\documentclass[
		openright,titlepage,numbers=noenddot,headinclude,
	 	footinclude=false,cleardoublepage=empty,
		dottedtoc, 
		BCOR=5mm,paper=a4,fontsize=10pt, 
		ngerman,american, 
		floatperchapter, 
		]{scrreprt} 
                
\PassOptionsToPackage{utf8}{inputenc} 
\usepackage{inputenc}

\PassOptionsToPackage{eulerchapternumbers,listings, pdfspacing, subfig,beramono,parts}{classicthesis}

\newcommand{\myTitle}{Capturing the Many-Time Physics of Non-Markovian Quantum Stochastic Processes\xspace}

\newcommand{\myName}{Gregory Anthony Liam White\xspace}

\newcommand{\myFaculty}{Faculty of Science\xspace}
\newcommand{\myDepartment}{School of Physics\xspace}
\newcommand{\myUni}{The University of Melbourne\xspace}
\newcommand{\myLocation}{Melbourne\xspace}
\newcommand{\myTime}{October 2023\xspace}
\newcommand{\myVersion}{version 4.2\xspace}

\newcommand{\ie}{i.\,e.}

\newcommand{\Eg}{E.\,g.} 

\newcounter{dummy} 
\providecommand{\mLyX}{L\kern-.1667em\lower.25em\hbox{Y}\kern-.125emX\@}

\usepackage{lipsum} 

\usepackage{babel}

\usepackage{csquotes}
\PassOptionsToPackage{%
backend=bibtex8,bibencoding=ascii,%
language=auto,%
style=numeric-comp,%
sorting=none, 
maxbibnames=3, 
doi=true,
url=false,
natbib=true 
}{biblatex}
\usepackage{biblatex}

 \usepackage{amsmath}

\PassOptionsToPackage{T1}{fontenc} 
\usepackage{fontenc}

\usepackage{textcomp} 

\usepackage{scrhack} 

\usepackage{xspace} 

\usepackage{mparhack} 

\usepackage{fixltx2e} 

\PassOptionsToPackage{smaller}{acronym} 
\usepackage{acronym} 

\PassOptionsToPackage{pdftex}{graphicx}
\usepackage{graphicx} 

\usepackage{tabularx} 
\usepackage{caption}
\usepackage{subfig}  

\usepackage{listings} 
\PassOptionsToPackage{hyperfootnotes=false,pdfpagelabels,bookmarks=false}{hyperref}
\usepackage{hyperref}  
\hypersetup{
colorlinks=true, linktocpage=true, pdfstartpage=3, pdfstartview=FitV,
breaklinks=true, pdfpagemode=UseNone, pageanchor=true, pdfpagemode=UseOutlines,%
plainpages=false, bookmarksnumbered, bookmarksopen=true, bookmarksopenlevel=1,%
hypertexnames=true, pdfhighlight=/O,
urlcolor=webbrown, linkcolor=RoyalBlue, citecolor=webgreen, 
pdftitle={\myTitle},
pdfauthor={\textcopyright\ \myName, \myUni, \myFaculty},
pdfsubject={},
pdfkeywords={},
pdfcreator={pdfLaTeX},
pdfproducer={LaTeX with hyperref and classicthesis}
}

\makeatletter
\@ifpackageloaded{babel}
{
\addto\extrasamerican{

}
\addto\extrasngerman{

}
}{\relax}
\makeatother

\usepackage{classicthesis} 
\usepackage[left=25mm, right=25mm, bottom=25mm]{geometry}

\usepackage[most]{tcolorbox}
\newcounter{example}[chapter]

\usepackage{xparse}
\usepackage{lipsum}

\definecolor{verylightgray}{gray}{0.98}

\def\exampletext{Example} 

\NewDocumentEnvironment{example}{ O{} }
{
\definecolor{colexam}{RGB}{77,154,207}
\newtcolorbox[use counter=example,number within=chapter]{examplebox}{%
    empty,
    title={\exampletext~\thetcbcounter: #1},
    coltitle=white,fonttitle=\bfseries\fontsize{10.5}{10.5}\selectfont,
    colback=verylightgray,
    before=\par\medskip\noindent,parbox=false,boxsep=3pt,left=3mm,right=3mm,top=6pt,bottom=4pt,breakable,pad at break=0mm,left skip=0.5cm,
    before upper=\csname @totalleftmargin\endcsname0pt, 
    enhanced,
    frame style={colexam,line width=0.5pt},
    boxed title style={empty,size=minimal,toprule=0pt,top=0pt,left=3mm,overlay={},boxrule=0pt,boxsep=0pt,colback=colexam, enlarge top by=2pt},
    overlay unbroken={},
    overlay first={},
    overlay middle={},
    overlay last={},%
    }
\begin{examplebox} }
{ \end{examplebox}\endlist}

\usepackage{mathrsfs}
\usepackage{algorithm} 
\usepackage{algpseudocode}
\usepackage{epigraph}
\usepackage{amsthm}
\usepackage{wrapfig}
\usepackage{longtable}

\usepackage{xcolor}

\newcommand{\kket}[1]{\ensuremath{\left| #1 \right\rangle\!\rangle}}
\newcommand{\bbra}[1]{\ensuremath{\left\langle\!\langle #1 \right|}}

\newcommand{\examplecontent}[1]{\setlength{\parindent}{0pt}{#1}\color{black}}
\newcommand{\pt}{$\Upsilon_{k:0}$}

\newcommand{\ket}[1]{\ensuremath{\left| #1 \right\rangle}}
\newcommand{\bra}[1]{\ensuremath{\left\langle #1 \right|}}
\newcommand{\frob}[1]{\ensuremath{\left|\right| #1 \left|\right|_F}}
\DeclareMathOperator*{\argmax}{\arg\!\max}
\DeclareMathOperator*{\argmin}{\arg\!\min}
\newcommand{\Tr}[0]{\ensuremath{\text{Tr}}}

\algnewcommand\algorithmicinput{\textbf{Input:}}
\algnewcommand\algorithmicoutput{\textbf{Output:}}
\algnewcommand\Input{\item[\algorithmicinput]}
\algnewcommand\Output{\item[\algorithmicoutput]}

\newtheorem{theorem}{Theorem}[section]

\theoremstyle{definition}
\newtheorem{definition}{Definition}[section]

\theoremstyle{remark}

\SetSymbolFont{operators}   {normal}{OT1}{cmr} {m}{n}
\SetSymbolFont{letters}     {normal}{OML}{cmm} {m}{it}
\SetSymbolFont{symbols}     {normal}{OMS}{cmsy}{m}{n}
\SetSymbolFont{largesymbols}{normal}{OMX}{cmex}{m}{n}
\SetSymbolFont{operators}   {bold}  {OT1}{cmr} {bx}{n}
\SetSymbolFont{letters}     {bold}  {OML}{cmm} {b}{it}
\SetSymbolFont{symbols}     {bold}  {OMS}{cmsy}{b}{n}
\SetSymbolFont{largesymbols}{bold}  {OMX}{cmex}{m}{n}

\SetMathAlphabet{\mathbf}{normal}{OT1}{cmr}{bx}{n}
\SetMathAlphabet{\mathsf}{normal}{OT1}{cmss}{m}{n}
\SetMathAlphabet{\mathit}{normal}{OT1}{cmr}{m}{it}
\SetMathAlphabet{\mathtt}{normal}{OT1}{cmtt}{m}{n}
\SetMathAlphabet{\mathbf}{bold}  {OT1}{cmr}{bx}{n}
\SetMathAlphabet{\mathsf}{bold}  {OT1}{cmss}{bx}{n}
\SetMathAlphabet{\mathit}{bold}  {OT1}{cmr}{bx}{it}
\SetMathAlphabet{\mathtt}{bold}  {OT1}{cmtt}{m}{n}

\font\greekcapstenrm=cmr10
\font\greekcapssevenrm=cmr7
\font\greekcapsfiverm=cmr5

\newfam\greekcapsfam
\textfont\greekcapsfam=\greekcapstenrm
\scriptfont\greekcapsfam=\greekcapssevenrm
\scriptscriptfont\greekcapsfam=\greekcapsfiverm

\let\tmpGamma=\Gamma \def\Gamma{{\fam\greekcapsfam\tmpGamma}}
\let\tmpDelta=\Delta \def\Delta{{\fam\greekcapsfam\tmpDelta}}
\let\tmpTheta=\Theta \def\Theta{{\fam\greekcapsfam\tmpTheta}}
\let\tmpLambda=\Lambda \def\Lambda{{\fam\greekcapsfam\tmpLambda}}
\let\tmpXi=\Xi \def\Xi{{\fam\greekcapsfam\tmpXi}}
\let\tmpPi=\Pi \def\Pi{{\fam\greekcapsfam\tmpPi}}
\let\tmpSigma=\Sigma \def\Sigma{{\fam\greekcapsfam\tmpSigma}}
\let\tmpUpsilon=\Upsilon \def\Upsilon{{\fam\greekcapsfam\tmpUpsilon}}
\let\tmpPhi=\Phi \def\Phi{{\fam\greekcapsfam\tmpPhi}}
\let\tmpPsi=\Psi \def\Psi{{\fam\greekcapsfam\tmpPsi}}
\let\tmpOmega=\Omega \def\Omega{{\fam\greekcapsfam\tmpOmega}}

\usepackage{amsmath}
\usepackage{amssymb}
\usepackage{enumitem}
\setlist{leftmargin=*}
\usepackage{float}
\usepackage{graphicx}
\usepackage{orcidlink}
\usepackage{lscape}

\graphicspath{{Draft_Figs/}}

\newcommand{\Description}[1]{\hangindent=0.5em\hangafter=0\noindent\footnotesize{#1}\par\normalsize\vspace{1em}} 

\newlength{\datebox}

\newcommand{\NewEntry}[3]{\noindent\hangindent=0.5em\hangafter=0 \parbox{\datebox}{\small \textit{#1}}\hspace{0.1em} #2 #3 
\vspace{0.15in}} 

\usepackage{etoolbox}
\makeatletter
\patchcmd{\scr@startpart}{\cleardoublepage}{\clearpage}{}{}
\makeatother

\begin{document}


\frenchspacing 

\raggedbottom 

\selectlanguage{american} 


\pagenumbering{roman} 

\pagestyle{plain} 


\begin{titlepage}

\begin{addmargin}[-0.5cm]{-0.5cm}
\begin{center}
\large

\hfill
\vfill

\begingroup

\color{Maroon}\spacedallcaps{Many-Time Physics in Practice}\\
\vfill
\spacedlowsmallcaps{Characterising and Controlling\\ non-Markovian Quantum Stochastic Processes} \\ \bigskip
\endgroup

\vfill
\hypersetup{hidelinks=true}
\emph{by}\\
\spacedlowsmallcaps{Gregory Anthony Liam White} \\
\orcidlink{0000-0001-6673-6676} \hspace{2mm} \href{https://orcid.org/0000-0001-6673-6676}{orcid.org/0000-0001-6673-6676}
\hypersetup{hidelinks=false}
\vfill
\vfill

Submitted in total fulfillment of the \\
requirements for the degree of \\
Doctor of Philosophy \\
\medskip 

\vfill
\myDepartment \\
\myUni \\ 
Australia
\bigskip

\myTime\ 

\vfill

\end{center}
\end{addmargin}

\end{titlepage} 













\setlength{\parindent}{0pt}
\cleardoublepage

\thispagestyle{empty}
\refstepcounter{dummy}

\pdfbookmark[1]{Dedication}{Dedication} 

\vspace*{6cm}






\linespread{2.0}\selectfont
\noindent
\emph{
I met a lot of things on the way that astonished me. Tom Bombadil I knew already; but I had never been to Bree. Strider sitting in the corner at the inn was a shock, and I had no more idea who he was than had Frodo. The Mines of Moria had been a mere name; and of Lothlórien no word had reached my mortal ears till I came there. Far away I knew there were the Horse-lords on the confines of an ancient Kingdom of Men, but Fangorn Forest was an unforeseen adventure. I had never heard of the House of Eorl nor of the Stewards of Gondor. Most disquieting of all, Saruman had never been revealed to me, and I was as mystified as Frodo at Gandalf's failure to appear on September 22.
}\\ 
\newline 
\newline 
\linespread{1.5}\selectfont

\qquad --- J. R. R. Tolkien, in a letter to W. H. Auden, 7 June 1955

\cleardoublepage

\refstepcounter{dummy}
\pdfbookmark[0]{Declaration}{declaration} 

\chapter*{Declaration} 

\thispagestyle{empty}

This is to certify that:

\bigskip

\begin{enumerate}[label=(\roman*)]
    \item the thesis comprises only original work towards the Doctor of Philosophy except where indicated in the preface;
    \item due acknowledgement has been made in the text to all other material used; and 
    \item the thesis is fewer than the maximum word limit in length, exclusive of tables, maps, bibliographies, and appendices.
\end{enumerate}

\bigskip
I authorise the Head of the School of Physics to make or have made a copy of this thesis to any person judged to have an acceptable reason for access to the information, i.e., for research, study, or instruction.
\bigskip
 
\noindent\textit{\myLocation, \myTime}

\smallskip

\begin{flushright}
\begin{tabular}{m{5cm}}
\\ \hline
\centering\myName \\
\end{tabular}
\end{flushright}

\cleardoublepage

\pdfbookmark[1]{Acknowledgements}{Acknowledgements} 




\begingroup

\let\clearpage\relax
\let\cleardoublepage\relax
\let\cleardoublepage\relax

\chapter*{Acknowledgements}

\begin{center}
\emph{In loving dedication to Yvonne Marie White}\\
(1956--2023)
\end{center}

It is an impossible task to adequately thank each person responsible for the academic works herein, or for friendship and guidance afforded throughout this PhD and throughout the pandemic. Nevertheless, the following is an incomplete list of people without whom this could not have happened.
First and foremost, my thanks go to my supervisor Kavan Modi for his years of diligent hard work and support.
Kavan has a knack for finding and conveying the most fascinating ideas to think about, and always made a point of always being available both to answer the basic questions and to push projects through their bottlenecks.
Next, to Charles Hill, who supervised the majority of this project. Charles played a large part in my formative academic years, and has both infectious excitement and is extraordinarily clever about a great many topics. It was always a joy to spend hours whiteboarding with Charles.
To my principal supervisor Lloyd Hollenberg, who took me on back in 2017, and has always commanded a phenomenally wide range of knowledge with enthusiasm. 
Lloyd has always fostered a supportive research environment, made sure that cutting edge resources were on hand, and backed his students in to the broader community. Lastly, to Matt Dolan, for his role as chair of my advisory committee. Matt has always given unwavering support, a sense of good humour, and a willingness to listen.

Navigating this PhD would have been impossible without essential friendships and support from those around me.
A core thank you to Fred Hiskens, Harish Vallury, Maddy Zurowski, and Innes Bigaran, and especially to Erin Grant and Alex Healey for staying the course alongside me from the start of master's to the end of PhD. Further including Neil Dowling and Isobel Aloisio for being great friends, great colleagues, instigators of enlightening chats about physics, and chief proofreaders.
I was always lucky to be mentored by and friends with people in the years ahead, especially featuring the wisdom and relentless know-how of Innes Bigaran; the whip-cracking of David Broadway; the cheerfulness and music recommendations of Scott Lillie; and the warmth and perspective of Robert de Gille. 

To the various people who helped me navigate administrative perils (and kept admirably patient with my being difficult) -- Rose Cooney, Mary Siciliano; as well as Anna Phan for helping out with everything IBM Quantum related.
A huge thanks growing group of quantum computing friends and colleagues in Gary Mooney, Aidan Dang, Sam Tonetto, Floyd Creevey, Michael Jones, Spiro Gicev, Tim Kay, and Maiyuren Srikumar. In addition, the de facto quantum sensing family in Sam Scholten, Nikolai Dontschuk, Ella Walsh, Di Wang, Julia McCoey, and Daniel McCloskey. From the top, the senior mentors in David Simpson, Liam Hall, and Jean-Philippe Tetienne. The quantum precinct has always had an air of camaraderie and welcoming, for which the listed people are chiefly responsible.

I was fortunate to have ample travel opportunity towards the end of my candidature, from which I made a great many friends and had excellent conversations. I am indebted to the various groups that hosted me, especially the hospitality of Petar Jurcevic, and the Sandia crew in Mohan Sarovar, Corey Ostrove, Kevin Young, and Robin Blume-Kohout.

To Mum and Dad, for sparking my scientific dreams at a very young age, as well as their endless sacrifice to give me the opportunities to make that happen. 
My siblings, Justine, Nicolette, and Tim for always supplying that love and support, as well as baby niece Annalisa, who has given us all so much joy! An immense thank you to my family for their instrumental part in bringing me to where I am today.
But most importantly, I would like to thank my wonderful fianc\'ee Nicole who, among many other things, has always been so patient and supportive of pursuing my physics passion; kept me sane during extended lockdowns; been tolerant of some more obsessive periods of research; and who has always kept me grounded with laughter and conversation. I can't wait for the years ahead.

\endgroup 



\cleardoublepage

\pdfbookmark[1]{Abstract}{Abstract} 

\begingroup
\let\clearpage\relax
\let\cleardoublepage\relax
\let\cleardoublepage\relax

\chapter*{Abstract}

The past few decades have seen an explosion of interest in many-body physics and its emergent phenomena. Researchers have been enticed not only by unique quantum properties -- such as entanglement, zero-temperature phase transitions, and complexity beyond classical systems -- but by the \emph{information-processing} capabilities of quantum matter. 

Every year, substantial theoretical and experimental progress is made towards the realisation of a genuinely new computational paradigm. But progress is fractal; to make headway is to unearth the next set of obstacles. Decades of work has so far overcome physical, mathematical, engineering, and information theoretic obstacles to produce the remarkable high-fidelity devices we see today. But these devices must be \emph{near perfect} to realise useful quantum algorithms. Indeed, advancements so far have precipitated sensitivity to a host of complex dynamical and control-based effects. Chief among these today are \emph{non-Markovian memory effects}, where interactions between a quantum system and its surrounding environment can give rise to multi-time correlations. This open quantum systems paradigm underlies all real-world systems, and contains exotic characteristics both in nature and in engineered quantum systems. 

Multi-time processes are endowed with the same features as many-body physics, including temporal entanglement and well-defined causal structures. We dub this \emph{many-time physics}, and in this thesis we both explore these phenomena and prescribe a methodology for its complete characterisation. Although more interesting than their Markovian counterparts, correlated quantum dynamics are problematic -- both in terms of adverse effects on controlled systems, and for their inscrutability using conventional tools of open quantum dynamics. Characterisation techniques have so far represented the front line in understanding and controlling quantum dynamics. But non-Markovianity runs contrary to the models typically employed. We address this issue and formally present a generalisation of quantum process tomography, called process tensor tomography (PTT). This establishes the ability to rigorously and systematically characterise non-Markovian open quantum systems, resolving many long-standing issues in the literature.

In the first part of this thesis, we present an original review of the literature and motivate the problem at hand. In the second part, we develop and demonstrate PTT. We describe in detail the experiment design, construct post-processing algorithms, and demonstrate the characterisation of multi-time statistics both numerically and experimentally on near-term quantum devices. In particular, we demonstrate this as a tool for obtaining in-depth diagnostics about the nature and origin of temporal quantum correlations, exploring many-time physics in various scenarios. Lastly, in the third part, we dedicate our efforts to efficiency and self-consistency. To this effect, we explore theoretically processes with sparse memory structures. We then leverage this to develop various efficient estimation techniques tailored for different settings. The result is a lightweight framework capable of reconstructing any non-Markovian open quantum process. We finally show how understanding the dynamics of any particular system can be fed forward into optimal control of that system. 

Our work encompasses fundamental results on the properties of quantum stochastic processes, as well as a full gamut of tools to access many-time physics properties, accompanied by instructive and non-trivial examples on various quantum processors. We find that non-Markovianity plays a significant role in the dynamics of modern day quantum hardware. This has consequential implications, both from a control and an error-decoding standpoint. Thus, the results we present are timely, and we argue could form the core of carefully developed software that forms the basis of fault-tolerant quantum computing. Further, looking beyond pure quantum computation, PTT provides a general framework with which to characterise and understand complex and naturally occurring open quantum systems.

\endgroup			

\vfill 

\cleardoublepage

\refstepcounter{dummy}

\pdfbookmark[1]{\contentsname}{tableofcontents} 

\setcounter{tocdepth}{2} 

\setcounter{secnumdepth}{3} 

\manualmark
\markboth{\spacedlowsmallcaps{\contentsname}}{\spacedlowsmallcaps{\contentsname}}
\tableofcontents 
\automark[section]{chapter}
\renewcommand{\chaptermark}[1]{\markboth{\spacedlowsmallcaps{#1}}{\spacedlowsmallcaps{#1}}}
\renewcommand{\sectionmark}[1]{\markright{\thesection\enspace\spacedlowsmallcaps{#1}}}

\clearpage

\begingroup 
\let\clearpage\relax
\let\cleardoublepage\relax
\let\cleardoublepage\relax


\refstepcounter{dummy}
\pdfbookmark[1]{\listfigurename}{lof} 

\listoffigures

\vspace{8ex}
\newpage



        
    



       

\refstepcounter{dummy}
\pdfbookmark[1]{Acronyms}{acronyms} 

\markboth{\spacedlowsmallcaps{Acronyms}}{\spacedlowsmallcaps{Acronyms}}

\chapter*{Acronyms}

\begin{acronym}[UML]
\acro{CJI}{Choi-Jamio\l kowski Isomorphism}
\acro{CP}{Completely Positive}
\acro{CPTP}{Completely Positive and Trace-Preserving}
\acro{DD}{Dynamical Decoupling}
\acro{GET}{Generalised Extension Theorem}
\acro{GKSL}{Gorini-Kossakowski-Sudarshan and Lindblad}
\acro{GME}{Genuine Multipartite Entanglement}
\acro{GST}{Gate Set Tomography}
\acro{IC}{Informationally Complete}
\acro{KET}{Kolmogorov Extension Theorem}
\acro{LI}{Linear Inversion}
\acro{LPDO}{Locally-Purified Density Operator}
\acro{MLE}{Maximum Likelihood Estimation}
\acro{MPO}{Matrix Product Operator}
\acro{MPS}{Matrix Product State}
\acro{MUUB}{Mutually Unbiased Unitary Basis}
\acro{NISQ}{Noisy Intermediate-Scale Quantum}
\acro{OSR}{Operator-Sum Representation}
\acro{POVM}{Positive Operator-Valued Measure}
\acro{PTM}{Pauli Transfer Matrix}
\acro{PTM}{Pauli Transfer Matrix}
\acro{PTT}{Process Tensor Tomography}
\acro{PVM}{Projection-Valued Measure}
\acro{QCVV}{Quantum Characterisation, Verification, and Validation}
\acro{QIP}{Quantum Information Processor}
\acro{QMI}{Quantum Mutual Information}
\acro{QPT}{Quantum Process Tomography}
\acro{QSP}{Quantum Stochastic Process}
\acro{QST}{Quantum State Tomography}
\acro{RB}{Randomised Benchmarking}
\acro{SDP}{Semi-definite Program}
\acro{SPAM}{State Preparation and Measurement}
\acro{TNI}{Trace Non-Increasing}
\acro{TP}{Trace-Preserving}
\end{acronym}


\endgroup 

\cleardoublepage

\pdfbookmark[1]{Foreword}{Foreword} 

\chapter*{Foreword} 



This thesis presents a set of theoretical and numerical tools for the general study of non-Markovian open quantum systems. We invest ourselves in the characterisation of many-time physics, insofar as temporal correlations can exhibit the same complex features as spatial correlations.
We specifically apply many of these tools to the investigation of non-Markovian noise on quantum devices, and attaining superior levels of control more broadly. The thesis directly incorporates the following primary publications, as well as ideas and skills developed in collaborative and thematically relevant projects. We acknowledge especially the technical contributions of Dr. Charles Hill and Dr. Felix Pollock in guiding many of the listed publications, as well as collaborative contributions from Dr. Petar Jurcevic.

\noindent
\spacedlowsmallcaps{Primary Publications}



\noindent
\emph{We visit the work of each of the following peer-reviewed manuscripts:}

\NewEntry{2020}{Demonstration of non-Markovian process characterisation and control on a quantum processor
\Description{
\newline
\cite{White-NM-2020}\quad \href{https://doi.org/10.1038/s41467-020-20113-3}{Nature Communications \textbf{11}, 6301} \qquad\href{https://arxiv.org/abs/2004.14018}{arXiv:2004.14018}
\\
\textbf{Gregory White}, Charles Hill, Felix Pollock, Lloyd Hollenberg, and Kavan Modi}}

\NewEntry{2021}{Performance optimisation for drift-robust fidelity improvement of two-qubit gates
\Description{
\newline 
\cite{white-POST}\quad \href{https://journals.aps.org/prapplied/abstract/10.1103/PhysRevApplied.15.014023}{Phys. Rev. Applied 15, 014023} \qquad \href{https://arxiv.org/abs/1911.12096}{arXiv:1911.12096}
\newline
\textbf{Gregory White}, Charles Hill, and Lloyd Hollenberg}}

\NewEntry{2022}{Non-Markovian quantum process tomography
\Description{
\newline
\cite{White-MLPT}\quad\href{https://doi.org/10.1103/PRXQuantum.3.020344}{PRX Quantum \textbf{3}, 020344} \qquad \href{https://arxiv.org/abs/2106.11722}{arXiv:2106.11722}
\newline
\textbf{Gregory White}, Felix Pollock, Lloyd Hollenberg, Kavan Modi, and Charles Hill}}

\NewEntry{2023}{Filtering crosstalk from bath non-Markovianity via spacetime classical shadows
\Description{
\newline
\cite{white2022filtering}\quad \href{https://link.aps.org/doi/10.1103/PhysRevLett.130.160401}{Phys. Rev. Lett. \textbf{130}, 160401} \qquad \href{https://arxiv.org/abs/2210.15333}{arXiv:2210.15333} 
\newline
\textbf{Gregory White}, Kavan Modi, and Charles Hill}}

\clearpage
\emph{in addition to the work presently under review:}

\NewEntry{2021}{From many-body to many-time physics
\Description{
\newline
\cite{white2021many}\quad \href{https://arxiv.org/abs/2107.13934}{arXiv:2107.13934} 
\newline
\textbf{Gregory White}, Felix Pollock, Lloyd Hollenberg, Charles Hill, and Kavan Modi}}

\NewEntry{2023}{Unifying non-Markovian characterisation with an efficient and self-consistent framework
\Description{
\newline
\cite{white2023efficient}\quad \href{https://arxiv.org/abs/2312.08454}{arXiv:2312.08454} 
\newline
\textbf{Gregory White}, Petar Jurcevic, Lloyd Hollenberg, Charles Hill, and Kavan Modi}}

\emph{the works in preparation:}

\NewEntry{2023}{A spacetime Markov order tensor network ansatz
\Description{
\newline
\cite{white2023stmo}
\newline
\textbf{Gregory White}, Charles Hill, and Kavan Modi}}

\vspace{-0.25in}
\emph{and the following perspectives:}

\NewEntry{2021}{Gate set tomography is not just hyperaccurate, it's a different way of thinking
\Description{
\newline
\cite{White2021gatesettomographyis}\quad \href{https://doi.org/10.22331/qv-2021-10-05-60}{Quantum Views 5, 60}
\newline
\textbf{Gregory White}}}

\NewEntry{2022}{Characterization and control of non-Markovian quantum noise
\Description{
\newline
\cite{white_characterization_2022}\quad \href{https://doi.org/10.1038/s42254-022-00446-2}{Nat Rev Phys \textbf{4}, 287}
\newline
\textbf{Gregory White}}}

\spacedlowsmallcaps{Secondary Publications}

Many of the ideas and tools that appear throughout this thesis have featured in the following thematically relevant works as co-author. In the interest of space, time, and without the necessary background, we do not explicitly incorporate these works, but they do inform the context and discussions of our work, as well as being referenced in various sections.

\NewEntry{2019}{Quantum bath control with nuclear spin state selectivity via pulse-adjusted dynamical decoupling
\Description{
\newline 
\cite{lang2019quantum}\quad\href{https://doi.org/10.1103/PhysRevLett.123.210401}{Phys. Rev. Lett. 123, 210401}\qquad \href{https://arxiv.org/pdf/1904.00893.pdf}{arXiv:1904.00893} 
\newline
Jacob Lang, David Broadway, \textbf{Gregory White}, Liam Hall, Alistair Stacey, Lloyd Hollenberg, Tanya Monteiro, and Jean-Philippe Tetienne}}

\clearpage

\NewEntry{2021}{Prospects for nuclear spin hyperpolarisation of molecular samples using nitrogen-vacancy centres in diamond
\Description{
\newline
\cite{tetienne2021prospects}\quad\href{https://doi.org/10.1103/PhysRevB.103.014434}{Phys. Rev. B 103 (1), 014434} \qquad \href{https://arxiv.org/abs/2008.12417}{arXiv:2008.12417}
\newline
Jean-Philippe Tetienne, Liam Hall, Alexander Healey, \textbf{Gregory White}, Marc-Antoine Sani, Frances Separovic, and Lloyd Hollenberg}}

\NewEntry{2021}{Generation and verification of 27-qubit Greenberger-Horne-Zeilinger states in a superconducting quantum computer
\Description{
\newline
\cite{mooney2021generation}\quad\href{https://doi.org/10.1088/2399-6528/ac1df7}{J. Phys. Commun. 5 095004} \qquad \href{https://arxiv.org/abs/2101.08946}{arXiv:2101.08946}
\newline
Gary Mooney, \textbf{Gregory White}, Charles Hill, and Lloyd Hollenberg}}

\NewEntry{2021}{Hyperpolarisation of external nuclear spins using nitrogen-vacancy centre ensembles
\Description{
\newline
\cite{healey2021pol}\quad\href{https://doi.org/10.1103/PhysRevApplied.15.054052}{Phys. Rev. Applied 15, 054052}\qquad \href{https://arxiv.org/abs/2101.12325}{arXiv:2101.12325}
\newline
Alexander Healey, Liam Hall, \textbf{Gregory White}, Tokuyuki Teraji, Marc-Antoine Sani, Frances Separovic, Jean-Philippe Tetienne, and Lloyd Hollenberg}}

\NewEntry{2021}{Whole-device entanglement in a 65-qubit superconducting quantum computer
\Description{
\newline
\cite{mooney2021whole}\quad\href{https://doi.org/10.1002/qute.202100061}{Adv. Quantum Technol. 2100061} \qquad \href{https://arxiv.org/abs/2102.11521}{arXiv:2102.11521} 
\newline
Gary Mooney, \textbf{Gregory White}, Charles Hill, and Lloyd Hollenberg}}

\NewEntry{2021}{Process Tomography on a 7-Qubit Quantum Processor via Tensor Network Contraction Path Finding
\Description{
\newline
\cite{dang2021process}\quad\href{https://arxiv.org/abs/2112.06364}{arXiv:2112.06364}
\newline
Aidan Dang, \textbf{Gregory White}, Lloyd Hollenberg, and Charles Hill}}

\NewEntry{2023}{Sampling complexity of open quantum systems
\Description{
\newline
\cite{aloisio-complexity}\quad \href{https://link.aps.org/doi/10.1103/PRXQuantum.4.020310}{PRX Quantum \textbf{4}, 020310}\qquad \href{https://arxiv.org}{arXiv:2209.10870} 
\newline
Isobel Aloisio, \textbf{Gregory White}, Charles Hill, and Kavan Modi}}

\NewEntry{2023}{Noise-robust ground state energy estimates from deep quantum circuits
\Description{
\newline
\cite{vallury2022noiserobust}\quad
\href{https://link.aps.org/doi/10.1103/PRXQuantum.4.020310}{Quantum 7, 1109}\qquad
\href{https://arxiv.org/abs/2211.08780}{arXiv:2211.08780} 
\newline
Harish Vallury, Michael Jones, \textbf{Gregory White}, Floyd Creevey, Charles Hill, and Lloyd Hollenberg}}




\pagenumbering{arabic} 



\ctparttext{The first three chapters of this thesis provide the necessary background for the original research presented later. Chapter~\ref{chap:QC-overview} reviews the fundamentals of quantum computing, current state-of-the-art research, and key challenges faced by hardware platforms, such as exotic quantum dynamics. Chapter~\ref{chap:OQS} introduces the mathematical framework needed to describe and understand open quantum system dynamics, as well as techniques for characterising these dynamical systems. Chapter~\ref{chap:stoc-processes} delves into multi-time settings, highlighting the role of temporal correlations in open system dynamics and providing an overview of both classical and quantum stochastic processes. 
In particular, we cover the process tensor framework, a formalism used to describe quantum stochastic processes. This will form the central topic of this thesis.}
\part{Background coverage} 
\chapter{Quantum Information Processing: Theory and Experiment}
\label{chap:QC-overview}
\epigraph{\emph{We shouldn't be asking 'where do quantum speed-ups come from?' we should say `all computers are quantum … where do classical slow-downs come from?'}}{Charles Bennett}
\noindent\colorbox{olive!10}{%
	\begin{minipage}{0.9555\textwidth} 
		\textcolor{Maroon}{\textbf{Chapter Summary}}\newline
		Quantum information processing, encompassing computation, communication, sensing, and cryptography, represents a new paradigm offering significant advancements through the application of quantum rather than classical physics. Designing controllable quantum systems remains challenging despite recent progress. This chapter introduces quantum information processing, focusing on proposed technologies for quantum computers and main thesis results. Hardware platforms face noise and scalability issues, with each having unique pros and cons. Often, underlying physics differs from standard error models -- complicating benchmarking and error correction, which is crucial for realising fault-tolerant quantum devices. We examine specific examples of exotic dynamics before discussing operational solutions and explore recent breakthroughs in the field. Some content serves as context for new results rather than direct thesis use. 
		\par\vspace{\fboxsep}
		\colorbox{cyan!10}{%
			\begin{minipage}{\dimexpr\textwidth-2\fboxsep}
				\textcolor{RoyalBlue}{\textbf{Further Reading}}\newline 
				In the introductory chapters, this cyan box will be used to suggest a handful of helpful and relevant resources. This is to provide both a broader context and more specific technical details. In later chapters, it will be used to highlight some key results of that chapter.
				Both prosaic and insightful perspectives on the field of quantum information processing can be found from Preskill~\cite{Preskill2018quantumcomputingin,Preskill2022ThePO,preskill2023quantum} and Deutsch~\cite{PRXQuantum.1.020101}.
				Ref.~\cite{crooks-on-gates} constitutes both a thorough and practical introduction to quantum computing, and Ref.~\cite{qec-textbook} one of the author's personal favourite references on noise and quantum error correction. 
		\end{minipage}}
\end{minipage}}
\clearpage
\pagestyle{scrheadings} 

\section{Introduction}


The foundations of quantum theory were well and truly established by the middle of the 20th Century~\cite{dirac1981principles,von2018mathematical}. So what remains? Thomas Kuhn famously argued that revolutions in science -- and in physics in particular -- are born from a tension in existing models; subsequent crisis in the field; and, finally, resolution and a shift in the paradigm of understanding~\cite{Kuhn1962,Hacking1981-HACSR}. 
Indeed, this has been true for a great many pillars of modern science and modern life: from the genesis of quantum mechanics and Einstein's relativity~\cite{allori2015quantum,weinberg1998revolution}, to the understanding of Earth's plate tectonics and the genetics of DNA~\cite{marx2013emergence,politi2018scientific}.
But this lens is too restrictive. Paradigm shift can also come from methodological breakthroughs\footnote{
Consider a recent flurry of breakthroughs in the biosciences due to CRISPR~\cite{barrangou2017decade}; the transition from behavioural psychology to cognitive psychology enabled by brain scanning technology~\cite{tryon2014}; or fundamental observations about the universe in observational astronomy~\cite{berry2019short}.\vspace{9pt}}, syntactic maturity\footnote{
One sees this in the power of Feynman diagrams for the understanding of high energy physics~\cite{kaiser2005physics}.
\vspace{9pt}}, and interdisciplinary crossover\footnote{The interdisciplinary field of computational neuroscience emerged in the 1980s and has since led to many breakthroughs in our understanding of the brain~\cite{boone2016cognitive}. Similarly, the integration of evolutionary biology and molecular genetics in the mid-20th century led to the discovery of the molecular basis of genetic variation and the role of natural selection in driving evolution~\cite{goldenfeld2007biology}. 
}. Even though little tension remains in quantum theory under everyday conditions, the onset of quantum information arguably constitutes the largest paradigm shift in the past half-century of physics and, we contend, is owed to all three progenitors.

Often the lofty goal of fully fault-tolerant quantum information processing is posed as a means to an end. Encoding information into quantum degrees of freedom allows for the exploitation of superposition and entanglement as a valuable resource in computation and communication. 
Society could stand to benefit immensely from this fundamentally different set of quantum problem-solving tools~\cite{divincenzo1998quantum,cleve1998quantum,montanaro2016quantum}; simulation of quantum systems~\cite{bauer2020quantum,jordan2012quantum}; precision metrology~\cite{szczykulska2016multi}; and biomedical breakthroughs~\cite{marx2021biology}. 
But beyond that, there is also an inextricable relationship between technological breakthrough and conceptual insight~\cite{goldenfeld2007biology}. 
This extends in two directions, the lead-in and the follow-up. The latter constitutes everything we might discover with the technology. A fully fledged fault-tolerant quantum computer might, for example, allow us to settle controversial issues such as interpretations of the axioms of quantum mechanics~\cite{bong2020strong,carroll2021energy}, or establish the true nature of quantum gravity~\cite{almheiri2015bulk,pastawski2015holographic}. 
The lead-in, however, encompasses the cumulative infrastructure of thought required to make the technology happen. A quantum computer would seem unimaginable to the early pioneers of quantum mechanics, despite obeying the exact same familiar set of physical laws. On the one hand, any person sent forward in time a century hence would of course be dazzled by the accomplishments of engineering and technology. But there has also been fundamental progress made on the nature of information and its role in physical systems. \par

The early connecting foundations between physics and information theory can be seen in John Wheeler's famous \emph{it-from-bit} essay, or in modern parlance, it-from-\emph{qubit}~\cite{wheeler2018information}. 
The driving contention here is that physics and information have an inseparable relationship, that ``information is physical''. 
Jaynes showed, for example, how the theory of statistical mechanics should be seen through the lens of Bayesian inference rather than physical law~\cite{PhysRev.106.620}. The problem of Szilard's engine~\cite{szilard1929entropieverminderung}, resolved by Bennett~\cite{bennett1982thermodynamics} with an application of Landauer's principle~\cite{keyes1970minimal}, collectively showed that information processing depends intrinsically on the underlying device physics.
We might understand, then, that the equations of motion governing a dynamical system are in a precise sense performing a computation. Moreover, that the active role we play in measuring that system is creating information of its own. From here, the conceptual leap was to imagine the creation of physical systems with computation as the end goal rather than the by-product, as famously posed by Feynman and Benioff~\cite{feynman2018simulating,benioff1980computer}. 
Setting the context of physical laws for computation, Feynman~\cite{feynman1982modeling}'s and Deutsch~\cite{deutsch1985quantum}'s momentous realisations were that a Turing machine with quantum -- rather than classical -- underpinnings could potentially solve problems more efficiently, in stark contradistinction to the tenets of the strong Church-Turing thesis~\cite{sep-church-turing}.

%

If the First Quantum Revolution was driven by crisis, the Second Quantum Revolution is driven by potential: the potential to use quantum mechanics by \emph{design}, rather than purely a framework of understanding~\cite{quant-tech-1,PRXQuantum.1.020101}.
This promise has instantiated a massive experimental effort, channelling technology and methodological progress into the shifting of a scientific paradigm. 
Building off a bedrock of condensed matter physics, the field now strives to realise quantum devices that significantly surpass their classical counterparts.
However, it remains uncertain which physical principles will best underpin quantum information encoding.
The range of hardware approaches to this end is vast and varied. Popular routes, such as superconducting circuits with both qubit and bosonic encodings, and trapped ion qubits, have made great strides in the past decade and constitute some of the highest quality control capabilities over any quantum physical system~\cite{grimm2020stabilization,Arute2019,PRXQuantum.2.020343}. But the long-term prospects and hurdles to scalability remain unclear~\cite{blinov2021comparison,bravyi2022future,bruzewicz2019trapped,brown2021materials}.
Meanwhile, alternative approaches such as neutral atoms systems and photonic quantum computers boast their own intrinsic properties that may be more suitable~\cite{Henriet2020quantumcomputing,Bourassa2020BlueprintFA}. Finally, in the event that respective challenges of scalability or noise are otherwise insurmountable, the nascent silicon spin approaches and theoretical topologically-encoded Majorana qubits could constitute long-term solutions, each front-loading many of their major challenges~\cite{mkadzik2022precision,aasen2016milestones,van2016path}.
The long road to useful quantum computing is fraught with obstacles but marked by innovation and exciting milestones.
It remains to be seen whether this journey involves a linear progression through engineering advances, or whether new physical breakthroughs are required, uncovering novel emergent physics at scale.

As well as significant technological progress, the field of quantum information has benefitted greatly from interdisciplinary overlap. In fact, it is a field born of interdisciplinarity. 
Turing's profound original insight was to conceptualise computation as a sequence of discrete actions that could be executed by a machine.
This, of course, had far-reaching technological implications for society. But it additionally changed the way we understand the nature of problem-solving.
Nearly 80 years later, complexity theory is one of the great drivers of problem-solving in modern science. Physical laws can be understood from the abstraction to computation~\cite{percus2006computational,998639,kreinovich2006towards,arora2009computational,Badii1997-BADCHS}, indeed the latter may one day inform the former~\cite{aaronsonNPcomplete}. We have seen a recent example of this in the milestone \texttt{MIP$^*$=RE} result, showing a strong separation between Bell inequalities for commuting correlations and tensor-product correlations -- that commuting bounds cannot even be approximated by finite dimensional systems~\cite{ji2021mip}.
This crossover between fields can help answer the question not only of how could humanity conceivably build a fault-tolerant quantum computer, but what are some critical applications of such a device? Even if we reach the experimental pinnacle, there is no guarantee that this question will be adequately answered~\cite{seunghoon2022,montanaro2016quantum}. \par 

So why does it appear as though computation is fundamental to quantum theory? One aspect to this might be the invasiveness of measurement.
Much of classical physics stood by as a passive observer to nature. Laplace supposed that a demon equipped with infinitely precise information about the boundary conditions of the universe and Newton's Three Laws could have access to all past and future events~\cite{laplace1995philosophical}. Quantum mechanically, we know this is cannot be the case -- even in principle. To extract information from a quantum system is to disturb it. When viewed as a series of discrete steps, the evolution of a quantum system can be understood as a sequence of apparatuses that take states from one point in time to the next. This view emphasises the role of control in quantum mechanics, and suggests that control should be considered a fundamental aspect of the theory. This idea of a `user interface', as Hardy calls it~\cite{hardy_operator_2012}, is in accord with the \emph{operational} view of quantum theory, where experimenters make choices of measurements and collect measurement outcomes. In other words, the very connection between us and our quantum systems is inherently a computation. 




\section{Thesis Outline and Main Results}


The road to fault-tolerant quantum computing is uniquely poised to make significant contributions to both our understanding of reality and our technological capabilities. In this thesis, we delve into the two interconnected perspectives: the study of fascinating physics and the development of quantum computers. Nascent quantum computers provide an advanced testbed for performing physics experiments and conversely, understanding their natural complex dynamics is crucial to attaining better control.
One of the core complications in the development of quantum technologies is that it is difficult to isolate a system from its environment. Failure to do so introduces a host of pernicious effects, broadly described as \emph{noise}. 

Initial scepticism of quantum computation wondered whether exponential gains in computational power might come at the expense of exponential effort to combat environmental decoherence. 
However, Shor's seminal work showed that coherence could in principle be maintained for arbitrarily long times with only polynomial physical overhead~\cite{PhysRevA.52.R2493}.
But principles of quantum error correction are often hinged on requirements of either spatial or temporal locality. If qubits interact strongly with their neighbours, or their environment, then noise can be correlated and less correctable. 

Participation in a computation, as the user, changes the system in non-deterministic ways. If that system also strongly interacts with its environment, then actions on the system can influence the environment and vice-versa to influence the system at a later point in time. Such dynamics are termed non-Markovian, and induce complex temporal correlations. In this thesis, we are primarily concerned with determining this connection for arbitrary open quantum systems. How does the control of an experimenter on their system influence the \emph{future} dynamics of that system, as dictated by the environment? This has significant implications for the ability to correct errors. We seek to understand how we can learn this connection efficiently, and how to use the information to improve control and error suppression in such devices.

To achieve this, we employ state-of-the-art quantum devices to study this spatiotemporal structure in the context of naturally occurring complex dynamics. We maintain a tight focus on noise, while also exploring new directions in which the physics of open quantum systems emerge. Specifically, examining this problem more broadly, we study temporal quantum correlations in a regime of \emph{many-time} physics, as inspired by traditional many-body physics. 
This approach opens up numerous possibilities for studying the structure of complex temporal correlations in experimentally relevant settings.


\begin{itemize}
	\item Chapters~\ref{chap:QC-overview}--\ref{chap:stoc-processes} constitute an original overview of the literature, as well as an introduction to the basic material required for the remainder of the thesis. 
	We focus on \emph{open} quantum systems: ones whose coupling to an inaccessible environment leads to irreversible stochastic dynamics. Chapter~\ref{chap:OQS} covers the standard theory of \emph{Markovian} open quantum systems -- and how to characterise them -- where coupling is weak and memory effects are neglibible. In Chapter~\ref{chap:stoc-processes}, we review the more general \emph{non-Markovian} theory of quantum stochastic processes, with a particular focus on process tensors as objects for the study of higher order probabilitiy theories. Some of the content in Chapter~\ref{chap:OQS} features in Refs.~\cite{White2021gatesettomographyis} and~\cite{white-POST}. Meanwhile some parts of Chapter~\ref{chap:stoc-processes} also feature in Ref.~\cite{white_characterization_2022}.
	\item In Chapter~\ref{chap:process-properties} we analyse the properties of process tensors. Specifically, this includes linear constraints of multi-time processes made by physical reasoning, and made in relation to the nature of various experimental probes. These details are then used for the remainder of the thesis in the estimation of quantum stochastic processes. We further establish firm relationships between types of open quantum dynamics, the classical and quantum structure of temporal correlations, and typicality of processes. Section~\ref{sec:PT-pauli} from this chapter contains work that appears in Refs.~\cite{White-MLPT} and~\cite{white2021many}.
	\item In Chapter~\ref{chap:PTT} we introduce a general operational protocol for characterising non-Markovian dynamics. This formally generalises the notion of quantum process tomography to the multi-time setting, and allows us to study arbitrarily structured temporal correlations. We develop the necessary algorithms to post-process data, with experimental demonstrations to high accuracy on IBM Quantum devices. This chapter is largely based off the works conducted in Refs.~\cite{White-NM-2020} and~\cite{White-MLPT}. 
	\item This broadly opens up the realisation of what we term \emph{many-time physics} in Chapter~\ref{chap:MTP}, the diagnostics and study of temporal quantum correlations. In this chapter, we explore quantum stochastic processes in a practical setting and how to determine crucial properties such as genuine multi-time entanglement. From this vantage point we consider questions of learnability of dynamical sampling problems, and separating out causal sources for non-Markovian memory effects. This chapter is formed predominantly from content in Refs.~\cite{white2021many} and~\cite{white2022filtering}, and contains ideas that also appear in Ref.~\cite{aloisio-complexity}.
	\item The tools developed up to this point are for fully general quantum stochastic processes, and hence have an exponential scaling. We thus turn in Chapter~\ref{chap:efficient-characterisation} to study the structure and compression of non-Markovian descriptions and characterisation efficiently. Here, we adapt generalised notions of Markov order to study processes with finite memory structures. We also use tensor network representations to characterise these systems in practice. The work of Section~\ref{sec:MO} appears in Ref.~\cite{White-MLPT}; Section~\ref{sec:stmo} forms the basis of Ref.~\cite{white2023stmo}; and Section~\ref{sec:tensor-networks} forms the first half of Ref.~\cite{white2023efficient}.
	\item In Chapter~\ref{chap:universal-noise}, we consider and develop a universal framework for the study of spatiotemporal quantum correlations. In particular, we incorporate erroneous memory effects in both process and control to self-consistently estimate all possible relevant effects from both probe and process. This chapter forms the basis of the second half of Ref.~\cite{white2023efficient}.
	\item Finally, in Chapter~\ref{chap:NM-control} we show how the tools developed are well-placed to be applied to the suppression of correlated noise in near-term quantum devices.
	We explore several avenues from which process tensor estimation can be used in optimal control settings and noise-aware circuit compilation. This achieves superior device control, and provides a path forwards by which higher fidelity quantum devices can be achieved through a combination of characterisation and feed-forward. Parts of this chapter feature in Refs.~\cite{White-NM-2020},~\cite{White-MLPT}, and~\cite{white2023efficient}.
\end{itemize}


We emphasise that we have aimed to construct this thesis in the form of a narrative structure rather than a necessarily chronological one. This is with the aim of expositional clarity in mind, rather than a transition from elementary to sophisticated results. 


\section{Quantum Mechanical and Quantum Computational Preliminaries}

The pedagogy of quantum mechanics usually begins with closed, or isolated quantum systems. These are the pure points of a more general theory, which we will introduce in Chapter~\ref{chap:OQS}. Although this thesis does not perform explicit quantum algorithms, our work is contextualised around quantum computers and quantum devices more broadly. In a closed system, states are represented as unit vectors on a state space $\mathcal{H}$ of dimension $d$. 
\begin{equation}
	|\psi\rangle\in \mathcal{H} \cong \mathbb{C}^d \quad \rightarrow \quad |\psi\rangle = \sum_{j=0}^{d-1}\alpha_i |i\rangle = \begin{pmatrix}
	\alpha_0\\
	\alpha_1\\
	\vdots\\
	\alpha_{d-1}
	\end{pmatrix}.
\end{equation}
In quantum computing syntax, we always deal with \emph{qubits}, or two-level systems. 
Composites of qubits are formed by a tensor product of their state spaces $\mathcal{H}_{12}\cong \mathcal{H}_1\otimes \mathcal{H}_2$. A composite state $|\psi_{12}\rangle \in\mathcal{H}_{12}$ is said to be separable if it can be written as a product state $|\psi_{12}\rangle = |\psi_1\rangle\otimes|\psi_2\rangle$, otherwise, it is \emph{entangled}. 
Generalising to $n$ qubits, a composite state vector can be written
\begin{equation}
	\ket{\psi} = \sum_{i=1}^{2^n}\alpha_i|i\rangle,\quad \alpha_i\in\mathbb{C},\quad \sum_{i}|\alpha_i|^2 = 1.
\end{equation}
We typically express qubits in the $Z$-basis: a linear combination of $\ket{0}$ and \ket{1}, and hence the $n$-qubit state vector $|i\rangle$ is a binary combination of the form $\{|0\rangle,|1\rangle\}^{\otimes n}$. 

A Hilbert space $\mathcal{H}\cong \mathbb{C}^d$ comes equipped with a Euclidean norm or inner product
\begin{equation}
	\langle \phi | \psi\rangle = \sum_{i,j=0}^{d-1}\beta_j^{\ast} \alpha_i \langle i | j \rangle = \begin{pmatrix}
		\beta_0^\ast & \beta_1^\ast & \cdots & \beta_{d-1}^\ast \end{pmatrix} \begin{pmatrix}
			\alpha_0\\
			\alpha_1 \\
			\vdots \\
			\alpha_{d-1}
		\end{pmatrix} = \sum_{i,j}\beta_j^{\ast} \alpha_i
\end{equation}

Basis elements of a state space are associated with physical outcomes of an observable.
In particular, from the coefficients of a wave function we get the Born rule. If one makes a measurement to determine the state of a particular system, measurement of state $\psi$ returns outcome $i$ with probability $\mathbb{P}$ given as
\begin{equation}
	\mathbb{P}(\psi = i) = |\langle i|\psi\rangle|^2 = |\alpha_i|^2.
\end{equation}

The set of linear operators $A : \mathcal{H}_{\text{in}}\rightarrow\mathcal{H}_{\text{out}}$ form their own vector space, the set of bounded linear operators from $\mathcal{H}_{\text{in}}$ to $\mathcal{H}_{\text{out}}$: denoted by $\mathcal{B}(\mathcal{H}_{\text{in}}, \mathcal{H}_{\text{out}})$. In the case where $\mathcal{H}_{\text{out}}\cong \mathcal{H}_{\text{in}}\equiv \mathcal{H}$, this will simply be written as $\mathcal{B}(\mathcal{H})$. 
This vector space comes equipped with its so-called Hilbert-Schmidt inner product $\langle A, B \rangle_{\text{HS}} := \Tr[A^\dagger B]$.
There are three classes of operator relevant to quantum mechanics: unitary operators, Hermitian operators, and projectors. Unitary operators $U$ satisfy $U^\dagger U = UU^\dagger = \mathbb{I}$, for appropriately-sized identity matrix $\mathbb{I}$. Unitary operations describe reversible, coherent manipulations of a quantum state: $|\psi\rangle \mapsto U|\psi\rangle = |\psi'\rangle$, or an $n$-qubit unitary, $U\in \mathbb{C}^{2^n\times 2^n}$.
Hermitian operators $O$ satisfy $O^\dagger = O$. In quantum mechanics, these are suggestively referred to as \emph{observables}, corresponding to physical quantities whose value may in principle be measured with respect to a quantum state.

Of special importance are the set of Hermitian and unitary $n-$qubit Pauli matrices $\mathbf{P}^{n} := \{\mathbb{I},X,Y,Z\}^{\otimes n}$, where 
\begin{equation}
	\mathbb{I} = \begin{pmatrix}
		1 & 0 \\ 0 & 1
	\end{pmatrix}; \qquad X = \begin{pmatrix}
		0 & 1 \\ 1 & 0 
	\end{pmatrix};
	\qquad Y = \begin{pmatrix}
		0 & -i \\ i & 0
	\end{pmatrix};
	\qquad Z  = \begin{pmatrix}
		1 & 0 \\ 0 & -1
	\end{pmatrix}.
\end{equation}

The set $\mathbf{P}^{n}$ forms a basis of all $n$-qubit Hermitian observables. That is, for any $O\in \mathcal{B}(\mathcal{H})$ with $O^\dagger = O$, we have 
\begin{equation}
	O = \frac{1}{2^n}\sum_{P\in \mathbf{P}^n} \Tr[O\cdot P] P.
\end{equation}
We will frequently use this basis to decompose Hermitian operators throughout the thesis, and for convenience occasionally normalise the set such that $\Tr[P_iP_j]=\delta_{ij}$, $P_i,P_j\in \mathbf{P}^n$. We have already seen the normalised eigenvectors of the Pauli $Z$ operator, which constitutes our canonical basis $\{|0\rangle,|1\rangle\}$. The eigenvectors of $X$ and $Y$ we will respectively denote by $\{|+\rangle, |-\rangle\}$ and $\{|i+\rangle, |i-\rangle\}$. The Pauli matrices also generate transformations. $\mathbf{P}^n$ forms a basis of the Lie algebra $\mathfrak{su}(2^n)$, which exponentiates to the special unitary group $SU(2^n)$ from which reversible transformations of a qubit register are made.

The evolution of a closed system is governed by a Hamiltonian $H(t)$, which may generally be time-dependent. For specific dynamics, the time evolution of a system is dictated by the Schr\"odinger equation: 
\begin{equation}
	i\hbar\frac{\text{d}|\psi\rangle}{\text{d}t} = H(t)|\psi\rangle,
\end{equation}
whose solution can be expressed by a unitary operator $U(t)$ via 
\begin{equation}
	|\psi(t)\rangle = U(t)|\psi(0)\rangle,\quad \text{where} \quad U(t) = \mathcal{T}\exp \left(\int_{0}^t \text{d}t' H(t')\right),
\end{equation}
where $\mathcal{T}$ is the time-ordering operator.

We will usually forego a continuous time expression of evolution, opting instead to use a digitised or discrete gate-based description of system evolution. Here, a system is driven via a series of discrete blocks, which themselves are generated by an underlying Hamiltonian. In particular, we will often consider the general case where some effective system evolution has taken place, as expressed by the unitary $U$ -- but we are not interested in \emph{solving} such dynamics, per se.\par 
As well as the above Pauli operators, we have some canonical single and two-qubit quantum gates we will employ 
\begin{equation}
	\begin{split}
	&H\text{ (Hadamard) } = \frac{1}{\sqrt{2}}\begin{pmatrix}
		1 & 1 \\ 1 & -1
	\end{pmatrix}; \quad R_z(\theta) = \begin{pmatrix} 
		1 & 0 \\ 0 & \text{e}^{i\theta}\end{pmatrix}; \quad R_x(\theta) = \begin{pmatrix}
			\cos\frac{\theta}{2} & -i\sin\frac{\theta}{2}\\
			-i\sin\frac{\theta}{2} & \cos\frac{\theta}{2}
		\end{pmatrix};\\ 
		&\quad \\
		&R_y(\theta) = \begin{pmatrix}
			\cos\frac{\theta}{2} & -\sin\frac{\theta}{2}\\
			\sin\frac{\theta}{2} & \cos\frac{\theta}{2}
		\end{pmatrix};\quad
		\text{CNOT} = \begin{pmatrix}
			1 & 0 & 0 & 0 \\
			0 & 1 & 0 & 0 \\
			0 & 0 & 0 & 1 \\
			0 & 0 & 1 & 0
		\end{pmatrix}; \quad \text{SWAP} = \begin{pmatrix}
			1 & 0 & 0 & 0 \\
			0 & 0 & 1 & 0 \\
			0 & 1 & 0 & 0 \\
			0 & 0 & 0 & 1
		\end{pmatrix}.
	\end{split}
\end{equation}
Finally, we have the conversion from quantum to classical information via readout. 
Measurements in pure-state quantum mechanics are represented by a \ac{PVM}. A \acs{PVM} is a set of Hermitian operators $\{|x\rangle\!\langle x|\}_{x=0}^{d-1} := \{\Pi_x\}$ such that $\sum_{x}|x\rangle\!\langle x| = \mathbb{I}$, $\Pi_x\Pi_y = \delta_{xy}\Pi_x$, and $\Pi_x^2 = \Pi_x$. Applying a measurement to a state $\ket{\psi}$ gives outcome $x$ with probability $|\langle x|\psi\rangle|^2$. After measurement, this becomes $\ket{\psi} \mapsto \frac{\Pi_x\ket{\psi}}{\sqrt{p_x}}$, leaving post-measurement state $|x\rangle$ (up to global phase). 
By the spectral theorem, any observable $O$ has eigendecomposition $O = \sum_{i}\lambda_i|i\rangle\!\langle i|$ into an orthonormal basis, and so defines a \acs{PVM}. 
Measurement of an observable gives outcome $\lambda_i$, the eigenvalue, as an outcome from projection $|i\rangle\!\langle i|$. The expectation value $\langle O\rangle$ of an observable with respect to state $|\psi\rangle$, then, is given by $\langle\psi|O|\psi\rangle = \sum_i\langle \psi|\lambda_i|i\rangle\!\langle i|\psi\rangle = \sum_i \lambda_i |\langle i|\psi\rangle|^2$, which is the weighted average of outcomes by the probability of finding $|\psi\rangle$ in state $|i\rangle$.
We will often consider projectively measuring in $Z$: $\{|0\rangle\!\langle 0|,|1\rangle\!\langle 1|\}$, $X$: $\{|+\rangle\!\langle +|,|-\rangle\!\langle -|\}$, and $Y$: $\{|i+\rangle\!\langle i+|,|i-\rangle\!\langle i-|\}$ bases, the \acs{PVM}s corresponding to the Pauli matrices.



\section{A Snapshot of Quantum Computing in the Early 2020s}


The field of quantum information is an exciting place to be. One that is characterised by rapid progress and frequent theoretical and experimental breakthroughs. However, there is still a level of uncertainty surrounding the ultimate success of this technology, and its potential impact on society. It would be presumptuous for this author to claim to offer definitive insights into this nascent, yet already vast field of quantum computation. Instead, there are numerous viewpoints and reviews available~\cite{PRXQuantum.1.020101,preskill2023quantum,Preskill2022ThePO,van2016path,quant-ladd} that cover a wide range of topics, including near-term and far-term algorithms~\cite{montanaro2016quantum,Bharti2021NoisyIQ,Cerezo2020VariationalQA}; various hardware platforms~\cite{gyongyosi2019survey,bruzewicz2019trapped,Kjaergaard2019,Henriet2020quantumcomputing,Flamini_2019}; as well as the speculative and the concrete aspects of quantum computing~\cite{aaronson2022structure,bennett1997strengths}; learning theory~\cite{carrasquilla2020machine,banchi2021generalization}; quantum advantage~\cite{lund2017quantum,harrow2017quantum}; and potential applications~\cite{cao2019quantum,orus2019quantum,NAP25196,outeiral2021prospects,cao2018potential}. In this section, we will provide a brief overview of current popular realisations of quantum hardware, review the recent state of the literature, and discuss what is yet to come.






\subsection{Current State-of-Play}

The quantum computing landscape has changed dramatically even since the present author began his candidature. Let us briefly summarise the state of the field, as encapsulated through some key results in the fleeting span of four years since the beginning of this undertaking. 

Looking back, this era of rapid advancement was kicked off by the famous Google `quantum supremacy' experiment, a marvellous achievement of science and engineering~\cite{Arute2019} whose (unintended, to those that recall) release was undoubtedly marked with much more fanfare than the subsequent results showing these circuits were, in fact, still classically simulable~\cite{pan2021simulating,pan2022solving,huang2020classical}. Nevertheless, the private and public imagination was captured and three more strong supremacy results were subsequently demonstrated, one on superconductors~\cite{wu2021strong}, and two photonic~\cite{zhong2020quantum,XanaduSupremacy}.

Fleets of quantum computers have now been constructed~\cite{bravyi2022future,PRXQuantum.2.020343,pompili2021realization,takeda2021quantum,mkadzik2022precision,hendrickx2021four,kues2019quantum,bruzewicz2019trapped,9805433,gyongyosi2019survey}, and their ever-increasing capabilities are marked by benchmarks showcasing their quantumness in just about every conceivable metric, including the creation of large entangled states~\cite{mooney2021generation,mooney2021whole,PhysRevX.9.031045}; increased coherence times~\cite{PhysRevX.9.031045,nguyen2019high,hu2019quantum,grimm2020stabilization}; and increased qubit numbers~\cite{wright2019benchmarking,ball2021first}.
Naturally, attempts have been made to collect these feats into single digestible benchmarks advertising device performance.
Most famous of which is the quantum volume~\cite{Cross-QV}, which, alongside the more recent mirror benchmarking circuits~\cite{proctor2022measuring} has become somewhat of a de-facto standard in measuring the capabilities of quantum computers~\cite{Jurcevic2021}. Quantum volume corresponds loosely to the classical complexity of simulating a quantum circuit with equal depth and width.
To give some indication of evolution, in the nascent stages of this work, the best device on the planet had a quantum volume of 8~\cite{Cross-QV}; today it is 32,768~\cite{quantinuum5qv} (one might more justifiably call this number an increase from 3 useful qubits to 15 useful qubits).


To speak even more technically, key developments in hardware capabilities have been made with 
open access pulse-level control~\cite{alexander2020qiskit}; advanced quantum assembly languages~\cite{morrison2020just,cross2022openqasm}; mid-circuit measurements and dynamic feed-forward~\cite{wright2019benchmarking,blinov2021comparison,Corcoles2021} capabilities; increased cryogenic size~\cite{9947181} and improved connectivity~\cite{webber2020efficient}.

Aside from the remarkable experiments in physics, it is the prospect of achieving fault-tolerant quantum computation that has generated excitement from the community.
A series of experimental milestones leave us tantalisingly close to the threshold where errors are suppressed in the addition of qubits, rather than amplified~\cite{google2023suppressing,sivak2022real,postler2022demonstration,saraiva2023dawn,Egan2020FaultTolerantOO,Chen2021ExponentialSO,Krinner2021RealizingRQ,RyanAnderson2021RealizationOR,google2023suppressing}.
Quantum error correction, in its own right, has flourished both with propositional change~\cite{Chamberland2020BuildingAF,Bartolucci2021FusionbasedQC,Bourassa2020BlueprintFA,Freedman2021SymmetryPQ}; breakthroughs in understanding~\cite{albert2020robust,brown2020fault,breuckmann2021quantum,Delfosse2020BeyondSF,Chamberland2020BuildingAF,Gullans2020QuantumCW,Webster2020UniversalFQ,Hastings2021DynamicallyGL,Cohen2021LowoverheadFQ,Tremblay2021ConstantoverheadQE} -- the introduction of good low-density parity check codes~\cite{panteleev2022asymptotically}, or the $XZZX$ surface code~\cite{BonillaAtaides2020TheXS} among some highlights -- and the development of better decoders~\cite{Chubb2021GeneralTN}. Moreover, myriad connections to physics have been made~\cite{Jahn2021HolographicTN, Liu2020ManyBodyQM,Movassagh2020ConstructingQC,haferkamp2022linear,Kesselring2022AnyonCA}.


A key theme throughout has been the emergence of learning theory as a topic of great interest. This queries the extent to which physical properties can be determined from a set of observations. On the classical side, contextualised in characterisation and benchmarking, this asks the extent to which quantum distributions may be reconstructed from a set of measurements~\cite{Harper2020,Eisert2020,Anshu2020SampleefficientLO,Brando2020FastAR,Helsen2020GeneralFF,Flammia2021AveragedCE,Hinsche2021LearnabilityOT,Huang2022FoundationsFL,Hinsche2022AST,Huang2022LearningMH,Yuen2022AnIS,Gong2022LearningDO,Chen2022TheLO,Fawzi2023LowerBO}. On the quantum side, we have much richer scope. Focus is narrowing on the highly relevant question: how can we use quantum devices to efficiently learn quantum properties~\cite{Huang2021,Huang2021InformationtheoreticBO, Chen2021ExponentialSB,Huang2022LearningTP,Onorati2023EfficientLO,huang2022provably,pesah2021absence,herrmann2022realizing}?
On both sides, the now-famous classical shadow tomography~\cite{huang-shadow} has cascaded into an entire subfield~\cite{elben2022randomized} -- not only of characterisation techniques~\cite{Helsen2021} -- but of remarkable insight into the connection between the generation of quantum data and the classical machine learning of said data to determine intricate quantum properties~\cite{Onorati2023EfficientLO,Huang2021InformationtheoreticBO, Chen2021ExponentialSB,huang2022provably, Huang2022LearningTP}. 

Access to increasingly sophisticated devices also has seen the beginning of a field of digitised quantum physics experiments whereby -- even in their noisy state -- quantum devices serve as a novel playground for exploring exotic physics~\cite{frey2022realization, mi2021information,randall2021many,else2020discrete,mi2022time} and chemistry~\cite{google2020hartree,PhysRevX.8.031022}.

Meanwhile, the topic of near-term quantum advantage remains far from settled~\cite{Bharti2021NoisyIQ,Cerezo2020VariationalQA,sharma2020noise,pesah2021absence,larose2019variational,cirstoiu2020variational,mcardle2019variational,jones2019variational}. We have seen the cutting of a wide swathe of near-term algorithms~\cite{Clinton2020HamiltonianSA} and even complexity-focused arguments that argue we need not wait until the far term to see useful quantum advantage~\cite{Chen2022TheCO,Liu2022NoiseCB}. But we now understand better than ever the negatives of noise on trainability of variational quantum algorithms, and this seems like a tall barrier to entry~\cite{Bravyi2019ObstaclesTV, Bittel2021TrainingVQ,StilckFrana2020LimitationsOO,Wang2020NoiseinducedBP}. Not least a role to play is that of error mitigation, the philosophy of trading sampling overhead for removal of stochastic errors; this has its upsides~\cite{Cai2022QuantumEM,Koczor2020ExponentialES,Piveteau2021ErrorMF} and its downsides~\cite{Takagi2021FundamentalLO,Quek2022ExponentiallyTB}.
Notwithstanding the difficulties of the near-term, the elusive nature of quantum advantage and avenues towards its realisation is much better understood now than it was four years ago~\cite{watts2019exponential,hangleiter2022computational,watts2023unconditional,yamakawa2022verifiable,Seddon2020QuantifyingQS,BenDavid2020SymmetriesGP,Aaronson2019OnTC,Huang2021InformationtheoreticBO,Cotler2021RevisitingDA,Chen2021ExponentialSB,Nezami2021PermanentOR,Aaronson2022HowMS,Szegedy2022QuantumAF,Schuld2022IsQA}, including blocked paths and no-gos~\cite{garg2020no,arunachalam2021quantum}.
Particularly, this is true for quantum chemistry problems~\cite{clinton2022towards,bauer2020quantum,cao2019quantum,mccaskey2019quantum,RevModPhys.92.015003,Childs2021,Bauer2020QuantumAF,Gharibian2021DequantizingTQ,McClean_2021,seunghoon2022}, but there have been more generally significant developments identifying where quantum advantage could come from, and exactly how much it could cost~\cite{Childs2020CanGP,Bravyi2019QuantumAW,Gidney2019HowTF,Liu2020ARA,Dalzell2020RandomQC,Maslov2020QuantumAF,Martyn2021GrandUO,huang2022provably,Huang2021,Pirnay2022ASQ,Dalzell2022MindTG}.

In the race to demonstrate utility of quantum devices, the field also faced its share of fierce classical competition, reminding one that classical algorithms and classical computers still remain extremely good~\cite{aharonov2022polynomial,bravyi2021classical,napp2022efficient}. The `dequantisation' approach, pioneered by Tang~\cite{Tang_2019} showed that the speed-up of known quantum algorithms need not be intrinsically quantum mechanical~\cite{Gilyen2020AnIQ,Gharibian2021DequantizingTQ,Shao2021FasterQA}. This comes amidst a whole host of other advances in classical computing within the quantum domain~\cite{Arrazola2019QuantuminspiredAI,Brando2019FasterQA,Gosset2020FastSO,Pashayan2021FastEO,Bravyi2021ClassicalAF,Gilyen2020AnIQ,Bravyi2021HowTS,Gharibian2021DequantizingTQ,Gao2021LimitationsOL,Shao2021FasterQA}.

Somewhat tangentially, these few years have also been dotted with remarkable insight into the nature of reality --
whether that be discovering the need for complex numbers in quantum mechanics~\cite{renou2021quantum}; confirming the existence of room temperature entangled states~\cite{anshu2022nlts,anshu2022construction}; or connecting quantum circuit complexity to black holes~\cite{haferkamp2022linear}. Much of this builds from the foundations of, and breakthroughs in, quantum complexity theory, one of the richest and most active fields of science today~\cite{ji2021mip,anshu2022nlts,anshu2022construction,natarajan2019neexp,Broadbent2019QMAhardnessOC,Alagic2019NoninteractiveCV,Aharonov2020StoqMAVM,Chen2022TheCO,Metger2023stateQIPS}. The insight of quantum complexity has been broadly mirrored by insight into physics, such as in the power of quantum computing, preparation of topological phases of matter, or the interior geometry of black holes, among more~\cite{yoganathan2019one,brandao2021models,haferkamp2022linear,bravyi2021complexity,anshu2022area,Harrow2019ClassicalAC,Bausch2019UncomputabilityOP,Anshu2020FromCC,huang2022provably,Eisert2021EntanglingPA,YungerHalpern2021ResourceTO,Onorati2023EfficientLO,Bouland2022QuantumP,Berta2022OnAG}.

\subsection{Overview of Different Hardware Platforms}



Most quantum information theory is highly non-specific to the underlying physics from which processing is realised. This leaves many different modalities available for quantum computing hardware. The widely agreed-upon properties are that the physical platform ought to satisfy DiVincenzo's five criteria: scalability; high fidelity; well-characterised resources; long coherence times; and realiable initialisation and measurement of the quantum constituents~\cite{divincenzo-criteria}.
The underlying physics need not even be two-level. Broadly speaking, qubits can be constructed either from a biased two-state system -- such as electron or nuclear spins, or photon polarisation -- or by engineering a two-level subspace from more complex macroscopic systems -- such as superconducting circuits. Qubits are ideally coupled in a deliberate fashion to one another, as well as to external control lines, but otherwise well-isolated from any other environmental properties. 
Although we will not explicitly visit the physics of quantum computing platforms, with key exceptions, it will serve as useful context to briefly discuss the underlying physics of several popular avenues for the realisation of quantum computers.




Due to the widespread commercial adoption of their approach, superconducting quantum computers are perhaps the most popular and most well-known platform for quantum computing. Simple resonant inductor-capacitor circuits are described by a Hamiltonian from which Cooper pair number $n$ and reduced flux $\phi$ -- or superconducting phase -- form a canonical conjugate pair. Incidentally, in this form, the Hamiltonian is identical to that of a 1D particle in a quadratic potential, whose solution is the archetypal quantum harmonic oscillator~\cite{Kjaergaard2019,krantz2019quantum}. The flux $\phi$ here is a generalisation of the position coordinate. This straightforward system does not serve as a viable qubit platform since all energy levels are evenly spaced, and so designated computational basis states could not be individually addressed without exciting other transitions. Instead, the linear inductor is replaced by a Josephson junction which is a non-linear circuit element. The inclusion of this component renders the system effectively an \emph{anharmonic} oscillator where the potential is sinusoidal. Hence, the different energy levels are forced to be meaningfully different from one another. In particular, the $|0\rangle\rightarrow \ket{1}$ transition is different from the $\ket{1}\rightarrow\ket{2}$ transition, allowing one to carve out a computational subspace from the bottom two energy levels.
The underlying physics of these circuits is dominated by the choice of either a high Josephson energy $E_J$ or a high charging energy $E_C$, required to add each electron to the island. A regime of $E_J \gg E_C$ implies high flux noise, whereas $E_C \gg E_J$ generates high charge noise, which is more challenging to address. Consequently, a popular architecture converged on by the community is known as the \emph{transmon qubit}. Here, the junction is shunted with a large capacitor to substantially increase Josephson energy. 

Transmon qubits fit into two broad categories with respect to qubit couplings: flux-tunable and fixed frequency. Tunable transmons, which is the architecture used by Google in their supremacy experiment~\cite{Arute2019}, have the advantage that qubit-qubit couplings are switched on by irradiation from an external magnetic field. Consequently, short two-qubit gates can be implemented with this method, and interaction effectively switched off outside. The drawback is the introduction of a significant level of flux noise. Fixed frequency transmons, in contrast, maintain connected qubits by a fixed coupling bus. This enables qubits to be much more stable, experiencing less stochastic noise, but at the cost of slower gates and experience always-on Hamiltonians with entangling $ZZ$ terms between connected qubits~\cite{PhysRevA.101.052308,malekakhlagh2020first}.

The ability to lithographically fabricate relatively macroscopic superconducting circuits makes them scalable in the sense that engineering is precise, but need not be atomically precise. These reduced demands meant that superconducting qubits very quickly took off in the construction of moderate numbers of qubits. But because of their macroscopic nature superconducting qubits experience a large amount of noise from their environments and possess relatively short coherence times -- usually on the order of tens to hundreds of microseconds~\cite{Kjaergaard2019}. Moreover, the question of long-term viability hangs large. Superconductors are macroscopic systems which need to be kept at cryogenic temperatures. Groups have, for example, been working to address these problems with capacity to create much larger fridges~\cite{9947181} and to transduce quantum information and communicate it between fridges~\cite{magnard2020microwave}, but envisaging the sheer size of an operation to maintain tens of millions of fault-tolerant qubits might give one pause.


Another highly mature platform for quantum computation has been based around ions contained in a Paul trap~\cite{bruzewicz2019trapped}. The ion species in question -- usually $^{171}\text{Yb}^+$, $^{40}\text{Ca}^+$ or $^{88}\text{Sr}^+$ -- are maintained by a set of DC and radio frequency voltages in a one-dimensional crystal structure. Different atomic electronic states are chosen to host a computational subspace. For Calcium, the hyperfine energy levels are used as $4S_{1/2,m_j=-1/2} := \ket{0}$ and $3D_{5/2,m_j = -1/2} := \ket{1}$. The metastable excited state can have lifetimes on the order of seconds, in stark contrast to superconducting systems~\cite{brown2021materials}. As such, ions do not suffer substantially from stochastic noise due to an external environment. Ion traps are contained in tightly sealed vacuum and, although vacuum systems require cryogenic temperatures, many of the components operate reliably at room temperature. Single qubit gates are typically driven resonantly by optical and magnetic fields, having demonstrated fidelities above 99.995\% in optically driven gates~\cite{bermudez2017assessing}. 
Trapped ion qubits also enjoy the substantial advantage of having all-to-all interactions within traps of around 20 or so, which can be extended between traps. Two-qubit gate operations are realised by illuminating the respective ions, enabling the photon momentum to transfer motion to the ions in a manner which is selective of the qubit state. Precisely controlling the ion motion is critical to achieving high fidelity control, and typically it is noise from the ion motion which contributes most to single and two-qubit error rates. Two-qubit gates, however, have still been realised to a fidelity of near 99.9\%~\cite{ballance2016high}. 

Much of the challenge in improving ion trap technology, therefore, lies in improvement of control instruments and materials~\cite{brown2021materials}. Additionally, it remains an open question how these qubits will be best scaled. Two principal paths to scalability have been proposed: (i) a large number of small trap sites, with ions shuttled from site-to-site, and (ii) a smaller number of traps with a larger number of ions, and photonic interconnects between the different traps. But gate speeds may be a longer-term sticking point in comparing technologies, taking on the order of a few microseconds for single qubit gates and tens to hundreds of microseconds for two-qubit gates~\cite{bruzewicz2019trapped}. This can be orders of magnitudes slower than superconductors, for example.

Relatedly, a large pocket of the quantum physics community is dedicated to exploring neutral atoms as a means by which quantum information may be processed~\cite{Henriet2020quantumcomputing,saffman2019quantum}. Specifically, any atom in a state with a highly excited electron (with principal quantum number in the tens to hundreds) is referred to as a Rydberg atom~\cite{wu2021concise}. Highly excited electrons display greatly exaggerated properties when compared to ground state atoms. For example, extremely large electric dipole moments can facilitate strong interactions with external fields or nearby particles. These interactions can be highly controllable, making Rydberg atoms an ideal candidate for quantum many-body simulation and the realisation of exotic phases of matter~\cite{semeghini2021probing}. The atoms possess many desirable properties: they are identical, have very long coherence times, and are amenable to a versatile range of interactions -- as opposed to the usual fundamental gate set limitations. Moreover, since atoms only weakly interact, a very large number can be fit together in a single trap, making them a key candidate scalability-wise. The identical nature of atoms is a double-edged sword. On the one hand, this means that only few control components are required in comparison to the total qubit number, since they are identically addressable. The cost, however, is a stringent requirement for precision addressing to ensure the correct atom is targeted. 






A highly sought-after alternative to materials-based quantum information processing is to encode qubit states into any number of degrees of freedom in photons -- such as polarisation, spatial distribution, path, or time-bin~\cite{o2009photonic,Flamini_2019}. Photonics are already seen as a highly desireable platform for quantum communication and quantum cryptography, not least due to ability to transmit quantum information over large distances quickly~\cite{RevModPhys.84.621,Flamini_2019,slussarenko2019photonic}. Initially, it seemed as though a photonic platform for quantum computing would require non-linear optical components deemed to be experimentally infeasible. Despite single-qubit operations being relatively straightforward (merely change the angle of a beam-splitter), implementing a CNOT gate appeared out of reach. However, it was shown in a 2001 breakthrough that with the addition of a number of auxiliary photons, CNOT gates could be implemented non-deterministically in photons~\cite{knill2001scheme}. Although this non-determinism adds a large resource overhead, photons do not face many of the stochastic environment issues that plague other platforms. 
One may alternatively interact photons with an atom cavity for deterministic operations, at the expense of incurring extra errors~\cite{PhysRevA.78.032318}.
In either case, as a trade-off, photonic hardware requires extremely high quality single photon sources~\cite{slussarenko2019photonic}. 

There, of course, remain a great number of other avenues being explored for the realisation of a \ac{QIP}. Examples include solid state silicon qubits~\cite{kane1998silicon}, which promise scalability and ease of integration into global silicon manufacturing chains; nitrogen-vacancy colour centres in diamond, which offer room temperature coherence~\cite{gulka2021room}; continuous variable oscillator qubits and encoded Gottesman-Kitaev-Preskill states, which boast some of the only break-even quantum error correction experiments~\cite{ofek2016extending,sivak2022real}; topological Majoranas, which are currently unrealised but are naturally protected from noise~\cite{aasen2016milestones}; among others~\cite{blais2020quantum,gross2017quantum,loss1998quantum,quant-ladd,jones1998implementation}.

\subsubsection*{Cloud Access}

Although the experimental requirement to construct quantum hardware is significant, access to these devices is somewhat democratised by various companies and institutions providing access to nascent quantum computers as accessed through cloud-run services. These devices all have a significant amount of noise on them, and are typically referred to as \ac{NISQ} computers. 
Cloud quantum computing access was pioneered with a free model by IBM~\cite{PhysRevA.94.032329}. Anybody with a computer and internet access can run basic quantum programs on these multi-million dollar devices. A paid model is now employed across the industry, with Amazon, IonQ, Oxford Quantum Circuits, Rigetti, Quera, Xanadu, and D-Wave offering access~\cite{gonzalez2021cloud}.

The present author has been fortunate enough to have extensive access to premium IBM Quantum superconducting devices over the course of this work. Information about the specifics of these devices can be found in the Appendix. The shared property is that they are all superconducting transmon devices, all operate with the same heavy honeycomb connectivity -- qubits are connected via a coupling bus and interact through the cross-resonance interaction~\cite{malekakhlagh2020first}. The fixed frequency architecture thus means that connected qubits experience a constant $ZZ$ interaction.
Different devices have different foci. Some specialise in readout fidelity, some are characterised by new materials and higher coherence times, and some boast large qubit numbers as their distinguishing feature. Each device will be indicated in the text with its name in italics, for example \emph{ibmq\_melbourne}.



\subsection{The Long Game}

Despite recent work showing immense progress towards the dream of fault-tolerant quantum computing, the realisation of such a device is still expected by most to be many years -- or even decades -- away~\cite{de2021materials,van2016path}. 
From an engineering perspective, each technology faces their challenges. 
Despite remarkable milestones and increasingly high fidelity hardware, some problems worryingly persist. Gate fidelities across most platforms have seemed to plateau.



Why are things so far away? One could consider the challenges of the building blocks themselves; the challenges of scaling those building blocks from a fabrication perspective; and the classical challenges required. There is, of course, interplay between the different hurdles. Fabrication defects, for example, could be compensated for with more sophisticated control schemes. 

We have discussed some hurdles facing present-day platforms for quantum computing. Summarily, the devices we currently have are not good enough to be below the fault-tolerant threshold. In a minimal sense, this requires better coherence times, which can be seen as a fabrication issue. In superconducting qubits, for example, the exploration of various metal alloys has yielded great improvements in coherence times~\cite{place2021new,wang2022towards} and improved understanding of amorphous defects in the fabrication process will better aid reducing environment effects~\cite{muller2019towards,}. Silicon processors, for example, have already reaped great benefits from isotopic enrichment removing nearby nuclear spins as a source of decoherence~\cite{yoneda2018quantum,takeda2021quantum}.
On the flip side, platforms which are limited more by control seek control-based solutions. A promising path forward in trapped ion addressability, for example, is the integration of classical control and measurement in the ion traps themselves through a mechanism such as on-chip delivery of light~\cite{brown2021materials}. 


Even with high quality qubits in hand, the scaling of qubit number is an engineering problem for which some platforms are more well-suited than others. Phosphorus donors in silicon have faced immense challenges thus far in the form of charge noise, for which it may seem as though other modalities are far leading the race. But once these initial hurdles are cleared, the path to tens of millions of qubits is far clearer in a silicon chip than thousands of cryogenic fridges. 
It is possible that the road to fault-tolerance is realised by a slow progression, step-by-step checking off each item on this list. Alternatively, it may be a brand-new approach that sees us the remainder of the way. Even if this should be possible, the required advances in classical computation for real-time decoding of errors in quantum error correction is itself eye-watering~\cite{battistel2023real}.

What we have discussed so far are the more conventional challenges faced by quantum computers: increasing the physical coherence of systems and producing more of them. 
One largely unaddressed question on the subject of quantum error correction, however, is the issue of \emph{complex} noise in quantum devices. We will frequently use `complex' to describe noise with a more intricate character than the usual identical, independently-distributed error models -- such as depolarising noise. It is unknown whether, for example, errors could emerge at scale. The emergent physics of hundreds of coherent quantum systems interacting with one another is, of course, unexplored. Once the bounds of classical computation are saturated, we are in uncharted territory with respect to the dynamical behaviour.

But even at a tractable level, complex correlated noise can be problematic. Quantum error correcting codes are predicated on a simple, uncorrelated error model. If real noise violates these assumptions, then more physical resources are required -- or correction may even be unachievable.
Correlated error simply brings about more total error, stemming from dynamics which will be harder to characterise, harder to understand, and harder to control. 
Exotic noise is already known to exist on every hardware platform we have discussed. 
This ranges from passive crosstalk between connected qubits in fixed-frequency transmons~\cite{PhysRevLett.129.060501}, coherent interactions with two-level system defects in the surrounding amorphous materials~\cite{muller2019towards}, and non-exponential dephasing~\cite{wilen2021correlated}. 
Trapped ions, in contrast, have more control-based problems and sources of electric field noise near surfaces~\cite{RevModPhys.87.1419}.
Imprecise addressing of ions constitutes one non-trivial introduction of spatially correlated noise~\cite{ParradoRodriguez2021crosstalk}, but even further one must be wary of temporally correlated noise from lasers with quasistatically fluctuating power spectra.
Rydberg atoms face similar control issues, as well as the potential for absorbing spontaneous emissions from nearby atoms during the measurement process~\cite{wu2021concise}.
Silicon meanwhile faces more complex crosstalk in Heisenberg interactions between qubits, as well as dipole coupling to highly coherent nearby nuclear spins~\cite{kuhlmann2013charge,yoneda2022noise,rojas2023spatial}.

Some work examining the effects of correlated noise on error-correcting codes has been conducted~\cite{correlated-qec,PhysRevA.83.052305,Clader2021,PhysRevA.69.062313,PhysRevA.56.1177,ahsan2018performance,PhysRevLett.97.040501}.
Although the impact is highly dependent on the underlying models, the end result is always that correlated noise results in higher effective thresholds than uncorrelated counterparts. Despite some optimism on achieving fault-tolerance via scaling current means under idealised models, the reality of complex noise may require substantially larger physical overheads than currently estimated.
The upshot appears to be that greater characterisation is needed. This can be used both to understand and remove the complex dynamics at both the fabrication and the control level, as well as to feed forward that information to quantum error correction decoders for more sophisticated error decoding.

In this thesis, we adopt the philosophy of near-term quantum devices as a novel testbed for probing interesting physics. We centrally discuss correlated quantum dynamics in the practical context of understanding and mitigating quantum noise. However, we are more generally motivated by a desire to understand complex temporal correlations and non-Markovian open quantum systems~\cite{deVega2017}. Our results, therefore, are not limited to studying noise on quantum computers. We hope instead that this will open up a broad avenue for both the study and the simulation of the most general setting in quantum dynamics.



\chapter{Open Quantum Systems and Operational Quantum Dynamics}
\label{chap:OQS}
\epigraph{\emph{Quantum phenomena do not occur in a Hilbert space. They occur in a laboratory.}}{Asher Peres}
\noindent\colorbox{olive!10}{%
	\begin{minipage}{0.9555\textwidth} 
		\textcolor{Maroon}{\textbf{Chapter Summary}}\newline
		Noise -- and other deleterious effects -- typically stems from either an interaction between a closed system and its environment, or the addition of classical stochasticity from control instruments. 
		This is known as the open quantum systems formalism, and can be thought of as the mixed interior to the space traced out by closed systems quantum mechanics.
		In this chapter, we introduce some mathematical machinery required to model dynamics of this type. We predominantly take an \emph{operational} viewpoint. That is, one which is in terms of measurable input and output contents to a quantum map, and hence agnostic to the underlying physics. This type of formalism is amenable to direct experimental reconstruction, and we review the basics of both quantum state and quantum process reconstruction. 
		\par\vspace{\fboxsep}
		\colorbox{cyan!10}{%
			\begin{minipage}{\dimexpr\textwidth-2\fboxsep}
				\textcolor{RoyalBlue}{\textbf{Further Reading}}\newline 
				The set of lecture notes in Ref.~\cite{lidar2019lecture} constitutes a first-principles and pedagogically-motivated introduction to open quantum system dynamics. This supplements more extensive discussions in the textbook Ref.~\cite{lidar_brun_2013}, for which noise is placed into quantum computing and quantum error correction contexts.
				For a comprehensive discussion of open quantum dynamics, quantum maps, and their representations, see Ref.~\cite{Milz2017,Flory2010}. For a specifically graphical calculus presentation of the above, Ref.~\cite{wood2011tensor} introduces and demonstrates a comprehensive graphical language.
				Finally, Ref.~\cite{intro-GST} presents a thorough introduction to quantum state tomography, quantum process tomography, and gate set tomography. The latter includes numerical motivation for self-consistent reconstruction of quantum dynamics.
		\end{minipage}}
\end{minipage}}
\clearpage
\section{Introduction}

The postulates of quantum mechanics are, in principle, sufficient to describe all reality at a microscopic, non-relativistic level. Indeed, the dynamics of closed systems can be completely captured by the Schr\"odinger equation without reference to size. 
But we are not Laplace's Demon: the contents of a closed system which are accessible and from which we can acquire knowledge is vanishingly small when compared to the total system. Not least of the consequences of this is the emergence of classicality~\cite{Zurek1991}. Although we do not yet understand where the line is drawn between definitely quantum and definitely classical realms, non-quantum phenomena such as irreversibility and stochasticity are known to emerge from massive, poorly isolated quantum systems~\cite{spohn1978irreversible}. 
But without total information, how then do we have any hope of describing the evolution of our tiny fragment of this vast space?
Open systems paradigms are not uniquely quantum. Almost every physical system -- whether it be the weather, a coin flip, or a biological process -- is ignorant to some extent of its overarching deterministic physics. 
To address this, there is a vast literature on stochastic theory to describe the probabilistic evolution of systems~\cite{breuer2016,breuer2002theory,rivas2012open,198297,lindblad1975completely,davies1970operational}. 

The origins of stochastic theory in quantum mechanics can be traced chiefly back to von Neumann with the introduction of the density matrix and its connection to dynamics~\cite{von2018mathematical}. Stochastic dynamics was studied phenomenologically from the 1950s and 1960s~\cite{van1957derivation,van1957approach,PhysRev.134.A1429,weidlich1965coherence,risken1966fokker,REDFIELD19651,PhysRev.70.460}, but it was not until Nakajima and Zwanzig's seminal work until an exact master equation was derived, consistent with any environment~\cite{nakajima1959quantum,zwanzig1960ensemble}. Other, inexact but eminently useful, master equations were similarly derived by Bloch~\cite{PhysRev.70.460}; Redfield~\cite{REDFIELD19651}; Lindblad~\cite{1976CMaPh..48..119L}; Gorini, Kossakowski, and Sudarshan~\cite{gorini1976completely}. 
Quantum maps were introduced by Sudarshan, among others, in 1961~\cite{jordan1961dynamical,PhysRev.121.920}, and later rediscovered and popularised by Kraus~\cite{kraus1971general}. Much of the structure of such maps was exposed mathematically by Choi~\cite{choi_1972} and Stinespring~\cite{stinespring1955positive,CHOI1975285}. These considerations form the basis of quantum probability theory, for which some crucial players have been Lindblad~\cite{lindblad1975completely}, Davies and Lewis~\cite{davies1970operational}, Jamio\l kowski~\cite{JAMIOLKOWSKI1972275}, and Accardi \emph{et al.}~\cite{198297}. 
Many of the concepts discussed both in this chapter and the next are quantum generalisations of classical probability theory. Probability vectors, transition matrices, master equations, and stochastic processes all have a well-formalised quantum analogue which has been remarkably successful at describing reality. 

The theory of open quantum systems plays a role in optimal quantum control and the development of quantum technologies~\cite{PhysRevLett.104.090501,koch2016controlling,schmidt2011optimal,PhysRevLett.102.080501}; quantum optics~\cite{breuer2002theory,Sieberer_2016,daley2014quantum}; quantum thermodynamics~\cite{kosloff2013quantum}; quantum metrology~\cite{PhysRevLett.112.120405,PhysRevLett.109.233601}; and quantum biology~\cite{mcfadden2018origins,lambert2013quantum}, among other areas~\cite{attal2006open}. 
One could consider open quantum systems as being the axiomatic formulation of quantum mechanics, for which pure-state quantum mechanics lies as a special case on the boundaries. Without discussing the philosophy of such things, here we provide a brief overview of the theory of open quantum systems at a level which is practically relevant to the remainder of the thesis. We shall consider the introduction of stochastic effects at the single and two-time level.
This introductory chapter is separated into three main components. First, we introduce open quantum systems as a concept and much of the mathematical machinery for dealing with them. Next, we look at open quantum system theory from an operational perspective and abstract all dynamics into a series of linear maps. Finally, we consider the problem of \ac{QCVV} --- or shorter, tomography. This asks how we can reconstruct or estimate different components of our open quantum systems in practice. This mathematical framework assumes the generation of no significant system-environment correlations, or Markovianity. We will also visit the extent to which the framework breaks down in a non-Markovian setting. We will both employ, and build on, many of the mathematical and algorithmic tools introduced in this chapter throughout the remainder of the thesis.

\section{Open Quantum Systems}

We start by defining the formalism of open quantum systems. An \emph{open} quantum system is one who is the substituent of a larger space. The picture we will typically deal with is in factor spaces: a system $S\equiv\mathcal{H}_S$ coupled to an environment $E\equiv\mathcal{H}_E$ such that $\mathcal{H}_{\text{total}} = \mathcal{H}_S\otimes\mathcal{H}_E$. But note that the total Hilbert space need not factor, and $\mathcal{H}_S$ may instead be some vector subspace of $\mathcal{H}_{\text{total}}$. A collective state $|\psi\rangle\in\mathcal{H}_{\text{total}}$ evolves unitarily according to the Schr\"odinger equation, but without access to the entire system-environment setting, the entanglement across the two systems manifests itself as randomness at the level of the system. 
Descriptions of open quantum systems are designed to deal with the dynamics of some sub-part of a larger composite system, where the environmental degrees of freedom are themselves evolving in some unknown manner. They are also designed to express the effects of classical randomness on quantum evolution.


\subsection{Density Operators}
A natural first step to introducing stochasticity in quantum theory is to consider mixed ensembles of states. The formalism introduced so far makes statistical predictions about a collection of identically prepared quantum states -- that is to say, a pure ensemble. Applying a $H$ gate to a $\ket{0}$ state produces an equal superposition $1/\sqrt{2}(\ket{0} + \ket{1})$, for which we say there is a 50\% probability of measuring $\ket{0}$ and a 50\% probability of measuring $\ket{1}$ in the $Z$ basis. But this statement only makes sense in the context of many identically prepared $\ket{0}$ states, to which we apply the control operation $H$. What if instead, we have a mixed ensemble? Instead of feeding all $\ket{0}$ states into the control operation, suppose we take a classical mixture of 60\% $\ket{0}$ and 40\% $\ket{+}$ states. Evaluating the same expectations for each instance, we find equal probability of $\ket{0}$ and $\ket{1}$ for the first set of states, and 100\% probability of measuring $\ket{0}$ in the second instance. Stochastically mixing these outcomes together gives a 70\% chance of measuring $\ket{0}$ and 30\% chance of measuring $\ket{1}$ across the ensemble. \par 

To be precise, consider the preparation of an ensemble $\{\ket{\psi_1},\ket{\psi_2},\cdots,\ket{\psi_N}\}$ with corresponding classical probabilities $\{p_1,p_2,\cdots, p_N\}$. Each $\ket{\psi_i}$ lives on a $d-$dimensional Hilbert space $\mathcal{H}$. There is no requirement that $N\leq d$, and hence the $\{\ket{\psi_i}\}$ need not be orthogonal. The fractional probabilities, however, are normalised such that $\sum_i p_i = 1$. Suppose we make a measurement on the ensemble to evaluate some observable $O$ with eigenbasis $\{\ket{\gamma_i}\}$ and corresponding eigenvalues $\{g_i\}$. This translates to the question ``what is the average measured value of $O$ when many measurements are carried out across the ensemble?'' As sketched above, the solution is contained in the ensemble average of $O$ evaluated on each $\ket{\psi_i}$. Hence:
\begin{equation}
\label{eq:ensemble-av}
	\begin{split}
	\langle O \rangle &= \sum_{i=1}^N p_i \bra{\psi_i} O\ket{\psi_i}\\
	&= \sum_{i=1}^N \sum_{j=1}^d p_i |\langle \gamma_j |\psi_i\rangle|^2 g_j.
	\end{split}
\end{equation}
We have two probabilistic weightings here: the classical probability $p_i$ of obtaining state $\ket{\psi_i}$, and the quantum mechanical probability $|\langle \gamma_j|\psi_i\rangle|^2$ of finding $\ket{\psi_i}$ in eigenstate $\ket{\gamma_j}$. Re-writing Equation~\eqref{eq:ensemble-av} in a general basis $\{\ket{a}\}$, we have:
\begin{equation}
\begin{split}
	\langle O \rangle &= \sum_{i=1}^N p_i \sum_{a}\sum_{a'} \langle\psi_i|a\rangle\!\langle a|O|a'\rangle\!\langle a'|\psi_i\rangle,\\
	&= \sum_a\sum_{a'}\langle a' |\left(\sum_{i=1}^N p_i |\psi_i\rangle\!\langle\psi_i|\right)|a\rangle\langle a|O|a'\rangle.
\end{split}
\end{equation}
The term in parentheses is now independent of the given observable, which motivates the definition of the \emph{density matrix} or \emph{density operator}\footnote{$\rho$ is indeed an operator, since it acts as a linear mapping $\rho : \ket{v}\in\mathcal{H}\mapsto \ket{v'}\in\mathcal{H}$. However, since we rarely employ the object in this context, we will employ the terminology `density matrix'.}
\begin{equation}
	\label{eq:DM-def}
	\rho := \sum_{i=1}^N p_i |\psi_i \rangle\!\langle \psi_i|.
\end{equation}
The density matrix contains all the information both about the ensemble probabilities and the quantum states therein. Computationally however, it is much simpler to deal with, as the above expectation value reduces:
\begin{equation}
\begin{split}
	\langle O\rangle &= \sum_a\sum_{a'}\langle a'|\rho|a\rangle\!\langle a|O|a'\rangle\\
	&= \Tr[\rho O].
\end{split}
\end{equation}
It is important to note that although density matrices contain all the physically extractable information about their ensembles, they do not uniquely define an ensemble. This can be seen, for instance, in the observation that $0.5 (|0\rangle\!\langle 0| + |1\rangle\!\langle 1|) = 0.5 (|+\rangle\!\langle +| + |-\rangle\!\langle -|)$. Since decompositions are not unique, one should take care when ascribing physical interpretations.\par 

Density matrices satisfy several key properties:
\begin{enumerate}
\item \underline{Unit trace:} This follows from normalisation of probabilities,
\begin{equation}
	\Tr[\rho] = \sum_{i=1}^N\sum_a p_i \langle a|\psi_i\rangle\!\langle\psi_i|a\rangle = \sum_{i=1}^Np_i\langle \psi_i|\psi_i\rangle = 1.
\end{equation}
\item \underline{Hermitian:} From Equation~\eqref{eq:DM-def} it follows that $\rho$ is Hermitian, since it is defined in terms of real, positive numbers and rank-one projectors.
\item \underline{Positive semidefinite:} For all vectors $\ket{v}\in\mathcal{H}$, we have $\langle v|\rho|v\rangle = \sum_{i=1}^N p_i |\langle \psi_i|v\rangle|^2 \geq 0$.
\end{enumerate}
It follows from conditions (1) and (3) that $\rho$ has positive eigenvalues $0\leq \lambda_i\leq 1$. We will henceforth identify the eigenvalues of density matrix $\rho$ with probabilities of an ensemble in the eigenbasis, and denote them by $\{p_i\}$.

\subsection{Composite Systems}

The system-environment distinction is particularly important both in this thesis and in all discussions of non-Markovianity. Indeed, non-Markovian dynamics loosely boil down to the generation of temporal correlations by random variables whose values have been marginalised over. Therefore, the presence of temporal correlations at all depends entirely on which variables one is able to condition over, and which are inaccessible. That is to say, one person's environment may be another person's system, and hence one person's non-Markovianity may be another person's closed and Markovian dynamics. 

Let us consider two systems: $S$, our system and $E$, its environment. We can assume that the total composite system evolves according to the Schr\"odinger equation, and let $\mathcal{H}_S = \text{span}(\{|i\rangle_S\})$ as well as $\mathcal{H}_E=\text{span}(\{|\mu\rangle_E\})$, where $\{|i\rangle_S\}$ ($\{|\mu_E\rangle\}$) is a size-$d_S$ ($d_E$) orthonormal basis. The combined space is hence $\mathcal{H} = \mathcal{H}_S\otimes \mathcal{H}_E$ with size $d_Sd_E$ basis $\{|i\rangle_S\otimes |\mu\rangle_E\}$. Much like before, we can write any density matrix on the combined Hilbert space as 
\begin{equation}
	\rho^{SE} = \sum_{ij\mu\nu} \lambda_{ij\mu\nu}|i\rangle_S\langle j|\otimes |\mu\rangle_E\langle\nu|.
\end{equation}

The partial trace over an environment is a linear operator $\mathcal{H}\rightarrow\mathcal{H}_S$ defined as 
\begin{equation}
	\begin{split}
	\Tr_E[\rho^{SE}] &= \sum_\gamma\langle\gamma|_E\left(\sum_{ij\mu\nu} \lambda_{ij\mu\nu}|i\rangle_S\langle j|\otimes |\mu\rangle_E\langle\nu|\right)|\gamma \rangle\\
	&= \sum_{ij\mu\gamma}\lambda_{ij\mu\gamma}|i\rangle_S\langle j| \langle\gamma|\mu\rangle_E\\
	&= \sum_{ij\mu}\lambda_{ij\mu\mu}|i\rangle_S\langle j| \equiv \sum_{ij}\bar{\lambda}_{ij}|i\rangle_S\langle j|.
	\end{split}
\end{equation}

Physically, what this represents is a sum over ignorance: a weighted average of the system corresponding to different environment eigenstates. 
Of course, we have used $SE$ terminology here, but the same applies for arbitrary composite Hilbert spaces. 
A reduced density matrix is the general object describing what we have physical access to, it represents all the information we can extract from a system at any single time.
Importantly, there is no way to physically distinguish ensembles of different elements from one another. One cannot, for example, know whether mixedness of a density matrix is due to proper mixtures from classical stochasticity, or improperly mixed via entanglement with an external system. 
Mixedness can in general be measured by the \emph{purity} of the state, which can be expressed in terms of eigenvalues $\{p_i\}$:
\begin{equation}
	\gamma := \Tr[\rho^2] = \sum_i p_i^2\quad \text{where}\ \frac{1}{d} \leq \gamma \leq 1.
\end{equation}
The lower bound is saturated for maximally mixed states $\rho = \mathbb{I}/d$, and the upper bound for pure states. 


\subsection{Time Evolution}

The time evolution of the density matrix subject to some Hamiltonian dynamics is given by the Liouville-von Neumann equation
\begin{equation}
	\frac{\text{d}\rho}{\text{d}t} = -i[\mathcal{H},\rho],
\end{equation}
which is exactly deterministic Schr\"odinger evolution, with the additional feature that we can start with initially mixed states. The closed evolution of some large quantum state is typically intractable to perform, and so under various assumptions the evolution of an open system is usually performed. 

The most general Markovian master equation was discovered by \ac{GKSL} independently~\cite{gorini1976completely,1976CMaPh..48..119L}. This includes the effects of dissipation, which is responsible for general non-unitarity:
\begin{equation}
	\frac{\text{d}\rho}{\text{d}t} = -i[\mathcal{H},\rho] + \sum_\alpha \gamma_\alpha \left(L_\alpha \rho L_\alpha^\dagger - \frac{1}{2}\left\{L_\alpha^\dagger L_\alpha,\rho\right\}\right).
\end{equation}
The \acs{GKSL} equation can be derived using numerous methods (for a review, see Ref.~\cite{lidar2019lecture}). It is not applicable to any open dynamics (for this we need a memory kernel, as discussed in Chapter~\ref{chap:stoc-processes}), and so hence requires a series of assumptions to be made. These typically amount to slowness of correlations between system and environment; dissipation of system excitations within the environment; and negligence of fast-oscillating terms when compared to the system timescale. Respectively, these are known as the Born, Markov, and rotating wave approximations. 
Nevertheless, the \acs{GKSL} equation has been immensely useful in the study of open quantum systems for decades hence. It is only now with the development of highly sensitive quantum technologies that a more general treatment of memory effects and the breakdown of these assumptions has precipitated.

\subsection{Ensemble Quantities}

A density matrix description allows for an ensemble description of quantum states, for which classically motivated tools from information theory can be described. In particular, this allows us to determine quantifiers not only of the quantum properties of the state, but the properties of the distribution itself. We have already seen this in state purity, $\Tr[\rho^2]$, which is an ensemble quantity. Let us briefly introduce several further quantities of interest that we will apply throughout the remainder of the thesis. 


The von Neumann entropy quantifies the departure of a density matrix from a pure state. In particular, this measures the Shannon entropy of the spectrum of the state. That is to say, the uncertainty of the quantum state as a stochastic mixture of pure states.
\begin{equation}
	S(\rho) := -\Tr[\rho\log\rho] = -\sum_{i=1}^{d} p_i\log p_i
\end{equation}
The von Neumann entropy pertains specifically to the properties of a single state. To compare different states, one typically employs the quantum relative entropy $S(\rho\mid\mid\sigma)$, which generalises the Kullback-Leibler (KL) divergence\footnote{Unlike metrics, divergences are not symmetric in their arguments.} $S(P\mid\mid Q)$, for states $\rho,\sigma$ and distributions $P,Q$. 
In the limit where the number of samplings from a distribution goes to infinity, a frequency distribution is guaranteed to match a given probability distribution. Relative entropy governs the rate at which these two objects coincide, as formalised by Sanov's Theorem. Given an experiment with $n$ outcomes described by the probability distribution $Q$, let the experiment be repeated $N$ times. 
We are interested in the probability that a frequency distribution $P$ will be observed. Let $E$ be the set of probability distributions for $n$ outcomes, then for large $N$, the probability $\mathbb{P}$ that a frequency distribution belonging to $E$ will be obtained is 
\begin{equation}
	\mathbb{P}(E) \sim \text{e}^{-N S(P_\ast \mid\mid Q)},
\end{equation}
where $P_\ast$ is the distribution in $E$ that minimises $S(P\mid\mid Q)$.
Roughly speaking, this tells us the number of experiments that will be required to rule out a model for some distribution to a given probability. The higher the relative entropy, the fewer experiments required to ascertain that we have modelled the distribution erroneously. 

The quantum relative entropy can be defined in terms of matrix logarithms, but, once more, is equivalently the KL divergence of the spectrum of two states $\rho$ and $\sigma$, with eigenvalues $\{p_i\}$ and $\{q_i\}$ respectively. That is, 
\begin{equation}
	\label{eq:relative-entropy}
	S(\rho\mid\mid\sigma) := \Tr[\rho(\log\rho - \log\sigma)] = \sum_{i=1}^{d}p_i\log\frac{p_i}{q_i}.
\end{equation}
A straightforward way to see the asymmetry in distributions here is to consider an unknown coin which is known to either be fair, or have heads on both sides. To distinguish between the two, one starts flipping. If the coin is fair, then one can soon say with certainty that it is not the unfair coin once tails shows up. However, if the coin is unfair, then it will take an infinite number of flips to rule out the fair possibility. The asymmetric quantities are hence
\begin{equation}
	S(Q_{\text{fair}} \mid\mid Q_{\text{unfair}}) = \log N\quad \text{and} \quad S(Q_{\text{unfair}}\mid\mid Q_{\text{fair}}) = \infty.
\end{equation}
Unbounded relative entropies occur when $P$ has support outside $Q$, and the denominator in Equation~\eqref{eq:relative-entropy} vanishes. 

For some $\rho^{AB}\in \mathcal{B}(\mathcal{H}_A\otimes\mathcal{H}_B)$, \ac{QMI} quantifies the shared information of the two composite systems.
\begin{equation}
	I(A : B) := S(\rho^A) + S(\rho^B) - S(\rho^{AB}) = S(\rho^{AB} \mid\mid \rho^A\otimes \rho^B)
\end{equation}
The second equality here emphasises that mutual information measures via relative entropy the error of the assumption to take a state as the product of its marginals. 

\subsection{Positive Operator-Valued Measures}
\label{ssec:POVMs}
In the same way that we generalised the notion of pure states to permit a reference to subsystems including the complete Hilbert space, so too can we generalise projective measurements. Specifically, we can ask what a projective measurement on a given Hilbert space does to a subsystem of that space, without referring to the whole system. To start, we can consider a generalised notion of measurement purely in terms of outcomes for some measurement apparatus.

\begin{definition}
On a finite-dimensional Hilbert space $\mathcal{H}$, a \emph{\ac{POVM}} is a collection of operators $\{E_m\}$ with the following properties:
\begin{equation}
	\begin{split}
		&E_m^\dagger = E_m \text{ (Hermitian),}\\[10pt]
		&\langle \psi|E_m|\psi\rangle \geq 0 \: \forall \: m, |\psi\rangle \text{ (Positive)},\\[10pt]
		&\sum_m E_m = \mathbb{I} \text{ (Partition of unity)}.
	\end{split}
\end{equation}
We call each component $E_m$ of the \acs{POVM} an \emph{effect}. 
\end{definition}

Each \acs{POVM} element $E_m$ is assigned to an outcome $m$ with probability $\langle\psi|E_m|\psi\rangle$, or $\Tr[E_m \rho]$ for mixed states. This represents the classical information extracted from a quantum state. In contrast with \acs{PVM}s, however, \acs{POVM}s do not \emph{uniquely} define a post-measurement state. \acs{PVM}s are consequently a subclass of the more general \acs{POVM}. 
Elements of a \acs{POVM} can be defined in terms of measurement operators $\{M_m\}$ via $E_m = M_m^\dagger M_m$. Since the $E_m$ are positive, then non-unique $M_m=\sqrt{E_m}$ can always be found. The post-measurement state is given by $|\psi\rangle\mapsto \frac{M_m|\psi\rangle}{\sqrt{p_m}}$, where $p_m = \langle\psi|M_m^\dagger M_m|\psi\rangle$. Importantly, the state has no longer necessarily `collapsed', in that repeated applications of a \acs{POVM} need not yield the same outcomes. One application of \acs{POVM}s is in optimal state discrimination. For non-orthogonal states, one can never say with certainty which state they have. \acs{POVM}s, however, allow a discriminator to proclaim ignorance so as to avoid providing the wrong answer.

As a consequence of Naimark's dilation theorem, all \acs{POVM}s can be realised as the reduction of a projective measurement on a larger Hilbert space. That is, starting from some $\rho=\rho_A\otimes\rho_B\in\mathcal{B}(\mathcal{H}_A\otimes\mathcal{H}_B)$, let $\{\Pi_m\}$ be a \acs{PVM} acting on the collective space. Then $\mathbb{P}(\rho = m) = \Tr[\Pi_m \rho] = \Tr_A[\Tr_B[\Pi_m\rho]]$ which can be equivalently written $\Tr_A[E_m\rho_A]$. 


\section{Graphical Calculus Preliminaries}
\begin{figure}[!ht]
	\centering
	\includegraphics[width = \linewidth]{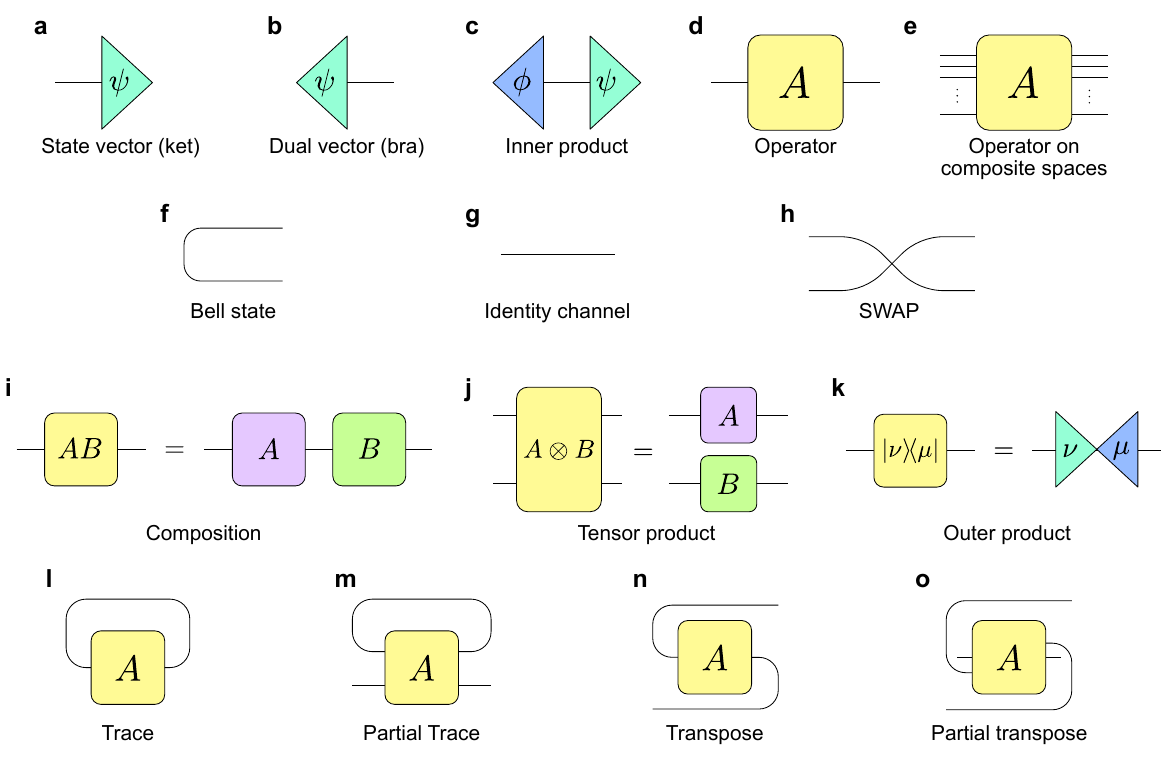}
	\caption[Graphical depiction of basic tensors and constructive expressions]{Graphical depiction of basic tensors and their relations. State vectors \textbf{a} and their dual vectors \textbf{b} are represented by triangles with a left or right-facing wire. Contraction of indices is represented by the connection of wires, such as the inner product in \textbf{c}. Operators $A$ are depicted by a box with left and right facing wires, as in \textbf{d} -- or with multiple wires for composite spaces, as in \textbf{e}. Simple wires have canonical relations: \textbf{f} A wire bent around a corner represents a Bell state $|\Phi^+\rangle$; \textbf{g} a single horizontal wire represents an identity operator or matrix; \textbf{h} two wires crossing over is a SWAP operation on two Hilbert spaces. Operators can also be created and manipulated in straightforward fashions: \textbf{i} composition $AB$ of two operators $A$, $B$ is given by connecting the right wire of $A$ with the left wire of $B$; \textbf{j} two operators $A$, $B$ stacked vertically is taken to represent a tensor product $A\otimes B$; \textbf{k} arranging the vertices of a state vector and dual vector depicts the outer product operation and constructs a new operator.
	Finally, expressions on operators may be given by manipulating their wires:
	\textbf{l} trace of an operator $\Tr[A]$ is taken by connection all left and right wires; \textbf{m} partial trace of an operator $\Tr_1{A_{12}}$ is taken by connecting left and right wires on only a subspace of an operator; \textbf{n} the transpose of an operator $A^{\text{T}}$ is represented by bending left wires around to the right and vice versa; \textbf{o} the partial trace of an operator $A^{\Gamma_1}$ is obtained by bending only left and right wires on a subspace.
	}
	\label{fig:graphical-calc}
\end{figure}

A popular expression of calculation in the quantum information community has been with the use of tensor network diagrams and graphical calculi to convey otherwise unwieldy multidimensional tensor algebra. This is already commonly seen with the use of quantum circuits to represent unitary manipulations of pure states. 
We will not employ formal graphical calculi, but rather use it to illustrate important concepts and adhere roughly to the set of rules as follows.
Tensors -- of any rank -- are represented by geometric shapes, with free indices represented by wires. The rank of the tensor is hence given by the number of wires, and whether these correspond to a Hilbert space or a dual vector space is indicated by the opposing direction of those wires. Contraction (summation) of indices is then given by the joining of co- and contravariant wires. State vectors $\ket{\psi}$ are represented by triangles with left-pointing wires, dual vectors $\bra{\psi}$ by triangles with right-pointing wires, and more general operators are depicted as rectangles with left and right-facing wires. For pictorial convenience we will occasionally have wires pointing up/down rather than left/right but from context these should be clear. A series of quantum objects, as well as manipulations of those objects in a graphical calculus sense is shown in Figure~\ref{fig:graphical-calc}.

\section{Operational Quantum Dynamics}
The time evolution of any non-relativistic closed quantum system can be perfectly described by a Hamiltonian detailing the energy configuration of the system. If one moved to a dissipative open environment setting, then a Lindbladian would be sufficient to generate and solve the dynamics. Moving even further, in a fully general open quantum systems setting, one could solve the Nakajima-Zwanzig master equation and fully describe the time evolution of a system. Each of these approaches is a powerful method (when chosen appropriately) to study the dynamics of a given system, especially numerically. 
In contrast, rather than setting the dynamics in motion, one may describe it more in the abstract operational framework in terms of state inputs and measurement outcomes. 
The former approach is more reductionist -- understand the physics and build the dynamics from there. The latter approach is more emergentist -- let the dynamics happen, and study its properties.
But they also carry associated baggage, one typically must a priori have a model for the physical system, they can be difficult to reconstruct. 
This section, and indeed the majority of this thesis, will be concerned with an operational approach to quantum dynamics. 

\subsection{Quantum Maps and Representations}
One broad consequence of the linearity of quantum mechanics is that the dynamical evolution of a system can be described by a series of linear maps on that system. Specifically, a density matrix at one time can be mapped to a density matrix at another time $\rho\mapsto\Lambda[\rho]$. 
Mappings of density operators $\Lambda$ are called \emph{superoperators}, as they act as operators on an operator space, taking an input $\rho\in \mathcal{B}(\mathcal{H})$ and produce an output $\rho'=\Lambda[\rho]\in \mathcal{B}(\mathcal{H}')$.
In different contexts these may inherit different names. Throughout this thesis, we will roughly keep to the following semantics: \emph{quantum maps} to describe mappings of density matrices generally; \emph{quantum channels and dynamical maps} to refer to uncontrolled -- and hence involve no measurements -- dynamics; and \emph{quantum operations} and \emph{quantum instruments} to refer to experimenter-chosen manipulation of a system, which can be probabilistic.

Formally speaking, a quantum map $\Lambda$ is a linear map $\Lambda:\mathcal{B}(\mathcal{H}_{\text{in}})\rightarrow \mathcal{B}(\mathcal{H}_{\text{out}})$. Here, the input state to the map $\rho\in\mathcal{B}(\mathcal{H}_{\text{in}})$ need not live on the same space as the output state $\rho'\in\mathcal{B}(\mathcal{H}_{\text{out}})$, although usually we will take them as such. 
From the linearity of mixing in stochastic theories, a quantum map must be linear. That is, we have 
\begin{equation}
	\Lambda[\alpha\rho_1 + \beta \rho_2] = \alpha\Lambda[\rho_1] + \beta\Lambda[\rho_2]\:\forall\:\rho_1,\rho_2.
\end{equation} 
It must also always produce positive probabilities, i.e. the density matrix of the output state must have only positive eigenvalues. Classically, this is a sufficient condition. Quantum mechanically, however, we must demand the stricter notion of \emph{complete positivity}. The \ac{CP} property says that if a quantum map is extended to act trivially on any extended ancilla space, then that map must also be positive\footnote{An example of a positive-but-not-completely-positive map is the matrix transpose. A transpose does not change the eigenvalues of a state, but a partial transpose renders an input state negative if the subsystems have non-bound entanglement~\cite{horodecki2009quantum}.}. That is, 
\begin{equation}
	(\Lambda\otimes \mathcal{I}_A)[\rho_{SA}]\succcurlyeq 0 \:\forall\:\rho_{SA}\in \mathcal{B}(\mathcal{H}_S\otimes \mathcal{H}_A); \:\forall\: \mathcal{H}_A.
\end{equation}
Finally, quantum maps may be either probabilistic or deterministic. Probabilistic maps may only be implemented non-deterministically, and rely on the collapse of a quantum wavefunction. For example, a projection onto the $\ket{0}$ state is a valid quantum map, but for an incoming $\ket{+}$ state, can only be applied with a probability of 0.5. In contrast, a $Z$ gate can be applied deterministically to this state. A deterministic quantum map must be \ac{TP}, that is, for all input states $\rho_{\text{in}}$, we must have 
\begin{equation}
	\Tr[\Lambda(\rho_{\text{in}})] = \Tr[\rho_{\text{in}}].
\end{equation}
Recalling that the eigenvalues of a density matrix are probabilities in the statistical ensemble, this is the statement that probabilities are preserved by deterministic operations. Quantum channels satisfying these properties are known as \ac{CPTP}. If the map is probabilistic, then we have the slightly less restrictive condition that the map must not increase the trace of the state:
\begin{equation}
	\Tr[\Lambda(\rho_{\text{in}})] < \Tr[\rho_{\text{in}}].
\end{equation}
Such maps are called \ac{TNI}.

\subsubsection*{Stinespring Dilation and Complete Positivity}

It is a consequence of the Stinespring dilation theorem that \acs{CP} maps can be imbued with the physical interpretation of originating from reduced unitary dynamics of some larger Hilbert space. Specifically, we suppose we have a system $S$ that evolves as part of a dilated Hilbert space $\mathcal{H}_{SE}$.
For ease of interpretation, we write $\mathcal{H}_{SE}$ as a factor space: $\mathcal{H}_{SE} \cong \mathcal{H}_S\otimes \mathcal{H}_E$, but in full generality $S$ may simply be a subspace of $\mathcal{H}_{SE}$, such as the lowest two energy levels of an anharmonic oscillator. $E$ is commonly taken to be an external `environment', but note that the delineation may not always be so clear. 

\begin{figure}[htbp]
	\centering
	\includegraphics[width=0.9\linewidth]{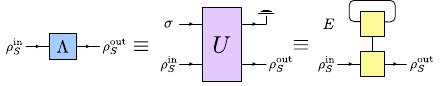}
	\caption[Graphical realisation of Stinespring's dilation theorem]{Graphical realisation of Stinespring's dilation theorem. Any \acs{CP} map on a quantum system can be expressed as a unitary evolution on a larger Hilbert space (the dilation) followed by a partial trace. The grounding symbol used throughout this thesis indicates a partial trace over inaccessible degrees of freedom.}
	\label{fig:stinespring-explanation}
\end{figure}

Stinespring's dilation theorem may be interpreted to say: for any \acs{CPTP} quantum map $\Lambda$, there exists an environment $E$ with initial state $\sigma$ and unitary dynamics $U$ such that
\begin{equation}
	\Lambda[\rho_S^{\text{in}}] = \Tr_{E}\left[U(\rho_{S}^{\text{in}}\otimes \sigma)U^\dagger\right].
	\label{eq:stinespring}
\end{equation}
Moreover, given the existence of some projective measurement $\Pi\in\mathcal{B}(\mathcal{H}_E)$, any \acs{CP}\acs{TNI} map may be realised by 
\begin{equation}
	\Lambda[\rho_S^{\text{in}}] = \Tr_{E}\left[(\Pi\otimes \mathbb{I}_S)\cdot U(\rho_{S}^{\text{in}}\otimes \sigma) \cdot U^\dagger (\Pi^\dagger\otimes \mathbb{I}_S)\right].
\end{equation}
This is expressed graphically in Figure~\ref{fig:stinespring-explanation}.

Stinespring dilation is very useful for ascribing an open quantum system picture to quantum mappings, but is not often used in practice with quantum maps. 

\subsubsection*{Kraus Operator-Sum Representation:}

\begin{figure}[htbp]
	\centering
	\includegraphics[width=0.9\linewidth]{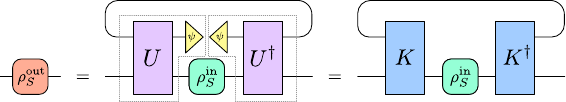}
	\caption[Graphical depiction of the Kraus operator-sum representation for quantum maps]{Graphical depiction of the Kraus \acs{OSR} for quantum maps. Kraus operators are defined by the partial trace performed going from Stinespring dilation back down to an open system.}
	\label{fig:kraus-explanation}
\end{figure}

The Kraus \ac{OSR} may be immediately derived from the Stinespring dilation of a quantum map. The initial state of the environment can be taken to be pure without loss of generality, so $\sigma = |\psi\rangle\!\langle\psi|$ and we let $\{\ket{\phi_i}\}$ be a complete set of states on $E$. Then, following on from Equation~\eqref{eq:stinespring}, we have 
\begin{equation}
	\begin{split}
		\rho_S^{\text{out}} &= \sum_i\langle\phi_i|U(\rho_S^{\text{in}}\otimes|\psi\rangle\!\langle\psi|)U^\dagger |\phi_i\rangle\\
		&= \sum_i \langle\phi_i| U|\psi\rangle \rho_S^{\text{in}}\langle\psi|U^\dagger |\phi_i\rangle\\
		&:= \sum_i K_i\rho_S^{\text{in}}K_i^\dagger,
	\end{split}
\end{equation}
This is illustrated graphically in Figure~\ref{fig:kraus-explanation}.
The last decomposition here is exactly the Kraus \acs{OSR}. It is a theorem of Kraus that a quantum map can be written in this form if and only if it is \acs{CP}. 

\subsubsection*{Superoperator Representation:}

\begin{figure}[htbp]
	\centering
	\includegraphics[width=0.9\linewidth]{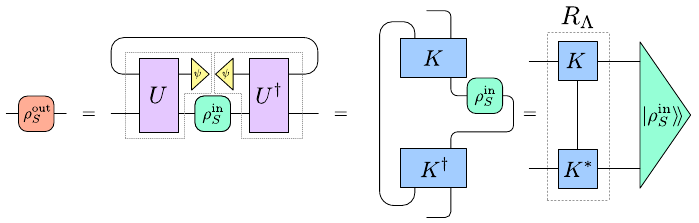}
	\caption[Graphical depiction of the superoperator representation for quantum maps]{Graphical depiction of the superoperator representation for quantum maps, obtained by vectorising the input state.}
	\label{fig:supop-explanation}
\end{figure}

Computations with quantum maps are most conveniently carried out in the superoperator representation. In this formalism, density matrices $\rho$ on a Hilbert space of dimension $d$ are represented as vectors $\kket{\rho}$ on a Hilbert-Schmidt space of dimension $d^2$. Quantum operations, then, are represented as $d^2\times d^2$ matrices which act on these Hilbert-Schmidt vectors. The natural inner product is the Hilbert-Schmidt inner product $\langle\!\langle A|B \rangle\!\rangle = \Tr[A^\dagger B]$. Its relationship to Stinespring dilation and the Kraus \acs{OSR} is depicted in Figure~\ref{fig:supop-explanation}.

Vectorisation is a reshaping transformation of matrices. One can convert an $n\times m$ matrix into a size $nm$ vector either by stacking the rows or the columns of the matrix into a single vector. Both operations are equivalent under a SWAP operation on the two spaces, and so it is merely a matter of consistency whichever one selects. In this thesis, we employ the row-vectorised convention, since it is naturally equivalent to the reshaping operation implemented in NumPy. A useful identity here is that left and right multiplication of an $r\times s$ matrix $X$ by $p\times r$ and $s\times q$ matrices $A$ and $B$ can be expressed as a single matrix-vector multiplication via the identity:
\begin{equation}
	\kket{A X B} = A\otimes B^{\ast} \kket{X}.
\end{equation}
Consequently, we can convert from the \acs{OSR} to superoperator form by vectorising 
\begin{equation}
	\begin{split}
		\rho_S^{\text{out}} &= \sum_i K_i\rho_S^{\text{in}} K_i^\dagger \\
		\implies |\rho_S^{\text{out}}\rangle\!\rangle &= \left(\sum_i K_i \otimes K_i^\ast \right) |\rho_S^{\text{in}}\rangle\!\rangle,
	\end{split}
\end{equation}
and hence the superoperator $R_{\Lambda}$ representation of $\Lambda$ is 
\begin{equation}
	R_{\Lambda} = \sum_i K_i \otimes K_i^\ast,
\end{equation}
where $(\cdot)^\ast$ indicates complex conjugation. It immediately follows that maps compose under this representation via matrix multiplication:
\begin{equation}
	R_{\Lambda_2\circ \Lambda_1} = R_{\Lambda_2}\cdot R_{\Lambda_1}.
\end{equation}
\noindent
This superoperator form is written with respect to the standard basis $\{\kket{i}\}$, but we will find it convenient to employ the Pauli basis to write maps in superoperator form. If we take the normalised $n$-qubit Pauli basis $\{\mathbb{I}/\sqrt{2}, X/\sqrt{2}, Y/\sqrt{2}, Z/\sqrt{2}\}^{\otimes n}$, then we can convert $R_{\Lambda}$ to the Pauli basis via
\begin{equation}
	(R_{\Lambda}^{\text{Pauli}}) = \sum_{ij} |i\rangle\!\rangle \!\langle\!\langle P_i|R_{\Lambda}|P_j\rangle\!\rangle\!\langle\!\langle j|.
\end{equation}
This leaves us with the straightforward interpretation that 
\begin{equation}
	(R_{\Lambda}^{\text{Pauli}})_{ij} = \Tr[P_j \Lambda(P_i)],
\end{equation}
the expectation value of the $j$th Pauli matrix with respect to the action of $\Lambda$ on the $i$th Pauli. This Pauli superoperator form is commonly termed the \ac{PTM} representation. It is commonly used in contrast to the standard basis because each matrix element is both real and readily interpretable as a mapping of Pauli eigenvectors.


\subsubsection*{Choi-Jamio\l kowski Isomorphism}
\label{ssec:CJI}

The final representation we introduce warrants special attention. It will be employed extensively throughout this thesis. 
The \ac{CJI} allows one to map results about states to processes and vice versa. 
Let $|\Phi^+\rangle = \sum_{i=1}^{d_S} |ii\rangle$ be the un-normalised maximally entangled state on the bipartite Hilbert space $\mathcal{H}_S \otimes \mathcal{H}_S$. Then the action of a quantum operation on one half of the state imprints all its information in the resulting state. Specifically, the Choi state $\hat{\Lambda}$ -- which we denote by a caret -- is given from the operation $\Lambda$ via
\begin{equation}
	\begin{split}
		\hat{\Lambda} &= (\Lambda \otimes \mathcal{I})[|\Phi^+\rangle\!\langle \Phi^+|] = \sum_{ij}(\Lambda \otimes \mathcal{I})[|ii\rangle\!\langle jj|]],\\
		&= \sum_{ij}\Lambda(|i\rangle\!\langle j|) \otimes |i\rangle\!\langle j|.
	\end{split}
\end{equation}
The first thing to note from the third equality is that the map is acting on a complete set of states in quantum parallelism, hence this correspondence uniquely determines the map\footnote{Indeed, the input state need not be a Bell state, it must simply be related to the Bell state by an invertible map. In other words, the reduced systems must have full Schmidt rank. Such input states which uniquely fix the mapping are known as \emph{faithful} states~\cite{PhysRevLett.91.047902}.}.
\begin{figure}[htbp]
	\centering
	\includegraphics[width=\linewidth]{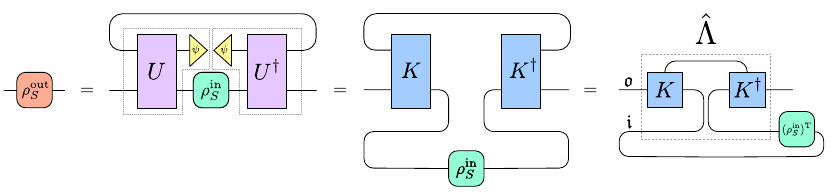}
	\caption[Graphical depicition of the Choi state representation of a quantum map]{Graphical depicition of the Choi state representation of a quantum map. This can be seen as an index reshuffling of the superoperator representation.}
	\label{fig:choi-explanation}
\end{figure}
We identify the left system of this bipartite space as being our `output' space, and the right as being the `input' for reasons which will become momentarily clear.
Whether the output space is on the left or right is entirely a matter of convention, in this thesis we will always use the association $\mathcal{H}_S\otimes \mathcal{H}_S\cong \mathcal{H}_{\text{out}}\otimes \mathcal{H}_{\text{in}}$, in accordance with the direction of matrix multiplication (inputs coming from the right and outputs travelling to the left). To shorten notation, we will always denote output spaces by the Gothic $\mathfrak{o}$ and input by $\mathfrak{i}$: $\mathcal{H}_{\mathfrak{o}}\otimes \mathcal{H}_{\mathfrak{i}}$.

The action of a map $\Lambda$ on input state $\rho_{\text{in}}$ is given by projecting the Choi state onto the transpose of the input state and tracing over the input space
\begin{equation}
	\label{eq:choi-act}
	\rho_S^{\text{out}} = \Tr_{\mathfrak{i}}[(\mathbb{I}_{\mathfrak{o}}\otimes \rho_{\mathfrak{i}}^\text{T})\cdot \hat{\Lambda}].
\end{equation}
We can see this by writing $\rho_{\text{in}}$ in the standard basis:
\begin{equation}
	\begin{split}
		\Tr_{\mathfrak{i}}[(\mathbb{I}_{\mathfrak{o}}\otimes \rho_{\mathfrak{i}}^\text{T})\cdot \hat{\Lambda}] &= \Tr_{\mathfrak{i}}\left[\sum_{ijkl} (\Lambda(|i\rangle\!\langle j|) \otimes |i\rangle\!\langle j|)\cdot(\mathbb{I}_{\mathfrak{o}}\otimes \rho_{kl}|l\rangle\!\langle k|)\right],\\
		&= \Tr_{\mathfrak{i}}\left[\sum_{ikl} \rho_{kl}(\Lambda(|i\rangle\!\langle l|) \otimes |i\rangle\!\langle k|)\right],\\
		&= \sum_{ml}\rho_{ml} \Lambda[|m\rangle\!\langle l|] = \rho_S^{\text{out}},
	\end{split}
\end{equation}
where we see that the last equality follows from the linearity of the map. The graphical depiction of the Choi state is shown in Figure~\ref{fig:choi-explanation}.


Composition of quantum maps is most easily performed in the superoperator representation, but is a valid transformation in all forms. For Choi states in particular, composition is defined by a \emph{link product} $\star$.

For $\hat{\Lambda}_1\in \mathcal{B}(\mathcal{H}_{AB})$ and $\hat{\Lambda}_2\in \mathcal{B}(\mathcal{H}_{BC})$,
\begin{equation}
	\hat{\Lambda}_2 \star \hat{\Lambda}_1 = \Tr_B[(\hat{\Lambda}_2\otimes \mathbb{I}_A) \cdot (\mathbb{I}_C\otimes\hat{\Lambda}_1^{\Gamma_B})]
\end{equation}
This is (i) a tensor product on the unshared spaces $A,C$, (ii) partial transpose of one map on the shared space $B$, and (iii) matrix multiplication over the shared space. The link product will be a convenient representation of the concatenation of higher order maps to represent multi-time processes. 

Finally, we note that there is a powerful association between the CJI and vectorisation, which can be seen by the observation that $|\Phi^+\rangle \equiv |\mathbb{I}\rangle\!\rangle$. Combining the Kraus \acs{OSR} with the definition of the Choi state, it follows that 
\begin{equation}
	\begin{split}
		\hat{\Lambda} &= \sum_j (K_j\otimes \mathcal{I}) |\Phi^{+}\rangle\!\langle \Phi^+|(K_j^\dagger \otimes\mathcal{I})\\
		&= \sum_j (K_j\otimes \mathcal{I}) |\mathbb{I}\rangle\!\rangle\!\langle\!\langle \mathbb{I}|(K_j^\dagger \otimes\mathcal{I})\\
		&= \sum_j |K_j\rangle\!\rangle \! \langle\!\langle K_j|.
	\end{split}
\end{equation}



\textbf{Physical Constraints}

Returning now to the physical constraints introduced above, we can evaluate these for a given quantum map under a given representation. By construction, each of these maps is linear. We can check the additional physical conditions via (i) \underline{\acs{CP}}: can be written in the Kraus \acs{OSR}, or the Choi state has strictly non-negative eigenvalues $\hat{\Lambda}\succcurlyeq 0$; (ii) \underline{\acs{TNI}}: Kraus operators $\sum_i K_i^\dagger K_i \preccurlyeq \mathbb{I}$, Choi state $\Tr_{\mathfrak{o}}[\hat{\Lambda}] \preccurlyeq \mathbb{I}_{\mathfrak{i}}$, or superoperator in the Pauli basis $(R_{\Lambda}^{\text{Pauli}})_{0i} = \delta_{0i}$, where equalities here designate \acs{TP} and inequalities \acs{TNI}. 
An additional property one might consider is whether the quantum map is unital. That is, whether the identity is a fixed point of the mapping. Operations satisfying $\Lambda[\mathbb{I}] \neq \mathbb{I}$ are called \emph{non-unital}, and physically correspond to the purification or unmixing of a state.



\textbf{Examples}

Maps represented by unitary operations $u$ form a subpart of the pure points on the space of operations. That is to say, they are closed system transformations of a state. In the Kraus \acs{OSR}, $u$ is the only Kraus operator: $\rho\mapsto u\rho u^\dagger$. In the \acs{PTM} form, unitality and trace-preservation of a unitary map ensures $R_{i0} = \delta_{i0}$, and $R_{0i} = \delta_{i0}$, whereas the submatrix given by $R_{ij}, \ 1\leq i,j\leq d^2$ is an orthogonal matrix.
The Choi state of a unitary operator is a maximally entangled pure state, as straightforwardly seen from the fact that a unitary rotation will preserve the pre-existing entanglement in the Bell state. The fact that unitaries are both rank one and \acs{TP} means that they are deterministic both in the classical and the quantum sense.

Certain noisy channels can be realised as probabilistic mixtures of unitary operations. The depolarising channel for example $\Lambda^{\text{Depol}}[\rho] :=  (1-p) \rho + p \mathbb{I}/d$ keeps a state with probability $1-p$, or sends it to the identity with probability $p$. Equivalently, this applies an identity operation with probability $1-\frac{3p}{4}$ or $X,Y,$ or $Z$ gate with equal probability $\frac{p}{3}$. A dephasing channel retains populations, but destroys coherence: $\Lambda^{\text{Dephase}}[\rho] := (1-p)\rho + p\sum_i \Pi_i \rho \Pi_i$ for \acs{PVM} ${\Pi_i}$. This is equivalent to applying identity gate with probability $(1-\frac{p}{2})$ or $Z$ gate with probability $\frac{p}{2}$. A famous example of a noisy process which does \emph{not} lie in the span of unitaries is the amplitude damping channel $\Lambda^{\text{AD}}[\rho] := K_0\rho K_0^\dagger + K_1\rho K_1^\dagger$, where $K_0 = |0\rangle\!\langle 0| + \sqrt{1-p}|1\rangle\!\langle 1|$ and $K_1 = \sqrt{p}|0\rangle\!\langle 1|$. This models thermal relaxation of a system to its environment. The PTM and Choi representations of these channels are given below, for concreteness.
\begin{enumerate}
	\item Depolarising Channel 
	\begin{equation}
			R_{\Lambda^{\text{Depol}}}  = \begin{pmatrix}
				1 & 0 & 0 & 0 \\
				0 & 1-p & 0 & 0 \\
				0 & 0 & 1-p & 0 \\
				0 & 0 & 0 & 1-p
			\end{pmatrix}
			\qquad 
			\hat{\Lambda}^{\text{Depol}} = \begin{pmatrix}
				1 - \frac{p}{2} & 0 & 0 & 1-p\\
				0 & \frac{p}{2} & 0 & 0 \\
				0 & 0 & \frac{p}{2} & 0\\
				1-p & 0 & 0 & 1 - \frac{p}{2}
			\end{pmatrix},
	\end{equation}
	\item Dephasing Channel 
	\begin{equation}
		R_{\Lambda^{\text{Dephase}}}  = \begin{pmatrix}
			1 & 0 & 0 & 0 \\
			0 & 1-p & 0 & 0 \\
			0 & 0 & 1-p & 0 \\
			0 & 0 & 0 & 1
		\end{pmatrix}
		\qquad 
		\hat{\Lambda}^{\text{Dephase}} = \begin{pmatrix}
			1 & 0 & 0 & 1-p\\
			0 & 0 & 0 & 0 \\
			0 & 0 & 0 & 0\\
			1-p & 0 & 0 & 1
		\end{pmatrix},
	\end{equation}
	\item Amplitude Damping 
	\begin{equation}
		R_{\Lambda^{\text{AD}}}  = \begin{pmatrix}
			1 & 0 & 0 & 0 \\
			0 & \sqrt{1-p} & 0 & 0 \\
			0 & 0 & \sqrt{1-p} & 0 \\
			p & 0 & 0 & 1-p
		\end{pmatrix}
		\qquad 
		\hat{\Lambda}^{\text{AD}} = \begin{pmatrix}
			1 & 0 & 0 & \sqrt{1-p}\\
			0 & p & 0 & 0 \\
			0 & 0 & 0 & 0\\
			\sqrt{1-p} & 0 & 0 & 1-p
		\end{pmatrix}.
	\end{equation}
\end{enumerate}

We lastly introduce measure-and-prepare operations. These are formally quantum instruments whose action is to measure a quantum state with some \acs{POVM} $\{E_i\}$, record the outcome, and then re-prepare a new state $\sigma$: $\Lambda^{\text{MP}}_i := \Tr[E_i\rho]\sigma$. 
Quantum instruments whose Choi states are product states, and whose corresponding \acs{CPTP} marginal are also product states are termed \emph{causal breaks}. This is because these impose a past-future independence, that is, there exists no correlations that may persist across $\mathcal{H}_{\mathfrak{o}}\otimes\mathcal{H}_{\mathfrak{i}}$. The corresponding Choi states in this instance are $\hat{\Lambda}_i^{\text{MP}} = \sigma \otimes E_i^{\text{T}}$.

\section{Quantum Characterisation, Verification, and Validation}

Quantum theory is not just about formalism. Every idea expressed, every mathematical principle must be testable and reconstructable up to some arbitrarily small statistical error. This translates into an experimental setting through Born's rule. Starting from a mathematical model, Born's rule allows one to make predictions about reality which may then be tested. From a model, a probability distribution can be constructed. The frequencies of this probability distribution can be measured in practice and compared to the prediction, and model accordingly updated until it matches experimental data as well as possible. 
Tomography in essence is an operational approach, it takes emergent dynamics from a system and aims to reconstruct those dynamics based on measurement outcomes. 
The act of performing quantum tomography is the act of taking measurements that in principle ought to reconstruct a certain operator, and then estimating that operator. There are many factors that may obfuscate that translation as we will discuss. The simplest derailment here is that a set of frequencies will never truly reproduce its underlying probability distribution, and so the operator that one obtains from those frequencies may not be correct or even physical. There is immense freedom in taking experimental data and developing a physical model that can reasonably reproduce or explain that data. \ac{QCVV} is the broad umbrella of procedures used to reconstruct, scrutinise, and benchmark properties of quantum computers.

In order for quantum devices to continue to improve in quality, robust benchmarking techniques that characterise their behaviour are essential. Once we understand how a device behaves, we can isolate out issues and determine the best manner to go about mitigating error. In this section, we detail the relevant characterisation techniques which have been previously developed, and have been implemented in this project. Broadly speaking, they aim to estimate to varying degrees of abstraction on a real device a quantum state, a quantum gate, or a generalised quantum process. For each technique in this section we are required to run a certain number of experiments. By \emph{experiments}, we mean the particular circuit which has been set up to estimate a physical probability. In order to reduce the statistical uncertainty in each estimate, we allocate a certain number of \emph{shots} to each experiment -- that is, repetitions of the same experiment. Typically, complete benchmarking experiments scale exponentially in the size of the system, but there are many approaches for which one can make this scaling more favourable. For example, randomised benchmarking style approaches aim to estimate only a single property of a quantum gate set, which is the average process fidelity~\cite{PhysRevA.77.012307,Wan2021ARQ,DRB,PhysRevLett.119.130502,PhysRevLett.109.240504}. Alternatively, one may adopt particular ans\"atze about the structure of a particular state or process that allows it to be represented sparsely~\cite{cramer2010efficient,PhysRevLett.111.020401,Lanyon2017,Baumgratz2013}.

\subsection{State and Process Tomography}
Two standard assets of the \acs{QCVV} toolkit are \ac{QST} and \ac{QPT}. Each designates a series of experiments to run on a \acs{QIP} as well as some post-processing with the goal to respectively reconstruct a density matrix, or a \acs{CPTP} map. We will briefly overview these two procedures. Fundamentally, each approach is the same: take a set of frequencies from experiment, and construct a model to explain those frequencies. We will begin by outlining the in-principle reconstruction of density matrices and quantum maps, and then overview some practical considerations.

We start with the general reconstruction of some state $\rho\in \mathcal{B}(\mathcal{H})$ of dimension $d$. Estimating this state requires a series of measurements in an \ac{IC} basis. A minimally \acs{IC} basis is a \acs{POVM} $\mathcal{J} = \{E_m\}$ with $d^2$ effects whose elements span the underlying Hilbert-Schmidt space. A general method to test informational completeness is to create a matrix $J$ with each row given by an effect
\begin{equation}
	\begin{pmatrix}
		\langle\!\langle E_1| \\
		\langle\!\langle E_2| \\
		\vdots \\
		\langle\!\langle E_m|\\
	\end{pmatrix},
\end{equation}
and checking that this matrix is full rank. Measuring $\rho$ in basis $\mathcal{J}$ generates series of probabilities (omitting sampling error for the moment):
\begin{equation}
	\label{eq:born-tom}
	p_i = \Tr[E_i \rho],
\end{equation}
each of which acts as a linear constraint on the state. From $\mathcal{J}$, one can define a dual set $\{\Delta_j\}$ satisfying $\Tr[E_i\Delta_j] = \delta_{ij}$. It then follows that $\rho$ can be written from the above information
\begin{equation}
	\rho = \sum_{i=1}^{d^2} p_i \Delta_i. 
\end{equation}
This can be easily checked by noting that a \acs{POVM} effect $\Pi_x$ outside $\mathcal{J}$ can be written as a linear combination $\sum_{i=1}^{d^2}\alpha_i E_i$ (by the \acs{IC} property of $\mathcal{J}$). Hence, 
\begin{equation}
	p_x = \Tr[\Pi_x\rho] = \Tr\left[\sum_i \alpha_i E_i \sum_j p_j\Delta_j
	\right] = \sum_{i}\alpha_i p_i. 
\end{equation}

This can be alternately and equivalently expressed by vectorising $\rho$ and writing everything as a matrix-vector equation. Collecting each outcome from Equation~\eqref{eq:born-tom} into a vector $\vec{b}$, we can express the problem as 
\begin{equation}
	J |\rho\rangle\!\rangle = \vec{b}.
\end{equation}
Since matrix $J$ is full rank, the solution to this is 
\begin{equation}
	|\rho\rangle\!\rangle = J^{-1}|b\rangle\!\rangle.
\end{equation}

\acs{QPT} can be seen in many ways as an extension to \acs{QST}. We can clearly see this via the \acs{CJI}: given that processes can be represented as states, then the act of doing tomography on a quantum process is mathematically equivalent to the act of doing tomography on a quantum state. The difference lies in both the experimental setup, and the physical constraints. \acs{QST} is reconstructing a single $\rho\in\mathcal{B}(\mathcal{H})$ -- the output of a single state at a single time. In \acs{QPT}, however, we would like to reconstruct some $\hat{\Lambda} \in \mathcal{B}(\mathcal{H}_{\mathfrak{i}}) \otimes \mathcal{B}(\mathcal{H}_{\mathfrak{o}})$, for which we now also need an \acs{IC} basis on $\mathcal{B}(\mathcal{H}_{\mathfrak{i}})$. This input to the map can simply be the preparation of any $d^2$ linearly independent states $\mathcal{P} =\{\rho_j^{\mathfrak{i}}\}$. To reconstruct the map, then, one simply needs to feed each $\rho^{\mathfrak{i}}\in \mathcal{P}$ into $\Lambda$, and then measure in $\mathcal{J}$ to estimate the corresponding output set $\{\rho_i^{\mathfrak{o}}\} \equiv \{\Lambda(\rho_i^{\mathfrak{i}})\}$. Suppose, for $\mathcal{P}$ we define another dual set $\{\omega_j\}$ such that $\Tr[\rho_i\omega_j] = \delta_{ij}$, then the map becomes
\begin{equation}
	\hat{\Lambda} = \sum_{i} \rho_i^{\mathfrak{o}}\otimes \omega_i^{\text{T}} = \sum_{ij} p_j^{\mathfrak{o}}\Delta_j\otimes \omega_i^{\text{T}},
\end{equation}
which is consistent with Equation~\ref{eq:choi-act} by design.

Once more, we can express this problem as a matrix-vector equation for \acs{IC} inputs $\{\rho_i^{\mathfrak{i}}\}$, \acs{IC} measurements $\{E_i\}$, and observed probabilities $p_{ij}$:
\begin{equation}
	\label{eq:QPT-LI}
	\begin{pmatrix}
		\langle\!\langle E_1|\otimes \langle\!\langle \rho_1^{\mathfrak{i}}|\\
		\langle\!\langle E_1|\otimes \langle\!\langle \rho_2^{\mathfrak{i}}|\\
		\vdots \\ 
		\langle\!\langle E_1|\otimes \langle\!\langle \rho_{d^2}^{\mathfrak{i}}|\\
		\vdots \\
		\langle\!\langle E_{d^2}|\otimes \langle\!\langle \rho_{d^2}^{\mathfrak{i}}|\\
	\end{pmatrix} |\hat{\Lambda}\rangle\!\rangle = \begin{pmatrix}p_{11}\\ p_{12}\\ \vdots \\ p_{1d^2} \\ \vdots \\ p_{d^2d^2}\end{pmatrix}
\end{equation}
for which the feature matrix is inverted to determine $|\hat{\Lambda}\rangle\!\rangle$, and hence $\Lambda$. This method of reconstruction is called a \ac{LI} estimate. 

\textbf{Post-Processing Methods}

Quantum tomography, like any reconstructive method, suffers from the fact that frequencies are only estimates of probabilities, not probabilities themselves. Suppose we draw $N$ samples from a probability distribution to obtain a series of outcomes $\{x_1, x_2, \cdots, x_n\}$. The total number of draws for each outcome defines a set of frequencies 
\begin{equation}
	n_x = N_x / N,
\end{equation}
which can at best estimate $p_x$ up to an error of $\mathcal{O}(1/\sqrt{N})$. As a result of this, the set of equations given in Equation~\eqref{eq:QPT-LI} need not be consistent with one another, and further, linear inversion of Equation~\eqref{eq:QPT-LI} need not produce a physical model for $\Lambda$: it could estimate a trace-increasing map with negative eigenvalues. 

One often seeks, instead, to find a physically reasonable estimate which is \emph{plausible} from the data. \ac{MLE} is a commonly used tool, both for its computational convenience and asymptotic limit statements. The procedure here is to find whichever quantum object is both physical and maximises a likelihood function. For example, with \acs{QST}, evaluated at some point $\rho$ across \acs{POVM} $\{E_x\}$, this is 
\begin{equation}
	\mathcal{L}(\rho) = \prod_{x}\Tr[E_x \rho]^{n_x}.
\end{equation}

There are, however, some known problems with \acs{MLE}. 
\acs{MLE} aims to find the quantum state which is most consistent with the data. In the case that linear inversion is physical, then, the estimate coincides with the \acs{MLE} estimate~\cite{scholten2018behavior}. In other words, this is a frequentist perspective of estimation; i.e., frequencies are treated as probabilities. This can be a dangerous practice, because it treats unobserved outcomes as zero-probability events. In the case where the \acs{LI} estimate is not physical, the \acs{MLE} estimate will always be rank deficient~\cite{RBK2010}. Ideally, one should never be certain that an unobserved event will never occur. This should be a statement qualified by the total number of experiments run. 
Nevertheless, the risk of using \acs{MLE} is small for relatively large numbers of experiments, and the computational cost of chasing more optimal procedures is large. Thus, \acs{MLE} remains the de-facto standard for performing quantum tomography of all genres, but some progress has been made in pursuing more Bayesian approaches~\cite{PhysRevApplied.17.024068,RBK2010,blume2010hedged,PhysRevLett.116.090407,Granade_2016,PhysRevA.85.052120}. Experimentally, \acs{QPT} has been performed on all platforms, but the prohibitive scaling has limited instances to only 2-3 qubits~\cite{doi:10.1063/1.1785151,govia2020bootstrapping,PhysRevA.64.012314,PhysRevLett.93.080502,mkadzik2022precision,Tinkey_2021,bialczak2010quantum,PhysRevB.82.184515}.

\textbf{Assumptions}

All models have faults, and neither \acs{QST} nor \acs{QPT} are an exception. In their standard forms, both procedures establish a set of relationships between linear operators and experimentally observed outcomes. But this takes for granted perfect knowledge of the operators themselves. In \acs{QPT} (and problems are similarly faced in \acs{QST}), reconstruction of the process assumes a particular matrix form for the \ac{SPAM}, $\{\rho_i^{\mathfrak{i}}\}$ and $\{E_i\}$ which may not be physically reasonable.
When \acs{SPAM} errors are not negligible, \acs{QPT} produces process estimates considerably far away from the true maps \cite{intro-GST}. This is particularly an issue since the primary source of error in current quantum computers are \acs{SPAM} errors. We discuss a solution to wrongly-chosen frames in the following section. 

Another trouble with the technique of \acs{QPT} is more fundamental to the formalism itself. In the early 1990s, when experimental setups caught up to the theory developed of operational dynamics, they began reconstructing maps that were not completely positive \cite{DAriano2010} -- and it was unclear why. Recall that Stinespring's dilation theorem showed that any \acs{CPTP} map can be represented as the contraction of unitary dynamics on a larger space. Pechukas explained these non-\acs{CP} results by showing that this was only true in the case where the system and its environment were initially uncorrelated -- a restrictive assumption in practice \cite{Pechukas1994, Alicki1995, Pechukas1995, Modi2012a}. Moreover, he showed that a map whose argument is the system state $s$ is both \acs{CP} \emph{and} linear if and only if there are no initial system-environment correlations. 
Even if this is not the case for the initial state, we see how this is problematic for our general picture of operational dynamics. If we start with some initially uncorrelated state, then it will generally evolve to some entangled state with the environment at a later time. We can then no longer describe its further dynamics with \acs{CPTP} linear maps. More concisely: dynamics with a system-environment interaction cannot be \acs{CP}-divided without giving up linearity. We will return to an operational resolution of this issue in Section~\ref{process-tensor}.



\subsection{Gate Set Tomography}

The resolution to the first \acs{QPT} issue, just introduced, is addressed by \ac{GST}. \acs{GST} aims to be a self-consistent characterisation method that relies on no \emph{a priori} knowledge of a quantum device or its operations. 
This section provides a brief overview of its methodology -- for a comprehensive guide to the techniques involved, see \cite{intro-GST,RBK2017}. Here, we operate in the \acs{PTM} representation of quantum channels. 
The only experimentally accessible quantities in a laboratory are measurements of a quantum state. For example, after a sequence of quantum operations $G$, many measurements are taken in order to form an estimate of 
\begin{equation}
	\langle\!\langle E|G|\rho\rangle\!\rangle
\end{equation}
for some preparation Hilbert-Schmidt vector $|\rho\rangle\!\rangle$ and some measurement effect $\langle\!\langle E|$. The aim here is to estimate a set of quantum objects from these observed frequencies.
Let us first define a \emph{gate set} $\mathcal{G} = \{|\rho\rangle\!\rangle , \langle\!\langle E_0|,\langle\!\langle E_1| , G_1, G_2,\cdots ,G_K\}$, which includes a chosen set of control operations, as well as the ability to initialise and read out the state. Here, $|\rho\rangle\!\rangle$ is the initial fiducial state, and $\langle\!\langle E_i|$ is an effect of a \acs{POVM}, considered to be two-outcome for simplicity. We will drop the subscript without loss of generality. From this set, one can define a \emph{fiducial} set $\mathcal{F} = \{F_1,\cdots,F_N\}$, where each $F_i$ is a composition of gates from $\mathcal{G}$. The purpose of the fiducial set is to generate an informationally complete set of preparations $|\rho_j\rangle\!\rangle$ and measurements $\langle\!\langle E|F_j$. The \acs{IC} requirement ensures that $|\mathcal{F}|\geq d_S^2$ for fixed $|\rho\rangle\!\rangle$.
The principle here will be to generate a set of probabilities from $\mathcal{G}$ that should fix each observable parameter in the gate set. These are the expectation values 
\begin{equation}
	p_{ijk} = \langle\!\langle E| F_iG_kF_j|\rho\rangle\!\rangle.
\end{equation}
Absorbing $\kket{\rho}$ and $\bbra{E}$ respectively into $F_j$ and $F_i$, we can define new matrices $A$ and $B$ such that $A_{ir} = \langle\!\langle E|F_i|r\rangle\!\rangle$ and $B_{sj} = \langle\!\langle s|F_j|\rho\rangle\!\rangle$. This implies that $p_{ijk} = (AG_kB)_{ij}$, or that the measurement of these probabilities corresponds to estimating the components of the matrix $\tilde{G}_k:=AG_kB$. In the special case where $G_1$ is chosen to be the null gate (do nothing for no time), we fix 
\begin{equation}
	g = \tilde{G}_1 = AB,
\end{equation}
which is the Gram matrix $\langle\!\langle F_i| F_j\rangle\!\rangle$. For each other $G_k$, if one left-multiplies by the inverse of $g$, one obtains 
\begin{equation}
	\label{eq:gst-LI}
	g^{-1}\tilde{G}_k = B^{-1}A^{-1}AG_k B = B^{-1}G_kB.
\end{equation}
Hence,
\begin{equation}
	\bar{G}_k = g^{-1} \tilde{G}_k
\end{equation}
is an estimate of the entire gate set up to a similarity transformation by matrix $B$. Note that if we take 
\begin{equation}
	\begin{split}
		&|\rho\rangle\!\rangle \mapsto B'|\rho\rangle\!\rangle \\
		& G_k \mapsto B'^{-1}G_k B'\\
		&\langle\!\langle E |\mapsto \langle\!\langle E | B,
	\end{split}
\end{equation}
then the transformed probabilities 
\begin{equation}
	p'_{ijk} = \langle\!\langle E|B B^{-1} F_i B B^{-1} G_kB B^{-1}F_jB B^{-1}|\rho\rangle\!\rangle = \langle\!\langle E| F_iG_kF_j|\rho\rangle\!\rangle,
\end{equation}
are identical to the $p_{ijk}$. Hence, a gateset can only be defined up to an unobservable gauge $B$. In other words, the linear inversion estimate in Equation~\eqref{eq:gst-LI} cannot be uniquely fixed, even in principle. In practice, once an estimate is obtained up to a gauge freedom, the matrix $B$ is then varied to optimise for finding the gauge that takes an estimated gate set as close as possible to a target gate set.

\textbf{Heisenberg Scaling}

Self-consistency, as discussed, is the first paradigmatic feature of \acs{GST}.
The second is that it can be carefully designed to amplify gate errors for more accurate estimation. On the road to fault-tolerant quantum computing, this is important. Quantum error correcting threshold metrics are typically designated by diamond distance -- the worst case error rate -- whereby coherent errors of order $\eta$ make $\mathcal{O}(\eta)$ contributions to the measure. In contrast, coherent gate errors can be suppressed by randomised benchmarking: their contribution to the reported error rate is $\mathcal{O}(\eta^2)$, dominated instead mostly by stochastic noise~\cite{sanders2016gatefid}. Randomised benchmarking cannot tell the whole story about device quality.


To amplify coherent errors, a set of gate compositions $\mathcal{G} = \{g_0,g_1,...,g_n\}$ is generated. The elements of this set are termed \emph{germs}, and each comprises a sequence of operations from the gateset. Gates contain many free parameters whose visibility depends on the input state, sequence of operations, and basis in which a measurement is made. Germs are chosen from an extensive search such that a possible error in each gate parameter may be amplified and made detectable in at least one germ. To overcome statistical error, each germ is repeated $L$ times for many different values of $L$. When every possible noisy parameter is made detectable, the germs set is termed \emph{amplificationally complete}~\cite{RBK2017}. For all values of $i,j,k,$ and $L$, the experimental data are then collected as
\begin{equation}
	\label{eq:GST-probs}
	p_{ijk}^L = \langle\!\langle E| F_i g_j^L F_k|\rho\rangle\!\rangle.
\end{equation}
Heisenberg-type scaling is estimation with sampling precision that goes down like $1/N$, rather than the usual $1/\sqrt{N}$. This is central to a \acs{GST} extension, called long-sequence \acs{GST}. The basic premise is as follows. Suppose a gate $G$ coherently over-rotates by some angle $\theta$. In a single application, this term shows up linearly, and estimation is shot-noise-limited, $\epsilon = \mathcal{O}(1/\sqrt{N})$. For precisions of even $10^{-3}$, this becomes unwieldy. Instead, however, we may consider characterisation of $G^L$, across some integer $L$ number of repetitions. 
Now, the over-rotation may be by up to $L\theta$, measured to within $\epsilon$, so $\theta$ is measured to within a precision of $\epsilon/L$. Here, $L$ is usually chosen to take a range of values so as not to understate the case where $\theta \approx 2\pi/L_0$ for a single $L_0$.
Repeating a gate, however, does not necessarily compound its error. 
Tilt errors, for example, can cancel themselves out after a few repeated applications. A way to address this is to compose gates together in different configurations such that the different parameters are structurally amplified.
The collection of experiments designated by \acs{GST} guarantees not only that all measurable parameters are exposed, but that all gate parameters are amplified for precise estimation. The end result is that if $N'$ total experiments are conducted, the precision of the \acs{GST} estimate of the repeated gates in the gate set scales as $1/N'$, rather than $1/\sqrt{N'}$. The total number of experiments is highly dependent on the gate set, including the ability to perform circuit reduction tricks and combine germ parameters. Roughly, it requires $d_S^4$ experiments per gate, per germ, and per germ repetition. For example, a typical single-qubit \acs{GST} circuit might use 6 fiducial circuits, 11 germs, and up to 13 powers. This makes $6\times 6 \times 11 \times 13 = 5148$ different measurement settings~\cite{nielsen-gst}.

\subsection{Breakdown of Markovianity and Miscellaneous Characterisation Techniques}


Frameworks for dynamics in the literature can be broadly classed under dynamical map or master equation formalisms. The former captures only two-time correlations. However, in the presence of strong system-environment correlations, multi-time effects may emerge, as we shall see in the proceeding chapter. This type of dynamics is known as non-Markovian, and is the primary subject of this thesis. Crucially, in the presence of non-Markovian noise, dynamical maps will fail to describe multi-time processes when composed together~\cite{rivas-NM-review,PhysRevA.83.052128}. This applies clearly to \acs{QPT}, but also to any \acs{QCVV} procedures born out of the umbrella of quantum channels, such as \acs{GST}, randomised benchmarking, and Hamiltonian tomography~\cite{nielsen-gst, PhysRevA.77.012307,PhysRevLett.113.080401,Wang_2015,Eisert2020}. The latter generally is a function of at most three-time correlations and be can be reduced to a family of dynamical maps~\cite{pollock-tomographic-equations}.

All characterisation techniques assume an underlying model to be fit. Here, we will see a flavour of how, in particular, attempts to characterise dynamics with strong system-environment coupling can break down.
Stinespring's dilation theorem guarantees that for any \acs{CP} quantum map, there exists a dilation to unitary dynamics on an enlarged space. However, the converse need not follow. That is, reduced dynamics of some larger space need not be \acs{CP} at the system level. Specifically, Pechukas showed that if there exists initial correlations between the system and the environment, then subsequent unitary dynamics can give rise to not-completely-positive maps at the system level.
This is known as the initial correlation problem, and is our first sign that this operational framework will be insufficient to deal with non-Markovian processes in general~\cite{Modi2012,Modi2012a,laine2011witness}. At time $t_0$, even if the $SE$ state factors: $\rho^{SE}_0 = \rho^S_0\otimes \rho^E_0$, if the unitary dynamics are sufficiently strong, they will lead to a generically correlated state at some later time $t_1$. Consequently, the dynamics cannot be described from $t_1$ onwards. This property is known as the breakdown in \acs{CP} divisibility and is a common proxy for non-Markovian dynamics~\cite{rivas-NM-review,PhysRevA.83.052128}. We emphasise that it is a sufficient, but not necessary, condition. We will explore these ideas (and their resolution) further in Chapter~\ref{chap:stoc-processes}.

In the case where these approaches are insufficient to characterise features of non-Markovian noise, ad-hoc extensions techniques exist to detect a departure from the Markov assumptions. These detect some parts of the non-Markovian character but are not generalisable or predictive. Their application is typically as witnesses or for shallow diagnostics, but cannot rigorously measure the memory or be employed systematically to control the system. Common examples include memory kernels for master equations~\cite{PhysRevA.93.052111}, or statistical tests to establish causal connections between environmental factors or gate choices, and system-level dynamics~\cite{PhysRevX.9.021045, veitia2020macroscopic, veitia2018testing, helsen2019spectral, Sarovar2020detectingcrosstalk}. Statistical tests have also been employed to quantify the confidence with which breakdown of Markovianity can be described~\cite{nielsen-gst}. GST, for example, comes equipped with a way to quantify model violation, which, if high, indicates that other techniques need to be employed~\cite{White-NM-2020,White-MLPT,Rudinger-context-dependent,Sarovar2020detectingcrosstalk,proctor2020detecting}. 
This hypothesis tests whether a fail to fit experimental data can be explained purely by finite sampling, or a more general breakdown of the model.
For large model violation, it may no longer be possible to operationally characterise gates on a system, motivating the need to move beyond this.
Similar statistical tests may be applied to flag the effects of non-Markovianity, none of which are completely systematic. 
Conventional measures of non-Markovianity are well-motivated, but typically only describe a subset of non-Markovian processes. That is to say, they are sufficient but not necessary measures~\cite{PhysRevLett.101.150402, PhysRevLett.103.210401,PhysRevLett.105.050403, PhysRevA.83.052128, vacchini_non-markovian_2013, rivas-NM-review, breuer2016, deVega2017, Li2018}. 

In this section, we have only touched on the vast literature of benchmarking quantum devices. 
In the examination of open quantum dynamics, we can split estimation of processes into a series of desired properties. Consider a dynamical map. We wish to (a) estimate it as per \acs{QPT}; (b) post-process so that the estimate is physical and plausible; (c) have the estimate be self-consistent; (d) be able to amplify parameters; (e) capture non-Markovian dynamics across multiple times; and (f) for the procedure to be scalable. 
\acs{QPT} combined with \acs{MLE} for example, takes care of (a) and (b). \acs{GST} is aimed at (a), (b), (c), and (d), but neither is valid for non-Markovian processes. 
We will return to various parts of this zoo of characterisation methods at relevant points throughout. In particular, we emphasise one glaring gap, which is the inability at present to systematically characterise non-Markovian quantum dynamics. It is the raison d'\^etre of this thesis to rectify this gap. In Chapter~\ref{chap:PTT} we combine (a), (b), and (e) to formalise a non-Markovian quantum process tomography; Chapter~\ref{chap:efficient-characterisation} incorporates (f) to model processes sparsely; and Chapter~\ref{chap:universal-noise} closes the gap, adding in (c) to make the procedure wholly self-consistent.

\chapter{Quantum stochastic processes}
\label{chap:stoc-processes}
\epigraph{\emph{From where we stand the rain seems random. If we could stand somewhere else, we would see the order in it.}}{Tony Hillerman, Coyote Waits}
\noindent\colorbox{olive!10}{%
	\begin{minipage}{0.9555\textwidth} 
		\textcolor{Maroon}{\textbf{Chapter Summary}}\newline
		Stochastic processes -- such as the weather, the stock market, or at the casino -- plague the certainty of everyday life, and represent a series of events proceeding non-deterministically over time.
		At the quantum level in particular, systems are very sensitive to interactions with nearby inaccessible or complex features. If the environment carries forward information from the system, this can lead to temporally correlated -- or \emph{non-Markovian} -- dynamics on the system level. 
		This chapter aims to provide a detailed and pedagogical introduction to the mathematical foundations of quantum stochastic processes, specifically addressing non-Markovian dynamics. Despite the challenges posed by the non-commutativity of quantum observables and the invasive nature of quantum measurement, recent developments in the theory of quantum stochastic processes have made significant progress in addressing these issues. Specifically, we introduce the process tensor framework to mathematically describe such processes. As a foundation, the machinery introduced in this chapter will underpin the remainder of the thesis.
		\par\vspace{\fboxsep}
		\colorbox{cyan!10}{%
			\begin{minipage}{\dimexpr\textwidth-2\fboxsep}
				\textcolor{RoyalBlue}{\textbf{Further Reading}}\newline 
				As a comprehensive overview to different physical and operational facets of quantum non-Markovianity, Ref.~\cite{Li2018} constitutes a thorough exploration and analysis. In particular, this establishes a hierarchy to many of the sufficient-but-not-necessary components of non-Markovian quantum effects. Many conventional aspects of non-Markovianity from a master equation perspective can be found in Refs.~\cite{rivas-NM-review,de2017dynamics,breuer2016}.
				For a more specific tutorial on the motivated construction of quantum stochastic processes and multi-time statistics, see Ref.~\cite{Milz2021PRXQ}. 
		\end{minipage}}
\end{minipage}}
\clearpage
\section{Introduction}
In the previous chapter, we saw the insufficiency of the dynamical map formalism to deal with arbitrary multi-time processes. In particular, the inabilitity to describe correlated or non-Markovian dynamics even in principle. Classically, non-Markovian processes are extremely well-understood. The question of non-Markovianity is simply a statement as to whether obtaining a particular outcome at an earlier time for a non-deterministic dynamical system affects the later statistics for some random variable. The fundamental issue with generalising this notion to the quantum regime is the non-commutativity of observables in quantum systems. In plainer words: the very act of measuring a quantum system collapses its state, and so of course future statistics are affected. That control is invasive is central to the problem. 
The problem is further exacerbated when one attempts to define so-called Kolmogorov consistency in the quantum setting~\cite{Milz2020kolmogorovextension}. Classically, marginalising over nuisance variables is equivalent to choosing not to observe that system. Quantum mechanically, this is not the case. Measuring and forgetting destroys coherence of a system. 

Ad-hoc extensions to the Markovian regime, such as non-Markovian master equations and sufficient-but-not-necessary conditions for non-Markovianity are commonplace.
These commonly include the inability to divide dynamics into a series of \acs{CP} dynamical maps~\cite{wolf2008dividing}; checking whether a dynamical map can be generated by a Lindbladian~\cite{PhysRevLett.101.150402}; and witnessing information backthrough through quantum data-processing inequalities~\cite{laine2011witness}.
However, it was only very recently that an operational resolution to the problem was established. In this chapter, we shall review this resolution, known as the process tensor framework. Process tensors formally generalise the notion of a classical stochastic process to the quantum setting. We will hence first briefly describe classical stochastic processes. We will omit formal notions here such as $\sigma$-algebras and measurable sets to instead focus on some of the intuitive and key aspects of classical stochasticity. 
The motivation here is to understand important facets of stochastic processes without the added baggage of quantum mechanical properties. 
This will serve as a platform from which we may then consider quantum generalisations to these classical notions, as well as the extent to which process tensors are able to resolve many of these long-standing problems.

\section{Classical Stochastic Processes}
Stochastic processes are ubiquitous in nature. These represent a series of events on some system proceeding non-deterministically over time. At any given time, one can ask the question: what state is my system in? This can amount to measuring the ambient temperature outside, looking up the latest share prices, or checking the outcome of dice rolls. A stochastic process contains all of the information about sequences of measurement outcomes, as well as rule that assigns probabilities to each of those sequences.

The machinery required for this can be given in three parts: single-time statistical states $\mathbb{P}$, two-time transition matrices $\Gamma$, and multi-time stochastic processes $\mathbb{P}_{\mathbf{T}_k}$. 
A statistical state $\mathbb{P}_X$ is a vector of size $d$ for a $d$-outcome random variable $X$. Each element $i$ of the vector elicits the probability of the random variable having outcome $x_i$, which we denote by $(\mathbb{P}_X)_i = \mathbb{P}(X = x_i)$, or, for brevity $\mathbb{P}(x_i)$. 
A transition matrix or stochastic matrix is a mapping of probability vectors, indicative of transition probabilities. Suppose we sample from the random variable $X$, obtaining outcome $x_i$. We then sample from random variable $Y$. The probability to obtain outcome $y_j$ is given conditionally by $\mathbb{P}(y_j\mid x_i)$. Correspondingly, the \emph{total} probability to obtain $y_j$ is $\sum_{x_i}\mathbb{P}(y_j\mid x_i)\mathbb{P}(x_i)$. Expressed more compactly, we have the relationship 
\begin{equation}
	\mathbb{P}_Y = \Gamma_{Y:X} \mathbb{P}_X.
\end{equation}
$\Gamma_{Y:X}$ is the mechanism behind which one probability distribution transforms into another. We will typically employ this in the context of mapping a distribution at time $t_i$ to $t_{i+1}$.
Most generally, a stochastic process is a collection of random variables which are indexed by some set. That is, each element in the collection can be uniquely specified by the index set. Typically, this set has the interpretation as a set of ordered times $\mathbf{T}_k := \{t_0, t_1,\cdots,t_k\}$ on which the system is observed. 
\begin{equation}
	\{\mathbb{P}(X_k = x_k, X_{k-1} = x_{k-1},\cdots, X_0 = x_0)\},
\end{equation}
where sometimes the time indexing is made explicit via
\begin{equation}
	\mathbb{P}(x_k,t_k;x_{k-1},t_{k-1};\cdots;x_0,t_0).
\end{equation}
These joint probability distributions tell us everything there is to know about a set of dynamics, and can physically be interpreted as a probability distribution over trajectories. 


A key property of classical stochastic processes is the ability to validly map from a stochastic process on a set of times to that same underlying process defined across a smaller set of times. Specifically, let $\mathbf{T}_\ell = \{t_0,\cdots,t_{\ell-1}\}$. The stochastic process $\mathbb{P}_{\mathbf{T}_\ell}$ contains the statistical information for fewer times, in some subset $\mathbf{T}_k\subseteq \mathbf{T}_\ell$ by marginalising over outcomes 
\begin{equation}
	\mathbb{P}_{\mathbf{T}_k} = \sum_{\mathbf{T}_\ell \backslash \mathbf{T}_k}\mathbb{P}_{\mathbf{T}_\ell} =: \mathbb{P}_{\mathbf{T}_\ell}^{\mid \mathbf{T}_k},
\end{equation}
where $\mathbf{T}_\ell \backslash \mathbf{T}_k$ constitutes all events at times in $\mathbf{T}_\ell$ which are not found in $\mathbf{T}_\ell$, and the final equality a new notation for marginalisation. Classical stochastic processes whose probabilities are defined on a set of times can always be obtained from those on a superset of times by marginalisation. These are called \emph{Kolmogorov consistency conditions}, and, if these hold for all finite sets of time, a continuous-time underlying stochastic process is guaranteed by the following.

\begin{theorem}[\ac{KET}]
	Let $\mathbb{T}$ be a set of times. For each finite $\mathbb{T}_k$, let $\mathbb{P}_{\mathbf{T}_k}$ be a $k$-step joint probability distribution. There exists an underlying stochastic process $\mathbb{P}_\mathbf{T}$ that satisfies $\mathbb{P}_{\mathbf{T}_k}=\mathbb{P}^{\mid\mathbf{T}_k}$ for all finite $\mathbf{T}_k\subseteq\mathbf{T}$ if and only if $\mathbb{P}_{\mathbf{T}_k} = \mathbb{P}_{\mathbf{T}_\ell}^{\mid \mathbf{T}_k}$ for all $\mathbf{T}_k\subseteq \mathbf{T}_\ell\subseteq\mathbf{T}$.
\end{theorem}

For continuous (and hence infinite) $\mathbf{T}$, the process will not in general be experimentally accessible. But the \acs{KET} provides a connection between the underlying physical reality and what is experimentally practical. A breakdown in \acs{KET} is an indication that marginalising over a distribution -- that is, measuring and discarding the outcome -- is inequivalent to doing nothing. The fundamental assumption of the \acs{KET} is hence that interrogating a system does not, on average, disturb the state of that system. Herein lies the difficulty of defining a quantum stochastic process: measurements are necessarily disturbing.

Stochastic processes express a general conditional relationship between all variables at all different times, and hence grow exponentially complex in the generic case. But in many physically reasonable cases, the structure of the process can be much sparser than this, allowing the description to be greatly simplified. In the simplest case, we have only correlations persisting one time back into the past. These are known as \emph{Markovian} processes, defined as follows.

\begin{definition}[Markov process]
	A stochastic process $\mathbb{P}_{\mathbf{T}_k}$ is said to be \emph{Markovian} if it satisfies 
	\begin{equation}
		\label{eq:classical-markov}
		\mathbb{P}(x_j,t_j\mid x_{j-1},t_{j-1};\cdots;x_0,t_0) = \mathbb{P}(x_j,t_j\mid x_{j-1},t_{j-1})\ \forall \ t_j,
	\end{equation}
	otherwise, it is \emph{non-Markovian}.
\end{definition}
Markovian processes are often said to be \emph{memoryless}, and from the above definition it is clear why: the future statistics of the system do not depend on any previously observed outcomes of states of the system except at the most recent time. 

A non-Markovian process need not rely on its entire history. A stochastic process can exhibit so-called \emph{Markov order}. Formally, a process satifying 
\begin{equation}
	\mathbb{P}(x_j,t_j\mid x_{j-1},t_{j-1};\cdots;x_0,t_0) = \mathbb{P}(x_j,t_j\mid x_{j-1},t_{j-1};\cdots ; x_{j-\ell}, t_{j-\ell})\ \forall \ t_j,
\end{equation}
is said to have Markov order $\ell$. Processes such as these have a memory, but the memory size is fixed to only incorporate outcomes at the previous $\ell$ times.

An important related concept is that of master equations, which are integro-differential equations of motion for underlying stochastic processes. The general form here is 
\begin{equation}
	\frac{\text{d}}{\text{d}t} \mathbb{P}(X_t) = \int_{s}^t \mathcal{G}(t,\tau)\mathbb{P}(X_\tau)\text{d}\tau.
\end{equation}
This says that the time derivative of some probability distribution depends on the previous states of the system via some memory kernel $\mathcal{G}$, up to some time $s$, where $s$ can go back arbitrarily far. Master equations express the probabilities of some process as a function of time, but it is important to emphasise that they are not the same as the underlying stochastic process. Stochastic processes contain generic multi-time correlations, whereas a master equation contains information only about at most two-time correlations. 

Master equations with a time-local structure are said to be \emph{divisible}. Specifically, if a process can be described by a family of stochastic matrices which can be arbitrarily divided up as per
\begin{equation}
	\Gamma_{t:r} = \Gamma_{t:s}\Gamma_{s:r} \ \forall \ t>s>r,
\end{equation}
then it can be shown that the process has a time-local master equation, in the sense that the derivative of $\mathbb{P}$ depends only on the current time.

Divisibility is often conflated with Markovianity. It is tempting to think that since the memory kernel of a master equation is null, that this means there is no non-Markovian memory. However, since master equations only describe up to two-time correlations, this is instead a statement that there are no two-time correlations. It does not preclude the possibility of multi-time correlations in the process.

\begin{example}[Multi-time Correlations]
	\examplecontent{
	Suppose we have three sequential two-outcome events described by random variables $x_1, x_2$, and $x_3$. Let $x_1$ and $x_2$ independently have a 50\% probability of each outcome, and suppose $x_3$ is then given by $(x_1 + x_2)\mod\ 2$. There are no two-time correlations between any of the random variables, and hence the process is divisible, but all three are correlated; the only memory effects present are genuinely multi-time.}
\end{example}


In summary, we have:
\begin{itemize}
	\item \underline{Two-time correlated processes $\{\mathbb{P}_{\mathbf{T}_2}\}$}: stochastic processes describable under a two-time setting. This includes Markov processes -- which have two time correlations between only consecutive times -- as well as more general master equation processes -- which may include two-time correlations between any pairs.
	\item \underline{Markov order processes $\{\mathbb{P}_{\mathbf{T}_\ell}\}$}: for which $\ell$ is the smallest number such that $\{\mathbb{P}_{\mathbf{T}_2}\}$: for which $\mathbb{P}(X_k\mid X_{k-1},X_{k-2},\cdots,X_0) = \mathbb{P}(X_k\mid X_{k-1},X_{k-2},\cdots,X_{k-\ell})$.
	\item \underline{General non-Markovian processes $\{\mathbb{P}_{\mathbf{T}_2}\}$}: for which 
	\begin{equation*}
		\mathbb{P}(X_k\mid X_{k-1},X_{k-2},\cdots,X_0)\neq \mathbb{P}(X_k\mid X_{k-1}',X_{k-2}',\cdots,X_0').
	\end{equation*}
	\item \underline{Underlying stochastic process $\mathbb{P}_\mathbf{T}$}: whose existence is guaranteed by the \acs{KET} on $\mathbb{P}_{\mathbf{T}_k}$.
\end{itemize}

\section{Process Tensors as Quantum Stochastic Processes}\label{process-tensor}

We have already seen several of the statistical features required to elevate quantum theory to a theory of stochastic processes. Namely, the density matrix permits us some level of randomness in the state of our system so that it can be described as mixed, and quantum channels constitute mappings of these system states. Respectively, these generalise the notion of a probability vector and a stochastic transition matrix to the quantum setting. We have already seen how a wide range of dynamics can be adequately described with the dynamical map formalism. But Markovianity is a statement about correlations between all times, and dynamical maps can only ever provide two-time correlations. Similarly, the most general quantum master equation 
\begin{equation}
	\frac{\text{d}}{\text{d}t}\rho(t) = \overset{\text{Hamiltonian Dynamics}}{\overbrace{-i[H,\rho(t)]}} + \overset{\text{Dissipator}}{\overbrace{\mathcal{D}[\rho(t)]}}+\overset{\text{Memory}}{\overbrace{\int_s^t \mathcal{K}(t,\tau)[\rho(\tau)]\text{d}\tau}},
\end{equation}
describe the dynamics of a system as it is idle. But it fails to capture multi-time correlations and so cannot be a quantum generalisation of a classical stochastic process.

It is not \emph{a priori} clear why we are interested in moving to the multi-time setting when it comes to quantum systems. For many decades, master equations were the tool of choice for studying any open dynamics. 
Motivation for the multi-time setting has been substantially driven by experimental efforts to develop fault-tolerant quantum technologies. The digitisation of quantum evolution into a series of discrete steps represented by gates fits the natural structure of multi-time processes. Moreover, it is in this particular setting where multi-time correlated effects will rear their head, and choices of quantum gates will be highly inter-related in terms of their effective dynamics. In the far term, this connection becomes more explicit. Quantum error correcting codes are structured as a sequence of continuous measurements (or syndrome extractions) followed by classical feedback. This is exactly the collection of statistics on random variables at different times, or a quantum stochastic process. There is a deep symbiosis between higher order probability theories and quantum computation at both a practical and fundamental level waiting to be explored. 



\subsection{The Problem with Quantum Stochastic Processes}

We saw Kolmogorov consistency conditions as one of the fundamental properties of classical stochastic processes. In order to guarantee an underlying physical reality, stochastic processes need always to be obtainable from marginalisation of larger processes. But quantum mechanically, observables do not commute at different times. The seemingly innocuous requirement of non-invasiveness does not apply to the quantum setting. If we were to measure and forget the outcome (i.e. marginalise), this is equivalent to the application of a maximal dephasing channel to the system. A naive definition of quantum stochastic processes might be initially to say that we sample from a density matrix at each time. But clearly, due to breakdown in Kolmogorov consistency, this cannot be correct.
This problem is not uniquely quantum -- any classical theory with invasive measurements will run into trouble when attempting to define a stochastic process. The difference however is that we \emph{cannot} have a complete quantum theory without invasiveness.

As we shall see, this issue is a problem with formalism and not something more fundamental. Put plainly, a quantum theory of stochastic processes must include a control aspect to it. For this, we need to revisit our generalised quantum instruments, or trace non-increasing maps. An instrument allows one to obtain both the probability to obtain different outcomes from measurements, and to know the change of the state post-measurement. Loosely speaking, this machinery will allow us to determine statistical properties of a quantum state at different times, and to be cognisant of how the control itself affected the state.

\subsection{Process Tensors as an Operational Resolution}

By considering quantum instruments, we can first operationally resolve the issue of initial correlations. Suppose we start with a system-environment quantum state which is correlated. In an experiment that aims to determine the dynamics from initial time $t_0$ to final time $t_f$, one will apply an instrument $\mathcal{J}=\{\mathcal{A}_j\}$ on the system at the initial time to prepare it into a known state. Next, the system-environment evolution propagates the total state via the unitary map 
\begin{equation}
	\mathcal{U}_{t_f:t_0}[\rho]:= U_{t_f:t_0}\rho U_{t_f:t_0}^\dagger.
\end{equation}
This full process, with respect to the system alone, can be written as 
\begin{equation}
	\label{eq:SC-open}
	\rho_j(t_f) = \Tr_E[\mathcal{U}_{t_f:t_0}\circ (\mathcal{A}_j\otimes \mathcal{I}_E)[\rho_{SE}(t_0)]].
\end{equation}
The dilated description is exact, but for initially correlated states we will not be able to write down a physical map that takes us from states of the system at the start to states of the system at the end. Instead, let us consider that a linear mapping is any general linear relationship between experimentally controllable inputs and measureable outputs. We can see this by observing that action of the uncontrollable process here on the instrument itself $\mathcal{A}_j$ is linear. 
The change in mindset here is to instead consider \emph{the choice of the instrument} as the input to the mapping, while keeping the output state reconstruction the same. We can hence re-write this equation as 
\begin{equation}
	\rho_j(t_f) = \mathcal{T}_{t_f:t_0}[\mathcal{A}_j].
\end{equation}
This map $\mathcal{T}_{t_f:t_0}$ is known as a superchannel, which is a linear map acting on the operations $\mathcal{A}_j$ on which is it defined. At first this perspective change may seem strange given how engrained quantum channels are. But, at least in the context of quantum experiments, this is a more natural way of considering things. An experimenter does not have access to a quantum state in the lab \emph{per se}. Were they to perform process tomography, they would use a series of control knobs to aim to prepare a known (and uncorrelated) quantum state which is then fed into the evolution. A superchannel, then, is the recognition that the control operation is the true experimenter-chosen input to the dynamics, and that the state itself is merely a proxy for this. Indeed, Ref~\cite{Modi2012} proved that the mapping $\mathcal{T}_{t_f:t_0}$, as well as being linear, is both \acs{CP} and \acs{TP}. In this context, \acs{CP} means that the action of $\mathcal{T}_{t_f:t_0}$ on any \acs{CP} quantum map is guaranteed to output a positive quantum state -- even when the instrument on which it acts also extends to an arbitrarily large ancilla space. We do not explicitly show this here until we obtain the more general multi-time case. \acs{TP} is also well-defined in that any trace-preserving map will be mapped to a unit-trace object. This can be seen by noting in Equation~\eqref{eq:SC-open} is a concatenation of $\Tr_E$ and $\mathcal{U}_{t_f:t_0}$ (both trace-preserving), and $\rho_{SE}(t_0)$, which is a unit trace state. Hence, $\rho_j(t_f)$ is only subnormalised whenever $\mathcal{A}_j$ is trace non-increasing, with trace given by the probability of $\mathcal{A}_j$ occurring. 

The superchannel is the first step to the multi-time setting. We can see from Equation~\eqref{eq:SC-open} that it looks at three-point correlations: the ``measurement'' of $\rho_S(t_0)$, the subsequent preparation fed back into the process, and finally the state reconstruction at time $t_f$. Observe also that in the case where $\rho_{SE}(t_0) = \rho_S(t_0)\otimes\rho_E(t_0)$, we have a well defined quantum channel $\mathcal{E}_{t_f:t_0}[\rho] = \Tr_E[\mathcal{U}_{t_f:t_0}[\rho\otimes \rho_E(t_0)]]$ and hence the superchannel reduces to the concatenation of instrument with this subsequent \acs{CPTP} map 
\begin{equation}
	\mathcal{T}_{t_f:t_0}[\mathcal{A}_j] = (\mathcal{E}_{t_f:t_0}\circ\mathcal{A}_j)[\rho_S{(t_0)}].
\end{equation}
Recall that the issue of initial correlations has been the fundamental issue behind using quantum maps to describe non-Markovian processes, because any propagation to a later time will have new `initial' correlations. One can consider a valid \emph{marginal} \acs{CP} map by first using $\mathcal{A}_j$ to erase those correlations (through, for example, a measurement) and determining the subsequent evolution. We will return to this point.
With the superchannel as an operational resolution, we are hence in a position to introduce the fully-fledged framework for multi-time quantum stochastic processes. 

We introduce the following notations for multi-time quantum processes 
\begin{equation}
	\begin{split}
		\mathbf{T}_k &:= \{t_0,t_1,\cdots, t_{k-1},t_k\},\\
		\mathbf{J}_{k:0} &:= \{\mathcal{J}_0,\mathcal{J}_1,\cdots,\mathcal{J}_{k-1},\mathcal{J}_k\}\\
		\mathbf{x}_{k:0} &:= \{x_0,x_1,\cdots, x_{k-1},x_k\}\\
		\mathbf{A}_{\mathbf{x}_{k:0}} &:= \{\mathcal{A}_{x_0},\mathcal{A}_{x_1},\cdots, \mathcal{A}_{x_{k-1}}, \mathcal{A}_{x_k}\}.
	\end{split}
\end{equation}
$\mathbf{T}_k$ is the set of times on which the process is defined. At each $t_j\in \mathbf{T}_k$, the system is interrogated with an instrument $\mathcal{J}_j\in\mathbf{J}_{k:0}$ to obtain outcome $x_j\in \mathbf{x}_{k:0}$. The \acs{CP} map corresponding to this outcome is given by $\mathcal{A}_j\in\mathbf{A}_{\mathbf{x}_{k:0}}$. 
From here on, we will use Roman indices $j$ to denote each $t_j\in \mathbf{T}_k$. 

\begin{figure}[!b]
    \centering
    \includegraphics[width=0.8\linewidth]{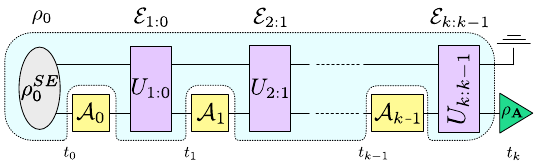}
    \caption[Circuit depiction of a process tensor]{Circuit depiction of a process tensor. A system is driven across $k$ time steps in tandem with uncontrolled interaction with an inaccessible environment. From a full open system perspective, a process tensor represents all of the uncontrollable dynamics -- initial state and subsequent $SE$ evolution, depicted in blue -- and maps system control operations to a final state.}
    \label{fig:comb-depiction}
\end{figure}

The setup is thus as follows: a joint state $\rho_0^{SE}$ at time $t_0$ is probed as in Equation~\eqref{eq:SC-open}, and then propagated to the next time $t_1$ by the system environment unitary $\mathcal{U}_{1:0}$, at which point the next instrument is employed. This sequence continues until the final time $t_k$, at which point it is measured with a POVM. 
No structure is imposed on this scenario, it is entirely general. Writing now the multi-time generalisation to Equation~\eqref{eq:SC-open}, we obtain the conditional probability distribution 
\begin{equation}
	\label{eq:PT-open}
	\mathbb{P}(\mathbf{x}_{k:0} \mid \mathbf{J}_{k:0}) = \Tr\left[\mathcal{A}_{x_k}\bigcirc_{j=0}^{k-1} \mathcal{U}_{j:j-1} \circ \mathcal{A}_{x_j} [\rho_0^{SE}]\right],
\end{equation}
where each $\circ$ is a composition. From this, we observe that the expression given can be written as multi-linear mapping from the sequence $\mathbf{A}_{k:0}$ to the output $\mathbb{P}(\mathbf{x}_{k:0}\mid \mathbf{J}_{k:0})$,
\begin{equation}
	\label{eq:pt-def-1}
	\mathbb{P}(\mathbf{x}_{k:0}\mid \mathbf{J}_{k:0}) =: \mathcal{T}_{k:0}[\mathbf{A}_{k:0}].
\end{equation}
The multi-linear mapping $\mathcal{T}_{k:0}$ is called the \emph{process tensor}. We identify several key features of process tensors from this initial rearrangement: (i) The mapping is entirely defined by Equation~\eqref{eq:pt-def-1}, which makes no assumptions about the form of the open quantum evolution. It can hence be an arbitrarily strong interaction. (ii) A process tensor is a partitioning of the controllable and uncontrollable: the mapping itself represents everything uncontrollable -- initial $SE$ state and subsequent $SE$ unitaries; its arguments are experimenter-chosen interventions, for which is is presumed an experimenter has full control. Thus, when referring to the underlying process itself, one only needs to refer to $\mathcal{T}_{k:0}$.
$\mathcal{T}_{k:0}$ is thus a generalisation of superchannels, which themselves were a generalisation of \acs{CPTP} maps. 


The above scenario describes a classical output at the final time through some POVM, but often, we will find it convenient to omit this and suppose we have access to the general quantum state where conditioned on a set of \acs{CP} maps $\mathbf{A}_{k-1:0}$. That is, 
\begin{equation}
	\rho_k(\mathbf{A}_{k-1:0}) = \mathcal{T}_{k:0}[\mathbf{A}_{k-1:0}].
\end{equation}
Specifically, throughout the thesis we will consider the use of $k$ instuments from time $t_0$ to $t_{k-1}$ followed by a final \acs{POVM} applied at $t_k$. 

\subsection{Quantum Stochastic Processes}

Let us return to the issue of Kolmogorov consistency. As we have seen, classical observables at different times commute. Hence, there is no difference between performing a measurement and applying no control to the system. The difference lies quantum mechanically, but which of the two options is more fundamental?
In spirit, what the \acs{KET} aims to do is to apply a consistency with respect to the `do-nothing' operation. That is, if we consider the underlying physics of some stochastic dynamics, marginalisation is simply a means to that end. 
It is more reassuring to think of the underlying process $\mathbb{P}_\mathbf{T}$ as some dynamical system evolving in its own right, rather than a hypothetical observer measuring-and-forgetting at infinitely many points in time.

With this philosophy in mind, one can define a set of generalised consistency conditions. For a selected set of times $\mathbf{T}_k$, one can condition a set of dynamics on the do-nothing operation by letting the process tensor act on the identity map $\mathcal{I}$. This is defined in the following notation.
\begin{equation}
	\mathcal{T}_{\mathbf{T}_\ell}^{\mid \mathbf{T}_k} := \mathcal{T}_{\mathbf{T}_\ell}\left[\bigotimes_{\alpha \in \mathbf{T}_\ell \backslash \mathbf{T}_k}\mathcal{I}_\alpha,\mathbf{A}_{\mathbf{T}_k}\right].
\end{equation}
It has been shown in Ref.~\cite{Milz2020kolmogorovextension} that process tensors satisfying a generalisation of the KET, taking into account this notion of a marginal process. Thus, process tensors formally generalise the notion of a classical stochastic process to the quantum setting. This is conveyed in the following theorem. 

\begin{theorem}[\ac{GET}]
	Let $\mathbf{T}$ be a set of times. For each finite $\mathbf{T}_k\subset \mathbf{T}$ let $\mathcal{T}_{\mathbf{T}_k}$ be a process tensor. Then there exists a process tensor $\mathcal{T}_{\mathbf{T}}$ that has all finite ones as marginals. I.e. $\mathcal{T}_{\mathbf{T}_k} = \mathcal{T}_{\mathbf{T}}^{\mid \mathbf{T}_k}$  if and only if all finite process tensors satisfy the consistency condition: $\mathcal{T}_{\mathbf{T}_k} = \mathcal{T}_{\mathbf{T}_\ell}^{\mid\mathbf{T}_k}$ for all finite $\mathbf{T}_k\subset \mathbf{T}_\ell \subset \mathbf{T}$. 
\end{theorem}

We can now understand process tensors as the most general object to characterise multi-time statistics in generic causal quantum processes. Note that acausal frameworks -- such as process matrices~\cite{Shrapnel_2018} -- exist, but these are designed to tackle a different set of problems.


\subsection{Process Tensor Representations}

\begin{figure}[!t]
    \centering
    \includegraphics[width=\linewidth]{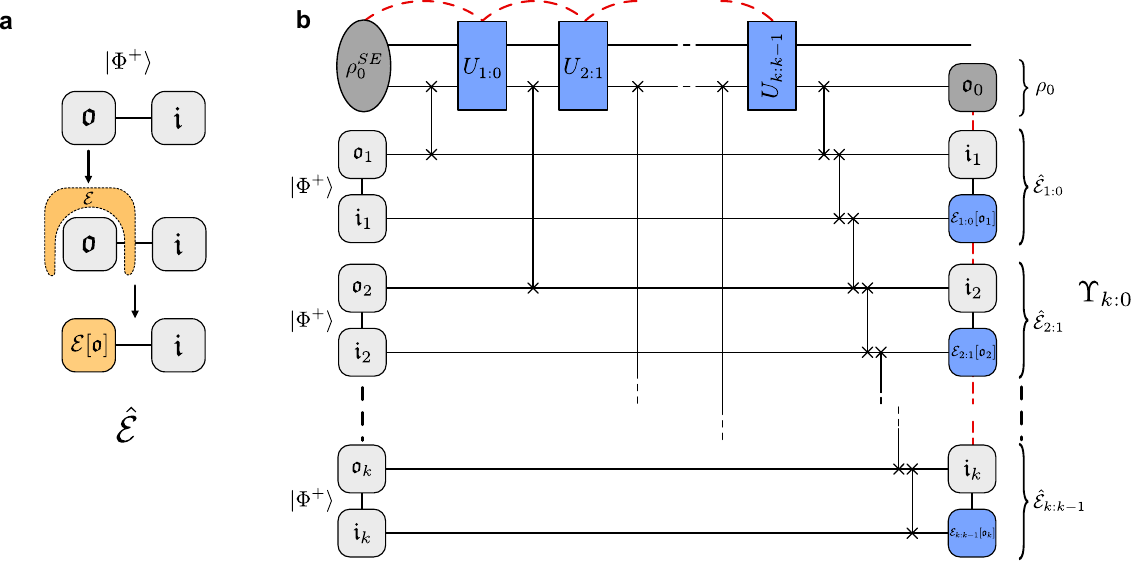}
    \caption[Circuit diagrams of the conventional and generalised Choi-Jamio\l kowski isomorphism]{Circuit diagrams of the conventional and generalised Choi-Jamio\l kowski isomorphism. \textbf{a} Two-time processes are represented by quantum channels; their Choi state is given by the channel acting on one half of a Bell pair. \textbf{b} Multi-time processes are represented by process tensors; their Choi state is given for a $k$-step process by swapping in one half of $k$ Bell pairs at different times. The non-Markovian correlations are mapped onto spatial correlations in the Choi state. The marginals of the process tensor are quantum channels, as well as the average initial state. Note that the final SWAP gates in the circuit diagram indicate a reordering of subsystems to align spatial and temporal conventions of this thesis.}
    \label{fig:choi_reps}
\end{figure}

This framework allows us to study quantum stochastic processes on a much more formal setting. But in a practical sense, the toolkit benefits greatly from a many-body state representation. 
Process tensors come equipped with a generalised version of the \acs{CJI}, and hence their own Choi form. In this picture, at each time, one half of a fresh Bell pair is swapped to interact with the environment. For a $k$-step process, feed one half of a Bell pair $|\Phi^+\rangle$ into the process, allow it to participate in the dynamics, and then swap a fresh Bell pair in at the next time. If this is repeated for each time, then the resulting state -- which we denote $\Upsilon_{k:0}$ can be shown to be element-by-element identical to the mapping $\mathcal{T}_{k:0}$. This relationship is extremely useful. What this statement shows is that temporal correlations can be mapped onto spatial ones, and understood in the exact same frameworks which have been developed under decades of many-body quantum information theory. 
A depiction of the conventional \acs{CJI} is shown in Figure~\ref{fig:choi_reps}a, along with the process tensor equivalent in Figure~\ref{fig:choi_reps}b.

The \emph{action} of a process tensor $\mathcal{T}_{k:0}$ is given by projecting the process tensor Choi state \pt{} onto the Choi state of a sequence of operations. 
This naturally generates a spatiotemporal version of Born's rule:
\begin{equation}
	\mathbb{P}(\mathbf{x}_k \mid \mathbf{J}_k) = \Tr[(\Pi_k \otimes \hat{\mathcal{A}}_{k-1}^{x_{k-1}\ \text{T}}\otimes \hat{\mathcal{A}}_{k-2}^{x_{k-2}\ \text{T}} \otimes \cdots \otimes \hat{\mathcal{A}}_{0}^{x_{0}\ \text{T}})\cdot \Upsilon_{k:0}].
\end{equation}
Similarly, this can be used to produce the action of the process tensor on any controlled sequence of operations (consistent with the time steps) by projection:
\begin{equation}\label{eq:PT-output}
    \rho_k(\mathbf{A}_{k-1:0}) = \text{Tr}_{\overline{\mathfrak{o}}_k}  \left[\Upsilon_{k:0}\left(\mathbb{I}_{\mathfrak{o}_k}\otimes \hat{\mathcal{A}}_{k-1}\otimes \cdots \hat{\mathcal{A}}_0\right)^\text{T}\right],
\end{equation}
where $\overline{\mathfrak{o}}_k$ is every index except $\mathfrak{o}_k$. This equation is inclusive of all intermediate $SE$ dynamics as well as any initial correlations, and illustrates how sequences of operations constitute observables of the process tensor. 

We use the notation $\mathfrak{o}_j$ to denote an output leg of the process at time $t_j$, and $\mathfrak{i}_j$ for the input leg of the process at time $t_{j-1}$. The collection of indices is therefore $\{\mathfrak{o}_k,\mathfrak{i}_k,\cdots,\mathfrak{o}_2,\mathfrak{i}_2,\mathfrak{o}_1,\mathfrak{i}_1,\mathfrak{o}_0\}$. These correspond to the marginals of the process, $\{\hat{\mathcal{E}}_{k:k-1},\cdots,\hat{\mathcal{E}}_{2:1},\hat{\mathcal{E}}_{1:0},\rho_0\}$. We will return to the interpretation of correlations between these different subsystems in Chapter~\ref{chap:process-properties} and Chapter~\ref{chap:MTP}.

A canny reader might notice that we have employed two different uses of the word `marginal' at this point. In the \acs{GET}, we used marginals of a process to mean projecting certain points in time onto the identity operation. In the Choi state context, however, we are using the term in the more conventional state-based setting, which is to take a partial trace over a set of subsystems. Given that we predominantly work with Choi states in this thesis, it is this latter case which we will retain as our convention. The former case, we will be explicit and refer to as a \emph{conditional} marginal, where this should be read as a process conditioned on the do-nothing operation. 

Each time has an associated output ($\mathfrak{o}$) and input ($\mathfrak{i}$) leg from the process. If the dynamical process is non-Markovian, then the system-environment interactions will distribute temporal correlations as spatial correlations between different legs of the process tensor. These correlations may then be probed using any number of established quantum or classical many body tools. In Figure~\ref{fig:choi_reps}b we have depicted these correlations in red. Correlations between consecutive $\mathfrak{i}/\mathfrak{o}$ legs (presented in black) are expected to be very high for nearly-unitary processes, and do not constitute a measure of non-Markovianity.

The Choi state of a process tensor can also be expressed in terms of the link product $\star$ of the dilated dynamics, as it was defined in Section~\ref{ssec:CJI}. That is, for Choi state $\hat{\mathcal{U}}$ of the open unitary propagator, we have
\begin{equation}
	\label{eq:pt-link}
	\Upsilon_{k:0} = \Tr_E[\hat{\mathcal{U}}^{SE}_{k:k-1}\star \hat{\mathcal{U}}^{SE}_{k-1:k-2}\star \cdots \star \hat{\mathcal{U}}^{SE}_{1:0} \star \rho_0^{SE}].
\end{equation}

\begin{figure}[!htb]
    \centering
    \includegraphics[width=\linewidth]{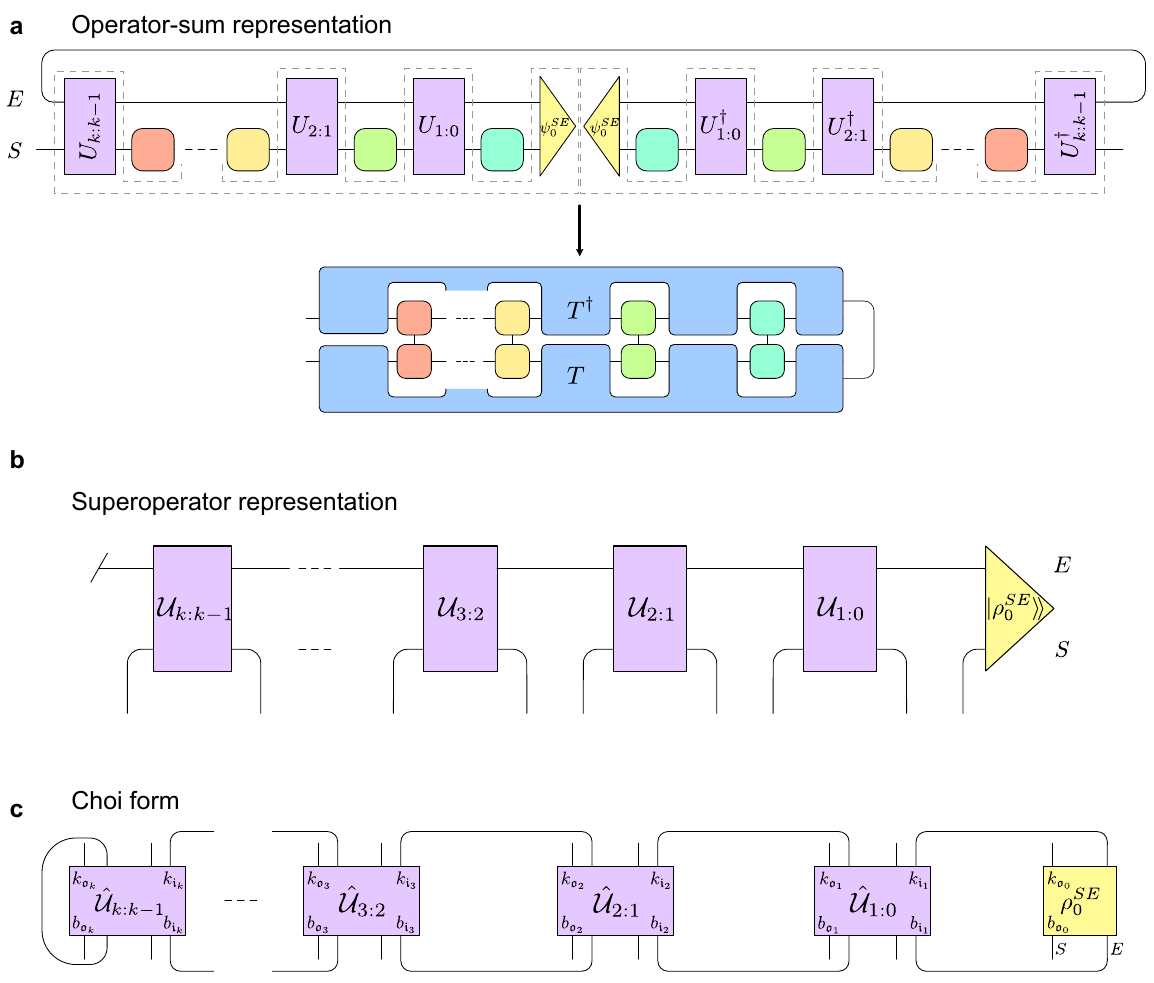}
    \caption[Graphical depiction of different process tensor representations]{Graphical depiction of different process tensor representations.
	\textbf{a} As a Kraus \acs{OSR}, guaranteeing complete positivity of the map.
	\textbf{b} In terms of the superoperator representation of the dilated unitaries. \textbf{c} In terms of the Choi representation of the dilated unitaries.}
    \label{fig:link_choi_reps}
\end{figure}


Much like process tensors generalise \acs{CPTP} maps to the multi-time setting, so too do they obey a generalised set of physical conditions, and have alternate matrix representations. It can be shown~\cite{Pollock2018a} that process tensors have a Kraus \acs{OSR} form, written via its action as 
\begin{equation}
	\mathcal{T}_{k:0}[\mathbf{A}_{k-1:0}] = \sum_l T_l\mathbf{A}_{k-1:0} T_l^{\dagger}.
\end{equation}
This implies that the linear map is completely positive, in that for any extension of the multi-time instrument to an identity tensor product on arbitrary ancilla space: $\mathbf{A}_{k-1:0}\mapsto \mathbf{A}_{k-1:0}\otimes \mathbf{I}_{k-1:0}^{A}$, the process tensor's action on this map remains positive. 
The generalisation of trace-preservation, is that the process must be \emph{causal} -- or satisfy a \emph{containment} property. The containment property can be written as 
\begin{equation}
	\Tr_{\mathfrak{o}_k}[\Upsilon_{k:0}] = \mathbb{I}_{\mathfrak{i}_k}\otimes \Upsilon_{k-1:0}.
\end{equation}
That is to say, that if we marginalise over the output at the last time of a process, the input leg $\mathfrak{i}_k$ is completely uncorrelated with the remainder of the process. The remainder of the process is then consistent with the process defined on $k-1$ steps. This amounts to generalised trace-preservation, because for any deterministic choice of maps, it guarantees that the final state will be normalised. Further, it can be described as causal because this relationship gives an arrow to the set of correlations. By definition, any choice of future operation or outcome does not affect the statistics of the process in the past.
We will explore this causal condition in greater detail in Chapter~\ref{chap:process-properties}.

We graphically depict the different process tensor representations in Figure~\ref{fig:link_choi_reps}. In Figure~\ref{fig:link_choi_reps}a, we show how the process tensor may be cast in an equivalent Kraus \acs{OSR} form. Figure~\ref{fig:link_choi_reps}b casts the open dynamics into their natural superoperator form. 
Finally, Figure~\ref{fig:link_choi_reps}c graphically depicts the link product in Equation~\eqref{eq:pt-link}, giving the process tensor Choi state in terms of compositions of $SE$ unitary Choi states. 
As well as the interpretive and practical benefits of a many-body state representation, this expression has the useful interpretation that when viewed as a \ac{MPO}, the minimal bond dimension can be seen as the size of the non-Markovian environment required to propagate memory. We will explore this idea further in Chapter~\ref{chap:efficient-characterisation}.

\subsection{Markov Conditions}

With this formalism established, we can now continue to the question of an operational condition for quantum (non) Markovianity. A process tensor's Choi state maps temporal correlations onto spatial ones. Thus, a Markovian process has a product state process tensor~\cite{Pollock2018}. Formally:
\begin{definition}[Quantum Markov Condition]
	A quantum stochastic process $\mathcal{T}_{k:0}$ is said to be \emph{Markovian} if its Choi state $\Upsilon_{k:0}$ satisfies 
	\begin{equation}
		\Upsilon_{k:0}^{\text{Markov}} = \bigotimes_{j=1}^k \hat{\mathcal{E}}_{j:j-1}\otimes \rho_0
	\end{equation}
	otherwise, it is \emph{non-Markovian}.
\end{definition}

It can be shown that this definition of Markovianity is a necessary and sufficient classification of Markov processes, and reduces to the familiar classical of Equation~\eqref{eq:classical-markov} in the appropriate limit. One can see this intuitively by noting that product states have no correlations between subsystems -- multi-point or otherwise. Further, one can quantify the degree of non-Markovianity in a process by quantifying the extent to which a process tensor's Choi state departs from a product state according to some measure~\cite{Pollock2018}.
\begin{definition}[Non-Markovian Measure]
	Any \acs{CP}-contractive quasi-distance $\mathcal{D}$ between the Choi state of a process tensor \pt{} and the closest Markov process $\Upsilon_{k:0}^{\text{Markov}}$ is a measure of the degree of non-Markovianity.
	\begin{equation}
		\mathcal{N} := \min_{\Upsilon_{k:0}^{\text{Markov}}}\mathcal{D}\left(\Upsilon_{k:0} \mid \mid \Upsilon_{k:0}^\text{Markov}\right)
	\end{equation}
	where a Markov process is defined as above. \acs{CP}-contractive here means $\mathcal{D}(\Lambda[X] \mid\mid \Lambda[Y]) \leq \mathcal{D}(X\mid\mid Y)$ for \acs{CP} $\Lambda$, and quasi-distance means satisfying all the properties of a metric except symmetry of the arguments. 
\end{definition}


In this thesis, we will frequently employ quantum relative entropy as our choice of $\mathcal{D}$. This takes a particularly nice form, in that the closest Markov process under this quasi-distance is exactly the tensor product of the marginals of the process tensor, as shown in Ref.~\cite{Pollock2018}. Hence, it has a closed form expression 
\begin{equation}
	\mathcal{N}_S(\Upsilon_{k:0}) = S(\Upsilon_{k:0}\mid\mid \bigotimes_{j=1}^k\hat{\mathcal{E}}_{j:j-1}\otimes \rho_0),
\end{equation}
where each $\hat{\mathcal{E}}_{j:j-1} = \Tr_{\bar{\mathfrak{o}}_j\bar{\mathfrak{i}}_j}[\Upsilon_{k:0}]$. Note that this is exactly a generalisation of the \acs{QMI} introduced in the previous chapter. We introduce this as $\mathcal{N}_S$ to emphasise that the degree of Markovianity warrants a choice of quasi-distance, but in future we will only employ relative entropy, and hence use $\mathcal{N}$ in all discussions of total non-Markovianity.\par 

\begin{wrapfigure}{R}{0.5\textwidth}
	\centering
	\includegraphics[width=0.8\linewidth]{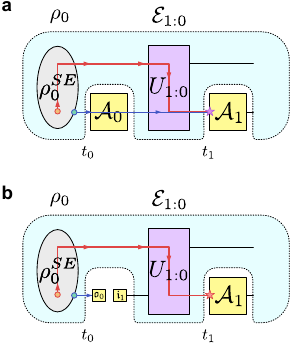}
	\caption[Circuit depiction of a causal break.]{Circuit depiction of a causal break. \textbf{a} Information from an initial time $t_0$ can persist to a later time $t_1$ either through the system, or through an inaccessible environment -- the two will be indistinguishable. \textbf{b} A causal break, represented mathematically by a product Choi state, prevents system information from propagating, and so only allows environment information to persist. }
	\label{fig:causal-break}
\end{wrapfigure}
More generally, one can witness the effects of non-Markovianity through \emph{causal breaks}. We can intuit this by considering that there are two ways for information to be transmitted from past to future: via the system or via the inaccessible environment, see Figure~\ref{fig:causal-break}. 
Suppose we apply an instrument at $t_0$ with outcomes $\{x_0\}$, an instrument at $t_1$ with outcomes $\{x_1\}$ and compare the joint probability distribution $\mathbb{P}(x_0,x_1)$ from its product distribution $\mathbb{P}(x_0)\mathbb{P}(x_1)$. Observing correlations here is not sufficient to claim non-Markovianity, because the information may persist through the system. 

Concretely, if we measure in the $Z$-basis at $t_0$ and then again in the $Z-$basis at $t_1$, it will not be surprising for these outcomes to be correlated. The first measurement forces a collapse, and so even if we have identity dynamics, the second outcome will be the same. 
Consider instead where at $t_0$ we apply the instrument $\hat{\mathcal{A}}_0^{(x_0)} := |0\rangle\!\langle 0| \otimes |x_0\rangle\!\langle x_0|$. That is, measure in the $Z$-basis and then, regardless of outcome, prepare the $|0\rangle\!\langle 0|$ state. Now if correlations are detected between outcomes, this is a definite flag of non-Markovianity because information was not allowed to persist through the system, and must have been transmitted through the environment.

A causal break at time $t_j$ is a control $\mathcal{A}_j^{(x_j)}$ such that the Choi state is a product: 
$\hat{\mathcal{A}}_j = \hat{\mathcal{A}}_j^{\mathfrak{i}_j}\otimes \hat{\mathcal{A}}_j^{(x_j) \ \mathfrak{o}_{j-1}}$ \emph{and} the marginal over outcomes is also a product: $\mathcal{A}_j = \sum_{x_0}\hat{\mathcal{A}}_j^{(x_j)} = \hat{\mathcal{A}}_j^{\mathfrak{i}_j}\otimes \hat{\mathcal{A}}_j^{\mathfrak{o}_{j-1}}$. 
Importantly, this implies that a fresh state is prepared at the input leg $\mathfrak{i}_j$ independent of the outcome at $\mathfrak{o}_j$. This notion really distils the essence of process tensors capturing non-Markovianity. Since process tensors are valid for any control sequence, they can detect correlations from all possible causal breaks. If no correlations persist across any causal break, the process is Markovian. The utility of process tensors -- and indeed, equivalent to the quantum Markov condition -- is the encoding of past-future correlations for \emph{all} choice of causal break. 


\subsection{Correlated Instruments: Testers}

For conceptual simplicity, we have only considered time-local sequences of quantum instruments. But, just as uncontrolled dynamics can carry classical or quantum memories, so too can multi-time instruments. These are known as \emph{testers} and are the dual object to process tensors.
A straightforward example of the former case is where control is made adaptive based on previous outcomes. Consider the following: an experimenter applies instrument $\mathcal{J}_0$ at time $t_0$, obtaining outcome $x_0$. This result is then fed forward to the next time, where instrument $\mathcal{J}_1^{(x_0)}$ is chosen based on the outcome at $t_0$ -- and so on. The result is a correlated instrument which can be described by a classical probability distribution. In the more general case, one can consider introducing a (quantum) ancilla space $A$. 
\begin{figure}[!b]
    \centering
    \includegraphics[width=0.7\linewidth]{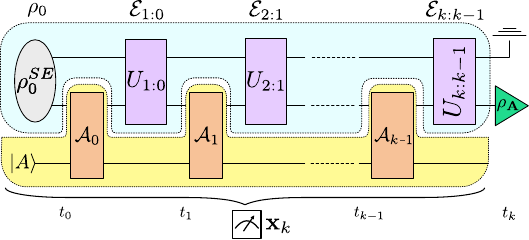}
    \caption[Circuit depiction of a tester]{Circuit depiction of a tester. A tester represents a multi-time instrument with memory. This can be visualised as a series of repeated interactions between $S$ and an ancilla system $A$, followed by a collection of measurements on $A$.}
    \label{fig:tester-depiction}
\end{figure}
At each $t_j$ an instrument is applied across the space $SA$ with outcome $x_j$, followed by \acs{POVM} at $t_k$ with outcome $x_k$. A tester $\mathbf{A}_{k:0}^{(\mathbf{x}_k)}$ has a Choi state that can generally be expressed in terms of a time-local linear basis:
\begin{equation}
	\hat{\mathbf{A}}_{k:0}^{(\mathbf{x}_k)} = \sum_{\mathbf{x}_k} \alpha_{\mathbf{x}_k} \hat{\mathcal{A}}_k^{(x_k)}\otimes \hat{\mathcal{A}}_{k-1}^{(x_{k-1})}\otimes \cdots \otimes \hat{\mathcal{A}}_0^{(x_0)},\: \alpha_{\mathbf{x}_k} \in \mathbb{R}.
\end{equation}
Although we have used the same notation $\mathbf{x}_k$ as with sequences of measurement outcomes, note that a general tester need not have each outcome tied to a specific time, nor does the number of outcomes necessarily equal the dimensionality of the system. One could consider measuring only at the end, or projectively measuring where $\text{dim}(A) > \text{dim}(S)$. One should simply think of this as a collection of measurement outcomes that dictates the effective probabilistic and correlated transformation on the system. 
Testers need not satisfy the same causal conditions as process tensors, since in general they can be probabilistic. But, just as we have the partition of unity in POVMs, and quantum instruments summing to a \acs{CPTP} map, the sum over tester outcomes must be a process tensor. That is, $\sum_{\mathbf{x}_k}\mathbf{A}_{k:0}^{(\mathbf{x}_k)}$ must be a valid process tensor.

\section{Notational Summary}

For convenience and for reference, we compile together many of the notational and convention choices for this thesis into Table~\ref{tab:notation}. Any departure from this standard style will be appropriately flagged in context. 

\begin{longtable}[c]{r|l}
	\label{tab:notation}
	\textbf{Notation}               & \textbf{Meaning/type/convention}         \\ \hline
	\endfirsthead
	\endhead
	$\rho_{j}^S$                    & Density matrix of system $S$ at time $t_j$ \\
	$\mathcal{H}_S$ & State space of a system \\
	$\mathcal{B}(\mathcal{H}_S)$ & Set of bounded operators on a Hilbert space \\ 
	$\mathcal{B}(\mathcal{B}(\mathcal{H}_S))$ & Set of superoperators on a Hilbert space \\ 
	$d_S$ & Dimension of system $S$ \\ 
	$|\rho\rangle\!\rangle$			& Row-vectorised density matrix			   \\
	$|\Phi^+\rangle$                & Normalised Bell state $\frac{1}{\sqrt{2}}(|00\rangle + |11\rangle)$ \\
	$\Lambda$                       & Arbitrary quantum channel                \\
	$\hat{\Lambda}$                 & Choi state of said quantum channel       \\
	$\mathcal{E}_{j:j-1}$           & Uncontrolled dynamical evolution from time $t_{j-1}$ to $t_j$         \\
	$\mathcal{A}_j$                   & Arbitrary control operation at time $t_j$              \\
	$\mathcal{B}_j^{\mu_j}$                   & The $\mu_j$th basis control operation applied at time $t_j$                 \\
	$\mathcal{J}$                   & Arbitrary quantum instrument             \\
	$\Pi$                           & Pure effect (projector)                  \\
	$E$                             & General effect                           \\
	$\mathcal{I}$                   & Identity map                          \\
	$\mathbb{I}$                    & Identity matrix                             \\
	$\mathcal{T}_{k:0}$             & Process tensor mapping across $k$ steps  \\
	$\Upsilon_{k:0}$                & Process tensor Choi state                \\
	$\mathfrak{i}_j$                & Input leg to a process at time $t_{j-1}$ \\
	$\bar{\mathfrak{i}}_j$          & Every input leg except $\mathfrak{i}_j$  \\
	$\mathfrak{o}_j$                & Output leg from a process at time $t_j$  \\
	$\bar{\mathfrak{o}}_j$          & Every output leg except $\mathfrak{o}_j$ \\
	$\mathbf{A}_{k-1:0}$            & Multi-time instrument, or tester across times $t_0$ to $t_{k-1}$         \\
	$\Upsilon_{k:0}^{(\mathbf{A})}$ & Process tensor conditioned on a tester $\mathbf{A}$  \\
	Time in a circuit diagram       & Left to right                            \\
	Time in a Choi state            & Right to left                           \\
	$\mathcal{N}(\Upsilon_{k:0})$	& Degree of non-Markovianity, as measured by the relative entropy
\end{longtable}

\section{Discussion}



The theory of quantum stochastic processes provides a powerful formalism with wide-ranging applicability, particularly in the rapidly developing field of quantum information processing. As all engineered quantum systems are inherently open to some degree, high fidelity control of these systems can be cast as a series of non-deterministic events. Achieving the low error rates necessary for successful quantum error correction depends on effectively navigating complex system-environment dynamics. To understand these dynamics, we must be able to cast them into the appropriate framework. It is the intention of this thesis to concretely study noisy quantum dynamics under the framework of process tensors.

Looking even further ahead, the process tensor formalism will be highly pertinent to realistic quantum error correction protocols. Quantum error correction involves making measurements on a quantum system and adaptively feeding forward this information. This procedure aligns with the process tensor's syntax, which accommodates multi-time measurement statistics and dynamic adaptation. Currently, quantum error correcting codes do very little to accommodate correlations between measurement outcomes. In an open system setting, then, we anticipate this to be a natural tool for implementing experiments of fault-tolerance. The aims of this thesis are to place analyses of open quantum systems in lock-step with the process tensor formalism. Although we examine and demonstrate this in the context of \acs{NISQ} devices, these tools will remain useful all the way up to a fault-tolerant setting.



\cleardoublepage 

\ctparttext{
We have established the incompatibility of conventional characterisation tools with non-Markovian dynamics and reviewed the process tensor framework for quantum stochastic processes. In this thesis's original research, we develop methods to generalise quantum process tomography to multi-time settings, which we call process tensor tomography. This allows for practical characterisation of arbitrary open quantum processes through experimental design, post-processing algorithms, and methods to address hardware limitations. As a result, we can effectively diagnose temporal quantum correlations, benefiting both noise characterisation in quantum devices and the study of many-time physics. We address these topics in Chapters~\ref{chap:process-properties}--\ref{chap:MTP}.
}
\part{Characterising non-Markovian quantum dynamics} 

\chapter{Properties of process tensors}
\label{chap:process-properties}
\epigraph{\emph{When you can measure what you are speaking about, and express it in numbers, you know something about it.}}{Lord Kelvin}
\noindent\colorbox{olive!10}{%
	\begin{minipage}{0.9555\textwidth} 
		\textcolor{Maroon}{\textbf{Chapter Summary}}\newline
		In this chapter, we analyse the properties of process tensors. This technical analysis provides insight into the structure of multi-time quantum correlations, and how they are formed, as well as information that can be extracted. As well as an original investigation in its own right, this sets the foundation to both inform and benchmark the results for the remainder of the thesis. We will make use of the tools and results here in later chapters.
		\par\vspace{\fboxsep}
		\colorbox{cyan!10}{%
			\begin{minipage}{\dimexpr\textwidth-2\fboxsep}
				\textcolor{RoyalBlue}{\textbf{Main Results}} 
				\begin{itemize}
					\item We translate the abstract requirement of causality into an explicit hyperplane defined by the solution to a series of homogenous linear equations. This concretely identifies the set of process tensors. 
					\item We consider various regimes of physical control, and determine which properties of process tensors may be extracted in these regimes. This experimentally pertinent question will form the backbone of characterisation work later in the thesis.
					\item We examine the structure of non-Markovian processes with strictly classical correlations, and identify physically relevant Hamiltonian classes that generate these simple non-Markovian processes.
					\item We develop a method to randomly sample from the flat distribution on the space of process tensors. We then employ this to investigate the complex properties of typical quantum processes.
				\end{itemize}
		\end{minipage}}
\end{minipage}}
\clearpage

\section{Introduction}
In our introductory chapters, we have seen how the generalised \acs{CJI} carries the remarkable insight that the space of multi-time quantum stochastic processes can be identified with a compact subset of quantum many-body states. As a consequence, the vast literature on many-body physics and its tools can be applied for similar understanding and scrutiny.
But this set has measure zero on the space of many-body states, and so properties of processes are highly atypical compared to regular quantum states -- it does not suffice to simply understand many-body physics and expect a direct translation. We must hence recalibrate our intuition to account for the extra structure imposed by causality in processes. 
This thesis is ultimately concerned with the methodology of learning process tensors in practice. As a precursor to this, we must first answer the important questions: how exactly do processes differ from states, and how exactly is the mapping composed? 
The present understanding of multi-time properties in the literature is only very nascent. This investigation will inform how to characterise processes, what properties we can expect, what different observables tell us about the process, and where different experimentally-available control partitions with respect to observables.

Another non-trivial translation to be made is in observables on processes. For many-body quantum states, this is well understood. A measurement apparatus -- described mathematically by a \acs{POVM} -- collects outcome statistics which may then be processed to estimate properties of the state, such as correlations between different subsystems. The requirements on this \acs{POVM} to extract all properties of a state are quite low; one might even consider the defining feature of a quantum device to have an informationally complete set of measurement operations. But observables on a quantum stochastic process are sequences of control operations, which themselves carry much more structure than their spatial equivalent. Unitary operations, for example, are deterministic projections of a subsystem of a process tensor, and has no single-shot spatial equivalent~\footnote{The spatial equivalent to a unitary operation is post-selected measurement in a Bell basis. Unitary observables can hence only be emulated in the spatial setting with a much larger number of experiments.}.
Control operations also carry the additional baggage that it is experimentally much more challenging to access all corners of a control vector space, and so the properties that may be learned with respect to a process is heavily limited in practice by the experimental hardware. \par 

Finally, we consider simple and complex properties of \acs{QSP}s. In the former case, we look at what exactly it takes from a dynamical perspective for non-Markovian correlations to be strictly classical. We identify an important class of Hamiltonians whose \acs{QSP} can be described using local control and classical probability distributions.
To aid our study of the latter case -- both here, and in later chapters -- we develop an algorithm to generate process tensors by generating random states and renormalising them to be causal. This is equivalent to sampling from the normalised Lebseque measure and provides an insight into the typical properties exhibited by processes. From this, for example, we see the contrast to classically correlated processes. Process tensors are highly likely to be complex, with genuinely quantum correlations. This computational foray into many-time physics is intended to set the tone for the remainder of this thesis. There is interesting physics to be discovered in multi-time \acs{QSP}s, and we focus on how to extract it.

\section{Process Tensors in the Pauli Basis}
\label{sec:PT-pauli}
To start our analysis, we wish to explore the numbers that make up a process tensor. A convenient choice of basis for Hermitian operators on binary dimensions is the $n$-qubit Pauli set $\mathbf{P}^n:=\{\mathbb{I},X,Y,Z\}^{\otimes n}$. The basis is Hermitian, and hence has only real expectation values; readily extensive, in that one can simply take tensor products of Paulis for larger subsystems; physically interpretable; and partitions neatly into unit-trace marginal terms, and correlation terms. 
The purpose of this section is to evaluate and analyse process tensors in the Pauli basis: what do causality constraints impose, and what do different instruments allow you to learn about different parts of the process? These results set the foundations of later sections, which are concerned with optimally reconstructing pre-existing process tensors from sets.

Consider first the Choi state representation of a process tensor, \pt{}, which is a $2k+1$-partite state, with $d_S$-dimensional subsystems. Our analyses here considers the case of $d_S=2$, but the results readily generalise to any generalised Hermitian basis for finite-dimensional systems. Using the Pauli basis, \pt{} can be expressed as:
\begin{equation}
	\Upsilon_{k:0} = \sum_{i=1}^{4^{2k+1}} \Tr[P_i\Upsilon_{k:0}]P_i,
\end{equation}
or, recalling that we label the marginal indices of \pt{} by interleaving $\mathfrak{o}_j$ and $\mathfrak{i}_j$ subsystems
\begin{equation}
	\Upsilon_{k:0} = \sum_{\mu_1,\mu_2,\cdots,\mu_{2k+1}} \Tr[P^{\mu_1}_{\mathfrak{o}_k}\otimes P^{\mu_2}_{\mathfrak{i}_k} \otimes \cdots \otimes P^{\mu_{2k+1}}_{\mathfrak{o}_0} \Upsilon_{k:0}]P^{\mu_1}_{\mathfrak{o}_k}\otimes P^{\mu_2}_{\mathfrak{i}_k} \otimes \cdots \otimes P^{\mu_{2k+1}}_{\mathfrak{o}_0} ,
\end{equation}
where in the former equation $P_i\in\mathbf{P}^{2k+1}$, and in the latter, $P_{\mu_i}\in \mathbf{P}^1$. For brevity, we will write the expectation values as $\langle P^{\mu_1}_{\mathfrak{o}_k}P^{\mu_2}_{\mathfrak{i}_k} \cdots  P^{\mu_{2k+1}}_{\mathfrak{o}_0}\rangle$, assumed to be with respect to some \pt{}. Additionally, any reduced expectation value is taken to implicitly mean identity elements on each other subsystem. \Eg{} $\langle P^{\mu_1}_{\mathfrak{o}_k}\rangle \equiv \langle P^{\mu_1}_{\mathfrak{o}_k}\mathbb{I}_{\mathfrak{i}_k}\cdots \mathbb{I}_{\mathfrak{o}_0}\rangle$. Finally, we will omit the tensor product symbol between Pauli matrices. \par 

Using Pauli bases is a convenient structure for studying process tensors both due to its orthonormality, and because the set partitions into one unit-trace and three traceless elements. The identity is responsible for identifying marginal terms, and the traceless elements represent the correlations in the state. 
Moreover, as we shall see, writing the process tensor in this capacity allows us to readily identify it as a positive matrix satisfying a set of linear constraints.

\subsection{Causality Conditions}
By virtue of the \acs{CJI}, all process tensors may be represented as quantum many-body states. However, the converse does not hold: random many-body states need not constitute valid processes. In fact, the set of process tensors has measure zero on the set of quantum states. The extra constraint introduced is that of \emph{containment}, or \emph{causality}. Explicitly:
\begin{equation}
	\label{eq:containment}
	\Tr_{\mathfrak{o}_k}[\Upsilon_{k:0}] = \mathbb{I}_{\mathfrak{i}_k}\otimes \Upsilon_{k-1:0}\quad \forall\: k.
\end{equation}
This mathematical condition has a variety of equivalent interpretations. It is a generalisation of trace-preservation, ensuring that a sequence $\mathbf{A}_{k-1:0}$ of deterministic instruments produces an output state $\rho_k(\mathbf{A}_{k-1:0})$ with unit trace. It is also a statement that marginalising over future measurement outcomes still retains a valid process for across the times prior to that. As a consequence, future instrument choices do not affect any of the past statistics.\par 

We wish to rewrite Equation~\eqref{eq:containment} in terms of constraints on our Pauli expectation values. Consider the first condition $\Tr_{\mathfrak{o}_k}[\Upsilon_{k:0}] = \mathbb{I}_{\mathfrak{i}_k}\otimes \Upsilon_{k-1:0}$. Let $\tilde{\mathbf{P}} := \{X,Y,Z\}$ be the traceless Pauli matrices, and observe that 
\begin{equation}
	\begin{split}
	\Tr_{\mathfrak{o}_k}[\Upsilon_{k:0}] &= \sum_{\mu_2,\cdots\mu_{2k+1}}\langle \mathbb{I}_{\mathfrak{o}_k}P^{\mu_2}_{\mathfrak{i}_k}\cdots P^{\mu_{2k+1}}_{\mathfrak{o}_0}\rangle P^{\mu_2}_{\mathfrak{i}_k}\cdots P^{\mu_{2k+1}}_{\mathfrak{o}_0}\\
	&= \sum_{\mu_2,\cdots\mu_{2k+1}}\langle P^{\mu_2}_{\mathfrak{i}_k}\cdots P^{\mu_{2k+1}}_{\mathfrak{o}_0}\rangle P^{\mu_2}_{\mathfrak{i}_k}\cdots P^{\mu_{2k+1}}_{\mathfrak{o}_0}\\
	&\overset{!}{=} \mathbb{I}_{\mathfrak{i}_k}\otimes \sum_{\mu_3,\cdots,\mu_{2k+1}}\langle P^{\mu_3}_{\mathfrak{o}_{k-1}}\cdots P^{\mu_{2k+1}}_{\mathfrak{o}_0}\rangle P^{\mu_3}_{\mathfrak{o}_{k-1}}\cdots P^{\mu_{2k+1}}_{\mathfrak{o}_0}.
	\end{split}
\end{equation}
From the final equality, using orthogonality of the Pauli matrices, we must have 
\begin{equation}
	\langle \mathbb{I}_{\mathfrak{o}_k}\tilde{P}_{\mathfrak{i}_k}^{\mu_2} P_{\mathfrak{o}_{k-1}}^{\mu_3}\cdots P_{\mathfrak{o}_0}^{\mu_{2k+1}}\rangle = 0.
\end{equation}
In plainer words, \pt{} must not have any correlation terms connecting the identity on $\mathfrak{o}_k$ with traceless Paulis on $\mathfrak{i}_k$.
Given that the containment condition is iterative -- that is, $\Upsilon_{k-1:0}$ in Equation~\eqref{eq:containment} must also be a valid process, we obtain in total
\begin{equation}
	\label{eq:pauli-causality}
	\begin{split}
			&\langle \mathbb{I}_{\mathfrak{o}_k}\tilde{P}_{\mathfrak{i}_k}^{\mu_2} P_{\mathfrak{o}_{k-1}}^{\mu_3}\cdots P_{\mathfrak{o}_0}^{\mu_{2k+1}}\rangle = 0\quad \text{for final time $t_k$,}\\
			&\langle \mathbb{I}_{\mathfrak{o}_k}\mathbb{I}_{\mathfrak{i}_k}\mathbb{I}_{\mathfrak{o}_{k-1}}\tilde{P}_{\mathfrak{i}_{k-1}}^{\mu_4} P_{\mathfrak{o}_{k-2}}^{\mu_5}\cdots P_{\mathfrak{o}_0}^{\mu_{2k+1}}\rangle = 0\quad \text{for penultimate time $t_{k-1}$,}\\
			&\qquad\vdots\\
			&\langle \mathbb{I}_{\mathfrak{o}_k}\mathbb{I}_{\mathfrak{i}_k}\cdots \mathbb{I}_{\mathfrak{o}_1}\tilde{P}_{\mathfrak{i}_1}^{\mu_{2k}}P_{\mathfrak{o}_0}^{\mu_{2k+1}}\rangle = 0\quad \text{until initial time $t_0$}.
	\end{split}
\end{equation}
A process tensor \pt{} is hence a quantum state with the extra structure that it must obey the linear equations given in Equation~\eqref{eq:pauli-causality}, of which there are
\begin{equation}
	\sum_{j=1}^k (d_S^2 - 1)d_S^{4j-2}
\end{equation}
many conditions. In other words, a process tensor \pt{} is an element of the intersection between the cone of positive semidefinite matrices, with the hyperplane defined by the above set of linear equations.
We will use this insight for the remainder of the thesis as it is useful both for imposing conditions of causality in model fits in Chapters~\ref{chap:PTT},~\ref{chap:efficient-characterisation}, and~\ref{chap:universal-noise}, as well as traversing process tensor landscapes in Chapter~\ref{chap:MTP}.

\subsection{Partitioning Observable Constraints By Available Control}
\begin{wrapfigure}{r}{0.45\textwidth}
	\centering
	\includegraphics[width=0.7\linewidth]{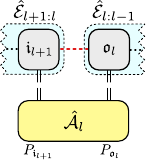}
	\caption[The projection of a process tensor onto some local control operation]{The projection of a process tensor onto some local control operation $\mathcal{A}_l$. We see that this evaluates expectation values on $\mathfrak{o}_l$ and $\mathfrak{i}_{l+1}$, and can be partitioned into Pauli tensor products on each of these legs.}
	\label{fig:pauli-expectations}
\end{wrapfigure}
The previous section considered Pauli basis coefficients of process tensors first in terms of physical reasoning. That is to say, that any collection of Bell pairs subject to a series of $SE$ isometries will always have Pauli coefficients set to zero as in Equation~\eqref{eq:pauli-causality}. We now move on to a consideration of which Pauli coefficients may be determined by which control implementations.
This is a highly pertinent question to explore before we embark on a tour of non-Markovian quantum process tomography. In the spatial setting, local and \acs{IC} observables are readily available in any typical experimental setting. Temporally, however, learning observables of a \acs{QSP} can be more restricted in terms of the available instruments. In particular, the ability to learn mid-circuit information about a quantum state is often a long, invasive, and technologically challenging operation. To this end, we first ask the question: what is in principle determinable about a process tensor with different levels of control? This amounts to examining the properties that control quantum channels may possess.\par

Recall that the action of a process tensor \pt{} on a multi-time instrument $\mathbf{A}_{k-1:0}$ and final \acs{POVM} $\{\Pi_k^{(i)}\}$ is
\begin{equation}
	p_{i|\mathbf{A}_{k-1:0}} = \Tr[(\Pi_k^{(i)}\otimes \hat{\mathbf{A}}_{k-1:0})\cdot \Upsilon_{k:0}].
\end{equation}
Consider first the action of time-local control operations. Figure~\ref{fig:pauli-expectations} shows how the legs of an operation $\hat{\mathcal{A}}_l$ are projected onto the marginals of a process tensor Choi state. We discuss a series of different control types.

\textbf{Unitary quantum channels:}
For gate-based quantum devices, the first obvious port-of-call are unitary channels
\begin{equation}
	\hat{\mathcal{A}}_{\text{u}} = (u\otimes \mathbb{I})\cdot |\Phi^+\rangle\!\langle \Phi^+| \cdot(u^\dagger \otimes \mathbb{I}),
\end{equation}
for unitary $u$.
\begin{equation}
	\begin{split}
	(u\otimes \mathbb{I})\cdot |\Phi^+\rangle\!\langle \Phi^+| \cdot (u^\dagger \otimes \mathbb{I}) &= \sum_{i,j}u(|i\rangle\!\langle j|)u^\dagger \otimes |i\rangle\!\langle j| ,\\
	\Rightarrow \Tr[{P}_\mathfrak{o}\otimes P_{\mathfrak{i}} \hat{\mathcal{A}}_{\text{u}}] &= \sum_{ijkl}\langle k|P_\mathfrak{o} u|i\rangle\!\langle j|u^\dagger|k\rangle \otimes \langle l|P_{\mathfrak{i}}|i\rangle\!\langle j|l\rangle,\\
	&= \sum_{ikl}\langle k|P_{\mathfrak{o}}u|i\rangle\!\langle l| u^\dagger|k\rangle\!\langle l|P_{\mathfrak{i}}|i\rangle\\
	&= \Tr[u P_{\mathfrak{i}}^{\text{T}}u^\dagger P_{\mathfrak{o}}].
	\end{split}
\end{equation}
We see from this that $\langle \tilde{P}_{\mathfrak{o}}\mathbb{I}_{\mathfrak{i}]}\rangle = \langle \mathbb{I}_{\mathfrak{o}}\tilde{P}_{\mathfrak{i}}\rangle = 0$ by the unitarity of $u$ and the tracelessness of $\tilde{P}$. 

A physical way to interpret this is that unitary maps -- and convex combinations thereof -- provide no information about the process marginals; they do not allow one to read out information about the state at $\mathfrak{o}_{l}$, nor do they allow one to know about the state entering the process at $\mathfrak{i}_{l+1}$. Instead, a sequence of unitary operations is equivalent to teleportation of the state through the process and applying operations to it along the way.
Thus, when restricted only to unitary operations on a single qubit, the observables $\mathcal{O}$ on $\Upsilon_{k:0}$ -- to which we have access -- 
take the form
\begin{equation}
	\label{eq:restr-obs}
	\begin{split}
		\mathcal{O} &= \sum_{i,\vec{\mu}}\alpha_{i,\vec{\mu}} P_i\otimes \bigotimes_{j=0}^k P_{\mu_j}, \qquad \text{where}\\
		&\alpha_{i,\vec{\mu}}\in \mathbb{R},\\
		&P_i\in \{\mathbb{I},X,Y,Z\},\\
		&P_{\mu_j}\in \{\mathbb{I\otimes I}\} \cup \{X,Y,Z\}\otimes \{X,Y,Z\}.
	\end{split}
\end{equation}
This observation is also true for generic unital -- or bistochastic -- quantum operations. 
Classically, Birkhoff's theorem says that all bistochastic matrices can be expressed as convex combinations of permutations (reversible transformations). The quantum version of this holds for $d_S=2$: that is, all bistochastic quantum channels are convex combinations of unitaries. But for $d_S>2$ it is known that this does not hold: there are bistochastic quantum channels which cannot be expressed as a mixed unitary channel~\cite{mendl2009unital}. They are, however, in the linear span of unitaries~\cite{PT-limited-control}.

\textbf{Non-unital channels:}
Non-unital channels are defined by their ability to unmix a quantum state. That is to say, that for some non-unital $\mathcal{A}_{\text{nu}}$, we have the property that $\mathcal{A}_{\text{nu}}[\mathbb{I}/d_S] \prec \mathbb{I}/d_S$. By definition, then, the expectation values $\langle \tilde{P}_{\mathfrak{o}}\mathbb{I}_{\mathfrak{i}}\rangle$ can be non-zero. 
Intuitively, this gives the ability to learn the marginal legs $\mathfrak{i}_{l+1}$ because the channel deterministically maps the fixed point of all input quantum states (the identity) into some non-trivial, and known direction. Hence, the non-unital channels give the experimenter some $\epsilon>0$ ability to prepare a chosen state for the next marginal of the dynamics, $\mathcal{E}_{l+1:l}$. 

\textbf{Trace-decreasing maps:}
For local control, the final class we are interested in are trace-decreasing maps $\mathcal{A}_{\text{d}}$. That is, elements of a quantum instrument. 
These are stochastic in the sense that the actual implementation cannot be chosen by the experimenter, and often constitutes the outcome of some sort of a weak-valued \acs{POVM}.
\acs{TP} quantum channels satisfy
\begin{equation}
	\Tr_{\mathfrak{o}}[\hat{\mathcal{A}}] = \mathbb{I}_{\mathfrak{i}} \qquad \Rightarrow \langle \mathbb{I}_{\mathfrak{o}} \tilde{P}_{\mathfrak{i}}\rangle = 0.
\end{equation}
In contrast, trace-decreasing maps have non-zero $\langle \mathbb{I}_{\mathfrak{o}} \tilde{P}_{\mathfrak{i}}\rangle$. For qubits, one can think of this as overlap between $X$, $Y$, and $Z$ projective measurements. This gives the final missing snapshot of information: observables that provide information about the marginals $\mathfrak{o}_{l}$. This constitutes the ability to read out the state in the middle of a process. 

\textbf{Spatial Locality:}
There is some extent to which we need to analyse control operations in moving to the multi-qubit setting. By this, we mean taking $d_S = 2^n$ for $n>1$ -- incorporating a process tensor with spatiotemporal correlations. Here, we can marginalise a process tensor \pt{} not only in time $\hat{\mathcal{E}}_{j:j-1}$ but also in space $\hat{\mathcal{E}}_{j:j-1}^{q_i}$. The question of tomographic completeness is further fine-grained, because the difficulty of implementing control operations can be highly dependent on locality. Suppose, for example, one wanted to consider a two-qubit process tensor across $k$ steps. Then local unitaries provide access only to observables of the form
\begin{equation}
	\begin{split}
		&\langle \tilde{P}_{\mathfrak{o}_j}^{(q_1)}\tilde{P}_{\mathfrak{o}_j}^{(q_2)}\tilde{P}_{\mathfrak{i}_j}^{(q_1)}\tilde{P}_{\mathfrak{i}_j}^{(q_2)}\rangle,\\
		&\langle \tilde{P}_{\mathfrak{o}_j}^{(q_1)}\mathbb{I}_{\mathfrak{o}_j}^{(q_2)}\tilde{P}_{\mathfrak{i}_j}^{(q_1)}\mathbb{I}_{\mathfrak{i}_j}^{(q_2)}\rangle,\\
		&\langle \mathbb{I}_{\mathfrak{o}_j}^{(q_1)}\tilde{P}_{\mathfrak{o}_j}^{(q_2)}\mathbb{I}_{\mathfrak{i}_j}^{(q_1)}\tilde{P}_{\mathfrak{i}_j}^{(q_2)}\rangle,\quad\text{and}\\
		&\langle \mathbb{I}_{\mathfrak{o}_j}^{(q_1)}\mathbb{I}_{\mathfrak{o}_j}^{(q_2)}\mathbb{I}_{\mathfrak{i}_j}^{(q_1)}\mathbb{I}_{\mathfrak{i}_j}^{(q_2)}\rangle.
	\end{split}
\end{equation}
The other traceless observables, such as $\langle \tilde{P}_{\mathfrak{o}_j}^{(q_1)}\tilde{P}_{\mathfrak{o}_j}^{(q_2)}\tilde{P}_{\mathfrak{i}_j}^{(q_1)}\mathbb{I}_{\mathfrak{i}_j}^{(q_2)}\rangle$ can only be obtained through entangling unitaries. For $n$ qubits, a local unitary control space can therefore set $10^n$ constraints, whereas generic $U\in SU(2^n)$ will measure $d_S^4 - 2d_S^2 + 2$ linearly independent observables. \acs{IC} local control, however, also constitutes \acs{IC} global control, and so with access to fully general quantum instruments, one need only manipulate the qubits locally in order to fully determine the entire process.

\textbf{No Initial Correlations:}
An extra assumption one might wish to employ is that of an initial state with no signalling to the rest of the process. Many quantum devices display excellent state preparation, in which the state $|0\rangle\!\langle 0|$ is created to a high fidelity. As alternative motivation, one might wish only to explore the physics of a well-defined process which starts from a point without initial correlations.
In this case, a $k-$step process may be written as $\Upsilon_{k:0} = \Upsilon_{k:1}\otimes \rho_0$. Since $\mathfrak{o}_0$ and $\mathfrak{i}_1$ are related by a tensor product, then only the local Pauli expectation values are required -- six for a single qubit, obtained with non-unital and trace reducing maps. In the restricted case, however, the ten required unitaries condenses down to four, since all $\langle \tilde{P}_{\mathfrak{i}_1}^j\tilde{P}_{\mathfrak{o}_0}^l\rangle$ terms are fixed by $\langle \tilde{P}_{\mathfrak{i}_1}^j\tilde{P}_{\mathfrak{o}_0}^j\rangle$ terms. This is true for all assumed tensor products between subsystems in the process tensor.

\begin{figure}
	\centering
	\includegraphics[width=0.8\linewidth]{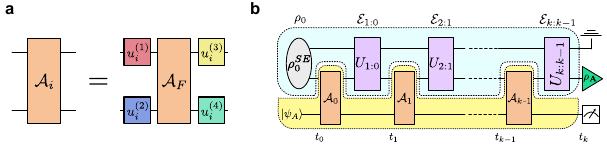}
	\caption[Circuit diagram of an example informationally complete tester ]{Circuit diagram of an example informationally complete tester. An ancilla qubit is taken to interact unitarily according to the expansion shown in \textbf{a} with the system qubit at a number of times, and is projectively measured at the end of the circuit, shown in \textbf{b}.}
	\label{fig:tester-tomographic}
\end{figure}

\begin{figure}
	\centering
	\includegraphics[width=\linewidth]{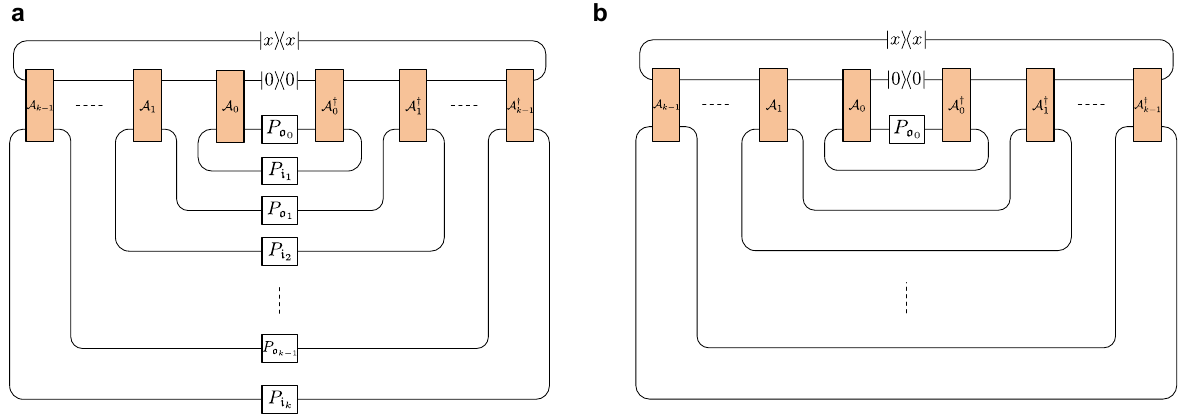}
	\caption[Informational completeness of a single qubit ancillary tester of multi-time statistics ]{Using a single ancilla system with a terminating two-outcome measurement and unitary interactions as an informationally complete probe. \textbf{a} Construction of the \acs{IC} ancillary tester. Each Pauli leg $P_{\mathfrak{o}_j}$ and $P_{\mathfrak{i}_j}$ feeds into a process tensor leg \pt{} (not depicted). \textbf{b} We show, for example, that such a tester can even determine the marginal initial state of the system.}
	\label{fig:testers-ic}
\end{figure}

\textbf{Testers:}
A further, conceptually different, avenue of control for learning quantum stochastic processes is that of correlated instruments, or testers. As discussed in Chapter~\ref{chap:stoc-processes}, a tester $\mathbf{\Theta}_{k-1:0}$ is a process tensor dual, it is a series of control operations with classical or quantum memory linking the different times. Since we have already shown the memoryless case of a tester to be \acs{IC},
we consider here the structure of a specific type of tester. 
The advantage of having an ancilla qubit to use as a tester is that this may be an alternative form of \acs{IC} control for systems without mid-circuit measurements. The level of control that this requires is a single native two-qubit gate, controllable local unitaries, and a projective measurement at the end of the circuit.
Suppose an experimentally available system constitutes an ancilla system $A$ with which $S$ interacts unitarily at each $t\in \mathbf{T}_k$. $A$ is then projectively measured at the end of the circuit, a scene depicted in Figure~\ref{fig:tester-tomographic}. Moreover, each $\mathcal{A}_i$ is locally equivalent to the same, fixed interaction, $\mathcal{A}_F$ see Figure~\ref{fig:testers-ic}a. The total open system dynamics in this case is:
\begin{equation}\label{eq:tester-multiproc}
	\rho^{(x)}_k\left(\Theta_{k-1:0}^{S}\right) = \text{tr}_{EA} [|x\rangle\!\langle x|_AU_{k:k-1}^{SE} \, \mathcal{A}_{k-1}^{SA} \cdots \, U_{1:0}^{SE} \, \mathcal{A}_{0}^{SA} (\rho^{SE}_0\otimes |\psi_A\rangle\!\langle \psi_A|)],
\end{equation}
where the $U_{j:j-1}^{SE}(\cdot) = [u^{SE}_{j:j-1}\otimes \mathbb{I}_A](\cdot)[u^{SE\dagger}_{j:j-1}\otimes \mathbb{I}_A]$, and $\mathcal{A}_j^{SA} = [a_j^{SA}\otimes \mathbb{I}_E](\cdot)[a_j^{SA}\otimes \mathbb{I}_E]$. In this sense, the neighbouring qubit acts as a probe of the system, performing joint measurements across each of the times of the \acs{QSP}. Consider the Pauli expectation values of our tester, as shown in the graphical diagram in Figure~\ref{fig:testers-ic}b. By choosing each $\mathcal{A}_j:=\mathcal{I}$, we readily obtain the full span of local unitaries the same as before. To show that $\Theta_{k-1:0}$ is indeed \acs{IC}, we again need the marginal trace-decreasing and non-unital observables $\langle \tilde{P}_{\mathfrak{o}_l}\mathbb{I}_{\mathfrak{i}_l}\rangle$ and $\langle \mathbb{I}_{\mathfrak{o}_l}\tilde{P}_{\mathfrak{i}_l}\rangle$, respectively. $A_F$ alone will be unable to generate this, since it only constitutes a single tester. But so long as $A_F$ can generate at least one non-unital or trace-decreasing observable, the local unitaries ensure that each Pauli can be transformed to another Pauli at will, and so a non-zero value in any given basis will suffice to show \acs{IC}. For concreteness: when we say that this particular tester is \acs{IC}, we mean that the span of testers given by the span of each local unitary, as it is defined, is \acs{IC}.\par 
By way of example, we will consider the special case where $A_F$ is a CNOT, which we denote by $C$. Note that the direction is $A\rightarrow S$ (control on $A$, target on $S$). For a single step, where $A$ is in any state, this interaction on its own is neither non-unital, nor trace-decreasing. The latter is manifestly true, because there is no classical output. The former follows since, for $\ket{\psi_A} = \cos\frac{\theta}{2}\ket{0} + \text{e}^{i\phi}\sin\frac{\theta}{2}\ket{1}$
\begin{equation}
	\mathcal{C}_S[\rho_S] = \Tr_A[C (|\psi_A\rangle\!\langle \psi_A| \otimes \rho_S) C^\dagger] = \cos^2\frac{\theta}{2} \rho_S + \sin^2\frac{\theta}{2} X\rho_S X,
\end{equation}
which is a mixture of unitaries, and hence, unital. Allowing for measurement on $A$ means we can postselect on outcome $x$, which purifies $S$. That is,
\begin{equation}
	\label{eq:AF-single}
	\mathcal{C}_S^{(x)}[\rho_S] = \Tr_A{[(|x\rangle\!\langle x|\otimes \mathbb{I}_S)\cdot C (|\psi_A\rangle\!\langle \psi_A| \otimes \rho_S) C^\dagger}] = \rho_S'^{(x)}\succcurlyeq \rho_S,
\end{equation}
with equality only when $\rho_S$ is stabilised by the CNOT -- if it is a mixture of $X$ eigenstates. Consequently, we have both trace-decreasing and non-unital properties. This particular example is simply an application of Naimark's dilation theorem (c.f. Section~\ref{ssec:POVMs}). \par 

The evaluation of individual Pauli expectations of a tester defined by multiple applications of $\mathcal{A}_i$ are shown graphically in Figure~\ref{fig:testers-ic}a. By Equation~\ref{eq:AF-single}, we obtain each of the $d_S^4$ expectation values $\langle P_{\mathfrak{i}_k}P_{\mathfrak{o}_{k-1}}\rangle$. 
To show that this tester contains non-zero expectation values on the earlier legs $\langle \mathbb{I}_{\mathfrak{i}_k} \mathbb{I}_{\mathfrak{o}_{k-1}}P_{\mathfrak{i}_{k-1}\cdots P_{\mathfrak{o}_0}}\rangle$, we need to be able to control the local unitaries such that $\mathcal{A}_{k-1}$ stabilises the population of $A$. That is, it should not alter the ancilla qubit's measurement probabilities. 

Consider first the problematic example where $\mathcal{A}_{k-1}$ is a CNOT controlled on $S$ and targeted on $A$. Inserting $\mathbb{I}_{\mathfrak{o}_{k-1}}$ is equivalent to tracing over the system before inserting it through the operation, which implements the maximal depolarising channel on $A$. Hence, when $A$ is projectively measured, a uniform distribution on $A$ will be obtained regardless of the state of $S$ or $A$ at earlier times. Hence, nothing is determined. However, if $\mathcal{A}_{k-1}$ is a CNOT controlled in the opposite direction, then the state of $S$ can only change the phase of $A$. In other words, the final $Z-$measurement on $A$ commutes through $\mathcal{A}_{k-1}$ and acts as a non-unital and trace-decreasing operation at the previous time. This can then iterate to find expectations at all earlier times, despite only measuring the ancilla once at time $t_{k-1}$. This is shown, for example, in Figure~\ref{fig:testers-ic}b.


%



\textbf{A note on normalisation:} Discussions of the \acs{CJI} often include an abuse of nomenclature (of which this thesis is not immune) with respect to Choi \emph{states} versus Choi \emph{matrices} or \emph{operators}. The primary consideration between the two is its choice of normalisation. When representing a quantum map $\mathcal{E}$ in its Choi form $\hat{\mathcal{E}}$, a choice of unit trace places the map in a proper quantum state form -- for which standard tools in quantum information can be applied. Conversely, we have seen that a \acs{TP} condition imposed requires $\Tr_{\mathfrak{o}}[\hat{\mathcal{E}}] = \mathbb{I}_{\mathfrak{i}}$. It follows then that $\Tr[\hat{\mathcal{E}}] = d_S$. This latter choice of normalisation takes on the matrix or operator terminology, because it acts as the bona fide mapping of some input $\rho$. \par 

The same considerations apply with respect to process tensors. The causality conditions in Equation~\eqref{eq:causality} demand the concatenation of $k$ identities $\mathbb{I}_{\mathfrak{i}_k}$ following a partial trace over $\mathfrak{o}_k$. This means that as an operator, a process tensor should be normalised to $\Tr[\Upsilon_{k:0}] = d_S^{2k}$. This can also be seen to follow from noting that a process tensor constitutes a collection of $k$ possibly correlated \acs{CPTP} maps, each of which have trace $d_S$. Throughout the thesis, process tensors are the object of our discussion, particularly with respect to correlations therein. To reduce confusion in future discussion (and for algorithmic simplicity), we adopt the convention compromise:
\begin{equation}
	\label{eq:norm-convention}
	\begin{split}
		\Tr[\Upsilon_{k:0}] &= 1,\\
		\Tr[\hat{\mathcal{A}}] &= d_S^{2}.
	\end{split}
\end{equation}
Squaring the normalisation on Choi matrices for control operations has several advantages. It maintains \pt{} as a valid state for us to study, and prevents the need to have a trace that scales with the number of steps. But by increasing the trace of our input control operations, we still retain desirable properties of trace preservation and causality. 

\section{Classically Correlated Processes}
By the state-process equivalence, processes can carry non-trivial physics through correlations in time as mediated by environmental memory. In considering process tensor properties, we are interested in establishing a boundary of sorts between simple processes and more complicated ones. Specifically, an important question we will come to in the study of correlated noise is whether the noise can be categorised as classical or quantum. By classical, we mean here not necessarily that the process only interacts with classical fields in its dynamics. Indeed, we will see that completely quantum environments might be called classical. What we mean is that the hypothetical quantum memory could be entirely simulated classically. That is, the correlations can be entirely explained by classical random variables. This constitutes a set of dynamics without any quantumness in its temporal correlations. This concept is illustrated as follows is as follows:
\begin{definition}
	If a process tensor \pt{} can be expressed in terms of quantum trajectories
	\begin{equation}
		\label{eq:p-sep}
		\Upsilon_{k:0} = \sum_{i=1}^{2^{2k}} p_i \left(\bigotimes_{j=1}^k \hat{\mathcal{E}}_{j:j-1}^{(i)} \otimes \rho_0^{(i)}\right),
	\end{equation}
	where each $\hat{\mathcal{E}}_{j:j-1}^{(i)}$ is a valid \acs{CPTP} map, then this is said to be \emph{process separable}, and the correlations are classically simulable.
\end{definition}
\begin{wrapfigure}{R}{0.3\textwidth}
	\centering
	\includegraphics[width=0.9\linewidth]{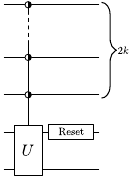}
	\caption[Circuit depiction of a classically correlated process. ]{Circuit depiction of a classically correlated process. }
	\label{fig:p-sep-multiplex}
\end{wrapfigure}
To unpack this slightly, we choose the terminology to be process separable rather than just separable because the Choi state is not completely separable -- that is, entanglement can exist between $\mathfrak{o}_j,\mathfrak{i}_j$ pairs. Entanglement across $\mathfrak{o}_j,\mathfrak{i}_j$ simply communicates the fact that process marginals do not destroy coherence of the incoming states, but that there is no non-Markovian temporal entanglement. We say that the correlations are classically simulable because no quantum resources are required in the environment. Although Equation~\eqref{eq:p-sep} is non-Markovian, it is a convex combination of Markovian processes. Thus, it fits into the class of dynamics known as quantum trajectories because the dynamics can essentially be simulated as a convex combination of system control protocols. On a shot-by-shot basis, one could draw an item $i_0$ from the classical distribution $\{p_i\}$, and then implement the dynamics $\bigotimes_{j=1}^k\hat{\mathcal{E}}_{j:j-1}^{(i_0)}\otimes \rho_0^{(i_0)}$. \par 

A useful way to think about processes of this type is that we can always dilate them to an environment implementing multiplex gates. Multiplex gates are generalisations of controlled-unitaries that implement unitaries on a target dependent on the bit string of the control register. For example, a multiplex gate with two control qubits can be written in block-diagonal form:
\begin{equation}
	\begin{pmatrix} 
		u_{00} & & & \\
		& u_{01} & & \\
		& & u_{10} & \\
		& & & u_{11}
	\end{pmatrix},
\end{equation}
where the unitary $u_{ij}$ is applied to the target qubits conditioned on the control qubits being in state $|ij\rangle$. Now, any process tensor in the form of Equation~\eqref{eq:p-sep} can be written with $2k$ environment qubits and $2$ ancilla qubits. In each step, $2k$ of them are controls of a multiplexed gate targeting an $SU(4)$ gate on the ancillas and the system. The ancillas are then reset to any pure state. This is depicted in Figure~\ref{fig:p-sep-multiplex}. The purpose of the ancillas is to allow the possibility that the $\hat{\mathcal{E}}_{j:j-1}^{(i)}$ be non-unital, and the reset to ensure that no quantum memory is carried forward between each step. Let us first eliminate the ancilla qubits, and have the target unitaries act only on the system. Note that as a result, the individual $\hat{\mathcal{E}}_{j:j-1}^{(i)}$ are mixed unitary channels. Let us denote a multiplex gate between times $t_{j-1}$ and $t_j$ by $M_{j:j-1}$.


We can relate this to a particular class of dynamics by converting processes of this mixed unitary type to a Hamiltonian model. Consider first a restriction of the controlled unitary to being a rotation about $Z$, $R_Z$.

\begin{wrapfigure}{l}{0.5\textwidth}
	\centering
	\includegraphics[width=0.9\linewidth]{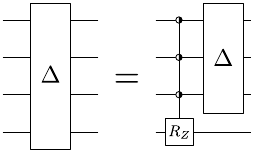}
	\caption[Decomposition of a diagonal gate into a multiplexed $Z$-rotation and another diagonal gate. ]{Decomposition of a diagonal gate into a multiplexed $Z$-rotation and another diagonal gate. }
	\label{fig:diag-to-rz}
\end{wrapfigure}
We make use of the fact that one may decompose a diagonal gate $\Delta$ on $n$ qubits to the composition into a composition of a multiplexed $R_Z$ on $n$ qubits followed by a diagonal gate (in the $Z$-basis) on the $n-1$ target qubits, as in Figure~\ref{fig:diag-to-rz}. 
The remaining $\Delta$ on $E$ commutes with any subsequent multiplex and can be brought to the end of the circuit and removed by the partial trace. Hence, any diagonal $SE$ interaction generates a process tensor of the form in Equation~\eqref{eq:p-sep}. We can add more structure to this by introducing the following definition~\cite{cubitt2016complexity}.
\begin{definition}
	Suppose we have a Hamiltonian $H_{SE}\in \mathcal{B}(\mathcal{H}_{S}\otimes \mathcal{H}_E)$. If 
	\begin{equation}
		H_{SE} = (u_S\otimes u_E) D_{SE} (u_S^\dagger\otimes u_E^\dagger),
	\end{equation}
	for diagonal $D_{SE}$, we say the Hamiltonian is \emph{locally diagonalisable}.
\end{definition}
It is easy to check that this definition implies
\begin{equation}
	\label{eq:local-H-structure}
	H_{SE} = \alpha A\otimes B + \beta A\otimes \mathbb{I} + \gamma \mathbb{I} \otimes B + \delta \mathbb{I}\otimes \mathbb{I},
\end{equation}
for real $\alpha, \beta, \gamma, \delta$, and Hermitian $A\in \mathcal{B}(\mathcal{H}_S), B\in \mathcal{B}(\mathcal{H}_E)$. We can show that dynamics generated by locally diagonalisable Hamiltonians also produce process-separable process tensors by noting that:
\begin{equation}
	\begin{split}
		U_{j:j-1}^{SE} &= \exp\left(-i(t_j - t_{j-1}) H_{SE}\right)\\
		&= (u_S\otimes u_E) \exp \left(-1 (t_j - t_{j-1})D_{SE}\right) (u_S^\dagger\otimes u_E^\dagger)\\
		&= (u_S\otimes u_E) \Delta_{j:j-1} (u_S^\dagger\otimes u_E^\dagger)\\
		&= (u_S\otimes u_E) M_{j:j-1}(R_Z) (u_S^\dagger\otimes u_E^\dagger)
	\end{split}
\end{equation}
Hence, for sequential $U_{SE}$:
\begin{equation}
	\begin{split}
		U_{j+1:j} \cdot U_{j:j-1} &= (u_S\otimes u_E) \Delta_{j+1:j} \Delta_{j:j-1} (u_S^\dagger\otimes u_E^\dagger)\\
		&= U_{j+1:j} \cdot U_{j:j-1} = (u_S\otimes u_E) M_{j+1:j}(R_Z) M_{j:j-1}(R_Z)(u_S^\dagger\otimes u_E^\dagger),
	\end{split}
\end{equation}
The $u_S$ applies a similarity transformation to each $R_Z$ in the multiplex. Meanwhile, each $u_E$ cancels out except for the very first and last. The first can be absorbed into the definition of the environment state, and the last absorbed into the partial trace. Hence, a concatenation of $k$ $SE$ evolutions generated by locally diagonalisable Hamiltonians produces a process tensor which is process-separable, as in Equation~\eqref{eq:p-sep}. Note also that any operation applied to $S$ does not change this structure, and so we can have the system-only part of $H_{SE}$ be arbitrary too, i.e. $H_S = \gamma I_E\otimes C$ for some Hermitian $C$. 

This provides us some insight into the complexity of different open quantum systems. Specifically, we can draw a distinction between non-Markovian processes whose memory can be simulated with classical resources only, versus those that need some quantumness. An interesting consequence of this is that $SE$ interactions that generate entanglement between a system and its environment is not sufficient to generate temporal entanglement, temporally entangled processes constitute a more restrictive class than the state equivalence. Below are some examples of $SE$ interactions (independent of environment state) that generate classically correlated processes:

\begin{itemize}
	\item Quantum Ising models $H_{SE} = \sum_{ij\in \mathcal{C}} J_{ij} Z_{i}\otimes Z_{j} + \sum_i Z_i$, modelling for example $ZZ$ coupling found in transmon devices with connectivity $\mathcal{C}$.
	\item A sequence of CNOT gates between system and environment, by virtue of the fact that CNOTs can be generated by a Hamiltonian $H_{CNOT}\propto (\mathbb{I} - X)\otimes (\mathbb{I} - Z)$. 
	\item Kitaev's toric code Hamiltonian~\cite{KITAEV20032}, with respect to any given qubit on a lattice: $H = -\sum_{\nu} A(\nu) - \sum_p B(p)$, where $A(\nu) = \sigma_{\nu,1}^x\sigma_{\nu,2}^x\sigma_{\nu,3}^x\sigma_{\nu ,4}^x$ and $B(p) = \sigma_{p,1}^z\sigma_{p,2}^z\sigma_{p,3}^z\sigma_{p,4}^z$. Here, $\nu$ are indices of a vertex on the lattice and $p$ are indices of a plaquette. $A$ and $B$ terms always commute since vertices and plaquettes share an even number of sites, hence the interaction is locally diagonalisable and leads only to classical temporal correlations. For clarity here, the system is a single qubit with respect to the rest of the array as an environment. 
\end{itemize}

Another avenue of significance here for later chapters is that it provides us with a counterfactual. If we can detect temporal entanglement from a process, then we can rule out dynamics of the form~\eqref{eq:local-H-structure}. It also provides us some indication of how to construct processes with temporal entanglement.





\section{Random Multi-Time Processes}
A powerful tool used in quantum information theory is the generation of random states to aid analysis. Numerical methods of this sort can be used for a range of applications such as examining questions of typicality in quantum physics, testing new methods, and determining the extent to which different classes of states can possess different properties. In this section, we propose a natural ensemble of process tensors and a method by which one may sample from it. Our method is constructive: generate a random state, and then project it onto the space of causal states in a way that preserves both rank and complete positivity. One might naturally consider random processes from the perspective of generating random multi-time dynamics on an environment and then taking a trace. We argue an alternate approach for several reasons: 
\begin{enumerate}
	\item \underline{Computational convenience}: by avoiding dilation, one can limit the dimensionality and study larger problems,
	\item \underline{Added structure}: As we will discuss we can impose extra constraints onto the type of processes that we produce, including, for example, bistochastic processes,
	\item \underline{Access to wider properties}: We find that the dilated perspective does not succeed at sampling as widely from non-Markovian properties.
\end{enumerate}
It is not our purpose to analyse the method of generation of random process in depth, but rather to develop a robust method that samples evenly from the flat distribution. We will then use this to look directly at properties of processes in the present chapter; in Chapter~\ref{chap:PTT}, we shall benchmark tomographic techniques on randomly sampled process tensors; and in Chapter~\ref{chap:MTP}, we test a variety of different many-time physics techniques on random process tensors as well.

We adapt a procedure inspired by the results of Ref.~\cite{Bruzda_2009} with respect to quantum channels.
For a system of dimension $d_S$ over $k$ steps, let us denote the process dimension $d_S^{2k+1}$ by $n$ and the rank of the process tensor by $r$. The first step is to generate a state from the Ginibre ensemble. Select a matrix $X\in \mathbb{C}^{(n\times r)}$ where each element of the matrix is a normally distributed complex number to define
\begin{equation}
\rho_k = \frac{XX^\dagger}{\Tr[XX^\dagger]}.
\end{equation}
This is a unit-trace, Hermitian, PSD matrix, and hence a valid quantum state with rank $r$. But it does not satisfy causality, and hence is not a valid process. Let us identify $\rho_k$ with the $k$-time process tensor, that is
\begin{equation}
\rho_k \in \mathcal{B}(\mathcal{H}_{\mathfrak{o}_k})\otimes \mathcal{B}(\mathcal{H}_{\mathfrak{i}_k}) \otimes \cdots \otimes \mathcal{B}(\mathcal{H}_{\mathfrak{o}_0}).
\end{equation}
We denote the marginals of $\rho_k$ as
\begin{equation}
\rho_{\mathfrak{i}_k,k-1} := \Tr_{\mathfrak{o}_k}[\rho]	
 \qquad \text{and} \qquad
 \rho_{k-1} := \Tr_{\mathfrak{o}_k\mathfrak{i}_k}[\rho].
\end{equation}
Now, let us define
\begin{equation}
	\breve{\rho}_{\mathfrak{i}_k,k-1} := \left(\sqrt{\rho_{\mathfrak{i}_k,k-1}}\right)^{-1} \qquad \text{and} \qquad \Lambda_{k}:=\rho_k.
\end{equation}
Since $\rho_k$ is PSD, $\rho_{\mathfrak{i}_{k},k-1}$ has a well-defined matrix square root. We will return to the existence of its inverse, but for now, assume this inverse to uniquely exist. Then let us define
\begin{equation}
\label{eq:lambda-def}
\begin{split}
	\Lambda_{k} &:= (\mathbb{I}_{\mathfrak{o}_k\mathfrak{i}_k} \otimes \sqrt{\rho_{k-1}})\cdot (\mathbb{I}_{\mathfrak{o}_k} \otimes \breve{\rho}_{\mathfrak{i}_k,k-1})\cdot \rho_{k} \cdot (\mathbb{I}_{\mathfrak{o}_k} \otimes \breve{\rho}_{\mathfrak{i}_k,k-1}) \cdot (\mathbb{I}_{\mathfrak{o}_k\mathfrak{i}_k} \otimes \sqrt{\rho_{k-1}}).
\end{split}
\end{equation}
This matrix has rank $r$. Moreover, because $A^{1/2}BA^{1/2}$ is PSD for any PSD $A,B$, we also have $\Lambda_k$ is PSD. Now we can see that if we take a partial trace over $\mathfrak{o}_k$ we obtain
\begin{equation}
\label{eq:lambda-causal}
\Tr_{\mathfrak{o}_k}[\Lambda_{k-1}] = (\mathbb{I}_{\mathfrak{i}_k}\otimes \sqrt{\rho_{k-1}})\cdot \breve{\rho}_{\mathfrak{i}_k,k-1} \cdot \rho_{\mathfrak{i}_{k}, k-1} \cdot \breve{\rho}_{\mathfrak{i}_k,k-1} \cdot (\mathbb{I}_{\mathfrak{i}_k}\otimes \sqrt{\rho_{k-1}})
=\mathbb{I}_{\mathfrak{i}_k}\otimes\rho_{k-1}.
\end{equation}

$\Lambda_k$ is hence locally causal, in that tracing over the final output reduces to an identity on its input leg, but the remaining marginal $\rho_{k-1}$ does not iteratively satisfy the same conditions. Fortunately, we can iterate our process by setting, for $1\leq j \leq k$:
\begin{equation}
\begin{split}
	\Lambda_j &:= (\mathbb{I}_{\mathfrak{o}_k\mathfrak{i}_k \dots \mathfrak{o}_j\mathfrak{i}_j} \otimes \sqrt{\rho_{j}})  \cdot (\mathbb{I}_{\mathfrak{o}_k\mathfrak{i}_k \dots \mathfrak{o}_j} \otimes \breve{\rho}_{\mathfrak{i}_{j+1},j})\cdot \Lambda_{j+1} \cdot (\mathbb{I}_{\mathfrak{o}_k\mathfrak{i}_k \dots \mathfrak{o}_j} \otimes \breve{\rho}_{\mathfrak{i}_{j+1},j}) \cdot (\mathbb{I}_{\mathfrak{o}_k\mathfrak{i}_k \dots \mathfrak{o}_j\mathfrak{i}_j} \otimes \sqrt{\rho_{j}}).
\end{split}
\end{equation}
The subsequent transformations do not affect the relation in Eq.~\eqref{eq:lambda-causal}, and hence the end result is a random rank $r$ process tensor.

The above derivation presupposed the existence of the matrix inverse of $\rho_{\mathfrak{i}_k,k-1}$. This matrix is defined through the marginalisation of $\rho_k$ over $\mathfrak{o}_k$, which has size $d_S$. This means $\text{rank}(\rho_{\mathfrak{i}_k,k-1}) = \text{min}(r d_S, n/d_S)$. Thus, if $r < n/d_S^2$, then $\rho_{\mathfrak{i}_k,k-1}$ is singular and hence cannot be inverted. This is problematic if we want to examine low-rank processes. To resolve this, we consider an `inner' iteration of the algorithm.
Let us instead set 
\begin{equation}
	\breve{\rho}_{\mathfrak{i}_{k},k-1} = (\sqrt{\rho_{\mathfrak{i}_{k},k-1}})^{+},
\end{equation}
where $(\cdot)^+$ denotes the Moore-Penrose pseudoinverse. This transformation takes the inverse on the image (Im) of $\rho_{\mathfrak{i}_{k},k-1}$, but leaves the kernel (Ker) untouched. The process is locally causal on the support of the marginal $\rho_{\mathfrak{i}_{k},k-1}$. Consider its eigendecomposition:
\begin{equation}
	\rho_{\mathfrak{i}_{k},k-1} = \sum_{i=1}^{r < n/d_S^2}p_i|\psi_i\rangle\!\langle \psi_i|,
\end{equation}
where $\{|\psi_i\rangle\}$ is an orthonormal basis of $\text{Im}(\rho_{\mathfrak{i}_{k},k-1})$. Let $\{|\phi_i\rangle\}$ be an orthonormal basis of $\text{Ker}(\rho_{\mathfrak{i}_{k},k-1})$
we have
\begin{equation}
\breve{\rho}_{\mathfrak{i}_{k},k-1} = \sum_{i=1}^{r} \frac{1}{\sqrt{p_i}}|\psi_i\rangle\!\langle\psi_i|.
\end{equation}
We see then that 
\begin{equation}
	\label{eq:deficient-projection}
  \breve{\rho}_{\mathfrak{i}_{k},k-1} \cdot \rho_{\mathfrak{i}_{k},k-1} \cdot \breve{\rho}_{\mathfrak{i}_{k},k-1}
= \sum_i |\psi_i\rangle\!\langle \psi_i| = \mathbb{I}_{\mathfrak{i}_{k},k-1} - \sum_{j=1}^{d_S^{2k}-r}|\phi_j\rangle\!\langle \phi_j|.
\end{equation}

The process is hence causal on the support of $\rho_{\mathfrak{i}_k,k-1}$. Let $\Lambda_k^{(0)}:= \Lambda_k$ and similarly for $\rho_{\mathfrak{i}_k,k-1}^{(0)}$ and $\breve{\rho}_{\mathfrak{i}_k,k-1}^{(0)}$. Then recursively define 
\begin{equation}
    \Lambda_{k}^{(l+1)} := (\mathbb{I}_{\mathfrak{o}_k\mathfrak{i}_k} \otimes \sqrt{\rho_{k-1}^{(l)}})\cdot (\mathbb{I}_{\mathfrak{o}_k} \otimes \breve{\rho}_{\mathfrak{i}_k,k-1}^{(l)})\cdot \Lambda_{k}^{(l)} \cdot (\mathbb{I}_{\mathfrak{o}_k} \otimes \breve{\rho}_{\mathfrak{i}_k,k-1})^{(l)} \cdot (\mathbb{I}_{\mathfrak{o}_k\mathfrak{i}_k}^{(l)} \otimes \sqrt{\rho_{k-1}^{(l)}}).
\end{equation}
At each iteration, the spectrum of $\Lambda_{k-1}^{(l+1)}$ will be different to the last (starting from a random state), and its rank will be unchanged. Moreover, because $\rho_{k-1}^{(l)}$ does not act on the $\mathfrak{i}_k$ subsystem, the state will remain causal on $\bigcup_{l}\text{Im}(\rho_{\mathfrak{i}_k,k-1}^{(l)})$. Combining these two facts, we have that eventually the algorithm converges such that
\begin{equation}
    \text{Tr}_{\mathfrak{o}_k}\left[\Lambda_{k}^{(l)}\right] = \mathbb{I}_{\mathfrak{i}_k}\otimes \Lambda_{k-1}^{(l)}
\end{equation}
where $\Lambda_k$ still has rank $r$.


The number of iterations is at worst $d_S^{2k}-r$, but in practice we find fewer than this are required. Note, however, that if $r\ll d$ the computation may become numerically unstable. This process is summarised in Algorithm~\ref{alg:random-pts}.


\begin{algorithm}[t]
\caption{Generating Random Process Tensors}
\label{alg:random-pts}
\begin{algorithmic}[1]
\Input{System dimension $d_S$, no. steps $k$, rank $r$, maximum iterations $M$.}
\Output{Random $k$-step process tensor $\Upsilon_{k:0}$ with environment size $r$.}
\State $j\gets0,$ $n \gets {d_S}^{2k+1}$
\State $X\gets$ \text{RandomGinibreMatrix($n,r$)}
\State $\rho \gets {XX^\dagger}/{\text{Tr}[XX^\dagger]}$
\State $\Upsilon \gets \rho$
\State $\Upsilon_{\text{c}}$, $\Upsilon_{\text{r}}$ \Comment{Cancellation and remainder terms}
\While{$\Upsilon \not\subset \mathcal{V}_{\text{c}}$ \text{ and } $j\leq M$}
\State $\Upsilon = \Upsilon / \text{Tr}[\Upsilon]$
    \For{$i \gets 1 \text{ to } k$}
        \State $\Upsilon_{\text{c}} = \text{Tr}_{[1:2i+1]}[\Upsilon]$
        \State $\Upsilon_{\text{r}} = \text{Tr}_{[1:2i+2]}[\Upsilon]$
        \State $\Upsilon_{\text{c}} = (\sqrt{\Upsilon_{\text{c}}})^{+}$
        \State $\Upsilon_{\text{r}} = \sqrt{\Upsilon_{\text{r}}}$
        \State $\Upsilon = (\mathbb{I}_{2i+1}\otimes \Upsilon_{\text{c}})\cdot \Upsilon \cdot (\mathbb{I}_{2i+1}\otimes\Upsilon_{\text{c}})$
        \State $\Upsilon = (\mathbb{I}_{2i+2}\otimes \Upsilon_{\text{r}})\cdot \Upsilon \cdot (\mathbb{I}_{2i+2}\otimes\Upsilon_{\text{r}})$
        \State $ j = j+1$
    \EndFor
\EndWhile
\State\Return $\Upsilon_{k:0} \gets \Upsilon$
\end{algorithmic}
\end{algorithm}

\textbf{Extra Structure:}
Our method for generating random process tensors is in essence a method to generate random Wishart matrices and project these onto the space of causal processes. We have seen that the space of process tensors is a convex set, hence if we wish to impose extra (convex) structure, then we can perform alternating projections onto different convex sets. The result is guaranteed to converge onto the intersection of the two sets. One example of this is that we can impose the condition of global unitality, defined as follows.
\begin{definition}
	\label{def:global-unital}
	If a process tensor \pt{} satisfies
	\begin{equation}
		\Tr_{\mathfrak{i}_k}[\Upsilon_{k:0}] = \mathbb{I}_{\mathfrak{o}_k} \otimes \Upsilon_{k-1:0}\quad \forall \:k,
	\end{equation}
	we say that process is \emph{globally unital}. Note that global unitality includes the special case of marginal unitality: that each \acs{CPTP} marginal to the process satisfies $\Tr_{\mathfrak{i}_j}[\hat{\mathcal{E}}_{j:j-1}] = \mathbb{I}_{\mathfrak{o}_j}$. 
\end{definition}
This condition is reminiscent to that of causality, and in fact the pair together produce a process which is bistochastic in its collective inputs and outputs. Dynamics of this type rule out observables of the form $\langle P_{\mathfrak{o}_k}\mathbb{I}_{\mathfrak{i}_k}P_{\mathfrak{o}_{k-1}}\mathbb{I}_{\mathfrak{i}_{k-1}}\cdots P_{\mathfrak{o}_1}\mathbb{I}_{\mathfrak{i}_1}\rangle$.

Taking once more some randomly generated $\rho$, if we define 
\begin{equation}
	\rho_{U_k} := (\sqrt{\rho_{\mathfrak{o}_k\mathfrak{o}_{k-1}\mathfrak{i}_{k-1}\cdots\mathfrak{o}_0}})^{-1},
\end{equation}
with $\rho_{R_k}$ as before. We can define a new matrix 
\begin{equation}
	\begin{split}
		\Gamma_k^{(0)} &:= \rho_{U_k}\cdot \rho \cdot \rho_{U_k}\\
		\Gamma_k &:= (\mathbb{I}_{\mathfrak{o}_k\mathfrak{i}_k} \otimes \rho_{R_k})\cdot \Gamma_k^{(0)}\cdot (\mathbb{I}_{\mathfrak{o}_k\mathfrak{i}_k} \otimes \rho_{R_k}),
	\end{split}
\end{equation}
where, in the first line, there is an implicit tensor product with $\mathbb{I}_{\mathfrak{i}_k}$. We thus have 
\begin{equation}
	\Tr_{\mathfrak{i}_k}[\Gamma_k] = \mathbb{I}_{\mathfrak{o}_k}\otimes \rho_{\mathfrak{o}_{k-1}\mathfrak{i}_{k-1}\cdots\mathfrak{o}_0}.
\end{equation}
If we iterate the above projection for all $k$, we obtain a state which is globally unital. Alternating causal and unital projections of a Wishart random matrix is thus guaranteed to produce a process tensor which is bistochastic in its collective inputs and outputs~\cite{Kukulski_2021}. Note that generic projections are not guaranteed to generate a flat distribution on the intersection of arbitrary sets, and should be treated only as a heuristic tool.

\subsection{Spectral Properties}
Here, we briefly analyse the spectral properties of randomly generated process tensors. This structure can be better seen by transforming the process tensor Choi state \pt{} to superoperator form, $\Phi_{k:0}$. 
\begin{figure}[!b]
	\centering
	\includegraphics[width=\linewidth]{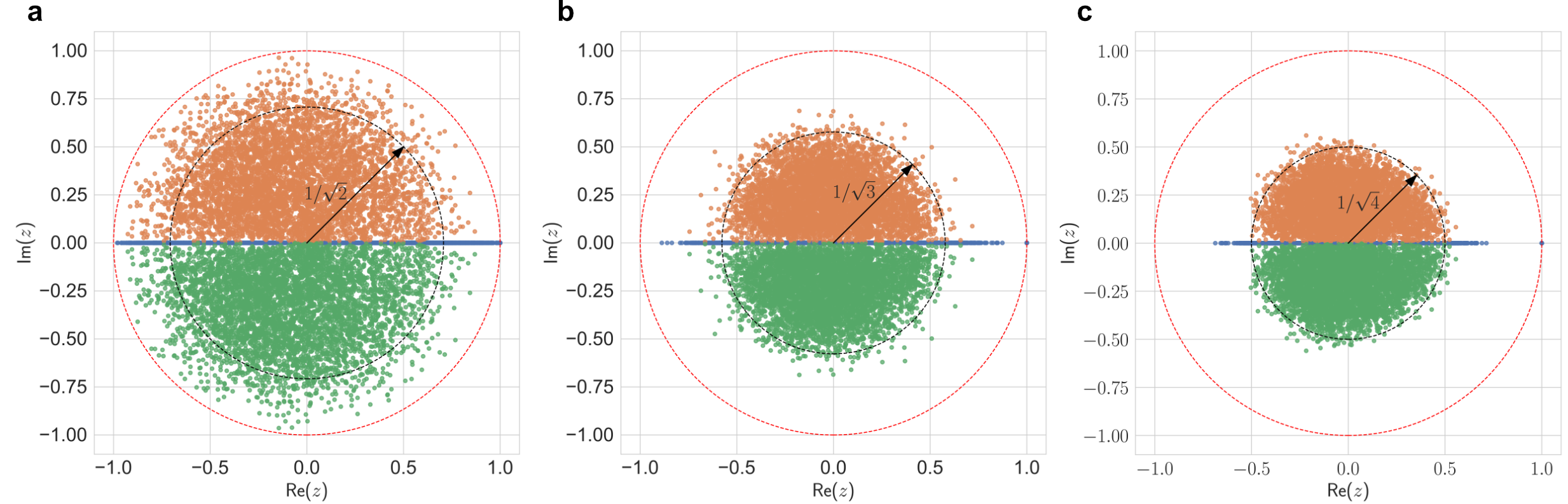}
	\caption[Eigenvalue distribution of 5000 random two-step process tensors ]{Eigenvalue distribution of 5000 random two-step process tensors. Plotted alongside the unit disc and the Girko disc $1/\sqrt{r}$. \textbf{a} $r=2$, \textbf{b} $r = 3$, and \textbf{c} $r = 4$. Orange, blue, and green colouring are eigenvalues with positive, zero, and negative imaginary components, respectively.}
	\label{fig:eval_dist}
\end{figure}
Recall, since process tensors are \acs{CP}, then a given process tensor $\mathcal{T}_{k:0}$ has operator-sum decomposition, as described in Chapter~\ref{chap:stoc-processes}
\begin{equation}
	\rho[\mathbf{A}_{k-1:0}] = \mathcal{T}_{k:0}[\mathbf{A}_{k-1:0}] = \sum_{l=1}^r T_l \hat{\mathbf{A}}_{k-1:0} T_l^\dagger.
\end{equation}
The Choi state is then defined as 
\begin{equation}
	\Upsilon_{k:0} = \sum_{l=1}^r |T_l\rangle\!\rangle \langle \!\langle T_l|,
\end{equation}
and superoperator form consequently as 
\begin{equation}
	\Phi_{k:0} = \sum_{l=1}^r T_l^\ast\otimes T_l.
\end{equation}
This is equivalent to a reshuffling operation $(\cdot)^R$,
\begin{equation}
	(|k_{\mathfrak{o}_j}\rangle\!\langle b_{\mathfrak{o}_j}| \otimes |k_{\mathfrak{i}_j}\rangle\!\langle b_{\mathfrak{i}_j}|)^R = | k_{\mathfrak{o}_j}\rangle\!\langle k_{\mathfrak{i}_j}| \otimes |b_{\mathfrak{o}_j} \rangle\!\langle b_{\mathfrak{i}_j}| \:\forall\: j,
\end{equation}
such that $\Phi_{k:0} = \Upsilon_{k:0}^R$. For quantum channels $\mathcal{E}$, the spectrum is contained within the unit disc $\{z\in \mathbb{C} : |z|\leq 1\}$, and there is a leading eigenvalue $\lambda_1 = 1$~\cite{Kukulski_2021}. The spectrum is also symmetric about the real axis. We can view process tensors as a \acs{CPTP} map from the space of inputs in a multi-time process to the space of outputs. As a consequence, its spectrum obeys the same conditions. In the limit of large dimension, random quantum channels drawn from a uniform measure have the same first two cumulants as the Gaussian distribution, with negligible higher order contributions. The trailing eigenvalues therefore tend to concentrate in the so-called \emph{Girko disc} of $1/\sqrt{r}$, recalling that $r$ is the rank, or number of Kraus operators. We see this too with process tensors. Expressed in superoperator form, Figure~\ref{fig:eval_dist} shows both the leading eigenvalues in the unit disc, the symmetry about $\text{Re}(z)$, and the clustering of eigenvalues in the Girko disc. From this we see good evidence that our method to sample processes is indeed according to the flat distribution.


\begin{figure}[!t]
	\centering
	\includegraphics[width=\linewidth]{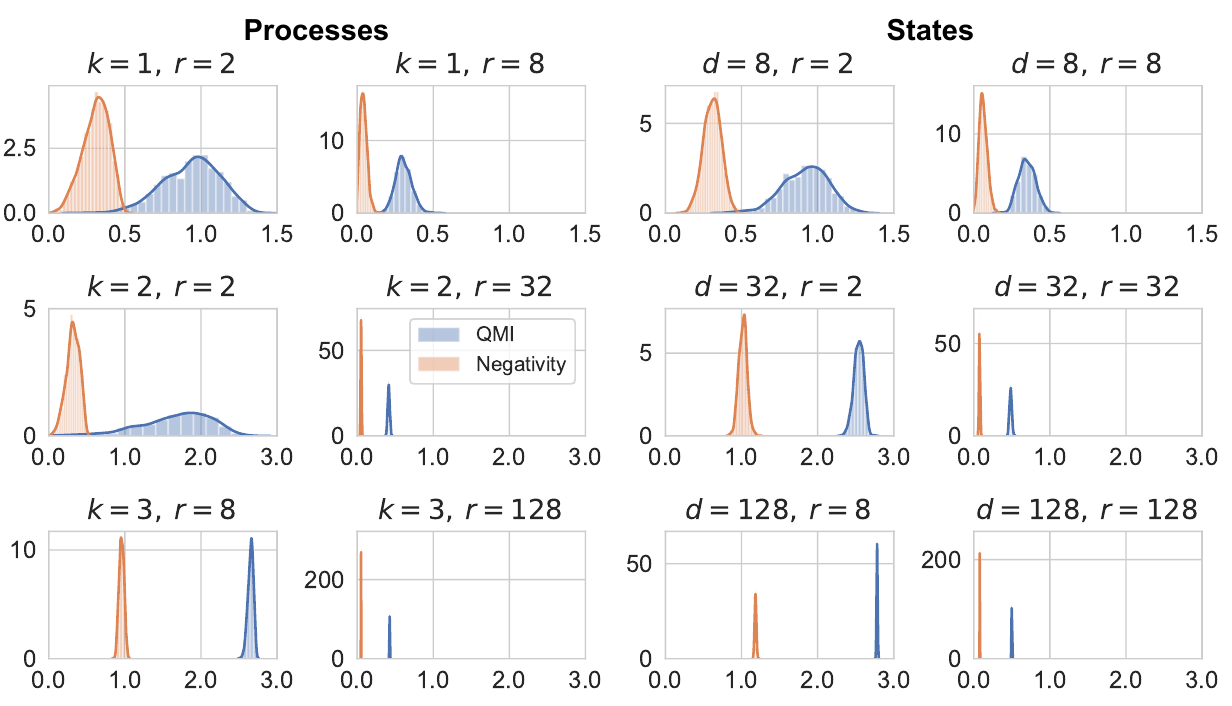}
	\caption[Correlation histograms for randomly generated process tensors of high and low rank ]{Histograms of total correlation and bipartite entanglement for 1000 randomly sampled processes and states of high and low rank. We see the same general trend, that typical processes can have highly complex temporal correlations much like typical states. However we note the widening of these distributions, and lower typical mean. This is most clearly seen when comparing the $d=32$, $r = 2$ cases for states and processes.}
	\label{fig:process-properties}
\end{figure}

\subsection{Sampling Typical Process Tensors}

We will employ our method to generate random $k-$step process tensors of rank $r$ throughout this thesis, but to begin we will showcase some numerics that demonstrate some properties of typical process tensors. In addition to the technique itself, what we wish to impress is that multi-time processes can be as complex as many-body quantum states. This is part of a recurring theme to explore notions of many-time physics. Figure~\ref{fig:process-properties} shows a series of histograms displaying the populations of different properties of both states and processes. For $k=1,2,$ and $3$, we draw random low-rank and high-rank process tensors, as well as equivalent dimension states. We compute the generalised quantum mutual information $\mathcal{N}(\Upsilon_{k:0})$,
as well as negativity across the respective bipartitions 
\begin{equation}
	\begin{split}
		&\{\{\mathfrak{o}_1,\mathfrak{i}_1\}, \{\mathfrak{o}_0\}\},\\
		&\{\{\mathfrak{o}_2,\mathfrak{i}_2\}, \{\mathfrak{o}_1,\mathfrak{i}_1,\mathfrak{o}_0\}\},\quad\text{and}\\
		&\{\{\mathfrak{o}_3,\mathfrak{i}_3,\mathfrak{o}_2,\mathfrak{i}_2\}, \{\mathfrak{o}_1,\mathfrak{i}_1,\mathfrak{o}_0\}\}.
	\end{split}
\end{equation}
We observe several key features. First, note that in the typical cases, both quantum and total correlation between states and processes are commensurate. 
Causality constraints do not significantly limit the quantum properties that a process may exhibit.

\begin{figure}[!b]
	\centering
	\includegraphics*[width=\linewidth]{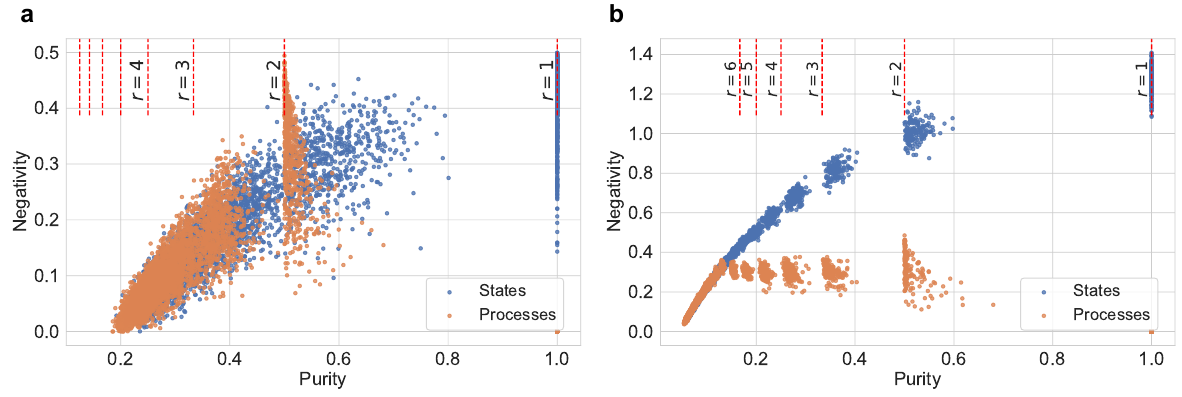}
	\caption[Relationship between purity and negativity in randomly generated states and processes ]{Relationship between purity and negativity in randomly generated states and processes at different ranks, \textbf{a} for $k=1,d=8$ and \textbf{b} for $k=2,d=32$. We see the effects of causality more clearly here. Typical processes do not follow the same purity-negativity trend as typical states: the features are more tightly distributed, and tend to zero where $\gamma = 1$.}
	\label{fig:process-purity-neg}
\end{figure}

It is also instructive to observe the general trend of how entanglement behaves as a function of purity in typical cases. As a general rule, the more pure a random quantum state is, the more entangled it is. However, a maximally pure process is necessarily a Markovian one for fixed steps. Rather, then, than total correlation scaling roughly as a function of purity, processes maximise their correlation at an intermediate purity -- where the environment is not too large to scramble, but not so pure as to lack an environment that mediates correlations at all. We show numerically the relationship between temporal entanglement and purity of a process for single and two-step process tensors in Figure~\ref{fig:process-purity-neg}.

\begin{figure}[!h]
	\centering
	\includegraphics*[width=0.45\linewidth]{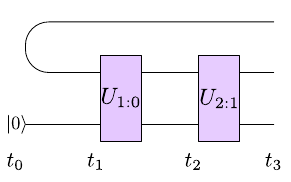}
	\caption[Example of a unital process with bipartite temporal entanglement ]{An example of a two-step process which is unital, and also exhibits bipartite entanglement between $\hat{\mathcal{E}}_{2:1}$ and $\hat{\mathcal{E}}_{1:0}$. Although this provides a example that unital processes need not only be separable, it is highly atypical and numerically we find that randomly generated unital dynamics do not generally produce temporal entanglement. }
	\label{fig:entangled-unital}
\end{figure}

Finally, one might ask the role of unital dynamics in the generation of temporal entanglement. Non-unitality is typically associated with quantumness of the dynamics, but in the unital case it can be harder to discern. When drawing random bistochastic process tensors, we observe no bipartite entanglement for single or two-step processes. This is not because entanglement in unital processes is impossible: consider the $k=2$ example depicted in Figure~\ref{fig:entangled-unital}. Subsequent interactions with the same maximally mixed state are unital, because the environment contains no coherence, but the process will be entangled for any choices of $U_{1:0}$ and $U_{2:1}$ that overlap with a SWAP gate.

This implies that the structure of unital quantum processes has measure zero on the space of unital process tensors. Surprisingly, we find this not to be the case once we move to $k=3$. Of course, mathematically unlikely does not necessarily mean physically unlikely. Still, this provides some impetus for probing processes at a time scale of three steps rather than two. 

\section{Discussion}

In this chapter, we have developed and analysed many properties of quantum stochastic processes. Our overarching aim has been to provide the reader with a foundational understanding of some of the key concepts that will be discussed in the remainder of the thesis.

One important concept we have highlighted is the spacetime structure of quantum theory. 
A quantum system is propagated through time by a series of apparatuses that manipulate it from one time to the next. These apparatuses, or user-controlled operations, allow us to extract information from the spacetime structure of the system. The type of information that can be extracted is dependent on the extent to which the system can be controlled. As we will see in the remainder of the thesis, this information is important both from a practical perspective (for example, in the reconstruction of quantum states) and for gaining physical insights into the behaviour of quantum systems.

We have considered two extremes of quantum processes. On the one hand, we have shown that entire classes of quantum dynamics give rise to simple processes, meaning that their memory can be replaced by a classical agent. On the other hand, we have shown that complex structures are generic, meaning that they are not confined to a measure-zero corner of Hilbert space.

There are several areas that could benefit from the study of quantum stochastic processes, including the understanding of noise in quantum devices and the behaviour of open quantum systems. This research is important for the development of fault-tolerant quantum technologies. Additionally, studying the emergence of spatiotemporal structures through open quantum systems can provide insights into the complexity of physical processes and the connections between quantum chaos and quantum thermodynamics.

We will explore these topics in more detail, looking at the role of quantum noise in open quantum systems and the implications of temporal quantum correlations. By examining the distribution of entanglement through complex environments, we can gain a better understanding of the fundamental properties of quantum theory and its applications in the real world.


\chapter{Characterisation of non-Markovian quantum processes with process tensor tomography}
\label{chap:PTT}
\epigraph{\emph{Memory is the seamstress, and a capricious one at that. Memory runs her needle in and out, up and down, hither and thither.}}{Virginia Woolf, Orlando}
\vspace{-0.5cm}
\noindent\colorbox{olive!10}{%
	\begin{minipage}{0.9555\textwidth} 
		\textcolor{Maroon}{\textbf{Chapter Summary}}\newline
		In this chapter, we formally generalise the practice of quantum process tomography to the multi-time regime, expanding to the much broader class of non-Markovian dynamics. The resulting estimate captures all multi-time correlations from memory effects, and serves as a reliable mapping from experimenter-chosen controls to observed circuit outcomes in arbitrary open quantum systems. This resolves a long-standing open problem about how to systematically characterise temporal correlations in a quantum setting. 
		\par\vspace{\fboxsep}
		\colorbox{cyan!10}{%
			\begin{minipage}{\dimexpr\textwidth-2\fboxsep}
				\textcolor{RoyalBlue}{\textbf{Main Results}}
				\begin{itemize}
					\item We introduce process tensor tomography, a means of estimating arbitrary quantum stochastic processes on controllable quantum devices.
					\item We develop a linear inversion reconstruction and demonstrate this on IBM Quantum devices, achieving average reconstruction fidelities of 99.9\%.
					\item We construct post-processing algorithms to obtain the maximum likelihood process tensor estimate for given dynamics. This reduces resource costs, improves accuracy, and produces a physically meaningful estimate. We analyse performance and assumptions of the approach in a series of synthetic and experimental tests.
					\item We develop techniques for the implementation of arbitrary quantum instruments given access to ancilla qubits, and use these to obtain full reconstruction of multi-time statistics.
				\end{itemize}
		\end{minipage}}
\end{minipage}}
\clearpage

\section{Introduction}
Central to the theme of progress in quantum computing has been the development and application of \acs{QCVV} procedures~\cite{Eisert2020,Endo2018,Ferracin2019,white-POST,Harper2020,Jurcevic2021,RBK2017}. These techniques model and identify the presence of errors in \acs{QIP}s. These errors may have different origins, such as coherent control noise, decoherence, crosstalk, or \acs{SPAM} errors.
The theoretical machinery for open quantum system dynamics is well-oiled in low-coupling cases, but strong environmental interactions can lead to non-trivial dynamical memory effects that are difficult to understand, much less control. 
The recent advent of high performance quantum information processors has precipitated greater sensitivity to complex dynamical effects.
In particular, it is clear that device behaviour must be understood under a relaxed Markov assumption~\cite{Haase2018, IBMStateFid,Sarovar2020detectingcrosstalk}. 
The resulting non-Markovian dynamics includes more general errors that may be temporally correlated or dependent on broader environmental context~\cite{Rivas2014, Li2018, fattah}.
Characterisation techniques of quantum devices such as randomised benchmarking and \acs{GST} have so far represented the front line in understanding and addressing noise~\cite{PhysRevLett.78.390, gst-2013, RBK2017, PhysRevA.87.062119, white-POST}.
However, constructing a digestible picture of non-Markovian behaviour has proven difficult, and, as we have seen, violates the error model assumed in these methods.
Chiefly, this is because quantum correlations can forbid the division of dynamical processes into arbitrary steps of \acs{CP}, linear maps~\cite{Pechukas1994}.
If information back-flow from the environment can occur, then noisy effects can be influenced by past factors; this detail can no longer be `forgotten'.

\par 
For device control, this is problematic.
The circuit model of quantum computation is predicated on identical gates implemented at different times having identical actions.
Markovian errors multiply out and propagate in predictable ways.
However, non-Markovian noise gives rise to adverse effects that are much more challenging to tame. 
For example, correlated errors can spread across the device, and have been shown to lower thresholds of quantum error correcting codes~\cite{correlated-qec,PhysRevA.99.052351}.
Similarly, context-dependent gates allow for poorly understood forms of dynamical errors not describable by a Markov model. 
This is one of the largest obstacles to near-term \acs{QIP}s: non-Markovian noise must be either eliminated or, as some have suggested, harnessed into a resource~\cite{non-Markovian-control,Pineda2016,Kumar2018,Bylicka2013,berk2021extracting}. 
\par
It therefore behoves us to develop \ac{QCVV} procedures capable of capturing the pernicious effects that arise as a consequence of complex background processes. Such a procedure should satisfy several desiderata:
\begin{enumerate}
	\item It should be concerned with multi-time quantum dynamics -- a system considered for at least three chosen times across some continuous window.
	\item It should not be tailored to any specific physical model -- rather than, say, fitting parameters to a particular master equation, it should be fully agnostic to the physics under consideration.
	\item It should come equipped with certification or convergence guarantees -- \ie{} a way to fully specify the model in a finite number of experiments.
	\item It should be predictive in some capacity -- the characterisation should be able to extrinsically predict the behaviour of experiments which were not used as data for the characterisation, to an arbitrary degree of accuracy.
	\item Finally, the procedure should be applicable for generic $SE$ interactions across arbitrary timescales.	
\end{enumerate}
In this chapter, we develop such a procedure and demonstrate it across an extensive range of experimental and numerical results.
Because the process tensor framework is able to mathematically describe arbitrary non-Markovian processes, it serves as a natural template to the characterisation of multi-time processes. Our procedure is hence concerned with estimating process tensor representations for given dynamics, and we term it \ac{PTT}. We justify the claim that this constitutes a complete generalisation of quantum process tomography to the multi-time, non-Markovian domain. Therefore, this resolves a long-standing open problem of experimentally characterising non-Markovian quantum dynamics. \par 

Although we introduce this predominantly in the context of quantum noise, we emphasise that the technique generically reconstructs any quantum stochastic process, and will be pertinent to all addressable quantum systems. Noise on quantum devices has the desirable property that it is naturally occurring, the devices in question are already highly programmable, and the upside for understanding these dynamics is particularly great.
Specifically, we consider the generic case of an open quantum system coupled to any arbitrary environment across a number of time steps. The experimenter has some access to the system, and can run sequences of controllable experiments. As we saw in Chapter~\ref{chap:stoc-processes}, sequences of operations constitute observables on a process tensor via the spatiotemporal Born rule. The challenge, therefore, is to (i) run a tomographically complete set of experiments to fix a process tensor, and (ii) process the relevant data to then estimate that process tensor.


First, we develop a linear inversion protocol that uses the process tensor as a framework to reconstruct a given set of dynamics. This is the simplest and least computationally-intensive method of estimation. We demonstrate on a set of simulated and experimental data the ability to accurately predict observables of non-Markovian processes. It has the drawbacks that it provides no reliable intrinsic information about a process, only dynamical predictions. It is also highly sensitive to finite sampling error in the estimates.
We rectify these issues by developing methods to determine optimal basis choice for the purpose of characterisation. We also design set of algorithms designed to produce physically meaningful maximum likelihood estimates of process tensors from given dynamics. The resulting characterisation captures a sequence of possibly-correlated \acs{CPTP} maps across a given window for any non-Markovian dynamics. Importantly, the post-processing supplies the physical process tensor (positive and causal) that maximises the likelihood function for a given set of experimental data. We show across a range of simulated and experimental data that this both greatly improves accuracy, and reduces resource overhead in its experimental demands.

\section{Problem Setup}
\begin{wrapfigure}{r}{0.6\textwidth}
	\centering
	\includegraphics[width=0.95\linewidth]{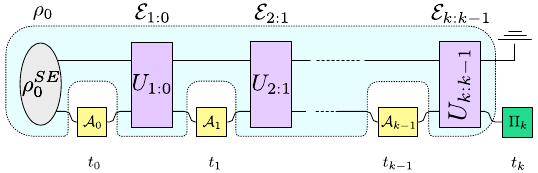}
	\caption[Circuit depiction of the process tensor tomography setup. ]{Circuit depiction of the process tensor tomography setup. A sequence of gates $\mathcal{A}_i$ at times $t_i$ drive a system among interleaving $SE$ interactions, followed by a final terminating measurement.}
	\label{fig:qspimage}
\end{wrapfigure}

The subject of our characterisation is a quantum stochastic process. \acs{QSP}s are defined by four properties: the system $S$ under scrutiny, the environment $E$ with which $S$ interacts, the time window $\mathbf{T}_k:= \{t_0,t_1,\cdots,t_k\}$ across which it evolves, and the corresponding $SE$ unitaries $\{U_{1:0},U_{2:1},\cdots,U_{k:k-1}\}$ governing the evolution of the system from one time to the next\footnote{Although, note that different $SE$ unitaries can lead to the same effective dynamics on the system}. First and foremost, in performing a non-Markovian generalisation of \acs{QPT}, the experimenter must first choose the \acs{QSP} that they would like to characterise -- just as a two-time process $\mathcal{E}$ must be first selected in \ac{QPT}. Consider, as an example, the circuit diagram shown in Figure~\ref{fig:qspimage}. 

Any continuous-time \ac{QSP} can be discretised into a number of time-steps $\mathbf{T}_k$ (for example, in the context of a quantum circuit). The overarching aim here is to determine all of the multi-time correlations that exist across $\mathbf{T}_k$, and hence these times constitute locations at which the system is probed. 
Since the environment and the interacting unitaries are open for selection in this context, these can be naturally occurring dynamics for the study of noise in a quantum device. Alternatively, these can be engineered for the purpose of studying simulated open quantum systems as an intended application of quantum hardware.\par 

The $k$-step process is driven by a sequence $\mathbf{A}_{k-1:0}$ of control operations, each represented mathematically by \acs{CP} maps: $\mathbf{A}_{k-1:0}:=\{\mathcal{A}_0,\mathcal{A}_1,\cdots,\mathcal{A}_{k-1}\}$. As a result of both the open dynamics and the control operations, the system is brought at the end of the circuit to state $\rho_k(\mathbf{A}_{k-1:0})$, conditioned on this choice of interventions. These controlled dynamics have the form:
\begin{equation}\label{eq:multiproc}
	\rho_k\left(\textbf{A}_{k-1:0}\right) = \text{Tr}_E [U_{k:k-1} \, \mathcal{A}_{k-1} \cdots \, U_{1:0} \, \mathcal{A}_{0} (\rho^{SE}_0)],
\end{equation}
where $U_{k:k-1}(\cdot) = u_{k:k-1} (\cdot) u_{k:k-1}^\dag$. Recall that Equation~\eqref{eq:multiproc} can be used to define a mapping from past controls $\mathbf{A}_{k-1:0}$ to future states $\rho_k\left(\textbf{A}_{k-1:0}\right)$, which is the process tensor $\mathcal{T}_{k:0}$:
\begin{equation}
	\label{eq:PT}
	\mathcal{T}_{k:0}\left[\mathbf{A}_{k-1:0}\right] = \rho_k(\mathbf{A}_{k-1:0}).
\end{equation}
The object of the characterisation, therefore, is to determine the mapping $\mathcal{T}_{k:0}$ which is valid for arbitrary sequences of instruments. This constitutes everything that can be determined about the \ac{QSP}, as it has been defined, and as has been discussed in Chapter~\ref{chap:stoc-processes}. 



\section{Linear Inversion Process Tensor Tomography}
\label{sec:LI-PTT}
We start first with a linear inversion reconstruction of the process tensor representation of some \ac{QSP}. `Linear inversion' in this context refers to the fact that any linear operator is uniquely defined by the input-output relations on a complete basis. Abstractly speaking, a Hermitian linear operator $A$ on a Hilbert-Schmidt space $\mathcal{B}(\mathcal{H})$ produces physically observable output $p$ from some operator $M$ on the dual space via the Hilbert-Schmidt inner product: 
\begin{equation}
	p = \Tr[AM^\dagger ].
\end{equation}
Vectorising the operators $|A\rangle\!\rangle:=\text{vec}(A)$ and $\langle\!\langle M|:=\text{vec}(M)^\dagger$, the same inner product can be expressed as
\begin{equation}
	p = \langle\!\langle M | A\rangle\!\rangle.
\end{equation}
We might interpret in this case $M$ to be the input choice to operator $A$ with corresponding output $p$. Now, supposing instead we have a list of inputs $\{M_i\}_{i=1}^N$ and outputs $\{p_i\}_{i=1}^N$, we can collect each row vector $\langle\!\langle M_i|$ into a single feature matrix, and each output into a single feature vector, this relationship can be extended to read
\begin{equation}
	\begin{pmatrix} \langle\!\bra{M_1} \\
		\langle\!\bra{M_1}\\
		\vdots \\
		\langle\!\bra{M_N}\\
	\end{pmatrix} \cdot \ket{A}\!\rangle = \begin{pmatrix} p_1\\p_2\\\vdots\\p_N\end{pmatrix},
\end{equation}
or, equivalently
\begin{equation}
	\mathcal{M}\cdot|A\rangle\!\rangle = \vec{p}.
\end{equation}
As long as $\mathcal{M}$ has full row rank ($N\geq \text{dim}(\mathcal{B}(\mathcal{H}))$), the matrix can be (pseudo-)inverted to uniquely determine the operator $A$:
\begin{equation}
\label{eq:LI-principle}
|A\rangle\!\rangle = \mathcal{M}^+ \vec{p}.
\end{equation}
In the case where $N >  \text{dim}(\mathcal{B}(\mathcal{H}))$, the set of equations may not be consistent. Applying the pseudoinverse allows one to find the least-squares minimiser of this set of inconsistent equations. In the remainder of this section, we apply these tenets specifically to the problem of determining a process tensor mapping for a QSP.

\subsection{Derivation}

\begin{figure}[h!]
	\centering
	\includegraphics[width=\linewidth]{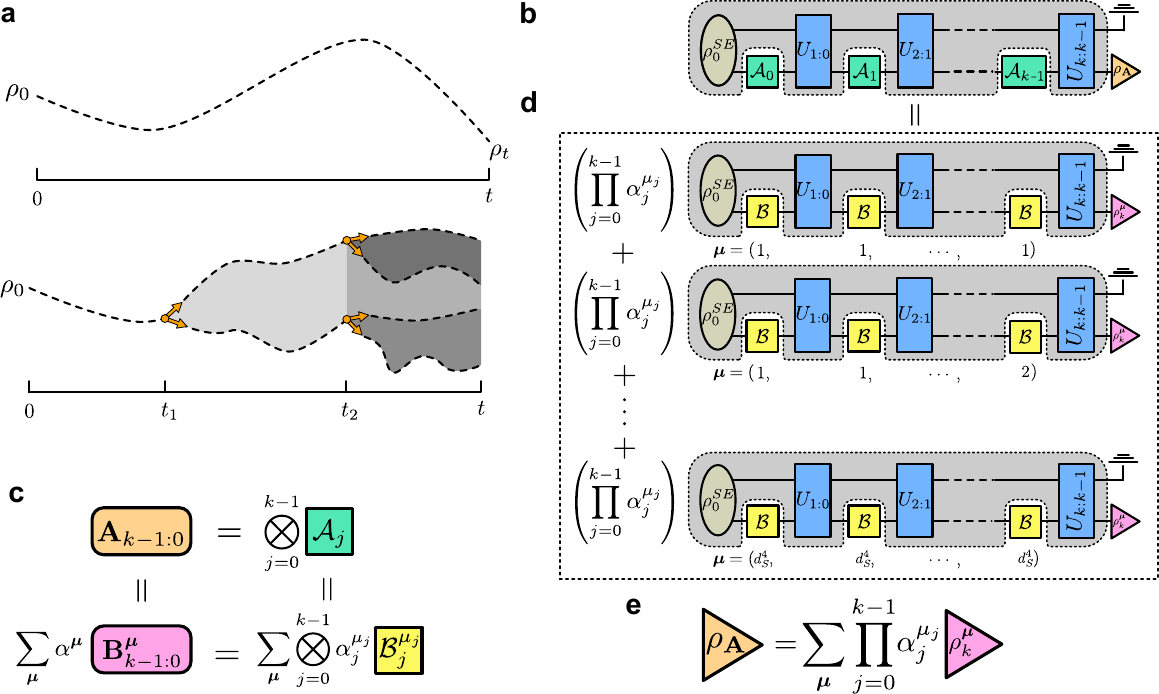}
	\caption[An illustrative summary of linear inversion process tensor tomography ]{An illustrative summary of process characterisation. \textbf{a} The state of an open system over time follows a trajectory through state space until some final time at which the state is probed (top). By applying control operations at times $t_1$ and $t_2$, an experimenter can anchor and change the trajectory, which can be inferred via a linear combination of trajectories corresponding to basis operations (bottom).
		\textbf{b} A circuit model showing a sequence of operations $\{\mathcal{A}_j\}$ interleaved with $SE$ interactions, resulting in a final state $\rho_{\mathbf{A}}$.
		\textbf{c}
		A sequence of operations $\mathbf{A}_{k-1:0}$ can be expressed as a tensor product of independently chosen operations $\mathcal{A}_j$ at each time step. These can then be individually decomposed into a chosen basis $\{\mathcal{B}_j^{\mu_j}\}$ together giving a basis of sequences $\{\mathbf{B}_{k-1:0}^{\vec{\mu}}\}$.
		\textbf{d}
		A process can be fully characterised by measuring the output state for a complete set of basis operations at different times. Then, an arbitrary process can be expressed as a linear combination of each basis process; because of the linear construction, the intermediate evolution is completely preserved in the description of the arbitrary process. 
		\textbf{e} The final state density matrix for the process $\mathbf{A}_{k-1:0}$ can be expressed by tracing over all of the intermediate operations, contracting to a coefficient expansion for the measured density matrices in the basis processes. This is the same density matrix as in \textbf{b}.}
	\label{fig:PT_summary}
\end{figure}

The set of possible sequences of \ac{CP} maps $\mathbf{A}_{k-1:0}$ forms a product vector space, built up from the spaces of temporally local operations; in particular, the Choi representation $\hat{\mathbf{A}}_{k-1:0} = \bigotimes_{j=0}^{k-1}\hat{ \mathcal{A}}_j$ when the operations at each time are chosen independently. 
As such, the process tensor is completely characterised by its input-output relations on a complete basis of control operations, just as a quantum channel is unambiguously defined by its input-output relations on a complete basis of states.
Each time-local operation may be expanded into a basis $\{\hat{\mathcal{B}}^{\mu_j}_j\}$ such that any \ac{CP} map can be expressed as $\hat{\mathcal{A}}_j = \sum_{\mu_j=1}^{d^4} \alpha_j^{\mu_j} \hat{\mathcal{B}}^{\mu_j}_j$. The subscript $j$ allows for the possibility of a different basis at each time, meanwhile the superscript $\mu_j$ denotes the elements of that particular set. The complete spatio-temporal basis is
\begin{equation}
	\{\hat{\mathbf{B}}_{k-1:0}^{\vec{\mu}}\} = \left\{\bigotimes_{j=0}^{k-1}\hat{\mathcal{B}}_j^{\mu_j}\right\}_{\vec{\mu}=(1,1,\cdots,1)} ^{(d^4,d^4,\cdots,d^4)}
\end{equation}
with vector of indices $\vec{\mu}$. To construct the process tensor, therefore, it suffices to measure the output $\rho_k^{\vec{\mu}} := \rho_k(\mathbf{B}_{k-1:0}^{\vec{\mu}})$ for each $\vec{\mu}$, see Figure~\ref{fig:PT_summary}c. 
Then the process tensor's action is defined by
\begin{gather}
	\label{DM-expansion}
	\rho_k(\mathbf{A}_{k-1:0}) = \sum_{\vec{\mu}} \alpha^{\vec{\mu}} \ \rho_k^{\vec{\mu}} \quad \text{with} \quad
	\rho_k^{\vec{\mu}} :=
	\mathcal{T}_{k:0}[\mathbf{B}^{\vec{\mu}}_{k-1:0}].
\end{gather}
In other words, to reconstruct the process tensor we need to experimentally estimate $\rho_k^{\vec{\mu}}$ for all $\vec{\mu}$, this is depicted in Figure~\ref{fig:PT_summary}d.

In general, to select a basis, it suffices to simply choose a collection of $d_S^4$ random instruments. This collection will be linearly independent with unit probability. However, a sufficient basis will often deviate greatly from an optimal basis. In general, the quality of the characterisation will depend on the conditioning of the basis~\cite{QPT-projection}. That is to say, the mutual overlap between different basis vectors. We shall first consider the ideal tomographic case (with no sampling noise) and return to a discussion of optimal basis choices at a later point.

To construct the tomographic representation of $\mathcal{T}_{k:0}$, we make use of the dual set $\{\Delta_j^{\mu_j}\}$ such that $\Tr[\mathcal{B}_j^{\mu_j}\Delta_j^{\nu_j}] = \delta_{\mu_j\nu_j}$. Then, the Choi state $\Upsilon_{k:0}$ of the process tensor $\mathcal{T}_{k:0}$ is given by
\begin{equation}
	\label{choi-pt}
	\Upsilon_{k:0} = \sum_{\vec{\mu}} \rho_k^{\vec{\mu}}\otimes \mathbf{\Delta}_{k-1:0}^{\vec{\mu}\ \text{T}},
\end{equation}
where $\{\mathbf{\Delta}_{k-1:0}^{\vec{\mu}}\} = \{\bigotimes_{j=0}^{k-1}\Delta_j^{\mu_j}\}$ satisfies $\Tr[\mathbf{B}_{k-1:0}^{\vec{\mu}}\mathbf{\Delta}_{k-1:0}^{\vec{\nu}}] = \delta_{\vec{\mu}\vec{\nu}}$. 

\textbf{Constructing a dual set}

The procedure to construct the dual operators is as follows, reproduced from Ref.~\cite{Pollock2018a}. For a complete set of linearly independent operations $\{\mathcal{B}_i\}$ whose Choi matrices are $\{\hat{\mathcal{B}}_i\}$, with dimension $d:=\text{dim}(\mathcal{B}(\mathcal{B}(\mathcal{H}_S))) = d_S^4$. We can compile the basis into a single matrix $\mathsf{B}$.
Write each $\hat{\mathcal{B}}_i = \sum_j b_{ij}\Gamma_j$, where $\{\Gamma_j\}$ form a Hermitian, self-dual, linearly-independent basis satisfying $\text{Tr}[\Gamma_j\Gamma_k]=\delta_{jk}$. 
In our case, we select $\{\Gamma_j\}$ to be the standard basis, meaning that the $k$th column of the matrix $\mathsf{B} = \sum_{ij}b_{ij}\ket{i}\!\bra{j}$ is $\hat{\mathcal{B}}_k$ flattened into a $1$D vector. 
Because the $\{\hat{\mathcal{B}}_i\}$ are linearly independent, $\mathsf{B}$ is invertible. 
Let the matrix $\mathsf{F}^\dagger = \mathsf{B}^{-1}$ such that $\mathsf{B}\cdot \mathsf{F}^\dagger = \mathbb{I}$.
This means that the rows of $\mathsf{F}^\dagger$ are orthogonal to the rows of $\mathsf{B}$.
The dual matrices can then be defined as $\Delta_i = \sum_j f_{ij}\Gamma_j$, ensuring that $\text{Tr}[\hat{\mathcal{B}}_i\Delta_j] = \delta_{ij}$.
We will discuss the relationship of $d$ to $n$ in later sections. The basis may be undercomplete, for which the dimension $d$ of the space is less than the order $n$ of the matrices. Alternatively, the basis may be overcomplete, giving up linear independence and taking $n>d$. We may also have overcompleteness on a subspace of the full vector space such that $n>d$, but $\text{dim}(\text{span}(\{\hat{\mathcal{B}_i}\})) < d$. In all instances, therefore we construct $\mathsf{F}^\dagger$ as the Moore-Penrose or the right inverse of $\mathsf{B}$.


\par 
As shown in previous chapters, the action of the process tensor with respect to a sequence of operations is found by projecting the process tensor onto the transposed Choi state of this sequence. The expansion coefficients from Equation~\eqref{DM-expansion} are implicitly calculated here, but not directly used.
Below, we explicitly step through this computation, which works by construction. 
We start with the mapping $\mathcal{T}_{k:0}[\mathbf{A}_{k-1:0}]$, which we would like to produce the final state $\rho_k(\mathbf{A}_{k-1:0})$. Here, we notationally collect the sequence of operations to one input space abbreviated as `in'. First,
\begin{equation}
	\label{eq:PT-action}
\mathcal{T}_{k:0}\left[\mathbf{A}_{k-1:0}\right] = \text{Tr}_{\text{in}} \left[\left(\mathbb{I}_{\text{out}}\otimes \hat{\mathbf{A}}_{k-1:0}\right)^\text{T}
        \Upsilon_{k:0}\right]
\end{equation}
Decomposing the set of operations $\mathbf{A}_{k-1:0}$, and substituting the tomographic expression~\eqref{choi-pt} yields
\begin{equation}
\text{Tr}_{\text{in}}\left[\left(\bigotimes_{i=0}^{k-1}\hat{\mathcal{A}}_i^\text{T} \otimes \mathbb{I}\right)\sum_{\vec{\nu}} (\mathbf{\Delta}^{\vec{\nu}}_{k-1:0})^\text{T}\otimes \rho_k^{\vec{\nu}}\right],
\end{equation}
for which the sequence linearly expands into its chosen basis:
\begin{equation}
\text{Tr}_{\text{in}}\left[
        \sum_{\vec{\mu}}
        \alpha^{\vec{\mu}}
        \bigotimes_{i=0}^{k-1} 
        \hat{\mathcal{B}}_i^{\mu_i \text{T}}
        \sum_{\vec{\nu}} 
        \bigotimes_{j=0}^{k-1}
        \Delta_j^{\nu_j \text{T}} \otimes \rho_k^{\vec{\nu}}\right].
\end{equation}
We then have an inner product between the basis elements and the chosen duals,
	\begin{align}
	&	\text{Tr}_{\text{in}}\left[
		\sum_{\vec{\mu},\vec{\nu}} 
		\alpha^{\vec{\mu}}
		\bigotimes_{i,j=0}^{k-1} \{\hat{\mathcal{B}}_i^{\mu_i \text{T}} \Delta_j^{\nu_j \text{T}}\}\otimes \rho_k^{\vec{\nu}}\right],\\
		&=\sum_{\vec{\mu},\vec{\nu}} 
		\alpha^{\vec{\mu}}
		\prod_{i,j=0}^{k-1}  \,
		\text{Tr}\left[
		\hat{\mathcal{B}}_i^{\mu_i} \Delta_j^{\nu_j}\right]
		\rho_k^{\vec{\nu}}.
	\end{align}
By construction, the duals are orthonormal to the basis elements, and this expression hence reduces to:
\begin{align}
	&\sum_{\vec{\mu},\vec{\nu}}
	\alpha^{\vec{\mu}}
	\prod_{i=0}^{k-1}  \, \delta_{\vec{\mu}\vec{\nu}} \, \rho_k^{\vec{\nu}},\\
	&=\sum_{\vec{\mu}} \alpha^{\vec{\mu}}\rho^{\vec{\mu}}_k,\\
	&=\rho_k(\textbf{A}_{k-1:0}).
\end{align}
If desired, the direct calculation of each expansion coefficient is therefore given by 
\begin{gather}
	\label{eq:coeff-calc}
\begin{split}
    \alpha^{\vec{\mu}} =& \text{Tr}\left[\hat{\mathbf{A}}_{k-1:0}\mathbf{\Delta}_{k-1:0}^{\vec{\mu}}\right],\\
    =& \text{Tr}\left[
    \bigotimes_{i=0}^{k-1} \hat{\mathcal{A}}_{i}
    \Delta^{(\mu,i)}\right],\\
    =& \prod_{i=0}^{k-1}
    \text{Tr}\left[
     \hat{\mathcal{A}}_{i}
    \Delta_i^{\mu_i }
    \right] = \prod_{i=0}^{k-1} \alpha_i^{\mu_i}.
    \end{split}
\end{gather}

One might note at this point computational differences between \pt{} as an object and its action $\mathcal{T}_{k:0}[\mathbf{A}_{k-1:0}]$. \pt{} is a matrix of order $d_S^{2k+1}$, and hence Equation~\eqref{eq:PT-action} requires multiplying out matrices of this size and taking a partial trace, with requirements scaling like $\mathcal{O}(d_S^{6k})$. However, we may also observe from Equation~\eqref{eq:coeff-calc} that the process tensor action needs only the linear combinations of measured $\rho_k^{\vec{\mu}}$ and inner products between basis operations and their duals. The cost of this is hence $\mathcal{O}(d_S^{12})$.  

\subsection{Restricted Process Tensor}
To reconstruct the process tensor, a minimal complete basis for the process tensor requires $d_S^4$ operations spanning the superoperator space $\mathcal{B}(\mathcal{B} (\mathcal{H}_S))$ at each time-step.
One mathematically convenient basis is an \ac{IC} \ac{POVM}, followed by a set of \ac{IC} preparation states which are independent of the measurement outcomes. However, this level of control is difficult to attain in quantum hardware -- not least because it requires electronics that can read out classical information and then send a series of instructions to manipulate the state in a timeframe much shorter than the qubit's coherence time. 
The approach we take in this work aims to be as cognisant as possible of experimental realities and constructing working frameworks in non-idealised settings. Throughout the course of this PhD, immense experimental progress has been made on the front of so-called dynamic quantum circuits with mid-circuit measurement capabilities and conditional feed-forward but the resulting control is still noisy, slow compared to background dynamics, and not universally available. If the objective of the characterisation is to study simulated quantum processes, this noise may be small compared to the effects under study, and thus permissable. However, if the aim is to benchmark the background noise of the device itself, then control noise commensurate with background noise will obfuscate any potential effects. \par



\par
For a typical \acs{NISQ} device, all intermediate operations are limited to unitary transformations, and a measurement is only allowed at the end -- or if mid-circuit measurement with feed-forward is possible, it is typically much slower than $SE$ dynamics. Nevertheless, it is possible to work within the experimental limitations and implement an informationally incomplete set of basis operations. This constructs what is known as a \emph{restricted} process tensor~\cite{PT-limited-control} and has full predictive power for any operation in a subspace of operations. As we discussed in the previous chapter, unitary operations capture Pauli expectation values of the form $\langle \tilde P_{\mathfrak{i}_j}\tilde{P}_{\mathfrak{o}_{j-1}}\rangle$. 
That is, these objects are well-defined as maps over the span of their incomplete basis, but do not form positive operators and do not uniquely fix a process tensor's Choi state.\par

We expand on the notion of a restricted process tensor here by offering an analogous quantum state perspective. The measurement of $\rho_k$ subject to some sequence of operations $\mathbf{B}^{\vec{\mu}}_{k-1:0}$ is akin to measuring a many-body observable on $\Upsilon_{k:0}$, as per Equation~\eqref{eq:PToutput}. The reconstruction of $\Upsilon_{k:0}$ then lies on the hyperplane defined by
\begin{gather}
	\text{span}(\{\Pi_i,\mathbf{B}^{\vec{\mu}}_{k-1:0}\}).
\end{gather}
When the set of operations is tomographically incomplete, $\Upsilon_{k:0}$ is non-uniquely fixed, thus termed `restricted'. Consider, for example, in the state case, if only $Z$ and $X$ measurements were performed on each subsystem. Then a consistent, non-unique state could be estimated with the correct expectation values for any real Hermitian operator. However, the $Y$ expectations would be a free parameter (up to positivity of the state).\par 

In the special case of unitary control, the Choi states representing each intervention are rank one projections onto a maximally entangled state of dimension $d_S^2$. Expanding these entangled measurements into a Hermitian basis yields only non-zero coefficients on the non-local terms. That is, in the Pauli basis for example, we have $\text{Tr}(\hat{\mathcal{B}}_l^{\mu_l}\cdot \mathbb{I}^{(\mathfrak{i}_{l+1})}\otimes P_j^{(\mathfrak{o}_l)}) = \text{Tr}(\hat{\mathcal{B}}_l^{\mu_l}\cdot P_i^{(\mathfrak{i}_{l+1})}\otimes \mathbb{I}^{(\mathfrak{o}_l)}) = 0\:\forall \:i,j$. Unitaries are consequently fully orthogonal to the span of non-unital (where the maximally mixed state is mapped to something more pure) and trace-decreasing (non-deterministically applied) maps. Linear inversion reconstruction of the process tensor, then, omits these local expectation values. \par 

A restricted process tensor is not a model restriction as such, but an observational restriction. Its properties are fully consistent with the discussed facets of full process tensors, but the mapping is only valid across the span of observed data. This means, for example, that it will provide a recipe for complete control (under the restriction) which is fully inclusive of non-Markovianity. However, the actual strength of the memory can only be inferred rather than directly measured. For example, in the absence of measurement causal breaks, correlations between past and future measurement statistics cannot be established. Moreover, any measure relying on a full eigendecomposition of the state (such as quantum mutual information, for example) is similarly out of reach. The relevant analogy then is performing \acs{QST} without measuring in all possible bases. The state will not be fully determined, but the information provided through Born's rule to predict the future will still be valid, so long as predictions are made within linear combinations of the measured bases.
Working within the constraints of \acs{NISQ} devices, we need to account for $d_S^4 - 2d_S^2 + 2$ unitary operations at each time-step. For a qubit, this amounts to $N=10$ unitary gates per time step. However, any estimation procedure will come with sampling error, leading to both an incorrect and unphysical representation of the map.

We emphasise that restricted process tensor is not a single definition but rather a concept that filling the superoperator space is often an unreasonable expectation and that for the purpose of characterisation, one should think about classes of restriction -- especially when we come to multi-qubit processes. In this Thesis, we will use `restricted' in context to mean to the class of unitary operations, since this is the most common experimental limitation. However, note that other restrictions may be pertinent, such as to projective measurements in photonic devices for example.

\subsection{Experiment Design}
We have introduced most of the formalism behind \acs{PTT} to this point. To condense some of these ideas, we explicitly outline the experimental procedure here and draw parallels to the practice of \acs{QPT} for pedagogical purposes.

\begin{figure*}[ht!]
	\centering
	\includegraphics[width=\linewidth]{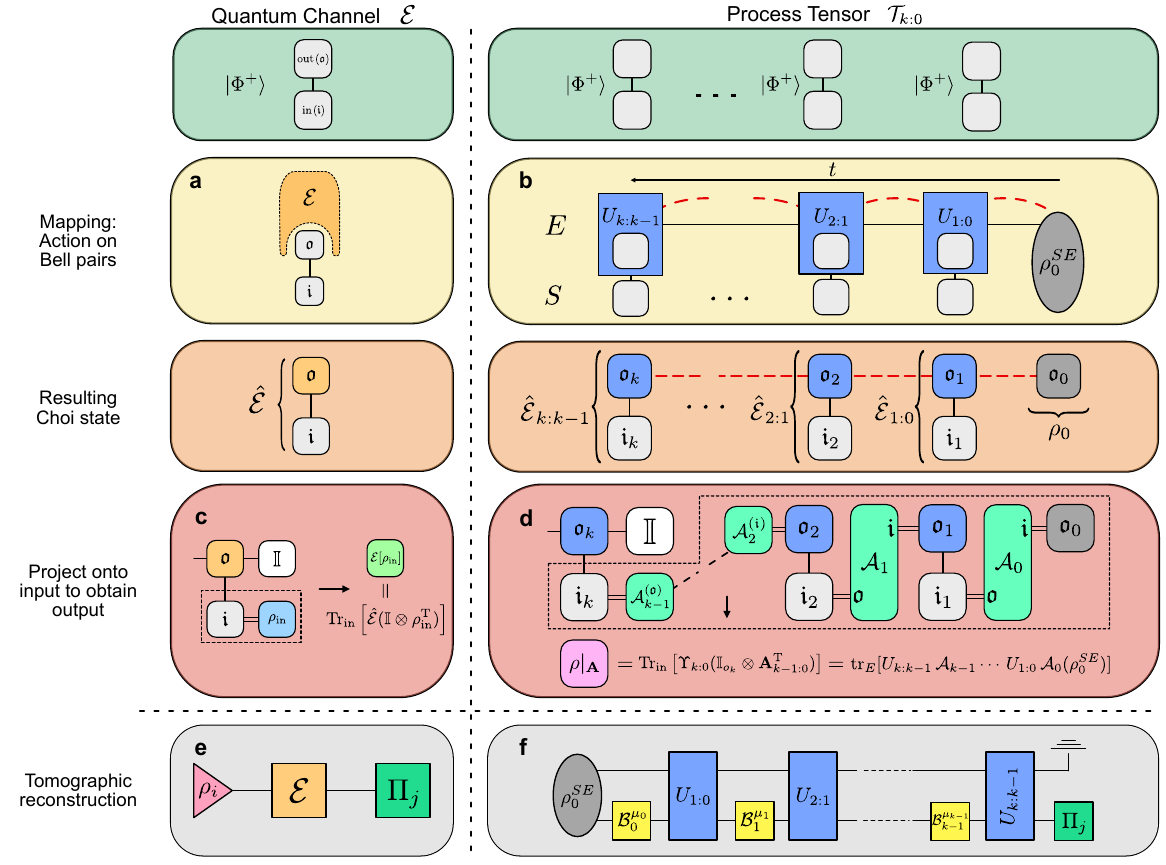}
	\caption[Operation, manipulation, and characterisation analogues between quantum process tomography and process tensor tomography ]{Operation, manipulation, and characterisation analogues between quantum process tomography and process tensor tomography. \textbf{a} The Choi-Jamiolkowski isomorphism represents a quantum process $\mathcal{E}$ by the density matrix $\hat{\mathcal{E}}$ using $\mathcal{E}$'s action on one half of a maximally entangled state. \textbf{b} In the generalised \acs{CJI}, $SE$ unitaries act on one half of a maximally entangled state per time-step. Correlations between times (denoted in red dashed lines) are then mapped onto spatial correlations between each output leg of a Bell pair. The result is a collection of (possibly correlated) \acs{CPTP} maps, as well as the average initial state. \textbf{c} The outcome of $\mathcal{E}$ on some $\rho_{\text{in}}$ is obtained by projecting the Choi state onto $\mathbb{I}\otimes \rho_{\text{in}}^\text{T}$ and tracing over the input space. \textbf{d} The outcome of a process conditioned on a sequence of operations $\mathbf{A}_{k-1:0}$ is obtained by projecting the Choi state of $\Upsilon_{k:0}$ onto the Choi state $\bigotimes_{i=0}^{k-1}\mathcal{A}_i$ and tracing over the input. Each $\mathcal{A}_i$ maps the output state of the $i$th \acs{CPTP} map to the input leg of the $(i+1)$th \acs{CPTP} map. \textbf{e} To reconstruct $\hat{\mathcal{E}}$ experimentally, prepare a complete set of states $\{\rho_i\}$, apply $\mathcal{E}$, and reconstruct the output state with an IC-POVM $\{\Pi_j\}$. \textbf{f} To reconstruct $\Upsilon_{k:0}$, measure each final state of the system subject to a complete basis of \acs{CP} maps $\{\mathcal{B}_{j}^{\mu_j}\}$ at each time. Note that the blue unitaries here are symbolic of any $SE$ interactions, and not gates that need performing.}
	\label{fig:PTT-explanation}
\end{figure*}

\begin{enumerate}
	\item Select the structure of open dynamics to probe, as in Figure~\ref{fig:PTT-explanation}b. That is, choose a system across timescale $\mathbf{T}_k=\{t_0,t_1,\cdots, t_k\}$. The intermediate $\{U^{SE}_{j:j-1}\}$ may then be either engineered interactions, or they may be the naturally noisy processes on the device.
	\item Choose a basis $\{\mathbf{B}_{k-1:0}^{\vec{\mu}}\}$ with which to probe the system. This may be time local, in that each $\hat{\mathbf{B}}_{k-1:0} = \bigotimes_{j=0}^{k-1}\hat{\mathcal{B}}_{j}^{\mu_j}$. Ideally this should be \acs{IC}, but may be a restricted basis, in that available control constrains the subspace on which the process may be probed. 
	\item Choose also an \acs{IC}-\acs{POVM} $\{\Pi_k^{\mu_k}\}$ with which to measure the state at $t_k$.
	\item Run a sequence of circuits with each basis element $\mathbf{B}_{k-1:0}^{\vec{\mu}}$ and measure in the \acs{IC}-\acs{POVM}. For example, if the basis is time-local, this means running circuits structured as $\mathcal{B}_0, \mathcal{B}_1,\cdots,\mathcal{B}_{k-1},\Pi_k$ for each $\mathcal{B}_0\in \{\mathcal{B}_0^{\mu_0}\}$, each $\mathcal{B}_1\in \{\mathcal{B}_1^{\mu_1}\}$, up to each $\mathcal{B}_{k-1}\in\{\mathcal{B}_{k-1}^{\mu_{k-1}}\}$ and each $\Pi_k \in \{\Pi_k^{\mu_k}\}$ -- see Figure~\ref{fig:PTT-explanation}f. If the cardinality of each $\{\mathcal{B}_j^{\mu_j}\}$ is the same, $N_B$ and $|\Pi_k^{\mu_k}|=N_M$ then the total number of circuits is $N_B^k \cdot N_M$.
	\item Repeat each circuit $\mathcal{O}(\frac{1}{\varepsilon^2})$ times for desired uncertainty $\varepsilon$.
\end{enumerate}
In the fully general case where $\mathbf{B}_{k-1:0}^{\vec{\mu}}$ is a $(k+1)-$outcome tester, then
\begin{equation}
	\mathbf{B}_{k-1:0}^{\vec{\mu}} = \sum_{j=0}^k\mathbf{B}_{k-1:0}^{\vec{\mu}(x_j)},
\end{equation}
and the frequencies with which outcomes $x_j$ in each experiment are observed provide an estimate for the probabilities
\begin{equation}
	\text{Pr}(x_k,t_k; x_{k-1},t_{k-1},\cdots,x_0 | \mathbf{B}_{k-1:0}^{\vec{\mu}}),
\end{equation}
obtained via, 
\begin{equation}
	\label{eq:ST-born-rule}
	p_{i,\vec{\mu}} = \text{Tr}\left[(\Pi_i \otimes  \mathcal{B}_k^{\mu_{k-1}\text{T}}\otimes \cdots \otimes \mathcal{B}_0^{\mu_0\text{T}})\Upsilon_{k:0}\right],
\end{equation}
also depicted in Figure~\ref{fig:PTT-explanation}d. These frequencies suffice to uniquely estimate the process tensor representation of the dynamics, and hence this concludes the experimental requirements. To reconstruct the process tensor then requires post-processing, which we continue to outline.\par 

We emphasise here the parallels between a \acs{QPT} experiment and a \acs{PTT} experiment, depicted in Figure~\ref{fig:PTT-explanation}. In \acs{QPT}, a single quantum channel $\mathcal{E}$ is probed, for which background (engineered or natural) dynamics must be chosen on times $t_0,t_1$. An \acs{IC} basis of inputs $\{\rho_\alpha\}$ and \acs{IC}-\acs{POVM} $\{\Pi_\beta\}$ is then chosen. The required circuits is then just each combination of all input states with output measurements, and sufficiently many repetitions to adequately estimate
\begin{equation}
	p_{\alpha \mid \beta} = \text{Tr}[(\Pi_\beta \otimes \rho_\alpha^{\text{T}})\cdot \hat{\mathcal{E}}].
\end{equation}

\subsection{Reconstruction Fidelity}
In the linear inversion regime, the process tensor's action on basis sequences will result in the experimentally observed density matrices by construction. Note that by `experimentally observed', we mean the density matrices as reconstructed by \acs{QST}. Since a process tensor is a linear operator, its action on linear combinations of basis sequences should be exactly the linear combinations of observed basis actions. This reiterates the idea of Figure \ref{fig:PT_summary}d: the $SE$ evolution between each operation is the same for all intermediate operations. In a linear combination, these arbitrarily strong dynamics are entirely accounted for. By tracing over the input space, we have the following relationship between the state conditioned on an arbitrary sequence of operations $\mathbf{A}_{k-1:0}$ and the states after each measured basis sequence:
\begin{equation}
	\label{eq:PT-response}
	\rho_k(\mathbf{A}_{k-1:0}) = \sum_{\vec{\mu}}\alpha^{\vec{\mu}} \rho_k(\mathbf{B}_{k-1:0}^{\vec{\mu}}),
\end{equation}
which is equivalent to Equation~\eqref{DM-expansion}.
Hence, we may (in principle) determine the system's exact response to any sequence of operations in the presence of non-Markovian interaction.
We use this as the figure of merit for the quality of characterisation. That is, we compute the fidelity over random sequences of operation between the state predicted by the process tensor -- Equation~\eqref{eq:PT-response} -- and what is realised on the device. 
We term this \emph{reconstruction fidelity}.

Formally, the Uhlmann fidelity $F$ between two process tensors $\mathcal{T}^{(1)}_{k:0}$ and $\mathcal{T}^{(2)}_{k:0}$ for a given sequence of interventions $\mathbf{A}_{k-1:0}$ is given by
\begin{align}
	&    F_{(1,2)}\left[\mathbf{A}_{k-1:0}\right]= F\left(\mathcal{T}^{(1)}_{k:0}[\mathbf{A}_{k-1:0}], \mathcal{T}^{(2)}_{k:0}[\mathbf{A}_{k-1:0}]\right),\notag \\
	& \text{where}\quad F(\rho,\sigma) = \text{Tr}\left[\sqrt{\sqrt{\rho}\sigma\sqrt{\rho}}\right]^2,
\end{align}
where $\mathcal{T}^{(1)}$ is taken to be the reconstructed process and $\mathcal{T}^{(2)}$ to be the real process, i.e., the experimental outputs. Then, the average reconstruction fidelity is an estimate of
\begin{equation}
	\mathcal{F} := \int \text{d}\mathbf{A}_{k-1:0} \  F_{(1,2)}[\mathbf{A}_{k-1:0}].
\end{equation}
We can use this to estimate the quality of our reconstruction. The outputs to a real process is simply the state reconstruction conditioned on the sequence of gates $\{\mathcal{A}_0, \cdots, \mathcal{A}_{k-1}\}$. This integral can be estimated by performing sequences of randomly chosen operations and comparing the fidelity of the predictions made by $\mathcal{T}_{k:0}$ with the actual outcomes measured. We use this as a metric for the accuracy with which a process has been characterised. We will employ this figure of merit throughout this entire thesis as an indicator of process reconstruction in a wide variety of contexts.\par

\subsection{Demonstration on NISQ Devices}


We look now to the practical determination of the process tensor in experiment. 
Specifically, we first wish to investigate the extent to which we can accurately operate our methods on real quantum hardware. With validation assured, we shall then explore facets of both engineered and naturally-occurring non-Markovian dynamics. 
The experiments carried out in this work used cloud-based IBM Quantum superconducting quantum devices. We first evaluated predictive capabilities of process tensor over a host of different experiments on the IBM Quantum devices \emph{ibmq\_johannesburg}, \emph{ibmq\_boeblingen}, \emph{ibmq\_poughkeepsie}, and \emph{ibmq\_valencia}. We are able to show high fidelity process characterisation of non-Markovian dynamics in a practical setting.
Ideally, complete process tensor construction would be achieved with the full span of \acs{CP} maps.
However, we reiterate at this point in time the lack of mid-circuit measurement functionality.
\begin{figure}[h]
	\centering
	\includegraphics[width=\linewidth]{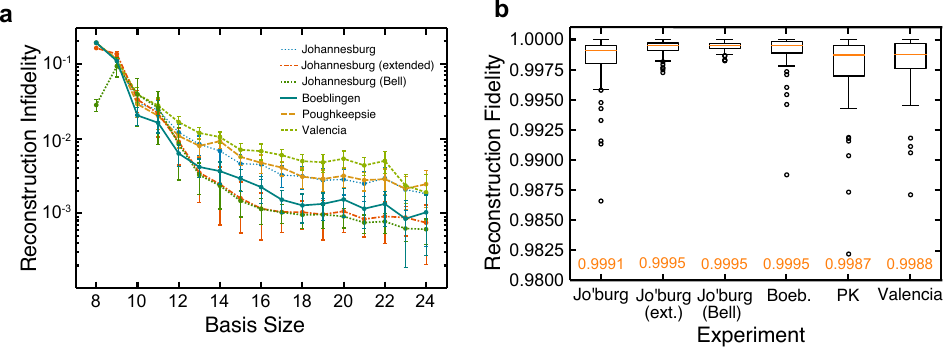}
	\caption[Reconstruction fidelities for linear inversion process tensor tomography performed on a series of IBM Quantum devices ]{Reconstruction fidelities for linear inversion process tensor tomography performed on a series of IBM Quantum devices. For each basis size, we compare the process tensor predictions with experimentally reconstructed density matrices for predictions that lay outside the basis set. \textbf{a} The average infidelity in reconstruction between the states predicted by the process tensor and the experimentally measured state.
		This includes a 95\% confidence interval, computed using the bootstrapping method described in~\cite{bootstrap-method}. The experiments compare the predictions of a basis $n$ process tensor with the experimental outcomes of the $4\times(28-n)\times(28-n)$ experiments from outside the basis set.
		In the notation of the Process Characterisation subsection, our basis is $\mathcal{P}\otimes\mathcal{U}^{(n)}\otimes\mathcal{U}^{(n)}$. \textbf{b} The distribution of fidelities of the predictions made by a basis-24 process tensor over a range of experiments. The top and bottom of the boxes are respectively the 25th and 75th percentiles, the whiskers are 1.5 times the inter-quartile range, and the orange lines are the medians of the distribution, with this last figure also provided in orange to four decimal places. Each black dot outside the boxes constitutes an outlier from the inter-quartile ranges. Note the shortening of device names for convenience in labelling. Further information about data collection and respective devices can be found in Appendix~\ref{chap:exp-details}.}
	\label{PT-fidelity-summary}
\end{figure}
For now, this rules out non-unital and trace-decreasing maps on superconducting devices, affording only unitary control, i.e., we do not have a complete basis of operations. With these limitations, we can still characterise processes in a restricted setting to obtain $\mathcal{T}_{k:0}^{(r)}$ as defined over the span of unitary operations, but not that we do not obtain access to \pt{}. This work deals only with single qubit process tensors, for which the unitary dimension is 10.\par 

We reconstruct and test the four-time restricted process tensor $\mathcal{T}_{3:0}^{(r)}$ for a single qubit process on IBM Quantum devices. To do so, we first reconstruct the final quantum states $\rho^{ijk}_3$. This state depends on the past controls, i.e., the initial preparation $\mathcal{P}^i_0 \in \mathcal{P}$ and the subsequent unitaries $\mathcal{U}^j_1 \in \mathcal{U}$ and $\mathcal{U}^k_2 \in \mathcal{U}$. The restricted process tensor is then obtained using Equation \eqref{DM-expansion}.

The set $\mathcal{U}$ contains 28 random unitaries, where the first $n$ elements $\mathcal{U}^{(n)}$ are used to reconstruct $\mathcal{T}_{3:0}^{(r)}$. Each smaller basis $\mathcal{U}^{(n)}$ is a subset of the larger bases. Randomly chosen unitaries are almost surely linearly independent, and are selected so as not to systematically preference any part of superoperator space.
The remaining $28-n$ elements are contracted with the reconstructed $\mathcal{T}_{3:0}^{(r)}$ to obtain predictions $\sigma_3^{ijk}$. 
We then compute the reconstruction fidelity
\begin{gather}
	\mathcal{F}_{ijk} := \left[\text{tr}\sqrt{\sqrt{\rho_3^{ijk}}\sigma_3^{ijk}\sqrt{\rho_3^{ijk}}}\right]^2
\end{gather}
to gauge the accuracy of the prediction.\par

In theory, a minimal complete basis ($n=10$) is all that is required for a restricted process tensor.
In practice, however, we find that sampling error and, to a lesser extent, gate error, introduces inconsistencies in the linear equations described in Equation~\eqref{DM-expansion}, amplifying reconstruction errors. Here, single-qubit gate errors are $\mathcal{O}(10^{-4}$--$10^{-3})$, and no error mitigation has been applied.
The Moore-Penrose pseudoinverse finds the coefficients minimising the least-squares error between overdetermined and inconsistent linear equations.
Consequently, adding in new basis elements will suppress the noise in the fidelities of prediction.
We find a surprisingly large improvement.
To further minimise bias in the noise, we also order our basis from least to most overlap with the rest of the set, as determined by the Hilbert-Schmidt inner product. 
This basis re-ordering improved predictive fidelity by 20\%.\par 
We summarise the average reconstruction fidelity between prediction and experiment of each basis in Figure~\ref{PT-fidelity-summary}a.
The `Johannesburg (extended)' experiment refers to process tensor experiments with idle time increased by a factor of 32. 
Meanwhile, `Johannesburg (Bell)' is the result of creating a Bell pair, and then acting the unitaries on one half. The intention of these is to probe different dynamics of the system: the former to add a longer time-scale, and the latter to test an initially correlated state. Standard \acs{CP} maps cannot describe the reduced dynamics of initially entangled states with the environment~\cite{arXiv:1011.6138, PhysRevLett.114.090402}, and so this evaluates a regime in which the process tensor is in principle more applicable.
The results both demonstrate the effects of basis size on process tensor performance, and showcase its ability to characterise processes.
Adding in new basis elements offers substantial improvement in comparison to a minimal complete basis.
Most of the error in reconstruction is statistical. The effects of this can be observed in the three highest fidelity experiments, `Johannesburg (extended)', `Johannesburg (Bell)', and `Boeblingen'. The first two produce more mixed final states, whose density matrices are naturally closer together, and the third is performed with 4096 shots per circuit, compared with 1600 for the remainder.
For a more fine-grained view, Figure~\ref{PT-fidelity-summary}b shows box plots of the predictive fidelity distribution of a size-24 basis on each experiment.
At this size, the median fidelity of characterisation is well within shot noise.
Here, we have shown how to extract useful and accurate predictions, and how unbiased and overcomplete basis sets are necessary for complete practical determination of the process tensor. 

\subsubsection*{Comparison with Markov models}

Fixed frequency transmon qubits -- such as the ones constructed by IBM Quantum -- have an always-on $ZZ$ coupling between connected qubits in the architecture. 
These dynamics provide a useful test-bed for the performance of the process tensor in a non-Markovian system when compared to a Markovian model for the process. 
As we saw in Chapter~\ref{chap:OQS}, \acs{GST} is a comprehensive tomographic procedure for estimating process matrices representing gate operations, preparations and measurements.
The maximum likelihood estimate of a gate set employs a Markov model, where repetitions of the gate are taken to be matrix powers.
\par
\begin{figure}
	\centering
	\includegraphics[width=0.5\linewidth]{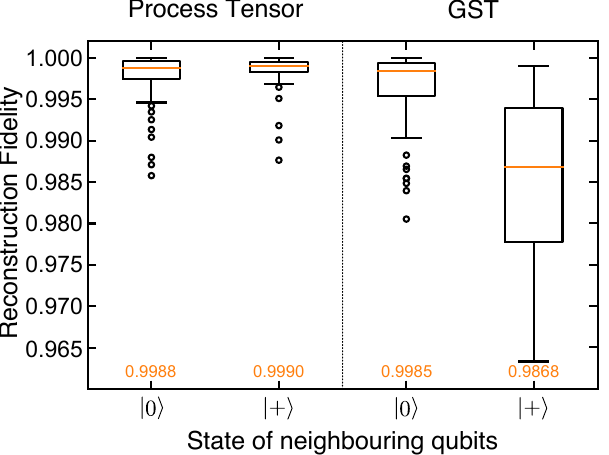}
	\caption[Comparison of process tensor tomography with gate set tomography in a non-Markovian setting ]{Comparison of process tensor tomography with gate set tomography in a non-Markovian setting. We benchmark the accuracy with which different techniques can predict the outcome of a given process for 64 circuits. When nearby qubits are initialised as $\ket{0}$, the median fidelity from \acs{GST} is similar to the process tensor in each scenario. When the neighbouring qubits are in state $\ket{+}$, however, \acs{GST} suffers from a fidelity drop of about $1.2\%$. This is a demonstration of how a technique like the process tensor could complement existing characterisation techniques in realistic non-Markovian settings.}
	\label{fig:valencia_boxplots}
\end{figure}

We performed two experiments under two different scenarios on the \emph{ibmq\_valencia} 5-qubit quantum device. 
The first is identical to the process tensor experiments the Experimental Implementation subsection using the set $\mathcal{U}$.
In addition, using \acs{GST} we characterised all 28 unitary operations in $\mathcal{U}$, the 4 preparations in $\mathcal{P}$, as well as the the initial state and the final measurement. 
The estimates for each map were multiplied out to produce a Markovian prediction for the final density matrix. 
Both the process tensor and \acs{GST} experiments were conducted first with neighbouring qubits initialised in the $\ket{0}$ state, and then again initialised in the $\ket{+}$ state.
Figure~\ref{fig:valencia_boxplots} shows the distribution of the reconstruction fidelities for both the process tensor and GST.
With a coherent environment, \acs{GST} performs about $1.2\%$ worse.
The process tensor tends to perform better in cases where the final state density matrices are more mixed, because this necessarily suppresses any directional bias in the noise.
We emphasise that our comparison of the outcomes of the two techniques is not framed competitively.
Indeed, they are qualitatively different: while \acs{GST} estimates the stationary maps of a given (presumed composable) gateset, the process tensor characterises all possible outcomes in a set process.
Figure~\ref{fig:valencia_boxplots} observes the breakdown of a Markov model, and benchmarks the process tensor against the state-of-the-art as a complementary tool to describing processes. 

The \acs{GST} experiments were completed using the \texttt{pyGSTi} quantum processor performance package~\cite{pygsti}.
Following the procedures outlined in the documentation, with background given in~\cite{intro-GST,RBK2017}, we characterised the 28 random unitaries as well as the 4 preparation gates in 8 groups of 4 gates. 
The software package designates the circuits required, and carries out the maximum likelihood reconstruction of the gates with the constraint of complete positivity and trace preservation. 
The gate sequences were repeated in powers of 2: $1, 2, 4, 8, 16,$ and $32$ times.
Included in this estimate are the state preparation and measurement vectors, $\kket{\rho}$ and $\bbra{E}$.
The process tensor and \acs{GST} experiments were conducted in the one calibration period for the device in a window of approximately 5 hours.
The gates were characterised in different groups for computational convenience, however this means that the final estimate for each group cannot necessarily be mixed. 
There exists a gauge freedom in \acs{GST}, in which measurement outcomes $\bbra{E} G\kket{\rho}$ is invariant under the transformation $\bbra{E} \mapsto \bbra{E}B$, $\kket{\rho} \mapsto B^{-1}\kket{\rho}$ and $G\mapsto B^{-1}GB$. In the \acs{GST} estimate, this gauge is optimised to bring the gate set as close as possible to the target set. However, in principle, each of the sets characterised will be in slightly different gauges. In order to estimate the effects of this, we computed the reconstruction fidelities with respect to the \acs{SPAM} vector estimates of each gate set estimate. Of the different gauges, the one with the maximum average difference between the data points in any of these distributions and the for the $\ket{+}$ neighbour given in Figure \ref{fig:valencia_boxplots} is $5.9\times 10^{-3}$, which is similar in magnitude to sampling error and does not significantly affect the comparison. This suggests that the absolute performance of the \acs{GST} estimates could be marginally better than what is shown.

\subsection{Optimal Basis Selection}
The effects of basis overlap in quantum tomography on the reconstruction have been discussed both with respect to conventional \acs{QST} and \acs{QPT}, and more recently with respect to the process tensor. 
In particular, the process tensor has shown itself to be highly sensitive to any overlap in its control basis. 
With access to all 16 dimensions of superoperator space, a mutually unbiased basis can be constructed in the form of a symmetric \acs{IC}-\acs{POVM} followed by an update~\cite{PhysRevA.70.052321}. 
However, in the limited case of a unitary-only basis, the ideal method is less straightforward. 
A randomly chosen unitary basis has been shown to adversely affect the reconstruction fidelity by as much as 30\%. 
Selecting a basis with mutual overlap here would be ideal, akin to the notion of a symmetric \acs{IC}-\acs{POVM} in conventional quantum state tomography. 
However, it has been shown that a \ac{MUUB} do not exist in dimension 10 (the dimension for single qubit channels)~\cite{Nasir2020}. 
Because of this limitation, we numerically search for a basis which minimises its average overlap with the remainder of the set to obtain something approximately unbiased. This procedure is performed as follows:\par 
We parametrise these gates using a standard single-qubit unitary parametrisation: 
\begin{gather}
	u(\theta,\phi,\lambda) = \begin{pmatrix}
		\cos(\theta/2) & -\text{e}^{i\lambda}\sin(\theta/2) \\
		\text{e}^{i\phi}\sin(\theta/2) & \text{e}^{i\lambda+i\phi}\cos(\theta/2) 
	\end{pmatrix}.
	\label{eq:unitary-paramch6}
\end{gather}
For two unitaries $u$ and $v$, let $\mathcal{U}$ and $\mathcal{V}$ be their superoperator equivalent, according to some representation. The overlap between the two channels is given by the Hilbert-Schmidt inner product: 
\begin{equation}
	\langle A, B\rangle_{\text{HS}} := \Tr[A^\dagger B].
\end{equation}
Importantly, this quantity is independent of representation, allowing us to select a form most desirable for computation. To this effect, we use the row-vectorised convention for states. Here, operations are given by $\mathcal{U} = u\otimes u^\ast$. The inner product between two unitaries parametrised as in~(\ref{eq:unitary-paramch6}) is then:
\begin{gather}
	\begin{split}
		\langle \mathcal{U},\mathcal{V}\rangle_{\text{HS}} &= \Tr[\mathcal{U}^\dagger \mathcal{V}]\\
		&= \Tr[(u\otimes u^\ast)^\dagger \cdot (v\otimes v^\ast)] \\
		&= \Tr[(u^\dagger\cdot v)\otimes (u^T\cdot v^\ast)]\\
		&= \Tr[u^\dagger v]\cdot \Tr[u^\dagger v]^\ast\\
		&= \left|\Tr[u^\dagger v]\right|^2
	\end{split}
	\label{HSIP-unitary}
\end{gather}
If we write $u=u(\theta_1,\phi_1,\lambda_1)$ and $v = v(\theta_2,\phi_2,\lambda_2)$, then (\ref{HSIP-unitary}) can be straightforwardly written (after some simplification) as
\begin{gather}
	\begin{split}
		& \Tr[u^\dagger v] = \cos\frac{\theta_1}{2}\cos\frac{\theta_2}{2} + \text{e}^{i(\phi_2 - \phi_1)}\sin\frac{\theta_1}{2}\sin\frac{\theta_2}{2} + \text{e}^{i(\lambda_2-\lambda_1)}\sin\frac{\theta_1}{2}\sin\frac{\theta_2}{2} + \text{e}^{i(\lambda_2 + \phi_2 - \lambda_1 - \phi_1)}\cos\frac{\theta_1}{2}\cos\frac{\theta_2}{2}\\
		& \Rightarrow \langle \mathcal{U},\mathcal{V}\rangle_{\text{HS}} = 
		4\cos^2\left(\frac{1}{2}(\lambda_1 - \lambda_2 + \phi_1 - \phi_2)\right)\cos^2\left(\theta_1 - \theta_2\right)
	\end{split}
\end{gather}
This simple expression for the inner product of any two single-qubit unitaries allows us to construct an objective function for the straightforward mutual minimisation of overlap between all ten elements of the basis set. Let 
\begin{gather}
	\mathscr{U}(\vec\theta,\vec\phi,\vec\lambda) = \{\mathcal{U}_i\}_{i=1}^{10} \equiv \{(\theta_i,\phi_i,\lambda_i)\}_{i=1}^{10}
\end{gather}
be our parametrised basis set. A basis set with the least mutual overlap can then be found by minimising the sum of the squares of each unitary with the remainder of the set. This minimises both the average overlap and the variance of overlaps with the remainder of the set. That is, by computing:
\begin{gather}
	\begin{split}
		\argmin_{(\vec\theta,\vec\phi,\vec\lambda)} \sum_{i=1}^{10}\sum_{j>i}\left(\langle \mathcal{U}_i,\mathcal{U}_j\rangle_{\text{HS}}\right)^2 = \sum_{i=1}^{10}\sum_{j>i}16\cos^4\left(\frac{1}{2}(\lambda_i - \lambda_j + \phi_i - \phi_j)\right)\cos^4\left(\theta_i - \theta_j\right)
	\end{split}
\end{gather}
One such ideal set can be found in Table~\ref{tab:muub-params}. This is the basis set used for the data obtained in the main text. Its overlaps with respect to the Hilbert-Schmidt inner product are listed in Table~\ref{tab:muub-overlaps}.
\begin{table}[h!]
	\centering
	\begin{tabular}{@{}lp{1.5cm}p{1.5cm}p{1.5cm}p{0cm}@{}}
		\hline
		& $\theta$ & $\phi$ & $\lambda$  &  \\ \hline
		$\mathcal{U}_1$    & 1.1148   & 1.5606    & 0.8160  &  \\
		$\mathcal{U}_2$    & -2.1993  & -2.0552   & -0.3564 &  \\
		$\mathcal{U}_3$    & 0.9616   & -0.8573   & 1.2333  &  \\
		$\mathcal{U}_4$    & 2.2655   & -2.7083   & 0.3154  &  \\
		$\mathcal{U}_5$    & -0.1013  & -0.5548   & -1.1472 &  \\
		$\mathcal{U}_6$    & 1.8434   & 0.8074    & -1.1772 &  \\
		$\mathcal{U}_7$    & -2.2036  & 1.9589    & 2.4002  &  \\
		$\mathcal{U}_8$    & -1.2038  & -0.2023   & 1.2355  &  \\
		$\mathcal{U}_9$    & 2.1791   & 3.2836    & 2.3524  &  \\
		$\mathcal{U}_{10}$ & -1.3116  & 2.3082    & 0.2882  &  \\ \hline
	\end{tabular}
	\caption{A set of thirty parameter values (given in radians) which constitute a set of ten unitary gates with minimal average mutual overlap.}
	\label{tab:muub-params}
\end{table}

\begin{table}[]
	\centering
	\resizebox{0.9\textwidth}{!}{%
		\begin{tabular}{@{}lllllllllll@{}}
			\toprule
			$\text{Tr}[\mathcal{U}_i^{\text{T}} \mathcal{U}_j]$ & $\mathcal{U}_1$ & $\mathcal{U}_2$ & $\mathcal{U}_3$ & $\mathcal{U}_4$ & $\mathcal{U}_5$ & $\mathcal{U}_6$ & $\mathcal{U}_7$ & $\mathcal{U}_8$ & $\mathcal{U}_9$ & $\mathcal{U}_{10}$ \\ \midrule
			$\mathcal{U}_1$ & 1 &  &  &  &  &  &  &  &  &  \\
			$\mathcal{U}_2$ & 0.19688 & 1 &  &  &  &  &  &  &  &  \\
			$\mathcal{U}_3$ & 0.19688 & 0.11111 & 1 &  &  &  &  &  &  &  \\
			$\mathcal{U}_4$ & 0.16758 & 0.19688 & 0.19688 & 1 &  &  &  &  &  &  \\
			$\mathcal{U}_5$ & 0.16758 & 0.19688 & 0.19688 & 0.16758 & 1 &  &  &  &  &  \\
			$\mathcal{U}_6$ & 0.19688 & 0.11111 & 0.11111 & 0.19688 & 0.19688 & 1 &  &  &  &  \\
			$\mathcal{U}_7$ & 0.03286 & 0.19688 & 0.19688 & 0.16758 & 0.16758 & 0.19688 & 1 &  &  &  \\
			$\mathcal{U}_8$ & 0.16758 & 0.19688 & 0.19688 & 0.16758 & 0.03286 & 0.19688 & 0.16758 & 1 &  &  \\
			$\mathcal{U}_9$ & 0.19688 & 0.11111 & 0.11111 & 0.19688 & 0.19688 & 0.11111 & 0.19688 & 0.19688 & 1 &  \\
			$\mathcal{U}_{10}$ & 0.16758 & 0.19688 & 0.19688 & 0.03286 & 0.16758 & 0.19688 & 0.16758 & 0.16758 & 0.19688 & 1 \\ \midrule
			Average & 0.24907 & 0.25146 & 0.25146 & 0.24907 & 0.24907 & 0.25146 & 0.24907 & 0.24907 & 0.25146 & 0.24907 \\ \bottomrule
		\end{tabular}%
		}
	\caption{Hilbert-Schmidt overlap between each element of the numerically constructed (approximate) \acs{MUUB}. We find this to be the most uniformly overlapping unitary basis possible, thus optimal for \acs{QST}.}
	\label{tab:muub-overlaps}
\end{table}

\section{Maximum Likelihood Estimation Process Tensor Tomography}

In the previous section, we derived and demonstrated a protocol for which experimental data could be used to capture non-Markovian multi-time statistics via direct linear inversion. But achieving high-fidelity characterisation was expensive, requiring highly overcomplete bases.
A major gap in the process tensor tomography toolkit so far is its lack of integration with standard tomography estimation tools like \acs{MLE}, whose underlying principle is to find a physical model estimate that maximises the probability with the observed data. Due to the intricate affine conditions of causality, this integration is nontrivial in general. The complexity of the procedure further grows when applied to restricted process tensors, e.g., when control operations are restricted and/or -- as we shall see in Chapter~\ref{chap:efficient-characterisation} -- when a finite Markov-order model is imposed. Our integration will naturally accommodate all of these variations.
We now close this gap and present a \acs{MLE} construction for \acs{PTT}, which helps to put this on the same footing as other tomographic techniques. The \acs{MLE} procedure estimates the physical quantum map most consistent with the data, according to some desired measure, while respecting the constraints discussed in Chapters~\ref{chap:stoc-processes} and~\ref{chap:process-properties}. The circuits required for tomography, depicted in Figure~\ref{fig:ml-flow}a, are the same as for linear inversion. That is, the scaling is the same. However, linear inversion typically requires an overcomplete basis to naturally average over inconsistencies. Meanwhile, \acs{MLE} treats the data such that a minimal tomographically complete basis suffices for accurate results.\par

In a practical setting with finite sampling error, it is best to set up tomographic protocols without bias in the basis vectors~\cite{PhysRevLett.105.030406}. This is especially true in high-dimensional spaces where even small errors may become significantly magnified. In the previous section, we found that a minimal single-qubit ($N_{\text{min}}=10$) unitary basis incurs substantial error in reconstructing the process. We thus resorted to an overcomplete basis of $N_{\text{oc}}=24$, leading to very high fidelity reconstruction for the process tensor, within shot-noise precision. However, this was very resource-demanding, since:
\begin{equation}
	\mbox{number of experiments} \sim \mathcal{O}(N_{\text{oc}}^k)
\end{equation}
for $k$ time-steps. \par 
Our focus here is to reduce the requirements of \acs{PTT} reconstruction while obtaining accurate estimates. To do so, in the next section we first integrate the maximum likelihood optimisation for \acs{QST}. This has the advantage that we will no longer need an overcomplete basis and thus reduce the base from 24 to 10. In order to do this, we ensure that the \acs{MLE} accommodates the affine conditions of the process tensor; devise a numerical method to generate an approximately unbiased basis, which minimises reconstuction errors; and develop a projection method to ensure the physicality of the process tensor. Of course, the \acs{MLE} alone is not enough if we also cannot reduce the exponent $k$. In Chapter~\ref{chap:efficient-characterisation}, we integrate the \acs{MLE} tools with more generic memory truncation methods to reduce the number of reconstruction circuits.
Along the way, we demonstrate that both \acs{MLE} and its subsequent variants are practically implementable by applying these ideas to superconducting quantum devices. \par


\subsection{Derivation}

An estimate for the map is coupled with metric of goodness (the likelihood) which quantifies how consistent the map is with the data. The cost function is then minimised while enforcing the physicality of the map.
The stored data vector in \acs{PTT} is the object $n_{i,\vec{\mu}}$, which are the observed measurement probabilities for the $i$th effect of an \acs{IC}-\acs{POVM}, subject to a sequence of $k$ operations $\bigotimes_{j=0}^{k-1}\mathcal{B}_j^{\mu_j}$. As is typical in \acs{MLE} tomography, this data is fit to a model for the process, $\Upsilon_{k:0}$, such that
\begin{equation}
	\label{PT-vector-est}
	p_{i,\vec{\mu}} = \text{Tr}\left[(\Pi_i \otimes  \mathcal{B}_k^{\mu_{k-1}\text{T}}\otimes \cdots \otimes \mathcal{B}_0^{\mu_0\text{T}})\Upsilon_{k:0}\right].
\end{equation}
\begin{wrapfigure}{r}{0.5\textwidth}
	\centering
	\includegraphics[width=0.95\linewidth]{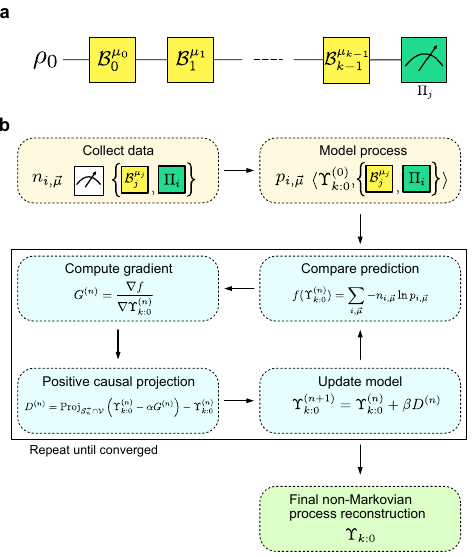}
	\caption[Circuit structure and logical workflow for performing process tensor tomography ]{\textbf{a} Circuit structure for \acs{PTT}. An arbitrary state is fed in, the experimenter acts with all combinations of different basis elements at different times, and a final measurement is recorded. \textbf{b} Logical flow of \texttt{pgdb} in the context of our \acs{MLE}-\acs{PTT} procedure. We maximise the likelihood of the model through iterative gradient descent and the physical projection of Section~\ref{ssec:projection} until some convergence condition is achieved. }
	\label{fig:ml-flow}
\end{wrapfigure}
These predictions are then compared to the observed frequencies, $n_{i,\vec{\mu}}$. The \emph{likelihood} of $\Upsilon_{k:0}$ subject to the data  -- that is $\text{Pr}(n_{i,\vec{\mu}}\mid \Upsilon_{k:0})$ is given by $\mathcal{L} = \prod_{i,\vec{\mu}} (p_{i,\vec{\mu}})^{n_{i,\vec{\mu}}}$. The cost function of \acs{MLE} algorithms is then the log-likelihood, i.e.,
\begin{equation}
	\label{ML-cost}
	f(\Upsilon_{k:0}) = -\ln \mathcal{L} = \sum_{i,\vec{\mu}} -n_{i,\vec{\mu}}\ln p_{i,\vec{\mu}},
\end{equation}
whose minimisation is the maximiser of the likelihood. A key part of the appeal to \acs{MLE} is that the cost function~\eqref{ML-cost} is convex, providing rigorous performance guarantees to the fit. \par 
An extensive selection of different semidefinite program packages exist in the literature for the log-likelihood minimisation in \acs{QST} and \acs{QPT} under the appropriate constraints.
In our construction of the \acs{MLE}-\acs{PTT} procedure, we employ and adapt the algorithm from Ref.~\cite{QPT-projection}, used for \acs{QPT}. This algorithm is termed `projected gradient descent with backtracking' (\texttt{pgdb}). We selected this both for its simplicity, and because it has been benchmarked as both faster and more accurate than other \acs{MLE}-\acs{QPT} algorithms. In this approach, the log-likelihood is minimised using conventional gradient descent, but at each iteration, a projection is made on the step direction to keep the map physical. The main steps are summarised in Figure~\ref{fig:ml-flow}b.


In Ref.~\cite{QPT-projection}, the relevant projection -- onto the intersection of the cone of \acs{CP} maps with the affine space of \acs{TP} channels  -- is performed using a procedure known as Dykstra's alternating projection algorithm~\cite{Birgin2005}.
We offer two key advancements here for \acs{PTT}. First, we determine the exact affine space generated by causality conditions on process tensors, such that the physical constraints are mathematically elucidated. Then, we adapt and introduce a conic projection technique from optimisation literature in order to project onto the space of completely positive, causal processes \cite{conic-projection}. We find this projection method to far outperform Dykstra's alternating projection algorithm in the problem instances, a fact which may be of independent interest for \acs{QPT}.
We detail each of these aspects in the following subsection. 
Finally, we benchmark the performance \acs{MLE}-\acs{PTT} on on superconducting quantum devices. These devices, as mentioned above, are limited to unitary control in the middle and a measurement at the end. This is insufficient to uniquely reconstruct the complete Choi state of a process. As such, our \acs{MLE} procedure yields a operationally well-defined restricted process tensor which has been completed into a full process tensor. One might consider a `family' of process tensors generated by the intersection of positive causal matrices with the affine space of observed experimental data. This yields all possible process tensors whose restriction to unitary operations is consistent with the observed data. Therefore, further information about the full dynamics may be inferred even with limited data.
We focus on the (non-unique) properties of the restricted process tensor family in Chapter~\ref{chap:MTP}, and the performance of the restricted process tensor in the present work.



\subsection*{Projecting onto the space of physical process tensors}
\label{ssec:projection}
Let us first describe in detail the physical conditions imposed on process tensors, as well as the approach used for projections onto the space of physical process tensors. Generally, this projection can be described as a problem of conic optimisation: finding the closest point lying on the intersection of a cone with an affine subspace. The affine constraints differ in each category: unit-trace for state tomography, trace-preservation for process tomography, and causality for the process tensor. 
Fundamentally, however, these techniques are applicable to all forms of quantum tomography.

Let $\Upsilon_{k:0}$ be the Choi form for a $k-$step process tensor (we will occasionally switch to $\Upsilon$ for brevity if the number of steps is not pertinent), and let $\kket{\Upsilon} := \text{vec}(\Upsilon)$, where we once more employ the row-vectorised convention \cite{gilchrist-vectorisation}. We discuss the mathematical demands of positivity and causality first, their individual projections, and then their simultaneous realisation. \par
Recall from Chapter~\ref{chap:stoc-processes} that, similar to a quantum channel, complete positivity of a process tensor is guaranteed by positivity of its Choi representation,
\begin{gather}
	\Upsilon_{k:0} \in \mathcal{S}_n^+,
\end{gather}
where $\mathcal{S}_n^+$ is the cone of $n\times n$ positive-semidefinite (PSD) matrices with complex entries. For $k$ time-steps, $n=2^{2k+1}$. The Euclidean projection is computed with a single eigendecomposition. Diagonalising $\Upsilon$ gives $\Upsilon = UDU^\dagger$ where $D = \text{diag}(\lambda_1, \lambda_2,\cdots,\lambda_n)$ is real. Then the projection onto $\mathcal{S}_n^+$ is:
\begin{gather}
	\label{PSD-proj}
	\text{Proj}_{\mathcal{S}_n^+}(\Upsilon) = U \text{diag}(\lambda^+_0,\cdots,\lambda^+_n)U^\dagger
\end{gather}
with $\lambda^+_j:=\max\{\lambda_j,0\}$. 

The Choi state must also obey causality, a generalisation of trace preservation. This is non-trivial to enforce, and ensures that future events should not influence past statistics. In the \acs{CJI} picture of Figure~\ref{fig:PTT-explanation}b, there should be no correlations between the final input leg and the rest of the process when the final output leg is traced out. This is also a statement of containment of the process tensor: that the process over a subset of the total period is contained within the larger process tensor:
\begin{equation}
	\label{containment}
	\text{Tr}_{\mathfrak{o}_k}\left[\Upsilon_{k':0}\right] = \mathbb{I}_{\mathfrak{i}_k}\otimes \Upsilon_{k'-1:0},
\end{equation}
iterated for all values of $k'$ from 1 to $k$. This statement is equivalent to causality in that the past stochastic process is unaffected by averaging over all future operations.\par 
We approach the problem of causality enforcement in the Pauli basis. Examining the Choi state, this condition places constraints on the values of these expectations. Let $\mathbf{P}:=\{\mathbb{I},X,Y,Z\}$ once more denote the single-qubit Pauli basis, $\mathbf{P}^n$ its $n-$qubit generalisation, and $\widetilde{\mathbf{P}}:=\{X,Y,Z\}$.
Focusing on the $\mathfrak{i}_{k'}$ subsystem, Equation~(\ref{containment}) can be enforced if all Pauli strings connecting the identity on the left subsystems with $\widetilde{\mathbf{P}}$ on the $\mathfrak{i}_{k'}$ subsystem have coefficients of zero. If this condition is imposed iteratively for all input legs on the process tensor, then Equation (\ref{containment}) will hold for all $k'$.
For example, in a two-step single qubit process (represented by a five-partite system), we have:
\begin{equation}
	\label{containment-example}
	\begin{split}
		&\langle \mathbb{I}_{\mathfrak{o}_2} P_{\mathfrak{i}_2}P_{\mathfrak{o}_1}P_{\mathfrak{i}_0}P_{\mathfrak{o}_0}\rangle = 0\: \forall \: P_{\mathfrak{i}_2}\in \widetilde{\mathbf{P}}; \: P_{\mathfrak{o}_1},P_{\mathfrak{i}_0},P_{\mathfrak{o}_0} \in \mathbf{P}\\ 
		&\langle \mathbb{I}_{\mathfrak{o}_2}\mathbb{I}_{\mathfrak{i}_2}\mathbb{I}_{\mathfrak{o}_1}P_{\mathfrak{i}_1}P_{\mathfrak{o}_0} \rangle = 0 \: \forall\: P_{\mathfrak{i}_0}\in\widetilde{\mathbf{P}}; P_{\mathfrak{o}_0}\in\mathbf{P}.
	\end{split}
\end{equation}
A simple way to enforce this condition is with the projection of Pauli coefficients. 
In particular, let $\mathcal{P}$ be the elements of $\mathbf{P}^{2k+1}$ whose expectations must be zero from equations~\eqref{containment} and~\eqref{containment-example}.
We can write this as a single affine constraint in the matrix equation:
\begin{gather}
	\label{affine-constraint}
	\begin{pmatrix} \langle\!\bra{\mathcal{P}_0} \\
		\langle\!\bra{\mathcal{P}_1}\\
		\vdots \\
		\langle\!\bra{\mathcal{P}_{m-2}}\\
		\langle\!\bra{\mathbb{I}}
	\end{pmatrix} \cdot \ket{\Upsilon}\!\rangle = \begin{pmatrix} 0\\0\\\vdots\\0\\d\end{pmatrix},
\end{gather}
where $d$ is the normalisation chosen for the Choi matrix (in this work, $d=1$). \par 

Letting this set the context for our discussion of the projection routine, consider a full rank constraint matrix $A\in\mathbb{C}^{m\times n^2}$, variable vector $\upsilon$, and fixed right-hand side coefficient vector $b$.
Let $\mathcal{V}$ be the affine space:
\begin{gather}
	\mathcal{V} = \{ \upsilon\in \mathbb{C}^{n^2} | A\upsilon = b\}
\end{gather}
The projection onto $\mathcal{V}$ is given by
\begin{gather}
	\label{affine-proj}
	\text{Proj}_{\mathcal{V}}(\upsilon_0) = \left[\mathbb{I} - A^\dagger (AA^\dagger)^{-1} A\right]\upsilon_0 + A^\dagger(AA^\dagger)^{-1}b.
\end{gather}

In general, however, a projection onto $\mathcal{S}_n^+$ and a projection onto $\mathcal{V}$ is not a projection onto $\mathcal{S}_n^+\cap \mathcal{V}$. The conic and affine constraints are difficult to simultaneously realise. 
One approach to this is to use Dykstra's alternating projection algorithm, as performed in \cite{QPT-projection} for \acs{QPT}. This applies a select iterative sequence of~\eqref{affine-proj} and~\eqref{PSD-proj}. Although this method is straightforward and has guaranteed convergence, we find it unsuitable for larger-scale problems. For large gradient steps the convergence can take unreasonably many steps. More importantly, however, each step of the gradient descent requires many thousands of applications of \eqref{affine-proj}. Although much of this expression can be pre-computed, the complexity grows strictly with $n$, rather than the number of constraints. Moreover, the matrix inverse requirement can reduce much of the advantage of having a sparse $A$.\par  
In our \acs{MLE}-\acs{PTT}, instead of Dykstra's alternating projection algorithm, we integrate a variant of the technique introduced in~\cite{conic-projection} and discussed further in~\cite{conic-handbook}. This method is targeted at the general problem intersecting cones and affine spaces. It regularises the projection into a single unconstrained minimisation, such that only eigendecompositions and matrix-vector multiplications by $A$ are necessary, avoiding the need for~\eqref{affine-proj}. Note that to guarantee uniqueness of the projection as well as convergence of projected gradient descent in general, the closest physical process tensor at each step is found in terms of Euclidean distance. For further detail on this, see Refs.~\cite{hauswirth2016projected,conic-projection}. Because the projection is the only component of \texttt{pgdb} that we change with respect to \cite{QPT-projection}, we explicitly walk through the steps in the following paragraph. We also benchmark this direct conic projection routine on normally distributed random matrices with respect to Dykstra's alternating projection algorithm for the case of \acs{QST}, \acs{QPT}, and \acs{PTT} in Figure \ref{fig:projection_comparison}. The scaling for each method is similar (dominated by the cost of eigendecompositions), but the absolute savings are of two orders of magnitude. \par 
In each respective regime of tomography, the increased number of constraints increases the amount of time, on average, for the projection to complete. However, we find substantial improvements in both the run-time and in the number of eigendecomposition calls between the direct conic projection in comparison to Dykstra's. This is especially necessary for the fitting of process tensors where the difference between the two can be the difference between a run-time of days, or of fractions of a minute. We include \acs{QST} here for completeness, however, note that the fixed projection of eigenvalues onto the canonical simplex with a single diagonalisation is more appropriate \cite{simplex-projection}.\par

We now explicitly step through the direct conic projection method. For a given $\upsilon_0$, we wish to find the closest (in Euclidean terms) $\upsilon \in \mathcal{S}_n^+ \cap \mathcal{V}$. That is, to compute 
\begin{gather}
	\label{primal}
	\argmin_{\upsilon \in \mathcal{S}_n^+ \cap \mathcal{V}} \|\upsilon - \upsilon_0\|^2.
\end{gather}
Note that when we talk about the vector $\upsilon$ being PSD, we mean that its matrix reshape is PSD. The dual approach introduces the Lagrangian, which is a function of the primal variable $\upsilon \in \mathcal{S}_n^+$ and dual variable $\lambda\in \mathbb{R}^m$ (for $m$ affine constraints): 

\begin{figure}
	\centering
\includegraphics[width=\linewidth]{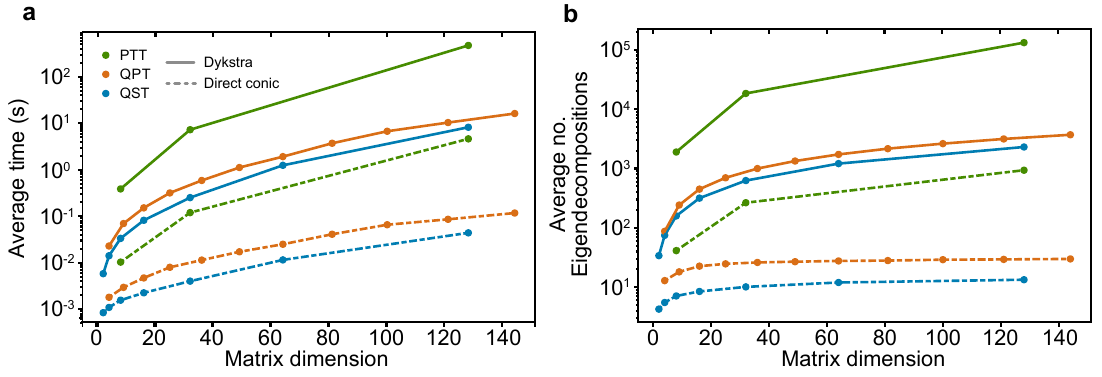}
	\caption[A comparison between projection methods imposing trace-preserving and positive physical conditions ]{A comparison between projection methods imposing physical conditions on 500 normally distributed random matrices. \textbf{a} Average time taken for a single projection for both Dykstra's alternating projection algorithm, and the direct conic projection. We compare conditions set by \acs{QST}, \acs{QPT}, and \acs{QST}. \textbf{b} Average number of eigendecompositions for each of the above. This dominates the runtime of each method.}
	\label{fig:projection_comparison}
\end{figure}

\begin{gather}
	\label{lagrangian}
	\mathcal{L}(\upsilon;\lambda) = \|\upsilon - \upsilon_0\|^2 - \lambda^\dagger (A\upsilon - b).
\end{gather}
Since $\upsilon$ is PSD, it is Hermitian, meaning that the matrix-vector product $A\upsilon$ is always real. This avoids the need for recasting the complex problem into real and imaginary pairs. \par 

The vector $\upsilon$ which minimises $\mathcal{L}$ for a given $\lambda$ provides a lower bound to the solution to the primal problem. We introduce the dual concave function
\begin{gather}
	\label{dual-func}
	\theta(\lambda) := \min_{\upsilon \in \mathcal{S}_n^+} \mathcal{L}(\upsilon;\lambda)
\end{gather}
whose maximum is exactly the solution to~\eqref{primal}. It is shown in Ref.~\cite{PSD-dual} that the minimum~\eqref{dual-func} is uniquely attained by $\upsilon(\lambda) = \text{Proj}_{\mathcal{S}_n^+}(\upsilon_0 + A^\dagger \lambda)$ \cite{PSD-dual}, and can hence be recast (up to a constant) as
\begin{gather}
	\label{dual-func-simplified}
	\theta(\lambda) = -\|\upsilon(\lambda)\|^2 + b^\dagger \lambda.
\end{gather}
It can further be shown that \eqref{dual-func-simplified} is differentiable on $\mathbb{R}^m$ with gradient
\begin{gather}
	\label{dual-func-grad}
	\nabla \theta(\lambda) = -A\upsilon(\lambda) + b.
\end{gather}
Thus, the solution to the projection problem \eqref{primal} becomes an unconstrained minimisation problem of \eqref{dual-func-simplified} with respect to $\lambda$, opening the door to a wealth of tested optimisation packages to be applied. The solution to the projection is then $\text{Proj}_{S_n^+}(\upsilon_0 + A^\dagger \lambda_{\text{min}})$. Specifically, in this work we select the L-BFGS algorithm to perform this minimisation, as we found it to give the fastest and most reliable solution \cite{lbfgs-alg}. Although we did not implement it here, it is also possible to compute the Clarke-generalised Jacobian of~\eqref{dual-func-grad}, allowing for exact second order optimisation techniques to be used \cite{projection-clarke-jacobian, spectral-derivatives}.
Note also that the difficulty of this minimisation is sensitive to the condition number of $A$. Thus, we find the best approach to be to always frame affine constraints in the Pauli basis to ensure a uniform spectrum.\par 

Another favourable reason to apply this conic projection method is in the arbitrary application of affine constraints. In the following chapter, we show how this can be used for searching (and thus bounding quantities of) manifolds of states consistent with an incomplete set of data. Introducing a feature matrix as part of the affine constraint with observed probabilities permits this exploration. Without the faster method, we found that this was infeasible to perform.\par
Using this modification of \texttt{pgdb} with an updated projection routine, we are able to implement \acs{PTT} both in simulation and in real data. 
We have thus formalised \acs{MLE}-\acs{PTT}, and are now in position to benchmark the performance of \acs{MLE}-\acs{PTT} tomography against the linear inversion method and look at applications. The results, summarised in the following sections, suggest full characterisation of quantum non-Markovian dynamics in an object which is both mathematically minimal, and which obeys all of the physical constraints of a quantum stochastic process.

\subsection*{Maximum likelihood, cost evaluation, and gradient}
Full details and benchmarking of the \texttt{pgdb} algorithm for \acs{QPT} can be found in Ref.~\cite{QPT-projection}. Here, we provide the pseudocode in this context, which forms the basis for our implementation of \acs{MLE}-\acs{PTT}.

\begin{algorithm}[H]
	\caption{\texttt{pgdb}}
	\begin{algorithmic}[1]
		\State $j=0,n=d_S^{2k+1}$
		\State \text{Initial estimate}:
		$\Upsilon_{k:0}^{(0)} = \mathbb{I}_{n\times n}/n$
		\State \text{Set metaparameters: }$\mu=3/2n^2,\gamma=0.3$
		\While{$f(\Upsilon_{k:0}^{(j)}) - f(\Upsilon_{k:0}^{(n+1)}) > 1\times 10^{-6}$}
		\State $D^{(j)} =\text{Proj}_{S_n^+\cap \mathcal{V}}\left(\Upsilon_{k:0}^{(j)} - \mu \nabla f(\Upsilon_{k:0}^{(j)})\right) - \Upsilon_{k:0}^{(j)}$
		\State $\beta = 1$
		\While{$f(\Upsilon_{k:0}^{(j)}) + \beta D^{(k)}) > f(\Upsilon_{k:0}^{(j)}) + \gamma \beta \left\langle D^{(j)}, \nabla f(\Upsilon_{k:0}^{(j)})\right\rangle$}
		\State $\beta = 0.5\beta$
		\EndWhile
		\State $\Upsilon_{k:0}^{(j+1)} = \Upsilon_{k:0}^{(j)} + \beta D^{(j)}$
		\State $j = j+1$
		\EndWhile
		\State\Return $\Upsilon_{k:0}^{(\text{est})} = \Upsilon_{k:0}^{(j+1)}$
	\end{algorithmic}
\end{algorithm}
$\text{Proj}_{S_n^+\cap\mathcal{V}}(\cdot)$ here is the projection subroutine as just described. Although we have fixed the gradient step size here to be the same as in~\cite{QPT-projection}, we find this to be slightly problem-dependent in terms of its performance. The reason is that the larger the step, the less physical $\Upsilon_{k:0}^{(j)} - \mu \nabla f(\Upsilon_{k:0}^{(j)})$ tends to be, increasing the run-time of the projection subroutine. In general, we find that decreasing $\mu$ to favour the runtime of the projection is overall favourable to the performance of the algorithm.\par

The process tensor action described in Equation~\eqref{eq:PT-action} is pedagogically useful, however in practice, we compute the action of some process tensor $\mathcal{T}_{k:0}$ on a sequence of control operations $\mathbf{A}_{k-1:0}$ via the projection of its Choi state onto as in Equation~\eqref{PT-vector-est}. So long as the input operations are always tensor product (omitting the case of correlated instruments), this can be performed fast as a tensor network contraction. In this form, computation of the cost and the gradient is significantly sped up in comparison to multiplying out the full matrices.\par 
Writing the Choi state of a process tensor $\Upsilon_{k:0}$ explicitly with its indices as a rank $2(2k+1)$ tensor, we have $2k+1$ subsystems alternating with outputs from the $j$th step ($\mathfrak{o}_j$) and inputs to the $(j+1)$th step ($\mathfrak{i}_j$), i.e.
\begin{equation}
	\Upsilon_{k:0} \equiv (\Upsilon_{k:0})_{k_{\mathfrak{o}_k},k_{\mathfrak{i}_k},\cdots,k_{\mathfrak{o}_0}}^{b_{\mathfrak{o}_k},b_{\mathfrak{i}_k},\cdots,b_{\mathfrak{o}_0}},
\end{equation}
where $b$ is shorthand for bra, and $k$ is shorthand for ket.
The basis operation at time step $j$ has indices (we write its transpose) $(\mathcal{B}_j^{\mu_j})_{b_{i_{j+1},b_{o_j}}}^{k_{i_{j+1}},k_{o_j}}$, meanwhile the \acs{POVM} element $\Pi_i$ is written $(\Pi_i)_{k_{o_k}}^{b_{o_k}}$. Consequently, the full tensor of predicted probabilities for all basis elements is given by
\begin{equation}
	\label{TN-cost}
	p_{i,\vec{\mu}} = \sum_{\substack{k_{\mathfrak{o}_k},k_{\mathfrak{i}_k},\cdots,k_{\mathfrak{i}_1},k_{\mathfrak{o}_0} \\ b_{\mathfrak{o}_k},b_{\mathfrak{i}_k},\cdots,b_{\mathfrak{i}_1},b_{\mathfrak{o}_0}}} (\Upsilon_{k:0})_{k_{\mathfrak{o}_k},k_{\mathfrak{i}_k},\cdots,k_{\mathfrak{o}_0}}^{b_{\mathfrak{o}_k},b_{\mathfrak{i}_k},\cdots,b_{\mathfrak{o}_0}}(\Pi_i)_{k_{\mathfrak{o}_k}}^{b_{\mathfrak{o}_k}} (\mathcal{B}_{k-1}^{\mu_{k-1}})_{b_{\mathfrak{i}_{k}},b_{\mathfrak{o}_{k-1}}}^{k_{\mathfrak{i}_{k}},k_{\mathfrak{o}_{k-1}}}(\mathcal{B}_{k-2}^{\mu_{k-2}})_{b_{\mathfrak{i}_{k-1}},b_{\mathfrak{o}_{k-2}}}^{k_{\mathfrak{i}_{k-1}},k_{\mathfrak{o}_{k-2}}}\cdots (\mathcal{B}_{k-1}^{\mu_{k-1}})_{b_{\mathfrak{i}_{1}},b_{\mathfrak{o}_{0}}}^{k_{\mathfrak{i}_{1}},k_{\mathfrak{o}_{0}}}
\end{equation}
We use the quantum information Python library \texttt{quimb} \cite{quimb} to perform this, and all future tensor contractions straightforwardly. The cost function is then evaluation as in Equation \eqref{ML-cost} in the same way: through an element-wise logarithm of $p_{i,\vec{\mu}}$ followed by contraction with the data tensor $n_{i,\vec{\mu}}$. Since the cost function is linear in $\Upsilon_{k:0}$, computing the gradient $\nabla f/\nabla \Upsilon_{k:0}$ is simply $\nabla p_{i,\vec{\mu}}/\nabla \Upsilon_{k:0} : (n/p)^{i,\vec{\mu}}$, which expands to:
\begin{equation}
	\frac{\nabla f}{\nabla \Upsilon_{k:0}} = \sum_{i,\vec{\mu}} \left[(\mathcal{B}_{k-1}^{\mu_{k-1}})_{b_{\mathfrak{i}_{k},b_{\mathfrak{o}_{k-1}}}}^{k_{\mathfrak{i}_{k}},k_{\mathfrak{o}_{k-1}}}(\mathcal{B}_{k-2}^{\mu_{k-2}})_{b_{\mathfrak{i}_{k-1},b_{\mathfrak{o}_{k-2}}}}^{k_{\mathfrak{i}_{k-1}},k_{\mathfrak{o}_{k-2}}}\cdots (\mathcal{B}_{0}^{\mu_{0}})_{b_{\mathfrak{i}_{1},b_{\mathfrak{o}_{0}}}}^{k_{\mathfrak{i}_{1}},k_{\mathfrak{o}_{0}}}\right] \frac{n_{i,\vec{\mu}}}{p_{i,\vec{\mu}}}
\end{equation}
i.e. Equation \eqref{TN-cost} without the inclusion of $\Upsilon_{k:0}$. \par 

\subsection{Demonstration on NISQ Devices}
\begin{figure}[t]
	\centering
	\includegraphics[width=0.75\linewidth]{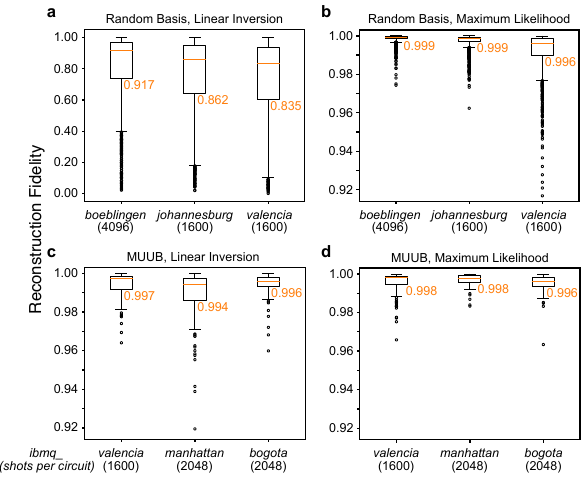}
	\caption[Reconstruction fidelity of various three-step process tensor procedures when using a minimal complete basis.]{Reconstruction fidelity of various three-step process tensor procedures when using a minimal complete basis. Each data point represents a different randomly generated unitary sequence. The top and bottom of the boxplots are 75 and 25 percentiles, orange line is the median (figure also printed), whiskers are 1.5 times the interquartile range, and any remaining data points are outliers beyond this. We compare reconstructions with \textbf{a} a randomly generated basis and linear inversion, \textbf{b} a randomly generated basis processed by maximum likelihood, \textbf{c} a \acs{MUUB} processed with linear inversion, and \textbf{d} a \acs{MUUB} processed with maximum likelihood.
	The results showcase high-fidelity, physical process tensors with minimal resources, in contrast with Section~\ref{sec:LI-PTT} where the random basis data was taken.}
	\label{fig:RF_boxplot}
\end{figure}

We probe each of these characteristics by looking at three-step \acs{PTT} on IBM Quantum devices. Using combinations of basis choice, and \acs{LI}/\acs{MLE} post-processing, we compute the reconstruction fidelities for random sequences of unitaries. The boxplots showing these distributions are shown in Figure~\ref{fig:RF_boxplot}. Each data point constitutes a different sequence of random unitary gates. The number is then the fidelity between the density matrix predicted through action of the reconstruction process tensor on these unitary mappings, and the actual density matrix reconstructed through \acs{QST} on the device after executing that specific unitary sequence. The use of \acs{MUUB} alone finds substantial improvement in characterising the process. We continue to use these optimal parameter values as our unitary control basis for the remainder of this work. Reconstruction is improved further by \acs{MLE}-\acs{PTT}, in which we see not only an increase in median reconstruction fidelity, but there are far fewer outliers in the distribution. Compared to the random basis, linear inversion case, reconstruction fidelity increases greatly to within shot noise. This is essential for both validating process characterisation, and optimal control of the system.

In Section~\ref{sec:LI-PTT}, much of the linear inversion characterisation noise was overcome with the use of an over-complete basis -- up to 24 unitaries. Since \acs{PTT} is exponential in the size of the basis, the employment of maximum-likelihood methods as shown here can offer a significant reduction in experimental requirements. Thus, in addition to offering an algorithm imposing physicality constraints, we show how to make the technique more practical to implement. The scaling of \acs{MLE} is therefore
\begin{equation}
	\label{eq:mle-scaling}
	\mbox{number of experiments} \sim \mathcal{O}(N_{\text{mle}}^k),
\end{equation}
where $N_{\text{mle}}$ can now be 10 regardless of the specific basis choice.
As well as comparing processing methods, we also juxtapose our approximate \acs{MUUB} with the minimal randomly-chosen unitary basis, where $N_{\text{muub}} = 10$. This, too, sees a drastic improvement of the method: though it is not guaranteed to produce a physical process tensor, we see that much higher quality predictions are possible without any additional effort in the linear inversion approach.

\section{Full Process Tensor Tomography}
If one had access to \acs{IC} quantum control, then performing full \acs{PTT} is a straightforward application of our methods presented so far in this chapter. By this, we mean the ability to uniquely and unambiguously recover an estimate for any process tensor representation in an experimental setting. We hence omit going through this procedure in detail.

Instead, we shall address a practical issue. Given access to only noisy quantum devices, either with or without mid-circuit measurement capabilities, how does one provide themselves with \acs{IC} control? 
The tenets are identical to before, but with two practical changes. First, we need to construct our basis of instruments and have them well-characterised. The basis $\{\mathbf{B}_{k-1:0}^{\vec{\mu}}\}$ is no longer constructed by idealised means, but by first performing experiments and reconstructing their Choi states as described in the previous section. Second, our data no longer consists of 2-outcome measurements at the end of each circuit, but rather $d_S^k$ different outcomes for the various instrument sequences.\par 

In our experiments on IBM Quantum, we employed both devices with and without mid-circuit measurement capabilities. 
At a brief summary, we use a nearby ancilla qubit with the means to probe our system. As we saw in Chapter~\ref{chap:process-properties}, unitary interactions with a secondary qubit, followed by projective measurement on that qubit suffices for \acs{IC} control. 
In the case where a device does not have mid-circuit measurements, the measurement can be deferred to the end, and one ancilla per step is required. In the case where mid-circuit measurements are available, a single ancilla can be measured and reused.
In both cases, the measurements are too noisy and too slow to use on the system directly for a reliable reconstruction. We resolve this by constructing a different basis.
We will show here how restricted process tensors can manipulate the interaction with the neighbouring qubit, creating an arbitrary well-characterised effective map. These maps can then be used to probe $SE$ dynamics with a full process tensor; hence, this `bootstraps' access to a full basis.

We will employ our bootstrapped set of instruments as part of further analysis in Chapter~\ref{chap:MTP}. We have obtained results for full \acs{PTT} on IBM Quantum devices to similarly high reconstruction fidelities of the full multi-time statistics. Since we use these tools in the following chapter, for brevity we omit these results. A numerical analysis of full \acs{PTT} is performed in Section~\ref{sec:ptt-analysis}.

\subsection{Manufacturing Informationally Complete Quantum Instruments}
\label{sec:IC-control}
A central obstacle so far to complete non-Markovian process characterisation has been the absence of an informationally complete set of controls in practice. Accessible devices typically only have unitary control of a qubit, which hinders actual measurement of different temporal correlations. In this section, we consider the design of quantum instruments that may be used to probe non-Markovian correlations in practice. A primary reason why even functional mid-circuit measurements on current devices are undesireable on a system is that such measurements take a long time to implement, on the range of microseconds in superconducting devices. Hence, if the system itself is measured, then the estimate will necessarily omit any continued dynamics for that particular window. 
We can circumvent this problem by introducing an ancilla qubit. The central idea is that \acs{IC} control of a qubit may be realised through a two-qubit unitary control operation with the aid of an ancilla qubit, and a projective measurement. 
By the principle of deferred measurement, once the system has interacted with its ancilla, it does not matter when that ancilla is measured. The system can continue to participate in any $SE$ dynamics, and the relevant time scale becomes the $SA$ unitary interaction. However, in practice, this level of control will also be unideal. We want to be able to implement control of this style, but it must be (i) short compared to the surrounding dynamics, and (ii) a known quantum operation. Introducing the ancilla allows this to some extent, but we must also have some method of knowing the exact structure of the probe we are constructing. 

\par

A previous key observation has been that non-Markovianity on a system can be an extra resource~\cite{Bylicka2013}. Functionally, non-Markovian dynamics (if properly controlled) gives one access to a larger range of controls than the equivalent closed system. Employing this philosophy, we use the tools we have so far developed with respect to restricted process tensors to bootstrap the construction of our own quantum instruments, satisfying both of the above criteria. Consider the following situation between a single system qubit $S$ and an ancilla qubit $A$. A quantum map on $S$ is defined by the interaction:
\begin{equation}
	\mathcal{E}_x[\rho_S] = \text{Tr}_A[|x\rangle\!\langle x|_A \cdot U_{SA}\cdot  (|0\rangle\!\langle 0|_A \otimes \rho_S) \cdot U_{SA}^\dagger].
\end{equation}
\begin{figure}[htbp]
	\centering
	\includegraphics[width=0.65\linewidth]{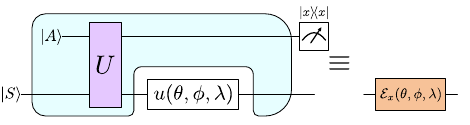}
	\caption[Realisation of a quantum instrument as a restricted process tensor ]{Realisation of a quantum instrument as a restricted process tensor. A parametrised quantum instrument can employ an ancilla qubit to achieve informational completeness as depicted. At a local level, this harnesses non-Markovianity to achieve extended control. We can hence bootstrap control availability in principle by first estimating the illustrated restricted process tensor.}
	\label{fig:instrument-comb}
\end{figure}
As discussed in the previous chapter, for most $U_{SA}$, this control operation can be \acs{IC} with respect to $S$. But ideally we would like not to have to vary $U_{SA}$, since decompositions of different arbitrary $SU(4)$ gates can vary wildly in terms of the physical gates. Alternatively, one could select a single fixed two-qubit unitary $U_F$ and maintain a parametrised single-qubit unitary on the system that would allow the desired flexibility. For example, $U_F$ might implement have non-zero $\langle X_{\mathfrak{o}}\mathbb{I}_{\mathfrak{i}}\rangle $ but $\langle Y_{\mathfrak{o}}\mathbb{I}_{\mathfrak{i}}\rangle  = \langle Z_{\mathfrak{o}}\mathbb{I}_{\mathfrak{i}}\rangle = 0$. A local unitary would appropriately convert the $Z$ eigenstates into $X$ and $Y$, generating informational completeness. We can then recast this situation in terms of a parametrised control instrument
\begin{equation}
	\mathcal{E}_x(\theta,\phi,\lambda)[\rho_S] = \text{Tr}_A[|x\rangle\!\langle x|_A (I_A\otimes u_S(\theta,\phi,\lambda))\cdot U_{SA} \cdot (|0\rangle\!\langle 0|_A \otimes \rho_S) \cdot U_{SA}^\dagger\cdot (I_A \otimes u^\dagger_S(\theta,\phi,\lambda)].
\end{equation}
The parametrisation allows us to consider not just a fixed set of instruments that we characterise, but rather a continuous manifold of structured operations. The family of parametrised instruments is \acs{IC}, but cannot directly realise any general operation as in Ref.~\cite{PhysRevLett.111.130504}.
Notice that this situation is starting to look like a process tensor, as depicted in Figure~\ref{fig:instrument-comb}. The $SA$ interaction is non-Markovian, and we can view the local parametrised unitary as being a leg of a restricted process tensor. We have already shown that we can reliably determine output estimates for arbitrary sequences of unitaries. We can apply this again to determine the dynamics of $\mathcal{E}_x(\theta,\phi,\lambda)[\rho_S]$, and learn this channel for any choice of parameter values. The knowledge of the control operation can then be fed forward to reliably determine a full process tensor.

\subsection{Demonstration on NISQ Devices}
\subsubsection*{Non-Unitary Control}

\begin{wrapfigure}{r}{0.5\textwidth}
	\centering
	\includegraphics[width=0.9\linewidth]{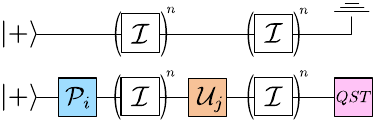}
	\caption[Circuit diagram enacting non-unitary gates with naturally occurring noise ]{To enact non-unitary gates of our choosing, we conduct a similar experiment. This time, however, there are four basis preparation operations to begin with, and \acs{QST} only on the single qubit. This is so that we can optimise the action of the gate over a complete basis of inputs.}
	\label{fig:nu-circuit}
\end{wrapfigure}

We validate and demonstrate these ideas again on IBM Quantum devices. Early progress on this demonstrated that non-unitary control could be obtained with the naturally idle noise of the device. 
Both the system qubit and its neighbour were both initialised in the $\ket{+}$ state. We sought to use the process tensor to control the always-on interaction between the two qubits without actually learning it. The circuit diagram is shown in Figure~\ref{fig:nu-circuit}.

For the purpose of implementing our own chosen non-unitary operations, we created a one-step basis-24 process tensor on a single qubit whose neighbour was in the $\ket{+}$ state: approximately $800$ ns of idle time after $\mathcal{P}$ preparations, followed by $\mathcal{U}^{(24)}$, followed by another $800$ ns and then \acs{QST}.
We then generated a set of random non-unitary operations with unitarity ranging from $1/3$ to $1.0$.
These are denoted by $\mathcal{N}(\alpha,\eta)$, where
\begin{gather}
	\begin{split}
		\mathcal{N}(\alpha,\eta) = \sqrt{\eta}\mathcal{E}(\alpha) + \sqrt{1-\eta}Y\mathcal{E}(\alpha),\\
		\text{and}\quad \mathcal{E}(\alpha) = (R_X(\alpha)R_Y(\alpha)R_Z(\alpha)).
	\end{split}
\end{gather}

\begin{wrapfigure}{l}{0.45\textwidth}
	\centering
	\includegraphics[width=0.9\linewidth]{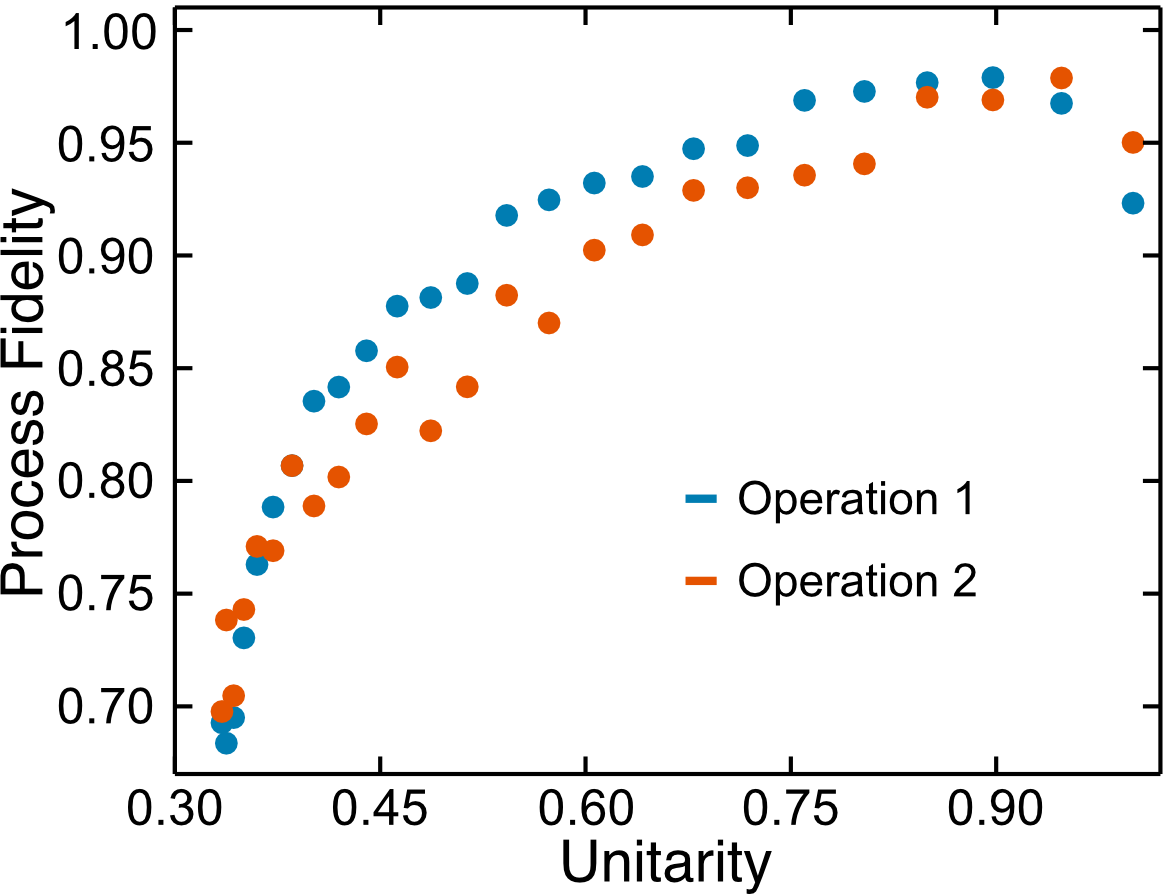}
	\caption[Demonstration of extended non-unitary control via a non-Markovian process characterisation. ]{Demonstration of extended non-unitary control via a non-Markovian process characterisation. }
	\label{fig:bootstrapped-non-unitary}
	\end{wrapfigure}

The two operations shown in Figure~\ref{fig:bootstrapped-non-unitary} are two different randomly generated values for $\alpha$. The unitarity of the operations is then varied by varying $\eta$ from $0$ to $0.5$ in the above equation.
Using these operations as a target map, we numerically found the gate parameters minimising the trace distance between the target outputs of the non-unitary map and the process tensor predictions for a set of four inputs.
That is, we applied the minimisation:
\begin{gather}
	\begin{split}
		\argmin_{\theta,\phi,\lambda} &\frac{1}{2} \left(||\tau_X - \rho_X||_1 + ||\tau_Y - \rho_Y||_1\right. \\ 
		&+ \left.||\tau_Z - \rho_Z||_1 + ||\tau_{I-Z} - \rho_{I-Z}||_1 \right),
	\end{split}
\end{gather}
where each $\rho_j$ is the ideal output of $\mathcal{N}(\alpha,\eta)$ acting on the $X, Y, Z, $ and $\mathbb{I}-Z$ eigenvectors, and each $\tau_j$ is the $\mathcal{T}_{2:0}\left[\mathcal{P}_j, U(\theta,\phi,\lambda)\right]$ predicted density matrices. 
Then, using the optimal values of $\theta, \phi, $ and $\lambda$, we performed quantum process tomography and compared the process tensor of our implementation $\mathcal{N}'(\alpha,\eta)$ with the ideal $\mathcal{N}(\alpha,\eta)$.\par 

The process fidelity of these non-unitary maps compared to their targets is plotted as a function of unitarity in Figure~\ref{fig:bootstrapped-non-unitary}.
It reaches up to $97\%$, showing that we can extend the control capabilities of the device by using the process tensor and a nearby coupled qubit. This target gate is achieved for a given interaction of the system with its neighbour. Since interaction time is not varied, the maximum achievable non-unitarity is fixed, which is why the process fidelity decreases when gates with a lower unitarity are targeted.

This shows a way forward in which extended control regimes could be used for the implementation of non-unital and trace-decreasing maps which are necessary for the reconstruction of the full process tensor.
Critically, for this to work, we do not need to perform control operations on the neighbouring qubit beyond its initialisation. For further detail about this implementation, see the methods section.\par 
This simple framework is widely applicable to many forms of quantum control. 
In particular, it allows for either mitigating or controlling non-Markovian noise without first understanding it at a microscopic level. 
Broadly, the user need only specify a desired outcome, without studying the means to achieve it.

\subsubsection*{Informationally-Complete Control}

Having demonstrated results showing control over non-unitary interactions shows that we may in principle control non-Markovian dynamics as an extra resource.
However, this does not actually help us tomographically, because the unitary mixtures still lie in the span of unitaries. To access the remaining six dimensions of single-qubit superoperator space we need to construct and characterise some bespoke circuits. In particular, we drive the native $ZX$ interaction between a system qubit and its neighbour to realise the extended control.

%
As an alternative, we leverage the use of an ancilla qubit and drive a short entangling interaction, followed by projective measurements on the ancilla. This generates an effective map on the system which can be partially controlled through a local unitary operation. The system-ancilla circuit we design uses the native cross-resonance interaction as well as local operations to achieve large values for each $(R_\mathcal{E})_{i0}$ and $(R_\mathcal{E})_{0j}$ such that a well-conditioned basis may be designed.

In \acs{PTT}, it is necessary that the instruments be well-characterised. To achieve this, we design a process tensor whose mapping is from the local pulse parameters to the effective operation on the system. The idealisation of this is as follows. An ancilla qubit is prepared in the state $(\ket{0} + i\ket{1})/\sqrt{2}$, meanwhile the system is prepared in one of the four states $\{\ket{+}\!\bra{+}, \ket{i+}\!\bra{i+}, \ket{0}\!\bra{0},\ket{1}\!\bra{1}\}$. We then drive the interaction $R_{XZ}(\gamma):=\text{exp}(-\frac{i\gamma}{2} X_A\otimes Z_S)$ -- which is the native cross-resonance interaction on IBM Quantum devices -- for some $\gamma$. An $S$ gate is then applied to the ancilla, followed by a unitary on the system parametrised in the standard \texttt{Qiskit u3} parametrisation.


A second $U_{XZ}(\gamma)$ chirp is then applied, followed by projective $Z$ measurement on the ancilla as well as \acs{QST} on the system. The outcome of the ancilla measurement $\ket{x}\!\bra{x}$ is then attached to the reconstructed state. The end result is a parametrised quantum instrument on the system $\Lambda_x(\theta,\phi,\lambda)$.  The circuit presented is designed such that a set of parameter values exists to amplify each of the six (three non-unital and three trace-decreasing) entanglement-breaking directions. The theoretical matrix elements corresponding to a $\ket{0}\!\bra{0}$ measurement on the ancilla, with $\gamma=\pi/4$ are:
\begin{equation}
	\begin{split}
		(R_{\Lambda_0})_{x0} &= \frac{\sqrt{2}}{2}\cos\phi\sin\theta - 2\cos\frac{\theta}{4}\sin^3\frac{\theta}{4}\sin\phi,\\
		(R_{\Lambda_0})_{y0} &= \frac{1}{4}\cos\phi\left(2\sin\frac{\theta}{2} - \sin\theta\right) + \frac{\sqrt{2}}{2}\sin\theta\sin\phi,\\
		(R_{\Lambda_0})_{z0} &= \frac{1}{8}\left(-1 + 5\cos\theta\right),\\
		(R_{\Lambda_0})_{0x} &= \sin\frac{\theta}{2}\left(\sqrt{2}\cos\lambda\sin^2\frac{\theta}{4} + \cos^2\frac{\theta}{4}\sin\lambda\right),\\
		(R_{\Lambda_0})_{0y} &= \frac{1}{2}\sin\frac{\theta}{2}\left[\left(1+\cos\frac{\theta}{2}\right)\cos\lambda - \sqrt{2}\left(1 - \cos\frac{\theta}{2}\right)\sin\lambda\right],\\
		(R_{\Lambda_0})_{0z} &= \frac{1}{8}(3+\cos\theta).
	\end{split}
\end{equation}
\begin{figure}[!t]
	\centering
	\includegraphics[width=0.7\linewidth]{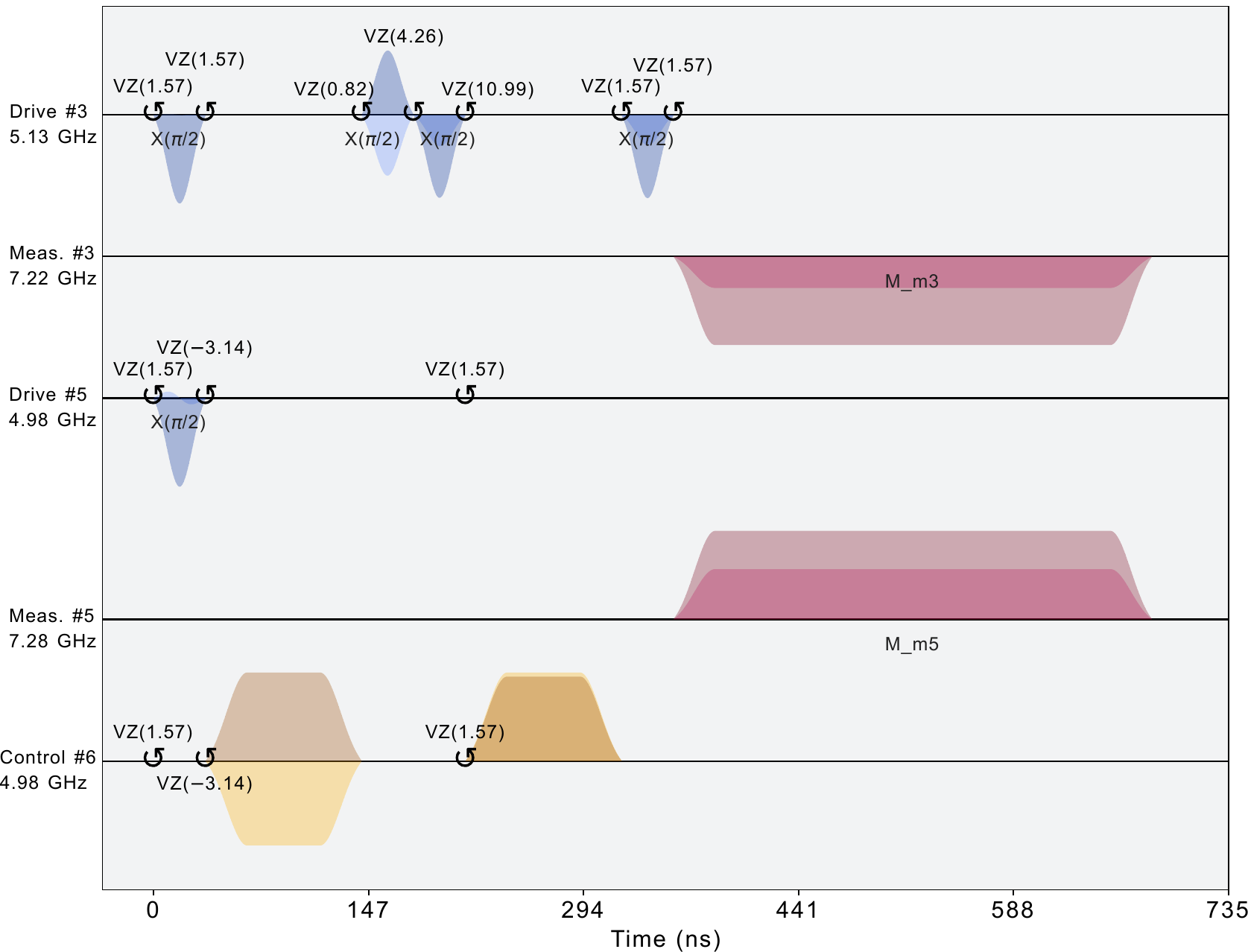}
	\caption[Sample pulse schedule for instrument creation using cross-resonance interactions ]{Sample pulse schedule for instrument creation on \emph{ibm\_perth}. We manually drive short Gaussian square cross-resonance pulses between the system qubit and the ancilla qubit. The local system operations -- depicted through a series of physical pulses and frame shifts on drive channel 3 -- constitute the restricted process tensor used for instrument creation.}
	\label{fig:instrument-schedule}
\end{figure}
\begin{figure}[h!]
	\includegraphics[width=\linewidth]{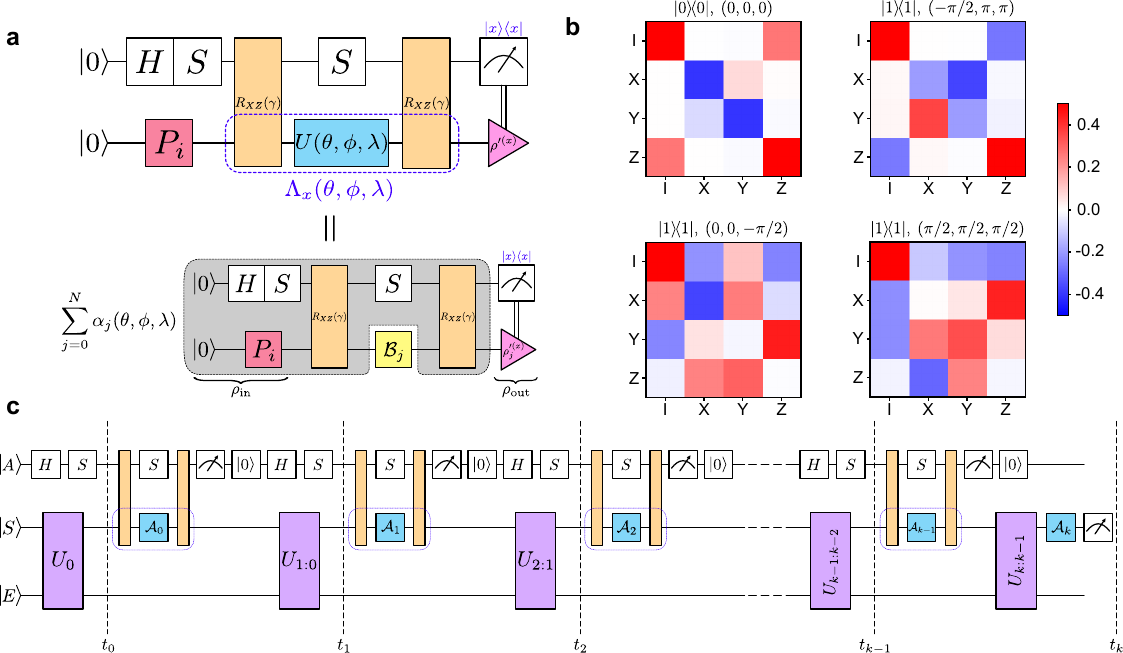}
	\caption[Instrument design for complete reconstruction of multi-time statistics ]{Instrument design for complete multi-time statistics. \textbf{a} Using a structured system-ancilla interaction, we construct a restricted process tensor with local unitaries to realise the desired mapping. This gives an exact characterisation of the effective instrument $\Lambda_x(\theta,\phi,\lambda)$ for any choice of parameter values. \textbf{b} Some sample basis instruments implemented using this method on \emph{ibm\_perth} written in the Pauli basis, showing \acs{IC} controllability including non-unital and trace-decreasing directions. \textbf{c} Many-time physics is then studied using the circuit structured as shown. The different instruments applied capture temporal quantum correlations from an arbitrarily strong $SE$ interaction.}
	\label{fig:instrument-creation}
\end{figure}
\noindent
This suffices for a complete basis. When implemented in practice, however, this instrument will be noisy: the cross-resonance interaction is imperfect and generates additional local terms in the Hamiltonian as well as slight decoherence. Moreover, the measurements themselves made on each qubit have some associated error. These issues may be well accounted-for as long as the instrument is characterised. However, employing \acs{QPT} will only be valid for a single set of local parameter values -- and it would be highly expensive to repeat the procedure for every desired local gate. Instead, we perform restricted \acs{PTT} with a basis of unitary operations on the system for an informationally complete set of measurements. This requires 40 circuits. Let the resulting two-step process tensor be denoted by $\Upsilon_I$. Note that this process tensor maps to the two-qubit $SA$ output which corresponds to the classical state $\ket{x}\!\bra{x}$ on $A$ and the quantum state $\rho^{(x)}$ on $S$. 
Let $\hat{\mathcal{P}}_i$ be the Choi state of the preparation operations $\{H, SH, \mathbb{I}, X\}$,
and finally, let $\hat{\mathcal{U}}(\theta,\phi,\lambda)$ be the Choi state of the local unitary. The Choi state of the parametrised instrument corresponding to outcome $x$ can be written as
\begin{equation}
	\label{eq:instrument-choi}
	\hat{\Lambda}_x(\theta,\phi,\lambda) = \sum_{i=0}^3 \rho_i'^{(x)}(\theta,\phi,\lambda)\otimes \omega_i^\text{T},
\end{equation}
where each $\rho'^{(x)}(\theta,\phi,\lambda)$ is the output state on the system level subject to some choice of local parameter values. Now, since $\Upsilon_I$ maps from these two local operations, we have
\begin{equation}
	\text{Tr}_\text{in}\left[\Upsilon_I\cdot(\mathbb{I}_S\otimes\hat{\mathcal{U}}(\theta,\phi,\lambda)^\text{T} \otimes \hat{P}_i^{\text{T}})\right] = \ket{0}\!\bra{0}\otimes \rho_i'^{(0)}(\theta,\phi,\lambda) + \ket{1}\!\bra{1}\otimes \rho_i'^{(1)}(\theta,\phi,\lambda).
\end{equation}
This allows us to re-write Equation~\eqref{eq:instrument-choi} in terms of our reconstructed process tensor:
\begin{equation}
	\label{eq:PTT-instrument}
	\hat{\Lambda}_x(\theta,\phi,\lambda) = \sum_{i=0}^3\text{Tr}_{\text{in},A} \left[\Upsilon_I\cdot\left(\ket{x}\!\bra{x}\otimes \mathbb{I}_S\otimes\hat{\mathcal{U}}(\theta,\phi,\lambda)^\text{T} \otimes \hat{P}_i^{\text{T}}\right)\right]\otimes \omega_i^{\text{T}}
\end{equation}
This outline is the idealised version of the procedure. In reality, we first perform \acs{GST} to obtain estimates of both the system input states $\rho_i$ and the system and ancilla \acs{POVM}s. This ensures that the resulting \acs{PTT} to estimate $\Upsilon_I$ and \acs{QPT} on the instrument is self-consistent. Note that the corresponding reconstruction is unaffected by the gauge freedom in \acs{GST} since this does not affect the linear relationship between basis operations.
Using Equation~\eqref{eq:PTT-instrument} we then have an accurate representation of the effective instrument acting on the system for any choice of $\theta$, $\phi$, and $\lambda$. Finally, noting that we may reset the ancilla qubit after each projective measurement to reuse it, this permits us the necessary control for tomographically complete observation of a multi-time process. The ideal circuit structure is given in Figure~\ref{fig:instrument-creation}, and allows one to be cognisant of any imperfections in the cross-resonance drive. 
Figure~\ref{fig:instrument-creation}b shows some representative basis instruments from applying this procedure on \emph{ibm\_perth}, which we employ further in Chapter~\ref{chap:MTP}. Because this method requires manipulation of a system with initial correlations, this procedure would be otherwise not possible without employing using of a process tensor to bootstrap the construction. Having demonstrated that these instruments permit us complete control of the system, we can use them for complete process characterisation as illustrated in Figure~\ref{fig:instrument-creation}c.

In the case where mid-circuit measurement capabilities are not available, we can still use ancilla qubits as instruments, but the measurement needs to be deferred until the very end of the circuit. This has two limitations: the first is that there now needs to be one ancilla qubit for each instrument, they cannot be reused. Practically, then, this limits us to 2--3 steps. Secondly, the ancilla qubits themselves may interact with the system (influencing the process tensor), or they may thermalise, changing the measurement populations. This means that the processes cannot be on too long a time-scale. We circumvent the interactions by applying dynamical decoupling protocols to the ancilla qubits. This type of process tensor is restricted in scope, but does demonstrate that even devices without any mid-circuit measurements can still achieve tomographically complete control. Note that in principle a single ancilla qubit suffices to be used as a tester across an arbitrary number of steps -- as suggested in the previous chapter -- but in practice the characterisation of such a tester is out of reach and so we are faced with the same issue of instrument error. We will revisit and rectify this problem in Chapter~\ref{chap:universal-noise}.

To summarise the important points and procedure:

\begin{enumerate}
	\item \acs{NISQ} devices typically do not have clean and fast control beyond unitary control. Even for devices with mid-circuit measurement capabilities, the duration of the instrument is slow when compared to the background dynamics.
	\item Ancilla qubits may be used to access non-unital and trace-decreasing directions. By interacting a system with a neighbour and projectively measuring that neighbour, the duration of the control is reduced to the duration of the $SA$ interaction, which can be fast. This projective measurement can either happen immediately, or at the end of the circuit depending on limitations.
	\item In order to characterise the instruments, we can manipulate the interaction with a local unitary and measure the subsequent responses with restricted-\acs{PTT}. This then tells us the effective quantum map for any choice of local parameter values. 
\end{enumerate}

\section{Compressed Sensing}

In all characterisation protocols it is imperative to apply the minimal effective model to describe the dynamics in question. Not only is this due to a preference in simplicity, but the fewer parameters, the more accurately one can estimate them, and the fewer experiments required to do so. So far, we have derived the characterisation of non-Markovian processes in full generality. This requires $\mathcal{O}(d_S^{4k})$ numbers to describe. In particular, we have described the estimation of generically full-rank process tensors. Recall from Chapter~\ref{chap:stoc-processes} the interpretation of process tensor rank $r$: it is the minimum of either the dimension of the process, or the number of environment eigenvectors coupling non-trivially to the system. 

The history of compressed sensing is long, and far pre-dates quantum information~\cite{bobin2008compressed,rilling2009compressed,boufounos20081}. In the context of quantum information, however, Gross \emph{et al.} first proposed the application of a rank deficient model 
to resource reduction in quantum state tomography~\cite{PhysRevLett.105.150401,flammia2012quantum}. If a state $\rho$ of dimension $d$ has rank $r$, then $\rho$ has an eigendecomposition $\rho = \sum_{i=1}^r p_i|\psi_i\rangle\!\langle \psi_i|$, which can be expressed in $\mathcal{O}(rd)$ numbers. They show that only $\mathcal{O}(rd\log^2 d)$ measurement settings are required. Here, we establish a method for \acs{PTT} based on compressed sensing for reconstructing processes with an approximately low dimension environment. 

In fact, we already have all of the tools we need to implement this. Instead of applying Equation~\eqref{PSD-proj} at each iteration, we can project onto $\mathcal{S}_{n,r}^+$ -- the set of rank $r$, $n\times n$ positive semidefinite matrices. 
\begin{equation}
	\text{Proj}_{\mathcal{S}_{n,r}^+}(\Upsilon) = U\text{diag}(\lambda_0^+,\cdots,\lambda_{r-1}^+,0,\cdots,0)U^\dagger
\end{equation}
Note that this problem is no longer convex, but satisfies a local form of regularity that guarantees local linear convergence~\cite{lewis2009local}. Concretely, this translates into taking much smaller steps in the \acs{MLE} projected gradient descent. Heuristically, we find a good step meta-parameter to be about an order of magnitude smaller than the fully general case, or $\mu = 3/20n^2$. 

If one has a small non-Markovian memory, then one can relax the requirements of informational completeness in the number of basis elements. The total number of measurement settings is then $\mathcal{O}(r d_S^{2k}\log^2 d_S^{2k})$, which in practice is a large reduction. As well as the advantage of fewer required quantum resources, there are fewer parameters in the model. Hence, the precision-per-parameter is also improved. This method would best fit an online data-collecting setting. Simply select an initial (small) rank for the model $r_0$ and collect some validation data. Collect an initial number of tomographic experiments for basis size $N \approx r_0 d_S^{2k}$, fit the data. The resulting estimate can be cross-validated against the validation data. If insufficient, then an increased rank $r_1$ is chosen and an extra $(r_1 - r_0)d_S^{2k}$ measurement settings collected. 
The experimenter then steps this procedure forward until a desired reconstruction fidelity is achieved.

There is a freedom in selecting basis operations here. If operations are selected pseudo-randomly -- i.e., as the reduction of some complete basis -- then the expected convergence will be $1/\sqrt{N}$, like all Monte Carlo methods. However, \emph{adaptive} strategies, such as adaptive \acs{QST}~\cite{PhysRevLett.111.183601} or cross-interpolation algorithms~\cite{savostyanov2011fast} can provide a quadratic improvement. The philosophy of these approaches is to use informed updates after each measurement to select the following measurement. 


\section{Analysis}
\label{sec:ptt-analysis}

We conclude this chapter by briefly analysing the tools we have developed. In particular, we look at some numerics of their reconstruction behaviour across random processes. We also contextualise them among more standard \acs{QCVV} procedures. We discuss model restrictions and paths forward to more accurate and robust process tensor estimation.
\subsection{Simulation}
As we have discussed, the methods introduced are not contingent on the particular $SE$ interactions, hence the reconstruction quality should not depend exactly on how non-Markovian the dynamics are. Still, we numerically benchmark these approaches on random processes using the techniques introduced in the previous chapter to draw random processes from an ensemble. This substantiates the claim of the ability to capture the dynamics of any multi-time process.

\begin{figure}[t]
	\centering
	\includegraphics[width=\linewidth]{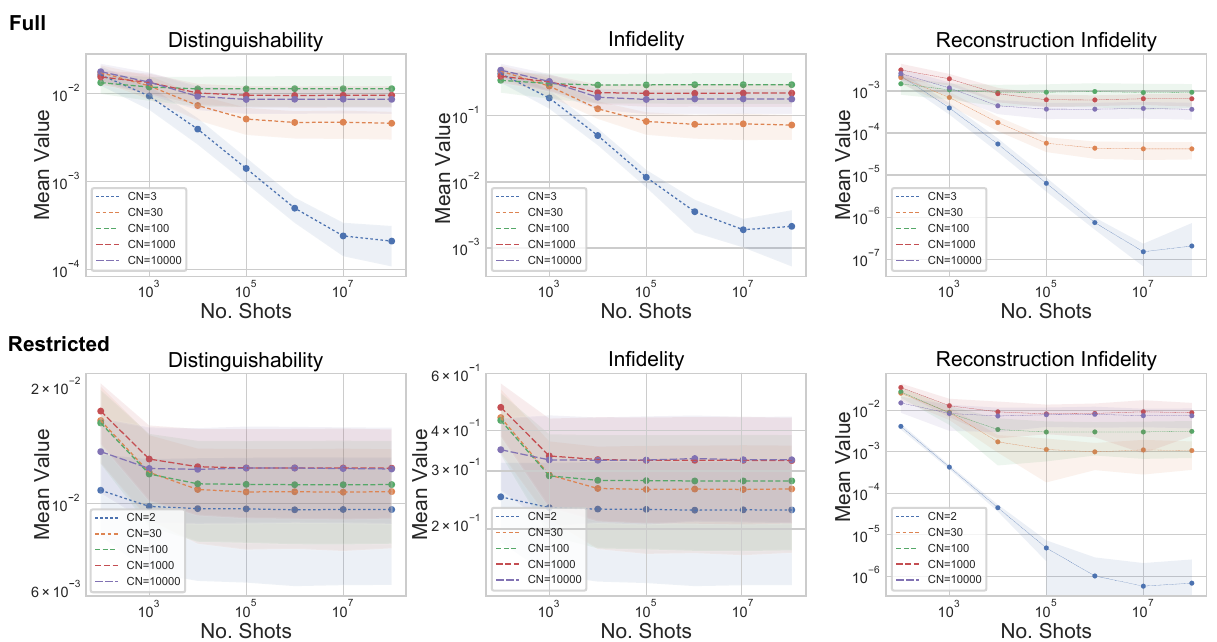}
	\caption[Analysis of shot noise and condition number in maximum likelihood estimation process tensor tomography ]{An analysis of maximum likelihood estimation process tensor tomography. We compare the effects of basis condition number (CN) and total shot number on the quality of the total reconstruction for both full (informationally complete) and restricted (unitary-only) process tensors. We examine the effects on distinguishability of the reconstructed process, infidelity of the characterisation, and reconstruction fidelity in the ability to predict the outcomes of new sequences.}
	\label{fig:mlpt-analysis}
\end{figure}

In Figure~\ref{fig:mlpt-analysis} we analyse convergence and accuracy properties of our \acs{MLE}-\acs{PTT} on randomly generated process tensors. Specifically, we tomographically reconstruct random two-step process tensors. First, we select a basis in which to measure, and a number of shots to collect. We then perform \acs{MLE}-\acs{PTT} to estimate the process tensor. We consider various properties for the quality of reconstruction: the scaled trace distance $\|\Upsilon_{\text{est}} - \Upsilon_{\text{true}}\|_1 / (2d_S^{2k+1})$, which relates to the average probability of distinguishing two processes~\cite{PhysRevA.71.062310}; the Uhlmann fidelity $F(\Upsilon_{\text{est}},\Upsilon_{\text{true}})$; and the reconstruction fidelity across random inputs. We consider these both for the case of a tomographically complete process tensor, and a restricted unitary-only process tensor. Note that in the latter case, the estimate is non-unique. Hence, there is no expectation that collecting more data will increase convergence between the estimated object and the true object, beyond the overlap on restricted data. This is reflected in the plateau of distinguishability and infidelity. However, since the mapping is valid on the subspace of the span of unitary operations, the reconstruction fidelity continues to decrease. 

The chosen bases are chosen according to the condition number of that basis. Condition number is defined as the ratio of highest and lowest singular values for the matrix where each basis vector is stored as a row. This is a measure of the stability of the linear inversion problem, and we contrast up to a condition number of 10\ 000. In the limiting cases, the full and restricted process tensors respectively have a symmetric \acs{IC}-\acs{POVM} and an approximate \acs{MUUB} as the chosen bases. The difference between appropriately chosen bases is stark, and in the asymptotic setting can mean an improvement of reconstruction estimate by several orders of magnitude.

\subsection{Hierarchy of Quantum Tomography}
\label{ssec:hierarchy}
Having establised a procedure for generalised multi-time tomography, it will be instructive to re-summarise the fundamentals of \acs{QST}, \acs{QPT}, and \acs{PTT}. This provides context to the work we have introduced, and clue to both shortcomings and paths forwards. These procedures build on each other; for a $d_S$-dimensional system, \acs{QST} requires a set of $\mathcal{O}(d_S^2)$ experiments, \acs{QPT} is most easily thought of as $d_S^2$ \acs{QST}s and requires $\mathcal{O}(d_S^4)$ experiments, and \acs{PTT} $\mathcal{O}(d_S^{4k})$ experiments, where $k=1,2,\ldots$ is the number of times steps~\cite{RBK2010,PhysRevA.87.062119, White-NM-2020}. 
\acs{PTT} is hence a formal generalisation of \acs{QPT} to the multi-time setting, a natural next object in the hierarchy of quantum tomography, in the sense that the information of each level is strictly contained within the characterisation of the next level. That is to say, \acs{QPT} can describe the reconstructed state of \acs{QST}, and \acs{PTT} can describe the dynamical map of \acs{QPT}. We present a summary of each map in Table~\ref{tab:tomography-summary}, as well as their physical requirements, and continue to flesh out here. 

\begin{table*}[t]
	\centering
	\resizebox{\textwidth}{!}{%
		\begin{tabular}{@{}l|l|l|l@{}}
			\toprule
			& Quantum State                  & Quantum Process                                                 & Process Tensor                                                                                                                     \\ \midrule
			Characterisation object & Density Operator $\rho$                  & Quantum Channel $\mathcal{E}$                                             & Process Tensor $\mathcal{T}_{k:0}$                                                                                                              \\
			Mapping                 & $\mathcal{H}_S\rightarrow \mathcal{H}_S$ & $\mathcal{B}(\mathcal{H}_{S})\rightarrow \mathcal{B}(\mathcal{H}_{S})$    & $\bigotimes_{i=1}^k \mathcal{B}(\mathcal{B}(\mathcal{H}_{S})) \rightarrow \mathcal{B}(\mathcal{H}_S)$                                                   \\
			Observed probabilities  & $p_i = \text{Tr}\left[\Pi_i\rho\right]$    & $p_{ij} = \text{Tr}\left[(\Pi_i\otimes \rho_j^\text{T})\hat{\mathcal{E}}\right]$         & $p_{i,\vec{\mu}} = \text{Tr}\left[(\Pi_i \otimes  \hat{\mathcal{B}}_{k-1}^{\mu_{k-1}\text{T}}\otimes \cdots \otimes \hat{\mathcal{B}}_0^{\mu_0\text{T}})\Upsilon_{k:0}\right]$ \\
			Positivity Constraint   & $\rho \succcurlyeq 0$                    & $\hat{\mathcal{E}} \succcurlyeq 0$                                              & $\Upsilon_{k:0} \succcurlyeq 0$                                                                                                              \\
			Affine Constraint       & $\text{Tr}\left[\rho\right] = 1$         & $\text{Tr}_{\text{out}}[\hat{\mathcal{E}}] = \mathbb{I}_{\text{in}}$ & $\text{Tr}_{\mathfrak{o}_k}\left[\Upsilon_{k:0}\right] = \mathbb{I}_{\mathfrak{i}_k}\otimes \Upsilon_{k-1:0}\quad \forall\quad  k$                           \\ \bottomrule
		\end{tabular}%
	}
	\caption{Detail on different levels of the quantum tomography hierarchy, pertinent to experimental reconstruction. \acs{QST} reconstructs a density operator, $\rho$, a positive matrix with unit trace representing the quantum state. \acs{QPT} reconstructs a quantum channel, $\mathcal{E}$ through its action on different states. This map must be both \acs{CP} and TP, conditions which, in Choi form, manifest themselves as positivity and affine constraints on the matrix. Finally, \acs{PTT} reconstructs a process tensor $\mathcal{T}_{k:0}$ through its action on different control operations. This object must have a positive matrix Choi form, and respect causality.
	The information of each column is strictly contained in the column to the right.}
	\label{tab:tomography-summary}
\end{table*}

The familiarity of the first two lays the groundwork for the latter. The treatment and practical concerns of each technique are similar with respect to real data. One key difference lies in the fact that due to the higher-dimensional superoperator basis, especially for \acs{PTT}, small errors can become magnified and require closer attention. In addition to an overview of tomography, we present the conceptual developments of \acs{PTT} in this section and analyse its particulars -- such as with respect to hardware control restrictions and \acs{SPAM} error.\par
Fundamentally, tomography is an exercise in reconstructing linear maps from experimental data. This can be accomplished by measuring the input-output relations on a complete basis for the input space. A disconnect between theory and experiment occurs when, in practice, the input vectors are faulty in some way, such as in a noise prepartion. Or, if the measured output frequencies differ from that of the real population (due to a noisy probe or finite sampling error)~\cite{RBK2010}. As well as producing an object that may disagree with experiment for inputs away from the characterisation, the resulting estimate might not even be physical. A variety of different methods may be employed to overcome some of these problems: the collection of more data, the elevation of inputs and outputs to the model~\cite{gst-2013,PhysRevA.87.062119}, employing an overcomplete basis in the characterisation~\cite{intro-GST}, and the treatment of the measured data to fit a physical model~\cite{Hradil2004}. These techniques are applicable, regardless of the model type, and we will discuss their utility in \acs{QST}. \par

\subsection{Model Considerations}

It is key to know the limitations of any \acs{QCVV} procedures, in particular how the estimate is affected by faulty control -- or \acs{SPAM} errors. In the context of \acs{PTT}, the usual notions of \acs{SPAM} need to be broadened. For example, quantum channels are assumed to act on ``known'' input states; when this assumption breaks it can be problematic for \acs{QPT}. In contrast, the initial state is a marginal to the process tensor, and any error is naturally estimated. A measurement probe, on the other hand, is required to read out information. Errors which are insensitive to the \acs{POVM} effect will absorb into the process. If they are common to all bases, the predictive capabilities will be unaffected. However, the estimated process tensor itself will then look slightly noisier. Accounting for either this or basis-specific noise can be straightforwardly achieved by using an estimate of the device \acs{POVM} in the model. Estimates may be obtained with consistent detector tomography outputs from procedures such as \acs{GST}~\cite{nielsen-gst}.

For \acs{PTT}, it is the input quantum operations that are assumed to be known, replacing `state preparation' in \acs{QPT}. In order of increasing consequence, violation of this assumption can occur in three ways: (i) with gate-independent error (such as decoherence), (ii) with gate-dependent coherent error, and finally, (iii) with significant $SE$ interaction during the finite time gate. 
Similar to the measurement case, \acs{PTT} is insensitive to independent error for the purposes of control, since it does not change the linear relation between basis elements. The second consideration is more problematic because it can lead to an inconsistent characterisation. This can be resolved in two ways: by \emph{a priori} characterising the gates themselves through \acs{GST}, or by using an overcomplete basis to average over the coherent error. Lastly, if a non-Markovian interaction occurs with coupling $\mathcal{O}(1/\tau_p)$ for control width $\tau_p$ then the process tensor model will break down. However, we expect this final possibility to be extremely rare for any functioning device -- but indeed could be circumvented with virtual gates, such as those described in Ref.~\cite{PhysRevA.96.022330}.

Because the input control must be high fidelity, we view \acs{PTT} predominantly as a useful tool for quantum devices clean enough to be sensitive to non-Markovian dynamics. In this work, the demonstrated results focus on single qubit unitary gates, for which the error is $\mathcal{O}(10^{-4})$.
A future extension to \acs{PTT} one might consider is a self-calibrating simultaneous estimation of both the process tensor and the input interventions, as in GST. Indeed, we pursue this direction in Chapter~\ref{chap:universal-noise}.
Though incorrect characterisations through gate errors are not implausible, single qubit gate errors for a typical \acs{NISQ} device are already smaller than the expected $1/\sqrt{N_{\text{shots}}}$ sensitivity, and most of this error is represented by decoherence during the small finite pulse width. 
Moreover, the control aspect may be self-consistently checked by comparing predictions made from estimates of the process tensor with random gate sequences on the real device, offering certification to the characterisation.

\subsection{Discussion}
In this chapter we have developed and demonstrated a fully general approach to non-Markovian quantum process tomography. This resolves a long-standing open problem in quantum characterisation, and comprehensively captures any open quantum dynamics to arbitrary accuracy, and is equipped with performance guarantees.
As well as an experimental procedure, we introduced two different ways to post-process the data: linear inversion and maximum likelihood estimation. The former is computationally cheap, whereas the latter circumvents issues with sampling error and outputs a physical process estimation which can then be interrogated. It is worth noting that there are other estimation procedures in the literature which have been well-studied, and that \acs{PTT} is fully consistent with any of these approaches.
\acs{MLE}, for example, is widely used for its convenience but is known to be slightly sub-optimal by most metrics, and can result in rank-deficient estimates~\cite{RBK2010,scholten2018behavior,ferrie2018maximum}. In practice, however, we find the approach to be robust and the quality of the reconstruction to be as high as we desire.
Bayesian mean estimation is a provably optimal procedure~\cite{Granade_2016}, but it is still not known how to construct generically efficient algorithms even in the state case. It is also an open question how to best select a prior distribution. Other approaches, such as hedged-\acs{MLE}~\cite{blume2010hedged}, minimax tomography~\cite{PhysRevLett.116.090407}, self-guided tomography~\cite{PhysRevLett.126.100402}, and a variety of others have advantages and are fully consistent with \acs{QST}. We do not explore their application here, but it would be straightforward to do so as a potential avenue for research. 


An obvious drawback to the characterisation is resource requirements. A generic non-Markovian process is exponentially complex to describe, and hence to characterise. This is because if the environment is complex enough, then at each step the intervention on the system can uniquely change the trajectory of the system. This is true even of classical stochastic processes, but the sample space becomes much larger in the quantum setting. Much like quantum states, the curse of dimensionality prohibits us from reconstructing anything beyond a few steps. Further, we are hindered by similar problems to \acs{QPT}, in that we assume the instruments used to probe the process are known.  
Fortunately, in the context of quantum noise, the extent to which correlations persist ought to be sparse, and the quality of control to be high. In Chapter~\ref{chap:efficient-characterisation}, we address these drawbacks by developing methods with which sparseness of the correlations can be leveraged to produce and estimate more efficient models. We also extend the protocol to be self-consistent, including the ability to simultaneously estimate control operations as well in Chapter~\ref{chap:universal-noise}. 
We have also focused at this point only on the extrinsic characterisation. That is, the ability to predict the behaviour of the dynamics. We focus on intrinsic properties, such as the structure of correlations and how to extract meaningful information about the many-time physics in Chapter~\ref{chap:MTP}.

\chapter{From many-body to many-time physics}
\label{chap:MTP}
\epigraph{\emph{We've come to replace your clock's old stubborn hands.}}{Andrew Savage, Tenderness}
\noindent\colorbox{olive!10}{%
	\begin{minipage}{0.955\textwidth} 
		\textcolor{Maroon}{\textbf{Chapter Summary}}\newline
		The \acs{CJI} captures temporal quantum correlations, equating them with those within a quantum state. This approach reveals rich physics of multi-time quantum stochastic processes and exotic quantum phenomena. However, extracting these features lacks practical methods. This chapter explores techniques for uncovering non-Markovian memory, drawing inspiration from many-body physics, with demonstrated effectiveness using real experimental data. We address technical challenges, limitations of control hardware, and informationally incomplete settings, all the while maintaining an experimental focus.
		\par\vspace{\fboxsep}
		\colorbox{cyan!10}{%
			\begin{minipage}{\dimexpr\textwidth-2\fboxsep}
				\textcolor{RoyalBlue}{\textbf{Main Results}}
				\begin{itemize}
					\item We develop methods to measure and bound total non-Markovian memory and temporal entanglement in a quantum stochastic process solely through sequences of unitary operations.
					\item We generalise the method of classical shadows to the spatiotemporal domain to efficiently extract observables from a process.
					\item We measure and benchmark the natural temporally correlated noise on \acs{NISQ} devices, finding it to be quantum in nature. 
					\item We develop a procedure to efficiently filter out the effects of qubit-crosstalk to determine the non-Markovianity of inaccessible baths.
					\item We demonstrate the accurate learning of multi-time sampling statistics of simulated open quantum systems on \acs{NISQ} devices. 
				\end{itemize}
		\end{minipage}}
\end{minipage}}
\clearpage
\section{Introduction}

Quantum information processors -- computing devices, sensors, and communicators -- have recently seen an explosion in both accessibility and interest.
Even more broadly, adjacent fields such as condensed matter physics and biology are becoming increasingly engaged with the direct effects of quantum mechanics in their respective areas~\cite{marx2021biology,mcfadden2018origins,roman2021quantum}.
Crucially, what these areas all have in common is that the need to understand the nature and dynamics of open quantum systems. But the ability to equip this nascent technology to study the dynamics of open quantum systems is still very much lacking. In this chapter, we show that this is a problem that can be solved through methodology and conceptual breakthrough, rather than a limitation of hardware or scope. This provides a new set of tools designed to understand the nature of temporal correlations both in generic open quantum systems and, importantly, in the context of noise on quantum devices. 

Quantum correlations are at the heart of most captivating microscopic phenomena, which are linked in stronger ways than any classical system. The microscopic correlations affect the macroscopic properties of many-body quantum systems, leading to exotic phases of matter. Discourse on this topic is typically dominated by spatial correlations exhibited at a single time, or where time is merely a parameter for investigating the evolution of spatial properties~\cite{horodecki2009quantum,lanyon2017efficient}.

In contrast, comparatively little has been studied on the subject of \emph{many-time} physics. As we have seen, a single quantum system may have rich dynamics, especially when it interacts with its environment. The relative sparseness of attention is in part due to the fact that it has been a challenge to put multi-time correlations on equal footing as spatial ones. Recent developments in the theory of quantum stochastic processes have closed this gap to map multi-time processes to many-body states~\cite{chiribella_memory_2008, Pollock2018a, 1367-2630-18-6-063032, PhysRevA.98.012328, Milz2021PRXQ} by means of the generalised Choi-Jamio\l kowski isomorphism~\cite{wilde_2013}. It is becoming evident that many-time physics is as vibrant as many-body physics; they are endowed with nontrivial temporal entanglement~\cite{milz21}, exotic causal properties~\cite{chiribella_quantum_2013, OreshkovETAL2012, Ringbauer2017, Carvacho2017, Milz2018, Ringbauer-npjQI}, rich statistical structures~\cite{Romero2020}, resources~\cite{berk, berk2021extracting, arXiv:2110.03233}, and a window into novel physics~\cite{hardy_operator_2012, cotler_superdensity_2017}.

In this chapter, we address the question: how can we access the fantastic phenomena listed above? By doing so, we open the door up to the realistic observation and study of temporal correlations on much the same footing as many-body states. To achieve this goal, we combine approaches well-known in many-body physics, as well as develop new techniques uniquely tailored for observables on temporal quantum states -- particularly with respect to different control paradigms. In the previous chapter, we focused heavily on an ability to capture the extrinsic properties of different aspects of non-Markovian dynamics. This is crucial for certifying the accurate estimation of exotic dynamics. That was the `how'; here we focus on the `what'. What interesting physics can we glean from these processes? What exactly can we say about complex system-environment interactions from system control alone?

We show that quantum computers, even in their current state, offer a timely opportunity to explore the richness of many-time physics. While these machines are too noisy to manipulate into the large entangled registers required for interesting quantum algorithms, they are riddled with complex noise~\cite{White-NM-2020,White-MLPT}, which we can access with high fidelity controls. Thus, we show that there is scope to explore interesting physics even with relatively small devices -- a single qubit probed across many times, for example, can exhibit many of the same properties as large quantum states. In this sense, highly nontrivial many-time physics can be more accessible than many-body physics for the near-term quantum processors, which opens a new rich area of investigations. 
\begin{figure*}[t]
	\centering
	\includegraphics[width=\linewidth]{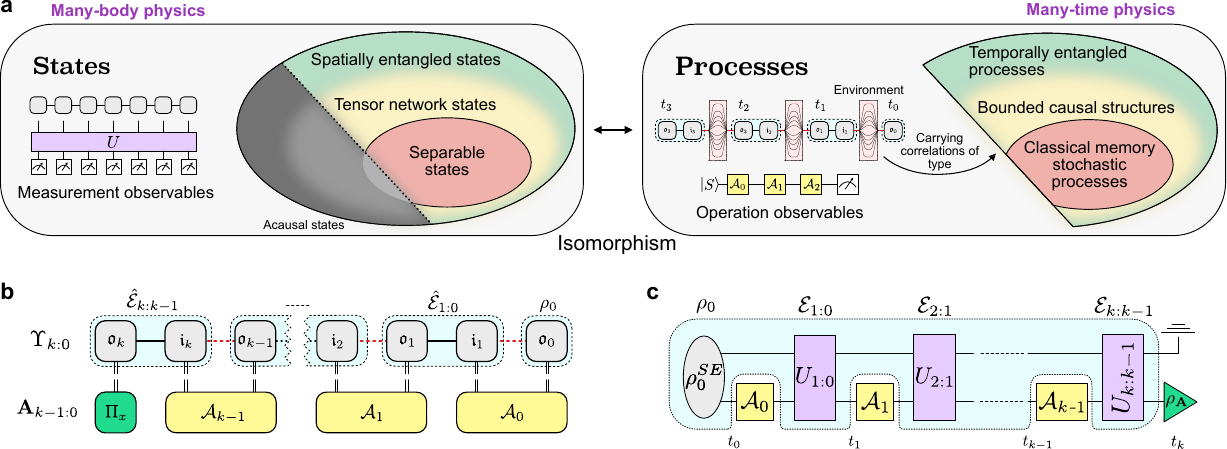}
	\caption[Illustrative summary comparing quantum states from quantum stochastic processes ]{Illustrative summary comparing quantum states from quantum stochastic processes. We highlight the potential for interesting physics to be found in temporal quantum correlations. 
		\textbf{a} A depiction of the space of processes as compared to a subset of states, linked by the Choi-Jamiolkowski isomorphism. Many of the exotic features of many-body physics can be linked to processes, thus generating an equivalent many-time physics. The space of processes is not isomorphic to all states, but rather states that obey a set of causality conditions.
		\textbf{b} A $k$-step process tensor Choi state. This object represents a sequence of possibly correlated \acs{CPTP} maps between different points in time, plus average initial state. Output legs $\mathfrak{o}_l$ are mapped by a control operation $\mathcal{A}_l$ to the next input leg $\mathfrak{i}_{l+1}$.
		\textbf{c} Circuit form of a process tensor with arbitrary $SE$ interactions. Operations at different times map to a final state conditioned on these choices. The process tensor represents everything shaded in blue, the uncontrollable system-enviroment dynamics. 
	}
	\label{fig:subsystems}
\end{figure*}
Non-Markovian dynamics are an essential property to understand when studying quantum devices. 
This type of noise can significantly impact the performance and accuracy of quantum computations.

Our results are divided up into two broad regimes: \textit{micro} and \textit{macro} properties. This is a split is as natural in many-time physics as for many-body physics. In the micro regime, restricted only to high fidelity unitary operations, followed by a final terminating measurement, we show how to witness genuine multitime entanglement, as well as devise tight bounds for various facets of non-Markovian memory. 

To access the macro regime, we look to generalise established techniques for efficient learning of state properties. In particular, we adapt the classical shadows~\cite{huang-shadow} protocol to its most general spacetime setting. As well as derivation of the procedure, we explore the limitations uniquely imposed by causality constraints on the learnability of process tensors in comparison to more general states. 
Using established techniques such as classical shadows and reconstruction methods for finitely correlated states, we show how to access properties of arbitrarily sized processes. 

We then show the high efficacy of these methods by extracting features of many-time physics in a series of quantum simulations of complex dynamics on IBM Quantum processors. These vary from in-depth small scale processes, to characterisation of a 20-step process. We supplement all our results with a plethora of numerical and experimental demonstrations. Applications of this work range from the diagnostics of temporally correlated noise in quantum technologies~\cite{altherr2021quantum, White-NM-2020}, to the observation and simulation of dynamical complexity in condensed matter systems~\cite{nitzan2003electron, PhysRevLett.124.043603, PhysRevLett.115.043601}, to quantum biology~\cite{lambert2013quantum, McGuinness2011} and quantum causal modelling~\cite{1367-2630-18-6-063032}.

\section{Diagnosing Process Properties With Restricted Control}

In Chapter~\ref{chap:PTT} we developed \acs{PTT}, which we rigorously benchmarked to accurately reconstruct quantum stochastic processes.
This procedure included the generic case of informationally complete control. In these instances, one could experimentally reconstruct a process tensor Choi state given a complete set of data. From this, it would then be straightforward to apply the usual quantum information toolkit to probe properties of the process, including entanglement, discord, and other multi-time correlations functions -- all allowing for the extraction of detail information about the system's dynamics. However, notwithstanding the interpretational aspect, this level of control is not always readily available in practical settings. The majority of our experiments were with respect to a restricted process tensor, demonstrating a high-fidelity characterisation of process tensors limited to the span of unitary operations. Even when \acs{IC} control is available, it is typically invasive, slow, and comes with errors. In practice, then, access to the full process tensor Choi state is typically unavailable. 

\begin{figure}[h!]
	\centering
	\includegraphics[width=0.7\linewidth]{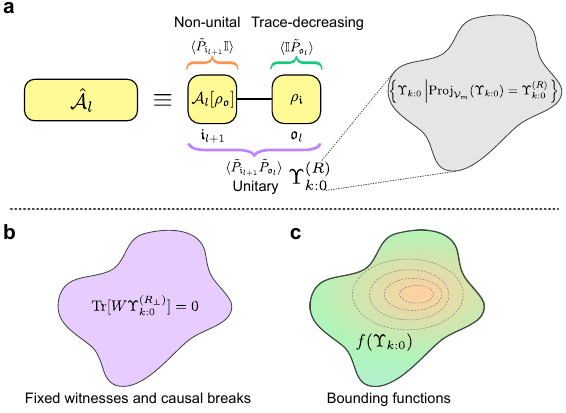}
	\caption[Illustration of various properies that can be determined from restricted process tensors ]{Property determination from restricted process tensors. \textbf{a}
		Observable interventions naturally partition into non-unital, trace-decreasing, and unitary controls. Unitary-only characterisation generates a family of process tensors consistent with this limited data. \textbf{b} Temporal entanglement may be certified by finding entanglement witnesses whose value is determined only by the observed data. Additionally, incoherent causal breaks also lie in the span of unitary operations and can be used to test non-Markovian correlation. \textbf{c} More informative functions of the process may be bounded by finding extrema across the process tensor family.
		}
	\label{fig:RPT}
\end{figure}

In this section, we will explore methods for determining the properties of non-Markovian noise in the regime of restricted control. This is an important problem, as it allows us to probe correlated noise on quantum computers, and more broadly extract further information about an environment in a metrological sense with, for example, quantum sensors. By developing methods that can be applied to systems with limited control, we can gain insight into the dynamics of these systems and improve their performance. Recall from Chapter~\ref{chap:process-properties} that we explored what different regimes of control allowed us to extract from the process, when partitioned into unitary, non-unital, and trace-decreasing maps. This distinction is summarised in Figure~\ref{fig:RPT}a; the span of unitary operations is equal to the span of non-local Pauli terms, and therefore marginals are not available in this regime of control. One might imagine -- also depicted in Figure~\ref{fig:RPT}a -- that the measurement of only these observables gives rise to a subspace of process tensors whose unitary expectation values are identical, but whose non-unital and trace-decreasing parts are free to vary within the constraints of positivity and causality. We call this a \emph{family} of process tensors and consider how one might learn meaningful information about the process from this space of possible data. 

To overcome control limitations, we propose three methods. First, we use the fact that the maximal depolarising channel lies within the span of unitary operations. These maximally noisy operations constitute a causal break on the system, no information may persist through the application. We can hence determine the conditional process tensor with this applied in the middle of a circuit. Any non-zero mutual information between the past and the future is a bound on the total amount of non-Markovian memory in the dynamics. Second, we consider temporal quantum entanglement as measured by entanglement witnesses, but we constrain the witnesses to lie in the span of unitary operations so as to not rely on any observables that are not experimentally captured. Finally, we consider spectral properties of the process tensor Choi state. Although we cannot fully estimate the process spectrum without \acs{IC} control, we have developed a method that allows us to identify extreme characteristics of process tensors within a restricted space, establishing tight bounds on the non-Markovian properties in practice.

A common thread among these three techniques is that these are bounds on non-Markovianity rather than direct measures. This will always be true for restricted control. Recall from Chapter~\ref{chap:process-properties} that unitary operations determine Pauli expectation values of the form $\langle \tilde{P}_{\mathfrak{i}_{l+1}}\tilde{P}_{\mathfrak{o}_l}\rangle$, and so the expectation values of local observables may vary, which in turn varies the total amount of correlation present. Nevertheless, we find that we are able to construct a variety of useful process diagnostics. By developing these methods, we can gain insight into the dynamics of quantum devices and improve their utility in practical settings.

The results presented here are not just methodological advances, but are conceptually important results about what properties of non-Markovianity may be determined solely with restricted control. In near-term computers, and even in far-term sensors, we expect this to be the overwhelmingly prevailing regime for control. It thus behoves us to explore the extent to which many-time physics can be explored even with this restricted control. 
We establish a concrete relationship between the ability to extract information about non-Markovian correlations, and the purity of the process. As one might expect, pure processes are able to be probed in much more detail than mixed ones.

\subsection{Mutual Information Across Random Unitary Causal Breaks}
We begin by defining a clear measure of information backflow accessible to restricted process tensors. 
A key property of process tensors is the ability to discern non-Markovian correlations in a way which is cognisant of the effects of control on the system. In particular, this relies on the ability to make a causal break on the system. If one applies a causal break, then properties of the state cannot persist from one time to the next. Hence, any detection of past-future correlations conditioned on an intermediate causal break must necessarily 
originate from the system-environment interactions. 
Thus, to fully account for the non-Markovianity in a system requires in-situ measurements, which break correlations between the system and its environment, and represent a clean barrier to any past-future dependence~\cite{Pollock2018}.
Barring access to these, a restricted process tensor can only infer aspects of the non-Markovianity.
Here, we introduce one such method to extract a lower bound on non-Markovianity.
\par
\begin{figure}[!b]
	\centering
	\includegraphics[width=0.5\linewidth]{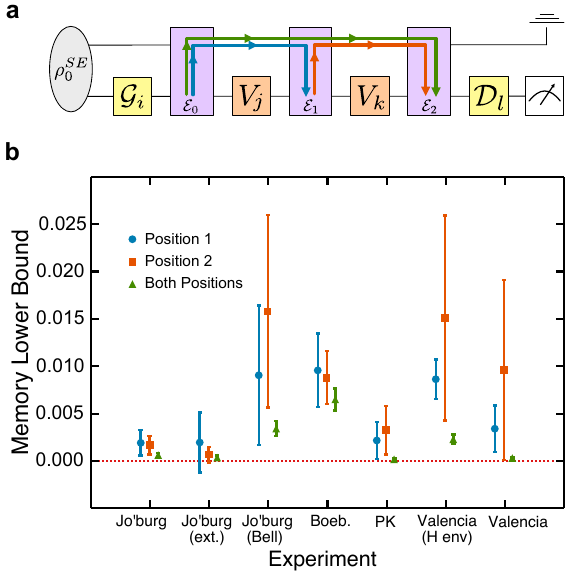}
	\caption[Bounding non-Markovian memory on IBM Quantum devices with incoherent causal breaks ]{Bounding non-Markovian memory on IBM Quantum devices with incoherent causal breaks. \textbf{a} The circuit depicting the process tensor. Quantum information can travel in and out of the system across one or many operations. Each gate is a place-holder for a larger set. Each $V_i$ is an arbitrary unitary operation that need not belong to the set $\mathcal{U}$. \textbf{b} The maximum CMI, which is a conservative lower bound for non-Markovian memory, through $\mathcal{R}$ for each process tensor experiment, with 95\% confidence intervals. This shows statistically significant non-zero memory in the device, which shows consistency in the timescale and the environmental interactions present.}
	\label{fig:mutual_information_circuits}
\end{figure}

Because the maximally depolarising channel, $\mathcal{R}[\rho] = \mathbb{I}/d_S \:\forall \: \rho$,
lies within the span of unitary operations, we can use it as an information barrier between time steps.
A non-zero mutual information between the input operation and final measurement is only possible when information has travelled into the environment and returned after $\mathcal{R}$ has been applied~\cite{taranto3}.
Figure~\ref{fig:mutual_information_circuits}a illustrates this idea for the processes we consider here, for a three step process, where $\mathcal{R}$ takes either the first operation position, the second, or both.
This tests the timing and duration of different memory effects. We describe the computation of this mutual information in the following, and then use it to estimate the mutual information in the experiments from Chapter~\ref{sec:IC-control} on IBM Quantum devices. Note that the effect of the depolarising channel is to erase the quantum state, and hence only non-unital memory effects may be detected here. 
\par 
Having estimated the process tensor enables us to numerically search for the encoding and decoding operations which give the largest lower bound to non-Markovianity along different paths.
Respectively, these are sets $\mathcal{G}$ and $\mathcal{D}$, the first of which contains two unitary operations applied with equal likelihood, and the second contains two orthogonal measurement effects.

The quantities we compute are the conditional mutual information (CMI) for each case:
\begin{align}
	\label{MI-1}
	&\argmax_{\mathcal{G},V_1,\mathcal{D}} I(G:D|\mathcal{G},V_1,\mathcal{R},\mathcal{D}),\\
	\label{MI-2}
	& \argmax_{\mathcal{G},V_2,\mathcal{D}} I(G:D|\mathcal{P},\mathcal{R},V_2,\mathcal{D}),\\
	\label{MI-3}
	&\argmax_{\mathcal{G},\mathcal{D}} I(G:D|\mathcal{P},\mathcal{R},\mathcal{R},\mathcal{D}),
\end{align}
where:
\begin{gather}
	I(G:D) = \sum_{g\in\mathcal{G}} \sum_{d\in\mathcal{D}} p_{(G,D)}(g,d)\log \left(\frac{p_{(G,D)}(e,d)}{p_G(g)p_D(d)}\right).    
\end{gather}

Each of the respective placements of $\mathcal{R}$ generates a conditional dynamical map $\hat{\mathcal{P}}_{3:0}^{(\mathcal{R},\mathcal{V}_1)}$, $\hat{\mathcal{P}}_{3:0}^{(\mathcal{V}_1,\mathcal{R})}$, and $\hat{\mathcal{P}}_{3:0}^{(\mathcal{R},\mathcal{R})}$. The accessible quantities in Equations~\eqref{MI-1}--\eqref{MI-3} then lower bound the quantum mutual information between the input and output spaces of these conditional dynamical maps, which in turn lower bounds the unconditional quantum mutual information of the whole process tensor. 
For each experiment, we summarise the memory lower bound in Figure~\ref{fig:mutual_information_circuits}b. 
Note that we include an extra experiment `Valencia (H env)', in which the neighbouring qubits are initialised into the $\ket{+}$ state.
In almost every case, we find non-zero CMI, flagging non-Markovianity within the device.
The extended Johannesburg experiment is the only case for which CMI overlaps zero in all three tests. Given that the effects are no longer observable on this longer timescale, this suggests that the memory has a finite lifetime which can be loosely upper-bounded by this experiment. This is further shown with the lower values where $\mathcal{R}$ is contracted in both positions.
The memory size is especially high for the experiments with coherent neighbours (`Joburg (Bell)' and `Valencia (H env)'), suggesting a passive crosstalk interaction might account for some of the environmental memory effects observed. Note, however, that $ZZ$ coupling is not sufficient to explain these effects, since this generates only unital dynamics. 

\label{ssec:coherent}

\subsection{Restricted Entanglement Witnesses}

In continuing analogy with properties of states, temporal entanglement may be identified via its process tensor Choi state through any of the number of mechanisms in entanglement literature.
However, absent a unique or complete estimate for the process tensor, it is unclear whether genuinely quantum temporal correlations can be recognised. Here, we show by construction that the data generated by sequences of unitary operations surprisingly suffice in many instances to detect both bipartite and \ac{GME} in time. 
We derive a restricted entanglement monotone constructed from unitary-only observables of the process. This lower bounds the generalised multipartite negativity, introduced in Ref.~\cite{PhysRevLett.106.190502}. 

Where measurements at different times are possible, temporal entanglement seems much like spatial entanglement: measurements on different subsystems are more strongly correlated than can be explained by any classical cause. Where the connection is less obvious, however, is where input legs are included and only unitary control is applied at different times followed by a final terminating measurement. This question is of keen relevance in the \acs{NISQ} era -- how much information can be extracted about temporal correlations without the technology to perform high fidelity causal breaks? Here, we add to at least one facet of the discussion by showing a somewhat remarkable fact to the affirmative: that temporal entanglement can indeed be both detected and measured with only unitary operations followed by a terminating measurement. 

We consider three situations: experiments conducted from a qubit interacting with an ancilla on a noisy quantum device; simulations of a quantum sensor subject to a spin-bath interaction; and random numerically-generated process tensors. The first demonstrates that current experimental hardware is capable of probing these many-time physics even with relatively simple control mechanisms. The second explores novel approaches to metrology in determining the nature of an environment~\cite{schuff2020improving}. The final examines how accurate this restricted monotone is for typical quantum processes. We find this to be highly dependent on the process purity, and that for pure processes we recover near-perfect entanglement characterisation. \par 

\subsubsection*{Background}

Consider a three-qubit state $\rho_{ABC}\in\mathcal{B}(\mathcal{H}_A\otimes\mathcal{H}_B\otimes \mathcal{H}_C)$. This state is called \emph{separable} with respect to a bipartition $A\mid BC$ if it can be written as a convex mixture of product states across that bipartition: 
\begin{equation}
	\rho_{A\mid BC}^{\text{sep}} = \sum_i p_i |\psi_i^{A}\rangle\!\langle \psi_i^{A}|\otimes |\phi_i^{BC}\rangle\!\langle \phi_i^{BC}|,
\end{equation}
and similarly across the second two bipartitions. A state which is not separable with respect to a given bipartition is said to be \emph{bipartite entangled} across that partition. If a state can be written as a convex mixture of separable states 
\begin{equation}
	\rho^{\text{bs}} = p_1\rho_{A\mid BC}^{\text{sep}} + p_2\rho_{B\mid AC}^{\text{sep}} + p_3\rho_{C\mid AB}^{\text{sep}}
\end{equation}
is said to be \emph{biseparable}. A state which is not biseparable is \emph{genuinely multipartite entangled} (GME), and is a property which is notoriously difficult to characterise except in extreme cases. 

It is well-known that separable states have a positive partial transpose (PPT). That is, let $(\cdot)^{\Gamma_A}$ be the partial transpose with respect to partition $A$. Then, for separable states the matrix $\rho_{ABC}^{\Gamma_A}$ will have only positive eigenvalues. Conversely, it is known that if $\rho_{ABC}^{\Gamma_A}$ has negative eigenvalues, then it is biparitite entangled across this partition. This is a sufficient criterion for bipartite entanglement, but is only also necessary in the case of qubit-qubit and qubit-qutrit systems\footnote{Entangled PPT systems have a weak form of entanglement known as \emph{bound entanglement}, from which no Bell states may be distilled using local operations and classical communication~\cite{horodecki2009quantum}.}. 

Ref.~\cite{PhysRevLett.106.190502} considers the characterisation of a slightly weaker form of \acs{GME}. They define PPT mixtures to be states which can be written 
\begin{equation}
	\rho^{\text{p mix}} = p_1 \rho_{A\mid BC}^{\text{PPT}} + p_2 \rho_{B\mid AC}^{\text{PPT}} + p_3 \rho_{C\mid AB}^{\text{PPT}}. 
\end{equation}
This is slightly stronger than the biseparability criterion, and so any state which cannot be written in this form is \acs{GME}. The advantage of this is that the set of PPT mixtures can be fully characterised through the use of a \ac{SDP} with efficiency and convergence guarantees; the drawback is that some weakly \acs{GME} states can be written as PPT mixtures. 

Entanglement is typically studied through the use of \emph{entanglement witnesses}. An entanglement witness $W$ is a Hermitian operator that takes a positive expectation value on all biseparable states, and a negative value on at least one entangled state. It is known that for every entangled state, one can find a witness $W$ that certifies its entanglement. However, discovery of witnesses in practice can be quite challenging due to the complicated structure of the set of biseparable states. This motivates the relaxation to finding converse states to the set of PPT mixtures. Ref~\cite{PhysRevLett.106.190502} shows that a witness $W$ of \acs{GME} can be found for a given set of partitions by the following \acs{SDP}:
\begin{equation}
	\begin{split}
	&\min \Tr[W\rho]\\
	&\text{s.t. }\Tr[W] = 1\quad \text{and, for all }M:\\
	&W = P_M + Q_M^{\Gamma_M};\: P_M\succcurlyeq 0,\: Q_M\succcurlyeq 0.
	\end{split}
\end{equation}
Here, where $M$ denotes a partition of $\rho$. $W$ and each $P_M$ are the free parameters of the optimisation, equivalently imposing that $(W - P_M)^{\Gamma_M}\succcurlyeq 0$ . If the minimum of this optimisation is negative, then $\rho$ is not a PPT mixture, and hence \acs{GME}.

\subsubsection*{Unitary Temporal Entanglement Witnesses}

We employ and modify the aforementioned entanglement witness, constraining it to lie within the span of unitaries. We then treat a similar \acs{SDP} which we can use to study entanglement even in restricted process tensors. 
Recall from Chapter~\ref{chap:process-properties} that observables of a process tensor generated by the span of sequences of unitary operations can be described as:
\begin{equation}
	\label{eq:restr-obs-restate}
	\begin{split}
		\mathcal{O} &= \sum_{i,\vec{\mu}}\alpha_{i,\vec{\mu}} P_i\otimes \bigotimes_{j=0}^k P_{\mu_j}, \qquad \text{where}\\
		&\alpha_{i,\vec{\mu}}\in \mathbb{R},\\
		&P_i\in \{\mathbb{I},X,Y,Z\},\\
		&P_{\mu_j}\in \{\mathbb{I\otimes I}\} \cup \{X,Y,Z\}\otimes \{X,Y,Z\}.
	\end{split}
\end{equation}
We denote the subset of observables of the form of Equation~\eqref{eq:restr-obs-restate} by $\mathfrak{U}$.
The following modified \acs{SDP} produces a restricted entanglement witness for processes:
\begin{equation}
	\label{eq:SDP-witness}
	\begin{split}
		&\min_{W\in\mathcal{W}} \text{Tr}[W\Upsilon_{k:0}],\\
		&\mathcal{W} = \{ W |\: \forall\: M : \exists \: P_M,Q_M\succcurlyeq 0: W = P_M + Q_M^{\Gamma_M} , \\
		&\text{Tr}[W] = 1, W\in \mathfrak{U}\}.
	\end{split}
\end{equation}
Here, $M$ denotes a partition of the subsystems of $\Upsilon_{k:0}$, and $\Gamma_M$ is the partial transpose with respect to that partition. The set of all available partitions will depend on the target of the entanglement witness -- i.e. an $n$-partite entanglement witness will cover $2^{n-1}$ unique partitions. If $W$ is found such that $\text{Tr}[W\Upsilon_{k:0}] < 0$ then $\Upsilon_{k:0}$ is genuinely multipartite entangled across the defined partitions. By replacing condition $\text{Tr}[W] = 1$ with $P_M,Q_M\preccurlyeq\mathbb{I}$ then the resulting minimisation defines a lower bound entanglement monotone, which we employ. In the bipartite case, this lower bounds the negativity.

When a process tensor $\Upsilon_{k:0}$ is estimated from restricted experimental data, a non-unique Choi state is constructed obeying conditions of complete positivity and causality. But so long as $W\in\mathfrak{U}$ is imposed, the witness is restricted to be an observable on $\Upsilon_{k:0}^{(R)}$ which \textit{is} uniquely fixed. Thus, we may solve the \acs{SDP} given in Equation~\eqref{eq:SDP-witness} to detect both bipartite and multipartite entanglement in time, even with extremely simple circuits such as sequences of unitary operations. Although this witness is not as powerful as a fully generic observable on $\Upsilon_{k:0}$, we find it sufficient to detect temporal entanglement in practice. This result extends the boundaries of what can be determined about the dynamics of an open quantum system, and should in particular find utility in the study of naturally occurring quantum stochastic processes. \par

\begin{figure}
	\centering
	\includegraphics[width=0.8\linewidth]{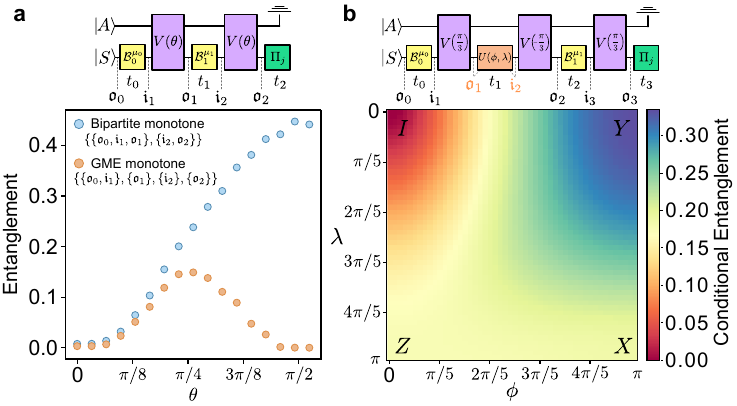}
	\caption[Implementing a unitary-restricted class of temporal entanglement witnesses for the detection of genuine multi-time entanglement ]{ Implementation of unitary-restricted temporal entanglement witnesses. \textbf{a} Results of a circuit designed to manufacture different temporal entanglement structures implemented on \emph{ibm\_cairo}. \textbf{b} We construct a three-step process tensor on \emph{ibm\_auckland} and show how deterministically changing the intermediate unitary gives control over the resulting temporal quantum correlations.}
	\label{fig:ent-witnesses}
\end{figure}

\textbf{Demonstration on NISQ Devices}

Our first demonstration consists of temporal entanglement experimentally verified through unitaries only on the device \emph{ibm\_cairo}. With one qubit as the simulated environment, and a two-step (three time) process tensor, we tune the $SE$ interaction between two qubits to $V(\theta) := \exp\left(-i\frac{\theta}{2}\sum_iP_i\otimes P_i\right)$, and determine both the detectable bipartite and genuinely multipartite entanglement as a function of $\theta\in [0,\pi/2]$. The start, middle, and end of these values respectively produce an identity, $\sqrt{\text{SWAP}}$, and SWAP gate. These interactions are designed to produce different temporal entanglement structures. Repeated $\sqrt{\text{SWAP}}$ gates in principle produce \acs{GME} across legs $\{\mathfrak{i}_1,\mathfrak{o}_1,\mathfrak{i}_2,\mathfrak{o}_2\}$~\cite{Milz2021PRXQ}. Meanwhile SWAPs ought to produce maximal bipartite entanglement between $\hat{\mathcal{E}}_{1:0}$ and $\hat{\mathcal{E}}_{2:1}$, since $\mathfrak{i}_1$ to $\mathfrak{o}_2$ is an effective identity channel -- but $\mathfrak{o}_1$ and $\mathfrak{i}_2$ are uncorrelated, hence no \acs{GME}. We realise and quantify these entanglement structures using our entanglement bound which relies only on unitary sequences. Results are shown in Figure~\ref{fig:ent-witnesses}a, demonstrating the ability to robustly detect entangled multi-time processes on real hardware even with limited control. In addition to varying the interaction, we also show that for a fixed process, one can `inject' complexity into the circuit~\cite{aloisio-complexity}. We reconstruct a three-step process on \emph{ibm\_auckland} and project the second time onto some parametrised unitary $U(\phi,\lambda)$. Then, for varied parameter values, compute the certifiable bipartite entanglement between $\{\mathfrak{o}_0,\mathfrak{i}_1,\mathfrak{o}_2\}$ and $\{\mathfrak{i}_3,\mathfrak{o}_3\}$ -- see Figure~\ref{fig:ent-witnesses}b. Thus, process correlations may be greatly manipulated even at the system level.
Applying a fixed operation on the system physically transforms one process into a new one. This new process can be taken to be the old process, where one step is projected onto the fixed operation. In our example of controllable entanglement, Figure~\ref{fig:ent-witnesses}b, we first estimate a three-step process tensor $\Upsilon_{3:0}$. A conditional two-step marginal is then found for parametrised unitary
\begin{equation}
	V(\phi,\lambda) = \begin{pmatrix}
		\cos(\phi/2) & -\text{e}^{i\lambda}\sin(\phi/2) \\
		\text{e}^{i\lambda}\sin(\theta/2) & \cos(\phi/2)
	\end{pmatrix}.
\end{equation}
Let $\hat{\mathcal{V}}(\phi,\lambda)$ be the Choi state of $V(\phi,\lambda)$, then
\begin{equation}
	\Upsilon_{2:0}^{\left(V(\phi,\lambda)\right)} = \text{Tr}_{\mathfrak{i}_2,\mathfrak{o}_1}\left[(\mathbb{I}_{\mathfrak{o}_3,\mathfrak{i}_3, \mathfrak{o}_2}\otimes \hat{\mathcal{V}}(\phi,\lambda)^\text{T}\otimes \mathbb{I}_{\mathfrak{i}_1,\mathfrak{o}_0})\cdot\Upsilon_{3:0}\right],
\end{equation}
which is the two-step conditional marginal process where $V(\phi,\lambda)$ is applied in position 1 of the circuit. For $\phi\in[0,\pi]$ and $\lambda\in[0,\pi]$ we compute the unitary entanglement lower bound Equation~\eqref{eq:SDP-witness} between legs $\{\mathfrak{o}_0,\mathfrak{i}_1\}$ and $\{\mathfrak{o_2},\mathfrak{i}_3,\mathfrak{o}_3\}$. This determines how much complex temporal correlations can be produced or modified by user-chosen controls.

\begin{figure}
	\centering
	\includegraphics[width=\linewidth]{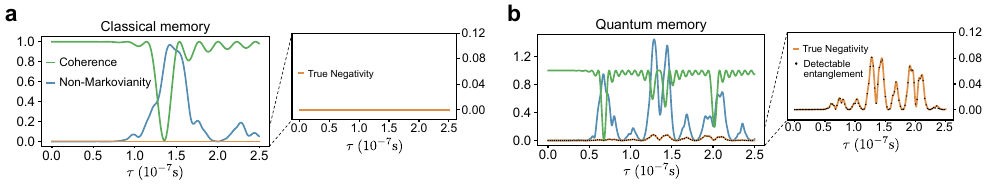}
	\caption[Scheme to integrate unitary-restricted temporal entanglement witnesses in the context of quantum sensors. ]{Implementation of unitary-restricted temporal entanglement witnesses in the context of quantum sensors. \textbf{a} A classical field generates non-Markovian correlations, but entanglement in the process tensor of a simulated quantum sensor remains zero. \textbf{b} When the multi-time correlations are genuinely quantum, we show that this may be detectable by the sensor with unitary-constrained control. }
	\label{fig:sensor-entanglement}
\end{figure}

\textbf{Application to Quantum Sensing}

In the second instance, we contrast a single spin coupled respectively to classical and quantum baths. Using \ac{DD} sequences as filter functions and sweeping through various frequencies, coherence dips correspond to locks on environmental frequencies -- the point at which interactions are amplified rather than suppressed. By constructing unitary-restricted process tensors within the filter, we show that quantumness of the interaction may be certified by measuring entanglement in the multi-time correlations. However, this method does not distinguish between sources which are quantum, but with interactions that can be classically emulated. We explored generators of classical temporal correlations in Chapter~\ref{chap:process-properties}. Here, we show how one in principle may operationally distinguish between different classes of Hamiltonian dynamics. Our procedure is entirely consistent with the capabilities of quantum sensors, and adds a different capacity to learn about the system-environment interaction through structured sequences on the system.

A straightforward quantum sensing protocol is to apply \acs{DD} sequences (we selected $XY4$~\cite{gullion1990new}) to the central spin, and sweep across a range of frequencies. This is achieved by varying the wait time $\tau$ between decoupling pulses. In our numerical simulation, the central spin is first placed in the $\ket{i+}$ eigenstate with a $R_x(\pi/2)$ rotation, a DD sequence applied, and then $R_x(-\pi/2)$ followed by measurement of the ground state population. This is the coherence curve presented in Figures~\ref{fig:sensor-entanglement}a and~\ref{fig:sensor-entanglement}b. 

\begin{figure}
    \centering
    \includegraphics[width=\linewidth]{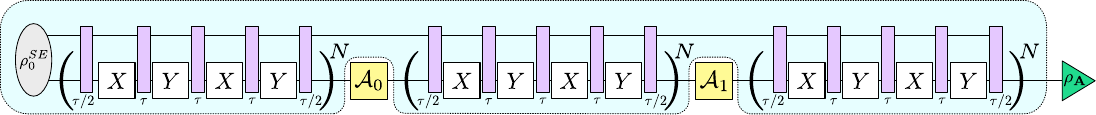}
    \caption[Circuit diagram illustrating a process tensor built around a dynamical decoupling sequence ]{A circuit diagram illustrating a process tensor built around a DD sequence. When on resonance, the DD sequence amplifies interactions with the environment. The process tensor can then be used to detect and characterise the temporal correlations generated.}
    \label{fig:sensing-PT}
\end{figure}

By inserting some probe operations within the filter sequence as per Figure~\ref{fig:sensing-PT}, we can also spectroscopically determine non-Markovianity when the sequence locks onto these frequencies. We propose the reconstruction of a restricted process tensor around the \acs{DD} sequence. The use of our unitary temporal entanglement witnesses can then determine whether the interaction is genuinely quantum or not.

The simulation takes a simple spin-spin Hamiltonian $H = H_S\otimes H_E$ in the rotating frame of the sensor. This has the form:
\begin{equation}
   H = \omega_L\sigma_z^{(E)} + H_I,
\end{equation}
where $\omega_L$ is the Larmor frequency of the environment spin -- we take this to be a nuclear spin of $1.85\times 10^6$ Hz); $\sigma_Z^{(E)}\in\mathcal{B}(H_E)$ is an operator on the nuclear spin (dropping the system identity for simplicity); and $H_I$ is the interaction Hamiltonian.   
For the examples of classical and quantum interactions, we choose interactions:
\begin{equation}
    \begin{split}
    H_I^{(\text{Classical})} &=  g\cdot Y\otimes Z\\
    H_I^{(\text{Quantum})} &= \frac{g}{2}(Z + X)\otimes (Z + Y)
    \end{split}
\end{equation}

The interaction $H_I^{(\text{Classical})}$ generates no entanglement in the process tensor because it is a two-qubit commuting Hamiltonian, as in Chapter~\ref{chap:process-properties}. Thus, after tracing over the environment, the interaction is locally equivalent to a stochastic mixture of $Z-$rotations, which is non-Markovian but separable. It could hence be replaced with a classical source of correlated randomness, such as a fluctuating magnetic field. Meanwhile, as verified by the entanglement witness (and the negativity of the process tensor), $H_I^{(\text{Quantum})}$ generates interactions that do not commute between different times. It hence verifiably generates quantum correlations in time, which could only be created by a genuinely quantum environment.

\textbf{Benchmarking on Randomly Generated Processes}

\begin{figure}[t]
	\centering
	\includegraphics[width=\linewidth]{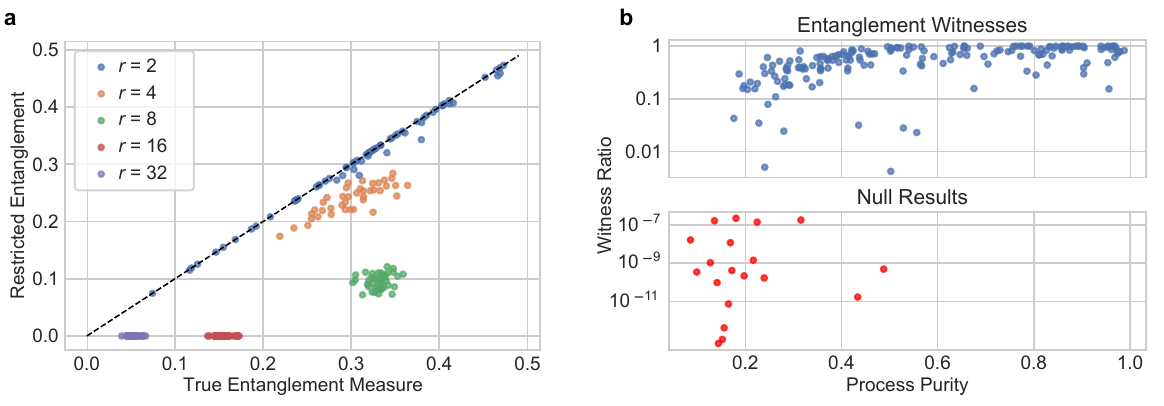}
	\caption[Restricted entanglement witnessing for randomly generated two-step process tensors ]{Results of restricted entanglement witnessing for randomly generated two-step process tensors. \textbf{a} We compare the true negativity of a given process with the negativity as bounded by unitary observables, generated across different ranks. \textbf{b} We plot the ratio of witnesses as a function of the puritity of different processes. The top panel shows all the instances where entanglement is sufficiently measured by unitary observables. The bottom panel shows that for highly mixed processes, we get some instances where entanglement cannot be observed in the restricted regime. }
	\label{fig:entanglement-benchmarking}
\end{figure}

Lastly, we benchmark our witness across a set of randomly generated two-step process tensors. We sample according to the method introduced in Chapter~\ref{chap:process-properties}.
We first compute the negativity across $\{\{\mathfrak{o}_2,\mathfrak{i}_2\},\{\mathfrak{o}_1,\mathfrak{i}_1,\mathfrak{o}_0\}\}$, and then solve the restricted \acs{SDP} to determine an entanglement bound as certified by the set of restricted observables. In Figure~\ref{fig:entanglement-benchmarking}a, we plot the restricted entanglement bound against the true negativity for random processes generated with increasing ranks, denoted by $r$. Interestingly, we see that up to rank eight process, restricted observables always suffice to witness temporal entanglement, and for rank two processes the resulting amount is almost identical to the true negativity. This can be seen that when the process is more mixed, it is harder to extract a signal with unitary operations without getting drowned out in the noise. Whereas a causal break purifies the process for any particular run, allowing correlations to be detected.

This effect is shown plotted in Figure~\ref{fig:entanglement-benchmarking}b, where we have generated processes via random Heisenberg interactions with an $n$-qubit environment $1\leq n \leq 5$. This produces a range of processes with purities that depend on the strength of the interaction. We plot the ratio between the unitary-witnessed entanglement, and the true entanglement value. For very low purities, unitary observables often fail to detect the entanglement, but it is significantly more reliable for even relatively pure processes.

\subsection{Bounding Process Properties}

The previous two subsections concerned witnesses of non-Markovianity which could be found entirely through the span of unitary operations, both in terms of total and of quantum non-Markovian correlations. But this is insufficient to describe more sophisticated measures of non-Markovianity with operational interpretations. Moreover, the bounds supplied were only lower bounds, and thus the tightness cannot be ascertained. 
Measures of non-Markovianity typically require access to the full Choi state, including its spectrum. This cannot be uniquely reonstructed in the tomographically incomplete regime. Here, we develop a workaround that allows for both the lower and upper bounding of properties of a non-uniquely reconstructed process tensor. From this, we show how it is possible to obtain various microscopic properties of quantum processes, including purity, fidelity, and generalised quantum mutual information. This permits the study of many-time physics on both quantum computers without mid-circuit measurement capabilities and, importantly, quantum sensors which may be placed in highly complex environments.



We circumvent the unitary limitation by constructing regions of plausibility and bounding process properties among these. Measuring the outcomes to sequences of unitaries naturally generates a `family' of process tensors which are consistent with the incomplete data. We develop an algorithm that searches this space of plausible process tensors to find bounds over objective functions of the user's choosing. Thus, even without a complete estimate of a multi-time process, one may answer the next best question: what range of dynamics are consistent with a limited set of observations?

\subsubsection*{Derivation}

Since the causality constraints take the same form as basis measurements of the process tensor (linear inhomogeneous equations), here, we extend this method to allow for straightforward optimisation of arbitrary objective functions over the spaces of processes consistent with limited experimental data. 
Consider a process tensor with decomposition $\Upsilon_{k:0} = \Upsilon_{k:0}^{(R)} + \Upsilon_{k:0}^{(R_\perp)}$,
where $\Upsilon_{k:0}^{(R)}$ is fixed by a restricted set of experimental observations, and $\Upsilon_{k:0}^{(R_\perp)}$ its free (modulo physical constraints) orthogonal complement. 
Let $\Upsilon_{k:0}^{(R)}$ be reconstructed through the final-state measurement conditioned on a set of basis sequences $\{\mathbf{B}_{k-1:0}^{\vec{\mu}}\} = \{\bigotimes_{i=0}^{k-1}\mathcal{B}_i^{\mu_j}\}_{\vec{\mu}=(1,1,\cdots,1)} ^{(d^4,d^4,\cdots,d^4)}$. Although this object has well-defined properties -- such as its action on the span of its basis elements -- one cannot use it to directly estimate correlations in $\Upsilon_{k:0}$.
For each sequence, and each measurement outcome, we have a set of equations
\begin{equation}
    \text{Tr}\left[\Upsilon_{k:0} \left(\Pi_i\otimes \mathbf{B}_{k-1:0}^{\vec{\mu}\text{T}}\right)\right] = p_{i,\vec{\mu}},
\end{equation}
which we choose to impose in our exploration of the process tensor family.
To fix these, we append the expectation of each basis sequence $\Pi_i\otimes\mathbf{B}_{k-1:0}^{\vec{\mu}}$ with its corresponding observed probability $p_{i,\vec{\mu}}$ which describes the affine space. That is, imposing the matrix-vector equation 
\begin{gather}
	\label{affine-constraint-bound}
		\begin{pmatrix} \langle\!\langle P_0| \\
		\vdots \\
		\langle\!\langle P_N|\\
		\langle\!\langle\mathbb{I}|\\
		\langle\!\langle{\Pi_{i_0}\otimes\mathbf{B}_{k-1:0}^{\vec{\mu}_0}}|\\
		\langle\!\langle{\Pi_{i_0}\otimes\mathbf{B}_{k-1:0}^{\vec{\mu}_1}}|\\
		\vdots \\ 
		\langle\!\langle{\Pi_{i_L}\otimes\mathbf{B}_{k-1:0}^{\vec{\mu}_N}}|\\
		\end{pmatrix} \cdot |\Upsilon_{k:0}\rangle\!\rangle = \begin{pmatrix} 0\\\vdots\\0\\1\\p_{i_0,\vec{\mu}_0}\\p_{i_0,\vec{\mu}_1}\\\vdots\\p_{i_L,\vec{\mu}_N}\end{pmatrix},
	\end{gather}
where $|\cdot\rangle \!\rangle$ is the row-vectorisation of an operator, and $\langle\!\langle\cdot|$ its complex conjugate.
Each $P_i$ here is an $n-$qubit Pauli operator whose expectation needs to be zero for causality to follow. It is important that the observed constraints come from a procedure that guarantees the feasibility of the problem, such as a maximum likelihood estimation first. If experimental data are directly used as the set of observed probabilities, there may not exist a positive operator consistent with that data -- i.e. the intersection between the affine space and positive semi-definite matrices may be empty. \par 

Rigorous characterisation of non-Markovianity necessarily requires measurement and feed-forward capabilities. This is because temporal correlations due to the environment must be separated out from correlations at the system level with the use of causal breaks. Given that many engineered and naturally occurring systems can only be manipulated with fast unitary operations followed by a projective measurement, it is necessary to consider what exactly can be inferred about the non-Markovianity given this limitation. We investigate this situation here. With unitary-limited control, the measurement apparatus for probing a process tensor is not tomographically complete. We first explore what this means explicitly, and subsequently determine what can still be inferred about the process from limited data.



Our aim is to infer a full process and quantify non-Markovian correlation therein from given limited data. The space of $k-$step process tensors is given through the CJI by the set of $2k+1-$partite density matrices satisfying causality conditions. Specifically, they are unit-trace positive matrices satisfying
\begin{equation}
	\label{eq:causality}
	\text{Tr}_{\mathfrak{o}_j}\left[\Upsilon_{j:0}\right] = \mathbb{I}_{\mathfrak{i}_j} \otimes \Upsilon_{j-1:0}
\end{equation} for each $j$.
Physically, causality conditions ensure that averaging over future operations does not affect the statistics of the past. Mathematically, they generate a hyperplane of dimension $\sum_{j=1}^k(d^2-1)d^{4j-2}$ plus one for normalisation~\cite{White-MLPT}. We denote this affine space $\mathcal{V}_c$. 

In Chapter~\ref{chap:PTT}, we developed a projection routine as part of the \acs{MLE} algorithm to impose $\Upsilon_{k:0}\in\mathcal{S}_n^+ \cap \mathcal{V}_c$. We extend this in two ways here: first, we expand the affine space to include observed data, and second, we consider a range of different information-theoretic quantities as our objective to find maxima and minima.
This opens up exploration of process tensor families consistent with the limited observations. The technique functions generically, however for concreteness we consider the restriction of unitary-only control sequences followed by a terminating measurement, as in Figure~\ref{fig:RPT}a. 

Suppose we have probed a process solely without an \acs{IC} basis. Then, after fitting a model to the data with MLE, we have a set of observations $p_{i,\vec{\mu}}$ corresponding to our restricted basis $\mathfrak{B}:=\{\mathbf{B}_{k-1:0}^{\vec{\mu}}\}:=\{\bigotimes_{j=0}^{k-1}\mathcal{B}_j^{\mu_j}\}$ and \acs{POVM} $\mathcal{J}:=\{\Pi_i\}$. We can construct a new affine space $\mathcal{V}_m$ from these observations, given by
\begin{equation}
	\begin{split}
		&\left\{\Upsilon_{k:0} \in \mathbb{C}^{n\times n}\left| \text{Tr}\left[\Upsilon_{k:0} \left(\Pi_i\otimes \mathbf{B}_{k-1:0}^{\vec{\mu}\text{T}}\right)\right] = p_{i,\vec{\mu}}\right\}\right.\\
		&\forall\:\Pi_i\in \mathcal{J},\:\mathbf{B}_{k-1:0}^{\vec{\mu}}\in\mathfrak{B}.
	\end{split}
\end{equation}
Consequently, the family of process tensors which are consistent with our observations is given by the intersection of the cone $\mathcal{S}_n^{+}$ and the affine space $\mathcal{V} := \mathcal{V}_c\oplus \mathcal{V}_m$ (Figure~\ref{fig:subsystems}b). As in Chapter~\ref{chap:PTT}, we optimise some objective function over the feasible set using the \texttt{pgdb} algorithm~\cite{White-MLPT, QPT-projection}. In short, we perform gradient descent with a chosen function, where at each iteration the object is projected onto $\mathcal{S}_n^+ \cap \mathcal{V}$. Using this technique we find minimum and maximum bounds for the values that any differentiable function can take for all process tensors consistent with observed data. The gradient of the function may be supplied analytically, computed through automatic differentiation libraries, or calculated numerically through the method of finite differences. 

\subsubsection*{Diagnosing Non-Markovian Noise on NISQ Devices}

We apply this framework in the interest of mapping out the quantum and classical temporal correlations present in naturally occurring noise. We constructed six three-step process tensors in different dynamical setups for a system qubit on the superconducting device \emph{ibmq\_casablanca}. These setups were designed to examine base non-Markovianity, as well as crosstalk influence. A ten-unitary basis was used, and each process tensor obtained with maximum-likelihood \acs{PTT}. Fixing these observations in the model, we then searched the family of consistent process tensors for a variety of diagnostic measures.

Specifically, we consider five information-theoretic quantities: \textbf{(i)} To measure total non-Markovianity, we employ quantum relative entropy, $S(\rho\mid\mid\sigma) := \text{Tr}\left[\rho(\log\rho - \log\sigma)\right]$, between $\Upsilon_{k:0}$ and the product of its marginals $\bigotimes_{j=1}^k\hat{\mathcal{E}}_{j:j-1} \otimes \rho_0$. This is a generalisation of \acs{QMI} beyond the bipartite scenario and is endowed with a clear operational meaning~\cite{Pollock2018}. \textbf{(ii)} To determine whether the non-Markovianity has genuine quantum features we quantify the entanglement in the process~\cite{Giarmatzi2021, milz21} by means of negativity, $\max_{\Gamma_B}\frac{1}{2}(\|\Upsilon_{k:0}^{\Gamma_B}\|_1-1)$, where $\Gamma_B$ is the partial transpose across some bipartition~\footnote{Entanglement across indices $\mathfrak{o}_j, \ \mathfrak{i}_j$ will trivially be large for nearly unitary processes. We care about the entanglement across indices $\mathfrak{i}_j, \ \mathfrak{o}_{j-1}$, which measures coherent quantum memory effects between time steps.}. \textbf{(iii)} The purity of the process, $\text{Tr} \left[ \Upsilon_{k:0}^2\right]$, measures both the strength and size of non-Markovianity through the ensemble of non-Markovian trajectories. \textbf{(iv)} Finally, quantifying both Markovian and non-Markovian noise over multiple time steps, we compute the fidelity between the ideal process and the noisy process:  $\text{Tr}[\Upsilon_{k:0}(\bigotimes_{j=1}^k\ket{\Phi^+}\!\bra{\Phi^+}\otimes \rho_{\text{ideal}})]$ with $\ket{\Phi^+} = (\ket{00}+\ket{11})/\sqrt{2}$. Our results are summarised in Table~\ref{tab:memory-bounds}. We contrast these results with the coarse bounds obtained in Figure~\ref{fig:mutual_information_circuits}, and find the measures of non-Markovianity to be more than an order of magnitude higher. In the former case, maximal depolarising channels are used as causal breaks. Although this is in line with restricted process tensor capabilities, it also renders the dynamics incoherent and scrambles almost all of the information. Instead, Table~\ref{tab:memory-bounds} offers an exact estimate, and is much more sensitive to the signal. 

\begin{table*}[t]
	\centering
	\resizebox{0.9\textwidth}{!}{%
		\begin{tabular}{@{}lcccc@{}}
			\toprule
			Setup &
			QMI (min, max) & 
			Negativity &
			Purity & Fidelity \\ \midrule
			\#1. System Alone & 
			(0.298, 0.304) & 
			(0.0179, 0.0181) &
			(0.8900, 0.8904) &
			(0.9423, 0.9427) \\
			\#2. $\ket{+}$ Nearest Neighbours (NN) &
			(0.363, 0.369) &
			(0.0255, 0.0259) &
			(0.7476, 0.7485) &
			(0.7239, 0.7244) \\
			\#3. Periodic CNOTs on NNs in $\ket{+}$ &
			(0.348, 0.358) &
			(0.0240, 0.0246) &
			(0.7796, 0.7811) &
			(0.7264, 0.7272) \\
			\#4. $\ket{0}$ NNs with QDD &
			(0.358, 0.373) &
			(0.0205, 0.0210) &
			(0.8592, 0.8612) &
			(0.8034, 0.8045)\\
			\#5. $\ket{+}$ Long-range Neighbours & 
			(0.329, 0.339) &
			(0.0209, 0.0214) &
			(0.8594, 0.8608) &
			(0.9252, 0.9260) \\
			\#6. $\ket{+}$ NN, delay, $\ket{+}$ next-to-NN & 
			(0.322, 0.329) & 
			(0.0209, 0.0213) &
			(0.8534, 0.8549) &
			(0.9077, 0.9083)\\
			\bottomrule
		\end{tabular}%
	}
	\caption{Property bounds on process tensors with different background dynamics. \acs{QMI} measures the non-Markovianity of the process, with conditional \acs{QMI} the minimal and maximal past-future dependence for first and last steps. Negativity measures unbound temporal bipartite entanglement. Purity indicates purity of the ensemble of open system trajectories, while fidelity captures the level of noise. }
	\label{tab:memory-bounds}
\end{table*}

Surprisingly, we find these bounds to be extremely tight. 
This appears to be due to the relative purity of the process, with ranks (to numerical threshold of $10^{-8}$) of between 44 and 68 for the $128\times 128$ matrix \acs{MLE} estimates, which can limit the object's free parameters~\cite{flammia2012quantum}. 
Process rank is also directly related to the size of environment, with each nontrivial eigenvector constituting a relevant bath degree of freedom. In practice, then, this both compresses requirements and estimates the effective dimension of the non-Markovian bath. 

Interestingly, \acs{QMI} is much larger than zero in all situations. Moreover, from the negativity across $\mathfrak{i}_2$ and $\mathfrak{o}_1$ we find temporal entanglement in all cases. Thus, not only are these noisy stochastic processes non-Markovian, they are bipartite quantum entangled in time. This includes Experiment \#1, suggesting non-Markovian behaviour is prominent beyond even the immediate effects of crosstalk. Crosstalk certainly increases correlation strength, as well as decreasing process fidelity -- an effect shown by switching neighbours into the $\ket{+}$ state and coupling to $ZZ$ noise -- but $ZZ$ noise alone cannot generate temporal entanglement~\cite{Milz2021PRXQ}. We remark that when a quadratic dynamical decoupling (QDD) protocol~\cite{QDD} is performed on neighbouring qubits in Experiment \#4, \acs{QMI} increases with respect to the other experiments, but with reasonably high fidelity. This may suggest that naive QDD reduces noise strength while increasing complexity, warranting further studies.

Although this procedure is in principle to bound generic processes, in practice, we find near complete determination for only relatively few numbers of measurements. An \acs{IC} three-step process tensor requires $12\:288$ circuits, but with only unitary gates this is $3000$, so the savings are substantial. To limit experimental overhead, one might consider increasingly measuring basis sequences until bounds achieved were satisfactorily tight. This supplements, for example, machine learning approaches to directly estimate the memory~\cite{guochu2020,Goswami2021} or full characterisations~\cite{Xiang2021}. 




\begin{figure}[htbp]
	\centering
	\includegraphics[width=0.5\linewidth]{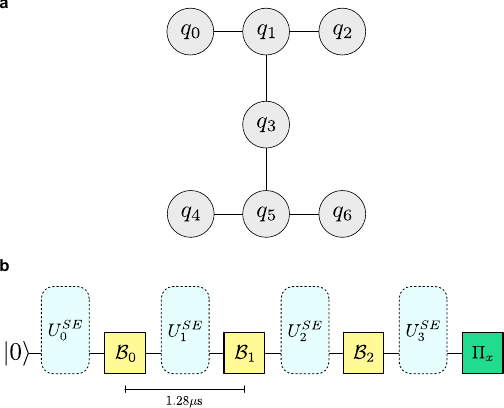}
    \caption[Experiments conducted on \emph{ibmq\_casablanca} to determine non-Markovian memory ]{\textbf{a} Qubit map of \emph{ibmq\_casablanca}, $q_5$ was the system qubit in all experiments. \textbf{b} Structure of the process tensor circuits for this device: a three-step process tensor constitutes three basis operations, followed by measurement. For all six experiments, we fixed the wait time at $1.28\:\mu$s and varied the background.}
	\label{fig:casablanca_layout}
\end{figure}

The layout of the device from Table~\ref{tab:memory-bounds} is shown in Fig~\ref{fig:casablanca_layout}a and the structure of the process tensor experiments in Fig~\ref{fig:casablanca_layout}b. $q_5$ was used as the system qubit for all experiments shown in Table~\ref{tab:memory-bounds}, and had a measured $T_2$ time of 123.65 $\mu$s during the period over which data was collected. For experiments \#1--\#6 the backgrounds were varied, and the wait time between each time step identically fixed to 1.28 $\mu$s. A ten-unitary basis was used, for a total of $3000$ circuits per experiment at $4096$ shots. The details of each background are as follows: 
\begin{enumerate}
    \item No control operations on any other qubits.
    \item $q_3$, $q_4$, and $q_6$ were each initialised into a $\ket{+}$ state at the beginning of the circuits and left idle.
    \item $q_3$, $q_4$, and $q_6$ were each initialised into a $\ket{+}$ state at the beginning of the circuits, and, in each wait period, four sequential CNOTs controlled by $q_3$ and with $q_1$ as a target were implemented. 
    \item $q_3$, $q_4$, and $q_6$ were all dynamically decoupled using the QDD scheme with an inner loop of 2 $Y$ gates and an outer loop of 2 $X$ gates. i.e. a sequence $Y$--$Y$--$X$--$Y$--$Y$--$X$--$Y$--$Y$.
    \item $q_0$, $q_1$, and $q_2$ were initialised into a $\ket{+}$ state and left idle. 
    \item $q_3$ was initialised into a $\ket{+}$ state, followed by a wait time of 320 ns, and then $q_1$ also initialised into a $\ket{+}$ state and left idle.
\end{enumerate}

To account for the effects of measurement errors on all process tensor estimates, we first performed \acs{GST} on the system qubit using the \texttt{pyGSTi} package~\cite{pygsti}. This was used to obtain a high quality estimate of the \acs{POVM} $\{\ket{+}\!\bra{+}, \ket{-}\!\bra{-}, \ket{i+}\!\bra{i+},\ket{i-}\!\bra{i-}, \ket{0}\!\bra{0},\ket{1}\!\bra{1}\}$, which was then used in the \acs{MLE} fit for each process tensor. \par 

\begin{figure}[!b]
	\centering
	\includegraphics*[width=0.7\linewidth]{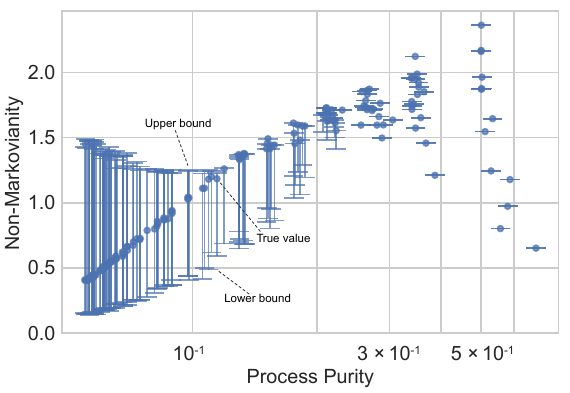}
	\caption[Bounds on generalised quantum mutual information for random restricted process tensors ]{Bounds given on generalised \acs{QMI} for randomly generated two-step process tensors, plotted against purity. For mixed processes, we see that the bounds are sufficient to detect non-Markovian correlations, but are reasonably wide. With increasing purity, the bounds tighten until unitary data appears sufficient to fully determine the memory of a process.}
	\label{fig:random-PT-bounds}
\end{figure}

\subsubsection*{Benchmarking on Randomly Generated Processes}

We saw in Table~\ref{tab:memory-bounds} that memory bounds obtained on real data in practice are tight. It is natural, then, to wonder if this is a generic property of processes and how the bounds behave for typical instances. To investigate this, we apply our bounding procedure for randomly generated process tensors, shown in Figure~\ref{fig:random-PT-bounds}. Once more, the process tensors are randomly sampled according to the procedure in Chapter~\ref{chap:process-properties}. Here, we generate two-step processes and determine their non-Markovianity in the form of generalised \acs{QMI}. We then optimise to find lower and upper bounds for the memory that are consistent with the set of restricted observables. We plot this against the purity of the generated process.
For processes of low purity, the bounds are wide, but well above zero. For higher purity processes, however, the bounds tighten significantly. This is consistent with the set of observations so far that unitary operations in practice suffice to completely determine the process when the dynamics are not too mixed. 
It would be an interesting future research direction to place guarantees on the size of the bounds that may be obtained from restricted processes.





\section{Spatiotemporal Classical Shadows}

In the previous section, we focused on examining the properties of stochastic processes from both theoretical and practical perspectives, integrating a microscopic viewpoint with a consideration of near-term control constraints. In this section, we aim to extend the state-of-the-art in learning properties of quantum states to quantum stochastic processes. This extension allows for more powerful methods to learn about non-Markovian quantum stochastic processes, even when dealing with large numbers of steps or qubits.

Recent seminal work by Huang et al. has introduced the concept of classical shadows, a technique that uses randomised measurements and post-processing to estimate $M$ quantum state observables in $\log M$ measurements, with additional scaling factors depending on the properties of the observable and the randomisation procedure~\cite{huang-shadow}. This concept has been expanded to include quantum process tomography and the estimation of gate set properties~\cite{Kunjummen2021,Helsen2021}. In this section, we demonstrate the natural application of classical shadows to multi-time, multi-qubit processes and address causally-imposed idiosyncrasies.
The issue of causality highlights a possible learnability gap between processes and states, particularly in physically plausible scenarios. That is, causal ordering typically implies that correlations are high weight. But classical shadows is mostly only able to capture low-weight properties.
We employ this method to investigate various properties of quantum stochastic processes and identify potential applications that can overcome the limitations imposed by the inherent structure of these processes.

By virtue of the state-process equivalence for multi-time processes~\cite{chiribella_memory_2008,Shrapnel_2018,Pollock2018a,PhysRevA.98.012328}, quantum operations on different parts of a system at different times constitute observables on a many-body quantum state. Consequently, properties of quantum stochastic processes can be similarly probed using state-of-the-art quantum state characterisation techniques. Classical shadow tomography~\cite{huang-shadow,elben2022randomized} is one such technique and already has many generalisations and applications~\cite{helsen2021estimating,hadfield2022measurements,huang2022provably,PhysRevLett.125.200501}. Measuring classical shadows allows for exponentially greater observables to be determined about a state, provided that those observables are sufficiently low-weight. But this restriction
means the technique has limitations for the study of temporal correlations in contrast to spatial ones, as we shall see. Namely, typical temporal correlations appears in high-weight observables. Hence, the full set of low-weight correlations in-principle detectable by classical shadows is shrunk considerably by conditions of causality on the process. The free operation in classical shadows -- the identity observable -- is a maximally depolarising channel. At a minimum, this limits detections of past-future correlations to those generated by non-unital dynamics, and partially scrambles the environment signal. We saw some of the impositions placed by unital dynamics on temporal correlations in Chapter~\ref{chap:process-properties}.
Despite this limitation, we identify and demonstrate some desireable applications of classical shadows to multi-time processes.

We first derive the extension of classical shadows applied to process tensors. Specifically, we show how one can achieve the same exponential improvement in determining spatiotemporal properties of process tensors through randomised measurements. We perform this first in the idealised case of perfect Clifford measurements and state preparations, and then in the more general case of noisy quantum instrument bases, such as those constructed in Chapter~\ref{chap:PTT}. There are many surprising limitations of classical shadows when applied to processes that we raise along the way. 
In subsequent sections, we then look to some applications of classical shadows in the process setting to determining key features of quantum stochastic processes.


\subsection{Background}

Given a fixed $n-$qubit state $\rho$, one sets a fixed ensemble of unitary rotations from which the state is rotated $\rho\mapsto U\rho U^\dagger$, and then projectively measured to generate a length $n$ bit string $|\tilde{b}\rangle \in \{0,1\}^n$. This series of steps generates a snapshot of the state which can be stored efficiently in classical memory: $U^\dagger |\tilde{b}\rangle\!\langle \tilde{b}|U$. 
Define a quantum channel $\mathcal{M}$ which is the averaging of both classically random unitary, and quantumly random measurement outcome 
\begin{equation}
	\label{eq:shadow-channel}
	\mathcal{M}(\rho) = \mathbb{E}\left[U^\dagger |\tilde{b}\rangle\!\langle \tilde{b}| U\right]\implies \rho = \mathbb{E}\left[\mathcal{M}^{-1}\left(U^\dagger |\tilde{b}\rangle\!\langle \tilde{b}| U\right)\right].
\end{equation}
$\mathcal{M}^{-1}$ is not \acs{CP}, but can still be applied to the measurement outcomes $U^\dagger|\tilde{b}\rangle\!\langle\tilde{b}|U$ in post-processing. Note that $\mathcal{M}$ is invertible for a tomographically complete set of measurements, but later we discuss tomographically incomplete cases where we invert only on the support of $\mathcal{M}$. A single classical snapshot of the state $\rho$ is given by $\tilde{\rho} = \mathcal{M}^{-1}\left(U^\dagger|\tilde{b}\rangle\!\langle\tilde{b}|U\right)$. In accordance with Equation~\eqref{eq:shadow-channel}, this snapshot reproduces the state in expectation. Repeating the procedure $N$ times results in a collection of $N$ independent, classical snapshots of $\rho$:
\begin{equation}
	\label{eq:shadow-def}
	\mathtt{S}(\rho;N) = \left\{\tilde{\rho}_1 = \mathcal{M}^{-1}\left(U_1^\dagger|\tilde{b}_1\rangle\!\langle\tilde{b}_1|U_1\right), \cdots, \tilde{\rho}_N = \mathcal{M}^{-1}\left(U_N^\dagger|\tilde{b}_N\rangle\!\langle\tilde{b}_N|U_N\right)\right\}.
\end{equation}
This array is called a \emph{classical shadow} of $\rho$. We will now state, but not show, some results on classical shadows. 

A sufficiently sized classical shadow is expressive enough to predict many different observables of the state. For an $N$ shot classical shadow, the procedure involves computing observables on each snapshot and averaging the data. Because the snapshots are efficiently storable, this processing is also efficient. In practice, to avoid corruption from outliers, the snapshots are chunked into $K$ different sets, the observables computed on those sets, and the median value taken (so-called median-of-means algorithm).
To estimate a set of $M$ observables $\{O_1,\cdots,O_M\}$ from a classical shadow $\mathtt{S}(\rho,N)$, average the $K$ groups of size $\lfloor N/K \rfloor$ 
\begin{equation}
	\tilde{\rho}_{(k)} = \frac{1}{\lfloor N/K\rfloor}\sum_{i= (k-1\lfloor N/K \rfloor + 1)}^{k\lfloor N/K\rfloor}\tilde{\rho}_i,
\end{equation}
Then compute 
\begin{equation}
	\tilde{o}_i(N,K) := \text{median}\{\Tr[O_i\tilde{\rho}_{(1)}],\cdots, \Tr[O_i\tilde{\rho}_{(k)}]\}.
\end{equation}
for each $O_i$. 
This procedure comes equipped with the following set of guarantees: for $M$ observables on an $n$-qubit system $\{O_i\}$ select accuracy parameters $\epsilon,\delta\in [0,1]$, set 
\begin{equation}
	K = 2\log(2M/\delta)\quad\text{and}\quad N = \frac{34}{\epsilon^2}\max_{1\leq i \leq M} \|O_i - \frac{\Tr[O_i]}{2^n}\mathbb{I}\|^2_{\text{shadow}},
\end{equation}
where $\|\cdot\|_{\text{shadow}}$ is an operator norm that depends on the measurement ensemble. Then a classical shadow with $NK$ snapshots accurately determines the observables through the median-of-means estimation 
\begin{equation}
	|\tilde{o}_i(N,K) - \Tr[O_i\rho]|\leq \epsilon\:\forall \:1\leq i \leq M
\end{equation}
with probability at least $1-\delta$. 

We can see from this that the protocol does not a priori depend on the properties under scrutiny. This flexibility is one aspect to the appeal of shadows: the measurement primitive is independent of the scientific question, and hence a wide range of fundamentally different questions can be asked about a quantum state through the same set of data. This spans the estimation of local observables, entanglement verification, entanglement detection, computation of Renyi entropies, and can be generalised much further. We will discuss the effects of the operator norm in a later section. The protocol most used involves sampling each unitary from the set of local Cliffords and then projectively measured. The single-qubit inverted channel $\mathcal{M}_1^{-1}(X) = 3X - \mathbb{I}$, and across $n$ qubits scales to $\mathcal{M}_{\mathcal{C}_n}^{-1} = \bigotimes_{i=1}^n\mathcal{M}_1^{-1}$. The operator norm for this choice of measurements scales as $4^k\|O\|_{\infty}^2$, for which $k$ is the locality of the observable -- the number of qubits on which the observable acts non-trivially. In total, $\mathcal{O}(\log(M)4^k/\epsilon^2)$ local Clifford measurements are required to determine $M$ observables with locality at most $k$.

\subsection{Derivation}
We now turn to the act of generalising the classical shadows procedure from states to process tensors. Consider a quantum device with a register of qudits $\mathbf{Q}:=\{q_1,q_2,\cdots,q_N\}$ across a series of times $\mathbf{T}_k:=\{t_0,t_1,\cdots, t_k\}$. We take the whole quantum device to define the system: $\mathcal{H}_S:=\bigotimes_{j=1}^N\mathcal{H}_{q_j}$. The device interacts with an external, inaccessible environment whose space we denote $\mathcal{H}_E$ (and interchangeably refer to as a bath).
The $k$-step open process is driven by a sequence $\mathbf{A}_{k-1:0}$ of control operations on the whole register, each represented mathematically by \acs{CP} maps: $\mathbf{A}_{k-1:0} := \{\mathcal{A}_0, \mathcal{A}_1, \cdots, \mathcal{A}_{k-1}\}$, after which we obtain a final state $\rho_k^S(\mathbf{A}_{k-1:0})$ conditioned on this choice of interventions. Note that where we label an object with time information only, that object is assumed to concern the entire register.
Recall that these controlled dynamics have the form:
\begin{equation}
	\label{eq:shadow-pt-output}
	\rho_k^S\left(\textbf{A}_{k-1:0}^{\mathbf{Q}}\right) = \text{Tr}_E [U_{k:k-1} \, \mathcal{A}_{k-1}^{\mathbf{Q}} \cdots \, U_{1:0} \, \mathcal{A}_{0}^{\mathbf{Q}} (\rho^{SE}_0)],
\end{equation}
where $U_{k:k-1}(\cdot) = u_{k:k-1} (\cdot) u_{k:k-1}^\dag$. 
Now let the Choi representations of each $\mathcal{A}_j$ be denoted by a caret, i.e. $\hat{\mathcal{A}}_j = \mathcal{A}_j\otimes \mathcal{I}[|\Phi^+\rangle\!\langle \Phi^+|] = \sum_{nm}\mathcal{A}_j[|n\rangle\!\langle m|]\otimes |n\rangle\!\langle m|$.
Then, the driven process Equation~\eqref{eq:shadow-pt-output} for arbitrary $\mathbf{A}_{k-1:0}$ uniquely defines a multi-linear mapping across the register $\mathbf{Q}$ -- a process tensor, $\Upsilon_{k:0}$ -- via the generalised Born rule~\cite{Pollock2018a,Shrapnel_2018}. We restate this equation from Chapter~\ref{chap:stoc-processes}, emphasising that now the instruments act on the entire qubit register.
\begin{equation}\label{eq:PToutput}
	p_k^S(\mathbf{A}_{k-1:0}^{\mathbf{Q}})= \text{Tr} \left[\Upsilon_{k:0}^{\mathbf{Q}}\left(\Pi_k^{\mathbf{Q}}\otimes \hat{\mathcal{A}}_{k-1}^{\mathbf{Q}}\otimes \cdots \otimes \hat{\mathcal{A}}_0^{\mathbf{Q}}\right)^\text{T}\right],
\end{equation}
\par 
At each time $t_j$, the process has an output leg $\mathfrak{o}_j$ (which is measured), and input leg $\mathfrak{i}_{j+1}$ (which feeds back into the process).
The two important properties that we stress are: (i) a sequence of operations constitutes an observable on the process tensor via Equation~\eqref{eq:PToutput}, generating the connection to classical shadows, and (ii) a process tensor constitutes a collection of possibly correlated \acs{CPTP} maps, and hence may be marginalised in both time and space to yield the $j$th \acs{CPTP} map describing the dynamics of the $i$th qubit $\hat{\mathcal{E}}^{(q_i)}_{j:j-1}$.

When marginalising across all but a handful of times or qubits, we will denote the remaining steps or registers by commas, i.e.
\begin{equation}
	\label{PT-marg}
	\begin{split}
		\hat{\mathcal{E}}^{(q_{i_0},q_{i_1})}_{j_0:j_0-1,j_1:j_1-1}\!&:=\! \text{Tr}_{ {\{\overline{q_{i_0}},\overline{q_{i_1}}\}}, \{\overline{t_{j_0}},\overline{t_{j_0-1}},\overline{t_{j_1}},\overline{t_{j_1-1}}\}}[\Upsilon_{k:0}];\\
		\Upsilon_{k:0}^{(q_i)} &= \text{Tr}_{\overline{q_i}}[\Upsilon_{k:0}],
	\end{split}
\end{equation}
where the overlines denote complement: $\mathbf{Q}\backslash\{q_i\}$.\par



\begin{figure}
	\centering
	\includegraphics[width=0.75\linewidth]{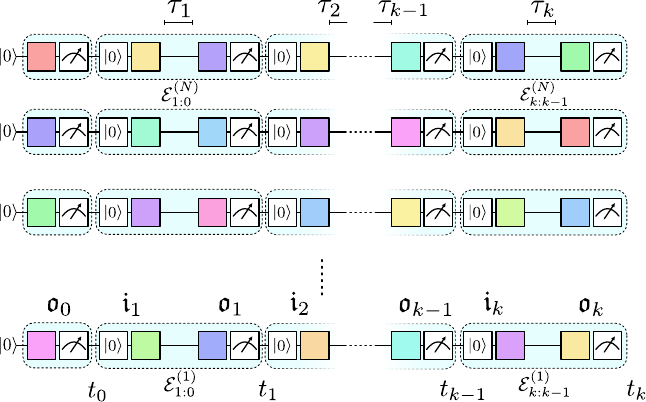}
	\caption[Circuit diagram depicting tomography of spacetime classical shadows. ]{Circuit diagram of our proposed procedure. Spatiotemporal classical shadows can be obtained by applying random Clifford operations to each qubit, projectively measuring, resetting, and then a random Clifford preparation. By repeating these instruments across the circuit with chosen wait times, the appropriate shadow post-processing may be used to determine spatiotemporal marginals of the process.}
	\label{fig:shadow-circ}
\end{figure}

\subsection{Procedure}\label{ssec:shadow-procedure}
We now derive the shadows procedure for processes with respect to two different ensembles of random instruments. In the first instance, we have (perfect) random Clifford measurements and preparations, and in the second we have the experimentally bootstrapped basis from Chapter~\ref{chap:PTT}.

\textbf{Clifford Ensemble}

To map these correlations on each qubit, the classical shadows procedure naturally extends as follows: at each $t\in \mathbf{T}_k$, on each $q\in \mathbf{Q}$, apply a random Clifford operation followed by a $Z-$measurement. This defines a \acs{POVM} on all qubits across all times $\{U_i^\dagger |x\rangle\!\langle x| U_i\}_{t_j}^{q_m}$.
The measurements considered have four defining features: the qubit $q$ on which they act, the time $t$ at which they are implemented, and the basis change $U$ applied prior to a measurement outcome $x$. To avoid notational overload, we will omit these final two labels when writing instruments where the context is clear.

Record both the the outcome of the measurement and the random unitary. Reset the qubit to state $\ket{0}$ and apply a random Clifford, recording this operation as well. The intended effect of this is to apply a randomised quantum instrument -- i.e. a random measurement with an independently random post-measurement state. The circuit diagram for this is given in Figure~\ref{fig:shadow-circ}. Note that any informationally complete set of instruments here is valid, but we choose measure-and-prepare operations for conceptual ease. \par 
The application of an instrument in each chosen location in space and time constitutes a single-shot piece of information about the process tensor, projecting the process tensor onto the sequence of interleaving measurements $\Pi$ and preparations $P$:
\begin{equation}
	\begin{split}
		\hat{\mathbf{\Pi}}_{\mathbf{T}_k} &=  \bigotimes_{l=0}^k \bigotimes_{j=1}^N (U_{\mathfrak{o}_{l}}^{j\ast} |x\rangle\!\langle x | U_{\mathfrak{o}_{l}}^{j\text{T}})\\
		\hat{\mathbf{P}}_{\mathbf{T}_{k}} &=\bigotimes_{l=1}^k \bigotimes_{j=1}^N(U_{\mathfrak{i}_l}^j |0\rangle\!\langle 0 | U_{\mathfrak{i}_l}^{j\dagger}),
	\end{split}
\end{equation}
with probability given in accordance with Equation~\eqref{eq:PToutput}, and noting that $\hat{\Pi} = \Pi^{\text{T}}$ for any \acs{POVM} element.
Since there is a tensor product structure, we can examine each measurement and each preparation at each time on each qubit separately. The preparations $P_l^{q_j}$ are all deterministic, and enact (in isolation) the quantum channel
\begin{equation}
	\mathcal{M}_{\mathcal{P}}(\sigma_{\mathfrak{i}_l}^{q_j}) = \mathbb{E}_{U_{\mathfrak{i}_l}^j\sim \mathcal{U}}[U_{\mathfrak{i}_l}^j|0\rangle\!\langle 0|U_{\mathfrak{i}_l}^{j\dagger}],
\end{equation}
whose inverse gives the classical shadow on the process input legs
\begin{equation}
	\hat{D}_{\mathfrak{i}_l}^{j} := \mathcal{M}_\mathcal{P}^{-1}(U^j_{\mathfrak{i}_l}|0\rangle\!\langle 0|U^{j\dagger}_{\mathfrak{i}_l}) = 3 U^j_{\mathfrak{i}_l}|0\rangle\!\langle 0|U^{j\dagger}_{\mathfrak{i}_l} - \mathbb{I}.
\end{equation}

For the measurements $\Pi_l^{q_j}$ we have the usual single qubit Clifford channel:
\begin{equation}
	\mathcal{M}_{\mathcal{D}}(\sigma_{\mathfrak{o}_l}^{q_j}) = \mathbb{E}_{U_{\mathfrak{o}_l}^j\sim \mathcal{U}, x\sim \text{Tr}[\Pi_l^{q_j}\sigma_{\mathfrak{o}_l}^{q_j}]}[U_{\mathfrak{o}_l}^{j\ast}|x\rangle\!\langle x|U_{\mathfrak{o}_l}^{j\text{T}}].
\end{equation}
Note that $|x\rangle$ on each qubit at each time is sampled according to the generalised Born rule in Equation~\eqref{eq:PToutput}, and depends generally on the operations that come before it. The inverse of this channel gives the shadow on the output legs:
\begin{equation}
	\hat{\Delta}_{\mathfrak{o}_l}^j:= \mathcal{M}_{\mathcal{D}}^{-1}(U_{\mathfrak{o}_l}^{j\ast}|x\rangle\!\langle x|U_{\mathfrak{o}_l}^{j\text{T}}) = 3U_{\mathfrak{o}_l}^{j\ast}|x\rangle\!\langle x|U_{\mathfrak{o}_l}^{j\text{T}} - \mathbb{I}.
\end{equation}

Hence, for a $k$-step process on $N$ qubits, the classical shadow is a reshuffling of
\begin{equation}
	\label{eq:process-shadow}
	\begin{split}
		\tilde{\Upsilon}_{k:0} &= \hat{\mathbf{D}}_{\mathbf{T}_k}^{\text{T}}\otimes \hat{\mathbf{\Delta}}_{\mathbf{T}_k}^{\text{T}},
	\end{split}
\end{equation}
to have the $\mathfrak{o}$ and $\mathfrak{i}$ legs alternating, and from which properties can be efficiently determined using the usual median-of-means estimation described at the start of this section, and in Ref.~\cite{huang-shadow}.

\textbf{Arbitrary Instruments}

Selecting Clifford measurements and re-preparations as the ensemble from which classical shadows are drawn is both conceptually simple, and allows for neat derivations of subsequent results. However, in practice, this basis will often be inaccessible for the contexts we treat. We must hence have a numerical procedure for determining classical shadows across arbitrary ensembles of instruments. 

Recall the tomographic representation of a process tensor, given a basis of operations at the $i$th time $\{\mathcal{B}_i^{\mu_i}\}$ with Choi states $\{\hat{\mathcal{B}}_i^{\mu_i}\}$. There exists a \emph{dual} set $\mathcal{D} := \{\Delta_j^{\mu_j}\}$ such that the process tensor may be written~\cite{Pollock2018a}
\begin{equation}
	\label{eq:choi-pt-full}
	\Upsilon_{k:0} = \sum_{i,\vec{\mu}} p_k^{i,\vec{\mu}}\omega_i\otimes
	\Delta_{k-1}^{\mu_{k-1}\ \text{T}}\otimes \cdots \otimes \Delta_1^{\mu_1\ \text{T}}\otimes \Delta_0^{\mu_0\ \text{T}}.
\end{equation}
When a linearly independent basis is chosen, the duals satisfy $\text{Tr}[\mathcal{B}_i\Delta_j] = \delta_{ij}$. Similarly, we have expanded the final state $\rho_k^{\vec{\mu}}$ in terms of the dual operators $\omega_i$ of the probing \acs{POVM} $\{\Pi_i\}$, i.e., $\rho_k^{\vec{\mu}} = \sum_{i} p_k^{i,\vec{\mu}} \omega_i$.
Here, $p^{i,\vec{\mu}}$ is the probability of the detector clicking with outcome $i$ subject to some sequence of operations $\mathbf{B}_{k-1:0}^{\vec{\mu}}$. Another way to say this is that Equation~\eqref{eq:choi-pt-full} is the linear inversion estimate of $\Upsilon_{k:0}$, as described in Chapter~\ref{chap:PTT}. 

Let $\mathcal{M}_i' : \Delta_i^{\mu_i} \mapsto \hat{\mathcal{B}}_i^{\mu_i}$, and let $\mathcal{M}' = \bigotimes_i\mathcal{M}_i'$. We see then that 
\begin{equation}
	\begin{split}
		\Upsilon_{k:0} &= \sum_{i,\vec{\mu}} p_k^{i,\vec{\mu}}\omega_i\otimes
		\Delta_{k-1}^{\mu_{k-1}\ \text{T}}\otimes \cdots \otimes \Delta_1^{\mu_1\ \text{T}}\otimes \Delta^{\mu_0\ \text{T}},\\
		&= \sum_{i,\vec{\mu}} p_k^{i,\vec{\mu}}
		\mathcal{M}_k'^{-1}(\Pi_i)\otimes \mathcal{M}_{k-1}^{-1}(\hat{\mathcal{B}}_{k-1}^{\mu_{k-1}})'\otimes \cdots \otimes \mathcal{M}_{1}^{-1}(\hat{\mathcal{B}}_{1}^{\mu_{1}})'\otimes \mathcal{M}_{0}^{-1}(\hat{\mathcal{B}}_{0}^{\mu_{0}})',\\
		&= \mathbb{E}\left[\mathcal{M}'^{-1}\left(\Pi_k\otimes \hat{\mathbf{B}}_{k-1:0}\right)\right].
	\end{split}
\end{equation}
We identify $\mathcal{M}'$ as being exactly the same channel as used to define classical shadows in Equation~\eqref{eq:shadow-channel}.
As a consequence, the single-shot version of Equation~\eqref{eq:choi-pt-full} is exactly a classical shadow. This connection points to a nice interpretation of classical shadow tomography: it is effectively a Monte-Carlo sampling of the tomographic representation of a quantum state. 
This also implies that if the basis is incomplete -- for example, restricted to the set of unitaries -- that classical shadows are still valid, but the state produced in expectation will be a restricted process tensor rather than a full one. 
We can now use our tools for computing dual sets to the problem of generating classical shadows on process tensors from arbitrary instruments.

To compute the classical snapshot from a sequence of instruments, first, let 
$N$ be the size of the (possibly under or overcomplete) basis. We can write each of the Choi $\hat{\mathcal{B}}_i^{\mu_i}$ as a row vector (adopting a row-vectorised convention) $\langle\!\langle\hat{\mathcal{B}}_i^{\mu_i}|$ and from this construct a single matrix $\hat{\mathcal{M}}'_i$:
\begin{equation}
	\hat{\mathcal{M}}'_i = \begin{pmatrix}
		\langle\!\langle\hat{\mathcal{B}}_i^{0}| \\ 
		\langle\!\langle\hat{\mathcal{B}}_i^{1}| \\ 
		\vdots\\
		\langle\!\langle\hat{\mathcal{B}}_i^{N-1}| \\ 
	\end{pmatrix}.
\end{equation}
Now, the matrix $\mathcal{M}_i'^{\ +\dagger}$ (right-inverse, conjugate transposed) contains the dual matrices $\{\Delta_i^{\mu_i}\}$ to $\{\hat{\mathcal{B}_i^{\mu_i}}\}$ in its rows, which are then trace normalised. 

In our forthcoming experiments on \acs{NISQ} devices, we use the parametrised set of bootstrapped instruments described in Chapter~\ref{chap:PTT}. Our basis set consists of the instruments generated by a local Clifford gate on the system, plus the projective measurement outcome on the ancilla for a total of 48 different basis elements. For each randomised sequence, a sequence of Cliffords are applied to the system (recalling that these are each sandwiched between fixed $SA$ interactions), and the outcome of the series of mid-circuit measurements on the ancilla qubit recorded, $x_kx_{k-1}\cdots x_0$. The duals to this basis of 48 instruments are computed, and the classical shadow of the process tensor represented from the snapshots
\begin{equation}
	\label{eq:bootstrapped-shadow}
	\hat{\Upsilon}_{k:0}^{\vec{x},\vec{\mu}} := \omega_{x_k} \otimes \Delta_{k-1}^{\mu_{k-1}(x_{k-1})\text{T}}\otimes \cdots \otimes \Delta_1^{\mu_1 (x_1)\text{T}}\otimes \Delta_0^{\mu_0(x_0)\text{T}}.
\end{equation}
This provides a computationally convenient method to compute the classical shadow for an arbitrary ensemble of quantum instruments, with efficient computation of observables as per the classical shadow routine. 


The scaling of the shadows procedure depends on the shadow-norm as it is defined below for projective measurements:
\begin{equation}
	\|O\|_{\text{shadow}} = \max_{\sigma : \text{state}} \left(\mathbb{E}_{U\sim\mathcal{U}}\sum_{\tilde{b}\in\{0,1\}^n} \langle \tilde{b}|U\sigma U^\dagger |\tilde{b}\rangle\!\langle \tilde{b}|U \mathcal{M}^{-1}(O)U^\dagger |\tilde{b}\rangle^2\right)^{1/2}.
\end{equation}
The $\langle \tilde{b}|U\sigma U^\dagger |\tilde{b}\rangle$ part of the expression is the probability of obtaining outcomes $\tilde{b}$ from state $\sigma$ in the basis of $U$, and the $\langle \tilde{b}|U \mathcal{M}^{-1}(O)U^\dagger |\tilde{b}\rangle$ is the expectation value of the snapshot with respect to the resulting outcome. Equivalently, for a given basis of mixed-state instruments, one can numerically compute the quality of the ensemble by solving the \acs{SDP} for the single step observable
\begin{equation}
	\|O\|_{\text{shadow} \ i} = \max_{\sigma :\text{state}}\left(\mathbb{E}_{\hat{\mathcal{B}}\sim\mathscr{B}_i} \Tr[\sigma \hat{B}^{\text{T}}] \Tr[\mathcal{M}_i'^{-1}(O)\hat{B}^\text{T}]\right)^{1/2},
\end{equation}
from which $\|O\|_{\text{shadow}} = \prod_{i=0}^k\|O\|_{\text{shadow}\ i}$. 

\subsection{Recovering Useful Observables and Process Marginals}

We have generalised classical shadows to learning properties of quantum stochastic processes.
It is important now to consider: what interesting information can we actually extract using this method? In
later sections, we consider several different applications thereof. 
Let us first discuss the constraint of requiring low-weight observables, and subsequently the application to finding process marginals. 

\textbf{Low Weight Observables}

For all locally drawn ensembles -- and many global ones -- the procedure of classical shadows scales with an operator norm that is exponential in the locality $\ell$ of the observables. That is, the number of qubits that some operator $O$ acts non-trivially on. For a single tensor product Pauli string, this is the number of traceless Pauli operators present. Consider observables of this form:
\begin{equation}
	O = P_{\mathfrak{o}_k}\otimes P_{\mathfrak{i}_{k}}\otimes\cdots \otimes P_{\mathfrak{o}_1}\otimes P_{\mathfrak{i}_1}\otimes P_{\mathfrak{o}_0}.
\end{equation}
The implication here is that for any feasible set of experiments, the desired operators to learn must be mostly $\mathbb{I}$, with a handful of $X$, $Y$, and $Z$. We recall that $\mathbb{I}$ in the Choi picture is either a partial trace (measure and discard) in the outputs, or a preparation of a maximally mixed state in the inputs. 
In quantum states, there is no a priori structure to the set of correlations between qubits. However, with processes, we have seen in Chapter~\ref{chap:process-properties} the effects of causality constraints. These ensure that there is a causal direction. But this is problematic, because information must travel in and out of the system -- and so any discarding of the system by low-weight observables equally erases important information. One concrete example is in a two-step unital quantum process from which we wish to extract two-point correlation functions. Causality forbids any $\mathfrak{i}_1-\mathfrak{i}_2$, and $\mathfrak{i}_1-\mathfrak{o}_0$ or $\mathfrak{i}_2-\mathfrak{o}_1$ correlations; unitality forbids $\mathfrak{o}_2-\mathfrak{o}_1$, $\mathfrak{o}_1-\mathfrak{o}_0$, and $\mathfrak{o}_2-\mathfrak{o}_0$ correlations. This leaves $\mathfrak{i}_1-\mathfrak{o}_2$ as the only non-trivial two-point correlator of the process. 

The problem more generally stems from the fact that dynamics build up non-trivial entanglement between a system and its environment. If any past leg of the system is rendered incoherent, this incoherence will destroy the system-environment entanglement, preventing future temporal entanglement. 
This unique structure of processes means we have to be somewhat careful and imaginative when applying the framework of classical shadows to the temporal domain.

\textbf{Process Marginals}

We find that one particularly useful application of classical shadows to quantum stochastic processes is simultaneous determination of process tensor marginals, particularly contiguous ones. This allows us to determine a subset of correlations between different times. More usefully, though, we shall explore different classes of processes that may be reconstructed from their marginals. 

A $k-$step process marginal across $n$ qubits is fully determined by a set of $\mathcal{O}(2^{2kn})$ weight $kn$ Pauli observables. 
Although collecting shadows suffices to estimate properties of the process marginals, they will not in general be able to reconstruct a physical estimate of the process marginals. 
To this effect we estimate enough observables to constitute informationally complete information about the process tensor, and then employ \acs{MLE}-\acs{PTT} to process the data and obtain a physical estimate. The advantages of a physical (positive, causal) estimate is that we make use of information theoretic tools that rely on the postitivity of the state.

\subsection{Discussion}


In this work, we extend the method of classical shadow tomography to the spatiotemporal domain, enabling the efficient identification of properties for multi-time, multi-qubit processes. We will demonstrate this procedure and its applications in subsequent sections.
The randomisation of our current ensembles produces property estimates that are efficient when perturbed around the identity, obtaining expectation values such as $\langle X \mathbb{I}\mathbb{I}\cdots \mathbb{I} Y\rangle$. However, causality constraints suggest that, in physically realistic scenarios, most non-trivial observables are likely to be highly non-local ones, involving coherent control at the system level. As a result, the applications we present are not only useful but also tailored to accommodate these restrictions.
An intriguing question to explore is whether there exist accessible ensembles that allow for sampling and obtaining perturbations around high-weight observables, for expectation values such as $\langle X \Phi^+\Phi^+\cdots\Phi^+Y\rangle$. This could enable more sensitive probing of long-range non-Markovian memory. Nevertheless, in the following sections, we will examine some applications of classical shadow tomography in the setting of quantum stochastic processes.

\section{Correlated Noise Characterisation}
\label{sec:correlated-noise}



Currently, there exists a dearth of \acs{QCVV} tools for the characterisation of correlated noise. There are approaches at a coarse-grained level, with for example, noise twirled into depolarising channels~\cite{Harper2020,flammiaACES}. But detailed information is missing here. Specifically, any coherent or correctable noise, or clues as to the physical origin of the correlations is washed away by the twirling practice. This information is crucial to inform better control of a device, the attention paid to different fabrication practices, and for the design of bespoke quantum error correction codes. 

In Chapter~\ref{chap:PTT}, we developed \acs{PTT} which can perform this task in principle, but the resource requirements are too large to be tractable over more than a small number of steps. If, rather than considering generic multi-time correlations, we are interested in collections of small correlated marginals of the dynamics, then the problem becomes much more feasible. We approach this problem using the classical shadows developed thus far. This allows us to estimate arbitary groups of process tensor marginals. The resulting estimates then contain all information about the structure of different correlations between those times. As discussed in the previous section, this is conditioned on tracing out the remainder of the observed steps and is hence only with respect to the class of non-unital dynamics. 

The method of classical shadows permits the determination of $k-$point correlation functions. Translating this to processes, we examine two questions: (i) what is the average case probability of temporally correlated Pauli noise at different times?, and (ii) given the application of some operation at an earlier time, how do the dynamics of the system change at a later point? This is measured in terms of both total non-Markovianity, and in tempoeral entanglement. 
The former gives a coarse view on temporal correlations, but is an important benchmark for the feasibility of quantum error correction. The latter is a fuller diagnosis, indicating the behaviour of the device, the complexity and nature of the noise, and may be used for a more sophisticated form of quantum error mitigation, such as probabilistic error cancellation.

\begin{figure}[t]
	\centering
	\includegraphics[width=\linewidth]{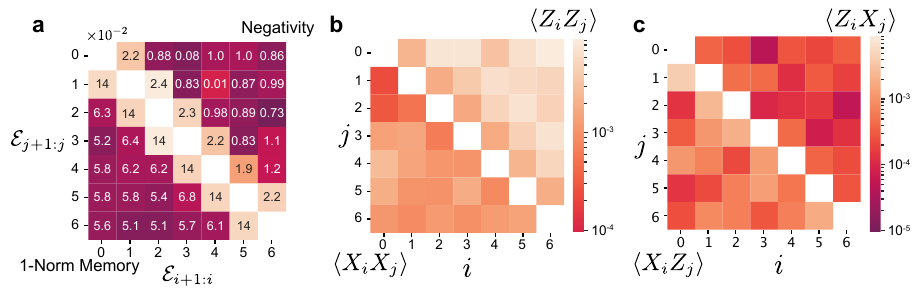}
	\caption[Heatmap capturing the two-map (four-body) correlated marginals on \emph{ibm\_perth} ]{The two-map (four-body) marginals of a system enduring naturally idle dynamics on \emph{ibm\_perth}. \textbf{a} We compute both the negativity between the two steps and the trace-distance between the maps and the product of their marginals. \textbf{b} We compute the joint overlap of correlated Pauli errors: $\langle \hat{P}_i\hat{P}_j\rangle - \langle \hat{P}_i\rangle\langle \hat{P}_j\rangle$ for $ZZ$ and $XX$ as well as \textbf{c} $ZX$ and $XZ$}
	\label{fig:correlated-noise-maps}
\end{figure}

We demonstrate this approach on an IBM Quantum device, obtaining these detailed diagnostics across many different time steps from only a relatively small number of shots.
Given also the scale-up in approach, the tools developed to bound restricted processes are no longer feasible here.
Instead, we require an \acs{IC} basis of quantum instruments. This is achievable by interacting the system with ancilla qubits, and projectively measuring. With mid-circuit measurement capabilities, only a single ancilla is required -- to be measured and reused within the circuit. Otherwise, one ancilla per time-step is necessary, and will need to be isolated post-interaction through e.g. dynamical decoupling. 
In this work, we perform experiments on devices with mid-circuit measurement capabilities. Note that the choice to use an ancilla is because these measurements typically have a long duration (1--5 $\mu$s). Deferring to an ancilla allows the system to continue participating in dynamics -- otherwise the window is lost if measuring the system directly. 
We employ the bootstrapped \acs{IC} basis using the procedure outlined in Chapter~\ref{chap:PTT}, summarised in Figure~\ref{fig:instrument-creation}. For the local unitary operation, random Clifford operations are applied to the system qubit between system-ancilla interactions.


On the device \emph{ibm\_perth}, we apply randomised instruments using our complete basis (Fig.~\ref{fig:instrument-creation}b) to characterise an idle process with random $SU(4)$ gates on neighbouring qubits. 
Across a seven step process, we compute the correlated dynamical maps from $t_i\rightarrow t_{i+1}$ and $t_j\rightarrow t_{j+1}$. These are given by
\begin{gather}\label{eq:correrr}
\hat{\mathcal{E}}_{j+1:j;i+1:i} :=
\text{Tr}_{\bar{i},\bar{j}}[\Upsilon_{k:0}],   
\end{gather}
where $\bar{i},\bar{j}$ indicates the trace over all process tensor legs except for $\{\mathfrak{i}_i, \mathfrak{o}_i,\mathfrak{i}_j,\mathfrak{o}_j\}$.
From these estimates, we compute some representative components of the noise. Specifically, Figure~\ref{fig:correlated-noise-maps}b displays the negativity between each pair of maps; the trace distance between the correlated pairs and tensor product of marginals; and overlap of these maps with Pauli error channels. Note that for marginals between non-neighbouring times, there is a maximal depolarising channel applied in the interim. This is comparable to the effect of the causal break in Figure~\ref{fig:mutual_information_circuits}. However, we see more sensitivity here to the exact multi-time correlations, owing to the use of an \acs{IC} probe to distinguish different aspects of the process.

Some of the results are consistent with device expectations, such as e.g. the prominence of correlated $Z$ errors compared to the other Paulis. $Z$ errors are typically much more significant in general due to their energy-conserving nature when compared to $X$ or $Y$.
Given that the values of these overlap terms is not commensurate with the total non-Markovianity, however, this suggests that correlated Pauli error models are not yet a good total description of the noise on current devices.
$10^7$ shots were collected, and Equation~\eqref{eq:bootstrapped-shadow} used to construct each of the corresponding classical shadows. This is then further postprocessed to determine \acs{IC} information about the marginals, and an \acs{MLE} estimate obtained. We note that although we considered seven times steps for this process, the procedure scales only logarithmically with the number of times and so in principle could be extended far further.

\emph{ibm\_perth} has the same physical layout as \emph{ibmq\_casablanca}, shown in Figure~\ref{fig:casablanca_layout}. On this device, $q_3$ was chosen as the system qubit, and $q_5$ the ancilla for implementation of instruments, since these two qubits had the shortest CNOT gate time. In the background process, we applied two random elements of $SU(4)$ between pairs $q_0-q_1$ and $q_1-q_2$, compiled down to the native gates of the device. This was intended to emulate a realistic background of driven dynamcis in typical states. The idle time on the system qubit per step was around $1.8\:\mu$s as a result.

\section{Filtering Crosstalk From Bath non-Markovianity}

\begin{figure}[t]
	\centering
	\includegraphics[width=0.75\linewidth]{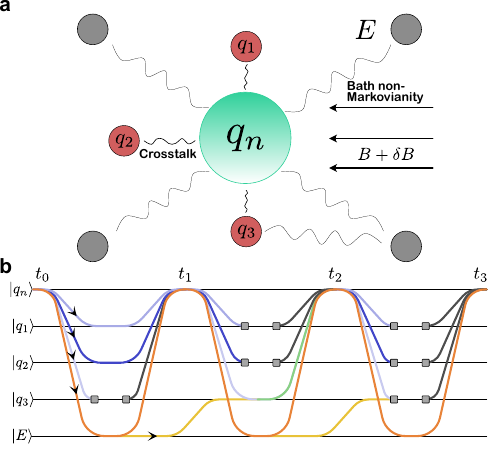}
	\caption[Depiction of crosstalk mechanisms from interacting qubits as well as inaccessible non-Markovian environment ]{System of interacting qubits and an inaccessible non-Markovian environment $(E)$. \textbf{a} A target qubit $q_n$ may interact via crosstalk mechanisms with other qubits, $\{q_1,q_2,q_3,\cdots\}$, in a quantum device, as well as defects in the bath and fluctuating classical fields $B+\delta B$. \textbf{b} The non-Markovian correlations for that system may be separated into different causal pathways by which the correlations are mediated. Causal breaks (depicted in grey) erase any temporal correlations from a given pathway, allowing one to infer the various contributions to total non-Markovianity from nearby qubits and environment.}
	\label{fig:crosstalk_idea}
\end{figure}

In the race to fault tolerant quantum computing, magnified sensitivity to complex dynamics in open quantum systems requires increasingly tailored characterisation and spectroscopic techniques~\cite{chalermpusitarak2021frame,ferrie2018bayesian,nielsen-gst,PhysRevX.9.021045,White-NM-2020,White-MLPT,youssry2020characterization,von2020two}. Correlated dynamics are one particularly pernicious class of noise, and can be generated from a variety of sources, including inhomogeneous magnetic fields, coherent bath defects, and nearby qubits, see Figure~\ref{fig:crosstalk_idea}a~\cite{paladino20141,mavadia2018experimental}. Concerningly, these effects are often omitted from typical quantum error correction noise models despite being ubiquitous in \acs{NISQ} hardware~\cite{Clader2021,White-NM-2020,white2021many, White-MLPT,correlated-qec,Harper2020}.

We have shown the means with which we can learn temporally correlated dynamics in detail using a combination of \acs{PTT} and classical shadow tomography. This determines only the effects of the dynamics at the system level. Here, we now focus on the question of \emph{where} the non-Markovianity is sourced from.
From an open system perspective non-Markovian effects due to a nearby bath or neighbouring qubits are dynamically equivalent. However, there is a conceptual distinction to account for: neighbouring qubits may be controlled. 
In this section we combine our advances in non-Markovian quantum process tomography with the framework of classical shadows to characterise spatiotemporal quantum correlations. Observables here constitute operations applied to the system, where the free operation is the maximally depolarising channel. Using this as a causal break, we systematically erase causal pathways to narrow down the progenitors of temporal correlations, illustrated in Figure~\ref{fig:crosstalk_idea}b. We show that one application of this is to filter out the effects of crosstalk and probe only non-Markovianity from an inaccessible bath. It also provides a lens on spatiotemporally spreading correlated noise throughout a lattice from common environments. We demonstrate both examples on synthetic data.
Owing to the scaling of classical shadows, we can erase arbitrarily many neighbouring qubits at no extra cost. Our procedure is thus efficient and amenable to systems even with all-to-all interactions.

Temporal -- or non-Markovian -- correlations are elements of error that are correlated between different points in time, as mediated by interactions with an external system~\cite{Pollock2018a,White-MLPT}. The specific mediator of these effects is both conceptually and experimentally relevant device information. Is it something we can control, or is it part of the inaccessible bath? Entangling crosstalk, such as the always-on $ZZ$ interactions in transmon qubits~\cite{PhysRevA.101.052308}, can generate temporal correlations. But whether the dynamics look non-Markovian depends on whether it is feasible or not to dilate the characterisation to multiple qubits -- which, typically, it is not. 
Since crosstalk and non-Markovianity can easily be conflated, it is crucial to find robust methods that can not only account for their behaviour, but distinguish them.

Here, we establish a systematic, concrete, and efficient approach to the two pragmatic questions: (1) if non-Markovian dynamics are detected across different timescales for a qubit, do they come from neighbouring qubits or a nearby bath? And (2) how can we determine when two qubits are coupled to a shared bath generating common cause non-Markovian effects, versus independent sources or direct interaction?
The solutions here have highly practical implications. Namely, whether curbing the correlated effects is achievable through more intricate control of the various subsystem, or whether more careful attention needs to be paid to defects in the fabrication process~\cite{winick2021simulating,wei2022hamiltonian}.
Up to this point, attempts to answer the first question have been mostly heuristic: detect non-Markovian behaviour, and then search through candidate explanations of the underlying physics to see whether this explains the phenomena~\cite{tripathi2022suppression,Sarovar2020detectingcrosstalk,krinner2020benchmarking}.
But this approach is not scalable, and highly model-dependent.
\acs{PTT}, as we have introduced it, is equipped to chacterise non-Markovian quantum stochastic processes and indeed to distinguish between purely spatial interactions versus temporal correlations generated by an external bath. However, in full generality the number of experiments required grows as $\mathcal{O}(d^{2kn+n})$ to find correlations across $k$ steps over $n$ qudits. 

The basic premise of our work is to apply the method of classical shadows to \acs{PTT}, resolving these problems. Then, instead of reconstructing the whole multi-time process for an entire quantum register, we can simply look at process marginals. 
This gives us access to maximal depolarising channels at no extra cost on the remainder of the qubits at each time. 
Classical shadow tomography provides access to a small number of low weight observables, with $\langle\mathbb{I}\rangle$ on the remainder of the subsystems. The case where $\langle\mathbb{I}_{\mathfrak{i}_{j+1}}\mathbb{I}_{\mathfrak{o}_j}\rangle$ is evaluated is equivalent to selecting an $\hat{\mathcal{A}}_j =\mathbb{I}_{\mathfrak{i}_{j+1}}\otimes\mathbb{I}_{\mathfrak{o}_j}\equiv \mathbb{I}_{\mathfrak{i}_{j+1}\mathfrak{o}_j}$. This is the Choi state of the maximal depolarising channel, up to normalisation.
These operations act as causal breaks on controllable systems. When suitably placed, these operations eliminate temporal correlations as mediated on the chosen Hilbert spaces, thus allowing non-Markovian sources to be causally tested. We illustrate this idea in Figure~\ref{fig:crosstalk_idea}b. The end result is the simultaneous determination of the bath-mediated non-Markovianity on all qubits.
Our approach hence only depends on the individual system size (in this work, qubits), and is a physics-independent way for us to test the relevant hypotheses. 
We are also able to simultaneously compute all $\binom{nk}{k}$ spacetime marginals (with resources growing exponentially in $k$), i.e. the generic process correlations between groupings of qubit-time coordinates. This extends the randomised measurement toolkit to the spatiotemporal domain~\cite{elben2022randomized}. Note that while this approach can also be used to characterise crosstalk, extensive work has already been conducted on this topic, and so we do not concentrate on these spatial correlations~\cite{Sarovar2020detectingcrosstalk,rudinger2021experimental, helsen2021estimating}. Instead, we focus on the ability to filter out crosstalk effects and study what is left over from the bath.

When non-Markovian correlations persist as mediated by the inaccessible bath, we designate this bath non-Markovianity (BNM). When the correlations are mediated from neighbouring qubits, we designate this register non-Markovianity (RNM).
Naturally, since the bath cannot be controlled by definition, BNM can be probed without RNM, but RNM effects cannot be isolated by themselves. Instead, one might consider the spatial process marginals alone to measure direct crosstalk~\cite{Sarovar2020detectingcrosstalk,rudinger2021experimental,helsen2021estimating}.
\par

\subsection{Erasing non-Markovian Pathways}
The procedure introduced in Section~\ref{ssec:shadow-procedure} suffices to estimate marginals of a process tensor with the familiar logarithmic scaling properties of classical shadows in the number of marginals. We consider two possible applications of spatiotemporal classical shadows, and supplement with numerical demonstrations.



Assuming now that we have a process tensor defined on a register $\mathbf{Q}=\{q_1,\cdots,q_N\}$, then we can fine-grain not just in time but also in space. We first note that without loss of informational completeness, each instrument $\hat{\mathcal{A}}_j$ can be factored into local operations $\bigotimes_{i=1}^N\hat{\mathcal{A}}_j^{(q_i)}$. The dynamical maps of the process can also be further marginalised in the same way to obtain $\hat{\mathcal{E}}_{j:j-1}^{(q_i)}$: the average \acs{CPTP} map taking the $i$th qubit from time $t_{j-1}$ to $t_j$. Although process tensors contain information about all multi-time correlations, for simplicity's (and efficiency's) sake, here, we consider only two map correlations. Therefore, spacetime correlations can be probed across the following pairs of marginals:

\begin{itemize}
	\item \underline{Purely spatial}: $\hat{\mathcal{E}}_{j:j-1}^{(q_n)}$ and $\hat{\mathcal{E}}_{j:j-1}^{(q_m)}$,
	\item \underline{Purely temporal}: $\hat{\mathcal{E}}_{j:j-1}^{(q_n)}$ and $\hat{\mathcal{E}}_{l:l-1}^{(q_n)}$,
	\item \underline{Spatiotemporal}: $\hat{\mathcal{E}}_{j:j-1}^{(q_n)}$ and $\hat{\mathcal{E}}_{l:l-1}^{(q_m)}$.
\end{itemize}

Pure spatial correlations constitute either direct crosstalk, or presence of an environmental common cause factor. As we show in the next section, purely temporal marginals constitute genuine bath non-Markovian correlations. And finally, spatiotemporal correlations indicate two qubits coupled to the same non-Markovian bath.

\begin{figure}[h!]
	\centering
	\includegraphics[width=\linewidth]{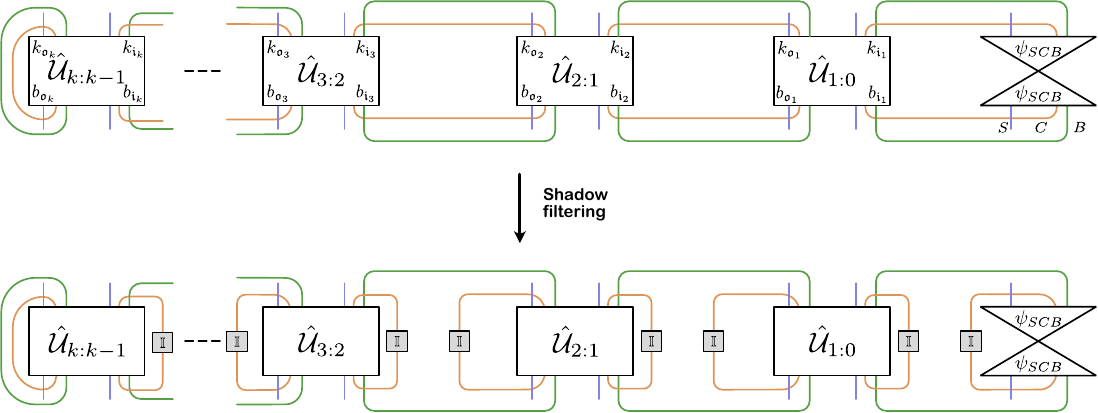}
	\caption[Graphical depiction distinguishing crosstalk from bath non-Markovianity via causal testing ]{Graphical depiction of causal testing. At the top, we represent the process tensor in terms of its dilated link product representation. Below, we show how the transformation implements depolarising channels, leaving the process as tensor product structure across $\mathcal{B}(\mathcal{H}_C)$.}
	\label{fig:PT-link}
\end{figure}

Here, we give a short graphical proof for the fact that applying depolarising channels to other qubits in the register at each time eliminates them as a potential non-Markovian source.
Recall from Chapter~\ref{chap:stoc-processes} that in the dilated open quantum evolution picture, a process tensor may be written as the \emph{link product} $\star$ of a series of $SE$ unitaries. 
\begin{equation}
	\label{eq:PT-link}
	\Upsilon_{k:0} = \text{Tr}_{E}\left[\bigstar_{i=1}^{k} \hat{\mathcal{U}}^{(SE)}_{i:i-1} \star \rho_0^{(SE)}\right],
\end{equation}
where $\hat{\mathcal{U}}_{i:i-1}$ is the Choi representation of the $i$th step unitary $U_{i:i-1}$. This lives on $\mathcal{B}(\mathcal{H})_S^{\mathfrak{i}}\otimes\mathcal{B}(\mathcal{H})_E^{\mathfrak{i}}\otimes\mathcal{B}(\mathcal{H})_S^{\mathfrak{o}}\otimes\mathcal{B}(\mathcal{H})_E^{\mathfrak{o}}$. Suppose the environment factorises into two spaces: $\mathcal{H}_E \cong \mathcal{H}_C \otimes \mathcal{H}_B$ -- the controllable qubits and the uncontrollable bath -- then each $\mathcal{U}_{i:i-1}$ has twelve indices: ket and bra for the input and output spaces for each of $S, C, B$. Equation~\eqref{eq:PT-link} can be written more intuitively using the graphical calculus formalism~\cite{wood2011tensor}, as shown at the top of Figure~\ref{fig:PT-link}.

Because depolarising channels are a product state over $\mathcal{B}(\mathcal{H}_{\mathfrak{o}})\otimes \mathcal{B}(\mathcal{H}_{\mathfrak{i}})$, they hence implement a causal break.
Note that any choice of instrument for which the Choi state is a product state (for example, a projective measurement and fresh preparation) constitutes a valid causal break, but a depolarising channel is the only one for which the locality of the observable does not grow, and hence can be efficient under the classical shadows procedure.

The same logic applies for the observation of correlations between some $\hat{\mathcal{E}}_{j:j-1}^{(q_n)}$ and $\hat{\mathcal{E}}_{l:l-1}^{(q_m)}$. Obtaining these two correlated marginals means that a causal break is applied at the $\mathfrak{o}$ leg of $\hat{\mathcal{E}}_{j:j-1}^{(q_m)}$ and at the $\mathfrak{i}$ leg of $\hat{\mathcal{E}}_{l:l-1}^{(q_n)}$. Hence, this erases any mixing terms between $\mathcal{H}_{q_n}$ and $\mathcal{H}_{q_m}$ in the unitaries. Consequently, non-zero values for the mutual information between these maps must be due to interactions involving $\mathcal{H}_B$ and $\mathcal{H}_{q_n}$ at the first time step, and then $\mathcal{H}_B$ and $\mathcal{H}_{q_m}$ at the second step. I.e., the two qubits shared the same bath. 



\subsubsection*{Distinguishing Between Passive Crosstalk and Bath Non-Markovianity}

First, we consider certifying when non-Markovian correlations originate via an inaccessible bath, or from neighbouring qubits in the register.
Certifying bath non-Markovianity means estimating $\Upsilon_{k:0}^{(q_i)}$ -- the marginal process tensor for a single qubit. Through the usual procedure of classical shadows, this can be simultaneously performed for all $q_i\in \mathbf{Q}$. The Choi state of the operations on the remainder of the qubits at each time will be $\mathbb{I}/d$, i.e. a maximally depolarising channel. Because this enacts a causal break -- the operation is a product state -- any information travelling from the system into the register cannot persist forwards in time. Hence, a suitable choice for computing the non-Markovianity in $\Upsilon_{k:0}^{(q_i)}$ will be a measure of correlations from an inaccessible bath alone.
\begin{figure}
	\centering
	\includegraphics[width=0.75\linewidth]{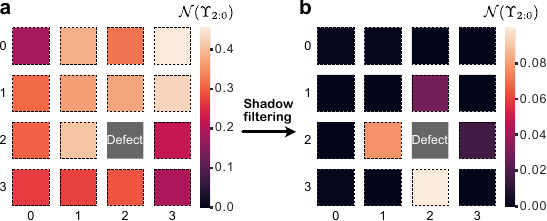}
	\caption[Numerical simulation demonstrating classical shadows to isolate environmental effects from local crosstalk ]{A numerical simulation demonstrating the proposed technique to isolate environmental effects. A grid of 15 qubits is simulated with crosstalk effects and an inaccessible non-Markovian defect. \textbf{a} Determining the non-Markovianity on each qubit individually (with the remainder idle) is not very informative, since each qubit looks non-Markovian due to passive crosstalk. \textbf{b} After learning all of the shadow marginals $\Upsilon_{2:0}^{(q_j)}$, the crosstalk is filtered out to reveal which qubits possess temporal correlations from the environment.}
	\label{fig:simulated-baths}
\end{figure}

We demonstrate this numerically in Figure~\ref{fig:simulated-baths}. Here, we have 15 qubits and one defect quantum system acting as the bath in a two-step process, and then compute $\mathcal{N}(\Upsilon_{2:0}^{q_i})$. The qubits each experience a random nearest-neighbour $ZZ$-coupling crosstalk, and the ones geometrically closest to the defect are Heisenberg-coupled to that system. Figure~\ref{fig:simulated-baths}a shows the standard fare: estimating the process tensor of each qubit and determining its non-Markovianity while the other qubits remain idle. However, the results are not so informative, because they do not distinguish between RNM and BNM effects, and so every qubit experiences temporal correlations. 
We show in Figure~\ref{fig:simulated-baths}b the results of a shadow marginal estimation. 
We see that, only the qubits which are actually coupled to non-Markovian bath defects have a non-zero $\mathcal{N}(\Upsilon_{2:0}^{(q_i)})$, and we can readily identify them.

\subsubsection*{Isolating Shared Baths}

A second important scenario we consider is where two qubits are correlated via common cause from a shared bath. For example, this might be experiencing the same stray magnetic field inhomogeneities or through a coupling to a common defect.
This is sometimes referred to as crosstalk, because the joint map $\hat{\mathcal{E}}_{j:j-1}^{(q_1,q_2)}$ does not factor to $\hat{\mathcal{E}}_{j:j-1}^{(q_1)}\otimes \hat{\mathcal{E}}_{j:j-1}^{(q_2)}$~\cite{Sarovar2020detectingcrosstalk}. However, we consider this a coarse description because neither system acts as a direct cause for each other's dynamics. Instead, they are subject to spatiotemporal correlations as mediated by the same non-Markovian bath -- i.e. the bath is the common cause for the shared correlations. The key, therefore, is to measure the relationship between the maps $\hat{\mathcal{E}}_{j:j-1}^{(q_1)}$ and $\hat{\mathcal{E}}_{j+1:j}^{(q_2)}$.
\begin{figure}[t]
	\centering
	\includegraphics[width=0.9\linewidth]{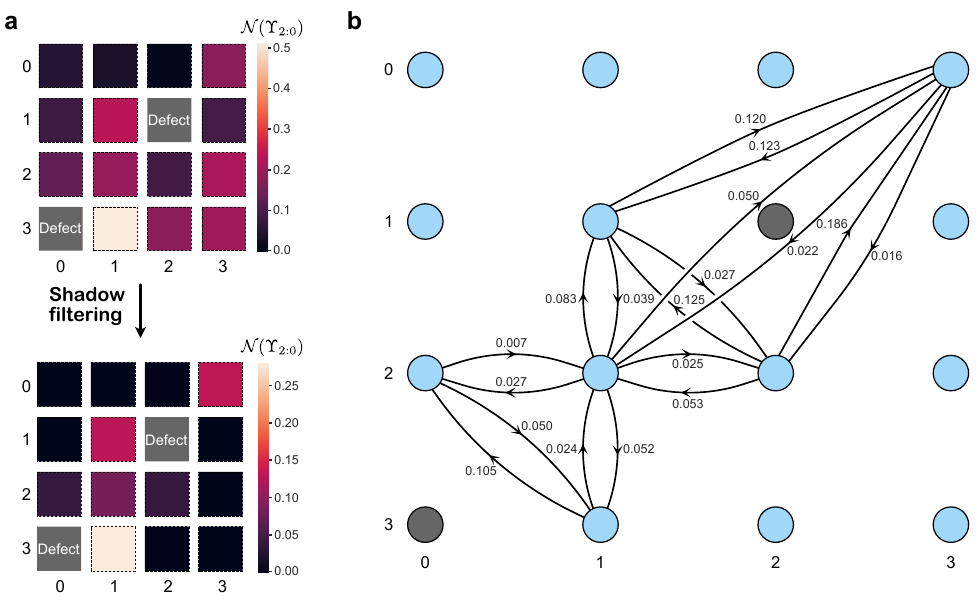}
	\caption[Numerical simulation computing spatiotemporal process marginals to local qubits with a common-cause shared non-Markovian bath ]{
		A numerical simulation demonstrating the proposed technique to determine qubits with shared baths. A grid of 14 qubits is simulated with crosstalk effects and two inaccessible non-Markovian defects. \textbf{a} Shadow filtering may be used to find qubits coupled to an inaccessible bath as before
		\textbf{b} By looking at the correlations between map $\hat{\mathcal{E}}_{1:0}^{(q_m)}$ with $\hat{\mathcal{E}}_{2:1}^{(q_n)}$, we can infer which qubits share common baths and the extent to which the defects redistribute quantum information.
		Note, the relationship lines between qubits are not direct crosstalk interactions, but bath-mediated spatiotemporal correlation.}
	\label{fig:cc-relations}
\end{figure}

We demonstrate this numerically in Figure~\ref{fig:cc-relations}. We have a similar setup to before, except this time with two bath defects. Performing a shadow filtering (Figure~\ref{fig:cc-relations}a) again reveals which qubits are coupled to the defects. However, in Figure~\ref{fig:cc-relations}b, we look at the spacetime marginals estimated from the shadow data. This fine-grained data indicates which qubits are commonly coupled to bath defects, versus independently coupled. The arrows from qubit $q_i$ to qubit $q_j$ indicate the \acs{QMI} in the process with marginals $\hat{\mathcal{E}}_{2:1}^{(q_j)}\otimes \Upsilon_{1:0}^{(q_i)}$.
The erasure of $q_j$ in the first step and $q_i$ in the second step eliminates the possibility of direct-cause correlations between the two qubits, leaving only the possibility of common-cause. In other words, non-zero values are a measure of non-Markovian correlations distributed by a shared bath between two qubits. This generates a more informative view of the interplay between qubits and their environment, breaking down the extent to which different systems share temporal correlations and which are fully independent from one another. 

In both of these simulated scenarios, causality conditions fix the expectations of $\sum_{j=1}^k(d^2-1)d^{4j-2}$ Pauli operators. For a two step process then, this leaves $M=820$ expectation values to estimate. From the bounds given in Ref.~\cite{huang-shadow}, we therefore have $N_{\text{shots}} = \mathcal{O}(\log(N_{\text{qubits}}\cdot M)/\epsilon^2 3^5)\approx 2\times 10^7$ measurements in the worst case, which is what we use in our numerical experiments. The median-of-means algorithm from Ref.~\cite{huang-shadow} is used to estimate an informationally complete set of observables for each process marginal, and the maximum likelihood estimation obtained using the methods developed in Chapter~\ref{chap:PTT}.

\subsection{Discussion}
We have introduced a scalable and conceptually simple method to distinguish between non-Markovian dynamics generated by nearby qubits in a quantum device, and those from an inaccessible bath. 
This contributes to the growing zoo of quantum characterisation, verification, and validation techniques, and yet satisfies a unique niche. Geometrically isolating non-Markovian sources across a device can inform various facets of the development process: the signals can warrant further investigation and inform the fabrication process; they can flag qubits to be given extra control attention; and they can be used in error-finding mechanisms for error-correction decoders.

This also extends the capabilities of the randomised measurement toolbox in the multi-time and multi-qubit domain. Despite the limitations of classical shadows for quantum stochastic processes, we have identified an important use case in efficient casual testing.
We have introduced this in full generality with respect to \acs{PTT}, however the techniques are generic: the only important point is that causal breaks are applied to neighbouring qubits between each steps. The same principles will broadly apply to other approaches to learning non-Markovian dynamics~\cite{PhysRevLett.124.140502,PhysRevLett.112.110401,nielsen-gst,PRXQuantum.2.040351}.


\section{Dynamically Sampling Non-Markovian Open Quantum Systems}
\label{sec:dynamic-sampling}

In recent years, there have been explosions of interest in using quantum devices to determine classically intractable properties. 
The seminal development of the randomised measurement toolbox provided not only an extremely practical \acs{QCVV} tool, but kicked off a flurry of research poised at the question: what are the fundamental limits to which we can learn properties of quantum states? 
One recent development has been the idea to take quantum data from a device in the form of randomised measurements, and process this with classical machine learning algorithms. It has been shown that this approach generates an exponential advantage in the learnability of quantum states.
Meanwhile, Ref.~\cite{aloisio-complexity} (co-authored by the present author) showed the ability for quantum stochastic processes to be efficiently simulable (in the sampling sense) on a quantum computer, but intractable for a classical computer. This opens up the possibility of realising quantum advantage not only with large entangled registers, but through small devices with complex environments.

In this section, we make a step towards this goal, 
by specifically considering the characterisation of multi-time sampling statistics in practice. This is hence a \emph{dynamic} sampling problem, whose complexity is studied in Ref~\cite{aloisio-complexity}.
We move onto the study of much larger processes, coarser observations, and evolution of marginals of a process. These fit under the broad umbrella of macroscopic process properties, and can tell us, for example, how microscopic properties change as a function of time-step. Given the concerted effort to simulate correlated quantum processes classically~\cite{strathearn_efficient_2018, Luchnikov2019L, Cygorek-2022}, this connection lays the groundwork for fully general quantum simulation of non-Markovian open quantum systems. As well as simulating the dynamics (system as a function of time), a quantum computer may also access the dynamical multi-time correlations, which encapsulates every extractable piece of information from the system.
We perform a demonstration that consolidates our tools and reconstructs a long non-Markovian process simulated on a quantum device. For this purpose, we reconstruct process tensors under a \emph{finitely correlated state} ansatz~\cite{PhysRevLett.111.020401}. States of this type may be reconstructed by only their marginals. 
Although our reconstruction still falls into the realm of classically simulable, the intended purpose is to show that dynamical sampling of multi-time statistics is possible even with near-term devices. This not only establishes that scalable non-Markovian characterisation is achievable, but presents a new aspect to simulation capabilities of quantum computers. 


\subsection{Demonstration on NISQ Devices}

Specifically, we consider one end of a Heisenberg-coupled spin-chain and estimate the 20-step/42-body process tensor Choi state in a finitely correlated state form. Ref.~\cite{PhysRevLett.111.020401} presents a method for reconstructing finitely correlated quantum states from their marginals. 
There, it was shown that a 1D state could be reconstructed from reduced states on an odd number $R$ neighbouring spins of dimension $d$, so long as the state satisfies certain left and right invertibility conditions. We briefly summarise the procedure, and outline its application in the context of process tensors. As an ansatz, a left and a right rank ($d^{2l},d^{2r}$ respectively) are chosen for an $n$-body state. The marginals for this state will then be $R$-body, where $R = l + r + 1$. This gives a total of $n - R + 1$ local marginals.
In a process tensor context, we have $k$ steps, or a $2k+1$-body state. If $l = r :=\ell$, then the problem requires estimation of all $\ell$ step marginals: $\{\Upsilon_{\ell:0}$, $\Upsilon_{\ell+1:1}$, $\cdots$, $\Upsilon_{k:k-\ell}\}$. With the input of classical shadows, these $k - \ell + 1$ marginals may be simultaneously estimated, leaving the characterisation scaling logarithmic in the number of steps. The reconstruction of a process tensor using this method is discussed further below.
We use this as a platform to experimentally reconstruct correlated multi-time statistics for large processes. Specifically, we use a quantum device to simulate a spin chain and then examine the properties of the quantum stochastic process at one end of the chain. The device used was \emph{ibm\_lagos}, where the spin chain was simulated with four qubits. An additional ancilla qubit was also used for the purpose of applying quantum instruments to the system. 

\begin{wrapfigure}{r}{0.5\textwidth}
	\centering
	\includegraphics[width=0.95\linewidth]{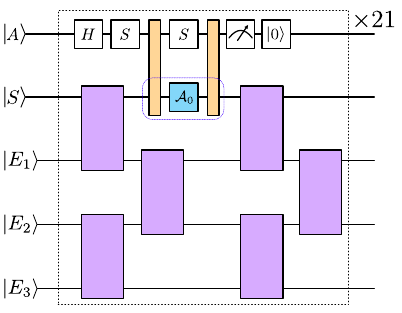}
	\caption[Circuit diagram collecting multi-time statistics as part of a spin chain ]{Circuit diagram collecting multi-time statistics as part of a spin chain. An exchange interaction is driven between neighbouring qubits across 21 repeated blocks, and ancilla qubit used to probe the system. A $R_{ZX}(\pi/4)$ operation is used to interact $S$ and $A$, followed by projective measurement on $A$. In each run the local unitary on $S$ is chosen to be a random Clifford operation, generating the classical shadow data.}
	\label{fig:spin-chain-circ}
\end{wrapfigure}

\begin{figure}[t]
	\centering
	\includegraphics[width=0.6\linewidth]{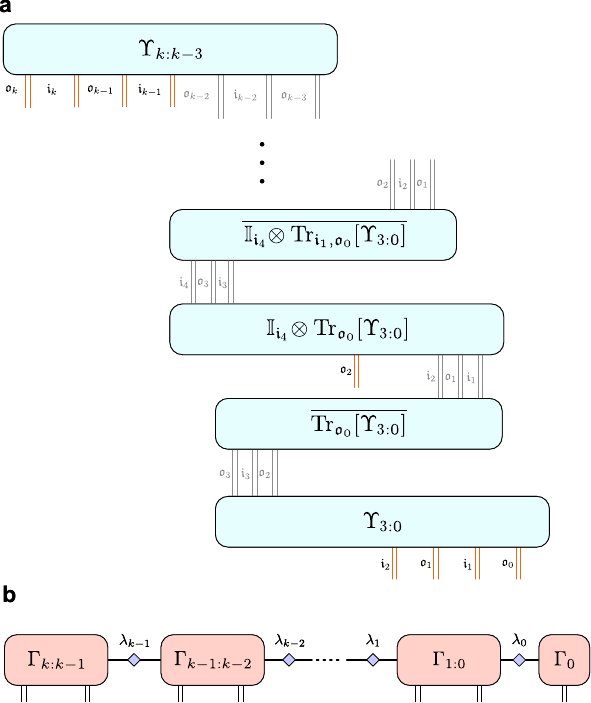}
	\caption[Graphical illustration of a reconstructed many-body process tensor from its marginals ]{Graphical illustration of a reconstructed many-body process tensor from its marginals. \textbf{a} We construct the process tensor as a finitely correlated state by stitching together $\ell$ step marginals estimated from classical shadows. Here is an example taking $\ell=3$. Note that the overline denotes a pseudoinverse of the matrix. Orange legs are free indices, and grey legs are contracted indices between process tensor marginals. \textbf{b} The resulting many-time representation after condensing down the above, and appropriately splitting sites at the start and finish. The $\lambda_j$ indicate summation over the indices given in the marginal pseudoinverses. }
	\label{fig:PT-MPO}
\end{figure}

At each step, an exchange coupling is simulated by the application of a random $XX + YY + ZZ$ rotation between each of the neighbouring qubits. Using the ancilla qubit, a random (bootstrapped, as per Chapter~\ref{chap:PTT}) instrument is applied to the system for a total of $4.4\times 10^7$ shots. Equation~\eqref{eq:bootstrapped-shadow} is then used to construct the classical shadow for each shot, and the data post-processed to produce the three step marginals $\{\Upsilon_{l:l-3}\}_{l=3}^{21}$. Then, by evaluating expectation values of the global chain in Figure~\ref{fig:PT-MPO}, we can generate a predicted probability distribution for time-series data -- the 21 measurements across the circuit. The overall circuit diagram is given in Figure~\ref{fig:spin-chain-circ}.

Because we have simulated a relatively small environment, and because noise on quantum devices is expected to render this mostly dissipative, we take three step marginals as our ansatz for reconstruction. This assumes that the left and right ranks at each site of the process tensor is no more than 64. 
With the three-step marginals estimated through repeated randomised measurements made on the ancilla qubit, we may now reconstruct the entire 20-step process tensor Choi state according to our finite correlation ansatz.

Full details of the theorem reconstructing a state from its marginals may be found in Ref.~\cite{PhysRevLett.111.020401}, however we reconfigure the statements here in the form of a graphical diagram in Figure~\ref{fig:PT-MPO}. 
In summary, the procedure involves computing pseudoinverses to connect together process tensor marginals. This further motivates the need for a robust marginal estimate, since inverses can only be computed with complete tomographic information about a state.
We start with $\Upsilon_{\ell:0}$. Moving one leg across, the next $2\ell+1$ body marginal is  $\text{Tr}_{\mathfrak{o}_{\ell+1},\mathfrak{o}_0}[\Upsilon_{\ell+1:0}]$. However, causality constraints (Equation~\eqref{eq:causality}) mean that this term reduces to $\mathbb{I}_{\mathfrak{i}_{\ell+1}}\otimes \text{Tr}_{\mathfrak{o}_0}[\Upsilon_{\ell:0}]$, which is information we already have. Moving one leg over, the next marginal is $\Upsilon_{\ell+1:1}$. This continues until we get to $\Upsilon_{k:k-\ell}$. Collectively, these form the set
\begin{equation}
		E:=\{\Upsilon_{k:k-\ell}, \mathbb{I}_{\mathfrak{i}_k}\otimes \text{Tr}_{\mathfrak{o}_{k-\ell-1}}[\Upsilon_{k-1:k-\ell-1}], \Upsilon_{k-1:k-\ell-1},\cdots,\mathbb{I}_{\mathfrak{i}_{\ell+1}}\otimes\text{Tr}_{\mathfrak{o}_0}[\Upsilon_{\ell+1:0}],\Upsilon_{\ell:0}\}.
\end{equation} 
The connections of these terms to construct the overall finitely correlated stote is depicted in Figure~\ref{fig:PT-MPO}a, with the effective many-time state depicted in Figure~\ref{fig:PT-MPO}b.

\begin{figure}
	\centering
	\includegraphics[width=\linewidth]{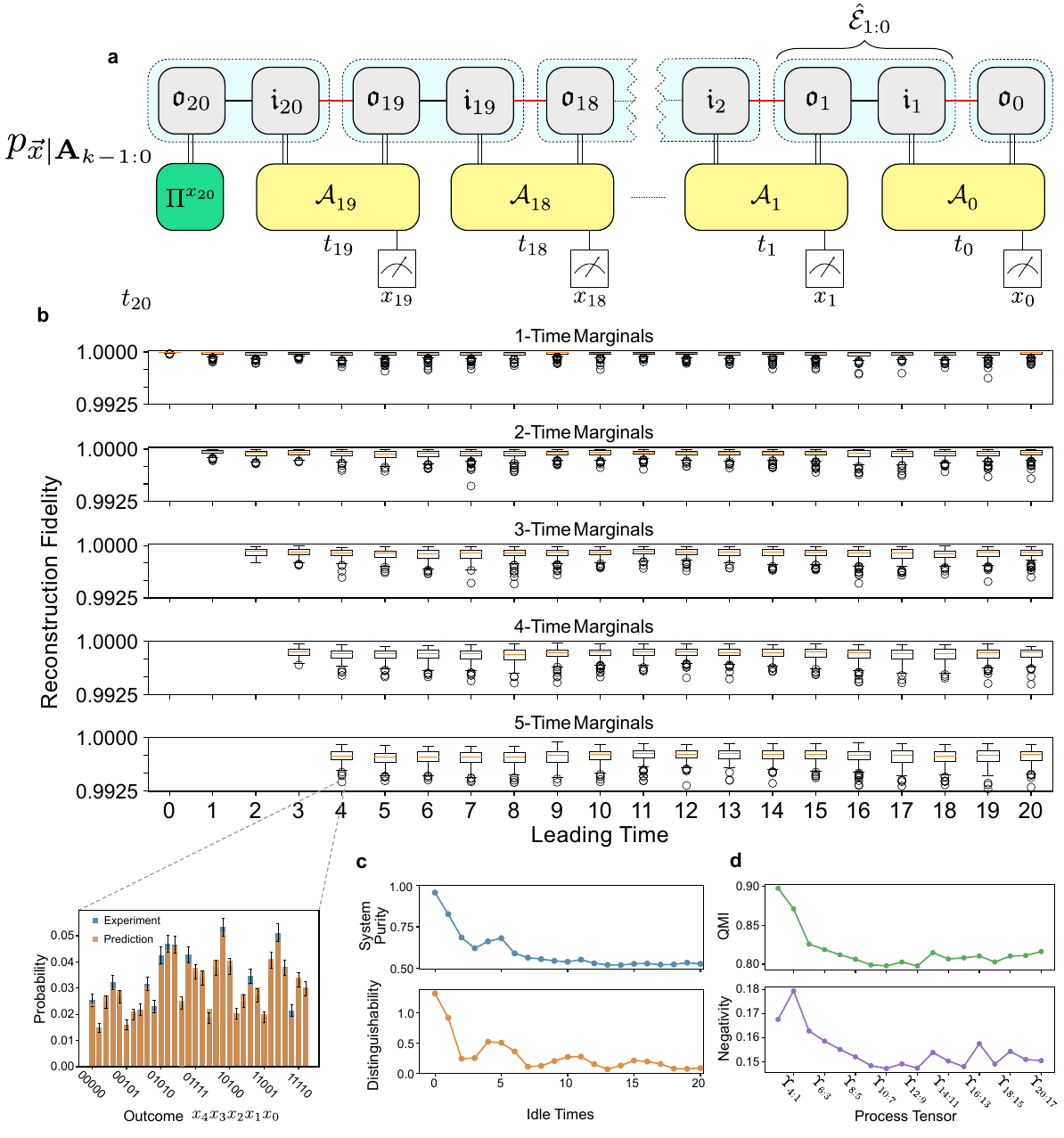}
	\caption[Experimentally reconstructing dynamic sampling statistics from a finitely correlated process tensor ]{
	\textbf{a} We characterise a 20-step process of a non-Markovian quantum environment on \emph{ibm\_lagos} and construct its finitely correlated state representation. \textbf{b} This characterisation is then validated against 170 sequences of random instruments: the reconstructed process tensor is used to predict different $l$-time marginal probability distributions of the instrument outcomes and compared against the experimental results.
	\textbf{c} Properties of the system when left idle as a function of time step, showcasing conventional measures of non-Markovianity. We observe the non-monotonic behaviour of both the purity of the system, and the trace distance between two states initialised as $Z$ and $Y$ eigenstates. \textbf{d} Properties of the 3-step marginal process tensors as a function of time, we measure both the total memory and the temporal quantum entanglement, demonstrating a genuinely complex non-Markovian process.}
	\label{fig:full-mpo-results}
\end{figure}


To validate that the estimate does indeed accurately capture environmental dynamics, 170 different sequences of random instruments are then run on the device at $16\:384$ shots. The outcomes of the instruments across the 21 different times are collected and used to generate different marginal probability distributions. The effect of marginalising over the outcomes at a given time is that the operation applied is a deterministic \acs{CPTP} channel. The marginal probability distributions for each sequence of operations is predicted by our process characterisation and compared via Hellinger fidelity to the actual results. Figure~\ref{fig:full-mpo-results}b summarises the distributions of these results for up to 5-time marginals across all 21 times, showcasing the ability to predict arbitrary components of the stochastic process to high accuracy.

The cross-validation of multi-time statistics is as follows. Consider one such test sequence of random instruments $\mathbf{J}_{20:0} = \{\mathcal{J}_{20}^{(x_{20})},\mathcal{J}_{19}^{(x_{19})},\cdots \mathcal{J}_0^{(x_0)}\}$. Note that the last instrument, $\mathcal{J}_{20}$, is actually a \acs{POVM} -- an instrument where the post-measurement state is thrown away.
Each instrument is selected by choosing a Haar random unitary to act on the system, and the ancilla qubit measurement outcomes recorded. Because of the characterisation procedure given in Chapter~\ref{chap:PTT}, each instrument's effective map is known, regardless of the local unitary. A single circuit consequently induces a joint probability distribution
\begin{equation}
	\label{eq:prob-dist}
	\mathbb{P}(x_{20},t_{20}; x_{19},t_{19};\cdots;x_0,t_0 | \mathbf{J}_{20:0})
\end{equation}
for the measurement outcomes at the different times. We first sample from this distribution for a given $\mathbf{J}_{20:0}$ by running the sequence on the device at $16\:384$ shots and recording each of the results. Naturally, at $2\times 10^6$ outcomes, this distribution is already too large to sensibly measure and compare. Hence, we consider marginals of the distribution. A marginal of length $\ell$ at a leading time $j$ is the marginal distribution of Equation~\eqref{eq:prob-dist} 
\begin{equation}
	\label{eq:prob-marg}
	\mathbb{P}(x_{j},t_{j}; x_{j-1},t_{j-1};\cdots;x_{j-\ell+1},t_{j-\ell+1} | \mathbf{J}_{j:0}).
\end{equation}

Figure~\ref{fig:full-mpo-results} summarises up to each contiguous 5-time marginals of a sequence of random instruments on \emph{ibm\_lagos}. The estimate of the finitely correlated state representation of $\Upsilon_{20:0}$ can produce a corresponding estimate for the distributions in Equation~\eqref{eq:prob-marg} by computing
\begin{equation}
	\label{eq:PT-probs}
	p_{x_jx_{j-1}\cdots x_{j-\ell+1} | \mathbf{J}_{j:0}} = \text{Tr}\left[\left(\mathbb{I}_{t_{20}:t_{j+1}}\otimes \mathcal{J}_j^{(x_j)\text{T}}\otimes \mathcal{J}_{j-1}^{(x_{j-1})\text{T}}\otimes\cdots \otimes \mathcal{J}_{j-\ell+1}^{(x_{j-\ell+1})\text{T}}\otimes \mathcal{J}_{j-\ell-2}^{\text{T}}\otimes \cdots \otimes \mathcal{J}_0^{\text{T}}\right)\cdot \Upsilon_{20:0}\right],
\end{equation}
which, for the resulting graphical diagram in Figure~\ref{fig:PT-MPO}, can be contracted efficiently. Note that where we have dropped the instrument superscript, this indicates that we have discarded the ancilla outcome, effecting a \acs{CPTP} map: $\mathcal{J}_i = \mathcal{J}_i^{(0)} + \mathcal{J}_i^{(1)}$. The boxplots of Figure~\ref{fig:full-mpo-results}b display the Hellinger distance between the quantum computer sample, and Equation~\eqref{eq:PT-probs} for each of the 170 sequences, for each marginal $1-5$, and for each time $0-20$.

In Figures~\ref{fig:full-mpo-results}c and \ref{fig:full-mpo-results}d we examine the non-Markovian properties of the process from different perspectives. Figure~\ref{fig:full-mpo-results}c measures the purity $\Tr[\rho^2]$ of the idle state of the system as a function of time -- with no control operations applied. The oscillations seen indicate decoherence and recoherence of the state, a strong characteristic of information backflow from the environment~\cite{rivas-NM-review}. 
We also demonstrate another aspect of two-time correlations, which is the non-monotonicity of trace distance between two differently initialised states~\cite{PhysRevA.83.052128}. We start $\rho_1 = |0\rangle\!\langle 0|$ and $\rho_2 = |i+\rangle\!\langle i+|$, and plot $\|\rho_1(t) - \rho_2(t)\|_1$ for each time step. This quantity oscillates in phase with the state purities, and exhibits a much stronger signal. Collectively, these demonstrate substantial environmental back-action. Thus, the IBM Quantum device is able to coherently simulate this non-Markovian dynamics, which we have demonstrated we have been able to fully characterise.

 Meanwhile Figure~\ref{fig:full-mpo-results}d conveys exact measures of non-Markovianity throughout the process. We see that both the generalised \acs{QMI}, and negativity across the middle cut of the three-step process marginals both decay to a non-zero constant value. Despite the nominal decay of the system, memory effects persist in the dynamics, as can be detected through mid-circuit measurements. The reason the memory persists in the multi-time statistics but not in the purity or distinguishability measures is two-fold: (i) these measures are not as general, and will always lower bound the multi-time quantity, and (ii) measuring the state of the system as a function of time is likely to lead to dissipation. Incorporating the effects of control operations can act as a randomiser of the state of the system, and effectively decoupling from the decoherence process.


Here, we have gone beyond the state-of-the-art in characterising quantum stochastic processes. We have shown how one may completely characterise arbitrarily many time-steps in a quantum process by using the available hardware to manufacture full system control. Through our demonstration we validate the ability to predict the statistics produced by arbitrary sequences of instruments with mid-circuit measurements. This demonstrates the first full-scale characterisation of a quantum stochastic process, and showcases how multi-time statistics can be learned in practice.
The memory of the process studied is genuinely complex, but not so large that our finitely correlated state model is unable to capture the necessary degrees of freedom. This further provides a clue on constructing compressed models to effectively determine non-Markovian dynamics. We explore this problem further in Chapter~\ref{chap:efficient-characterisation}.
Our approach and demonstration broadens the study of processes to a far more macro level, widening the scope to more intricate considerations such as thermalisation and equilibration.

\section{Discussion}
Although the study of many-body physics has uncovered a rich tapestry of structure in different corners of applied quantum mechanics, temporal quantum correlations remain relatively under-explored. In this work, we have motivated, demonstrated, and made accessible the study of many-time physics beyond the two-time correlators regime. Multi-time correlations will showcase the full potential of the field of many-time physics, including capturing dynamical phases of matter~\cite{heyl2018dynamical}.
\par

A primary achievement is in showing how rich non-Markovian effects can be studied with only unitary control and a final projective measurement. Rather than merely simulating non-Markovian quantum processes on fully controllable quantum computers, this allows observation of complex naturally occurring phenomena~\cite{PhysRevA.61.023603,jaksch2005cold,PhysRevLett.101.260404,PhysRevA.90.032106,caruso2009highly,lambert2013quantum}. One might imagine, for example, the use of quantum sensors in condensed matter or biological systems. Nitrogen-vacancy centres in diamond, for example, can be unitarily controlled with arbitrary waveform generators, projectively read out optically, and have been shown to be biocompatible~\cite{McGuinness2011}.\par 

Our results on IBM Quantum devices show that even idle qubits can exhibit surprisingly complex and temporally correlated dynamics. We have discovered that the non-Markovianity persists not simply as a classical set of correlations, but as temporal quantum entanglement. This prompts the need for further study of non-Markovian behaviour on \acs{NISQ} devices. Specifically, if not fabricated away, correlated noise ought to be able to be converted into clean channels through appropriate control sequences~\cite{berk2021extracting}.\par 

Finally, given the use of ancilla qubits, we have also shown how to arbitrarily capture many-time physics through repeated probes of a system. This naturally fits into models of quantum simulation, but provides strictly more information than simply observing evolution of a state.
From this, we can envisage the transduction of multi-time observables from a single non-Markovian quantum system into (classical or quantum) machine learning algorithms, which has recently been shown as a promising avenue for quantum advantage~\cite{Huang2021}.

%

\ctparttext{
We have formalised the process tensor tomography technique to characterise non-Markovian processes fully. However, several questions remain: Can we make the technique scalable and practical? How can we account for faulty components and background processes? What are the real-world applications of these methods for quantum devices? Quantum stochastic processes are exponentially large, making generic characterisation infeasible for all but small-scale cases. We address this by studying the structure of process tensors and leveraging both sparseness and finite quantum memory to develop efficient characterisation methods. We also create a method to estimate both background processes and faulty control operations, resulting in a self-consistent non-Markovian process characterisation procedure. Finally, we apply our tools to optimal quantum control, demonstrating how dynamic characterisation can inform device control to mitigate the effects of correlated quantum noise.
}
\part{Practical and Applied Considerations}
\chapter{Structuring and Characterising Multi-Time Processes Efficiently}
\label{chap:efficient-characterisation}
\epigraph{\emph{Blessed are the forgetful, for they get the better even of their blunders.}}{Friedrich Nietzsche}
\noindent\colorbox{olive!10}{%
	\begin{minipage}{0.9555\textwidth} 
		\textcolor{Maroon}{\textbf{Chapter Summary}}\newline
		Non-Markovian processes can be challenging to characterise due to their complex structure and resource-intensive requirements. This chapter aims to introduce novel approaches for compressing these models and reducing the resources needed for characterisation. Building on the foundational understanding of non-Markovian processes developed in previous chapters and leveraging the sparsity of noise in advanced devices, our approaches enable more efficient and general models for characterisation. Our methods not only allow for efficient learning and reconstruction of non-Markovian processes, but also shed light on the underlying structure of temporal correlations in these systems, enhancing our understanding of their nature and behaviour over time.
		\par\vspace{\fboxsep}
		\colorbox{cyan!10}{%
			\begin{minipage}{\dimexpr\textwidth-2\fboxsep}
				\textcolor{RoyalBlue}{\textbf{Main Results}}
				\begin{itemize}
					\item We develop a spacetime Markov order ansatz to processes: the modelling of processes whose memory persists approximately across only a finite cone in space and time. Reconstruction requires only the process marginals whose information carries the memory.
					\item We model multi-qubit process tensors with two-dimensional tensor networks and develop the required numerical tools to learn these representations of processes. In certain scenarios this is an exact and highly efficient approach to characterise spatiotemporal quantum correlations.
				\end{itemize}
		\end{minipage}}
\end{minipage}}
\clearpage

\section{Introduction}
In our discussions of non-Markovian processes, we have kept in the foreground the \acs{CJI} -- the remarkable equivalence between multi-time processes and many-body quantum states. This relationship says that the exotic physics enjoyed by sets of qubits can be similarly mapped to a single system across many times. Consequently, we have focused on characterising non-Markovian processes in complete generality. We have so far considered tomography of process tensors where in principle the environment responds in a highly chaotic way to the system. This is to say, that an exponentially complex history is tracked in time. \par

In fact, it is much more difficult for complex temporal correlations to emerge than spatial ones. As we have already seen, spatial entanglement with an environment is not sufficient to generate temporal entanglement. But causality conditions mean that for complex correlations to occur at all, information must be first coherently transferred from system to environment and then passed back again at a later point. Non-Markovian processes balance a fine line between having an effective environment which is small -- and hence, simple; and an environment which is too large -- and hence, dissipative and also simple. In practice, complex processes appear to have to be highly contrived. For the purposes of characterisation, then, this motivates us to leverage this memory sparsity to develop efficient techniques to characterise non-Markovian quantum processes. It is hardly ideal to have exponential scaling in process learning. In Chapter~\ref{chap:PTT}, this limited us to practical tomography of only up to three steps on a single qubit. Here, we go above and beyond this in both simulation and experiment, and construct and analyse methods to scalably implement process tensor tomography. In each respective section, we consider two conceptually distinct ways that a process might be sparse (i) where the memory is arbitrarily strong, but short-lived, and (ii) where the memory can be described by a small quantum environment. 

We respectively address both forms of sparse memory in making \acs{PTT} more scalable.
First, we consider and develop the notion of quantum Markov order. If the mutual information of a stochastic process does not grow unboundedly with time, then one can define a finite memory over which the joint statistics are kept. The necessary complexity of a process is then linear in the number of time steps and exponential in that memory size.
We then develop and apply powerful tensor network methods to the problem of spatiotemporal \acs{PTT}. In terms of classical computation, this is by far the most costly of the various methods, and comes equipped without the same convergence guarantees. But it is also extremely expressive, and can serve as a cheap way to represent and estimate process tensors whose effective environment is not too large.

It is our objective in this chapter to identify and leverage sparsity in a process. Some aspects of this source inspiration from similar approaches in \acs{QST}, and some come from analysing properties of stochastic processes. Each method considered here reduces the complexity of the problem from exponential to at most linear in the number of time steps and number of qubits. We have already seen some process model reductions. For example, we derived a compressed sensing procedure for \acs{PTT} in Chapter~\ref{chap:PTT} for estimating reduced-rank states. This is cheaper in the collection of data, but inherits only a quadratic reduction in the number of resources. We also considered the problem of low-rank and tomographically incomplete sets of measurements (with a slightly different application) in Chapter~\ref{chap:MTP}. This collection of approaches greatly helps to reduce the problem of representing and characterising non-Markovian processes by leveraging some aspect of their structure. However, each of these methods is still exponentially costly. 
Here, we consider processes with specific simple structures. 
We have seen already that non-Markovian quantum stochastic processes can be both as complex and as large as quantum states. If this is the case, then one will always need to expend an exponential number of resources to fully learn its nature. Conversely, if the temporal correlations are not complex, then we show here that we can fully determine the processes in only a polynomial number of experiments, making the problem far more tractable.

\section{Quantum Markov Order}

Even in the classical case, the price of characterising the joint statistics of a stochastic process in full generality is exponentially high. Often, however, this is unnecessary in practice as physical processes are sparse. This is because the memory must be carried by another physical system, whose size then bounds the size of the memory. Often in practice, the \emph{necessary} complexity of a process characterisation only grows modestly with the size of its memory if, after a certain amount of time, the history and the future are independent from one another. 

In such cases, the joint statistics are no longer required between those points in time. This motivates the idea of Markov order~\cite{Rosvall2014, pollock-tomographic-equations}: the number of previous time steps in the process which are relevant to the present. We first discuss classical and quantum notions of this property, and then translate it into the ability to reconstruct multi-time, and eventually multi-qubit processes with these finite memory structures.

\subsection{Background}

 Concretely, in classical theory, a stochastic process is described by the joint probability distribution of a sequence of events $\mathbb{P}(x_k, x_{k-1}, \dots, x_0)$, occurring at times $\{t_k, t_{k-1}, \dots, t_0\}$. A process with Markov order $\ell$ then conditionally separates the future $F_j=\{t_{j+\ell+1},\dots,t_{k}\}$ from the past $P_j=\{t_0,\dots,t_{j-1}\}$ given the knowledge of the state in the memory block $M_j=\{t_j, \dots, t_{j+\ell}\}$. That is, the above distribution takes the form
\begin{gather}\label{cl:cmo}
	\mathbb{P}(x_k, \dots, x_0) = 
	\sum_{M_j}
	\mathbb{P}(P_j|M_j) \
	\mathbb{P}(M_j) \
	\mathbb{P}(F_j|M_j).
\end{gather}
That is, in order to know the probability of an event at a given time, we only need to look at the past $\ell$ events. Anything beyond that will not affect the future. In the case where memory is indeed infinite but decays in time, we can turn the last equation into an approximate statement. Importantly, the complexity of the whole process goes as $d^k$, where $k$ is not bounded. Meanwhile, the complexity of a process with Markov order $\ell$ goes as $d^\ell$ with a fixed $\ell$. That is, the distribution in Eq.~\eqref{cl:cmo} is fully determined by knowing $\mathbb{P}(M_j)$.

It is possible to extend the notion of Markov order to quantum stochastic processes by replacing $\mathbb{P}$ with its quantum counterpart $\Upsilon$, roughly speaking. We now apply this idea to \acs{MLE}-\acs{PTT}, and derive a resource-efficient way to characterise even processes with very large numbers of steps. Concretely, integrating \acs{MLE}-\acs{PTT} with a Markov order of $\ell$ would reduce the exponential scaling in Equation~\eqref{eq:mle-scaling} to
\begin{equation}
	\label{eq:mo-mle-scaling}
	\mbox{number of experiments} \sim \mathcal{O}(k \cdot N_{\text{mle}}^\ell).
\end{equation}
To achieve this, we build upon the ideas established in Ref.~\cite{taranto1} and subsequently realised in Ref.~\cite{PhysRevLett.126.230401}.

\subsection{Multi-time Conditional Markov Order}
\label{sec:MO}
A summary of our approach is to divide a $k-$step process up into a number of smaller, overlapping process tensors. These smaller process tensors are designed to account for a truncated number of past-time correlations. We then use \acs{MLE} estimation to fit each of the memory process tensors according to a Markov order model chosen by the experimenter. The finite Markov order process tensor fitting, adaptive memory blocking, and action across sequences that we introduce here are all novel features of quantum Markov order, and offers a method by which non-Markovian behaviour on \acs{NISQ} devices can be feasibly characterised and controlled. We note in passing that without \acs{MLE}-\acs{PTT} a Markov order integration would not be possible when working with restricted process tensors. This is because in this regime the partial traces of the process tensor are not well-defined, which makes it difficult to split the process into parts.\par 

In the quantum realm, the matter of Markov order is more nuanced than for classical processes. A quantum stochastic process has either Markov order one (the output at any leg is affected only by the previous input) or infinite Markov order (the memory persists indefinitely)~\cite{taranto1, taranto3}. That is to say, a Choi state may only be written as a product state, or there will exist correlations between all points in time (though, saying nothing about the strength of these correlations). For practical purposes however -- especially in the context of quantum computing -- there exists the useful concept of \emph{conditional} Markov order, studied recently in Ref.~\cite{taranto1}. This operationally characterises the form of processes which may have low \acs{QMI} between past and future when conditioned on some memory block. What is missing is an ansatz that allows for the modelling and characterisation of such processes. We introduce such an ansatz here.


\subsubsection*{Structure of Quantum Markov Order}
To properly explain this statement about conditioning, we first re-emphasise that a process tensor represents a quantum stochastic process. As a consequence, if we gain extra information about the past -- for example, what operation was applied by the experimenter -- then we may update our description of the process when conditioned on that choice of operation. This is akin to the quantum state picture: if a measurement is made on one qubit as part of a many-body system, then the remaining state can be updated based on the outcome of the measurement. For a dynamical process, this intervention can, in full generality, be a quantum instrument. 
The conditional state of the process may then exhibit past-future independence.
We now introduce Markov order for a quantum stochastic process, as well as the related notion of instrument-specific conditional Markov order. We then make clear that processes conditioned on generic operations may only exhibit \emph{approximate} conditional Markov order. Finally, we explicitly walk through our calculations of tomographically reconstructing processes with an approximate conditional Markov order ansatz.\par 
To begin, we describe a $k-$step process with Markov order $\ell$. When $\ell=1$, the only relevant information to the \acs{CPTP} map $\hat{\mathcal{E}}_i$ is the state mapped at the output of the $\hat{\mathcal{E}}_{i-1}$ step, described by the $\mathfrak{o}_{i-1}$ leg of the process tensor. Consequently, there is no context to the previous gates. The choice of operation $\mathcal{A}_{i-1}$ is only relevant insofar as determining the output state for time $(i-1)$. This constitutes a Markovian process, and the dynamics are \acs{CP}-divisible. Otherwise, it is non-Markovian with $\ell = \infty$~\cite{taranto1}. Intuitively one may think of this as the statement that there is no way to consistently write a generic quantum state with strictly limited correlations -- for example where each subsystem might have nearest-neighbour correlations but zero correlations with any subsystem outside this.  \par
Although finite $\ell>1$ Markov order is well-defined for classical stochastic processes, where there is only one basis, there is no generic way to write a quantum state with correlations persisting to the last $\ell$ subsystems. However, future and past statistics may be independent of one another for quantum stochastic processes when conditioned on the choice of an intermediate instrument. This notion of conditional Markov order may be described as follows. Consider a process tensor $\Upsilon_{k:0}$, which we denote as $\Upsilon_{FMP}$ with the groupings for the past, the memory, and the future, respectively:
\begin{equation}
	\begin{split}
		P_j &=\{t_0,\cdots,t_{j-1}\},\\
		M_j &=\{t_{j}, \cdots, t_{j+\ell}\},\\
		F_j &= \{t_{j+\ell+1}, \cdots, t_k\}.
	\end{split}
\end{equation}

Let a sequence of operations $\mathbf{C}_{j+\ell:j}$, with $k > j+\ell$ act on the memory block of the process. For the moment, while discussing the basic properties of Markov order in quantum processes, we omit the $j$ and $\ell$. However, these will become important when propagating processes with a Markov order assumption. Thus, $\mathbf{C}_{j+\ell:j}$ will be expressed as $\mathbf{C}_{M}$ henceforth. Let $\{\mathbf{B}_{M}^{\vec{\mu}}\}$ be a minimal \acs{IC} basis for these times, which includes $\mathbf{C}_{M}$, and let $\{\mathbf{\Delta}_{M}^{\vec{\mu}}\}$ be its dual set, as defined in Chapter~\ref{chap:PTT}. The conditional process is given by
\begin{equation}
	\Upsilon_{FP}^{(\mathbf{C}_{M})} = \text{Tr}_{M}\left[\Upsilon_{FMP} \mathbf{C}_{M}^\text{T}\right],
\end{equation}
where $\Upsilon_X$ is the process tensor across the legs given by the set(s) $X$. If the past and the future are independent in this conditional process, then it can be written as
\begin{equation}
	\label{eq:conditional-independence}
	\Upsilon_{FP}^{(\mathbf{C}_{M})} = \Upsilon_{F}^{(\mathbf{C}_{M})} \otimes \Upsilon_{P}^{(\mathbf{C}_{M})},
\end{equation}
where
\begin{equation}
	\Upsilon_{X}^{(\mathbf{C}_{M})} = \text{Tr}_{M\overline{X}}\left[\Upsilon_{FMP} \mathbf{C}_{M}^\text{T}\right] \quad X\in\{F,P\},
\end{equation}
recalling that the overline indicates set complement.
Note that the condensed language used here is identical to the description used in Equation~\eqref{eq:PToutput}. If Equation \eqref{eq:conditional-independence} holds for all elements of $\{\mathbf{B}_{M}^{\vec{\mu}}\}$, then the process, by construction, can be written as
\begin{equation}
	\label{eq:cmo-pt}
	\Upsilon_{FMP} = \sum_{\vec{\mu}} \Upsilon_{F}^{(\mathbf{B}_{M}^{\vec{\mu}})} \otimes \mathbf{\Delta}_{M}^{\vec{\mu}} \otimes \Upsilon_{P}^{(\mathbf{B}_{M}^{\vec{\mu}})}.
\end{equation}

A fact of practical importance is that for all sequences of operations $\mathbf{A}_{M}\not\in \{\mathbf{B}_{M}^{\vec{\mu}}\}$, Equation~\eqref{eq:conditional-independence} cannot hold. This is because $\{\mathbf{B}_{M}^{\vec{\mu}}\}$ is informationally complete, meaning that some operation sequence from outside the set can be expressed as a linear combination
\begin{equation}
	\mathbf{A}_{M} = \sum_{\vec{\nu}} \alpha_{\vec{\nu}}\mathbf{B}_{M}^{\vec{\nu}}.
\end{equation}
Contracting this operation into the process then yields
\begin{equation}
	\begin{split}
		\Upsilon_{FMP}^{(\mathbf{A}_{M})} &= \text{Tr}_{M}\left[\Upsilon_{FMP}  \mathbf{A}_{M}^\text{T} \right]\\
		&= \text{Tr}_{M}\left[\Upsilon_{FMP} \left(\sum_{\vec{\nu}} \alpha_{\vec{\nu}}\mathbf{B}_{M}^{\vec{\nu}\text{T}}\right)\right]\\
		&= \sum_{\vec{\nu}} \alpha_{\vec{\nu}}\Upsilon_{F}^{(\mathbf{B}_{M}^{\vec{\nu}})}\otimes \Upsilon_{P}^{(\mathbf{B}_{M}^{\vec{\nu}})},
	\end{split}
\end{equation}
which is no longer a product state, and thus the future and the past are \emph{separable}, but not completely uncorrelated. \par
The complexity of characterising a process grows exponentially in $\ell$; we would prefer to drop the instrument-specific component, and employ a generic conditional Markov order model. Explicitly, in this model, we truncate all conditional future-past correlations, treating the conditional state as a product. i.e.,
\begin{equation}
	\label{eq:cmo-approx}
	\Upsilon_{FMP}^{(\mathbf{A}_{M})} \approx \Upsilon_{F}^{(\mathbf{A}_{M})} \otimes \Upsilon_{P}^{(\mathbf{A}_{M})}.
\end{equation}
The cost, or approximation, in doing so will be determined by the actual memory strength of the process over different times. One meaningful measure of this is the quantum mutual information (QMI) of the conditional state, the LHS of Equation~\eqref{eq:cmo-approx}. Of course, this information is inaccessible in our truncated characterisation. Instead, we continue to use the reconstruction fidelity, and experimentally estimate this model error in the ability of each Markov order to predict the behaviour of actual sequences of random unitaries~\cite{taranto3}.

\begin{figure*}[ht!]
	\centering
	\includegraphics[width=0.85\linewidth]{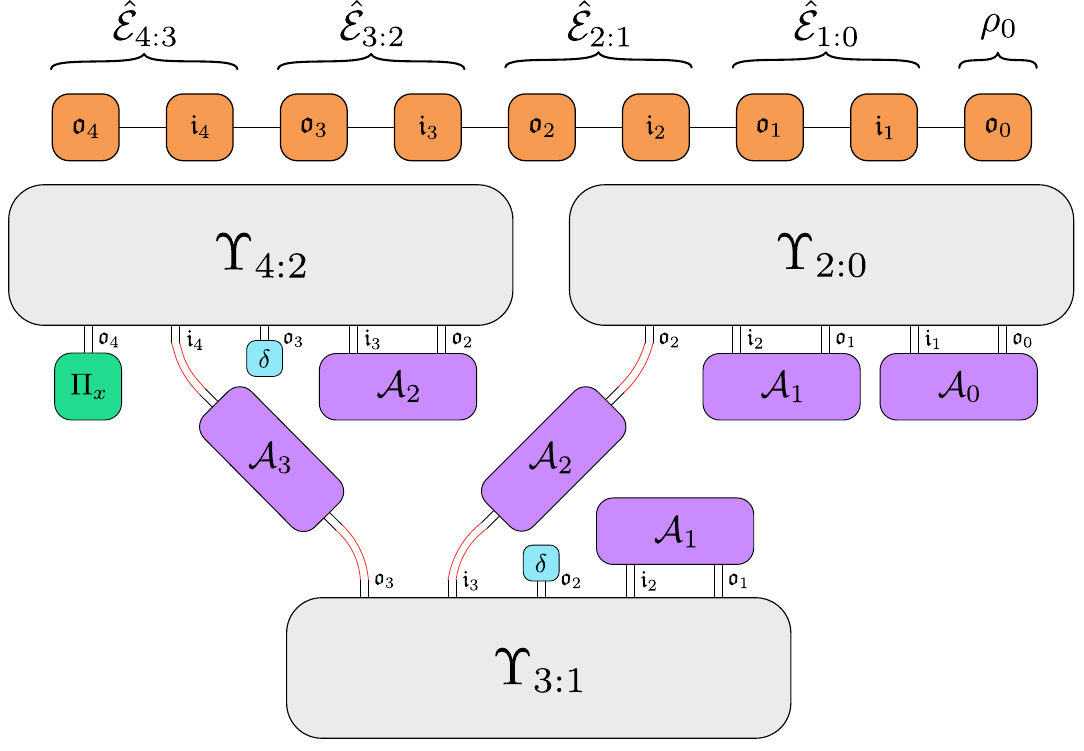}
	\caption[A contraction strategy for mapping multi-time gate sequences using a conditional Markov order ansatz ]{A contraction strategy for mapping multi-time gate sequences using a conditional Markov order ansatz. Here, we show how a four-step process can be modelled by two-step memory process tensors. The memory process tensors are stitched together by first contracting the relevant operations to their times to account for correlations. After tracing over the state output (denoted by $\delta$), the conditionally independent parts can be treated as tensor products, and thus stitched together with the latest common operation, where the output of the earlier process tensor is mapped to the input of the later one.}
	\label{fig:stitching-cmo}
\end{figure*}
\subsubsection*{Stitching Together Finite Markov Order Processes}\label{se:MLPTTMO}

We turn now to our work in extending the concept of conditional quantum Markov order to a quantum circuit context, and tomographic characterisation. Here, we are interested not only in dividing up the process into a single past, memory, and future, but to do this for all times in the process so that each mapping is dynamically updated. Then, for each time, the previous $\ell - 1$ operations are taken into account. Here, $\ell$ is chosen based on an expectation of the memory structure, and can be updated if the model does not adequately fit. The reconstruction is performed by iterating through the above computation and at each time step, dividing the circuit up into past, memory and future. Correlations due to the memory are taken care of via contraction of the relevant operations, leaving the future and the past conditionally independent. This requires a tomographically reconstructed process tensor for the relevant memory steps. These memory process tensors are then stitched together by the overlapping operation's map of the earlier output state. As an example, we depict the procedure in Figure~\ref{fig:stitching-cmo}.\par 
Our goal is to predict $\rho_k(\mathbf{A}_{k-1:0})$ by making use of the Markov order structure. At the first reduced process tensor, we have no past $P_0 = \{ \emptyset \}$, and the memory is given by the first $\ell$ operations. That is, we have to construct the full process tensor $\Upsilon_{M_0} = \Upsilon_{\ell:0}$, which contains all the conditional states $\rho_j(\mathbf{A}_{j-1:0})$ in $M_0$, i.e., $j \le \ell$. To go beyond time $t_\ell$, we need the conditional state $\rho_\ell(\mathbf{A}_{\ell-1:0})$ given by Equation~\eqref{eq:PToutput} (see also the first line of Equation~\eqref{conditional-PTs}): contracting $\Upsilon_{M_0}$ with $\mathbf{A}_{\ell-1:0}$. Importantly, this is the state propagated along with the sequences of operations. For this reason, the first memory process tensor is the only one for which the output state is not traced over, since we are not tracing over any alternative pasts. This means that it contracts one more local operation than the remainder. 

To get state $\rho_{\ell+1}(\mathbf{A}_{\ell:0})$, we move one step forward
with $P_1, \ M_1, \ F_{1}$. The relevant information is stored in $\Upsilon_{M_0}$ and $\Upsilon_{M_1} = \Upsilon_{\ell+1:1}$. For this process (and for all intermediate blocks in the process), there are three considerations: first, we must account for the memory through its action on the sequence $\mathbf{A}_{\ell-1:1}$ across $\Upsilon_{M_1}$ and trace over its output index at time $t_\ell$ (see the second line of Equation~\eqref{conditional-PTs}) since this state corresponds to a different, fixed past. Finally, the operation $\mathcal{A}_{\ell}$ connects $\Upsilon_{M_0}$ and $\Upsilon_{M_1}$ by mapping the state $\rho_{\ell}(\mathbf{A}_{\ell-1:0})$ to time $t_{\ell+1}$ since, as per our conditional Markov order assumption, once $M_1$ is accounted for, $F_1$ and $P_1$ are independent, i.e., their dynamics can be treated as a tensor product. See Figure~\ref{fig:stitching-cmo} for a graphical tensor network depiction. Note the distinction between here and Figure~\ref{fig:PTT-explanation}d. For a full four step process tensor, estimating a single expectation value involves contracting a tensor of matrix size $512\times 512$. With $\ell=2$ conditional Markov order, however, these require only three tensors with matrix size $32\times 32$.\par 
Following this recipe, we proceed forward in single steps, generating blocks of $P_{j}, \ M_{j}, \ F_{j}$ until we reach time $t_k$ at which point the final state may be read out. For clarity, the sequence of conditional memory process tensor states is given by:
\begin{equation}
	\label{conditional-PTs}
	\begin{split}
		&\rho_\ell(\mathbf{A}_{\ell-1:0}) = \text{Tr}_{\overline{\mathfrak{o}}_{\ell}}\left[\Upsilon_{M_0} \bigotimes_{i=0}^{\ell-1}\mathcal{A}_i^\text{T}\right],\\
		&\Upsilon_{j}^{(\mathbf{A}_{j-2:j-\ell})} := \text{Tr}_{\overline{j}}\left[\Upsilon_{M_{j-\ell}} \bigotimes_{i=j-\ell}^{j-2}\mathcal{A}_i^\text{T} \right].
	\end{split}
\end{equation}
The conditional state $\rho_\ell(\mathbf{A}_{\ell-1:0})$ has the free index $\mathfrak{o}_\ell$, corresponding to its output state. All others $\Upsilon_{j}^{(\mathbf{A}_{j-2:j-\ell})}$ have free indices $\mathfrak{i}_j$ and $\mathfrak{o}_j$ which, respectively, are contracted with the output and input legs of the operation $\mathcal{A}_j$ and $\mathcal{A}_{j+1}$. These are the operations which stitch together the different conditional memory process tensors, where the conditional independence means that the state can be mapped as though it were a tensor product. Finally, the last output leg $\mathfrak{o}_k$ is read out by some \acs{POVM}. \par 
We condense this $k$-step Markov order $\ell$ process in the object $\mathbf{\Upsilon}_{k:0}^\ell:= \{\Upsilon_{M_{k-\ell}}, \Upsilon_{M_{k-\ell-1}},\cdots,\Upsilon_{M_0}\}$. That is, the final state is defined by the collective action of each $\Upsilon_{M_j}$ as
\begin{align}
	\label{CMO-action}
	&\rho_k(\mathbf{A}_{k-1:0}) \approx \mathbf{\Upsilon}_{k:0}^\ell \ast \mathbf{A}_{k-1:0} \\ \notag
	& \qquad := \text{Tr}_{\overline{\mathfrak{o}}_k}\left[ \Upsilon_{k}^{(\mathbf{A}_{k-2:k-\ell})} \left(\bigotimes_{j=\ell}^{k-1}\Upsilon_{j}^{(\mathbf{A}_{j-2:j-\ell})} \mathcal{\mathcal{A}}_j^\text{T}\right)\right].
\end{align}
Note that since the same control operation may contract into multiple different memory process tensors, this action is no longer linear in $\mathbf{A}_{k-1:0}$.\par 
Recalling our earlier depiction of the process tensor in Figure~\ref{fig:PTT-explanation}, the dynamics can be described as a collection of correlated \acs{CPTP} maps $\{\hat{\mathcal{E}}_{j:j-1}\}$. In the \acs{CJI} picture, past operations are equivalently seen as measurements on these earlier states. Thus, the $\Upsilon_{j}^{(\mathbf{A}_{j-2:j-\ell})}$ are exactly the conditional memory states $\hat{\mathcal{E}}_{j:j-1}^{(\mathbf{A}_{j-2:j-\ell})}$. With correlations accounted for, they can be treated locally in time. Any process may be written exactly as a sequence of conditional \acs{CPTP} maps, but in full generality they depend on the whole past. Here, they only depend on the memory. The difference in complexity of characterisation is $\mathcal{O}(N^k)$ vs. $\mathcal{O}(N^\ell)$.
Putting it all together we have an equivalent form of Equation~\eqref{CMO-action}
\begin{align}\notag
	\rho_k(\mathbf{A}_{k-1:0}) \approx & \mathcal{E}_{k:k-1}^{(\mathbf{A}_{k-2:k-\ell})} \circ \mathcal{A}_{k-1} \circ \cdots \circ \mathcal{E}_{\ell+2:\ell+1}^{(\mathbf{A}_{\ell:2})} \circ \mathcal{A}_{\ell+1} \circ  \notag
	\mathcal{E}_{\ell+1:\ell}^{(\mathbf{A}_{\ell-1:1})} \circ \mathcal{A}_{\ell} [\rho_{\ell}(\mathbf{A}_{\ell-1:0})] \\
	\mbox{with} \quad &\rho_{\ell}(\mathbf{A}_{\ell-1:0}) = \mathcal{T}_{M_0}[\mathbf{A}_{\ell-1:0}].
\end{align}
To summarise, the process with a conditional Markov order $\ell$ ansatz $\mathbf{\Upsilon}_{k:0}^{\ell}$ is represented by a collection of memory process tensors $\left\{\Upsilon_{k:k-\ell}, \Upsilon_{k-1:k-\ell-1}, \cdots, \Upsilon_{\ell+1:1}, \Upsilon_{\ell:0}\right\}$. As discussed, it cannot be represented generically by a quantum state, but this collection of memory process tensors defines its action on a sequence of $k$ operations.
The contraction strategy for a series of control operations (with $k=4$ and $\ell=2$) is shown in Figure~\ref{fig:stitching-cmo}. In short:
\begin{enumerate}
	\item Contract the first $\ell$ operations into $\Upsilon_{\ell:0}$, producing the output state at time $t_\ell$,
	\item Contract operations $2$ to $\ell$ into $\Upsilon_{\ell+1:1}$,
	\item Trace over the output index of $\Upsilon_{\ell+1:1}$ at time $t_\ell$ (since this is not representative of the actual state of the system subject to all operations),
	\item Taking the $(\ell+1)$th operation to be conditionally independent of the first, this can be applied across the tensor product of the two process tensors into the indices for the output state at time $t_\ell$ and the input for time $t_{\ell+1}$,
	\item Repeat this pattern for the next $k - \ell - 2$ process tensors,
	\item Read out the final state at time $t_k$.
\end{enumerate}
The action of the reduced process tensors under a finite conditional Markov order model is best posed as a tensor network contraction, so that optimisation tasks can be performed quickly. We show this here. Note that the following extravagant working is equivalent to Figure \ref{fig:stitching-cmo}, but we write it out in full generality in order to make the indices explicit and replication more straightforward. First, the conditional reduced states of each of the process tensors must be taken by contracting the relevant control operations (including final measurement) into the process tensors. These are:
\begin{equation}
    \begin{split}
        &\left(\Upsilon_{\ell:0}^{(\mathbf{B}_{\ell-1:0}^{\vec{\mu}})}\right)_{k_{\mathfrak{o}_\ell}}^{b_{\mathfrak{o}_\ell}} =  \sum_{\substack{
			k_{\mathfrak{o}_{0:\ell-2}},k_{\mathfrak{i}_{1:\ell-1}}\\
			b_{\mathfrak{o}_{0:\ell-2}},b_{\mathfrak{i}_{1:\ell-1}}}}
			\left(\Upsilon_{\ell:0}\right)_{k_{\mathfrak{o}_{0:\ell}},k_{\mathfrak{i}_{1:\ell}}}^{b_{\mathfrak{o}_{0:\ell}},b_{\mathfrak{i}_{1:\ell}}}(\mathcal{B}_{\ell-1}^{\mu_{\ell-1}})_{k_{\mathfrak{i}_{\ell}},k_{\mathfrak{o}_{\ell-1}}}^{b_{\mathfrak{i}_{\ell}},b_{\mathfrak{o}_{\ell-1}}}\cdots (\mathcal{B}_{0}^{\mu_{0}})_{k_{\mathfrak{i}_{1}},k_{\mathfrak{o}_{0}}}^{b_{\mathfrak{i}_{1}},b_{o_{0}}},\\
        &\left(\Upsilon_{j:j-\ell}^{(\mathbf{B}_{j-2:j-\ell}^{\vec{\mu}})}\right)_{k_{\mathfrak{o}_j},k_{\mathfrak{i}_{j}}}^{b_{\mathfrak{o}_j},b_{\mathfrak{i}_{j}}} = 
		\hspace{-0.25cm}
		\sum_{\substack{k_{\mathfrak{o}_{j-\ell:j-\ell-2}},k_{\mathfrak{i}_{j-\ell+1:j-1}}\\
		b_{\mathfrak{o}_{j-\ell:j-\ell-2}},b_{\mathfrak{i}_{j-\ell+1:j-1}}}}
		\hspace{-0.75cm}
		\left(\Upsilon_{j:j-\ell}\right)_{k_{\mathfrak{o}_{j:j-\ell}},k_{\mathfrak{i}_{j:j-\ell+1}}}
		^{b_{\mathfrak{o}_{j:j-\ell}},b_{\mathfrak{i}_{j:j-\ell+1}}}
		(\mathcal{B}_{j-2}^{\mu_{j-2}})_{k_{\mathfrak{i}_{\ell-1}},k_{\mathfrak{o}_{\ell-2}}}^{b_{\mathfrak{i}_{\ell-1}},b_{\mathfrak{o}_{\ell-2}}}\cdots (\mathcal{B}_{j-\ell}^{\mu_{j-\ell}})_{k_{\mathfrak{i}_{j-\ell+1}},k_{\mathfrak{o}_{j-\ell}}}^{b_{\mathfrak{i}_{j-\ell+1}},b_{\mathfrak{o}_{j-\ell}}},\\
        &\left(\Upsilon_{k:k-\ell}^{(\mathbf{B}_{k-2:k-\ell}^{\vec{\mu}},\Pi_i)}\right)_{k_{\mathfrak{i}_k}}^{b_{\mathfrak{i}_k}} = 
		\hspace{-0.25cm}
		\sum_{\substack{ k_{\mathfrak{o}_{k-\ell:k}}, k_{\mathfrak{i}_{k-\ell+1:k-1}} \\
		b_{\mathfrak{o}_{k-\ell:k}}, b_{\mathfrak{i}_{k-\ell+1:k-1}}
		}}
		\hspace{-0.75cm}
		\left(\Upsilon_{k:k-\ell}\right)_{k_{\mathfrak{o}_{k-\ell:k}},k_{\mathfrak{i}_{k-\ell-\ell+1:k}}}
		^{b_{\mathfrak{o}_{k-\ell:k}},b_{\mathfrak{i}_{k-\ell-\ell+1:k}}}
		(\mathcal{B}_{k-2}^{\mu_{k-2}})_{k_{\mathfrak{i}_{\ell-1}},k_{\mathfrak{o}_{\ell-2}}}^{b_{\mathfrak{i}_{\ell-1}},b_{\mathfrak{o}_{\ell-2}}}\cdots (\mathcal{B}_{k-\ell}^{\mu_{k-\ell}})_{k_{\mathfrak{i}_{k-\ell+1}},k_{\mathfrak{o}_{k-\ell}}}^{b_{\mathfrak{i}_{k-\ell+1}},b_{\mathfrak{o}_{k-\ell}}}.
    \end{split}
\end{equation}
Then, the tensor of predicted probabilities $p_{i,\vec{\mu}}$ is obtained by stitching each conditional process tensor together with the overlapping control operations. That is:
\begin{equation}
\begin{split}
    &\sum_{\substack{k_{\mathfrak{o}_{\ell:k-1}}k_{\mathfrak{i}_{\ell+1:k}}\\ b_{\mathfrak{o}_{\ell:k-1}}b_{\mathfrak{i}_{\ell+1:k}}}}
    \left(\Upsilon_{k:k-\ell}^{(\mathbf{B}_{k-2:k-\ell}^{\vec{\mu}},\Pi_i)}\right)_{k_{\mathfrak{i}_k}}^{b_{\mathfrak{i}_k}}(\mathcal{B}_{k-1}^{\mu_{k-1}})_{k_{\mathfrak{i}_k},k_{\mathfrak{o}_{k-1}}}^{b_{\mathfrak{i}_k},b_{\mathfrak{o}_{k-1}}}\\
    &\left( \prod_{j=\ell+1}^{k-1}\left(\Upsilon_{j:j-\ell}^{(\mathbf{B}_{j-2:j-\ell}^{\vec{\mu}})}\right)_{k_{\mathfrak{o}_j},k_{\mathfrak{i}_{j}}}^{b_{\mathfrak{o}_j},b_{\mathfrak{i}_{j}}}(\mathcal{B}_{j-1}^{\mu_{j-1}})_{k_{\mathfrak{i}_j},k_{\mathfrak{o}_{j-1}}}^{b_{\mathfrak{i}_j},b_{\mathfrak{o}_{j-1}}}\right)(\mathcal{B}_{\ell}^{\mu_{\ell}})_{k_{\mathfrak{i}_{\ell+1}},k_{\mathfrak{o}_{\ell}}}^{b_{\mathfrak{i}_{\ell+1}},b_{\mathfrak{o}_{\ell}}}\left(\Upsilon_{\ell:0}^{(\mathbf{B}_{\ell-1:0}^{\vec{\mu}})}\right)_{k_{\mathfrak{o}_\ell}}^{b_{\mathfrak{o}_\ell}}.
    \end{split}
\end{equation}
Which is precisely the generalisation of the strategy presented in Figure \ref{fig:stitching-cmo}. When evaluated from left to right, this can be performed efficiently since every contraction is a rank-2 tensor with a rank-4 tensor. Note that in this instance, the index vector $\vec{\mu}$ does not run from $(0,0,\cdots,0)$ to $(d_S^4, d_S^4,\cdots,d_S^4)$ but rather for each block of memory, it contains all $d_S^{4\ell}$ combinations of basis elements, with all other operations fixed at $\mu_0$. There are therefore $(k-\ell+1)\cdot d_S^{4\ell}$ values taken by $\vec{\mu}$.
Although there are a linear number of model parameters here, we can employ the techniques of classical shadow tomography in Chapter~\ref{chap:MTP} to simultaneously estimate these non-overlapping marginals. The upshot is that the number of quantum experiments required to run scales only logarithmically with the number of time steps. This brings down the scaling of the protocol substantially further.


\subsubsection*{Circuits for $\Upsilon_{M_j}$}
With a framework established for constructing and operating a finite Markov order ansatz, we now explicitly detail how to tomographically reconstruct this model on a real device. For concreteness, we present it in this complete form, where the number of circuits scales linearly in $k$. But, as we have just detailed, applying classical shadows presents a completely different set of circuits that can be used to achieve logarithmic scaling overall.
\begin{figure}[ht]
	\centering
	\includegraphics[width=0.8\linewidth]{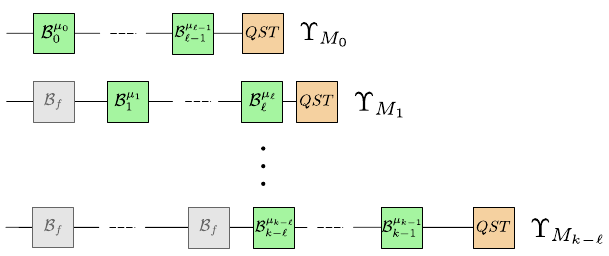}
	\caption[Circuits to construct a process tensor with conditional Markov order $\ell$ ]{Circuits to construct a process tensor with conditional Markov order $\ell$. For a $k$-step process, $\mathbf{\Upsilon}_{k:0}^{\ell}$, there are $k-\ell+1$ memory process tensors that need constructing -- each circuit represents the estimation of each of these, by varying all combinations of all indices from $i_0$ to $i_{\ell-1}$. In the other gate positions, a fixed operation $\mathcal{B}_f$ is applied.}
	\label{fig:cmo_circs}
\end{figure}
In order to estimate $\mathbf{\Upsilon}_{k:0}^\ell$, we must estimate each of the memory block process tensors. Recall that each $\Upsilon_{M_j} := \Upsilon_{\ell+j:j}$ is equivalent to $\Upsilon_{\ell+j:0}$ with the first $j$ times projected out onto some series of interventions. In order to experimentally reconstruct each $\Upsilon_{M_j}$, then, it suffices to fix the first $j$ operations in the circuit, and then perform a complete basis of operations in each position from $t_j$ to $t_{\ell+j}$ and estimate the associated $\ell$-step process tensor. The circuits required for each of these are illustrated in Figure~\ref{fig:cmo_circs}, with the fixed operation labelled $\mathcal{B}_f$. As well as sufficiently describing a Markov order $\ell$ model, these circuits contain all the information required for any lower-order Markov model if it is a full process tensor. Under the unitary-only restriction, there will be a few extra experiments required for any smaller memory blocks terminating earlier than $t_{\ell}$. We note here also that the maximum likelihood procedure of the previous Chapter is necessary for conditional Markov order models if the set of instruments is restricted to the unitaries. This is because Equation~\eqref{CMO-action} requires local partial traces, but unitary gates are equivalent to entangled measurements in the Choi picture. Hence, if a linear inversion restricted process tensor is constructed, the partial traces will not be well-defined.
\par 

In the action of $\mathbf{\Upsilon}_{k:0}^\ell$, the state generated by each $\Upsilon_{M_j}$ is traced over for all $j>0$. Consequently, the fixed operations that precede $M_j$ in the reconstruction circuits should, in principle, not affect the final outcome $\rho_k$. Since, however, the CMO ansatz is an approximation to the true dynamics, then the fixed past operations will, in practice, affect this approximation. 
In the generic case, there is no reason to suspect any operation will put forth a better or worse approximation, hence we arbitrarily set this operation to be the first element of the basis set each time. This choice may require closer attention in practical situations.

With this, we have described how to adaptively characterise a process with quantum and classical requirements only as large as the complexity of the noise (or, up to the error the experimenter is willing to tolerate). Since a process characterisation may be verified through the reconstruction fidelity, the best approach to this is to progressively build up and verify a more complex model until the desired precision has been reached. This will depend on the intended applications of the characterisation. We believe that this is the first procedure to methodically characterise finite Markov order quantum processes. We expect this to be greatly useful in stemming the effects of both correlated and uncorrelated noise on \acs{NISQ} devices, where the open dynamics is already clean enough so as to be mostly -- but not strongly -- non-Markovian. \par

\subsubsection*{Demonstration on NISQ Devices}

\begin{figure}[h]
	\centering
	\includegraphics[width=0.6\linewidth]{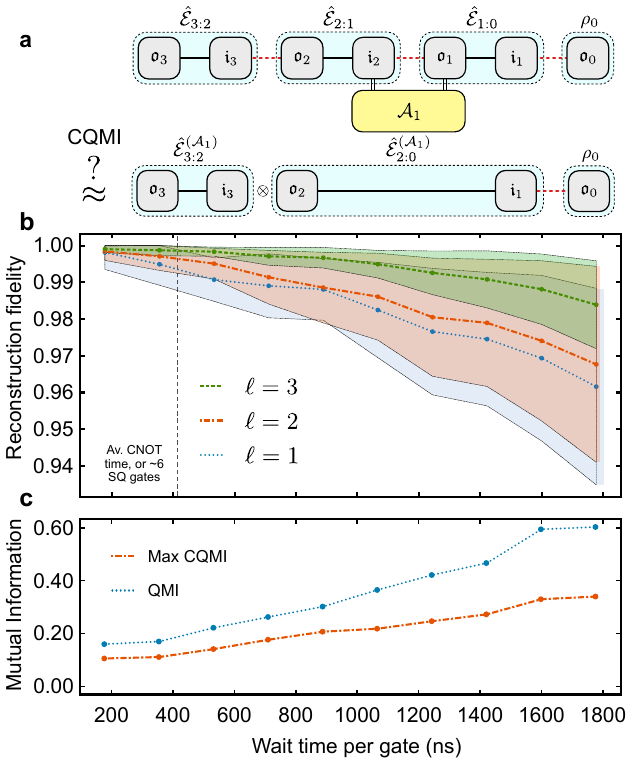}
	\caption[Build-up of temporal correlations with increasing interaction time on IBM Quantum devices]{Build-up of temporal correlations with increasing interaction time. \textbf{a} The CQMI quantifies the error in truncating correlations beyond $\ell=2$ when conditioned on intermediate operations. \textbf{b} A four step process is considered with increasing wait time after each gate. Using Markov orders $\ell=1$, $\ell=2$, and $\ell=3$, we quantify how each model reconstructs 100 random unitary sequences. Points indicate average fidelity, shaded regions indicate standard deviation. \textbf{c} \acs{QMI} and CQMI both increase with interaction time, respectively bounding the breakdown of $\ell=1$ and $\ell=2$ models.}
	\label{fig:cmo_reconstruction}
\end{figure}

We can combine some of the ideas introduced to give a more fine-grained measure of non-Markovian memory on quantum devices, both in terms of its length and its strength. Specifically, these simplified models of the process can be employed either to streamline control of the quantum system, or they may be used as a diagnostic tool by observing how well different restrictions describe the dynamics. This also validates our method of reconstructing processes with conditional Markov order efficiently.
To estimate non-Markovianity we construct increasingly complex models by accounting for increasing memory, and quantify how well they describe observed device dynamics under the measure of reconstruction fidelity. This measure is computationally convenient, scaling linearly in time-steps; has an immediately available interpretation; and may be performed up to acceptable approximation, or where costs become prohibitive. The breakdown of conditional Markov order is bounded by the maximum conditional quantum mutual information (CQMI) as described in Ref.~\cite{taranto3}. CQMI is taken for a three-step process tensor to be
\begin{equation}
	\max_{\mathcal{A}_1} S\left[\Upsilon_{3:0}^{(\mathcal{A}_1)} \mid\mid \hat{\mathcal{E}}_{3:2}^{(\mathcal{A}_1)}\otimes \Upsilon_{2:0}^{(\mathcal{A}_1)}\right],
\end{equation}
where $S[\rho\mid\mid\sigma] := \text{Tr}[\rho(\log\rho - \log\sigma)]$ is the von Neumann relative entropy. The conditional Markov order approximation is illustrated in Figure~\ref{fig:cmo_reconstruction}a.

We performed this procedure as an example on \emph{ibmq\_guadalupe} to observe non-Markovianity as a function of time. For a four step process, we constructed process tensors $\mathbf{\Upsilon}_{4:0}^\ell$ for $\ell\in \{1,2,3\}$. With a $\ket{+}$ state neighbour, the duration of each step was varied across ten different times, ranging from 180 ns up to 1800 ns. For each value of $t$, we executed 100 sequences of random unitaries $\{\mathbf{U}_{3:0}^i\}$ followed by state reconstruction. Then the action of $\mathbf{\Upsilon}_{4:0}^{\ell}$ on $ \mathbf{U}_{3:0}^i$ is used to predict the resulting states. The fidelity between predicted state and actual state is computed, and the distribution for each data point shown in Figure~\ref{fig:cmo_reconstruction}. We see here the build-up of memory effects; the timescales constitute relatively short-depth effective circuits, meaning these temporal correlations are likely to accumulate across practical circuits. Interestingly, the $\ell=3$ model performs significantly better than the other two, whereas $\ell=2$ is only marginally better at predicting the dynamics than $\ell=1$ \footnote{$\ell = 2$ and $\ell=1$ models can be constructed from subsets of the $\ell=3$ data, however on their own they minimally require $3\times 300 = 900$ and $4\times 30 = 120$ circuits per time, respectively.}. This suggests that most of the memory effects in these devices are higher order -- they persist across multiple times. This observation is substantiated by computing the QMI, as well as finding the operation which maximises the conditional \acs{QMI} for the three step process tensors. The ability to compute these measures comes from our compressed sensing approach to \acs{PTT}, detailed in Chapter~\ref{chap:MTP}. The increase of both of these non-Markovian measures is shown in Figure~\ref{fig:cmo_reconstruction}c. Because of the cumulative build-up, for long-time dynamical processes in real situations, mitigating the effects of these correlations would require either decoupling early, or fine-graining the process into many more time steps. These memory effects manifest themselves over a time frame of only a few CNOT gates, indicating that non-Markovian dynamics are likely a significant class of noise in regular circuits. 

Markovianity breakdown has been previously quantified in terms of model violation in gate sets, or in the loss of \acs{CP} divisibility~\cite{RBK2017,white-POST,PhysRevA.83.052128}. However these approaches only coarsely diagnose temporal correlations, and are not generic to the process. We have presented a systematic framework by which different levels of finite conditional Markov order may be tested on quantum devices with both a rigorous foundation and practical interpretation. 

\subsection{Spatiotemporal Quantum Markov Order}
\label{sec:stmo}
We have thus far been restricted in our analyses to many-time processes: quantum stochastic processes that represent a single system as it evolves in time. This conveys no sense of modularity. This is to say, real correlations spread out spatiotemporally, and if we coarse grain the physics into a \emph{single} system, then we lose access to these internal mechanisms. Indeed, the interplay between different physical subsystems is crucial for understanding and developing devices that can withstand the effects of environmental noise and control issues. Since quantum devices are already well into the multi-qubit regime, we thus turn to understanding multi-qubit quantum stochastic processes. Direct spatial correlations can emerge via two channels: a coherent interaction between two different subsystems, as described by an always-on Hamiltonian; or a common bath acting as a simultaneous common-cause variable for the two subsystems, such as superdecoherence processes. Spatio\emph{temporal} correlations, however, emerge only from the latter. Correlations are generated between instruments outcomes on different qubits at different times as information is mutually exchanged between qubits and the same shared environment. 
\begin{example}[Spatiotemporal correlations]
	\examplecontent{
		Suppose a qubit $q_1$ interacts with some defect in the environment $E$. $q_1$ is measured at time $t_1$ and the outcome $x_1$ is stored; the environment is now transformed when conditioned on that outcome. $q_1$ is then deterministically reset to the $\ket{0}$ state. Between $t_1$ and $t_2$, $E$ interacts with $q_2$. Because the state of $E$ depends on the outcome of $q_1$ at $t_1$, the interaction between $q_2$ and $E$ depends also on $x_1$. Consequently, when $q_2$ is measured at $t_2$, the outcome $x_2$ is correlated with $x_1$, even though $q_1$ was deterministically re-prepared at $t_1$. 
	}
\end{example}


Chapter~\ref{chap:PTT} provided all the necessary tools to perform spatiotemporal tomography in principle. However, if one wished to perform complete \acs{PTT} on more than one qubit, they would quickly realise that the fully generic situation is prohibitive. Even assuming nothing about the experimental capabilities, \acs{PTT} requires $\mathcal{O}(d_S^{4k+2})$ different circuits, which is so expensive that it is not worth even discussing $k>1$ when $d_S\geq 4$ in practice.
But experimental constraints even bring their own issues: a restricted process tensor across two qubits requires implementation of arbitrary $SU(4)$ gates which can have complicated decompositions. For larger numbers of qubits, the challenge is even greater. We must therefore broaden our approach to compressing the representations of dynamics. Even multi-time conditional Markov order compression does not suffice to make a multi-qubit experiment feasible. We must subdivide the memory block amongst qubits.
In this section we define and investigate the notion of spatiotemporal Markov order: quantum stochastic processes for which correlations between events decay in both space and time. This obeys many of the same principles as the temporal Markov order we have seen so far, but we must appropriately generalise the notions of past, memory, and future. Where appropriate, we will lean on techniques and exposition from the previous Section and emphasise only the key differences in translating to the spatiotemporal setting. 

One key difference in introducing a spatial dimension is that we no longer have a causal ordering that gives a direction to correlations: past to future. We previously leveraged this fact to feed past conditional outputs into future dynamics. However, if we wish to treat multiple qubits, then this approach will no longer be possible. Instead, we will employ recovery maps to approximate a dynamical map from its overlapping marginals. We will introduce these recovery maps in the context of well-studied approximate quantum Markov chains; motivate the idea of a causal cone in which one has non-Markovian memory; and finally derive a method from which one can reconstruct correlated spatiotemporal dynamics with a finite memory ansatz.



\subsubsection*{Quantum Markov Chains}


The study of classical Markov chains dates back to Andrei Markov's seminal work on the structured dependence of random variables~\cite{chung1967markov}. We have seen this notion extensively throughout this thesis so far. 
Given three classical random variables $X,Y,$ and $Z$ with joint probability distribution $\mathbb{P}_{XYZ}$, the three variables form a \emph{Markov chain} if $X$ and $Z$ are independent when conditioned on $Y$. One can then say that $\mathbb{P}_{XYZ} = \mathbb{P}_{XY}\mathbb{P}_{Z\mid Y}$. Alternatively, this is the statement that there exists a stochastic matrix $\Gamma_{Z\mid Y}$ such that $\mathbb{P}_{XYZ} = \mathbb{P}_{XY}\Gamma_{Z\mid Y}$. Mathematically, this means that the entire joint probability distribution can be reconstructed using only joint information about $XY$ and $YZ$. 

The natural quantum generalisation to a Markov chain was first characterised by Petz in Refs.~\cite{petz1986sufficient,petz2003monotonicity}. Approximations to these objects were then rigorously investigated by Fawzi, Renner, Sutter, among others~\cite{fawzi2015quantum,sutter2016universal,sutter2018approximate,gao2022complete}.
This satisfies the following definition. 
\begin{definition}
	A tripartite state $\rho_{ABC}$ on $A\otimes B \otimes C$ is called a \emph{quantum Markov chain} if there exists a recovery map $\mathscr{R}_{B\rightarrow BC}$ from $B$ to $B\otimes C$ such that 
	\begin{equation}
		\rho_{ABC} = (\mathcal{I}_A \otimes \mathscr{R}_{B\rightarrow BC}(\rho_{AB})).
	\end{equation}
\end{definition}
Here, $\mathscr{R}_{B\rightarrow BC}$ is a \acs{CPTP} map. The statement is then that the entire state may be recovered by acting on the middle subsystem. Moreover, in the same reference, Petz proved that quantum Markov chains are completely characterised by the entropic relation:
\begin{equation}
	\rho_{ABC} \text{ is a quantum Markov chain }\quad \iff \quad I(A:C\mid B) = 0,
\end{equation}
where 
\begin{equation}
	I(A:C\mid B) := S(AB) + S(BC) - S(ABC) - S(B)
\end{equation}
denotes the conditional quantum mutual information of the state. Petz also constructed a specific form for $\mathscr{R}_{B\rightarrow BC}$, known as the \emph{Petz recovery} or \emph{Petz transpose} map, in:
\begin{equation}
	\mathscr{T}_{B\rightarrow BC}(X_B) := \rho_{BC}^{\frac{1}{2}}\rho_B^{-\frac{1}{2}} X_B \rho_B^{-\frac{1}{2}}\rho_{BC}^{\frac{1}{2}}, 
\end{equation}
with inverse taken on the support of $\rho_B$ and identities implicit. Thus, not only may the state be reconstructed only by acting on the subsystem $B$, it is fully characterised by the two marginals $\rho_{AB}$ and $\rho_{BC}$ via 
\begin{equation}
	\rho_{ABC} = \mathscr{T}_{B\rightarrow BC}(\rho_{AB}).
\end{equation}

The above fully characterises quantum Markov chains. However, in the instance where
\begin{equation}
	I(A:C | B) =\varepsilon,\quad\text{for }\varepsilon >0,
\end{equation}
the state is called an \emph{approximate} quantum Markov chain. An interesting property of these classes of states is that unlike the classical case, it is known that approximate quantum Markov chains need not be $\varepsilon$-close to any exact quantum Markov chain (as measured by the trace distance).



Sutter introduced the rotated Petz recovery map in Ref.~\cite{sutter2018approximate}, defined as follows. 
\begin{equation}
	\mathscr{T}_{B\rightarrow BC}^{[t]}(X_B) = \rho_{BC}^{\frac{1+it}{2}}\rho_B^{-\frac{1+it}{2}} X_B \rho_B^{-\frac{1-it}{2}}\rho_{BC}^{\frac{1-it}{2}},\:\: t\in\mathbb{R}.
\end{equation}
This map satisfies the same properties as the original Petz recovery map. That is for $\rho_{ABC}$ satisfying $I(A:C\mid B) = 0$, we have $\rho_{ABC} = \mathscr{T}_{B\rightarrow BC}^{[t]}(\rho_{AB})$. Moreover, this can be used to define a sufficient criteria for approximate recoverability of approximate quantum Markov chains, as follows~\cite{sutter2016universal}. 
\begin{theorem}
	Let $\rho_{ABC}\in A\otimes B \otimes C$. Then
	\begin{equation}
		I(A:C\mid B)\rho \geq S_M(\rho_{ABC}\mid\mid \bar{\mathscr{T}}_{B\rightarrow BC}(\rho_{AB})),
	\end{equation}
	where the rotated Petz recovery map $\bar{\mathscr{T}}_{B\rightarrow BC}$ is defined as 
	\begin{equation}
		\label{eq:petz-universal}
		\bar{\mathscr{T}}_{B\rightarrow BC} = \int_{-\infty}^\infty \text{d}t \frac{\pi}{2}\left(\cosh(\pi t) + 1\right)^{-1} \mathscr{T}_{B\rightarrow BC}^{[t]}.
	\end{equation}
\end{theorem}

Note that $S_M$ here is no longer the quantum relative entropy, but rather the \emph{measured} quantum relative entropy~\cite{berta2017variational}, which is the supremum of relative entropy with measured inputs over all \acs{POVM}s:
\begin{equation}
	S_M(\rho\mid\mid\sigma) = \sup_{M\in\mathscr{M}} S(M(\rho) \mid\mid M(\sigma)),
\end{equation}
where $M(\cdot)$ takes the classical-quantum form $M(\rho) = \sum_x\Tr[E_x\rho] |x\rangle\!\langle x|$, where $E_x$ is a \acs{POVM}.
This relation provides a frame of reference for approximately reconstructing approximate quantum Markov chains. Reconstruction incurs an additive error as bounded by the conditional mutual information of the state.
We shall use this universal recovery map across \emph{spatial} quantum states to implement a more general framework of quantum Markov order reconstruction. First, let us define what we mean by spatiotemporal quantum Markov order.

\subsubsection*{Causal Cones}

In the single qubit setting, we had a single notion of Markov order. If an operation beyond the previous $\ell$ steps does not meaningfully update future statistics, then we can omit this from our description of the memory. Moving into the multi-qubit setting, we now have an expanded notion. If an event occurs outside the previous $\ell_t$ steps or the nearest $\ell_s$ neighbouring qubits and does not meaningfully update future statistics, then we will omit this from our description of the memory. 

Intuitively, this gives rise to a \emph{causal cone} -- a collection of events in space and time which are correlated with future outcomes. 
In reality, the cone will have a graduated decrease in the memory size, but in practice we truncate this at some level. 
We re-emphasise that these concepts take on meaning only for a prescribed set of dynamics. 
We consider a chain of qubits with open boundary conditions for simplicity, but the structure could be readily updated in accordance with, for example, the topology of a given device. 

Now that we are in the multi-qubit regime, to maintain an efficient characterisation it does not purely suffice to just have low non-Markovian memory. Indeed, a generic Markovian process across $n$ qubits might only scale linearly in the number of time steps, but if each map looks like a Haar random unitary, for example, then an efficient classical description of the process is equally hopeless. Thus, we must also consider the structure of spatial correlations in the dynamics and assume these to be low. 

The inclusion of spatial degrees of freedom complicates the matter in more ways than one. We cannot simply describe the dynamics of a single qubit as tracked by spatiotemporal events, because we need also to update the correlations of the surrounding qubits. Additionally, we need to incorporate purely spatial interactions between the different subsystems. 
The 2D process can be seen, for each time, and each qubit grouping, as being covered by a series of overlapping causal cones which act as the memory structure in a quantum Markov network which we now define. There naturally will be some freedom of choice in the ansatz here, about which we shall be explicit.

\begin{definition}
	Given a device with a register of qubits $\mathbf{Q} := \{q_1,q_2\cdots, q_n\}$, a \emph{causal cone} $\mathfrak{C}_{t_j}^{\vec{q}}$ is a memory structure ansatz for a quantum stochastic process at a single time $t_j$ across a set of qubits $\vec{q} = \{q_i, q_{i+1}, \cdots, q_{i+\ell_s}\}$. That is, it pertains to the dynamical map $\mathcal{E}_{j:j-1}^{\vec{q}}$ acting across qubits $\vec{q}$ at time $t_j$. 
	$\mathcal{C}_{t_j}^{\vec{q}}$ encompasses its front, $(t_j,\vec{q})$, as well as a set of qubit-time tuples $\{(t_{j-1}, \vec{q}^{(1)}), \cdots , (t_{j-\ell_t}, \vec{q}^{(\ell_t)})\}$ at which instruments $\{\mathcal{A}_j^{(q_i)}\}$ act. When conditioned on the instruments of $\mathfrak{C}_{t_j}^{\vec{q}}$, the marginal many-body dynamical map $\mathcal{E}_{j:j-1}^{\vec{q}}$ is \emph{independent} of all operations that occur on any qubit prior to time $t_{j-\ell_t}$. This implies a spatial locality $\ell_s$ for any direct interactions between qubits, and a finite temporal correlation $\ell_t$. 
\end{definition}

We call $\mathfrak{C}_{t_j}^{\vec{q}}$ a cone because for each step back, the set $\vec{q}^{(i)}$ gets larger. Assuming our 1D chain to be the geometric connectivity of a qubit device, it is supposed that geometrically non-local qubits are coupled to different baths to one another. Appealing to Lieb-Robinson arguments, it will take longer for information to reach further distances, thus motivating a conic shape. We note however that our analysis isn't specific to this shape, nor this 1D connectivity but rather come from general physical reasoning and simplicity of exposition. Each component could be modified to be made fit-for-purpose if necessary. For all discussions that follow, we take the cone to grow outwards by one qubit per step. Instruments will also all be local without loss of generality, since non-local instruments can be expressed as linear combinations of local ones. 

\begin{figure}[!t]
	\centering
	\includegraphics[width=\linewidth]{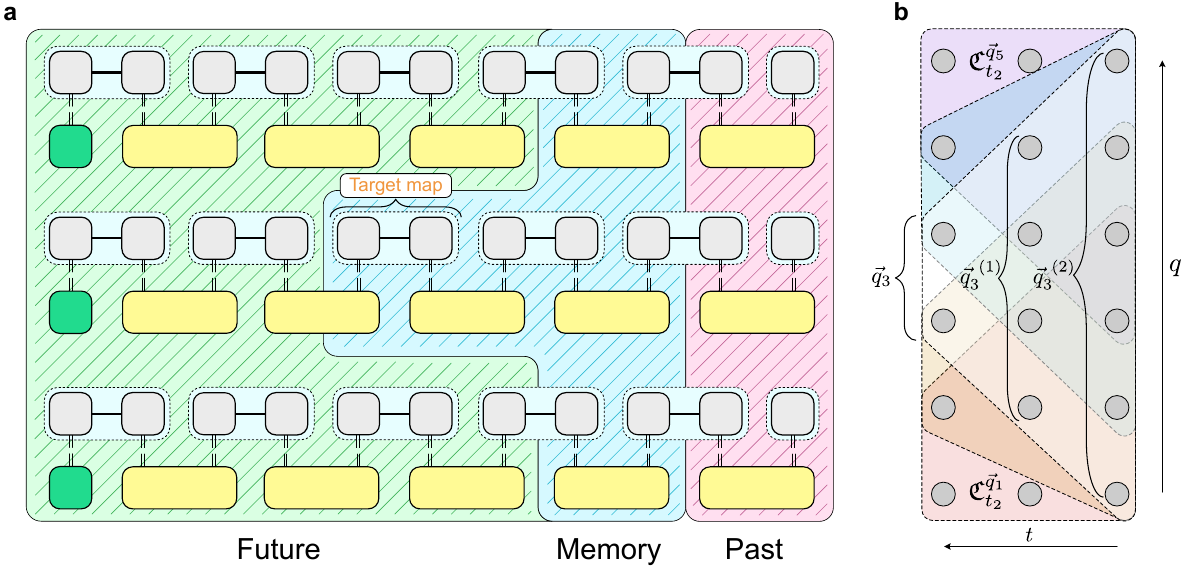}
	\caption[A depiction of a process with a spacetime Markov order ansatz ]{A depiction of a process with a spacetime Markov order ansatz. \textbf{a} For a target qubit at a target time, the process conditionally factors into a past block, a future block, and a memory block upon which the past and the future are conditionally independent. The memory block indicates the set of operations relevant to the marginal target map. \textbf{b} A stochastic process on a qubit register across several times is represented at the most recent time by different overlapping causal cones $\{\mathfrak{C}_{t_2}^{\vec{q}_1},\cdots,\mathfrak{C}_{t_2}^{\vec{q}_5}\}$, where $\ell_s = 2$. The fronts of the cones are denoted by groupings $\vec{q}_i$. Taking one step back, the relevant qubit groups in the cone are denoted by $\vec{q}_i^{(1)}$.}
	\label{fig:PMF-coverage}
\end{figure}

Continuing, let us define $\mathfrak{C}_{t_j\:P}^{\vec{q}}$, the cone's \emph{past}, and $\mathfrak{C}_{t_j\:F}^{\vec{q}}$, the cone's \emph{future}. The past with respect to $\mathfrak{C}_{t_j}^{\vec{q}}$ is the collection of tuples 
\begin{equation}
	\begin{split}
	\mathfrak{C}_{t_j\:P}^{\vec{q}}:=\{(t_{j-\ell_t - 1},\mathbf{Q}), (t_{t_{j-\ell_t - 2}},\mathbf{Q}), \cdots , (t_0,\mathbf{Q})\}.
	\end{split}
\end{equation}
That is, it is the collection of operations which act on all qubits at all times prior to the back of the causal cone.
The future with respect to $\mathfrak{C}_{t_j}^{\vec{q}}$ is thus the remaining qubit-time pairs: those contemporaneous to the cone, but on alternate qubits, and those strictly after $t_j$:
\begin{equation}
	\begin{split}
		\mathfrak{C}_{t_j\:F}^{\vec{q}}:=&\{(t_j,\vec{p}), (t_{j-1}, \vec{p}^{(1)}), \cdots , (t_{j-\ell_t}, \vec{p}^{(\ell_s)})\} \quad \text{where }\vec{p^{(i)}}:= \mathbf{Q}\backslash \vec{q}^{(i)},\quad \text{and}\\
		&\{(t_k,\mathbf{Q}), (t_{k-1},\mathbf{Q}),\cdots,(t_{j+1},\mathbf{Q})\}.
	\end{split}
\end{equation}

This set of definitions is illustrated in Figure~\ref{fig:PMF-coverage}. Note that we have introduced $\mathfrak{C}_{t_j}^{\vec{q}}$ as a collection of qubit-time pairs, but we will occasionally interchange this to also by proxy mean the operations that occur and these spacetime locations, as well as the marginal legs of the underlying process. From context, the distinction should be clear.


Let us unpack this definition further. Correlations at the \emph{front} of the cone across
\begin{equation}
	(t_j, \vec{q})
\end{equation}
are purely spatial ones, and could be present even for a Markovian process. These constitute crosstalk interactions between topologically connected qubits, common-cause variables, or a fixed experimenter-chosen circuit. We first maintain that these spatial correlations have weight $\ell_s$, which is small. According to spatial locality $\ell_s$, each marginal $\hat{\mathcal{E}}_{j:j-1}^{\vec{q}}$ has its own causal cone. 
The assertion here is that for any choice of past operations $\mathbf{A}_{j-1:0}$, the Choi state of the conditional dynamics $\hat{\mathcal{E}}_{j:j-1}^{(\mathbf{A}_{j-1:0})}$ is a quantum Markov chain satisfying the entropic relation 
\begin{equation}
	\label{eq:qmc-all-cones}
	I(q_{i-1}:q_{i+\ell_s} \mid \vec{q}) \approx 0,
\end{equation}
implying the state may be optimally reconstructed with the recovery map in Equation~\eqref{eq:petz-universal} containing marginals of size $\ell_s+1$. Equation~\eqref{eq:qmc-all-cones} is a statement about dependence of different qubits that is intended to hold for \emph{all} events in the different overlapping cones.
The limiting cases of $\ell_s=0$ and $\ell_s = n-1$ respectively imply zero crosstalk and all-to-all interaction scenarios. Moving one step backwards to $t_{j-1}$ we have the tuple
\begin{equation}
	(t_{j-1}, \vec{q}^{(1)}).
\end{equation}
Operations $\mathcal{A}_{j-1}^{\vec{q}^{(1)}}$ act at time $t_{j-1}$ across qubits $\vec{q}^{(1)}$, and update the status of the marginal $\mathcal{E}_{j:j-1}^{\vec{q}}$ through 
\begin{equation}
	\hat{\mathcal{E}}_{j:j-1}^{\vec{q}(\mathcal{A}_{j-1}^{\vec{q}^{(1)}})} = \Tr_{\bar{\mathfrak{o}}_j^{\vec{q}}\: \bar{\mathfrak{i}}_j^{\vec{q}}}\left[\Upsilon_{k:0}\cdot (\mathbb{I}_{{\mathfrak{o}}_j^{\vec{q}}\: {\mathfrak{i}}_j^{\vec{q}}}\otimes \bigotimes_{q\in\vec{q}^{(1)}}\hat{\mathcal{A}}_{j-1}^{q\:\text{T}})\right].
\end{equation}

This proceeds until the back of the cone:
\begin{equation}
	\hat{\mathcal{E}}_{j:j-1}^{\vec{q}(\mathcal{A}_{j-1}^{\vec{q}^{(1)}}\cdots\mathcal{A}_{j-\ell_t}^{\vec{q}^{(\ell_s)}})} = \Tr_{\bar{\mathfrak{o}}_j^{\vec{q}}\: \bar{\mathfrak{i}}_j^{\vec{q}}}\left[\Upsilon_{k:0}\cdot (\mathbb{I}_{{\mathfrak{o}}_j^{\vec{q}}\: {\mathfrak{i}}_j^{\vec{q}}}\otimes \bigotimes_{i=j-1}^{j-\ell_t}\bigotimes_{q\in\vec{q}^{(i+2)}}\hat{\mathcal{A}}_{i}^{q\:\text{T}})\right].
\end{equation}

At this point, we have reached the edge of our memory block. According to our 2D Markov order ansatz, then, we have 
\begin{equation}
	\Upsilon_{k:0}^{(\mathfrak{C}_{t_j}^{\vec{q}})} \approx \Upsilon_{k:0}^{(\mathfrak{C}_{t_j\: F}^{\vec{q}})}\otimes \Upsilon_{k:0}^{(\mathfrak{C}_{t_j\:P}^{\vec{q}})}.
\end{equation}

\subsubsection*{Reconstructing Memory Blocks -- Example}

Our Markov order ansatz asserts that a spacetime process is covered by a series of overlapping causal cones, each of which encodes the relevant memory structure with respect to a particular qubit set at a particular time. 
In principle, the \acs{PTT} procedure from Chapter~\ref{chap:PTT} suffices to estimate these process marginals $\Upsilon_{\mathfrak{C}_{t_j}^{\vec{q}}}$: these marginals are well-defined process tensors where input and output spaces $\mathcal{B}(\mathcal{H}_{\mathfrak{o}_i}),\mathcal{B}(\mathcal{H}_{\mathfrak{i}_i})$ are permitted to change total system dimension at each time. But even this structure is likely too large to estimate in practice without further compression. Consider, for concreteness, a causal cone with spatial extent $\ell_s = 1$ and temporal extent $\ell_t = 2$, i.e., two-body spatial interactions and a minimally non-Markovian memory. The front of this cone is a \acs{CPTP} map $\mathcal{E}_{2:1}$ across two qubits $q_2,q_3$; one step back is a \acs{CPTP} map $\mathcal{E}_{1:0}$ across four qubits $q_1,q_2,q_3,q_4$; and finally, the initial state $\rho_0$ across the same four qubits. The marginal process hence maps six different instruments to an output state and requires 
\begin{equation}
	\mathcal{O}(d_S^{4\cdot \left(\sum_{j=0}^{\ell_t-1} \ell_s+1 + 2j \right)  + 2\cdot(\ell_s+1)}) \approx r\cdot 2^{14}
\end{equation}
unique circuits to estimate it for rank $r$ process, which is already infeasible even in this small example. But reconstructing the entire marginal is overkill. Since the ansatz supposes $\ell_s = 1$, then spatially we expect only weight-two correlations. This means, for example, that we need not account for high-weight correlations between events $\mathcal{A}_0^{q_0}$ and $\mathcal{A}_0^{q_3}$. The state is a quantum Markov network with $I(q_4 : q_2\mid q_3) \approx 0$ and $I(q_2 : q_4 \mid q_3) \approx 0$. This is true for all cones.
We hence have three marginal regions:
\begin{equation}
	\begin{split}
		\mathfrak{C}_1&:=\{(t_2,q_2), (t_1, q_2), (t_0, q_1, q_2)\},\\
		\mathfrak{C}_{2}&:= \{(t_2,q_2,q_3), (t_1, q_2,q_3), (t_0,q_2,q_3)\},\\
		\mathfrak{C}_3 &:= \{(t_2,q_3),(t_1,q_3),(t_0,q_3,q_4)\},
	\end{split}
\end{equation}
from which we can recover the entire causal cone. We may start with the process marginal across $\mathfrak{C}_1$, which maps three instruments $\{\mathcal{A}_0^{q_1}, \mathcal{A}_0^{q_2},\mathcal{A}_1^{q_2}\}$ to a single qubit output $\rho_2^{q_2}$. The union $\mathfrak{C}_{12} = \mathfrak{C}_1\cup\mathfrak{C}_2$ can then be recovered using
\begin{equation}
	\Upsilon_{\mathfrak{C}_{12}} = \mathscr{R}_{\mathfrak{C}_1\rightarrow \mathfrak{C}_{12}}[\Upsilon_{\mathfrak{C}_1}].
\end{equation}
This requires the $\mathfrak{C}_2$, which is a map from four instruments $\{\mathcal{A}_{0}^{q_2},\mathcal{A}_0^{q_3},\mathcal{A}_1^{q_2},\mathcal{A}_1^{q_3}\}$ to two-qubit output $\rho_{2}^{q_2,q_3}$. And finally, make the recovery: 
\begin{equation}
	\Upsilon_{\mathfrak{C}_{123}} = \mathscr{R}_{\mathfrak{C}_2\rightarrow \mathfrak{C}_{23}}[\Upsilon_{\mathfrak{C}_{12}}].
\end{equation}
The maximum marginal now has $r\cdot 2^{10}$ unique circuits required to estimate. While this is still large, it is tractable. We return to the issue of further reducing this number in the second half of this chapter.


\subsubsection*{Stitching Together Overlapping Memory Blocks}

To generalise our reconstruction from microscopic features to the macroscopic features of an entire process with overlapping causal cones, one could iteratively apply recovery maps until the global state is fully expressed. Indeed, this approach will be efficient in its use of quantum resources, but the representation will inevitably be that of a 2D network. Expectation values of 2D networks cannot be classically evaluated efficiently~\cite{PhysRevResearch.2.013010}. We consequently seek other more efficient methods from which we can represent approximately the same information.

As stated above, we treat each conditional dynamical map $\mathcal{E}_{j:j-1}^{(\mathbf{A}_{j-1:0})}$ as a spatial quantum Markov chain. Thus, it can be approximately reconstructed by its $\ell_s$- and $(\ell_s + 1)$-body marginals. Moreover, since this is a 1D state, it can be efficiently treated classically, so long as the input state can also be efficiently represented. 
\begin{figure}[!t]
	\centering
	\includegraphics[width=\linewidth]{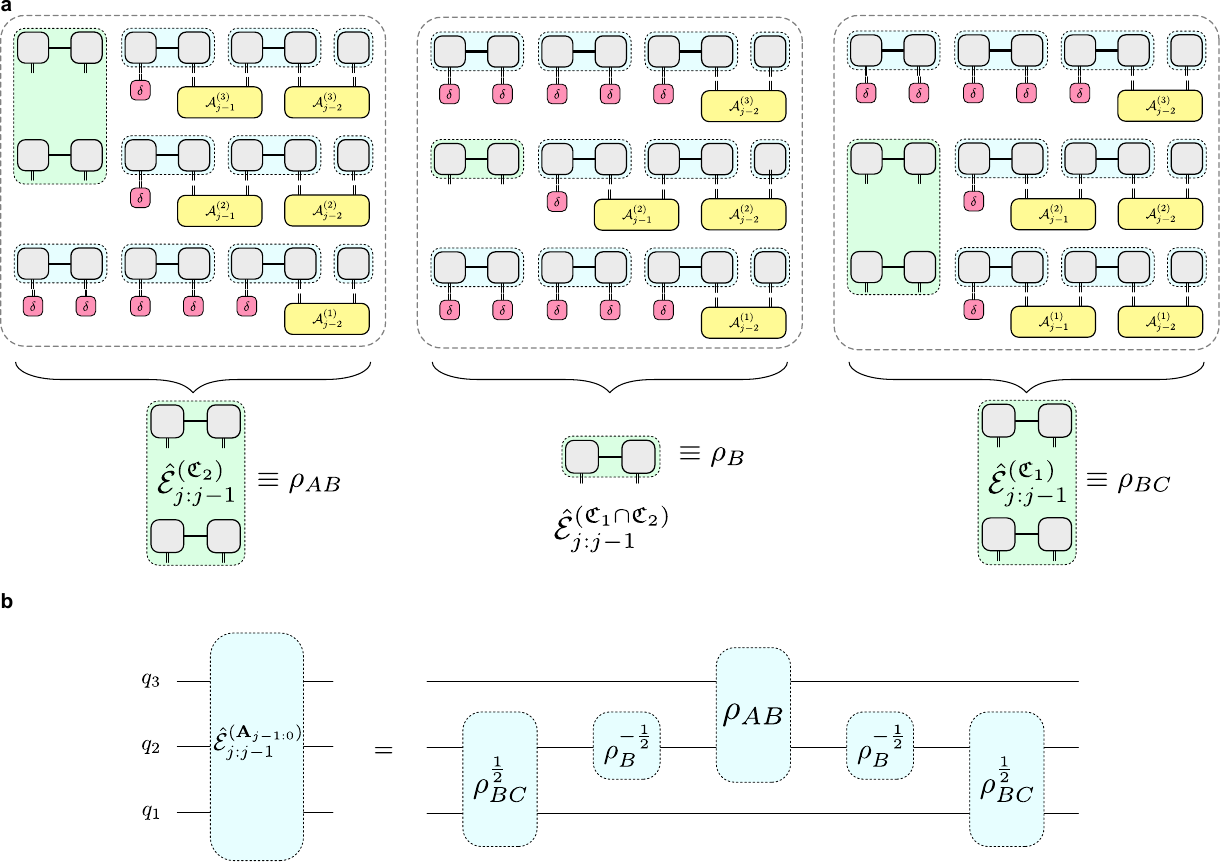}
	\caption[Reconstructing a single conditional dynamical map of a process from its overlapping marginals for $n=3$, $\ell_s = 1$ and $\ell_t = 3$ ]{Reconstructing a single conditional dynamical map of a process from its overlapping marginals for $n=3$, $\ell_s = 1$ and $\ell_t = 3$. \textbf{a} The three marginals required can be conditioned on their parts of the different causal cones $\mathfrak{C}_1$, $\mathfrak{C}_{2}$, as well as the intersection $\mathfrak{C}_{1}\cap\mathfrak{C}_{2}$. \textbf{b} Using the Petz recovery map, we show how the conditional marginal may be approximately reconstructed.}
	\label{fig:STMO-stitching}
\end{figure}
The marginal maps from which $\hat{\mathcal{E}}_{j:j-1}^{(\mathbf{A}_{j-1:0})}$ is recovered are themselves conditioned on operations in their own causal cones, rather than the entirety of $\mathbf{A}_{j-1:0}$. Explicitly, this means taking
\begin{equation}
	\begin{split}
		\left\{\hat{\mathcal{E}}_{j:j-1}^{(q_1,\cdots, q_{\ell_s})}, \hat{\mathcal{E}}_{j:j-1}^{(q_2,\cdots, q_{\ell_s})},\hat{\mathcal{E}}_{j:j-1}^{(q_2,\cdots, q_{\ell_s + 1})}, \cdots , \hat{\mathcal{E}}_{j:j-1}^{(q_{n-\ell_s},\cdots, q_{n-1})}, \hat{\mathcal{E}}_{j:j-1}^{(q_{n-\ell_s},\cdots, q_{n})}\right\}
	\end{split}
\end{equation}

as conditioned on the operations in their respective cones. 
The combined stitching procedure is then equivalent to that of Section~\ref{sec:MO}, where the initial state
\begin{equation}
	\rho_{\ell_t}(\mathbf{A}_{\ell_t-1:0}) = \Tr_{\overline{\mathfrak{o}}_{\ell_t}}\left[\Upsilon_{M_0}\bigotimes_{i=0}^{\ell_t -1 }\hat{\mathcal{A}}_i^{\text{T}}\right]
\end{equation}
can be computed from the quantum Markov network $\Upsilon_{M_0}$, and similarly for the subsequent $\Upsilon_{j}^{(\mathbf{A}_{j-2:2-\ell_t})}$. The initial state is then propagated along by the series of conditional process tensors on the memory blocks, while taking into account the impact of the $\mathcal{A}_j$. 
We emphasise that we still have a Markov order $\ell_t$ ansatz, and that spatial complexity is dealt with by employing Petz recovery maps to fully reconstruct each $\Upsilon_{M_j}$ as we have thus described. 
The procedure is collectively described by way of a simple single-step example in Figure~\ref{fig:STMO-stitching}.

We summarise this in reverse order to illustrate the top-down approach here:
\begin{enumerate}
	\item Temporally divide the dynamics into the composition of conditional dynamical maps:
	\begin{equation}
		\rho_k(\mathbf{A}_{k-1:0}) = \mathcal{E}_{k:k-1}^{(\mathbf{A}_{k-2:0})}\circ \mathcal{A}_{k-2}\circ \mathcal{E}_{k-1:k-2}^{(\mathbf{A}_{k-3:0})}\circ\cdots\circ \mathcal{A}_1 \circ {\mathcal{E}_{2:1}^{(\mathcal{A}_0)}}\circ\mathcal{T}_{1:0}[\mathcal{A}_0].
	\end{equation}
	\item Spatially divide each map as a quantum Markov chain of $\ell_s+1$ marginals:
	\begin{equation}
		\hat{\mathcal{E}}_{j:j-1}^{(\mathbf{A}_{k-1:0})} = \mathscr{R}_{(n-\ell_s)\cdots(n-1)\rightarrow (n-\ell_s)\cdots n}\circ \cdots\circ \mathscr{R}_{2\cdots\ell_s\rightarrow 2\cdots \ell_s+1}\left[\hat{\mathcal{E}}_{j:j-1}^{(q_1,\cdots,q_{\ell_s}),(\mathbf{A}_{j-2:0})}\right].
	\end{equation}
	\item Each recovery map will comprise the dynamical maps as marginals. According to the Markov order ansatz, these marginals will depend only on operations that make up their causal cones:
	\begin{equation}
		\begin{split}
			&\mathfrak{C}_{t_j}^{(q_1,\cdots ,q_{\ell_s})},\\
			&\mathfrak{C}_{t_j}^{(q_2,\cdots,q_{\ell_s })},\\
			&\mathfrak{C}_{t_j}^{(q_2,\cdots ,q_{\ell_s+1})},\\
			&\qquad \qquad \vdots\\
			&\mathfrak{C}_{t_j}^{(q_{n-\ell_s},\cdots ,q_{n-1})},\\
			&\mathfrak{C}_{t_j}^{(q_{n-\ell_s},\cdots, q_{n})}.
		\end{split}
	\end{equation}
	\item Each causal cone is also a quantum Markov chain with spatial weight $\ell_s$ which may itself be reconstructed in a series of recovery maps.
	This ensures that opposite spatial ends of the cone may update future dynamics, but only in a low-weight fashion rather than having high-weight spatiotemporal correlations.
	\item With the previous steps as a derivation of the marginals that make up the process, the final step is to characterise these marginals which may be performed using the results of Chapter~\ref{chap:PTT}.
\end{enumerate}

\clearpage
To summarise this entire ansatz, we have the following example.

\begin{figure}[!h]
	\centering
	\includegraphics[width=0.9\linewidth]{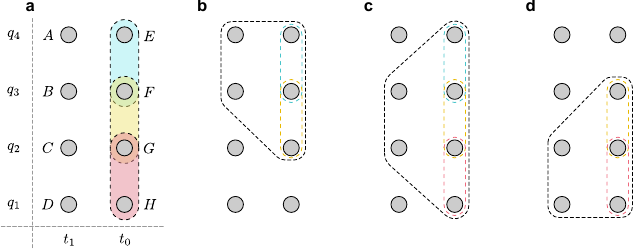}
	\caption[Illustration of a spacetime Markov order ansatz across four qubits and two times ]{Simplified illustration of the entire spacetime Markov order ansatz across four qubits and two times. \textbf{a} depicts the various conditional purely-spatial correlations for $\ell_s=2$. \textbf{b}, \textbf{c}, and \textbf{d} indicate the three causal cones that cover the process for the given subsystems. }
	\label{fig:STMO-walkthrough}
\end{figure}

\begin{example}[Four qubits, two times]
	\examplecontent{
	Start with a two-time process across four qubits where $\ell_s=1$, $\ell_t = 2$, and $\vec{q}^{(1)}$ expands by at most two. 
	Let us label the systems at each coordinate from $A$--$H$ as depicted in Figure~\ref{fig:STMO-walkthrough}a. To cover the process with a series of overlapping causal cones, we start with time $t_0$. Since $\ell_s = 1$, we have the entropic relations 
	\begin{equation}
		\label{eq:4q2t-entropies}
		\begin{split}
			I(E : G \mid F)&\approx 0\\
			I(F : H \mid G)&\approx 0.
		\end{split}
	\end{equation}
	The initial state is thus reconstructed by three two-body marginals.
	\begin{equation}
		\rho_{EFGH} \approx \mathscr{R}_{G\rightarrow GH}\circ \mathscr{R}_{F\rightarrow FG}[\rho_{EF}].
	\end{equation}
	Moving forward to $t_1$, there are three cones that cover the process. 
	(i) $ABEFG$, (ii) $BCEFGH$, and (iii) $CDFGH$. These are depicted in Figures~\ref{fig:STMO-walkthrough}b--d. Each of these respectively have the relations from Equation~\eqref{eq:4q2t-entropies} at $t_0$. So the cones are constructed with these marginals as a starting point:
	\begin{equation}
		\begin{split}
			\rho_{ABEFG} &\approx \mathscr{R}_{FB\rightarrow BFG}[\rho_{ABEF}]\\
			\rho_{BCEFGH} &\approx \mathscr{R}_{CG\rightarrow CGH}\circ \mathscr{R}_{BF\rightarrow BFCG}[\rho_{BEF}]\\
			\rho_{CDFGH} &\approx \mathscr{R}_{CG\rightarrow CGDH}[\rho_{FCG}].
		\end{split}
	\end{equation}
	Finally, using these three marginals, we can contract the various operations into their causal cones to obtain 
	\begin{equation}
		\begin{split}
			&\rho_{AB \mid EFG}\\
			&\rho_{BC\mid EFGH}\\
			&\rho_{CD\mid FGH}.
		\end{split}
	\end{equation}
	from which we obtain 
	\begin{equation}
		\rho_{ABCD} \approx \mathscr{R}_{C\rightarrow CD}\circ \mathscr{R}_{B\rightarrow BC}[\rho_{AB}],
	\end{equation}
	where we have dropped notation of dependencies of the marginals. 

	\color{black}
	}
\end{example}

\vspace{8cm}
\subsubsection*{Analysis}

We have presented an ansatz to approximately -- but efficiently -- represent any multi-time, multi-qubit quantum stochastic process of bounded memory. This opens up the exploration of quantum stochastic processes at scale, incurring only additive error in the spatiotemporal quantum mutual information. The next step will be to examine how well this ansatz suits real quantum processes.
Although we have used a model of memory effects being contiguous in time, there is nothing about our structure that imposes this. For example, one could imagine a general directed acyclic graph structure for arbitrary spatiotemporal correlations that could equivalently serve as the Markov blanket for the process.

The dimension of the necessary memory blocks will be bounded above by $d_S^{2\ell_s\ell_t}$, and there will be $\mathcal{O}((k-\ell_t+1)\cdot (n-\ell_s + 1))$ of them. The number of cones then grows linearly with the area of the process, and so the overall complexity is dominated by the exponential of the memory. 
In the next section, however, we will explore ways in which we can further compress the model of these memory blocks using sparseness of the process. 
We have seen in Chapter~\ref{chap:MTP} how these marginal blocks can be simultaneously characterised using classical shadows, implying that the total number of measurements required to estimate the process to precision $\delta$ will be $\mathcal{O}(\log kn/\delta^2)$. Assuming an average quantum Markov chain error of $\varepsilon$ growing additively, the size of the total approximation error will be $\mathcal{O}(kn\varepsilon)$. 

Recovery maps are in general \acs{CP}, which maintains positivity of the process. All marginals of a causal process tensor are themselves causal, implying that they generate \acs{TP} marginals of $\mathcal{E}_{j:j-1}$. Finally, the recovery of a single conditional map from the marginals is itself.
We can see this by noting that for some $\hat{\mathcal{E}}_{ABC}$, we have 
\begin{equation}
	\begin{split}
		\Tr_{\mathfrak{o}}[\hat{\mathcal{E}}_{ABC}] &= \Tr_{\mathfrak{o}_A,\mathfrak{o}_B,\mathfrak{o}_C}[\hat{\mathcal{E}}_{BC}^{\frac{1}{2}}\hat{\mathcal{E}}_{B}^{-\frac{1}{2}}\hat{\mathcal{E}}_{AB}\hat{\mathcal{E}}_{B}^{-\frac{1}{2}}\hat{\mathcal{E}}_{BC}^{\frac{1}{2}}]\\
		&= \mathbb{I}_{\mathfrak{i}_A} \otimes \Tr_{\mathfrak{o}_B\mathfrak{o}_C}[\hat{\mathcal{E}}_{BC}^{\frac{1}{2}} \hat{\mathcal{E}}_{B}^{-\frac{1}{2}}\hat{\mathcal{E}}_{B}\hat{\mathcal{E}}_{B}^{-\frac{1}{2}}\hat{\mathcal{E}}_{BC}^{\frac{1}{2}}]\\
		&= \mathbb{I}_{\mathfrak{i}_A} \otimes \mathbb{I}_{\mathfrak{i}_B} \otimes \mathbb{I}_{\mathfrak{i}_C},
	\end{split}
\end{equation}
following from the trace preservation of the marginal maps. Hence, the entire process is represented dynamically by a composition of conditional \acs{CPTP} maps, implying that the whole process is both \acs{CP} and causal, regardless of how coarse the Markov network approximations are. 

It is the topic of future research to carry out this spacetime Markov order approach in practice, both numerically and experimentally. It would also be interesting to first determine the structure of a good ansatz based on the expected dynamics of the system. For instance, one might appeal to Lieb-Robinson bounds to roughly determine in advance the expected spatial locality of the causal cones, as well as an expected rate of dissipation for the temporal dimension of the memory kernel.

\section{Tensor Networks}
\label{sec:tensor-networks}
The archetypal method for compressing quantum states into more efficient representations has been to employ the use of tensor networks. The philosophy behind tensor networks is that quantum states, which live on a series of tensor products of Hilbert and dual spaces, may be broken down across these spaces using well-known decompositions. Then, if these decompositions contain sparse features, they can be zeroed out to build more efficient representations. The most famous example is a \ac{MPS} which can exactly represent any pure quantum state~\cite{vidal-MPS}. If that state has exponentially decaying correlations in its topology, then the \acs{MPS} can represent it in only a polynomial number of parameters~\cite{Orus2014}.

Tensor networks are important for a variety of reasons~\cite{Cirac2020MatrixPS}. They highlight the necessary role played by entanglement in pure-state quantum computation~\cite{Eisert2021EntanglingPA,vidal-MPS,entanglement-jozsa}. They also play a crucial role in the advancement of classical computers for the simulation and emulation of quantum algorithms, furthering the barrier through which quantum supremacy may be realised~\cite{huang2020classical,china-sim,pan2021simulating}. 
More recently, tensor networks have also been realised as a useful tool to extend the ensembles on which classical shadows can operate~\cite{akhtar2022scalable,bertoni2022shallow}. Here, we will focus on the application of tensor networks to sparse characterisation, which has recently been achieved with \acs{QPT}~\cite{torlai2020quantum} and Hamiltonian tomography~\cite{wilde2022scalably}.

\subsection{Background}

We first review \acs{MPS}s as per their original introduction in Ref.~\cite{vidal-MPS}.
A pure state vector $\ket{\psi}$ on a system of $n$ qudits, each with orthonormal basis $\{\ket{s}\}_{s=1}^{d}$ may be expressed in the form 
\begin{equation}
	\label{eq:dense-state}
	\ket{\psi} = \sum_{s_1,\cdots,s_n=1}^d \alpha_{s_1s_2\cdots s_n} |s_ns_{n-1}\cdots s_2s_1\rangle,
\end{equation}
requiring exponentially many complex numbers $\alpha_{\vec{s}}$ to describe it. 

Consider a partitioning $A\mid B$ of the $n$ qudits. The Schmidt decomposition of $\ket{\psi}$ with respect to $A\mid B$ is given by 
\begin{equation}
	\label{eq:schmidt-decomp}
	\ket{\psi} = \sum_{\alpha = 1}^{\chi_A}\sigma_\alpha |\Phi_\alpha^{[A]}\rangle \otimes |\Phi_\alpha^{[B]}\rangle.
\end{equation}
The $\{\sigma_\alpha\}$ are the singular values of the decomposition, of which there are $\chi_\alpha$ many, a quantity known as the Schmidt rank. $\{\Phi_\alpha^{[A]}\}$ and $\{\Phi_\alpha^{[B]}\}$ are respective eigenvectors of the reduced density matrices $\rho_A$ and $\rho_B$. The entropy of entanglement $\mathcal{S}$ between partitions $A$ and $B$ is given by the von Neumann entropy of either of the reduced states 
\begin{equation}
	\mathcal{S}(\ket{\psi}_AB) = S(\rho_A) = -\Tr[\rho_A\log\rho_A] = -\sum_\alpha |\sigma_\alpha|^2\log(|\sigma_\alpha|) = S(\rho_B).
\end{equation}
Thus, the Schmidt rank is both a natural measure of entanglement across the bipartition, and describes the length of the list of numbers required to fully describe the decomposition in Equation~\eqref{eq:schmidt-decomp}. The derivation of \acs{MPS}s follows from repeated Schmidt decompositions in a 1D chain. Let us first consider the biparitition $A = \{q_1\}$ and $B = \{q_2,\cdots q_n\}$. We can write the total state as:
\begin{equation}
	\label{eq:mps-step-one}
	\begin{split}
		\ket{\psi} &= \sum_{\alpha_1=1}^{\chi_1}\sigma_{\alpha_1}^{[1]}|\Phi_{\alpha_1}^{[1]}\rangle|\Phi_{\alpha_1}^{[2\cdots n]}\rangle\\
		&= \sum_{s_1,\alpha_1}\Gamma_{\alpha{1}}^{[1]s_1}\sigma_{\alpha_1}^{[1]}|s_1\rangle|\Phi_{\alpha_1}^{[2\cdots n]}\rangle.
	\end{split}
\end{equation}
Here, $\Gamma_{\alpha_1}^{[1]s_1}$ is introduced as a change of basis from the Schmidt vectors to the standard basis: 
\begin{equation}
	|\Phi_{\alpha_1}^{[1]}\rangle = \sum_{s_1}\Gamma_{\alpha_1}^{[1]s_1}|s_1\rangle.
\end{equation}
To concatenate this procedure, we can move forward to $|\Phi_{\alpha_1}^{[2\cdots n]}\rangle$ across the bipartition $\{q_2\}\mid \{q_3\cdots q_n\}$. First write the eigenvector in terms of a local basis for $q_2$
\begin{equation}
	|\Phi_{\alpha_1}^{[2\cdots n]}\rangle = \sum_{s_2}|s_2\rangle |\tau_{\alpha_1 s_1^{[3\cdots n]}}\rangle,
\end{equation}
and re-write the $|\tau_{\alpha_1s_1}^{[3\cdots n]}\rangle$ in terms of the Schmidt vectors 
\begin{equation}
	\label{eq:mps-step-two}
	|\tau_{\alpha_1s_1}^{[3\cdots n]}\rangle = \sum_{\alpha_2=1}^{\chi_2}\Gamma_{\alpha_1\alpha_2}^{[2]s_2}\lambda_{\alpha_2}^{[2]}|\Phi_{\alpha_2}^{[3\cdots n]}\rangle.
\end{equation}
Now, Equation~\eqref{eq:mps-step-one} can be substituted into Equation~\eqref{eq:mps-step-two} to obtain
\begin{equation}
	|\psi\rangle = \sum_{s_1\alpha_1s_2\alpha_2}\Gamma_{\alpha_1}^{[1]s_1}\lambda_{\alpha_1}^{[1]}\Gamma_{\alpha_1\alpha_2}^{[2]s_2}\lambda_{\alpha_2}^{[2]}|s_1s_2\rangle|\Phi_{\alpha_1}^{[3\cdots n]}\rangle.
\end{equation}
Repeating these steps across all remaining bipartitions allows one to express the total state as 
\begin{equation}
	\label{eq:mps-decomp}
	\ket{\psi} = \sum_{s_1,\cdots s_n \alpha_1 \cdots \alpha_n}\Gamma_{\alpha_1}^{[1]s_1}\lambda_{\alpha_1}^{[1]}\Gamma_{\alpha_1\alpha_2}^{[2]s_2}\lambda_{\alpha_2}^{[2]}\cdots \Gamma_{\alpha_{n-2}\alpha_{n-1}}^{[n-1]s_{n-1}} \lambda_{\alpha_{n-1}}^{[n-1]}\Gamma_{\alpha_{n-1}}^{[n]s_n} |s_1\cdots s_n\rangle.
\end{equation}
In other words, each of the $2^n$ state coefficients can be expressed in terms of the tensors $\{\Gamma_{\alpha_{i-1}\alpha_{i}}^{[i]s_i}\}$ and vectors $\{\lambda_{\alpha_i}^{[i]}\}$. The former are referred to as sites, and the latter are referred to as bonds.
\acs{MPS}s can be hence seen as a repeated sparse reshaping of the dense vector described in Equation~\eqref{eq:dense-state}. The name `matrix product state' refers to the fact that to find coefficients of the \emph{state} in $|s_1,\cdots s_n\rangle$, one computes the \emph{matrix products} $\Gamma_{\alpha_{i-1}\alpha_{i}}^{[i]s_i}\cdot \lambda_{\alpha_i}^{[i]}\cdot \Gamma_{\alpha_{i}\alpha_{i+1}}^{[i+1]s_{i+1}}$. 
Defining $\chi:=\max\{\chi_i\}$, the number of parameters to describe the entire state is hence upper bounded by $(2\chi^2 + \chi)n$. As long as the state does not get too entangled -- i.e., $\chi$ is upper bounded by $\text{poly}(n)$, then this representation is efficient. This is known as the canonical form of an \acs{MPS}. Gauge freedoms exist in \acs{MPS} representations. First and foremost, one has a choice of whether to absorb the singular values into the left or right sites. Additionally, gauge pairs $XX^{-1}$ may be inserted within each bond, changing the local sites.
This form is useful not only for direct access of quantities like entanglement entropy, but readily also permits approximations in the form of truncation. That is, a state may be approximated either by thresholding all $\sigma_i^{[i]} < \epsilon$ to zero, or by keeping only the $\chi$ largest singular values. 
We will state, but not show, that the decomposition is local. This means that updating the state according to any $k$-local evolution requires only an update of the $k$ sites. 

The \acs{MPS} representation can be used to exactly represent any quantum state. However, in practice, they are typically only referred to in contexts where the Schmidt rank grows like $\text{poly}(n)$. One particularly important example is that low energy states of gapped local Hamiltonians can be efficiently represented by an \acs{MPS}. \acs{MPS}s satisfy a one-dimensional area law, which means that the entanglement entropy of a block of sites is bounded by a constant ($\chi$). Finally, they are finitely correlated, which means that the correlation functions of an \acs{MPS} decay exponentially with the separation distance. 

\acs{MPS}s form a foundation for many generalisations which we will employ, but not labour on introducing: (i) The above procedure can be repeated for operators (with standard bases $|s_i\rangle\!\langle s_i'|$), to derive the \ac{MPO}. (ii) By extension, density operators may be represented in this form, typically by representing a state on a larger space and then taking a partial trace. If one keeps this ancilla space, so as to have a bra MPS, a ket MPS, and the partial trace over the environment uncontracted, then the network is referred as a \ac{LPDO}. (iii) One can also take this representation to higher dimensions by performing Schmidt decompositions across bipartitions in each dimension. Note, however, that evaluating expectation values is only efficient in one dimension. In two-dimensions and above, this is a \texttt{\#P-complete} problem~\cite{PhysRevResearch.2.013010}. Collectively, we will designate all of these representations under the umbrella of tensor networks, where rank-deficient decompositions are leveraged to represent large vectors and matrices in a more efficient way. 

\subsection{Process Tensors as Locally Purified Density Operators}
In many physically relevant processes, the non-Markovian memory, or the effective environment ought not to grow too large. Indeed, the bond dimension of the \acs{MPO} representation of a process tensor has been shown to be a measure of non-Markovianity~\cite{Pollock2018a}.
However, most analyses of process tensor \acs{MPO}s have been in the form of numerical experiments. 
It is one thing to say that you can represent an object efficiently, but one of the chief difficulties in translating this to an experimental setting is actually estimating this object in a robust and physically sensible way. 
In this section, we focus on this problem. Given an experimental dataset, how can we perform a sparse version of \acs{PTT} and estimate a tensor network representation of our process? In contrast with Markov order models, we will see many examples where tensor networks can \emph{exactly} represent a process. This comes at the expense of much greater classical computational cost, and a lack of convergence guarantee in the estimation. However, we will present a method which details a powerful estimation procedure that we find effective in characterising multi-qubit, multi-time non-Markovian processes. Our approach is modular: we construct a tensor network ansatz for the process, employ a log-likelihood objective function, use autodifferentiation to obtain the gradients of the individual tensors, and finally perform a variant of stochastic gradient descent to find the maximum-likelihood model. The flexibility here is that the tensor network model for the process is fully generic and can be user-chosen. Indeed, in this section and in Chapter~\ref{chap:universal-noise}, we explore the versatility of this for different processes.


\subsubsection*{Representation} 

\begin{figure}
	\centering
	\includegraphics[width=\linewidth]{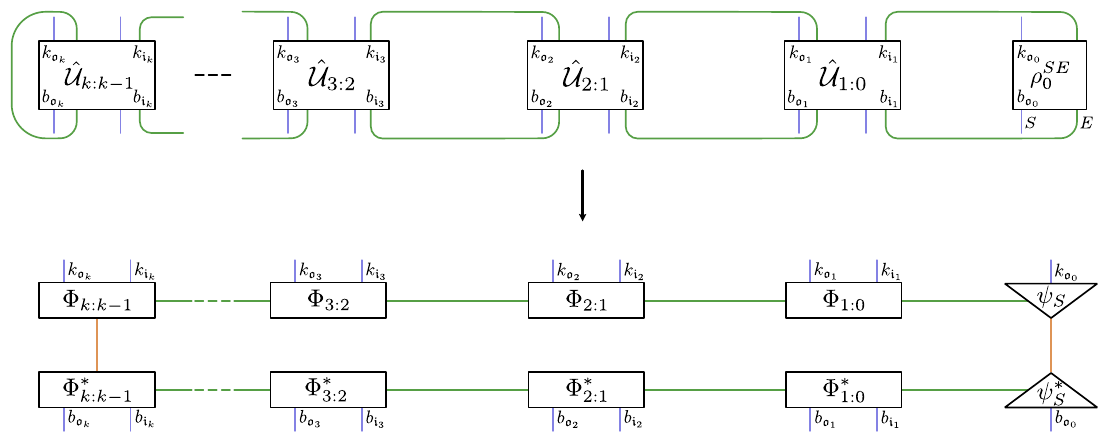}
	\caption[The link product representation of the process tensors shows how it can readily be expressed as an LPDO, where each step is pure except for the start and end.]{The link product representation of the process tensors shows how it can readily be expressed as an LPDO, where each step is pure except for the start and end. The size of the non-Markovian memory hence corresponds to the horizontal bond dimension, and the vertical bonds correspond to the total size of the environment, as sensed by the process.
	}
	\label{fig:mpo-ring}
\end{figure}

Process tensors are naturally equipped to be represented with \acs{LPDO}s. Apart from an initial (possibly) mixed state and the trace over the environment at the end, dynamics are unitary. Hence, only the start and end of a process require operator representations. We expand on the theoretical prescription given in Ref.~\cite{Pollock2018a}. A process tensor Choi state, in full generality, has \acs{MPO} representation 
\begin{equation}
	\label{eq:pt-mpo}
	\Upsilon_{k:0} = \sum (\Gamma_{k:k-1})_{b_{\mathfrak{o}_k}b_{\mathfrak{i}_k}}^{k_{\mathfrak{o}_k}k_{\mathfrak{i}_k}}
	\cdots (\Gamma_{1:0})_{b_{\mathfrak{o}_1}b_{\mathfrak{i}_1}}^{k_{\mathfrak{o}_1}k_{\mathfrak{i}_1}}(\Gamma_0)_{b_{\mathfrak{o}_0}}^{k_{\mathfrak{o}_0}} |k_{\mathfrak{o}_k}k_{\mathfrak{i}_k}\cdots k_{\mathfrak{o}_1}k_{\mathfrak{i}_1}k_{\mathfrak{o}_0}\rangle \! \langle b_{\mathfrak{o}_k}b_{\mathfrak{i}_k}\cdots b_{\mathfrak{o}_1}b_{\mathfrak{i}_1}b_{\mathfrak{o}_0}|,
\end{equation}
where, for $j\neq k,0$, the following are $d_E^2\times d_E^2$ matrices:
\begin{equation}
	(\Gamma_{j:j-1})_{b_{\mathfrak{o}_j}b_{\mathfrak{i}_j}}^{k_{\mathfrak{o}_j}k_{\mathfrak{i}_j}} = \langle b_{\mathfrak{o}_j} | U_{j:j-1}|b_{\mathfrak{i}_j}\rangle \otimes \langle k_{\mathfrak{o}_j}| U_{j:j-1}^\ast | k_{\mathfrak{i}_{j}}\rangle.
\end{equation}
The final step, then, is a length $d_E^2$ row vector
\begin{equation}
	(\Gamma_{k:k-1})_{b_{\mathfrak{o}_k}b_{\mathfrak{i}_k}}^{k_{\mathfrak{o}_k}k_{\mathfrak{i}_k}} = \sum_{\gamma}\langle b_{\mathfrak{o}_k} \gamma_E| U_{k:k-1}|b_{\mathfrak{i}_j}\rangle \otimes \langle k_{\mathfrak{o}_k} \gamma_E| U_{k:k-1}^\ast | k_{\mathfrak{i}_{k}}\rangle,
\end{equation}
and supposing the initial state has eigendecomposition
\begin{equation}
	\rho_0^{SE} = \sum_i p_i |\psi_i\rangle \! \langle \psi_i|,
\end{equation}
then we have a length $d_E^2$ column vector
\begin{equation}
	(\Gamma_0)_{b_{\mathfrak{o}_0}}^{k_{\mathfrak{o}_0}} = \sum_i p_i \langle b_{\mathfrak{o}_0}|\psi_i\rangle \otimes \langle k_{\mathfrak{o}_0}|\psi_i\rangle^\ast.
\end{equation}
This can also be seen if we recall the link product definition
\begin{equation}
	\Upsilon_{k:0} =  \Tr_E[\hat{\mathcal{U}}_{k:k-1}\star \hat{\mathcal{U}}_{k-1:k-2}\star \cdots \star \hat{\mathcal{U}}_{1:0}\star \rho_0^{SE}],
\end{equation}
sketched in Figure~\ref{fig:mpo-ring}.
We notice several features. First, only $\Gamma_0$ and $\Gamma_{k:k-1}$ are non-pure, as these respectively represent a generically mixed initial state, and the trace over the environment at the end of the dynamics. Each other intermediate step is pure. Second, the bond dimension grows with the size of the environment.
We can introduce two ancilla systems as proxies for the environment, one to purify $\Gamma_0$ and one to purify $\Gamma_{k:k-1}$. In the former case, we have 
\begin{equation}
	\begin{split}
		|\Psi_0^{SA}\rangle &= \sum_{\mu_0=1}^{\chi_{\mu_0}}\sqrt{p_{\mu_0}}|\psi_{\mu_0}\rangle\otimes |a_{\mu_0}^{(1)}\rangle,\\
	\end{split}
\end{equation}
which we can contract with the ancilla bond of its complex conjugate to yield the same local tensor. 
\begin{equation}
	\begin{split}
	(\Gamma_0)^{k_{\mathfrak{o}_0}}_{b_{\mathfrak{o}_0},\alpha_1} &= \sum_{\mu_0=1}^{\chi_{\mu_0}} [\Psi_0]^{k_{\mathfrak{o}_0}}_{\mu_0,\nu_1}[\Psi_0^\ast]^{\mu_0,\nu_1'}_{b_{\mathfrak{o}_0}} \\
	&= \sum_{\mu_0=1}^{\chi_{\mu_0}}\sqrt{p_{\mu_0}} \langle k_{\mathfrak{o}_0}|\psi_{\mu_0}\rangle\otimes \langle a_{\mu_0}^{(1)}|a_{\mu_0}^{(1)}\rangle\sqrt{p_{\mu_0}} \langle\psi_{\mu_0}|b_{\mathfrak{o}_0}\rangle^\ast\otimes \langle a_{\mu_0}^{(1)}|a_{\mu_0}^{(1)}\rangle\\
	&= \sum_{\mu_0}p_{\mu_0}\langle b_{\mathfrak{o}_0}|\psi_{\mu_0}\rangle \otimes \langle k_{\mathfrak{o}_0}|\psi_{\mu_0}\rangle^\ast
	\end{split}
\end{equation}
We say then that $\Gamma_0$ is locally purified, because the addition of the bond $\mu_0$ represents an ancilla system that encodes the mixedness of the reduced state $\Gamma_0$. The same procedure is repeated for $\Gamma_{k:k-1}$. This \acs{LPDO} process representation immediately gives rise to a ring-like structure, as in Figure~\ref{fig:mpo-ring}b. One great advantage to using a locally purified state rather than a generic \acs{MPO} as in Equation~\eqref{eq:pt-mpo} is that this form naturally encodes positivity of the state. As well as being a generally desirable physical property, this parametrisation naturally produces only positive, real probabilities when evaluated. Hence, it is well-behaved when considering the log-likelihood to find our optimal model. 

For a single-system, the process tensor for a quantum stochastic process can therefore be written in terms of this \acs{LPDO} parametrisation:
\begin{equation}
[\Upsilon_{k:0}^{\vec{\theta}}]^{\vec{k_{\mathfrak{o}}}\vec{k_{\mathfrak{i}}}}_{\vec{b_{\mathfrak{o}}}\vec{b_{\mathfrak{i}}}} = \sum_{\mu_k,\mu_0} \sum_{\vec{\nu},\vec{\nu}'} [\Phi_{k:k-1}]_{\mu_k,\nu_k}^{k_{\mathfrak{o}_k}k_{\mathfrak{i}_k}} [\Phi_{k:k-1}^\ast]^{\mu_k,\nu'_k}_{b_{\mathfrak{o}_k}b_{\mathfrak{i}_k}}\left(\prod_{j=1}^{k-1}
	[\Phi_{j:j-1}]_{\nu_{j+1}\nu_j}^{k_{\mathfrak{o}_j}k_{\mathfrak{i}_j}}[\Phi_{j:j-1}^\ast]^{\nu'_{j+1}\nu'_j}_{b_{\mathfrak{o}_j}b_{\mathfrak{i}_j}}\right) [\Psi_{0}]_{\mu_0,\nu_1}^{k_{\mathfrak{o}_0}} [\Psi_{0}^\ast]^{\mu_0,\nu'_1}_{b_{\mathfrak{o}_0}},
\end{equation}

where $\vec{\theta} = \{\Phi_{k:k-1},\{\Phi_{j:j-1}\},\Psi_0\}$. In this variational form, bond dimensions must be chosen \emph{a priori}. We have Kraus rank $\chi_{\mu_0},\chi_{\mu_k}$ reflective respectively of the rank of the initial $SE$ state and number of environment degrees of freedom traced at the end of the circuit. 
The $\chi_{\nu_j}$ are measures of non-Markovianity: they decree the effective size of the non-Markovian environment relevant to possible memory effects. Implicitly, we will always absorb eigenvalues and singular values into the left and right singular vectors of each site. This is simply a gauge choice that eases notational overload.

\begin{wrapfigure}{r}{0.65\textwidth}
	\centering
	\includegraphics[width=0.9\linewidth]{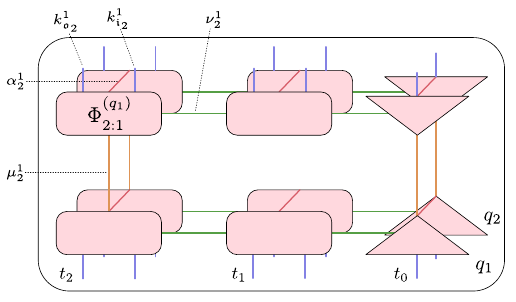}
	\caption[An indicative block of the 3D tensor network used to represent multi-qubit, multi-time quantum stochastic processes ]{An indicative block of the 3D tensor network we employ to represent multi-qubit, multi-time quantum stochastic processes. Green horizontal bonds represent non-Markovian memory; orange vertical bonds indicate local purification (Kraus bonds); red bonds into the page indicate qubit interactions, or crosstalk.}
	\label{fig:tn-3d-exp}
\end{wrapfigure}

Before proceeding to fit this model, we make one final adjustment. Each of these variational tensors are dense representations with local system sites of size $d_S^2$. But if we wish to choose a system whose Hilbert space structure is composite -- for example a chain of spins -- then we can further decompose this to take advantage of the sparseness of spatial correlations. For a system $S$ defined as a register of qubits $\{q_1,\cdots,q_n\}$, we can perform a series of singular value decompositions across the subsystems such that 
\begin{equation}
	\Phi_{j:j-1,\mu_j\nu_j}^{k_{\mathfrak{o}_j}k_{\mathfrak{i}_j}} = \sum_{\vec{\alpha}}\prod_{i=1}^n\Phi_{j:j-1, \alpha_j^{i-1}\alpha_j^i, \mu_j^i\nu_j^i}^{(q_i)k^i_{\mathfrak{o}_j}k^i_{\mathfrak{i}_j}}.
\end{equation}
This rather messy piece of index notation is all to say that we have a 3D tensor network representation of our process, summarised both in the following list and in Figure~\ref{fig:tn-3d-exp}.

\begin{itemize}
	\item A single site $\Phi_{j:j-1}^{(q_i)}$ is the local purification of the $j$th dynamical map on the $i$th qubit. $k_{\mathfrak{o}_j}^i$ and $k_{\mathfrak{i}_j}^i$ are the local site indices for the input and output spaces of the dynamical map represented for that particular qubit at that particular time step.
	\item The bonds in the $x$-direction are indexed by $\nu_j^i$ and encode temporal correlations on that qubit.
	\item The $y$-direction bonds are indexed by $\mu_j^i$, these are Kraus bonds and are only present for $j=0,k$.
	\item The $z$-direction bonds are indexed by $\alpha_j^i$: these encode spatial correlations between the qubits generated across a given time step.
\end{itemize}

The efficiency of this ansatz relies on the process obeying an area law growth. 
That is, the maximum bond dimension of the whole network should be bounded by a constant. 
Incidentally, although we have three bond dimensions here to make this a geometrically three-dimensional network because of the ring-like structure naturally encoded by process tensors, this is topologically only two-dimensional. 
Nevertheless, we must be careful. Two-dimensional tensor networks might encode a state efficiently -- hence requiring fewer quantum resources, but they cannot be contracted efficiently~\cite{Orus2014} -- and so the classical computation grows exponentially in the area of the network. For this reason, we limit our approach to either only few time steps and many qubits, or many qubits and few time steps. 





\subsection{Estimation} 
We are now in a position to lay out our tensor network estimation procedure for non-Markovian quantum stochastic processes. We have made some specific choices about the form of the gate operations here, but these do not affect the generality of the procedure. 
We are again focused on reconstructing restricted process tensors, hence the use of Clifford unitaries as our basis for characterisation. 
Predominantly, this is because we still face the same experimental hurdles as in Chapter~\ref{chap:PTT}; mid-circuit measurements are difficult to implement, and so this characterisation is the most applicable to any quantum hardware. However, the procedure is readily generalisable to any \ac{IC} basis if available. 
To perform \ac{PTT} with \ac{LPDO}s, we consider the standard setup of the previous Chapter. Our multi-time instrument will consist of sequences of unitary operations followed by a final terminating measurement. We take a time-local basis of operations $\{\mathcal{B}_j\}$, moreover for concreteness we take this basis to be the set of (for the time being) single-qubit Clifford operations. Our approach is inspired by generative models of neural networks in machine learning applications. At the cost of rigorous performance guarantees, we gain a significant reduction in required classical and quantum computational resources. 

\begin{wrapfigure}{l}{0.5\textwidth}
	\centering
	\includegraphics[width=0.9\linewidth]{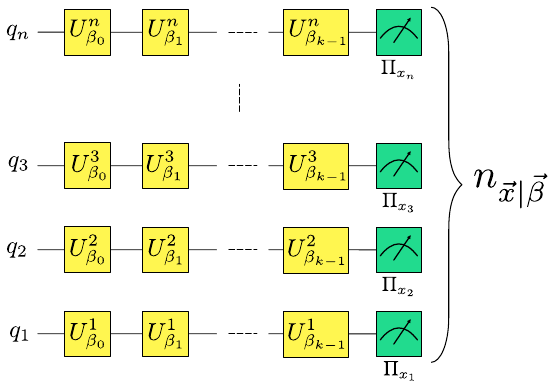}
	\caption{Circuit representation of the data collection procedure for tensor network characterisation of process tensors. }
	\label{fig:tn-circuit-char}
\end{wrapfigure}
As in fully general \ac{PTT}, one should decide on the QSP they wish to estimate, defined across a system $S$ and a series of times $\mathbf{T}_k$. At each time $t_j$, a random Clifford operation $\mathcal{U}_j^i$ is applied to qubit $i$, followed by a final projective measurement in a random Clifford basis, obtaining outcome $x_i$. Let us index the Clifford sequence by $\vec{\beta}$. The circuit for this is shown in Figure~\ref{fig:tn-circuit-char}.
The sequence of unitaries is represented by 
\begin{equation}
	\hat{\mathcal{B}}_{\vec{\beta}} = \bigotimes_{j=0}^{k-1} \bigotimes_{i=1}^n \hat{\mathcal{U}}_{\beta_j}^i,
\end{equation}
with outcome 
\begin{equation}
	\Pi_{\vec{x}} = \bigotimes \Pi_{x_i}.
\end{equation}


A single experiment run with $N$ shots hence returns a series of at most $N$ frequencies $\{n_{\vec{x} \mid \vec{\beta}_0}\}_{\vec{x} = (0,0,\cdots,0)}^{(1,1,\cdots,1)}$ for each of the observed outcomes. 
Correspondingly, the tensor network ansatz can make a prediction for each of the frequencies which depends on the local tensor parameters $\vec{\theta}$:
\begin{equation}
	p_{\vec{x} \mid \vec{\beta}_0}^{\vec{\theta}} = \Tr[\Upsilon_{k:0}^{\theta}\cdot (\Pi_{\vec{x}}\otimes \hat{\mathcal{B}}_{\vec{\beta}_0})].
\end{equation}

The experimental procedure is to run $M$ experiments and collect $N$ shots per experiment, forming the set of runs $\mathcal{R} = \{\vec{\beta}\}_{\vec{\beta} = \vec{\beta}_1}^{\vec{\beta_M}}$ with corresponding dataset $\mathcal{D} = {n_{\vec{x}\mid\vec{\beta}}}$. Our objective function $f$, which quantifies the goodness-of-fit of our model, is the same as that used in Chapter~\ref{chap:PTT}: the log-likelihood:
\begin{equation}
	\label{eq:tn-ll}
	f(\vec{\theta}) = \sum_{\vec{x}}\sum_{\vec{\beta}} -n_{\vec{x}\mid\vec{\beta}}\log p_{\vec{x} \mid \vec{\beta}}^{\vec{\theta}}.
\end{equation}
In practice, evaluating Equation~\eqref{eq:tn-ll} can be computationally arduous, and so at each iteration we randomly select a subset of $\mathcal{D}$ of size $M_{\text{batch}}$. This has the additional benefit of rendering the objective function to be stochastic, which can be useful for overcoming local minima in the optimisation space. 

\subsubsection*{Regularisation}

Our tensor network ans\"atz is defined to be positive, but it has no constraints to ensure it is causal. In Chapter~\ref{chap:PTT}, we rigorously encoded causality into our process tensors by projecting onto the linear space of causal quantum states at each step. This projected gradient descent came equipped with performance guarantees, namely that the final output of the maximum likelihood estimation algorithm would be guaranteed to lie on the intersection of the cone of positive semidefinite matrices with linear space defined by causality conditions. As we turn to larger problems, this approach becomes infeasible due to the exponential scaling of the projection. We encode positivity of the process into our \acs{LPDO} parametrisation, but we propose a more heuristic approach to maintain causality, which is that we regularise it in the objective function.

Recall from Chapter~\ref{chap:process-properties} that we can readily express the set of causality conditions $\Tr_{\mathfrak{o}_k}[\Upsilon_{k:0}] =\mathbb{I}_{\mathfrak{i}_k}\otimes \Upsilon_{k-1:0}\:\forall \:k$ as the requirement that a set of local Pauli expectation values must be zero. If we adopt the perspective that these causality requirements constitute data that we can feed our model, then we can combine these into the experimentally observed data at only slight extra computational expense. Remarkably, we find this to be an extremely effective approach to ruling out acausal estimates. 

We generate some number $M_{\text{causal}}$ of Pauli tensor products $P_{\text{c}} = P_{\mathfrak{o}_k}\otimes P_{\mathfrak{i}_k}\otimes \cdots\otimes P_{\mathfrak{o}_1}\otimes P_{\mathfrak{i}_1}\otimes P_{\mathfrak{o}_0}$ for which causality demands that $\Tr[P_{\text{c}}\cdot \Upsilon_{k:0}^{\vec{\theta}}] = 0$, as determined in Chapter~\ref{chap:process-properties}. We adopt the same notation, where $\mathbf{P}^{(n)}$ are the full set of $n$-qubit Pauli matrices, and $\tilde{\mathbf{P}}^{(n)}$ is the set of traceless $n$-qubit Pauli matrices. 
To generate a set of $M_{\text{causal}}$ Pauli causality constraints $\mathcal{C}:=\{P_{\text{c}}\}$, we repeat the following procedure:
\begin{enumerate}
	\item Select a number $j_0$ between 1 and $k$ (inclusive).
	\item Set $P_{\mathfrak{o}_j} = \mathbb{I}_{\mathfrak{o}_j}$.
	\item Set $P_{\mathfrak{i}_j}$ to be a random member of $\tilde{\mathbf{P}}^{(n)}$.
	\item Set all $P_{\mathfrak{o}_i}$ and $P_{\mathfrak{i}_i}$ to be equal to $\mathbb{I}$, for $j_0 < i \leq k$. 
	\item Set $P_{\mathfrak{o}_i}$ and $P_{\mathfrak{i}_i}$ to be (individually) random members of $\mathbf{P}^{(n)}$ for $0 < i < j_0$.
\end{enumerate}

Evaluating $\langle P_{\text{c}}\rangle$ with respect to a valid process tensor will always result in zero. Hence, we can regularise our tensor network model by evaluating this expectation value across randomly generated sets $\mathcal{C}$. This will bias our optimisation towards valid process tensors (which we may then later verify). 


\subsubsection*{Fitting the Model}

We are now in a position to set up the problem of fitting our tensor network model for a multi-time, multi-qubit quantum stochastic process to some experimental data. The objective function we choose is a sum of the (average) log-likelihood, and the causal regularisation:
\begin{equation}
	\label{eq:tn-objective}
	f(\vec{\theta}) = \sum_{\vec{x}}\sum_{\vec{\beta}} -\frac{1}{|\mathcal{D}|}n_{\vec{x}\mid\vec{\beta}}\log p_{\vec{x} \mid \vec{\beta}}^{\vec{\theta}} + \kappa \sum_{P_{\text{c}} \in \mathcal{C}} \Tr[P_{\text{c}}\cdot \Upsilon_{k:0}^{\vec{\theta}}]
\end{equation}
emphasising again that the respective sums are of size $M_{\text{batch}}$ and $M_{\text{causal}}$, and are chosen randomly at each evaluation. $\kappa$ here is a meta-parameter of the optimisation governing the strength of the regularisation. If it is too small, the optimal model may not be causal; if it is too large then it may slow down convergence. The first term in Equation~\eqref{eq:tn-objective} is called the cross entropy (equivalent to log-likelihood) and attains a minimum value at the data entropy. That is, when $p_{\vec{x}\mid \vec{\beta}} = n_{\vec{x}\mid \vec{\beta}} $ for each $\vec{x}$, $\vec{\beta}$.

Now our aim is to minimise Equation~\eqref{eq:tn-objective} with respect to $\theta$. We do this using the Adam optimiser, which has found remarkable success in optimising stochastic objective functions~\cite{kingma2014adam}. For tensor network semantics, we use the Python library \texttt{quimb}~\cite{quimb}. To obtain the gradients of the objective function with respect to the local tensors, $\nabla_{\vec{\theta}}f$, we use the library JAX for numerical autodifferentiation~\cite{jax2018github}. We additionally collect a smaller validation dataset $\mathcal{D}_v$ on which Equation~\eqref{eq:tn-objective} is evaluated. By computing the likelihood of the model with respect to this validation dataset, we can cross-validate the model to ensure that no-overfitting has occurred.

\textbf{Avoiding Premature Convergence} 

A consequence of adopting a completely positive parametrisation of our process tensor in its tensor network form is that the objective function is now quadratic, and hence the optimisation problem is no longer convex. This is a problem broadly applicable to machine learning and is often tackled by using stochastic optimisation methods such as stochastic gradient descent or Adam. Randomising the data partitions fed into the objective function can be a powerful approach to escape the effects of local minima. Moreover, a common supplementary technique is to begin the optimisation from randomly generated seeds to avoid getting stuck in the same local minima, and ideally find the globally optimal solution.
Although we do adopt these approaches, through a series of trial and error we still find that the problem of tensor network learning can be prohibitively slow, requiring a large number of iterations or cold restarts to converge. This is in general unideal, but it is particularly a problem if the purpose of the characterisation is to feed forward into calibration of a device. Further, one cannot say in general whether failure to converge is a deficiency of the chosen bond dimensions of the model, or simply whether not enough trial seeds have been examined. 
Consider instead that often a process is in the neighbourhood of something expected. By this we mean close to
\begin{equation}
	\Upsilon_{k:0}^{\text{(ideal)}} = \bigotimes_{j=1}^k|\Phi^+\rangle\!\langle \Phi^+| \otimes |0\rangle\!\langle 0|.
\end{equation}
Thus, rather than starting from a random seed, we let each 
\begin{equation}
	[\Phi_{j:j-1}]_{\nu_{j+1}}^{k_{\mathfrak{o}_j}k_{\mathfrak{i}_j}} = [|\Phi^+\rangle]_{\nu_{j+1}}^{k_{\mathfrak{o}_j}k_{\mathfrak{i}_j}} +[\tilde{R}_{j:j+1}]_{\nu_{j+1}}^{k_{\mathfrak{o}_j}k_{\mathfrak{i}_j}},
\end{equation}
where $|\Phi^+\rangle$ is a Bell state made into the appropriate tensor shape by padding out the extra axes with zeros, and $\tilde{R}$ is an equivalently shaped tensor with its entries chosen from a complex Gaussian distribution $\mathcal{N}(0,0.1)$. \par


In all simulated and real experiments, we find that this starting point is not only useful, it is essential to obtain a solution in a reasonable time. Moreover, we find convergence from this starting point in all cases. It is, however, not entirely reasonable to expect that one would always know the process around which to perturb. But we anticipate that the only scenario in which post-processing time is a significant factor is in the event that an experimenter wished to tune up their device to apply, for example, error suppression or correction protocols. In this event, the experimenter will always know their target channel -- the device is aimed at producing a clean, identity set of dynamics. If, however, the purpose of characterisation is to learn a completely unknown quantum stochastic process (for example, from a quantum sensor in an unknown environment), then the post-processing time is not such a significant factor and eventual convergence could be found using the myriad approaches in machine learning.

Another meta-parameter relevant to the problem of obtaining convergence is the number of shots per experiment. This, too, requires some fine-tuning. Too few shots-per-experiment and the landscape may be extremely noisy, preventing the fit to an adequate model. Too many shots-per-experiment, and we tend to find that the model overfits with respect to its data. The tuning here is not particularly fine, and in practice we often find that a large range of values results in an excellent model fit. Nevertheless, we examine this problem heuristically for different problem sizes to find a good practical guide.


\subsection{Demonstrations}
We have constructed both an efficient ansatz for representing multi-qubit quantum  stochastic processes and an algorithm to fit the ansatz to experiment data. We are now in a position to robustly test it. Some fine-tuning is required in the fitting process, and so we point out several best-usage policies which were discovered by trial and error. 

\textbf{Benchmarking Time Taken}

To start, we consider the time taken to contract expectation values of various process tensor networks. Linearly connected tensor networks -- such as \acs{MPO}s -- have efficient algorithms to evaluate, and hence the time taken grows approximately linearly in the number of steps for fixed qubit number. However, it is in general a \texttt{\#P-complete} problem to find the optimal contraction~\cite{PhysRevResearch.2.013010}. \texttt{opt\_einsum}~\cite{daniel2018opt} is a Python package designed for heuristically finding hypercontraction paths for tensor networks. After some exploration, we found that the \texttt{auto-hq} algorithm performed best with our specific topology. Although contraction costs grow exponentially for 2D tensor networks, using this approach we find it is feasible to fit tensor network ans\"atze up to around 4-5 qubits on a personal computer. 

\begin{figure}[htbp]
	\centering
	\includegraphics[width=\linewidth]{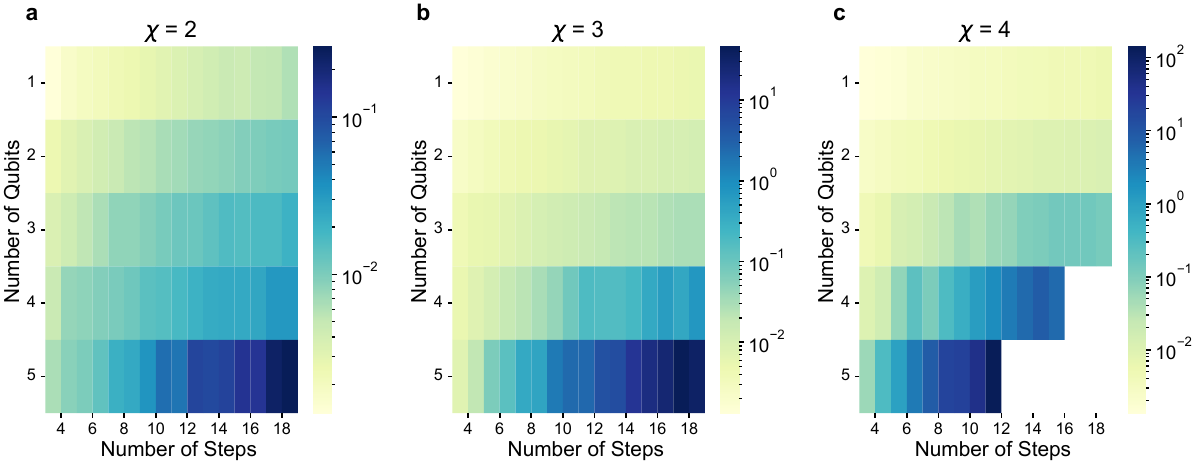}
	\caption[Benchmarking contraction times for different spatiotemporal tensor network expectation values ]{Benchmarking contraction times (given in seconds) for a single expectation value evaluation across different maximal bond dimensions $\chi$, number of qubits $n$, and number of steps $k$. The contraction time grows exponentially in the area of the network. Not all cells are filled in the $\chi=4$ instance due to memory limitations. }
	\label{fig:contraction-times}
\end{figure}

In Figure~\ref{fig:contraction-times} we benchmark the time it takes to perform a single contraction across different values of $k$, $n$, and $\chi$ on an 3.2 GHz 6-Core Intel Core i7 processor. This is performed using the Python library \texttt{quimb} for tensor network semantics, which deploys to the \texttt{numpy.einsum} function for computation. We can clearly identify the scaling behaviour between different network sizes and bond dimensions. Small numbers of times, qubits, and bond dimensions, are feasible on a personal computer, but estimating larger than this may take the use of GPUs or cluster computing.

In practice, while performing the optimisation. we use a compiled version of this code using the \texttt{JAX} library~\cite{jax2018github}, for which the individual contractions are approximately two orders of magnitude faster. Using the Adam~\cite{kingma2014adam} optimiser, we randomise over the data at each iteration of stochastic gradient descent. A good trade-off between memory requirements and convergence rates seems to be to use 1000 data points per iteration of the fit.

\subsubsection*{Demonstrations on Synthetic Data}

We now demonstrate this on synthetic data, scaling both number of qubits and number of time steps. To simulate non-Markovian behaviour, we couple a chain of qubits via a Heisenberg interaction with random couplings to a randomly initialised single-qubit environment. We also include random $ZZ$ couplings between each of the nearest-neighbour qubits in the chain. We are predominantly interested in controlling such non-Markovian systems, and so we will consider only restricted process tensors here. Note, however, that no modifications need to be included to estimate full process tensors. 

As part of our investigation, we look extensively into the effects of different metaparameters on both the quality of the optimisation and the rate of convergence. As we have already mentioned, we use the \texttt{auto-hq} path-finding algorithm to find appropriate contraction paths. 
A good choice of the regularisation metaparameter $\kappa$ seems to be $\kappa=1$: if this is too low, it can sometimes lead to unphysical process tensor estimates; too high, and it can slow down the convergence unreasonably. In our numerical experiments, this choice reliably produces process tensors with acausal Pauli expectation values suppressed to $10^{-7}$ or lower.

\begin{figure}[!htbp]
	\centering
	\includegraphics[width=\linewidth]{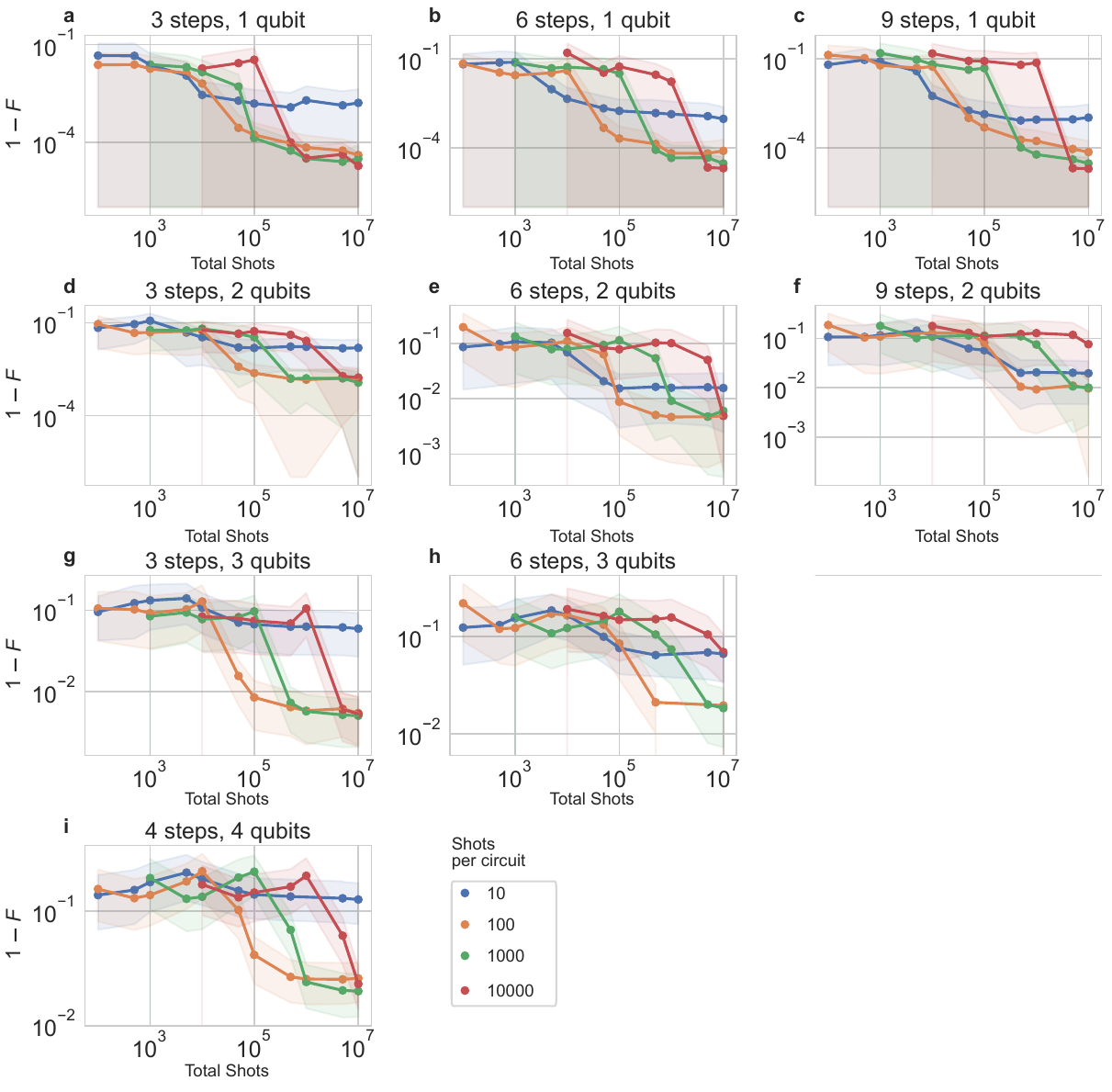}
	\caption[Benchmarking 2D tensor network fits from synthetic data]{Benchmarking 2D tensor network fits from synthetic data. For a variety of step numbers and qubit numbers, we fit tensor network models to qubits coupled via exchange interactions to a non-Markovian bath. We observe convergence properties as dependent on both the total number of shots, and the number of shots per circuit. We see a trade-off between quality and quantity of data, observing that for small numbers of shots, more circuits is preferable and vice versa. In all cases, we see orders of magnitudes fewer circuits required. }
	\label{fig:tn-convergence-graph}
\end{figure}

The next metaparameter we experimented with was the total number of shots per circuit. Given a total number of experiments run, one has the broad choice of collecting either numerous statistics on a few circuits, or few repetitions of many circuits. The trade-off here is quality of the data versus number of points in the space of parameter values. 
For similar problems applied to \acs{QPT}, Ref.~\cite{torlai2020quantum} suggest using single shot measurements across many random circuits. This is in the spirit of unsupervised machine learning, where it is envisaged that a rough optimisation landscape can help to avoid local minima, as well as avoid overfitting to produce an approximately generalisable estimate. We find two issues with this: the first is that for extremely stochastic data, the time taken to converge can be unreasonably long. Secondly, although for very few numbers of shots the final estimate can be marginally better than a tensor network which has overfit, it severely underperforms for even modest amounts of data.

We explore this key point in Figure~\ref{fig:tn-convergence-graph}, where we numerically demonstrate our process tensor network fits for the above dynamics. For a variety of different numbers of qubits and numbers of steps, we fit a process tensor estimate and then compute the reconstruction (in)fidelities for 300 randomly generated validation sequences at $16\ 384$ shots. We perform this for differently sized pools of data, ranging from $10^2$ up to $10^7$.
We highlight that in Chapter~\ref{chap:PTT}, many of the three-step process tensor experiments used $3\times 10^3\times 4096\approx 10^7$ shots. Hence, the upper limit here requires no more data from the quantum device; it is simply used much more efficiently. 
We see in Figure~\ref{fig:tn-convergence-graph} the effects of changing the shots-per-circuit metaparameter in data collection. Note that the \emph{total} amount of data is fixed entirely by the $x$-axis, and is the same for each numbers of shots-per-circuit. That is, a single circuit run $10\ 000$ times will be equivalently compared to 100 circuits each run 100 times. For lower numbers of total data, it appears preferable to choose a lower number of shots-per-circuit. These fits appear to converge to a much more generalisable estimate of the underlying process. 
Likely, this is because the landscape for higher quality data is much smoother, causing the optimisation to either get stuck in a local minimum or simply to overfit. However, at larger numbers of total shots $10^6$--$10^7$, it appears better to smooth that data, since the case of 10 shots-per-circuit tends to hit a plateau and fails to improve substantially. 
The upshot is that 100--1000 shots-per-circuit appears to be a quasi-optimal choice for moderate amounts of data.

Note the dependence of fit quality on number of qubits or number of times: for a given pool of data, the quality of the fit is relatively insensitive to the number of steps. 
This is less true for increasing the number of qubits, since central tensor blocks are not spatially repeated in the same way that they are temporally repeated. Further, increased size of the probability distribution -- as measured by number of measurement outcomes -- will naturally increase variance of the fit. Nevertheless, we see that even for a four qubit process, we still achieve an average reconstruction fidelity close to $0.99$, despite having a comparable number of data points supplied when compared with the three-step process tensors in Chapter~\ref{chap:PTT}. This reduces estimation from the previously infeasible $\mathcal{O}(10^{19})$ experiments, to the much more palatable $10^7$--$10^8$. 
Since we are estimating fewer parameters, we also achieve a greater accuracy than in the fully general version of \acs{PTT}, even for three-step processes, as benchmarked in Chapter~\ref{chap:PTT}.


\subsubsection*{Demonstrations on NISQ Devices}
\label{ssec:nisq-tn}

We now have a robust method for efficiently estimating process tensors with sparse memory structures in practice. We have tested this on synthetic data, but it remains to be seen how this approach performs in a real device setting. On three different IBM Quantum devices, we perform a total of four experiments to capture the underlying process. Each one of these is far larger than previously achievable: a single qubit captured across 18 time steps, two qubits across nine time steps, three qubits across six time steps, and four qubits across four time steps. 

We elected to use 512 shots per circuit in the characterisation. Although fewer shots might be slightly more efficient in terms of total quantum resources, there is a practical trade-off here. When submitting jobs to IBM Quantum, the bottleneck faced is in time taken to load data from a circuit onto the FPGA. Consequently, for example, 1000 circuits at a single shot will take orders of magnitude longer than a single circuit at a thousand shots. 512 shots, then, represents a good middle ground for the data collection. Using a total of $10^7$ shots, we reconstructed each of the above processes with a wait time of $800$ ns -- or about the duration of two CNOT gates -- between each step. This is well within the coherence times of each of the qubits used. 

\begin{figure}[!t]
	\centering
	\includegraphics[width=\linewidth]{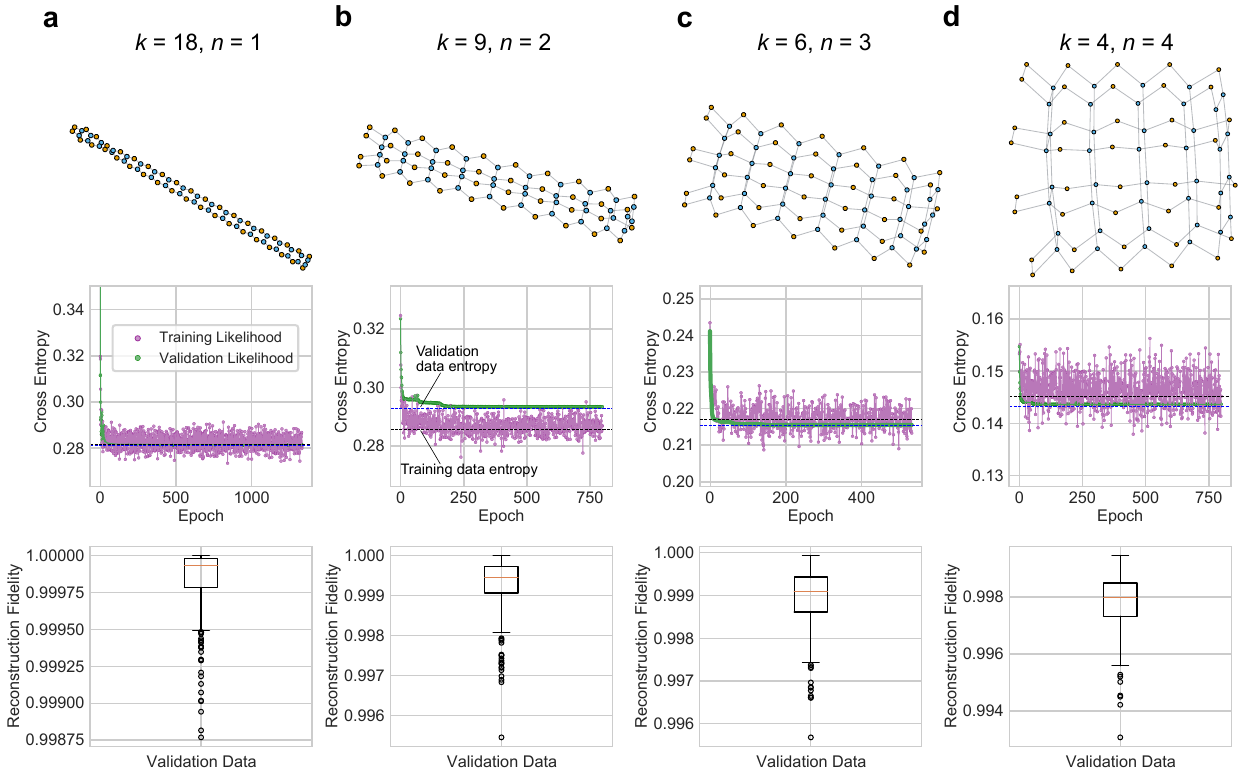}
	\caption[Benchmark estimating large spatiotemporal tensor networks on IBM Quantum devices]{Benchmark estimating large spatiotemporal tensor networks on IBM Quantum devices. We present the convergence of estimation as well as reconstruction statistics of a series of large tensor network reconstructions at a bond dimension of $\chi=2$. In particular, the validity of the model is given by the convergence of the validation cross entropy to the validation data entropy (given as a blue dashed line).
	\textbf{a} 18 steps on a single qubit on \emph{ibm\_cairo}; \textbf{b} 9 steps across two qubits on \emph{ibm\_perth}; \textbf{c} six steps across three qubits on \emph{ibm\_perth}; and \textbf{d} four steps across four qubits on \emph{ibm\_kolkata}. This encodes all Markovian, crosstalk, and naturally occurring non-Markovian dynamics from these instances to a high fidelity. }
	\label{fig:tn-ibm-benchmarks}
\end{figure}

The resulting estimates are given in Figure~\ref{fig:tn-ibm-benchmarks}. The convergence graphs show the cross entropy, or log likelihood, between the predictions made by the estimate and the observed data. We plot this for the estimation data (purple), as well as overlaid with the validation data (green). In dashed lines, we also indicate the entropy of each data set, which is the theoretical minimum of the cross entropy. Note that at each iteration, a random batch of size 1000 of the characterisation data is taken. These data points naturally vary from the entropy of the whole data set. Consequently, it is the convergence of the green curve to the dashed blue line which is the best indicator of goodness-of-fit. In each instance, we see a difference in log-likelihood of less than $5\times 10^{-3}$.

Remarkably, choosing $\chi=2$ is sufficient to capture all crosstalk, non-Markovian, and Markovian dynamics on these devices. In the respective cases, we achieve a median reconstruction fidelity of $1 - 10^{-4}$; $1 - 5\times 10^{-4}$; $1 - 10^{-3}$ and $1 - 2\times 10^{-3}$. 
The results here, when contrasted with simulation in Figure~\ref{fig:tn-convergence-graph} are significantly more accurate. Likely this is because in the simulated case, the dynamics had no extra incoherent noise added, and so there is a larger error faced in attempting to fit a state of higher purity. 

Note that we are using local unitary operations, as the error in each is $\mathcal{O}(10^{-4})$. 
If we wished to include imperfect non-local gates -- such as the native CNOTs -- then we could encode these into the vertical bonds of the model itself. We shall revisit the issue of erroneous control operations in Chapter~\ref{chap:universal-noise}. 
Here, we have demonstrated a highly efficient and highly versatile technique for capturing all models of naturally occurring dynamics. Moreover, we have shown that it functions in the practical test-bed of IBM Quantum devices. 


\subsection{Discussion}

In this chapter, we have studied several methods to leverage the sparseness of memory structures in quantum stochastic processes, enabling more efficient characterisation. In the fully general case, processes will always take exponentially many resources to reconstruct -- much like the complexity of quantum states. 
The first half of the chapter looked at the structure of processes with finite Markov order. This borrows from concepts in classical stochastic processes where the process memory comprises a fixed number $\ell$ of past events. Building on this concept of conditional Markov order, we determined how one can model ongoing sequences of processes by stitching together the overlapping memory blocks. We then proceeded to generalise this to the spacetime setting, where we used the concept of a causal cone to denote the finite memory kernel. We developed an equivalent procedure for stitching together these components. This allows one to approximately represent large many-body quantum stochastic processes with resources growing only logarithmically with the area of the process. It would be an interesting next direction to investigate finite memory structures in archetypal non-Markovian dynamics, such as resonant level models~\cite{PhysRevLett.83.3021}, acoustic phonon modes~\cite{PhysRevB.96.201201}, or anharmonic environments~\cite{PhysRevA.101.012101}.

In the second half of this chapter, we looked at a different type of compression through the use of tensor networks. The physical interpretation here is that rather than using the assumption of memory dissipating with time, instead we consider that the environment responsible for that memory may not be exponentially large. Equivalently, this states that the number of linearly independent updates of future dynamics based on past events may be smaller than the total dimension of the superoperator space. We have developed a method by which the total non-Markovian environment can be encoded into the bonds of tensor networks and then subsequently estimated to a high degree of accuracy.

We envisage a nice symbiosis between these two approaches. 2D tensor networks could, in principle, effectively capture reduced memory models of large process tensors. But evaluating expectation values is exponentially costly. Meanwhile, spacetime Markov order models have memory blocks which may be too large to fully reconstruct, but the overall stitching protocol to reconstruct the process is efficient, even for arbitrarily large processes in space and time. A compromise, then, would be to play to the strengths of both: use process tensor networks to represent the memory blocks, and then use the spacetime Markov order ansatz as an efficient method to build the larger process from these smaller pieces. 

The experimental results demonstrated here show that the number of resources required to characterise spatiotemporally correlated quantum noise need not be large in practice. This offers a huge reduction in the number of resources required to accurately capture these complex quantum effects. One can now apply all the tools of Chapters~\ref{chap:PTT} and~\ref{chap:MTP} to characterise and diagnose non-Markovian noise. We could, for example, use the restricted process tensor bounding procedures from Chapter~\ref{chap:MTP} to determine exact non-Markovian measures across many times.
Furthermore, in Chapter~\ref{chap:NM-control}, we shall see how this methodology can be applied to cancel out correlated noise and optimally control quantum devices.





\chapter{Calibration-Free Universal Quantum Tomography}
\label{chap:universal-noise}
\epigraph{\emph{I am terrified of all the things I feel but cannot see.}}{Jordan Dreyer, The Last Lost Continent}
\noindent\colorbox{olive!10}{%
	\begin{minipage}{0.955\textwidth} 
		\textcolor{Maroon}{\textbf{Chapter Summary}}\newline
		In many real-world experimental settings, characterising non-Markovian quantum stochastic processes is complicated by the lack of exact details about the operation being applied. To compound the problem, control electronics themselves may induce correlations from the experimenter rather than the environment. To overcome these challenges, this chapter aims to develop a generic framework for categorising and characterising quantum noise in such complex scenarios. While previous approaches relied on known control sequences and system-only manipulations, our proposed framework expands the model to require no a priori input information from the experimenter. We present a novel model and a series of experiments designed for efficient characterisation of completely general non-Markovian open quantum systems.
		\par\vspace{\fboxsep}
		\colorbox{cyan!10}{%
			\begin{minipage}{\dimexpr\textwidth-2\fboxsep}
				\textcolor{RoyalBlue}{\textbf{Main Results}} 
			\begin{itemize}
				\item We model a broader categorisation of quantum noise to explicitly include generic control imperfections, environment couplings, and control-environment interactions. We argue that this constitutes a nearly-universal framework of quantum noise.
				\item We implement a mathematical description of instruments which dynamically impact the surrounding environment, combining both process and control interactions.
				\item Building off the tensor network methods of the previous chapter, we show how to characterise the aforementioned model. The end result is a scalable and self-consistent characterisation protocol which is general to all gate-based quantum devices.
			\end{itemize}
		\end{minipage}}
\end{minipage}}
\clearpage

\section{Introduction}

In this thesis so far we have considered quantum stochastic processes in a highly idealised setting. The mathematical characterisation presented is designed to account for correlations between different events in space and time. But we have thus far avoided the implicit requirement that in order to study correlations between events, the events themselves must be known. In other words, we possess infinitely accurate control to probe a system. Such an idealisation may be approximately true in the case of single qubit unitaries on superconducting quantum devices, for which the gate fidelities are much lower than the statistical noise. But it is not even close to generically true, even for simple operations. In Chapter~\ref{chap:MTP}, we circumvented this issue for mid-circuit measurements by first characterising them, but this is not a generally simple or self-contained approach.
As we explored in Chapter~\ref{chap:OQS}, this problem motivated the development of \acs{GST}. \acs{QPT} carried out with invalid assumptions about the state preparation and measurement effects would reconstruct inaccurate and unphysical maps. \acs{GST} hence sought to rectify this problem by making the estimation procedure self-consistent: assume that \emph{nothing} is known, and nothing will be taken for granted. A sequence of experiments is designed to estimate both a set of gates, and the \acs{SPAM} apparatuses used for the characterisation.\par 

In this chapter, we adopt a similar philosophy to \acs{GST} and aim to bridge gaps in the consideration of non-Markovianity. Rather than unknown process and known control, we will take both to be unknown, and both to be estimated.
This forces us to make a choice about the mathematical structures representing the control. We explore this in detail. Our ultimate aim here is to construct an operational framework to describe the dynamics of any noisy controlled quantum device without resorting to dilation or hidden Markov models. The intuition here is that control -- in the abstract -- may be represented as a dual operation to process tensors. 
Thus, correlations generated from the control hardware itself (but not an always-on $SE$ interaction) may be encoded and accounted for. These are only present when the instrument is actually used. Naturally, Markovian control noise also falls under this framework -- for example through coherent overrotation. 
We firstly motivate a categorisation of all possible dynamical effects. This factors broadly into a coupling between system and environment; the effects of control operations on the system (including correlations within the controls); and the effects of control operations on the environment. We then develop a model to mathematically describe the above scenario, and finally present a method to estimate its components. The result is a fully self-consistent extension to \acs{GST}.

We will promote the structure of control operations to incorporate similar effects as non-Markovian processes. This allows much more expressive descriptions such as, for instance, the notion of correlated control operations. We then derive a self-consistent estimation protocol for this model -- a notion of non-Markovian \ac{GST}. We focus on clarity with respect to the effects we are characterising. We hope that even if the literature about quantum non-Markovianity is without semantic agreement, that we can ascribe a model that captures all effects which might \emph{conceivably} be called non-Markovian. 
Now in our categorisation we have an unknown background process, described by a process tensor. We also have a set of unknown control operations, wielded by the experimenter. There is a final missing piece in the framework, which is control-environment interplay. A control operation that enacts some change on the environment is currently undescribable under the process tensor framework. We design an expanded framework to incorporate these effects

Introducing the possibility of imperfect control brings about ambiguity as to what actually constitutes a set of non-Markovian quantum dynamics. Performing a literature search on notions of quantum non-Markovianity is a fraught task. One could come to many alternate conclusions about definitions, measures, and properties of non-Markovian quantum dynamics~\cite{Li2018}. 
Partially, this comes from the evolution of different subfields in quantum information over time, each different purposes and assumptions. It is also due to the fact that many of the formal difficulties in generalising classical notions of non-Markovianity to the quantum setting have only recently been resolved~\cite{Milz2021PRXQ}. 
As we saw in Chapter~\ref{chap:stoc-processes}, some of this is resolvable with a clear hierarchy. For example, information backflow from an environment as measured by non-monotonic distinguishability of different initial states constitutes a sufficient but not necessary condition of non-Markovianity. A formal description of quantum non-Markovianity for some quantum stochastic process relates to the non-factorisability of its process tensor marginals, and incorporates information backflow.

Practically speaking, however, the \acs{QCVV} community typically treats non-Markovianity to be any sequence of dynamical steps that cannot be represented by a series of \acs{CPTP} maps. Indeed, a commonly employed `measure' of non-Markovianity is the total model violation produced when attempting to fit some experimental data to a Markov model~\cite{nielsen-gst}. This definition is heuristically motivated, and broadly does not fit with the current formulation of process tensors.
Part of the issue is that until the present thesis, process tensors (and other variants) had not been considered in the context of quantum noise characterisation. Hence, practicalities -- such as the unknown status of control operations, or the existence of a wall clock -- had not been considered. The resolution here is clarity as to the exact structure of the model: which properties are known. One component is a lack of consensus about what constitutes an environment. It is ambiguous to say that non-Markovian dynamics are simply any form of correlated noise, because depending on which marginalised variable actually generates the correlations may offend different sensibilities. Time-dependent Markovian dynamics, for example, may \emph{look} non-Markovian when coarse grained. But does a wall clock count as an environment? It depends on who you ask. Again, the practicalities of characterisation are at odds with our idealised process tensor framework, and clarity of meaning is required here.\par 

Such a model is invariably complex -- too complex to estimate in full generality in a practical setting. We saw in the previous chapter the power of tensor networks to expressively resolve the scalability issue in \acs{GST}. We thus adapt these techniques to estimate our model. The result is a step towards a universal noise framework, capable of incorporating almost all physical effects experienced by noisy quantum devices. This is both practically useful and acts as a conceptual step forwards in tomographic models of quantum dynamics. 
A key property of the approach we choose is modularity. In all characterisation techniques, maintaining the minimal complete model is key. We discuss how, through the use of tensor networks, different parts of the model may be naturally compressed or expanded as per physical expectations.

\section{Categorising Quantum Noise}

This chapter presents an operational viewpoint which is designed to be highly expressive, capturing and controlling dynamical effects not treated elsewhere in the literature. Moreover, we endeavour to make the approach highly modular.
Before we get into the actual methodology of characterisation, we first discuss in this section our partitioning of quantum dynamical effects that might reasonably be construed as `noise'. The classification is, of course, subjective (and, as we discuss later, in some sense arbitrary). But we aim to meaningfully motivate our choices according to the philosophy of what an experimenter can and cannot control.
\begin{figure}[!b]
	\centering
	\includegraphics[width=\linewidth]{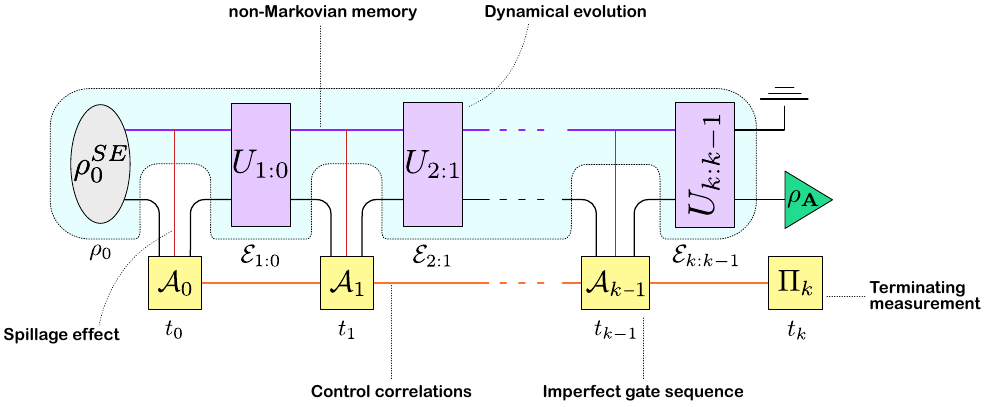}
	\caption[A circuit diagram illustrating different categories of complex noise ]{A circuit diagram schematic of the noise classification in this chapter. Broadly speaking, we separate out the effects of background process from that of experimenter-chosen control. Each component can have its own set of dynamics, including temporal correlations and non-trivial interplay between the two in the form of spillage.}
	\label{fig:noise-categorisation}
\end{figure}

Let us emphasise that from the perspective of quantum stochastic processes, there is no ambiguity on the matter. Non-Markovianity refers to a process whose dynamics do not factor into the product of dynamical maps. 
However, we are speaking in the context of quantum computing here, and must integrate somewhat into the language of the literature. Here, the presence of gate error is of utmost importance. In fact, the distinction between control and process error is often not made in practice. We aim to meet these two communities in the middle with extreme specificity about our contexts.


\subsection{Distinguishing Process and Control}


The resolution of the problem defining a quantum stochastic process came about from the insight that the underlying process and experimenter-implemented control need to be formally separated to account for the non-commutativity of observables at different times. Consequently, we resolve all driven dynamics into two objects: process tensors and their dual, testers. Recall from Chapters~\ref{chap:stoc-processes} and~\ref{chap:process-properties} that a tester is a multi-time instrument with the possibility of carrying memory.

One might argue that there is an epistemic difference between process tensors and testers, but not an ontological distinction. In this context, testers can be viewed as the process that we, as experimenters, have control over. This perspective assumes a certain state of knowledge about the process being implemented as a control operation.
However, when we relax these assumptions, the distinction between the two objects becomes less clear. Intuitively, the process is the piece of dynamics that occurs in each experiment, while control is what allows one experiment to be differentiated from another at the behest of the experimenter. Ultimately, the experimenter is able to effect changes on the dynamics, even if they don't necessarily know what those changes will look like. 
We start with this formal separation of processes as a foundation to categorise, model, and estimate non-Markovian open quantum systems in full generality.

To start with, suppose we have a quantum device represented by a series of qubits $\mathbf{Q}_N :=\{q_1,q_2,\cdots, q_N\}$.
We define the system $S\subseteq \mathbf{Q}_N$ whose state space is $\mathcal{H}_S = \prod_{q_i\in S} \mathcal{H}_{q_i}$.
Everything else, \emph{including the device qubits not in $S$}, is considered to be the environment $E$. This distinction is important for clarifying whether crosstalk-induced dynamics are Markovian or non-Markovian. The system is permitted to evolve across a window $[t_0,t_k]$, divided into $k$ steps to define the set $\mathbf{T}_k:=\{t_0,t_1,\cdots, t_k\}$. Across $\mathbf{T}_k$, the system is manipulated by a multi-time instrument $\mathbf{A}_{k-1:0}$, followed by a \acs{POVM} $\Pi_k$. The instrument may be time local (factor into a tensor product of operations), but it need not be. 

We start with our first definition, covering the underlying dynamics.
\begin{definition}
	A process error is when, without experimenter intervention, we have $S$ evolving according to some Hamiltonian $H$ which differs from the ideal dynamics prescribed by $H_{\text{ideal}}$. That is, it is an always-on interaction between an experimentally accessible system and its experimentally inaccessible environment. 
\end{definition}
Process errors are exactly stochastic noise as we have discussed so far in this thesis. It is stochastic noise that occurs no matter what the actions of an experimenter are. These effects are entirely encoded in the idealised model of a process tensor $\Upsilon_{k:0}$.
We next have the perhaps the most conventional notion of an error in a quantum device, errors due to instruments themselves. 
\begin{definition}
	A control error is an operation, or sequence of operations, designed to manipulate a system whose manipulation of the system deviates from the intended effect. The dynamics should be (a) \underline{controllable:} the effect should be able to be switched on and off by the experimenter, and (b) \underline{not always-on:} the effect should not be present in all experiments, otherwise it is part of the process. This latter condition is sometimes terms \emph{gate-dependent} noise.
\end{definition}
Control errors are dynamical effects on the system whose origin stems from control equipment. Thus, even if the control itself cannot be perfect, these effects can always be switched on or off at will be an experimenter. 

Finally, we consider another unintended, but distinct, consequence of imperfect control operations. 
\begin{definition}
	A superprocess error is a \emph{control operation} (may be turned on or off by the experimenter) which, (a) acts on an extended Hilbert space such that not only the system is manipulated, but part of the environment, and (b) does so in a way that modifies the part of the environment responsible for any present process error.
\end{definition}
The use of \emph{superprocess} terminology will become clear momentarily. Superprocess errors overlap with process and control errors. They are a subset of control errors, in that the physical origin is control equipment and the effect may be switched on or off by the experimenter. But since they also manipulate the broader environment with which the system interacts, these can modify the always-on interaction between $S$ and $E$, and hence change the nature of the process. These are not describable under the present process tensor framework.


\subsection{Subdividing Control and Dynamics}

We will now further subdivide these categories and offer some physical examples for concreteness. Note that these categories are very much overlapping, and the distinction we offer are to properties we believe are worth highlighting in particular. 

\subsubsection*{Markovian Error Channels}
We have seen a great many conditions (both operational and not) for which quantum dynamics might be said to be Markovian. Unequivocally accepted in the literature is the case where a series of control operations $\{\mathcal{A}_i\}$ and time steps $\{\mathcal{E}_{j:j-1}\}$ factorise, i.e., they can be composed: $\mathcal{A}_{j} \circ \mathcal{E}_{j:j-1}\circ \cdots \circ \mathcal{A}_1 \circ \mathcal{E}_{1:0}\circ\rho_0^S$. 
These have been well characterised in the literature~\cite{resch2021benchmarking,noise-coherence-2015,postler2022demonstration}, see especially Ref.~\cite{blume2022taxonomy}. This is consistent with our definition of both process tensor and tester represented as product states, and includes both gate dependent and independent noise. 

\begin{example}[Markovian Process]
	\examplecontent{
	Relaxation processes, quantified by $T_1$ decay, are irreversible. A qubit couples weakly to a cold, large, environment and exchanges energy 
	with the environment. But qubits are engineered to have resonant frequencies $\omega \gg k_B T$ for ambient temperature $T$, meaning that energy gain by the qubit is exponentially suppressed~\cite{krantz2019quantum}. Practically, then, energy leaks in one direction  until the qubit has relaxed into its lowest energy state. The coupling is fixed, and hence stationary. 
	}
\end{example}

\begin{example}[Markovian Control]
\examplecontent{
	A qubit is driven around its $x$-axis with pulse envelope $e(t)$ and coupling strength $h_d$ between the qubit and the drive field. The resulting time evolution operator is given by 
	\begin{equation}
		U(t)=\exp\left(\frac{i}{2}h_d \int_0^{t}e(t')\text{d}t' X\right).
	\end{equation}
	This equation can be solved to rotate about the $x$-axis by angle
	\begin{equation}
		\Theta(t) = -h_d\int_0^t e(t')\text{d}t'.
	\end{equation}
	If the coupling strength $h_d$ is not accurately known, the pulse envelope will not correspond to the desired $\Theta(t)$ and will be a coherent over or under rotation compared to the ideal gate.
}
\end{example}

\subsubsection*{Active and Passive Crosstalk}
Crosstalk denotes the presence of undesirable interactions between different qubits on a device. We first make the distinction that crosstalk only refers to the case where the qubits involved are entirely contained in $S$. Otherwise, if not accounted for, the qubits constitute an environment and will be treated differently. We make the further distinction between \emph{passive} and \emph{active} crosstalk. 

\begin{definition}
	For qubits $\{q_i\}\in S$, if the uncontrolled dynamics feature at least a weight-2 Pauli term in the Hamiltonian, this has the capacity to generate entanglement between different qubits, irrespective of the control operations. This is said to be \emph{passive crosstalk}.
\end{definition}

\begin{example}[Passive Crosstalk]
	\examplecontent{
	Fixed-frequency transmon qubits with fixed couplings enables entangling interactions between qubits connected to the same coupling bus~\cite{krantz2019quantum}. But the coupling leads to state-dependent frequency shifts of coupled qubits in the form of a static $ZZ$ term in the interaction Hamiltonian~\cite{PhysRevLett.129.060501}. The coupling dresses the energy levels of each qubit: if either qubit is in a $Z$-eigenstate, this resonance shift causes coherent $Z$-rotations in the other. If both qubits have non-zero transverse Bloch sphere components, this will lead to an entangling interaction. 
	}
\end{example}

\begin{definition}
	If a control operation $\mathcal{A}$ is intended to act on some qubit $q_j\in S$, but instead also manipulates other qubits within the system, this is said to be \emph{active crosstalk}.
\end{definition}

\begin{example}[Active Crosstalk]
	\examplecontent{
	In optically-addressed trapped ion devices, a widely employed method to implement entangling gates is known as the M\o lmer-S\o rensen (MS) gate~\cite{PhysRevLett.82.1835}. For a given string of qubits $q_1,q_2,\cdots,q_n$, the gate uses a common vibrational mode to mediate entanglement, provided that the laser illuminates each qubit in the string uniformly. 
	If the beam of the laser is not tightly controlled, then residual light can illuminate unintended neighbouring ions. This allows them to participate in the same MS interaction, driving unwanted entangling interactions whenever the MS gate is used~\cite{ParradoRodriguez2021crosstalk}.
	}
\end{example}

\subsubsection*{Process non-Markovianity}

\begin{definition}
	The always-on interaction between a system $S$ and its environment $E$ defines a quantum stochastic process, whose representation is given by a process tensor \pt{}. These dynamics are said to be \emph{process non-Markovian} if \pt{} cannot be expressed as a product of its marginals. That is:
	\begin{equation}
		S(\Upsilon_{k:0}\mid\mid \bigotimes_{j=1}^k \hat{\mathcal{E}}_{j:j-1}\otimes \rho_0) > 0.
	\end{equation}
\end{definition}
This is exactly the definition of non-Markovianity that we have employed throughout this thesis so far, introduced in Chapter~\ref{chap:stoc-processes}. That is, we have temporal quantum correlations generated by an environment with which a system interacts. The reason we have included the `process' modifier in the label is that now we are including control operations in the world view, we wish to be precise about the physical origin of the correlations. 

\begin{example}[Non-Markovian Process]
	\examplecontent{
	The electron and nuclear spins of a $^{31}$P donor in isotopically pure $^{28}$Si (which has nuclear spin 0) can be used as qubits. However, if the silicon is not isotopically pure, then both electron and nuclear spin may coherently interact with the extraneous silicon nuclei. This always-on interaction will generate temporal correlations in the dynamics of the systems. 
	}
\end{example}

\subsubsection*{Control non-Markovianity}

\begin{definition}
	For a system probed across multiple times by a multi-time instrument $\mathbf{A}_{k:0}$, we say that the hardware is \emph{control non-Markovian} if the (appropriately normalised) instrument does not factor into a product of its marginals. That is:
	\begin{equation}
		S(\hat{\mathbf{A}}_{k:0}\mid\mid \Pi_k\otimes \bigotimes_{j=0}^{k-1} \hat{\mathcal{A}}j) > 0,
	\end{equation}
\end{definition}
for trace-normalised $\hat{\mathbf{A}}_{k:0}$ and marginals. Note that this is regardless of whether the control correlations are deliberately created. 

\clearpage

\begin{example}[Non-Markovian Control]
	\examplecontent{
	Ion trap qubits are optically addressed by lasers for their single qubit gates. The control lasers may have quasistatically fluctuating power spectra. An unaccounted for power fluctuation (i.e., if the gate time is not altered) has the effect that the unitary will coherently over or under rotate about its rotation axis. If the power fluctuates on the timescale of the circuit, then each $\mathcal{A}_i$ may over/under rotate by the same amount, some random variable $\epsilon$ drawn from the power spectrum of the laser. The error is then coherently correlated across different gates. 
	}
\end{example}


\subsubsection*{Time-Dependent Markovianity}

The next category we consider is that of time-dependent Markovianity. Some authors in the literature consider this to be non-Markovian, because it breaks the tenets of stationarity, even if not the tenets of temporal locality. 
We start with a circuit clock $\{t_0,t_1,\cdots,t_k\}$ to designate the spacing between steps in a process. This clock is reset with every run of the circuit.
\begin{definition}
	A $k$-step process whose process tensor Choi state \pt{} can be expressed as 
	\begin{equation}
		\Upsilon_{k:0} = \bigotimes_{j=1}^{k} \hat{\mathcal{E}}_{j:j-1}\otimes \rho_0,\:\text{if } \exists\: j\neq l\text{ s.t. }\hat{\mathcal{E}}_{j:j-1} \neq \hat{\mathcal{E}}_{l:l-1}
	\end{equation}
	is said to have \emph{time-dependent process Markovianity}.
\end{definition}

Additionally, we have instruments which do not behave as expected:
\begin{definition}
	A gate $\mathcal{A}$ is said to be \emph{time-dependent control Markovian} if, for $k$ applications of $\mathcal{A}$, its multi-time instrument is
	\begin{equation}
		\begin{split}
		\hat{\mathbf{A}}_{k-1:0}  &= \bigotimes_{i=0}^{k-1}\hat{\mathcal{A}}_i,\:\text{but}\\
		\hat{\mathbf{A}}_{k-1:0}&\neq \bigotimes_{i=0}^{k-1}\hat{\mathcal{A}},
		\end{split}
	\end{equation}
	where $\hat{\mathbf{A}}_{k-1:0}$ is the repeated application of $\mathcal{A}$. That is, the tester is a product state (and hence carries no correlations) but the effect of the gate takes on a time dependence.
\end{definition}
A key distinction with control here is intent. Of course, if one were to apply a series of different perfect gates, then this constitutes time-dependent control Markovianity in a sense. To tighten this, we consider it to be the case that one \emph{intends} to repeat a gate, and the action of the gate changes over time, even if the effects are still time-local.


\begin{example}[Time-dependent Markovianity]
		We have a faulty $X$ gate whose properties we wish to determine. We perform a \acs{GST} experiment with $k$ sequential applications of the gate for various values of $k$. But the gate degrades over time due to the laser heating up: at $t_0$, a perfect $R_X(\pi)$ rotation is performed. At $t_1$, $R_X(\pi + \theta_1)$ occurs, where $\theta_1$ is drawn randomly from some distribution. At $t_2$, $R_X(\pi+\theta_2)$ occurs where $\theta_2$ is drawn from a different distribution that has no covariance with $\theta_1$ -- and so on. Attempts, then, to fit this the one gate to different $(X)^k$ expressions will fail.
		The situation is rectified by treating each circuit time as an independent gate, collected in $\{X_0, X_1, \cdots, X_{k-1}\}$. The dynamics is then accurately modelled by the composition of these. 
\end{example}

\subsubsection*{Process and Control Drift}

\begin{definition}
	\emph{Drift} is if: with respect to a wall clock, either the underlying model for control, or the underlying model for process/control, changes with time.
	Suppose we have $M$ experiments (shots), each taking place at some wall clock time $\{s_1,\cdots s_M\}$. The wall clock time distinguishes between different runs of an identical circuit. Both process \pt{} and control $\mathbf{A}_{k:0}$ may be indexed by the times at which experiments are run to collect data. If there is any variation within the sets $\{\Upsilon_{k:0}^{(s_j)}\}$ and $\{\mathbf{A}_{k:0}^{(s_j)}\}$, then either the process or the control is said to have drifted, respectively. 
\end{definition}

We make the distinction that time-dependent Markovianity is a change in frequency on the timescale of the circuit, whereas drift more generally is on the timescale of the experiments.

Depending on timescales, drift can be very difficult to characterise.
Almost by definition, it violates the primary foundation of quantum tomography, which is the assumed ability to be able to prepare identical states or processes a large number of times. 
Although it cannot always be characterised, it \emph{can} be described. 
Indeed, one could always consider the effects of drift to be the property of dynamics with frequencies slower than the repetition rate of a device. For example, a process tensor whose times extended to the wall clock, rather than purely the circuit clock. 
Approaches to deal with drift in the literature, therefore, are primarily heuristic. One powerful approach, introduced in Ref.~\cite{proctor2020detecting}, is to perform spectral analysis on time-series data. This can then resolve time-dependence in a manner agnostic to the experiment.

The difficulty in integrating drift into the process tensor framework is because drift is a practical effect that arises as a consequence of the time it takes to characterise. The process tensor is sensitive only to a circuit clock, rather than a wall clock. The following example emphasises how marginalising over a wall clock can induce an identical non-Markovian ensemble as a classical environment.
\begin{example}[Drift]
	\examplecontent{
		Consider the reconstruction of two separate two step processes on a single qubit.
		\begin{enumerate}
			\examplecontent{
			\item The system $S$ is coupled to a single qubit environment $E$, which is initialised in a $\ket{+}$ state. The process then undergoes a series of two CNOT gates controlled by $E$ and targetted on $S$. This is non-Markovian because $E$ is a random variable controlling the correlated evolution of the system. If $E=\ket{0}$, the system undergoes a series of identity evolutions. If $E=\ket{1}$, it will be subject to a series of $X$ gates. The dynamics are the same across the entire period that an experimenter reconstructs the process tensor.
			\item The system is isolated from its environment but driven with faulty control electronics. An experimenter aims to reconstruct a process tensor by collecting statistics across a ten-minute period. Unbeknownst to the experimenter, for the first five minutes, the dynamics are perfect. But after five minutes, the control electronics randomly switch to applying sequential $X$ gates.} 
		\end{enumerate}
		In this scenario, the two reconstructed process tensors will be identical:
		\begin{equation}
			\Upsilon_{3:0} = \left(0.5|\Phi^+\rangle\!\langle \Phi^+|\otimes |\Phi^+\rangle\!\langle \Phi^+| + 0.5 |\Psi^+\rangle\!\langle \Psi^+|\otimes |\Psi^+\rangle\!\langle \Psi^+|\right) \otimes \rho_0.
		\end{equation}
		That is, the process is (classically) non-Markovian. In the first instance, a nearby qubit is responsible for the correlations. In the second, although it might be described nominally as Markovian, once we marginalise over the wall clock time variable we obtain the same non-Markovian process. The problem here is that the process tensor framework does not account for the wall clock, we have assumed an identical and repeated process in our experiments. 
	}
\end{example}


Effects of drift are non-Markovian in the sense that we can reconstruct process tensors that have correlations between different points in time. An important distinction to make is that we are marginalising over a time variable rather than a space variable. We can evaluate a perfectly good process tensor model for the time-averaged characterisation. But as soon as we attempt to apply our model to predict dynamics, we will be conditioning on a single value for time, at which point the model will break down. 
This is a classic example of Allan variance: where collecting more data would make the model worse rather than better~\cite{galleani2009dynamic}.
The introduction of clocks somewhat obfuscates the ontological existence of processes in a single-shot setting. Treating drift and time-dependent Markovianity introduces an epistemic ignorance into the matter, we cannot in practice always condition on the parameter values of a clock. Here, then, time is a basic form of memory. 

The presence of drift can be witnessed by applying spectral analysis techniques to any sets of experimental data to look for difference~\cite{proctor2020detecting}. 
One can mitigate the effects of drift on sets of data -- for example by rastering experiments rather than running them serially -- but ultimately low-frequency drift is a fundamental hardware issue, and little can be done to mitigate its effect in software.
In the case of our process tensor experiments, we can perform confidence checks to ensure that drift has not affected the characterisation. 
The circuits used for characterisation will be collected across some set of wall clock times, $s_0, s_1,\cdots$. We also collect validation data to compute reconstruction fidelities, taken either at the start ($s_0$) or the end ($s_f$). The validation data, then, is conditioned on a single time. Hence, if there is drift, the process tensor model we estimate will fail to accurately describe the validation data. If it \emph{does} describe it to high accuracy, then we can be reasonably confident that low-frequency drift has not significantly affected the accuracy of the model. 
For extra certification, one might consider evaluating the same validation circuits at both the start and the end of the collection period. 


\subsubsection*{Environment Spillage}

We have so far considered two primary mechanisms for noise in quantum processors: possibly correlated background dynamics, and possibly correlated control operations. We lastly introduce a third possible mechanism, which we formally refer to as a \emph{superprocess} error, and colloquially as a \emph{spillage} error. This mechanism describes the situation where the control operation implemented by an experimenter drives a range of frequencies, including some which are resonant with environment transitions. The effect, then, can be not only to manipulate the system, but also the environment with which it interacts. Since we do not have any intentional access to the environment, we cannot simply dilate the dynamics to include it. Instead, we continue with our open systems philosophy and model how such effects can modify the dynamics with respect to our system. Specifically, we defined spillage effects as follows.

\begin{definition}
	If an intended control operation $\mathcal{A}_j\in \mathcal{B}(\mathcal{H}_S)$ at time $t_j$ excites some transition \emph{outside} $\mathcal{H}_S$, \emph{and} there exists a \acs{QPT} experiment for a future dynamics $\mathcal{E}_{j':j'j-1},\ t_{j'} > t_j$, which can detect the effects of $\mathcal{A}_j$, then we term this either a superprocess or a spillage error.
\end{definition}

At first glance, it might seem that spillage errors are the same as non-Markovian environment back-action. In both cases, we could have two instruments $\mathcal{A},\mathcal{B}$ applied at time $t_j$. Then, conditioned on these instruments, we have two different future dynamics at some time $t_{j'}$: $\mathcal{E}_{j':j'-1}^{(\mathcal{A})}$ and $\mathcal{E}_{j':j'-1}^{(\mathcal{B})}$. The distinction here is that spillage errors need not be linear in the control. That is, for control operation $\mathcal{C} = \alpha\mathcal{A} + \beta\mathcal{B}$ we may have 
\begin{equation}
	\mathcal{E}_{j':j'-1}^{(\mathcal{C})}  \neq \alpha\mathcal{E}_{j':j'-1}^{(\mathcal{A})} + \beta \mathcal{E}_{j':j'-1}^{(\mathcal{B})}.
\end{equation}
Hence, the above scenario cannot be described with process tensors alone. 

\begin{example}[Superprocess Errors]
	\examplecontent{
	A quantum system passively interacts with some two-level system (TLS) defect in the environment. An experimenter applies a control operation to $S$ by switching on a drive field with pulse envelope $e(t)$. The Fourier transform of the envelope reveals a spectrum with frequencies resonant with the TLS. Some population of the TLS is then driven from the ground $|g\rangle$ to the excited $|e\rangle$ state. The system is then evolving with respect to a \emph{different} environment, and hence no longer described by the same quantum stochastic process.
	}
\end{example}

Physically, this category of noise is describing the implementation of control whose action affects the environment. 
It is identical, in fact, to the case of active crosstalk but without the constraint that the affected subsystem is a part of $S$.
For example, this incorporates well-studied active \emph{leakage} errors, where the state of a qubit leaves the computational subspace as a consequence of active driving~\cite{varbanov2020leakage,strikis2019quantum}.
Note the distinction here: with non-Markovian quantum processes, an instrument modifies a system at some time $t_i$ and then, by virtue of a naturally occurring system-environment interaction, this affects the environment before $t_{i+1}$. Here, however, the control operation is taken to directly map the state of the environment. That is,
\begin{equation}
	\mathcal{A}_j\in \mathcal{B}(\mathcal{H}_S\otimes \mathcal{H}_E).
\end{equation}
In practice, we expect this to only be an incredibly small subspace of $E$. Nevertheless, we incorporate the whole environment in our descriptions, and then later compress. 

A useful way to discuss spillage in the language of process tensors is with \emph{superprocesses}. A superprocess~\cite{PhysRevA.81.062348} is a way of describing transformations on a particular open quantum evolution, and are typically used in the context of resource theories~\cite{berk,berk2021extracting}. A higher order map $\mathbf{Z}_{k:0}$ is a mapping from process tensors to process tensors, or equivalently its dual action is from control sequences to control sequences. Recall from Chapter~\ref{chap:stoc-processes} that in the superoperator representation, the multi-time expectation value of some sequence of control operations $\mathbf{A}_{k:0}$ with respect to a process \pt{} is 
\begin{equation}
	\langle\!\langle \hat{\mathbf{A}}_{k:0} | \Upsilon_{k:0}\rangle\!\rangle.
\end{equation}
A superprocess extends this notion to 
\begin{equation}
	\langle\!\langle \hat{\mathbf{A}}_{k:0} |\mathbf{Z}_{k:0}| \Upsilon_{k:0}\rangle\!\rangle,
\end{equation}
where there is an assumption that $\mathbf{Z}_{k:0}$ retains causal order of control operations on input and output spaces. In other words, superprocesses are completely positive and causality-preserving transformations of process tensors: $|\Upsilon_{k:0}\rangle\!\rangle \mapsto \mathbf{Z}_{k:0} |\Upsilon_{k:0}\rangle\!\rangle  = |\Upsilon_{k:0}'\rangle\!\rangle$. Spillage noise, as we have defined it, is a control action on both system and environment. We can view this equivalently then as a type of transformation on the process tensor. Without loss of generality, take the control operation $\mathcal{A}_0$ to be applied at time $t_0$. If $\mathcal{A}_0$ acts on system only, then the future dynamics are well described by $\Upsilon_{k:1}^{(\mathcal{A}_0)} = \Tr_{\mathfrak{i}_1\mathfrak{o}_0}[\hat{\mathcal{A}}_0^{\text{T}} \Upsilon_{k:0}]$. But if $\mathcal{A}_0$ also acts on the environment in some ill-defined fashion, then we still obtain a valid future process, but it is no longer defined by contraction of $\hat{\mathcal{A}}_0$. Instead, we denote this by $\Upsilon_{k:1}'^{(\mathcal{A}_0)}$, where 
\begin{equation}
	\label{eq:superprocess-expectation}
	|\Upsilon_{k:1}'^{(\mathcal{A}_0)}\rangle\!\rangle = \mathbf{Z}_{k:0}^{(\mathcal{A}_0)} |\Upsilon_{k:0}\rangle\!\rangle.
\end{equation}

\begin{figure}[!b]
	\centering
	\includegraphics[width=0.6\linewidth]{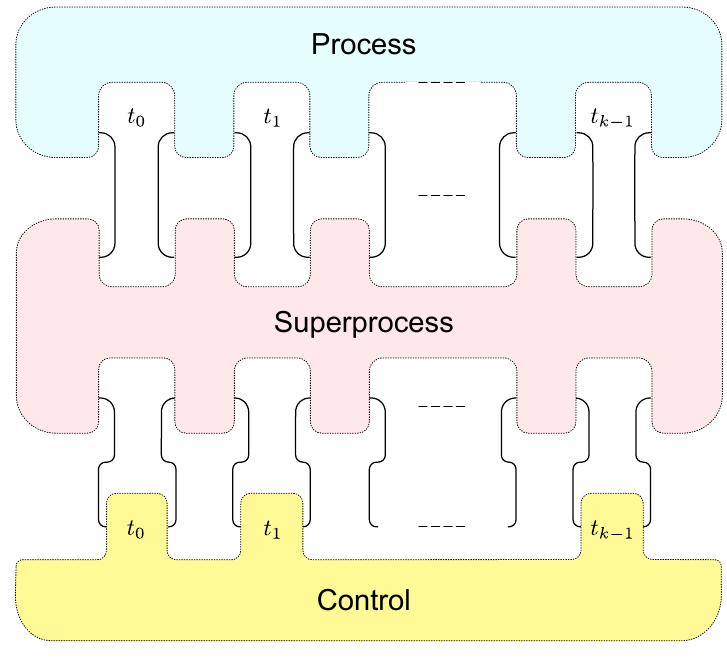}
	\caption[General model of driven quantum dynamics with arbitrary errors ]{General model of driven quantum dynamics with arbitrary errors. We conceptually partition process, control, and spillage effects from one another. These are respectively represented by process tensors, testers, and superprocess objects. This formalism captures a general setting of system-environment interaction, faulty correlated control, and active overlap between control and environment.}
	\label{fig:superprocess}
\end{figure}

Here, we have introduced $\mathbf{Z}_{k:0}^{(\mathcal{A}_0)}$ as the superprocess which (i) implements $\mathcal{A}_0$, and (ii) transforms $\Upsilon_{k:1}$:
\begin{equation}
	\bigotimes_{i=0}^k\mathcal{\mathcal{B}}(\mathcal{B}(\mathcal{H}_S))\rightarrow \bigotimes_{i=1}^k\mathcal{\mathcal{B}}(\mathcal{B}(\mathcal{H}_S) ).
\end{equation}
We illustrate the described picture in Figure~\ref{fig:superprocess}.
This is still extremely general. After all, the new process could be \emph{anything}. In principle, we could throw away our previous environment and replace it with something entirely different. 
To ensure that the transformation is not too significant, we once more incorporate the language of tensor networks to model a sparse set of spillage errors only. 

Note, however, that we do not actually \emph{require} the superprocess representation. Since the spillage effect is intrinsically tied to the control operation, we can always absorb $\mathbf{Z}_{k:0}^{(\mathcal{A}_0)}$ into $\mathbf{A}_{k:0}$ to create $|\mathbf{A}_{k:0}'\rangle\!\rangle := \mathbf{Z}_{k:0}^{(\mathcal{A}_0)} |\mathbf{A}_{k:0}\rangle\!\rangle $. The two models are physically indistinguishable, and so we opt for the smaller one.
We described the situation with superprocesses for conceptual reasoning, but in practice this shows that we do not need to explicitly \emph{model} the superprocess, as it is always contained in the tester.
In full generality, the bond dimension of $\mathbf{A}_{k:0}'$ would be the dimension of the future process tensor $\Upsilon_{k:j+1}$ itself, but in practice we can significantly compress things so long as the control only modifies a small part of the environment.

\section{Formalism for Self-Consistent Estimation}

We have explored the various categories of dynamics we wish to characterise in practice. We now derive a method to incorporate both experiment design and reconstruction of these general process and control dynamics. This is intended to be in the spirit of \acs{GST}: we take nothing for granted about the underlying gate set, and instead aim to self-consistently estimate the underlying dynamics. With this formalism in hand, we will then turn again to tensor networks to aid our practical compression and estimation in an experimental setting.





The language of \acs{GST} is to start with a gate set $\mathcal{G} = \{\kket{\rho},\bbra{\mathcal{J}}, G_0, G_1,G_2,\cdots,G_K\}$ which includes a reliably preparable initial state, a \acs{POVM}, and the set of control operations that an experimenter may implement. 
The set of gates contains two restrictions: it must include the special case of $G_0 = \{\}$, i.e. the do-nothing-for-no-time gate\footnote{The $G_0$ gate must not simply have its target be the identity, it must be a perfect implementation of the identity superoperator. Experimentally, this amounts to doing \emph{literally} nothing.}; and compositions of the gates must be able to be able to generate an \acs{IC} set of state preparations and measurements.
Rather than taking any one of these objects for granted, the experiment design seeks to simultaneously estimate the entire gate set self-consistently. 

The tenets of \acs{GST} are highly relevant to near-term quantum devices: \acs{SPAM} errors are significant, and gates not perfect. Thus, it is important to account for each of these noisy mechanisms individually in order to paint a holistic picture of device performance. But \acs{GST} has its drawbacks. Most notably, it adopts a totally Markovian model. Probabilities are obtained from the model by matrix multiplication, which is only an accurate representation if all dynamics (both process and control) are completely time local, an inadequate assumption in practice. Moreover, gates themselves as well as the surrounding process are combined into the one object, and no distinction is made between the two.
We have seen the ability of process tensors to encode temporal correlations, as well as to be estimated in practice. 
We wish now to further develop our formalism to incorporate the philosophy of \acs{GST} -- to assume nothing about the status of our control operations -- but to also characterise non-Markovian processes as we have so far done. The result is a robust estimate of the process tensor, as well as the possibly-noise interventions used to probe that process.

Let us clearly lay out our goals. In \acs{PTT}, we use a known basis of multi-time instruments $\{\mathbf{B}_{k-1:0}^\mu\}$ and terminating \acs{POVM} $\mathcal{J}$ to determine an underlying quantum stochastic process $\Upsilon_{k:0}$. Now, suppose that both instruments and measurements are taken to be unknown. We have an equivalent `gate set': $\{\Upsilon_{k:0}, \mathcal{J}, \mathbf{B}_{k-1:0}^{(1)}, \mathbf{B}_{k-1:0}^{(2)},\cdots,\mathbf{B}_{k-1:0}^{(K)}\}$ which we wish to estimate. This encodes both Markovian and non-Markovian \emph{process} errors into \pt{}, \emph{control} errors into $\{\mathbf{B}_{k-1:0}\}$, and measurement errors into $\mathcal{J}$. Our intent for this section is to first formally construct a fully general non-Markovian equivalent to \acs{GST}. This establishes a firm foundation on which we can consider self-consistent \acs{PTT}. However, this general formalism is simply too large to treat in practice, so we will chip away at the model to apply simplifications before eventually constructing a computationally efficient method to characterise such processes using tensor networks as our tool.

Using \acs{GST} as a foundation, we can derive a similar procedure that incorporates process tensors to account for any non-Markovian memory. To set the intuition, we can suppose that the resulting Choi state from a \acs{CJI} circuit could be used as an input state to a \acs{GST} experiment on $2k+1$ qubits. Before, we had a single fiducial state $|\rho\rangle\!\rangle$, which we could reliably prepare but not necessarily know; a two-outcome dual effect $\langle\!\langle E|$ that allows us to make observations; and a set of linear transformations $\{G_i\}$. We now make the identification of a $k$-step process whose dynamics we can reliably prepare -- but not necessarily know. The dual to a process is a multi-time instrument. 

Self-consistently determining multi-time control operations will require an \acs{IC} basis of process tensors $\{\Upsilon_{k:0}^\nu\}$. One could take this to mean the independent generation of a set of process tensors by manipulating the environment, which will be difficult in general. Alternatively, we can take a single fiducial process \pt{} and use control operations to transform it at the system level to generate an \acs{IC} set. This is equivalent to having a single preparable state $\rho$ and generating an \acs{IC} set of states by transforming the one state with gates $\{G_i\rho G_i^\dagger\}$. We also require an \acs{IC} set of testers. In keeping with the \acs{GST} analogy, we fix a single fiducial tester which shall be the `do-nothing' operation at each time $t_j$, followed by measurement with \acs{POVM} $\mathcal{J}$. This single tester can then be transformed by composing the do-nothing operation with some form of control.
The reason for these distinctions will become clear momentarily.


Both process and instruments can be transformed with superprocess transformations. This is, of course, a very general statement. Any change in the dynamics or the control instruments is indeed a superprocess mechanism. Typically, we use an instrument $\mathcal{A}_j$ to extract the state from a process at some $\mathfrak{o}_j$, transform it, and then feed into $\mathfrak{i}_{j+1}$. But suppose we fixed the instrument as a part of our process to define new process tensor $\Upsilon_{k:0}'$. The new output leg $\mathfrak{o}_j'$ is now the old process but with $\mathcal{A}_j$ applied. What we have done is rather than contract an operation into the process, we have \emph{composed} an operation with the process. For a Choi state, this is performed through the link product: 
\begin{equation}
	\Upsilon_{k:0}' = \Tr_{\mathfrak{o}_j}[\hat{\mathcal{A}}_{\mathfrak{o}_j'\mathfrak{o}_j}^{\text{T}_{\mathfrak{o}_j}}\Upsilon_{k:0}].
\end{equation}
In the extended \acs{PTM} representation, this is simply matrix multiplication:
\begin{equation}
	|\Upsilon_{k:0}'\rangle\!\rangle = \mathbf{Z}_{k:0}^{(\mathcal{A}_j)} |\Upsilon_{k:0}\rangle\!\rangle,
\end{equation}
and similarly for input legs. Note that if $\mathcal{A}_j$ is a trace-decreasing map, then the resulting process tensor must be trace-normalised in order to remain causal. This amounts to post-selecting on instrument outcomes in practice. 
From the states, gates, and measurements, we make the identification:
\begin{equation}
	\begin{split}
		|\rho\rangle\!\rangle &\rightarrow |\Upsilon_{k:0}\rangle\!\rangle, \\
		\langle\!\langle E | &\rightarrow \langle\!\langle \mathcal{J} \otimes \mathcal{I}\otimes\cdots\otimes \mathcal{I}|,\\
		\mathcal{G} = \{G_l\} & \rightarrow \mathscr{B} = \{\mathbf{B}_{k-1:0}^{\mu}\}.
	\end{split}
\end{equation}

We will drop the time subscripts temporarily to make indexing clearer. Also, let $\bbra{\mathcal{J}}$ be a proxy for $\langle\!\langle \mathcal{J} \otimes \hat{\mathcal{I}}\otimes\cdots\otimes \hat{\mathcal{I}}|$.
We define a fiducial set $\mathscr{F} = \{\mathbf{F}_1,\mathbf{F}_2,\cdots, \mathbf{F}_N\}$ to be drawn from $\mathscr{B}$. For simplicity, this is best chosen to be a deterministic subset such as sequences of unitary operations. In line with previous chapters, we denote the single-time components of each tester without the bold face, and with a subscript denoting the time.
Iterating over each of these sets generates the probabilities
\begin{equation}
	p_{ij\mu} = \langle \!\langle \mathcal{J}|\mathbf{F}_i\mathbf{B}_\mu \mathbf{F}_j| \Upsilon\rangle\!\rangle.
\end{equation}

\begin{figure}[!t]
	\centering
	\includegraphics[width = \linewidth]{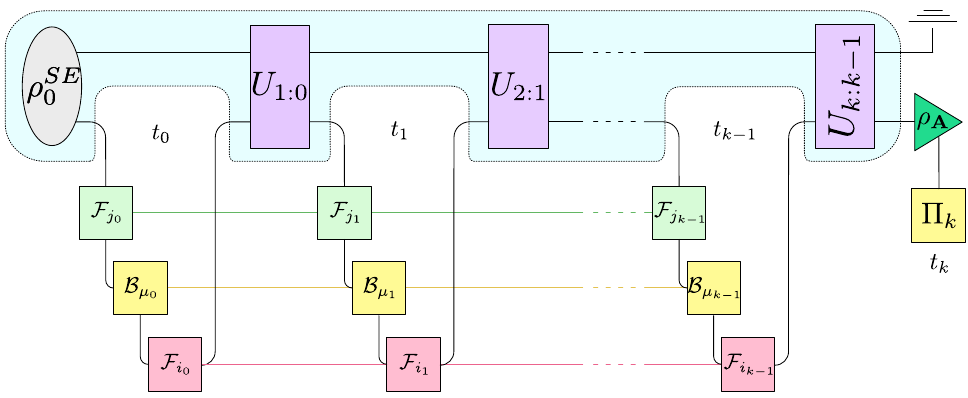}
	\caption[Circuit diagram for performing non-Markovian gate set tomography. ]{Circuit diagram for performing non-Markovian gate set tomography. Given a fixed, unknown process \pt{}, and fixed, unknown set of testers $\{\mathbf{B}_{k-1:0}^{\vec{\mu}}\}$, the underlying objects can be self-consistently determined (up to a gauge freedom) by forming combinations of \acs{IC} testers at each time step. Each fiducial $\mathbf{F}$ is drawn from $\mathscr{B}$. }
	\label{fig:nm-gst}
\end{figure}

Absorbing $\kket{\Upsilon}$ and $\bbra{\mathcal{J}}$ into $\mathbf{F}_j$ and $\mathbf{F}_i$ respectively, we can define new matrices $A$ and $C$ such that $A_{ir} = \langle\!\langle \mathcal{J} | \mathbf{F}_i|r\rangle\!\rangle$ and $C_{sj} = \langle\!\langle s | \mathbf{F}_j | \Upsilon\rangle\!\rangle$. This implies that $p_{ij\mu} = (A\mathbf{B}_\mu C)_{ij}$. That is, varying the fiducials and measuring the outcomes for a given $\mu$ is the measurement of the elements of the matrix $\tilde{\mathbf{B}}_\mu = A\mathbf{B}_\mu C$. In the special case where $\mathbf{B}_0$ is chosen to be the null gate at each time, we fix 
\begin{equation}
	g = \tilde{\mathbf{B}}_1 = AC,
\end{equation}
which is the Gram matrix of the fiducials $\bbra{\mathcal{J}}\mathbf{F}_i\mathbf{F}_j\kket{\Upsilon}$. For each other $\tilde{\mathbf{B}}_\mu$, left multiplying by the inverse of $g$ obtains 
\begin{equation}
	g^{-1} \tilde{\mathbf{B}}_\mu = C^{-1} A^{-1} A \tilde{\mathbf{B}}_\mu C = C^{-1} \mathbf{B}_\mu C.
\end{equation}
Hence, 
\begin{equation}
	\bar{\mathbf{B}}_\mu := g^{-1}\tilde{\mathbf{B}}_\mu
\end{equation}
is an estimate of each control operation up to similarity transformation by matrix $C$. Note that, like in \acs{GST}, this gauge freedom is an unavoidable consequence of taking no preferred reference frame. If we set 
\begin{equation}
	\label{eq:ptt-gauge}
	\begin{split}
		|\Upsilon\rangle\!\rangle &\mapsto C'|\Upsilon\rangle\!\rangle\\
		\mathbf{B}_\mu &\mapsto C'^{-1} \mathbf{B}_\mu C,\\
		\langle\!\langle \mathcal{J} | &\mapsto \langle\!\langle \mathcal{J} | C,
	\end{split}
\end{equation}
then the transformed probabilities 
\begin{equation}
	p_{ij\mu}' = \langle \!\langle \mathcal{J}|CC^{-1}\mathbf{F}_iCC^{-1}\mathbf{B}_\mu CC^{-1}\mathbf{F}_jCC^{-1}| \Upsilon\rangle\!\rangle,
\end{equation}
are identical to the $p_{ij\mu}$. This means that any reconstruction will \emph{always} be up to an unobservable gauge freedom across the set. In practice, gauges are often set by taking the transformation (as in Equation~\eqref{eq:ptt-gauge}) and performing a gauge optimisation. One has the freedom to choose gauges that obey certain properties -- such as maintaining \acs{CPTP} -- and also to select the gauge that takes the estimated gate set as close to the target gate set as possible. This is completely without loss of generality, and allows our estimates to look as familiar as possible with what we expect.

To obtain the two vectors, we note that we have the measurable estimate $g_{i0} = (AC)_{i0} = \langle\!\langle \mathcal{J}|\mathbf{F}_i|\Upsilon\rangle\!\rangle$. From this, we get the two equivalent relations 
\begin{equation}
	\begin{split}
		\langle\!\langle \mathcal{J} | \mathbf{F}_i |\Upsilon\rangle\!\rangle &= \sum_r\langle\!\langle \mathcal{J}|\mathbf{F}_i|r\rangle\!\rangle\!\langle\!\langle r|\Upsilon\rangle\!\rangle = A|\Upsilon\rangle\!\rangle := |\tilde{\Upsilon}\rangle\!\rangle, \\
		\langle\!\langle \mathcal{J} | \mathbf{F}_i |\Upsilon\rangle\!\rangle &= \sum_s\langle\!\langle \mathcal{J}|s\rangle\!\rangle\!\langle\!\langle s|\mathbf{F}_i|\Upsilon\rangle\!\rangle = \langle\!\langle \mathcal{J}|C := \langle\!\langle \tilde{\mathcal{J}}|.
	\end{split}
\end{equation}
and hence, the final estimate (up to gauge transformation) of each vector is:
\begin{equation}
	\begin{split}
		|\bar{\Upsilon}\rangle\!\rangle &= g^{-1}|\tilde{\Upsilon}\rangle\!\rangle = C^{-1}|\Upsilon\rangle\!\rangle,\\
		\langle\!\langle \bar{\mathcal{J}}| &= \langle\!\langle \tilde{\mathcal{J}}| = \langle\!\langle \mathcal{J}|C.
	\end{split}
\end{equation}

One might (rightly) point out that this formalism neglects the possibility of short-time correlations between the $\mathbf{F}_i$, $\mathbf{B}_\mu$, and $\mathbf{F}_j$, since we have assumed that they compose like tensor products within a given leg of the process tensor. On this issue, we make several remarks.
\begin{enumerate}[label=(\roman*)]
	\item The effect of ignoring these correlations in the formalism is equivalent to marginalising over them. This means, for example, that a given $\mathbf{B}_\mu$ is conditioned on the average case $\mathbf{F}_j$ which precedes it, rather than the specific one. In general, this practice can be dangerous, since if the two objects are strongly correlated then the conditional case can vary greatly from the average case. However, we do not expect pulse-level correlations to be the dominating problem here. Instead, it appears that the typical frequency of control fluctuations is on the time scale of dozens of gates, as evidenced by the large range of \acs{GST} experiments across different devices in the literature~\cite{white-POST,RBK2017, Dehollain_2016, kim_microwave-driven_2015}.
	\item One could not, in general, bootstrap correlated instruments to determine all information about each other. This can be seen with a simple parameter-counting argument: suppose at a single time we have the composition of three gates $\mathcal{F}_i \circ \mathcal{B}_\mu \circ \mathcal{F}_j$, except now they are generically correlated with one another. The collective sequence is thus described by a set of testers $\{\mathbf{A}_{2:0}^{ij\mu}\}$, whose \acs{IC} set size is $16^3$ with a total of $16^6$ unique parameters. Note that the action of composition here is now replaced by the projection onto two Bell states, so that the output of $\mathcal{F}_j$ feeds to the input of $\mathcal{B}_\mu$, and the output of $\mathcal{B}_\mu$ into the input of $\mathcal{F}_i$. 
	Let us presume that we have more capabilities than our present \acs{GST} experiment, we have the ability to perform \acs{QPT} on the input of $\mathcal{F}_j$ and the output of $\mathcal{F}_i$. Even in this idealised scenario, we only have $(4\times 4) \times 16^3 = 16^4$ linearly independent experiments we can perform. Hence, we lack full information to determine the entire object. 
	\item With undercomplete information, we \emph{could} always construct a plausible $\{\mathbf{A}_{2:0}^{ij\mu}\}$ to fit the data, but we would gain little extra insight about the instruments due to the large gauge freedoms. 
	\item With all this being said, this fully general case has little bearing on real life, since it is too onerous to characterise in complete generality. Hence, as we will see in the next section, we can resolve this issue by both constructing and estimating compressed models using tensor networks; these \emph{do} adequately account for both short and long time correlations in the instruments. 
\end{enumerate}

\begin{figure}[!t]
	\centering
	\includegraphics[width=0.8\linewidth]{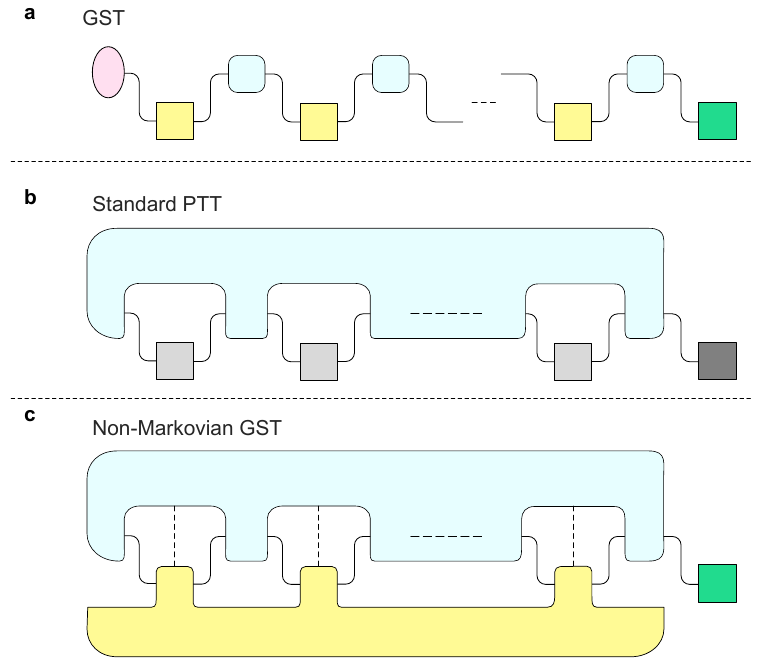}
	\caption[A comparison of various models for estimating dynamics Markovian and non-Markovian dynamics ]{A comparison of various models for estimating dynamics. Colour indicates an object which is estimated, grey indicates that object is taken to be known. \textbf{a} \acs{GST} estimates a set of time-local gates with some time-local background process. \textbf{b} \acs{PTT} from Chapter~\ref{chap:PTT} models a generically correlated background process, but assumes gates to be known. \textbf{c} The self-consistent protocol introduced in this chapter estimates all objects simultaneously, and allows for temporal correlations in both the background process and in the control instruments.}
	\label{fig:NMGST-buildup}
\end{figure}

\textbf{Time Local Bases and Time Local Processes}

We have derived a formulation of non-Markovian \acs{GST} in full generality. This accounts for the full swathe of noisy processes as we have introduced them in the previous Section. In particular, the use of multi-time instruments allows one to effectively model the possibility of control non-Markovianity, time-dependent Markovianity, and control spillage, as well as process non-Markovianity. If one were to expect -- or one wished to test a model of -- either process non-Markovian and control Markovian, or process Markovian and control non-Markovian, then this amounts respectively to treating $\{\mathbf{B}_{k-1:0}^\mu\}$ or \pt{} as a tensor product. Let us consider the former case, i.e. that $\{\hat{\mathbf{B}}_{k-1:0}^{\vec{\mu}}\} = \left\{\bigotimes_{i=0}^{k-1} \hat{\mathcal{B}}_{\mu_i}\right\}$. That is, all the sequences of operations are generated from the same set of $d_S^4$ time-local instruments, and the \emph{effects of an instrument is the same} no matter at which time it is applied. This substantially reduces the problem size. First, it allows the matrix $A$ to take a tensor product structure: 
\begin{equation}
	A_{\vec{i}\vec{r}} = \bigotimes_{j=0}^{k-1}\bigotimes_{l=0}^{k-1}\langle\!\langle \mathcal{J}_{j_l}|F_{j_l}|r_l\rangle\!\rangle.
\end{equation}
Hence, rather than measuring $i$ and $\mu$ from $1$ to $d_S^{4k}$, one only needs to measure each from 1 to $d_S^4$. Note that for full determination of $|\Upsilon\rangle\!\rangle$, one still has $j$ scaling like $d_S^{4k}$, since it is generally encodes arbitrary temporal correlations.
Another consequence of this is that it allows us to take the gauge freedom to also maintain a tensor product structure, since right multiplying by $g^{-1}$ selects $A$ as the gauge matrix rather than $C$, and we have already fixed this to be a tensor product.

\begin{figure}[h]
	\centering
	\includegraphics[width=\linewidth]{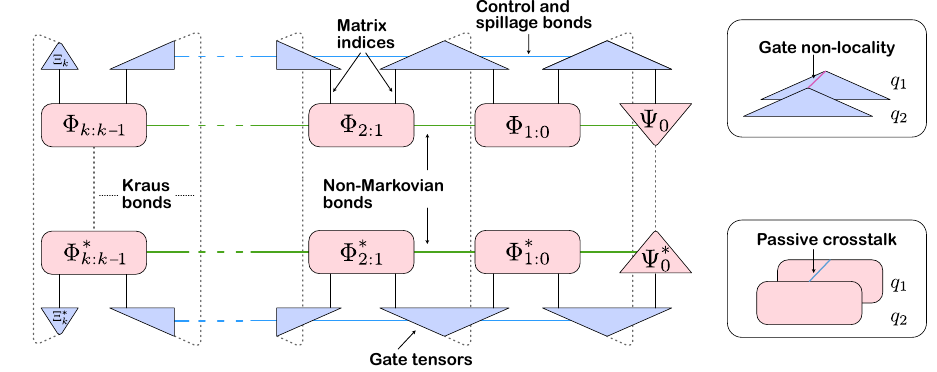}
	\caption[A tensor network diagram of locally purified process tensors represented alongside correlated control operations. ]{A tensor network diagram of locally purified process tensors represented alongside correlated control operations. This models all the different dynamical effects we have discussed in this chapter. }
	\label{fig:NM-GST-TN}
\end{figure}


\section{Implementing Self-Consistent Estimation}
In the previous section we presented a formalism to simultaneously capture various notions of control and process non-Markovianity, as well as the interplay between the two. For fully general processes, the complexity of the problem scales exponentially in the number of time steps $k$, the number of qubits, and the number of gates in the gate set. In Chapter~\ref{chap:PTT}, we were already limited to a single qubit at three time steps for the fully general case before being overcome by practical considerations. Here, as in Chapter~\ref{chap:efficient-characterisation}, we therefore apply tensor networks to the task from the outset. Not only does this compress the representation significantly, but allows for extreme modularity based on the expected physics of the system. We build on the approach from the previous chapter and develop a self-consistent method to estimate any quantum stochastic process as well as the noisy instrument used to probe it. We demonstrate the characterisation on a variety of synthetic and real data, showing how one may accurately capture a wide range of noisy quantum dynamics.

For a system $S$ composed of $N$ qubits characterised across times $\mathbf{T}_k$, we start with \pt{} representing the process, and $\{\hat{\mathbf{A}}_{k:0}^{(i)}\}_{i=1}^M$ representing the Choi states for the series of the $M$ multi-time control operations. Note that these testers include measurement outcomes in the index $i$, and are hence trace non-increasing. First, we cast these in variational tensor network form as \acs{LPDO}s, as depicted in Figure~\ref{fig:NM-GST-TN}. \pt{}, once again, can be written as 
\begin{equation}
	[\Upsilon_{k:0}^{\vec{\theta}}]^{\vec{k_{\mathfrak{o}}}\vec{k_{\mathfrak{i}}}}_{\vec{b_{\mathfrak{o}}}\vec{b_{\mathfrak{i}}}} = \sum_{\mu_k,\mu_0} \sum_{\vec{\nu},\vec{\nu}'} [\Phi_{k:k-1}]_{\mu_k,\nu_k}^{k_{\mathfrak{o}_k}k_{\mathfrak{i}_k}} [\Phi_{k:k-1}^\ast]^{\mu_k,\nu'_k}_{b_{\mathfrak{o}_k}b_{\mathfrak{i}_k}}\left(\prod_{j=1}^{k-1}
	[\Phi_{j:j-1}]_{\nu_{j+1}\nu_j}^{k_{\mathfrak{o}_j}k_{\mathfrak{i}_j}}[\Phi_{j:j-1}^\ast]^{\nu'_{j+1}\nu'_j}_{b_{\mathfrak{o}_j}b_{\mathfrak{i}_j}}\right) [\Psi_{0}]_{\mu_0,\nu_1}^{k_{\mathfrak{o}_0}} [\Psi_{0}^\ast]^{\mu_0,\nu'_1}_{b_{\mathfrak{o}_0}},
	\end{equation}
	where $\vec{\theta} = \{\Phi_{k:k-1},\{\Phi_{j:j-1}\},\Psi_0\}$. The control operations may similarly be locally purified, but since they do not stem from a unitary evolution from a dilated environment, they do not inherit the same ring-like structure. These are given as 
\begin{equation}
	[\hat{\mathbf{A}}_{k:0}^{\vec{\phi}_{i}}]^{\vec{b_{\mathfrak{o}}}\vec{b_{\mathfrak{i}}}}_{\vec{k_{\mathfrak{o}}}\vec{k_{\mathfrak{i}}}} = \sum_{\vec{\delta}} \sum_{\vec{\gamma},\vec{\gamma}'} 
	[\Xi_{k}]^{\delta_k,\gamma_k}_{k_{\mathfrak{o}_k}} [\Xi_{k}^\ast]_{\delta_k,\gamma'_k}^{b_{\mathfrak{o}_k}}
	\left(\prod_{j=1}^{k-1}
	[\Gamma_{j:j-1}]^{\gamma_{j+1}\gamma_j}_{k_{\mathfrak{i}_j}k_{\mathfrak{o}_{j-1}}}[\Gamma_{j:j-1}^\ast]_{\gamma'_{j+1}\gamma'_j}^{b_{\mathfrak{i}_j}b_{\mathfrak{o}_{j-1}}}\right),
\end{equation}
where $\vec{\phi}_{(i)} =  \{\Xi_{k}^{(i)},\{\Gamma_{j:j-1}^{(i)}\}\}$. Let us further subdivide the control index $i$ into $(j,x)$, where $j$ indicates the deterministically chosen multi-time instrument, and $x$ is its corresponding measurement outcome (or sequence of measurement outcomes). Then, when a quantum circuit is run with instrument $\mathbf{A}_{k:0}^{(j)}$, the probability of obtaining outcome $x$ is
\begin{equation}
	p_{x\mid j} = \Tr[\mathbf{A}_{k:0}^{(j)\text{T}}\Upsilon_{k:0}],
\end{equation}
modelled by our parametrised tensor networks as
\begin{equation}
	\label{eq:self-consistent-prob}
	p_{x\mid j}^{\vec{\theta}\vec{\phi_{j}}} = \sum_{\vec{k_{\mathfrak{o}}}\vec{k_{\mathfrak{i}}}\vec{b_{\mathfrak{o}}}\vec{b_{\mathfrak{i}}}} [\Upsilon_{k:0}^{\vec{\theta}}]^{\vec{k_{\mathfrak{o}}}\vec{k_{\mathfrak{i}}}}_{\vec{b_{\mathfrak{o}}}\vec{b_{\mathfrak{i}}}}[\hat{\mathbf{A}}_{k:0}^{\vec{\phi}_{i}}]^{\vec{b_{\mathfrak{o}}}\vec{b_{\mathfrak{i}}}}_{\vec{k_{\mathfrak{o}}}\vec{k_{\mathfrak{i}}}}.
\end{equation}

Employing Equation~\eqref{eq:self-consistent-prob} as our model probabilities, we use the same objective function as in previous chapters -- the negative log-likelihood averaged over the data set. We also use the same estimation technique as in Chapter~\ref{chap:efficient-characterisation}: Adam optimiser where the gradients are obtained using autodifferentiation performed by the JAX library~\cite{jax2018github}. 


A cornerstone of characterisation is that the resulting model must be physical. For process tensors, this is in the sense that we have outlined: it should be positive and causal. The former, we obtain from the \acs{LPDO} construction, and the latter we encode randomly into the objective function, as in Chapter~\ref{chap:efficient-characterisation}. The Choi states of multi-time instruments have the same positivity requirement (and imposition) but need not be causal. The natural intuition for this follows from the two-time case: quantum channels must be \acs{TP}, but quantum instruments may be trace non-increasing since the classical outcomes may be stochastically obtained. A sequence of measurements then, for example, will be an acausal multi-time instrument and trace will not be preserved. We might also consider the use of an ancilla tester through a simple example: suppose an ancilla qubit is placed into equal superposition and interacts with the system through only diagonal interactions. The ancilla is projectively measured at some early time $t_i$ and some later time $t_f$ in the same basis. This gives rise to four testers, one for each of the outcomes $\{0_i0_f,0_i1_f,1_i0_f,1_i1_f\}$. But, since the tester is defined by interactions that do not change the population of the ancilla qubit, the probability of obtaining outcomes $0_i1_f$ or $1_i0_f$ is zero. Hence, if we post-select on the later measurement outcome, we update the statistics of the earlier measurement.

The condition satisfied by testers must be that for a given $j$, if we sum over outcomes $x$, the result must be causal. This is necessary to preserve probabilities across the whole range of multi-time instruments. Much like the causality of the process tensor, we can regularise this by sampling the Pauli expectation values of 
\begin{equation}
	\hat{\mathbf{A}}_{k:0}^{(j)} = \sum_x \hat{\mathbf{A}}_{k:0}^{(j,x)}.
\end{equation}

Let us first write that $| \{\hat{\mathbf{A}}_{k:0}^{(j)}\}| = M_{\text{probes}}$.
Let $\mathbf{P}_c$ be the set of $2k+1$-qubit Pauli operators for which causality conditions dictate the expectation values must be zero for the respective gate set objects. That is to say, 
\begin{equation}
	\Tr[P\cdot \Upsilon_{k:0}] \overset{!}{=} 0 \quad \text{and} \quad \Tr[P \cdot \hat{\mathbf{A}}_{k:0}^{(j)}] \overset{!}{=} 0 \quad \forall \:j,P\in \mathbf{P}_c.
\end{equation}
At each iteration, we randomly generate $|D|$ subsets of $\mathbf{P}_c$ -- denoted by $\{\mathcal{C}_i\}_{i=0}^{|D|}$ -- as well as a subset $D\subset \mathcal{D}$ of the total dataset. The complete objective function for a given iteration is then
\begin{equation}
	f(\vec{\theta},\{\vec{\phi_j}\}) = \overset{\text{Model Log-Likelihood}}{\overbrace{\sum_{d\in D}-n_d\ln p_d^{\vec{\theta}\vec{\phi_{j}}} }}
	+ \overset{\text{Process Causality}}{\overbrace{\sum_{P\in \mathcal{C}_0} \Tr\left[P\cdot [\Upsilon_{k:0}^{\vec{\theta}}]\right]}}
	+ \overset{\text{Probe Trace Preservation}}{\overbrace{\sum_{d\in D}\sum_{P\in\mathcal{C}_d} \Tr\left[P\cdot \sum_x \hat{\mathbf{A}}_{k:0}^{\vec{\phi}_{(j,x)}}\right]}}.
\end{equation}

\textbf{Simplifications}

In general, we will not simultaneously employ each feature of the model as outlined here for reasons of computational convenience. Although accounting for the entire set of possibilities is not intractable, we find it more practical to design a model around what is physically reasonable for the expected physics. This is why the modularity of our approach is so useful. If we did not expect correlations in the control instrument, for example, we would simply set the bond dimension to be equal to 1, and the multi-time instrument would be completely time-local. 
\begin{figure}[!b]
	\centering
	\includegraphics[width=0.6\linewidth]{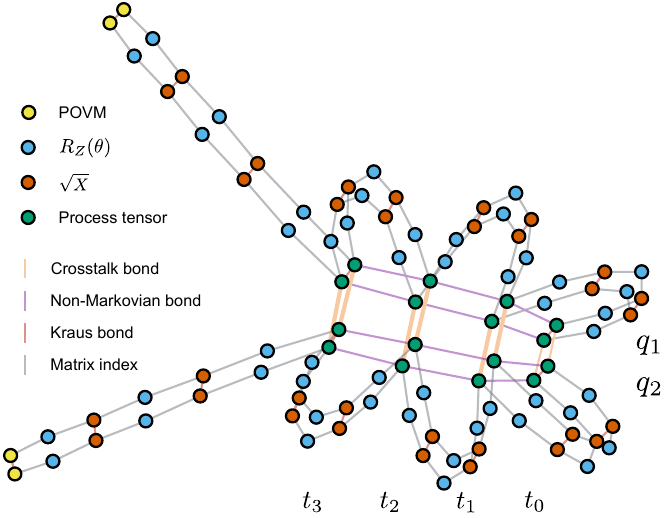}
	\caption[Tensor network decomposition of a three step, two qubit process partitioned into process and control models ]{Tensor network decomposition of a three step, two qubit process partitioned into process and control -- specific to IBM Quantum gate decompositions -- both of which are estimated. The background process partitions into non-Markovian (horizontal) and crosstalk (vertical) bonds as before. The single qubit control operations decompose into virtual $Z$-rotations and physical $X$-pulses, followed by final \acs{POVM}.}
	\label{fig:dragonfly_fig}
\end{figure}

We can also design hardware-specific simplifications of the model. In IBM Quantum devices, for example, single-qubit unitary gates decompose into a sequence of physical $X$ pulses and virtual $Z$ rotations:
\begin{equation}
	\label{eq:u3-decomp}
	u(\theta,\phi,\lambda) = R_Z(\phi + 3\pi)\circ R_X(\pi/2)\circ R_Z(\theta + \pi)\circ R_X(\pi/2)\circ R_Z(\lambda)
\end{equation}
and so it suffices to incorporate only this single physical gate into our model, taking the remainder to be perfect $Z-$rotations. The single-qubit rotations hence decompose into the tensor network shown in Figure~\ref{fig:dragonfly_fig}.
Assuming no control correlations then, the gate set is completely defined by $\{\Upsilon_{k:0}, \hat{R}_{X}(\pi/2)^{(q_i)}, \Pi_0^{(q_i)}, \Pi_1^{(q_i)}\}$. In fact, for many of our simulations and our demonstrations on IBM Quantum devices, this minimal model is the one we shall estimate, and we find it to be extremely accurate.

\textbf{Restricted Process Tensors}

In accordance with the availabilities of current technology and thematically with the remainder of this thesis, we consider predominantly the estimation of restricted process tensors here. By this we mean we have a sequence of unitaries operations followed by a final terminating measurement. This means that each tester $\mathbf{A}_{k:0}^{(j,x)}$ is only a two-outcome probe. The procedure is hence designed not to uniquely learn the properties of the process tensor, but to capture extrinsic behaviours of the dynamics. We emphasise, however, that the formalism is wholly compatible with the estimation of multi-time statistics, as per the experiments in Chapter~\ref{chap:MTP}.

\textbf{Statistical Precision}

We have claimed that this approach to estimating dynamics constitutes a generalisation of \acs{GST} to the non-Markovian setting. It is interesting then to ask whether we can achieve the same Heisenberg-type scaling in our estimation of model parameters, where estimation convergences with a precision of $1/N$ rather than $1/\sqrt{N}$. In short, it depends on the chosen model. \acs{GST} achieves $1/N_{\text{shots}}$ scaling in its accuracy because of repetitions in the gate sequences. If a gate $\mathcal{G}$ is repeated $L$ times, then certain parameters of $\mathcal{G}$ will be raised to the power $L$. A rotation by $\theta$ becomes a rotation by $L\theta$.

In the fully general case where we have a fixed process tensor and multi-time instruments, the scaling can only be at best $1/\sqrt{N_{\text{shots}}}$ -- no parameters are repeated and hence no parameters are amplified. However, if gates were taken to be time-local, then their matrix representations will be copied into each location of the instrument sequence tensor network, and the parameters will repeat. Similarly, if instead we took the central matrix of the process tensor in tensor network form, we could amplify this. Heisenberg precision is then possible, but it requires dynamics to repeat upon themselves so that coherent noisy can amplify. Correlations do not necessarily stand in the way of this.

We note this only for the sake of discussion. Rather than randomly, the engine of \texttt{pyGSTi} designs its gate sequences specifically to ensure that each parameter in the model is amplified by at least one circuit. Heuristically, this is necessary for Heisenberg precision. Design of an analogous engine here is beyond the scope of this thesis, though we believe the same principles should apply. 

\clearpage

\textbf{Estimating Full Process Tensors With Ancilla Assistance}

Restricted process tensor models are useful for control of non-Markovian systems, but they do not uniquely estimate the Choi state of a process and hence cannot be used to directly estimate non-Markovian properties. One solution we designed in Chapter~\ref{chap:PTT} was with the aid of an ancilla to drive an interaction, locally control it on the system, and projectively measure the ancilla at each step. This is then characterised and repeated to probe the system and perform full \acs{GST}. Although this procedure is useful, it is also cumbersome. Avoiding the prior characterisation, as well as reliance on mid-circuit measurements is desirable in order to have a frictionless procedure that makes fewer assumptions about the hardware.

Here, we develop one such procedure using our self-consistent estimation method. We showed in Chapter~\ref{chap:process-properties} that a tester $\Theta_{k-1:0}$ defined by a fixed interaction $\mathcal{A}_F$ between system $S$ and an ancilla qubit $A$ at each time, as well as local unitaries on both qubits and a final projective measurement of $A$ and $S$, constitutes an informationally complete probe of $S$ -- enough to learn its process tensor. However, on real hardware, two qubit gates $\mathcal{A}_F$ are likely to be noisy, and it cannot be assumed that the tester will be known. This can be resolved by estimating both \pt{} and $\Theta_{k-1:0}$. The end result is the use of a neighbouring qubit to repeatedly probe the system and learn any arbitrarily correlated dynamics. We shall demonstrate this principle further below. This unlocks the ability to estimate multi-time dynamics with extremely minimal assumptions in practice.

\section{Demonstrations}
To demonstrate this general framework for characterising correlated noise, we demonstrate it on a variety of synthetic data. 
We shall consider exotic noise models which are either operationally uncharacterisable by any techniques in the literature, or violate the model previously assumed with \acs{GST}. In particular, this means involving different forms of control noise. In each instance, we simulate a two-qubit, five-step process where each qubit is coupled to a common bath defect via a random Heisenberg interaction. Each characterisation uses 6000 circuits run at 1024 shots per circuit. The circuits are decomposed into a standard basis $\{R_Z(\theta), \sqrt{X}\}$ as per Equation~\ref{eq:u3-decomp}. We take the effects to apply to the physical $\sqrt{X}$ gate.
We have already established the ability of \acs{PTT} to capture these effects of process non-Markovianity. Here, we will see (i) how \acs{PTT} of the previous chapters breaks down when exposed to these classes of control noise, and (ii) how our universal approach is able to efficiently and effectively characterise all of these effects in tandem.



\subsection{Coherent Gate Error}
\begin{figure}[!t]
	\centering
	\includegraphics[width=0.85\linewidth]{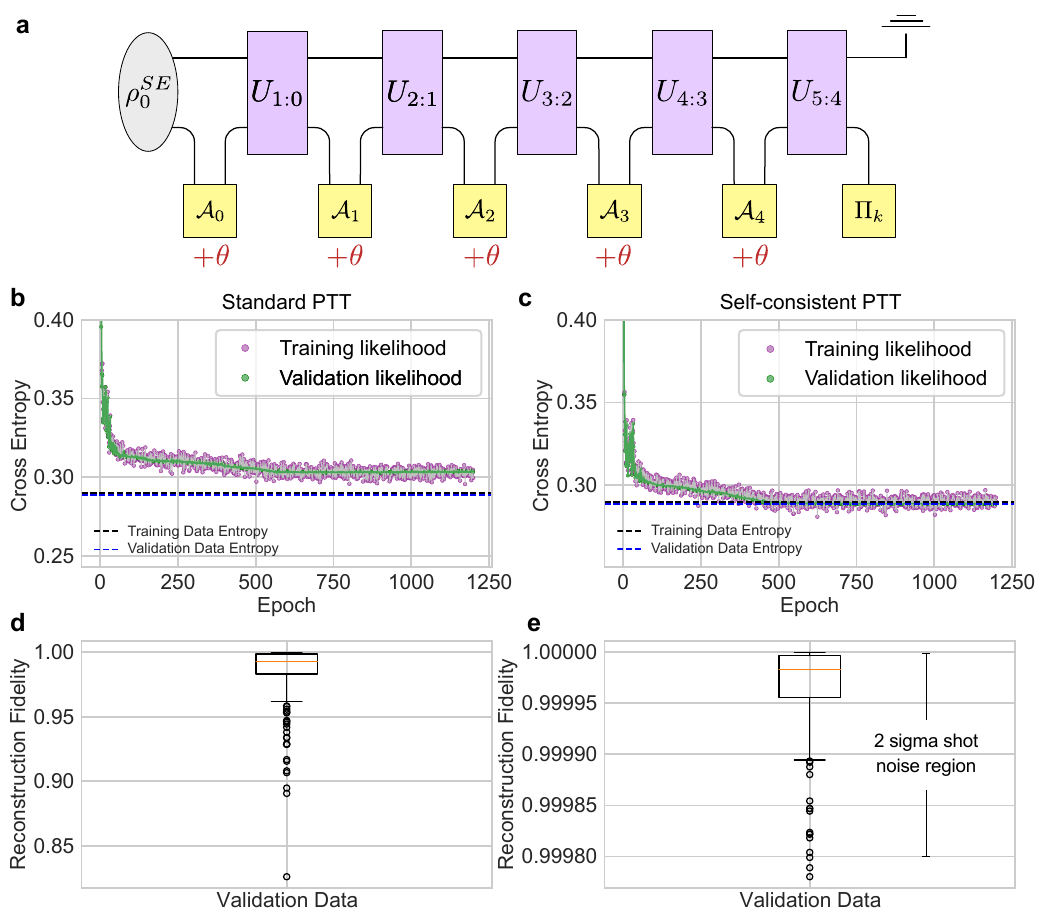}
	\caption[Simultaneous characterisation of coherent gate noise in a non-Markovian environment ]{Simultaneous characterisation of coherent gate noise in a non-Markovian environment. We compare standard \acs{PTT} (modelling only the process) with estimation of a self-consistent model. 
	\textbf{a} Circuit diagram of the simulated error model. Each $\sqrt{X}$ operation in the control decomposition is coherently rotated by $\theta = \pi/16$ in comparison to the ideal operation. This generates gate-dependent noise. Every detail here is unknown to the characterisation model.
	\textbf{b} and \textbf{c} indicate the convergence graphs for both the stochastic data fit and the quality with respect to validation data. \textbf{d} and \textbf{e} summarise reconstruction fidelities for each of the fit models: we see a significant improvement in the quality of the model by incorporating gate errors as well as background non-Markovianity. }
	\label{fig:coherent_error}
\end{figure}
We start with the most straightforward addition to our model with coherent over-rotations. Specifically, we decompose every gate in accordance with the decomposition given in Equation~\ref{eq:u3-decomp}. The virtual $Z$ gates are taken to be perfect, but the supposed $R_X(\pi/2)$ gate is really $R_X(\pi/2 + \pi/16)$. The effective gate then is a coherently transformed unitary. Importantly, because the $R_X(\pi/16)$ error does not commute through the $Z$ rotations, the effect of this error model is entirely gate-dependent. The reason this is important is that any gate-independent noise channels can factor outside and be incorporated into the process tensor estimate. A gate-dependent coherent model, however, changes the linear relationship between the various basis elements. This hence violates the model introduced in Chapters~\ref{chap:PTT} and~\ref{chap:efficient-characterisation}. 

We simulate and estimate the five-step process with controls as described above. The results are summarised in Figure~\ref{fig:coherent_error}. This displays both the convergence graphs of the estimation, and the reconstruction fidelities for a standard \acs{PTT} model as well as our self-consistent estimator. The graphs display the same convergence information as in Chapter~\ref{chap:efficient-characterisation} (Section~\ref{ssec:nisq-tn}). That is, the convergence of the green curve to the blue dashed line benchmarks the model's ability to capture the entire dynamics. There is a clear disparity between the two models shown. The former is unable to fit the data, and suffers from substantial error in predicting the dynamics for arbitrary sequences of unitaries thereafter. The latter fits the model correctly to a cross entropy of less than $10^{-5}$, with a median reconstruction fidelity of $1 - 10^{-5}$. This demonstration shows that we can characterise both the non-Markovian background process and noise in the control operations to a very high accuracy.


\subsection{Control non-Markovianity}
\begin{figure}[!t]
	\centering
	\includegraphics[width=0.85\linewidth]{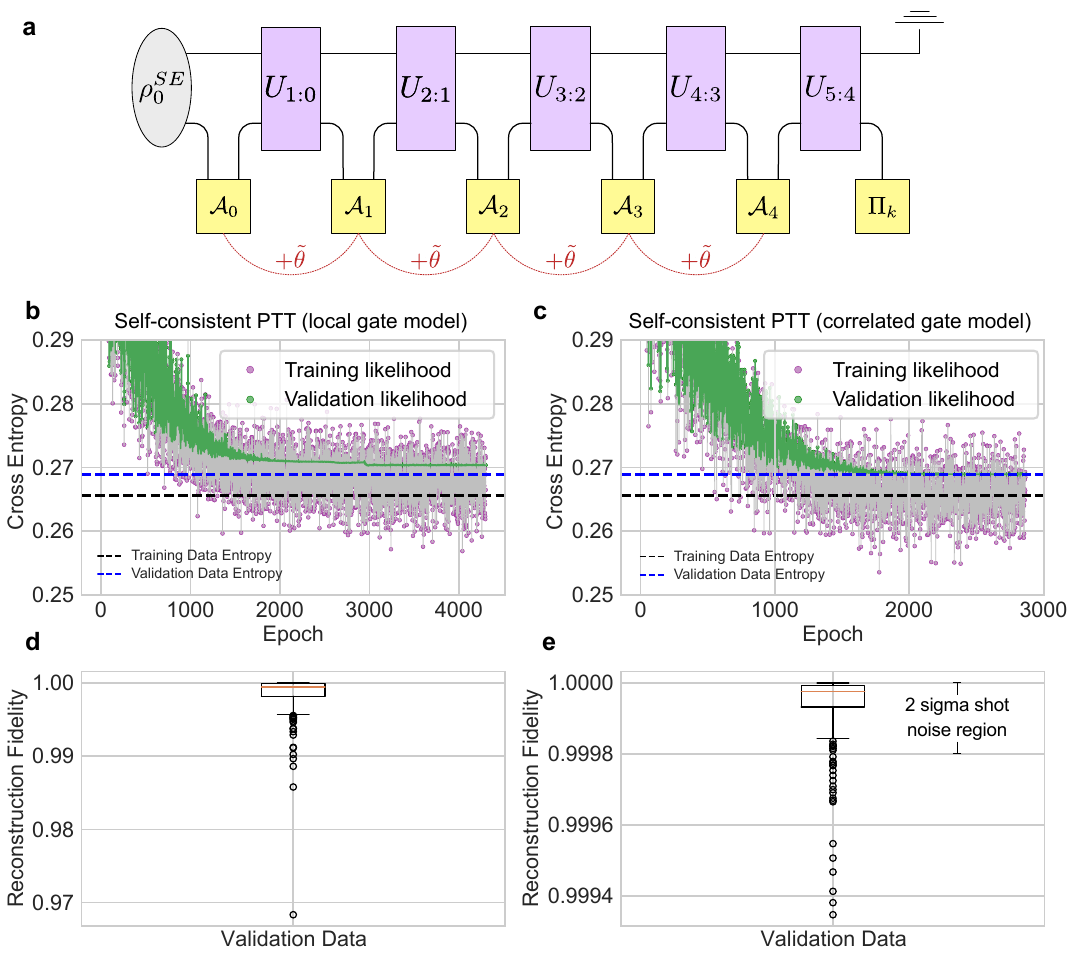}
	\caption[Characterisation of $1/f$ correlated gate noise in a non-Markovian environment ]{Characterisation of $1/f$ correlated gate noise in a non-Markovian environment. We compare the time-local self-consistent \acs{PTT} model with a fully fledged non-Markovian \acs{GST}, where both process tensor and testers are estimated. 
	\textbf{a} Circuit diagram of the simulated error model. Each gate in the circuit is coherently rotated by the same $\tilde{\theta}$, which is a stochastic variable resampled in each circuit. Every detail here is unknown to the characterisation model.
	\textbf{b} and \textbf{c} indicate the convergence graphs for both the stochastic data fit and the quality with respect to validation data. \textbf{d} and \textbf{e} summarise reconstruction fidelities for each of the fit models: we see a significant improvement in the quality of the model by estimating correlated gate errors as well as background non-Markovianity. }
	\label{fig:corr_coherent_error}
\end{figure}

The previous example showed that self-consistent \acs{PTT} is possible, and that our approach robustly estimates noise in the controls. Let us now turn to a more sophisticated error model, which cannot currently be captured by methods in the \acs{QCVV} literature.
Let us consider a circuit where every implementation of a gate $R_X(\pi/2)$ is shifted by some $\varepsilon$ so that the rotation is really $R_X(\pi/2 + \varepsilon)$. If $\varepsilon$ were fixed, then this would be the same coherent case as before. Suppose, however, that at the start of each circuit, $\varepsilon$ were drawn probabilistically from a distribution. Each gate rotates by the same amount, but the amount changes from circuit to circuit so that the coherent error is fully correlated. This scenario is depicted in Figure~\ref{fig:corr_coherent_error}a.

This models the effects of, for example, quasistatic laser noise. We implement this scenario where each $\varepsilon$ is sampled according to $1/f$ noise. $1/f$ noise is the phenomenon where a stochastic process has power spectrum $S(f) \propto 1/f^\alpha$; here we select $\alpha = 1$. In the context of signal processing, the power spectral density of a signal is the Fourier transform of its autocorrelation function. For $1/f$ noise, this implies that the autocorrelation function has a long-range dependence, meaning that the noise exhibits long-term memory, with correlations persisting over a wide range of time scales.
This archetypal spectrum models myriad physical processes in nature with long-time correlations~\cite{RevModPhys.53.497,west1989ubiquity,PhysRevE.54.2154}, including in almost all quantum hardware~\cite{paladino20141,aquino2023model,wilen2021correlated,PhysRevApplied.17.034074}. The overall process we sample from therefore has process non-Markovianity due to a stray two-level system, as well as instruments plagued with $1/f$ noise. This is the underlying phenomenological model.



Without tailoring anything to the underlying physics, we represent the processes with a generic process tensor network, and the controls with a generic tester tensor network, each with bond dimension $\chi = 2$. The results of this are depicted in Figure~\ref{fig:corr_coherent_error}. We compare the fully self-consistent estimate to the one from the previous results: self-consistent non-Markovian estimate with the controls modelled as time-local. This is intended to eliminate the possibility that the improvement is replicated simply by accounting for some coherent error. The delineation in this instance entirely stems from finding the best model estimate of the temporal correlations in the instrument, for which we see two orders of magnitude improvement in the reconstruction fidelities.

This thus demonstrates the ability to estimate compressed models representing physically realistic quantum correlated noise both in the background dynamics of the system, and in the control instruments of the experimenter.






\subsection{Spillage}
The third exotic noise model we consider is that of spillage, where the radius of the control operation leaks over into environment systems. This can typically happen in two ways: either the frequency spectrum of the control pulse is resonant with some part of the environment, and drives a stochastic transition; or, in the case of trapped ion qubits, for example, the laser addressing an individual ion may have some small width illuminating other ions, generating entanglement. 

\begin{figure}[!t]
	\centering
	\includegraphics[width=0.85\linewidth]{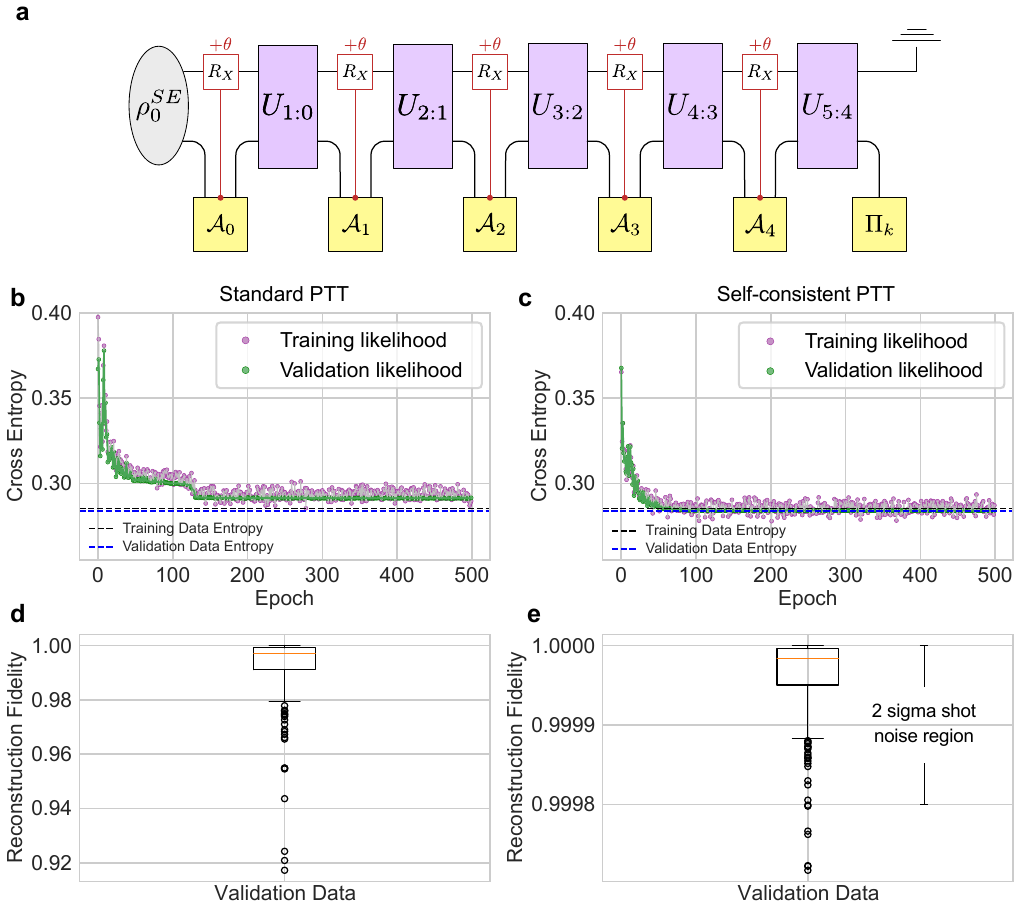}
	\caption[Simultaneous characterisation of spillage error in a non-Markovian environment ]{Simultaneous characterisation of spillage error in a non-Markovian environment. We compare standard \acs{PTT} (modelling only the process) with estimation of a self-consistent model. 
	\textbf{a} Circuit diagram of the simulated error model. At each application of control operations, a controlled $X$ rotation by $\theta = \pi/16$ is applied to the environment two-level system. 
	Every detail here is unknown to the characterisation model. 
	\textbf{b} and \textbf{c} indicate the convergence graphs for both the stochastic data fit and the quality with respect to validation data. \textbf{d} and \textbf{e} summarise reconstruction fidelities for each of the fit models: we see a significant improvement in the quality of the model by incorporating gate errors as well as background non-Markovianity. }
	\label{fig:spillage_error}
\end{figure}

In either case, the effect is the same: the act of performing a control operation modifies the environment state. 
One could dilate the system to account for this fact, but it would leave a large gauge freedom and be generally unsatisfactory. In effect, this expands the model to fit the data, but is not indicative of the physics. We have already argued with the use of superprocesses, however, that this particular type of driven dynamical effect is fully captured under the process tensor framework, it simply remains to estimate it. Further, we have already argued that the superprocesses used, while conceptually useful, are not strictly necessary for the estimation.

The model we consider for proof-of-principle once more has a non-Markovian environment coupled by exchange interaction. Following each $\sqrt{X}$ gate on the system, a controlled-$R_X$ gate is applied between system and environment, controlling a rotation by $\theta = \pi/16$.
This gate dependent error does not factor outside each of the gates, and so cannot be absorbed into the process tensor. It is hence incompatible with the \acs{PTT} models from Chapters~\ref{chap:PTT} and~\ref{chap:efficient-characterisation}. We illustrate this in Figure~\ref{fig:spillage_error}a.


To respectively account both for non-Markovian correlations and the superprocess transform, we require a process tensor with bond dimension $\chi = 2$ and tester with bond dimension $\chi = 4$. The former encapsulates the single-qubit environment, and the latter encapsulates the two-qubit gate operation. Though, we do not choose these values in advance, we start from a Markov model and expand until the cross validation is sufficiently well-explained. 
The results' comparison between standard \acs{PTT} and self-consistent \acs{PTT} are shown in Figure~\ref{fig:spillage_error}. Although the noise model is pathological, we are able to completely capture the effects in practice, and attain a complete convergence. Except for drift, which we examine in the discussion section, this closes the gap on every category of noise considered in this chapter. We argue, therefore, that we have demonstrated a near-universal protocol for performing quantum tomography.



\subsection{Ancilla-Assisted Full PTT}
As a last demonstration, we revisit the question of informational completeness and fully learning process tensor representations of dynamics. We have so far mostly only considered the fitting of restricted process tensors. Although the models are fully general, the experimental restriction is intended to be in line with the current reality of most quantum hardware. Further, bespoke unitary control is likely to form the foundation of carefully designed error suppression protocols, as we investigate in Chapter~\ref{chap:NM-control}. 

\begin{figure}[!t]
	\centering
	\includegraphics[width=0.6\linewidth]{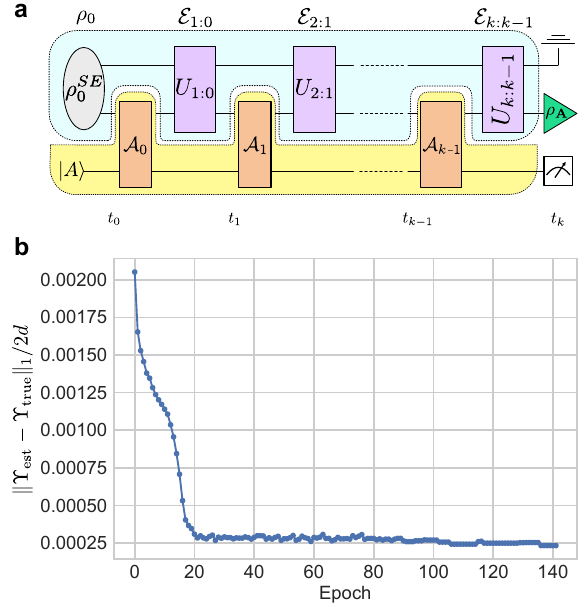}
	\caption[Self-consistently using noisy ancilla qubits to fully probe a multi-time quantum stochastic process. ]{Self-consistently using noisy ancilla qubits to fully probe a multi-time quantum stochastic process. \textbf{a} Repeated unitary interactions with a spectator qubit followed by a terminating measurement will generate an \ac{IC} set of correlated instruments which uniquely determine a process tensor. Although this will be noisy, this offers a far more experimentally accessible method to probe quantum stochastic processes. \textbf{b} Convergence graph from synthetic data. By self-consistently estimating both process tensor and tester, we can obtain an accurate picture of the dynamics, as evidenced by the convergence of estimate to the true process tensor. The final gap is a shot noise limitation.}
	\label{fig:tester-ptt}
\end{figure}

But suppose we want complete diagnostics of all temporal correlations, but we do not possess the ability to apply mid-circuit measurements. We can instead wield a spectator qubit to interact with our system for \ac{IC} control. 
We first recall several key facts from Chapters~\ref{chap:process-properties} and~\ref{chap:PTT}. In principle, tomographic completeness is achieved by having a full set of quantum instruments which can enact some weak measurement in every basis and re-prepare a post-measurement state in every basis. The challenge of this is that many systems do not possess dynamic measurement capabilities -- the ability to perform mid-circuit measurements and then conditionally feed back that measurement information. Moreover, the systems that \emph{do} possess this functionality are often noisy and, worse, slow with respect to background dynamics. 
We explored a set of \ac{IC} control operations in Chapter~\ref{chap:process-properties} of this nature: repeated interactions with a neighbour -- followed by a final projective measurement of that neighbour -- generates a tomographically complete set of testers. This is depicted in Figure~\ref{fig:tester-ptt}a. Unfortunately, although this circumvents the `slow' limitation of mid-circuit measurements, in general the interactions themselves will be noisy and unknown, and characterisation challenging.

Fortunately, the described situation falls under the umbrella of described capabilities introduced in the present chapter. If we have access to a device with more than one qubit, we can use that adjacent qubit as a probe of the primary one. 
The correlated probe will generally be subject to many different noise processes: decoherence and relaxation; interaction with its own non-Markovian environment; and crosstalk with the system of choice.
But all of these effects are naturally encoded in the tester structure and subsequently estimated.
Then, even if the correlated probe is noisy -- which it will almost certainly be -- we can still determine all temporal correlations in both system and the probe. 

We demonstrate this functionality in Figure~\ref{fig:tester-ptt}, where we simulate a three-step quantum stochastic process on two qubits. Each qubit is coupled via Heisenberg interaction to the environment two-level system, and has random $ZZ$ coupling between each other. The interaction of choice is a CNOT controlled by $S$ and targeted on $A$, sandwiched between random unitary operations on both system and ancilla. After randomly sampling 6000 circuits at 1024 shots per circuit, we estimate the process tensor of choice.
Figure~\ref{fig:tester-ptt}b shows the trace distance between the estimated process tensor and the true (numerically known) process tensor. We see convergence to this true process tensor up to sampling deviation.
This further validates the functionality of our methods by capturing the final piece of the puzzle: self-calibrating estimation of some process tensor, as well as the tester employed in characterisation. 

This technique now allows us to dynamically sample from processes in both a clean and efficient manner. This opens up further opportunities to explore the many-time physics of Chapter~\ref{chap:MTP} in both microscopic and macroscopic settings. Although we view this as important for the future development of fault-tolerant quantum technologies, this also makes exploring interesting temporal physics a possibility in a near-term setting. As well as experiments of this nature helping to better understand open quantum systems, we see this as playing a potential role in achieving quantum advantage on small-scale quantum devices.






\section{Discussion}

In this chapter, we have closed many of the gaps in the methodology we have introduced up to present. Specifically, we have developed a self-consistent characterisation procedure designed to incorporate fully general correlated process and correlated control errors. This extends the capabilities of \acs{PTT} by borrowing from the philosophy of \acs{GST}: to assume that nothing is known, and estimate every part of the model equally. 

The components of Markovian noise in control operations, such as \acs{SPAM}; decoherence; coherent gate error; and many more, are already well characterised by a variety of techniques. 
However, we introduce two new key contributions here: the ability to estimate elements such as gate noise and \acs{SPAM} error alongside a non-Markovian background interaction, and the ability to model and estimate correlated effects in the instruments themselves for large numbers of times and qubits. This addresses a significant gap in the \acs{QCVV} literature. 
However, there are still many questions that remain. First and foremost, we have predominantly examined these characterisation techniques from an extrinsic perspective in the ability to explain data. Indeed, this is useful for validating the model and, as we shall see in Chapter~\ref{chap:NM-control}, feeding forward control of the device. But we have not extensively considered the intrinsic properties. In particular, any self-consistent estimation will always be subject to some gauge freedom -- a rotation of each object in the gate set. It will be a natural next step of investigation to determine where precisely these gauge freedoms exist, what gauge-invariant properties can be determined from the estimates, and how an optimal gauge can be selected. 



We have found tensor network models to be incredibly useful to the effect of permitting efficient characterisation, but it would be an interesting future avenue to look beyond these. Specifically, one could consider a model of the process in terms of generators of the dynamics and the correlations therein, and then estimate a few physically reasonable parameters from this model
This may provide more physical insight into the nature of the underlying process and underlying controls. As it stands, tensor networks are not black box models. However, the properties of these objects are not instantly self-evident. 
It may be useful to find more transparent -- yet still efficient -- ways to represent the underlying process and the correlated instruments used to probe it.


\chapter{Techniques for Optimal Control of Non-Markovian Systems}
\label{chap:NM-control}
\epigraph{\emph{Do I dare \newline
Disturb the universe? \newline
In a minute there is time \newline
For decisions and revisions which a minute will reverse.}}{T. S. Eliot, The~Love~Song~of~J.~Alfred~Prufrock}
\noindent\colorbox{olive!10}{%
	\begin{minipage}{0.955\textwidth} 
		\textcolor{Maroon}{\textbf{Chapter Summary}}\newline
		Process tensors are mappings from sequences of gate operations to final states of a quantum stochastic process. In essence, they constitute a map that tells us the future: we can predict the final state of some dynamics conditioned on any sequence of operations. This makes the formalism a valuable tool for quantum optimal control -- pick a property of the state that one would like to achieve, and then extremise that property subject to sequences of gates. Here, we explore this idea. Using our various estimation methods for process tensors, we design protocols by which non-Markovian noise can be suppressed or eliminated through the careful choice of logical control operations. 
		\par\vspace{\fboxsep}
		\colorbox{cyan!10}{%
			\begin{minipage}{\dimexpr\textwidth-2\fboxsep}
				\textcolor{RoyalBlue}{\textbf{Main Results}}
				\begin{itemize}
					\item We show in practice how process tensors can be used to find optimal dynamical decoupling sequences from any open quantum dynamics, both in the single-qubit and multi-qubit case.
					\item We demonstrate how to find optimal decompositions of $SU(4)$ two-qubit generic gates in the presence of any noise.
					\item We provide a blueprint for how noise-aware compilation could be scaled using the tenets of Markov order: selecting or varying gates based on the choice of other gates in a system's causal light cone. 
				\end{itemize}
		\end{minipage}}
\end{minipage}}
\clearpage

\section{Introduction}
In this thesis we have considered the problem of characterising exotic models of quantum noise and open quantum systems. We have seen how these objects can be highly predictive and capture the essential extrinsic properties on both real and simulated quantum data. We have also investigated the extent to which the intrinsic characteristics of these estimations can aid with diagnostic properties from both generic open quantum systems and in the context of noise. In this chapter, we visit the final piece of the puzzle: how can we use these characterisations to feed forward for better control of quantum computers? We will make the connection between the estimation of process tensors and the circumvention of correlated noise on quantum computers in a variety of circumstances. Note that these results incorporate many of the different approaches from Part II of the thesis and represent a spectrum from preliminary to advanced results. Nonetheless, the principles are all equivalent. 

The characterisation given in \acs{PTT} can be useful for qualitatively different applications. Broadly speaking, these applications fall into two different camps: non-Markovian diagnostics, and non-Markovian optimal control. In the former, conventional many-body tools are applied to the Choi state to probe characteristics of the temporal correlations via correlations between the \acs{CPTP} marginals $\hat{\mathcal{E}}_{j:j-1}$. These characteristics can reveal a great deal about the noise: its complexity, the probability of Markov model confusion, the size of the environment, as well as its quantum or classical nature, as some examples~\cite{Pollock2018, giarmatzi_quantum_2018, White2021}. We explored these properties for both noisy quantum devices and in general open system settings in Chapter~\ref{chap:MTP}. 

Here, we focus on control. We have demonstrated a host of results indicating high predictive power in a variety of settings for arbitrary quantum operations. This capability straightforwardly translates to control applications.
We present some examples of how a process tensor characterisation can straightforwardly yield superior circuit fidelities on real \acs{QIP}s, and the extent to which a conditional Markov model can be used for this. Reconstruction fidelity validates the ability of the process tensor to accurately map a given sequence to its final state. This is especially applicable to near-term quantum devices whose control operations are high in fidelity but whose dynamics (non-Markovian or otherwise) are not under control. In the same way that a mathematical description of a quantum channel may be used to predict its behaviour on any input state, the process tensor can predict the output state of a process, subject to any sequence of input operations.
A mapping from unitary gates to outcomes allows an experimenter to ask `What is the optimal sequence of gates that best achieves this outcome?'. Two key features distinguishing this from regular quantum optimal control is firstly that after characterisation, all optimisation can be performed classically with confidence. Secondly, the process is fully inclusive of non-Markovian dynamics, allowing for the suppression of correlated errors.
Simply choose an objective function $\mathcal{L}$ of the final state. Then $\mathcal{L}\left(\mathcal{T}_{k:0}\left[\mathbf{A}_{k-1:0}\right]\right)$ classically evaluates $\mathcal{L}$ conditioned on some operation sequence $\mathbf{A}_{k-1:0}$ using the process characterisation. This can be cast as a classical optimisation problem to find the sequence of gates which best results in the desired value of $\mathcal{L}$. 

We explore and demonstrate various components of a toolkit here, with a particular focus on real results on IBM Quantum devices. We make it clear what should be possible in terms of using our characterisation framework to reduce the effects of correlated noise. Fundamentally, dealing with non-Markovian noise boils down to two key components: (1) controlling correlated errors to cancel with one another, and (2) accounting for serial context dependence (such as the effects of a particular gate on a neighbouring qubit) by adjusting future operations. 

\section{Cancelling Correlated Errors}

Non-Markovian noise is problematic because it generates complex dynamics which is both hard to capture, and introduces more overall error. But on the flip side, in contrast to Markovian noise, once these dynamics are well-understood, they can be manipulated to produce cleaner dynamics. We describe this with the following example. 

\begin{example}[Cancelling correlated errors]
	\examplecontent{
	Consider two locations in a circuit. In location \#1, a stochastic error $R_Z(\varepsilon_1)$ occurs, and similarly in location \#2 we have $R_Z(\varepsilon_2)$. If the covariance between $\varepsilon_1$ and $\varepsilon_2$ is zero, then we have two independent noisy channels decohering the system. However, if the two are maximally correlated, then $\varepsilon_1 = \varepsilon_2$. With this knowledge, one could place an $X$ gate before and after location \#1, performing the transformation $R_Z(\varepsilon_1)\mapsto XR_Z(\varepsilon_1)X = R_Z(-\varepsilon_1)$. When combined with location \#2, we end up with $R_Z(\varepsilon_2 - \varepsilon_1)$. If the two errors are maximally correlated, then they cancel out with one another. }
\end{example}

The above example distils the essence of how non-Markovianity can be controlled to produce a cleaner dynamics. In general, the situation will be more complex than this. Indeed, as we saw in Chapter~\ref{chap:MTP}, the noise observed in IBM Quantum devices cannot be totally classically explained, and does not purely decompose into correlated Pauli channels.
We first introduce some preliminary results as a proof of concept that characterisation can indeed be fed forward for optimal control. Next, we develop more sophisticated methods stemming from the efficient characterisation ideas in Chapters~\ref{chap:efficient-characterisation} and~\ref{chap:universal-noise}.

\subsection{Background}

The cancelling of correlated errors through stroboscopic control is broadly known as the practice of \acf{DD}, and so we will present our results in this framework. 
\ac{DD} is a powerful and well-known technique to suppress the effects of decoherence and other errors in quantum processes. 
One can view \acs{DD} as the consumption of non-Markovianity~\cite{berk2021extracting}. If a process is non-Markovian, that means instruments may be placed at different times to distil the process into a cleaner channel. Meanwhile, Markovian noise cannot be eliminated through active control methods and requires probabilistic error mitigation methods to remove its effects~\cite{Temme2017}. These, however, do not scale. 
Our characterisation techniques provide exactly the means to determine the non-Markovianity from the environment, and hence remove it. 



The principle of \acs{DD} is to use active control to manufacture symmetries in the system-environment interactions that can be employed as a decoherence-free subspace. Consider for instance the simple case where a qubit couples to a bath via $Z$-interaction terms,
\begin{equation}
	H_{SE} = Z \otimes B_Z.
\end{equation}
Left unchecked, this interaction will eventually decohere the system via the free evolution $f_t = \text{e}^{-i t H_{SE}}$. 
But suppose we can introduce a control term on the system, $H_S = \frac{\pi}{2}\delta(t-\tau_i)X$ which is a perfect $\pi$ pulse around the $X$ axis at designated times $\tau_i$. Because this control term is a delta function, no free evolution occurs when the pulse is turns on. Choosing $\tau_i = \tau$ and $2\tau$, let the system evolve for total duration $2\tau$. This amounts to evolution of  
\begin{equation}
	Xf_\tau X f_\tau = X \text{e}^{-i\tau H_{SE}} X \text{e}^{-i\tau H_{SE}} = \text{e}^{-i\tau XH_{SE}X} \text{e}^{-i\tau H_{SE}}.
\end{equation}
Now, since $X$ anti-commutes with $Z$, then we have $XH_{SE}X = -H_{SE}$ and so the total evolution is 
\begin{equation}
	Xf_\tau X f_\tau = \text{e}^{+i\tau H_{SE}}\text{e}^{-i\tau H_{SE}} = \mathbb{I}.
\end{equation}
We can interpret this to mean that from $t\in [0,\tau)$, the system undergoes some phase damping from the coupling; for $t\in (0,2\tau)$, the populations of the system are flipped, but it begins to recohere as the negated $H_{SE}$ cancels out the original evolution. Finally, at $t=2\tau$, the state of the system is flipped back and the total evolution is completely cancelled, meaning that the system is momentarily in its original state and completely decoupled from the bath.

In the more general case where $H_{SE} = \sum_i P_i\otimes B_i$, an $X$ gate will anticommute with (and hence cancel) $Z$ and $Y$ errors, but $X$ contributions will remain. The solution here is to have a recursive sequence where another loop of active control is designed specifically for the $X$ terms, for example by using $Y$ or $Z$ gates. 
With the same free evolution as before, $f_t = \text{e}^{-itH_{SE}}$, then following the application of an $X$ gate, between two evolutions by $\tau$, we have
\begin{equation}
	f_{2\tau}^{X} = f_\tau X f_\tau = \text{e}^{-i2\tau(X\otimes B_X) + H_E} + \mathcal{O}(\tau^2).
\end{equation}
Then, concatenating with $Z$ gates, 
\begin{equation}
	\begin{split}
		f_{4\tau}^{Z} &= Z f_{2\tau}^X Z f_{2\tau}^X\\
		&= ZX f_\tau X f_\tau Z X f_\tau X f_\tau\\
		&= Yf_\tau X f_\tau Y f_\tau X f_\tau.
	\end{split}
\end{equation}
The final result is an evolution of 
\begin{equation}
	f^{XY}_{4\tau} = \text{e}^{-i4\tau H_E} + \mathcal{O}(\tau^2),
\end{equation}
and hence the system is generically decoupled from the environment up to second order in the time scale. 
This gives a sense of how \acs{DD} works: use system control to anticommute with $SE$ terms. 

We can recognise this as non-Markovianity by observing that a free evolution by $\tau_1$ introduces a coupling error by some angle $\theta_1$. A subsequent evolution by $\tau_2$ introduces a second error by angle $\theta_2$. If $\tau_2$ is chosen to equal $\tau_1$, then the errors are equal in magnitude. By applying pulses to the system which anticommute with the interaction, one can ensure that a rotation by $\theta_1$ is followed by a second rotation of $-\theta_1$.
The upshot here is that correlated errors in some sense are preferable to uncorrelated ones. Although their collective effects are more deleterious, if one understands the nature of the correlation, then one knows their relative magnitudes and the errors can be actively manipulated to cancel one another out. 

There are two hurdles to \acs{DD} in practice. The first is that most \acs{DD} protocols derived in practice are intended to be universally optimal to some order. That is, they come equipped with guarantees to cancel out all terms in any Hamiltonian up to some level in, for example, a Magnus expansion. But one can get away with being far more efficient if they are willing to give up generality. 
The second problem is that non-Markovian effects -- generated by, for instance, bath self-correlation -- can cause serial context dependence of gates. This means that the first few choices of gates in a \ac{DD} sequence can adversely effect the next few choices, by virtue of that non-Markovianity. 
The motivation behind this chapter is, using techniques we have developed so far for non-Markovian noise characterisation, to feed the information forward into optimal control of non-Markovian settings. 

\subsection{Proof of Principle}

To begin, we shall present some introductory results which use characterisations of two and three-step process tensors to suppress small scale errors as they emerge on IBM Quantum devices. These establish that non-Markovian characterisation can indeed be fed forward to better control a quantum device. In the following sections we shall expand on many facets of these initial principles to develop robust and versatile control techniques.

\subsubsection*{Cancelling Two-Point Correlations}

In addition to non-Markovian characterisation and diagnostics, we now show that the process tensor can be a useful tool for quantum control.
With a direct map from control operations to experimental outcomes, the data can be used to find which gates optimally output a desired state in a parametrised circuit. 
This outcome could harness external couplings to that end, using only local operations to manipulate them. 
Having already captured the process, the need for hybrid quantum-classical optimisation is eliminated. 
The desired result could be the most entangled state, the highest fidelity equal superposition, or some member of a decoherence-free subspace.
The procedure naturally accounts for any mitigating background, such as environmental noise or crosstalk. 
It is a matter of simple numerical optimisation to find the sequence of operations achieving the closest possible state to the one we desire:
(i) Select an objective function $\mathcal{L}$ which computes some quantity on the output density matrix, subject to the sequence $\textbf{A}_{k-1:0}$ of operations performed.
(ii) Find:
\begin{gather}
	\label{best-op}
	\argmin_{\textbf{A}_{k-1:0}} \mathcal{L}\left(\mathcal{T}_{k:0}[\textbf{A}_{k-1:0}]\right).
\end{gather}
For unitaries, this is a straightforward minimisation over three parameters per time-step.\par
\begin{figure}[!htb]
	\centering
	\includegraphics[width=0.9\linewidth]{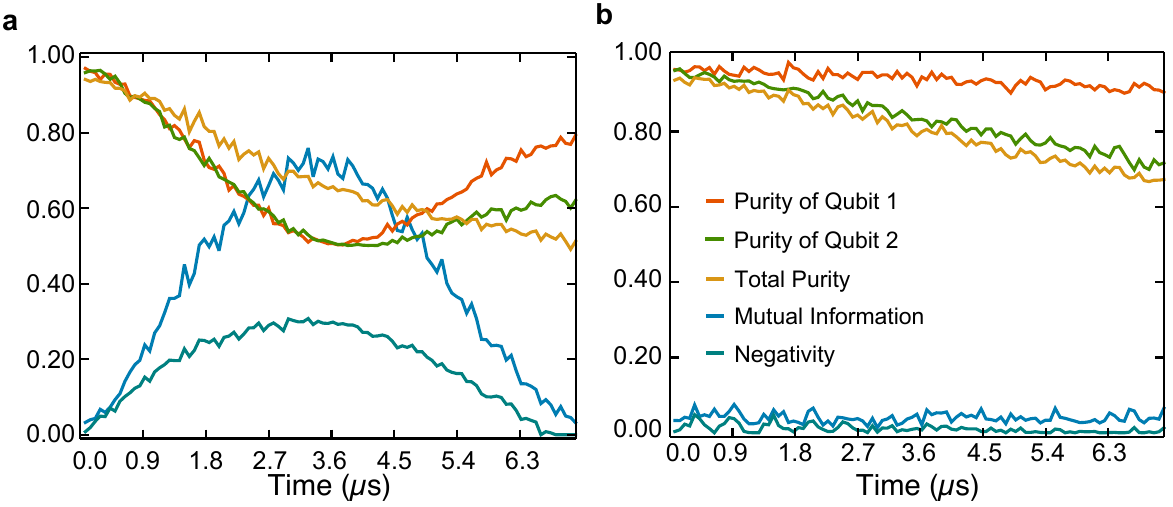}
	\caption[Demonstration of coherent control with non-Markovian noise ]{Coherent control with non-Markovian noise. Entanglement, mutual information, and purities extracted from the two-qubit density matrix after being initialised in the $\ket{++}$ state. \textbf{a} Both qubits are left idle and the natural evolution is tracked. \textbf{b} As a simple demonstrative application of the process tensor, we use the construction from Equation~\eqref{best-op} to find the optimal decoupling pulse. We periodically apply this gate to qubit 1. We see greatly improved coherences and almost complete elimination of entanglement between the two qubits, without actually characterising the nature of the interaction. }
	\label{coherence-plots}
\end{figure}

As example on an IBM Quantum device, we first consider two neighbouring qubits initialised in the $\ket{+}$ state on \emph{ibmq\_valencia}. 
Figure~\ref{coherence-plots}a shows the consequences of their natural coupling, extracted from the reconstructed two-qubit density matrix after some idle time.
The results, which summarise negativity, mutual information, and state purities, show genuine entanglement between the two qubits.
This form of dynamical behaviour will give rise to correlated errors in devices.
After detection of a non-trivial interaction, we can use Equation~\eqref{best-op} to decouple the qubits.
So-called `bang-bang' decoupling approaches have been thoroughly studied in the literature, but usually require \emph{a priori} knowledge of the system-environment interaction Hamiltonian~\cite{qec-textbook}.
Using a one-step process tensor to form outcomes, our objective function is $2-\gamma_1-\gamma_2$, where $\gamma_i$ is the purity of the reduced state: $\gamma_i = \text{tr}(\rho_i^2)$.
Performing the minimisation in Equation~\eqref{best-op}, we find the best decoupling operation. This turns out to be the gate 
\begin{gather}
	\begin{pmatrix}
		0.0051 & \text{e}^{-i\cdot(1.073)} \\
		\text{e}^{i\cdot(0.188)} & 0.0051\cdot\text{e}^{i\cdot(2.257)}
	\end{pmatrix}
\end{gather} 
which amounts to a rotation of approximately $\pi$ around the axis $(n_x,n_y,n_z) = (0.8076,0.5894,4.609\times 10^{-3})$.
We then repeat the experiment of Figure~\ref{coherence-plots}a, but periodically apply the decoupling operation approximately every 0.5 $\mu s$.
This yields the results in Figure~\ref{coherence-plots}b, wherein the purities of each qubit have been significantly increased, and the entanglement over time suppressed. Note that this is a demonstration of how the process tensor can be applied as an outcome-based control tool, rather than a rigorous benchmark of decoupling. We have not compared this to standard decoupling techniques, and the operation spacing times were arbitrarily chosen. 
\par
This simple framework is widely applicable to many forms of quantum control. 
In particular, it allows for either mitigating or controlling non-Markovian noise without first understanding it at a microscopic level. 
Broadly, the user need only specify a desired outcome, without studying the means to achieve it. 

Here, we more explicitly discuss our adaptive control methods using the process tensor. 
In each case, the system qubit and its neighbour were both initialised in the $\ket{+}$ state.
We sought to use the process tensor to control the always-on interaction between the two qubits without actually learning it.
In the first scenario, using operations only on qubit 1, we construct a single step process tensor with a size-24 basis, 256 ns of idle time on either side, and two-qubit state tomography at the end.
Altogether, this is $24\times 9 = 216$ experiments.
Strictly speaking only single qubit state tomography is required for the purpose of decoupling one qubit, however we created a mapping to the two-qubit output in order to best show these two qubits decoupled.
%


\subsubsection*{Optimising for State Preparation}

We can take this notion a step further. Given some noisy environment, we would like to prepare a given state according to that noise. Moreover, let this take place across several time steps. If a circuit has idle time available, then as well as cancelling correlated errors, this would also let us momentarily place the state into a decoherence-free subspace. Although this method is not scalable, it may better inform an understanding of the noise on the device and an in-principle optimal creation of a state.

To demonstrate the utility of this idea, we use the process tensor to improve the fidelity of IBM Quantum devices over multi-time processes. Note that this characterisation overcomes both Markovian and non-Markovian errors.
In particular, we apply many sequences of random unitaries to a single qubit and measure the final state. Interleaved between each operation is a delay time roughly equivalent to the implementation duration of a CNOT gate. 
We then compare the fidelity of this output state to the ideal output subject to those unitaries. That is, generate a set of ideal outputs:
\begin{equation}
	\rho_{\text{ideal}}^{ijk} = \mathcal{A}_3^k\circ \mathcal{A}_2^j \circ \mathcal{A}_1^i [|0\rangle\!\langle 0|],
\end{equation}
with a set of values $\mathcal{F}(\rho_{\text{ideal}}, \rho_{\text{actual}})$.
Mirroring these dynamics, we construct a process tensor whose basis of inputs is at the same time as the target unitaries. We supplement the reconstruction by using \acs{GST} to estimate the noisy device \acs{POVM}. The estimated \acs{POVM} is then used in the \acs{MLE} processing, rather than the ideal projective measurements. The purpose of this is to avoid inflating any circuit improvement. For example, a relaxation process during the measurement operation would be absorbed into the process tensor estimate and could be artificially overcome by increasing the $\ket{1}$ population. By accounting for measurement errors in \acs{PTT} and QST, we are considering only dynamics during the circuit as a more representative depiction of generic \acs{PTT} capabilities.
Finally, using the \acs{PTT} characterisation we determine which set of unitaries, $\mathcal{V}$ should be used (instead of the native ones, $\mathcal{A}$) in order to achieve the ideal output state. Let each unitary gate $V$ corresponding to the map $\mathcal{V}$ be parametrised in terms of $\theta$, $\phi$, and $\lambda$ (in the standard parametrisation we have so far given).

\begin{figure}[!t]
	\centering
	\includegraphics[width=\linewidth]{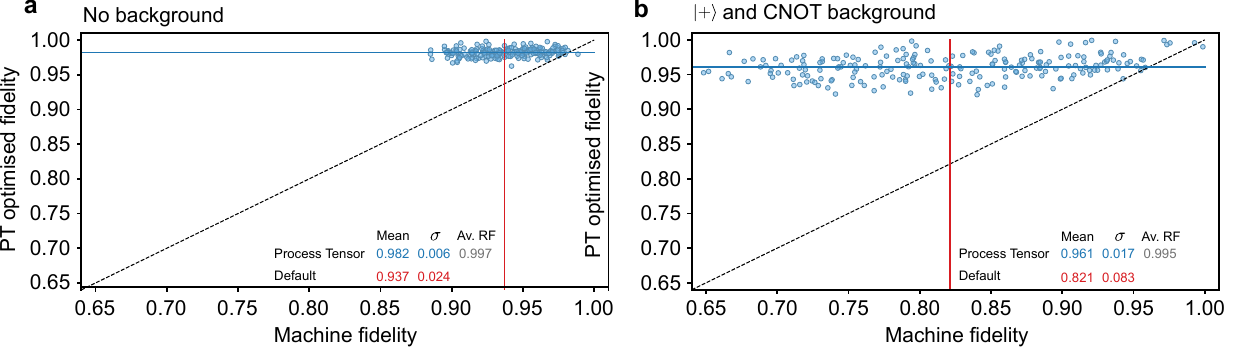}
	\caption[Results of using the process tensor to optimise multi-time circuits with different random single-qubit unitaries. ]{Results of using the process tensor to optimise multi-time circuits with different random single-qubit unitaries. The $x$-axis indicates the fidelity compared to the default machine fidelity. The $y$-axis is the fidelity of the sequence when the process tensor is used to optimise to the ideal case. \textbf{a} Here, we have three sequential unitaries with a wait time interleaved similar to that of two CNOT gates (0.71 $\mu$s) on the \emph{ibmq\_manhattan}.  \textbf{b} A similar setup is considered on \emph{ibmq\_bogota}, but with the neighbouring qubit in a $\ket{+}$ state and subject to four CNOT gates per time-step (2.5 $\mu$s).}
	\label{fig:circuit_improvement}
\end{figure}

The process prediction is then given with:
\begin{equation}
	\rho_{\text{predicted}}^{ijk}(\vec{\theta},\vec{\phi},\vec{\lambda}) = \mathcal{T}_{3:0}\left[\mathcal{V}^k_3, \mathcal{V}^j_2, \mathcal{V}^i_1\right].
\end{equation}
Then, for each combination $ijk$, we find
\begin{equation}
	\argmax_{\vec{\theta},\vec{\phi},\vec{\lambda}}\mathcal{F}(\rho_{\text{ideal}}^{ijk}, \rho_{\text{predicted}}^{ijk})
\end{equation}
and use these optimal values in sequences on the device.
The results across 216 random sequences are summarised in Figure~\ref{fig:circuit_improvement}a. The average observed improvement was 0.045, with a maximum of 0.10. In addition, the distribution of fidelities is much tighter, with the standard deviation of device-implemented unitaries at 0.0241, compared with our computed values at $0.00578$. We also repeated similar runs on \emph{ibmq\_bogota} intended to drive some characteristics of crosstalk: starting the neighbour in a $\ket{+}$ state followed by four sequential CNOTs to its other-side-nearest-neighbour between each unitary. Initialising the neighbour in a $\ket{+}$ state is intended to generate a passive entangling interaction between the two qubits due to the always-on $ZZ$ interaction found in superconducting transmons. We found that these native fidelities were much worse than on the \emph{ibmq\_manhattan}, despite possessing similar error rates -- implicating the effects of crosstalk. We emphasise that the noise encountered in all our results is naturally occurring from device fabrication, rather than a contrived environment. Nevertheless, the process-tensor-optimal fidelities in Figure~\ref{fig:circuit_improvement}b are nearly as high. This suggests a path forward whereby quantum devices may be characterised using \ac{PTT} and circuits compiled according to the correlated noise of that device. An obvious drawback of this is the characterisation requirements, and the fact that the compilation is specific to the state being created. We now investigate using a conditional Markov order model to perform a similar but more generalisable task.

\subsubsection*{Optimising Arbitrary Circuits with Finite Markov Order Process Tensors}
Using a complete process tensor model to optimise circuit sections may be feasible for a small number of gates, and, indeed, may be necessary for highly correlated noise. However, it is not desirable in a generic sense to characterise redundant information. Moreover, it is impractical to optimise over specific circuits in a state-dependent way when inputs may be reduced subsystems of a larger register. Here, we address both of these points. We target longer circuits with larger values of $k$ by using a truncated Markov model. In doing so, we both validate our conditional Markov order methodology, and demonstrate the need for approaching the problem of \acs{NISQ} noise with temporal correlations in mind. Further we also change our optimisation approach: instead of trying to create a specific state on a circuit-by-circuit basis, we numerically find the sequence of gates which most closely takes the effective process to be the identity channel. This allows for arbitrary addressing of non-Markovian noise without a priori knowing the input state.\par 
A five-step process is considered with delays of approximately 1.2$\ \mu$s after each gate. We characterise this process using conditional Markov order models of $\ell = 1$, $\ell=2$, and $\ell=3$ under three different cases: no operations on the background qubits, one nearest neighbour to the system initialised in a $\ket{+}$ state, and finally two nearest neighbours and one next-to-nearest neighbour in a $\ket{+}$ state. The first job took place on \emph{ibmq\_montreal} and the second and third on \emph{ibmq\_guadalupe}. The purpose of the latter two analyses is to encourage any (predominantly $ZZ$) interaction which realistically might occur between qubits in an algorithm. We then generate 100 sequences of 5 random unitary gates, followed by \ac{QST}. These sequences serve two purposes: first, we evaluate the reconstruction fidelity for the different Markov order models, and secondly we use these as our benchmark for adaptively improving the native fidelity of the device. \par 
\begin{figure}[t!]
	\centering
	\includegraphics[width=\linewidth]{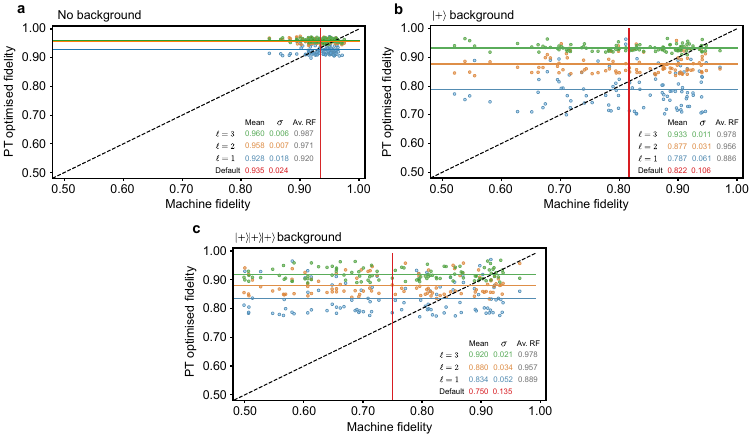}
	\caption[Results of using conditional Markov models to improve the fidelities of a five-step circuit on \emph{ibmq\_montreal} and \emph{ibmq\_guadalupe}. ]{Results of using conditional Markov models to improve the fidelities of a five-step circuit on \emph{ibmq\_montreal} and \emph{ibmq\_guadalupe}. We construct conditional Markov order models for $\ell=1$, $\ell=2$, and $\ell=3$ across five steps under a variety of background conditions. For randomised inputs, these models are used to optimise the next four operations. The four optimal operations are then applied to each input and the results compared to machine fidelity. The mean and standard deviation for each set of circuits is listed, as well as average reconstruction fidelity of the models. \textbf{a} No background operations are applied. \textbf{b} The nearest neighbour to the system is initialised in a $\ket{+}$ state. \textbf{c} Two nearest neighbours and one next-to-nearest neighbour from the system are initialised in a $\ket{+}$ state.}
	\label{fig:circuit_improvement_CMO}
\end{figure}
We select an \acs{IC} set of unitary gates $\{\mathcal{A}^i\}$ to be applied in the first circuit position, generating a set of ideal states $\{\rho_{\text{ideal}}\} := \{\mathcal{A}[\ket{0}\!\bra{0}\}$. That is, $i$ indexes a set of random states. We then parametrise the next four gates, again, in terms of $\vec{\theta}$, $\vec{\phi}$, and $\vec{\lambda}$. However, this time, the gates are the same for each input. Finally, using each $\mathbf{\Upsilon}_{5:0}^\ell$, we compute:
\begin{equation}
	\label{CMO-opt}
	\argmax_{\vec{\theta},\vec{\phi},\vec{\lambda}} \sum_i \left[\mathcal{F}(\rho_{\text{ideal}}^i,\rho_{\text{predicted}}^i)\right]^2.
\end{equation}
In plain words, we are finding the four gates which simultaneously best preserve all of our random input states. After running this optimisation for each Markov order model and each background, we then aimed to create the ideal output from the 100 random sequences. First, by creating the state with the first gate, then applying the four gates found from Equation~\eqref{CMO-opt}. Following \ac{QST} at the end, we compute the fidelity of each final state with respect to the ideal.
The purpose of this routine was two-fold: to determine whether active circuit improvements (akin to \acs{DD}) could be systematically found, even in the presence of non-Markovian noise, and to ascertain how the inclusion of higher order temporal correlations in the model could help achieve this task. Without randomising over the inputs, we found that the $\ell=1$ model would hide each state in a decoherence-free subspace until the last gate, which is not a generalisable strategy.
The results of these runs are shown in Figure~\ref{fig:circuit_improvement_CMO} for each sequence and each Markov model with both the mean circuit fidelities and reconstruction fidelities printed. 
With no activity on neighbouring qubits, we find a moderate amount of non-Markovian noise at this time scale. 
This is consistent with the findings of multi-time correlations across IBM Quantum devices in Chapter~\ref{chap:MTP}.
The optimal interventions improve the average circuit fidelity to a similar level for each given Markov order. Hence, the generic correctability for given circuit structures may saturate, regardless of the completeness of characterisation.
For the second and third situations, the dynamics are more complex and we see a clear separation between the different Markov orders. By accounting for these higher order temporal correlations, we are able to more substantially increase circuit fidelities, both in terms of the mean value, and in terms of the tightness of the distribution. Only in the last case do we find that a Markov model $\ell=1$ is able to achieve an improvement, further highlighting the need for our multi-time process characterisation on \acs{NISQ} devices.

For each of the conditional Markov order tests, a five step process with $\ell = 1$, $\ell = 2$, and $\ell = 3$ is considered. This amounts to reconstructing, respectively, 5, 4, and 3 memory block process tensors as detailed in Chapter~\ref{chap:efficient-characterisation}. The structure of the circuits is similar to the three step process tensor, however not all circuit elements are varied. For example, with $\ell=3$, this means reconstructing the three process tensors corresponding to circuit structure $\mathcal{U}_{\mu_0}-\mathcal{U}_{\mu_1}-\mathcal{U}_{\mu_2}-\Pi_i$; $\mathcal{U}_0 - \mathcal{U}_{\mu_0}-\mathcal{U}_{\mu_1}-\mathcal{U}_{\mu_2} -\Pi_i$; and $\mathcal{U}_0 -\mathcal{U}_0 - \mathcal{U}_{\mu_0}-\mathcal{U}_{\mu_1}-\mathcal{U}_{\mu_2}- \Pi_i$. For $\ell = 2$, a subset of the same data can be reused: fixing $\mu_0=0$ and varying $\mu_1$ and $\mu_2$, for example. The only additional information required is that a projective measurement needs to be made in position 2 of the circuit in order to determine the state at the end of the first $\ell=2$ memory block. A similar process follows for determination of $\mathbf{\Upsilon}_{5:0}^1$, with an extra memory block process tensor constructed with a projective measurement at position 1. This totals $3 \times (10\times10\times 10 \times 3) = 9000$ circuits for $\mathbf{\Upsilon}_{5:0}^3$, an extra $10\time10\times 3 = 300$ circuits for $\mathbf{\Upsilon}_{5:0}^2$, and an extra $10\times 3 = 30$ circuits for $\mathbf{\Upsilon}_{5:0}^1$.\par 
Since the state is being propagated along in our finite Markov order stitching procedure, it is important that it is well-characterised without measurement error. To this effect, we use \acs{GST} again to estimate our \acs{POVM}. This is more essential than before, since now our \acs{PTT} construction is contingent on inputting the correct form of the operation. This is also true of the unitary gates we apply, however single qubit error rates are $\mathcal{O}(10^{-4})$, compared with measurement errors of $\mathcal{O}(10^{-2})$, and so a far smaller assumption.

We have presented so far various demonstrations indicating that process tensor characterisations can provide superior control of a quantum device. However, these have mostly been proof-of-principle and not strictly generalisable. We seek a light-weight solution that can be readily implemented into near-term devices.

\subsection{Optimising Dynamical Decoupling Sequences}

Much like we built upon our characterisation protocols to be more efficient and more robust, so too can we focus ourselves on control tools that readily integrate into circuit compilation software. We first introduce our tensor network characterisation methods to the task. This permits a generalisation to arbitrary step numbers, and is cognisant of any errors in the electronics. 

Let us move away from the continuous time description and provide an operational description of \ac{DD}. Given a process $\Upsilon_{k:0}$, one can trace over the initial state to obtain a sequence of possibly correlated \acs{CPTP} maps 
\begin{equation}
	\Tr_{\mathfrak{o}_0}[\Upsilon_{k:0}] = \hat{\mathcal{E}}_{k:k-1;k-1:k-2;\cdots;1:0}.
\end{equation}
Applying a sequence of $k-2$ gates $\{D_1, D_2,\cdots,D_{k-1}\}$ generates the conditional dynamical map 
\begin{equation}
	\hat{\mathcal{E}}_{k:0}^{(D_1,D_1,\cdots,D_{k-1})} = \Tr_{\bar{\mathfrak{o}}_k \bar{\mathfrak{i}}_k}\left[\hat{\mathcal{E}}_{k:k-1;k-1:k-2;\cdots;1:0}\cdot \bigotimes_{i=1}^{k-1} \hat{\mathcal{D}}_i^{\text{T}}\right].
\end{equation}
If 
\begin{equation}
	\|\hat{\mathcal{E}}_{k:0}^{(D_1,D_1,\cdots,D_{k-1})} - \hat{\mathcal{I}}\| < \|\hat{\mathcal{E}}_{k:0}^{(I_1, I_2, \cdots , I_{k-1})} - \hat{\mathcal{I}}\|,
\end{equation}
for some norm $\|\cdot\|$, then the set of operations $\{D_i\}$ has (possibly imperfectly) decoupled the system from the environment: it has left the effective dynamics cleaner than doing nothing. If we find $\{D_i\}$ such that 
\begin{equation}
	\label{eq:dd-opt}
	\|\Tr_{\mathfrak{o}_0}[\Upsilon_{k:0}^{(D_0,D_1,\cdots,D_{k-1})}]- \hat{\mathcal{I}}\|
\end{equation}
is minimised, then we say a given set of gates is optimally decoupling. That is to say, the channel from time $t_0$ to time $t_k$ can be optimised to be as close to the identity as possible. One can choose different metrics for this; ideally the most effective would be the diamond norm, but this does not have a closed form and so in practice makes the optimisation difficult. Instead, we optimise for the trace distance between the conditional process tensor and the identity channel. 



We now propose to employ the full gamut of our characterisation tools for the purpose of error suppression. In particular, given \emph{any} set of dynamics across any given time scale, we wish to be able to determine the best \acs{DD} sequences for that dynamics without needing to possess a microscopic model of the underlying physics. What we propose is a highly useful and versatile tool. It is agnostic to the hardware, efficient to construct, and requires only an abstraction of the control implementation. 

We have seen so far how the results of Chapter~\ref{chap:PTT} gave us robust characterisation which could be parsed into superior control of a quantum device. But these fully general models are too tensor to practically characterise in general, one could only find 2-3 decoupling gates and at great expense. Instead, we use the methodology introduced in Chapter~\ref{chap:efficient-characterisation} for the estimation of tensor network forms of process tensors. 

An elegant property of the decomposition we use is that to the extent which $Z$ gates can be treated as `virtual' on near-term devices -- that is, up to the timing resolution of the classical hardware -- these pulses \emph{are} instantaneous and perfect. Any possible problem with the implementation of the physical gate, or the fact that it has finite width need not be treated theoretically, it is accounted for automatically. 

\begin{figure}[!t]
	\centering
	\includegraphics[width=\linewidth]{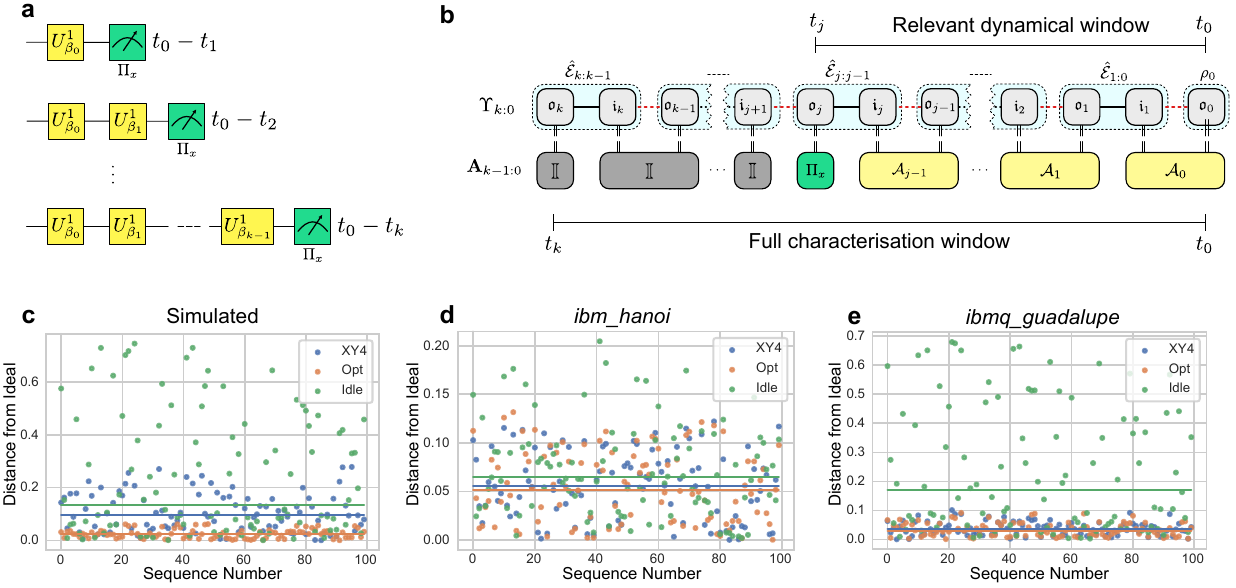}
	\caption[Protocol and results for optimisation of dynamical decoupling sequences. ]	{Protocol and results for optimisation of \acs{DD} sequences. To characterise a variety of differently sized windows, a maximum duration is selected, from which a particular window is used where necessary. \textbf{a} Characterisation takes place using the techniques of Chapter~\ref{chap:efficient-characterisation}, where random operations are applied. Now, however, the sequence is terminated at different points so differently sized windows can also be characterised. 
	\textbf{b} The process tensor for $t_0$ to $t_k$ is reconstructed, but when a relevant target window $t_0-t_j$ is selected, the remainder of the process can be traced out. The remaining window is then used to optimise a \acs{DD} sequence for that specific timing.
	\textbf{c--e} Simulated and experimental results using a nine-step process tensor to optimise \acs{DD} sequences.
	For the given window, a random state is created, then \acs{DD} sequence applied, and then compared with the perfect state at the end. We compare the case of completely idle dynamics, a standard $XY4$ sequence, and our noise-aware compilation. On the $y$-axis, we display the trace distance between ideal states and the noisy state. For each of idle dynamics, $XY4$-protected sequences, and optimal \acs{DD}, we plot the mean values as a horizontal line. \textbf{c} A simulated two-qubit system with random Heisenberg couplings. \textbf{d} Results from \emph{ibm\_hanoi}. \textbf{e} Results from \emph{ibm\_guadalupe}.}
	\label{fig:dd-results}
\end{figure}

It may seem that this approach is overly specific to the window $[0,t_k]$ that has been characterised. Often the idle time a qubit experiences in a circuit is highly variable. To account for this, one can characterise up to a fixed time $t_k$, but terminate early at different randomly selected times. That is, characterise circuits of the form given in Figure~\ref{fig:dd-results}a. This way, we obtain slightly more information than the standard restricted process tensor. In particular, we characterise a window for which the final state is known at any time up to $t_k$. Hence, when a circuit has idle time shorter than this, we can still optimise for the relevant window and trace over the remainder, as depicted in Figure~\ref{fig:dd-results}b. From this, we have a protocol which is both modular, and extendable to arbitrary numbers of times. 

We test this optimisation both in simulation and on IBM Quantum devices. We fix a nine step process. In the first operation, a random state is created. Then, the next eight operations constitute a \acs{DD} sequence. We take two repetitions of the $XY4$ sequence as the baseline with which we compare. This \acs{DD} sequence is universal in the sense that it removes all terms in the Hamiltonian to first order. If the interaction were purely coupled through $Z$ on the system, then it in principle cancels the interaction entirely. 

In our simulation, we take a qubit coupled via exchange interaction to another two-level system. This models potential defects in superconducting devices~\cite{muller2019towards} as well as nuclear spins and other qubits in solid-state hardware~\cite{He2019,yoneda2022noise}. A tensor network representation of the process tensor is constructed from $10^6$ shots to characterise the dynamics, and the optimisation in Equation~\eqref{eq:dd-opt} used to final the optimal \acs{DD} sequence. To benchmark how this sequence protects a broad spectrum of input states, 100 random states are created. The same \acs{DD} sequence is then applied to each one, and the final state at $t_9$ is compared to the ideal: $\|\rho_{\text{noisy}}^i - \rho_{\text{ideal}}^i\|_1$. We also perform the same procedure where (a) no \acs{DD} sequence is applied (idle), and (b) $XY4$ is applied twice. The results are shown in Figure~\ref{fig:dd-results}c. The mean values of each sequence are plotted as horizontal lines. We observe a clear separation between the different sequences. This validates the utility of using sequences designed to accommodate the particular noise model, rather than off-the-shelf \acs{DD}. Note that the set of states are still not perfect. We would expect that in the limit that one goes out to arbitrary numbers of pulses, this suppression goes to zero. 

In Figures~\ref{fig:dd-results}d, and~\ref{fig:dd-results}e, we repeat this demonstration respectively on \emph{ibm\_hanoi} and \emph{ibmq\_guadalupe}. Each device is subject to a nine step process with wait times of 1.6~$\mu$s. Although we see a separation between the three sequences, it is less stark than in the simulated setting. In particular, the optimal sequence we arrive at is very similar in performance to $XY4$. This may be indicative that the noise profile on IBM Quantum devices cannot be reduced much further, due to the lack of more complex noise terms. 
This warrants further investigation into finding optimal \acs{DD} sequences in different scenarios, and comparing with standard protocols to determine if a priori characterisation is required. It would also be worthwhile to extend the characterisation out to dozens of steps with minimal wait time, in order to determine what practical optimum is possible in controlled noise suppression.

We have developed and demonstrated a method for the suppression of noisy non-Markovian dynamics. This method is flexible, in that the one characterisation can be adapted to any desired time window. Moreover, as we saw in Chapters~\ref{chap:MTP} and~\ref{chap:universal-noise}, requires only relatively few shots to produce for any number of times. It is also fast, taking only a few seconds to perform the full optimisation.
Many \acs{DD} sequences are mathematically motivated without a clear notion of the true noise on the device. This provides a cheap and straightforward method through which bespoke \acs{DD} sequences can be applied to any given noisy setup, and are hence optimal in the number of pulses for that given setup.

Suppose that correlations persisted across different time-scales (i.e. short time and long time). Then a simple \acs{MPO} estimation would not suffice to determine these memory structures; the resulting optimal \acs{DD} sequence would only cancel out the short-time correlations rather than the long-time ones. There are two immediate approaches that one could consider to solve this problem. First, the short-time dynamics could be characterised and an optimal \acs{DD} sequence found. Then, this sequence could be fixed in the process, such that all fast errors were cancelled. Next, characterise a longer time-scale process tensor around the fixed sequence. After optimising for this \acs{DD}, one would then find the concatenation of an inner and an outer \acs{DD} sequence which respectively cancelled the short time and long time interactions. Alternatively, one could use more sophisticated tensor networks -- such as tree tensor networks or the multiscale entanglement renormalisation ansatz~\cite{PhysRevLett.101.110501} -- which are designed to capture different scales of correlation, estimate these and then optimise for a \acs{DD} sequence.



\section{Logically Improving Arbitrary Two-Qubit Gates}
In search for non-Markovian optimal control techniques, a key required property is generalisability. We cannot characterise each step of each quantum algorithm to run it optimally. Instead, we must focus on profiling the building blocks to quantum circuits so that the amount of required characterisation is fixed. In the previous section, we focused on the active suppression of non-Markovian noise during idle times. This is the first step to improved circuit outcomes: implementing an optimal identity operation.
But, of course, qubits are only idle a fraction of the time, and perhaps not at all if the device in question has good connectivity. We must therefore consider more broadly the problem of how one can design the best version of a quantum circuit in a noise-aware fashion. A generic choice for the layout of quantum circuits is in the implementation of arbitrary single and two-qubit gates. But most device hardware only has a limited native gate set, typically a single-qubit rotations and a CNOT. 

\begin{figure}[!htb]
	\centering
	\includegraphics[width=0.8\linewidth]{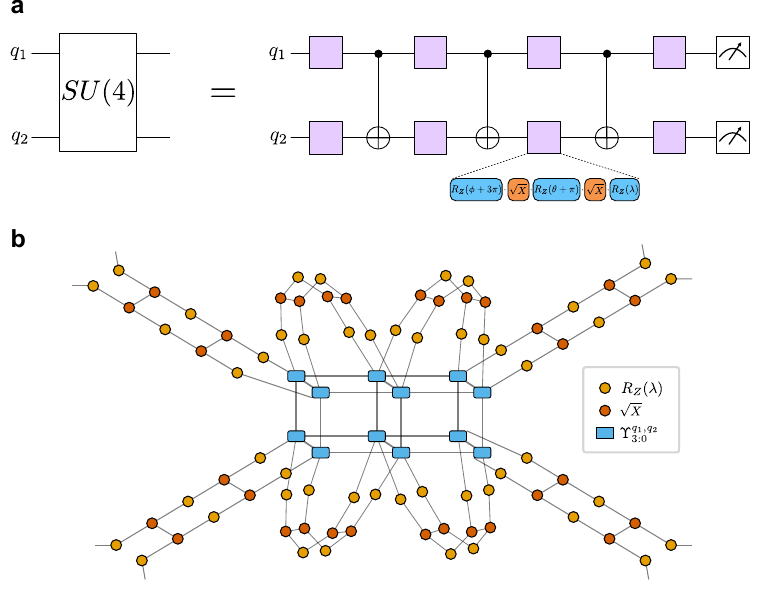}
	\caption[Scheme to characterise and improve arbitrary $SU(4)$ gate decompositions. ]{Scheme to characterise and improve arbitrary $SU(4)$ gate decompositions. We apply our non-Markovian-\acs{GST} tools from Chapter~\ref{chap:universal-noise} to optimise generic $SU(4)$ gates for the underlying device noise, accounting for crosstalk and non-Markovianity. \textbf{a} Any unitary operation in $SU(4)$ may be decomposed into three CNOT gates and eight local unitaries on each of the two qubits. For IBM Quantum devices, the local unitaries may be further decomposed into physical $\sqrt{X}$ gates and virtual $R_Z(\lambda)$ rotations. This naturally defines a three-step, two-qubit process tensor. \textbf{b} We show a tensor network representation of the two qubit gate. Once the complex noise is learned, we may then optimise the $R_Z$ parameters so that the effective channel is as close as possible to the two-qubit gate of our choosing. }
	\label{fig:su4-opt}
\end{figure}

Consider, then, the problem of constructing an arbitrary two-qubit gate in terms of CNOTs and local rotations. This decomposition is given in Figure~\ref{fig:su4-opt}a. We can recognise this gate decomposition as a three step, two qubit process tensor. Here, the CNOTs are a fixed part of the process, and the process accepts the single qubit gates as its input. We saw in Chapters~\ref{chap:efficient-characterisation} and~\ref{chap:universal-noise} that this is well within our capabilities to characterise. Let $\Upsilon_{3:0}^{q_i,q_j}$ be the two-qubit process tensor across three steps as represented by the given circuit. Each gate is decomposed into virtual $Z$ rotations and $R_X(\pi/2)$ pulses, as discussed in Chapter~\ref{chap:universal-noise}, and the resulting process tensor estimated using the tensor network methods therein. This gives us an estimate for the \emph{effective} noisy two-qubit gate for any values of local unitary operations:
\begin{equation}
	\begin{split}
		\mathcal{E}^{q_i,q_j}(\vec{\theta},\vec{\phi},\vec{\lambda}) &= (u_1 \otimes u_2) \circ \mathcal{E}_{2:1;1:0}^{q_i,q_j \ (u_3,u_4,u_5,u_6)} \circ (u_7 \otimes u_8),\quad \text{where}\\
		\hat{\mathcal{E}}_{2:1;1:0}^{q_i,q_j \ (u_3,u_4,u_5,u_6)} &= \Tr_{\mathfrak{o}_0,\mathfrak{o}_1,\mathfrak{i}_2,\mathfrak{o}_2,\mathfrak{i}_3}\left[\Upsilon_{2:0}^{q_i,q_j} \cdot \hat{u}_3\otimes \hat{u}_4\otimes \hat{u}_5\otimes\hat{u}_6\right].
	\end{split}
\end{equation}

In keeping with our optimisation theme, we may then classically optimise 
\begin{equation}
	\argmin_{\vec{\theta},\vec{\phi},\vec{\lambda}} 1 - \Tr(\hat{\mathcal{E}}^{q_i,q_j}(\vec{\theta},\vec{\phi},\vec{\lambda})\cdot \hat{U}_{\text{target}}^\dagger),
\end{equation}
to find the optimal decomposition of any two-qubit gate on our quantum device. The tensor network for this expression is shown in Figure~\ref{fig:su4-opt}b.

\begin{figure}[!t]
	\centering
	\includegraphics[width=0.95\linewidth]{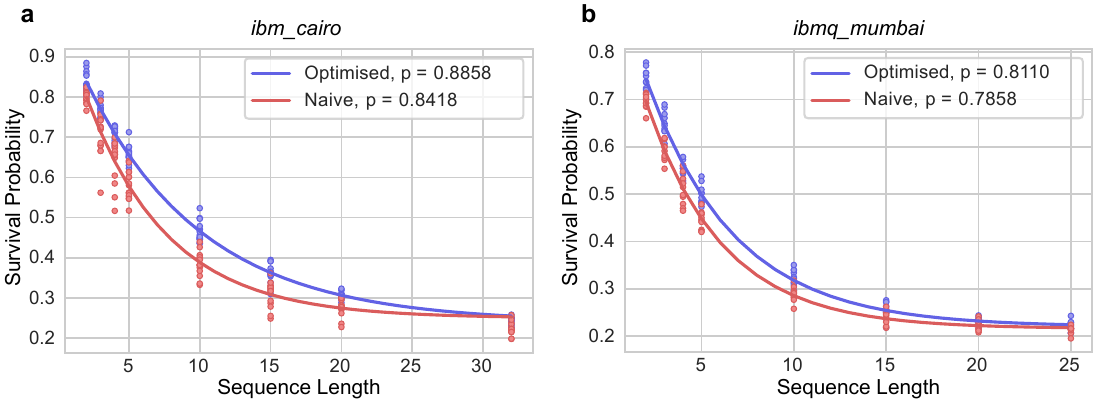}
	\caption[Predicted randomised benchmarking curves for $SU(4)$ gate optimisation on IBM Quantum hardware. ]{Predicted randomised benchmarking curves for $SU(4)$ gate optimisation on IBM Quantum hardware. Using a two-qubit, three-step characterisation with reconstruction error of $\approx 10^{-4}$, we compute the randomised benchmarking fidelity of both the native gate, and an optimised noise-aware recompilation on both \textbf{a} the device \emph{ibm\_cairo} and \textbf{b} the device \emph{ibmq\_mumbai}.}
	\label{fig:su4-rb}
\end{figure}

In Figure~\ref{fig:su4-rb} we demonstrate some preliminary results to this effect. We characterise the aforementioned process tensors on \emph{ibm\_cairo} and \emph{ibmq\_mumbai}. Each characterisation achieves a reconstruction error of $\lessapprox 10^{-4}$. We then sought to optimise the generic two-qubit gates as measured by \ac{RB}. This procedure is as follows: 
\begin{enumerate}
	\item From circuit depths 1 to $D_{\text{max}}$, generate Haar random two-qubit unitary operations $U_1, U_2, \cdots , U_{m}$, as well as inverse $U_f := (U_m\cdots U_2U_1)^{-1}$.
	\item For each $U_i$, use the conventional $SU(4)$ CNOT decomposition to express the unitary in terms of native single-qubit gates and CNOTs on the device.
	\item For each $U_i$, using the characterisation estimate $\mathcal{E}^{q_i,q_j}(\vec{\theta},\vec{\phi},\vec{\lambda})$, optimise to find the parameters such that the noisy gate best approximates the ideal unitary.
	\item Run the \ac{RB} sequence, and measure the final populations in the $|00\rangle$ state. 
	\item Repeat this procedure a number of times across each sequence length, and then fit a curve $A + B\beta^m$ to the averaged sequence fidelity. The average gate fidelity is then $p:= (3+\beta)/4$.
\end{enumerate}

Note that at this stage, the results are only a prediction taken from the characterisation. There are software engineering hurdles to overcome before we may construct a workflow that characterises the noise and optimises a gate before feeding back to the device. Currently, this process is too slow to feed back before the device is recalibrated or before its parameters drift. This is the subject of ongoing work, but we believe our present results to be a reliable indicator of how optimised gates will perform, since the reconstruction fidelities are much tighter than the difference between the native gate and the optimised gate. 
We additionally point out that the \ac{RB} results indicate an improvement in the average gate fidelity. Average gate fidelities, however, predominantly measure incoherent error. Coherent error of size $\theta$ contributes only $\mathcal{O}(\theta^2)$ to the average gate fidelity. 
Much of the improvement we expect to achieve is in the diamond norm between the ideal gate and actual, which is difficult to measure, but is the important figure of merit for achieving quantum error correction thresholds~\cite{aharonov1997fault}. 



\textbf{Approximating $SU(4)$ Gates}

Since we obtain $\Upsilon_{3:0}^{q_i,q_j}$ whilst measuring at each step, we also simultaneously estimate $\Upsilon_{2:0}^{q_i,q_j}$ and $\Upsilon_{1:0}^{q_i,q_j}$. In the \acs{DD} case, this allowed us to find optimal sequences for smaller windows of time. The advantage here is that we can explore a depth-approximation trade-off. Let $U$ be a Haar random unitary gate from $SU(4)$. Now, let $U_i$ be the best possible decomposition of $U$ into $i$ CNOTs and local unitary rotations. Moreover, let $\alpha_i = \|U_i - U\|$ be the distance between the $i$-CNOT decomposition and the target unitary. Clearly, $\alpha_3 = 0$, since any two-qubit gate can be perfectly decomposed with three CNOTs. Suppose, now, given a noisy device, we have $\tilde{U}$, the noisy implementation of $U$. Let $\epsilon_i = \|\tilde{U}_i - U_i\|$. In general, this means that $\|\tilde{U}_i - U\|$ will roughly depend on both $\alpha_i$, the approximation cost; and $\epsilon_i$, the noise cost. There is a tension in the following:
\begin{equation}
	\begin{split}
		&\alpha_3 \leq \alpha_2 \leq \alpha_1 \leq \alpha_0\\
		&\epsilon_3 > \epsilon_2 > \epsilon_1 > \epsilon_0.
	\end{split}
\end{equation}
An extra CNOT allows one to get closer to the mathematical ideal of $U$, but also introduces extra noise in the noisy decomposition. Whether $\alpha_i > \epsilon_i$ or $\epsilon_i > \alpha_i$ is highly dependent on the specific $U$. For example, a two-qubit unitary which is close to not entangling will not benefit from having a three CNOT decomposition. The extra CNOTs will only serve to introduce extra noise. Our two-qubit process tensor allows us to accurately assess this trade-off. We have already seen that we can optimally find the best $\tilde{U}_3$ for a given unitary, now we see that we can compute the smallest $\delta_i$, where $\delta_i = \|\tilde{U}_i - U\|$. This allows us to select between a 0, 1, 2, or 3 CNOT decomposition of the unitary. Hence, we are optimising both for the least noise, and the most favourable approximation error. 

This notion of approximating two-qubit gates in this fashion was explored in Ref.~\cite{Cross-QV}, but adopted the coarse approximation that each gate incurred the same error of a depolarising form. Here, where we know the noise to be more structured, we can use our process characterisation to concretely determine what the noise looks like and what the effects are of approximating. That is to say, that both $\alpha_i$ and $\epsilon_i$ will depend both on the noise of the device, and the unitary being compiled. 
Up to statistical fluctuation, we can accurately estimate both quantities to select the best trade-off.








\section{Future Directions: End-to-End Noise-Aware Compilation}

In this section, we do not present explicit results, but rather discuss a blueprint of how one might use the characterisation schemes we have developed to deal with the effects of correlated and non-Markovian noise on quantum devices. 
For concreteness, we consider a quantum volume circuit, which is the present de-facto standard for measuring capabilities of quantum computers~\cite{Cross-QV,9805433,Jurcevic2021}. The \emph{volume} moniker is intended to capture the largest square-shaped circuit that a quantum computer can reliably perform. Model circuits are structured with depth $d$ and width $m$ (where $m = d$) as a sequence $U = U_d\cdots U_2 U_1$, where each 
\begin{equation}
	U_i = U_i^{\pi_i(m'-1), \pi_i(m')}\otimes \cdots \otimes U_i^{\pi_i(1),\pi_i(2)}
\end{equation}
is a brickwork-type layering of two-qubit unitaries. Each layer is formed by selecting each $U_{i}^{a,b}$ on qubits $a$ and $b$ randomly according to the Haar measure on qubits $a$ and $b$. The qubits are chosen as a random uniform permutation of the $m$ indices.

\begin{figure}[htbp]
	\centering
	\includegraphics[width = 0.5\linewidth]{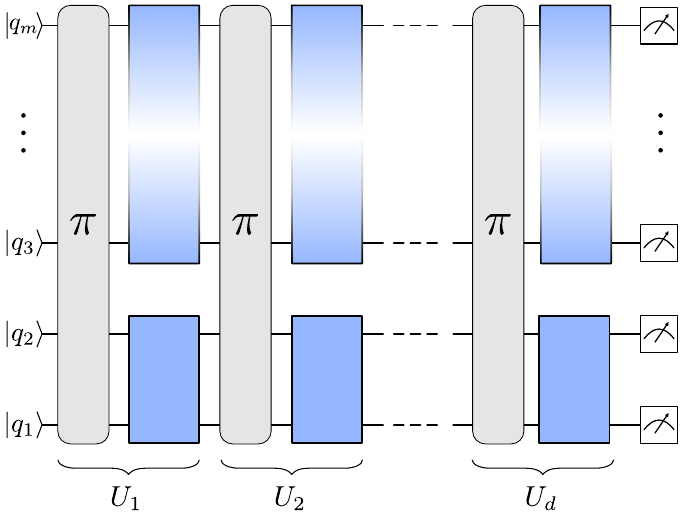}
	\caption[Diagram of a quantum volume circuit. ]{Diagram of a quantum volume circuit. Quantum volume circuits are designed to be robust tests of a device quality, interleaving layers of permutations ($\pi$) -- to probe device connectivity -- with generic brickwork patterned two-qubit $SU(4)$ gates, to achieve typicality.} 
	\label{fig:qv-circ}
\end{figure}

The specific evaluation of quantum volume is unimportant in this context, but to broadly summarise: numerous random circuits are generated, each with random intermediate permutations. The circuits are then run, and output distributions compared with the ideal case. 

\textbf{Atoms of a Circuit}

We can use the methods we have developed so far as the `atoms' that make up a quantum circuit. A quantum circuit compiled onto physical hardware will typically undergo several layers in that compilation. First, is that the chosen unitaries will be decomposed into the native single and two-qubit gates of the device. Next, to account for connectivity constraints, a series of SWAP operations will be layered in between gates to connect distant qubit with one another. 
Finally, the gates will be arranged into a timed pulse schedule, specific to the frequencies of single and two-qubit interactions. Among all this, various tricks will be applied to reduce the complexity of the circuit and make it more amenable to the physics of the device.

In the corresponding physical circuit, then, this essentially leaves single qubit unitaries on each qubit, two-qubit unitaries on connected qubits, and periods of idle time. We envisage two means with which non-Markovian characterisation can be employed to suppress errors. First, one can design optimal \acs{DD} sequences for each of the idle times, using the techniques we have introduced. The idle duration is immaterial, since we can select a $t_{\text{max}}$ and simultaneously characterise every time up to $t_{\text{max}}$. Each idle window on each qubit can be used to decouple that qubit from the surrounding environment and any crosstalk in a way that maximises the utility of resources. 
Next, rather than decomposing generic two-qubit unitaries in the naive way -- with a KAK decomposition -- one can readily use the techniques we have discussed to optimally decompose that gate to either circumvent or cancel the noise which is present. 
These approaches constitute basic and readily available control which we have demonstrated in an elementary form. Incorporating bespoke compilation into a compiler as we have discussed would be undoubtedly challenging software engineering work, but nevertheless could eliminate the harder task of fabricating these deleterious effects away.


\textbf{Accounting for Causal Cones}

Ideally, accounting for non-Markovian noise and cancelling correlated errors should not have to wait for an idle point in which to protect the qubit. It is preferable to have some active noise suppression throughout a circuit which accounts for context-dependence in the decision of actually applying the control. That is, once it has been made apparent the effects of gates at previous times and on neighbouring qubits, this characterisation can be fed forward to update future choice of operations at the circuit compilation stage. All optimisation could be done once ahead of schedule; it is highly parallelisable. At the compilation step, then, we simply have a series of decision problems to make within each causal cone. This type of adaptive control can be performed within the circuit itself. 

Using our notion of spacetime Markov order, described in Chapter~\ref{chap:efficient-characterisation}, we can consider relevant connected regions of spacetime within a quantum circuit. Finitely sized cones $\mathfrak{C}$ can be characterised as we have detailed. At first, one would optimise the atoms of the circuit as we have described. At the very beginning of a circuit, the gates will have been chosen appropriately. Then, move one step along. Conditioned on the previous gates in each of the overlapping causal cones, update the next gate such that the \emph{effective} dynamics is as close to ideal as possible. 


\textbf{Scalable Characterisation and Compilation}

\begin{figure}[htbp]
	\centering
	\includegraphics[width = \linewidth]{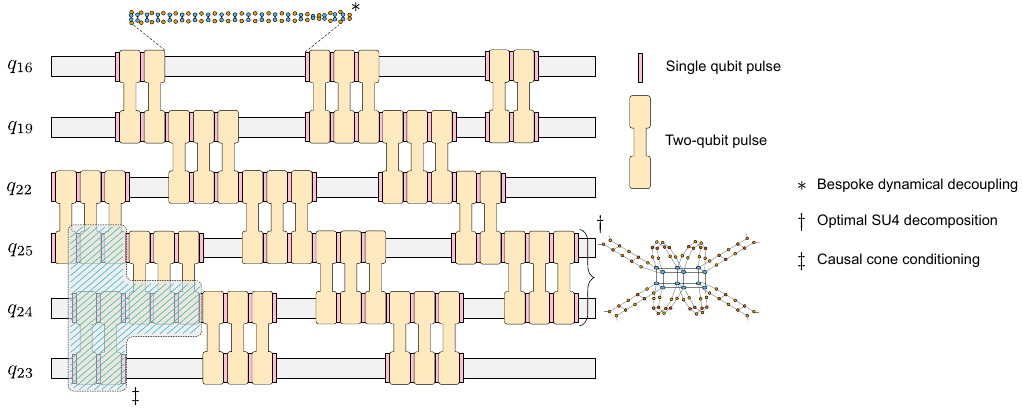}
	\caption[Depiction of targeted error suppression strategies to combat generic spatiotemporally correlated noise. ]{Depiction of targeted error suppression strategies to combat generic spatiotemporally correlated noise. The circuit drawing indicates the physical pulse compilation of part of a QV64 circuit on \emph{ibmq\_montreal}, accounting for connectivity and differences in gate duration. This is reproduced from Ref.~\cite{jurcevic2021demonstration}. We illustrate three potential strategies to target correlated noise using our characterisation method: (i) bespoke \acs{DD} sequences, specific to a qubit noise profile; (ii) optimal decomposition of two-qubit unitaries, choosing parameters to circumvent crosstalk and non-Markovianity; and (iii) updating future operations based on events within a past causal cone.}
	\label{fig:noise-aware-comp}
\end{figure}

Supposing that non-Markovian noise characterisation has taken place, we have outlined multiple ways in which a quantum circuit could be optimally compiled to either circumvent or cancel that noise. This covers active suppression via \acs{DD}, static suppression via decomposition of larger unitaries, and conditional suppression via selective updating of future gates. Figure~\ref{fig:noise-aware-comp} summarises these ideas in a practical setting. We have reproduced a pulse schedule from a quantum volume circuit compiled onto an IBM Quantum device in Ref.~\cite{jurcevic2021demonstration}. We see the need for all three features, there are idle periods in which decoupling can take place; gate decompositions; and, in a non-Markovian environment, there are causally connected regions by virtue of geometrically local qubits coupled to the same environment.

To characterise all the features as we have described sounds at first like a daunting task. We have gone to extensive efforts in this thesis to lighten the load, to the point where it takes $\mathcal{O}(10^6)$ shots to characterise a physically reasonable process. We have also seen that simultaneous characterisation experiences only a logarithmic overhead, rather than linear as one might expect. This key fact instils that characterisation need not be overly burdensome to understanding and removing complex noise. 

A single circuit can incorporate randomised single and two-qubit unitaries, designed for the connectivity of the device, and the maximum intended idle window length $t_{\text{max}}$. Data collection then begins with small numbers of shots and a random termination of the circuit after any of the different layers. 
Each process tensor: single qubit, two-qubit, and conic, will be captured under the characterisation circuit. One needs only to account for the features (i.e., unitary operations) of interest, and marginalise over the remainder of the information. 
Thus, each of these features will be characterised with only logarithmic scaling in both space and time, rendering the procedure efficient -- beyond the initial overheads.
Indeed, simultaneous characterisation is highly desirable, because marginalising the `background' data allows the characterisation to be with respect to the average-case dynamics, rather than any specific neighbouring state in particular.


\section{Discussion}

We have constructed a blueprint for how an efficient characterisation and noise compilation procedure could be constructed. We have argued that this can methodically and efficiently tackle the effects of non-Markovian noise. 
Once all correlated error is cancelled out to the best of our ability given the constraints, two properties remain: (i) any uncancelled correlated error, and (ii) any Markovian noise. Each of these can be eliminated to arbitrary accuracy using probabilistic quantum error mitigation and quantum error correction techniques. Respectively, this demands an increase in the number of circuit repetitions, and number of physical resources. Either an exponential time or polynomial space trade-off. The former requires bespoke characterisation, and the latter -- despite having some universal properties -- can benefit greatly from knowledge of the underlying noisy mechanisms. Our process characterisation provides both, and, with included information about non-Markovian errors, should always outperform any Markovian approach. 

Note that it is always possible to render errors incoherent by twirling through randomised compiling, but this does not eliminate the errors~\cite{PhysRevA.94.052325}. Our proposed procedure develops the foundations of how one might apply all the tools developed throughout this thesis to tackle the effects of correlated noise. We hope and anticipate that methods such as these, when carefully applied, could form the foundations of control software that enables fault-tolerant quantum computing.

\part{Conclusion}
\chapter{Final Remarks and Future Prospects}
\label{chap:conclusion}

\section{Thesis Summary}
Quantum information science has come a long way since the early days of pioneers such as Bell, Feynmann, Benioff, Bennett, Deutsch, Clauser, Aspect, Zeilinger, and many others. 
What started out as a niche interest in the nature of quantum correlations has flourished into a fully fledged quantum industry.
We have seen a remarkable improvement in experiments. In early days, experimenters would once search for any signal of non-classicality \emph{at all}. As the decades progressed, so too did the technology. From the generation of single photons and pursuance of ensemble NMR to highly modular and controllable quantum systems. This rapid progress has been matched by an infrastructure of theoretical understanding. In the 1970s, significant progress was made by \acs{GKSL} to understand Markovian open quantum systems. Then, fully incoherent noise was by far the most important experimental factor. The theory of open quantum systems marches mostly in lockstep with experimental breakthrough.

We are now entering the \emph{entanglement} or the \emph{complexity} frontier, as Preskill remarks~\cite{preskill2012quantum}. 
High fidelity quantum systems are commonplace. But the maintenance of more sophisticated system itself precipitates sensitivity to more complex background dynamics.
We must now turn to isolating, understanding, and suppressing the effects of increasingly intricate dynamical effects. 
Far from incoherent and overwhelming environments -- which is predominantly a matter for fabrication -- coherent and small effects are now laid bare to us and can be more adequately addressed with characterisation and control.

In this thesis, we have aimed to address this challenge. Broadly speaking, we have explored many angles to the problem of characterising and controlling the many-time physics of non-Markovian quantum stochastic processes. Chapters~\ref{chap:process-properties},~\ref{chap:PTT}, and~\ref{chap:MTP} examined this problem in detail and in full generality. How does one \emph{exactly} estimate all the properties of a multi-time process; how do we reconcile this problem with the available control in a quantum device; and how do we diagnose the various properties of that temporal structure?
We have developed and demonstrated robust approaches to answer each of these questions in tandem.

Naturally, full tomography of any quantum object is usually undesirable. Not only is this a cumbersome task, it is often redundant in the information it captures. To resolve this, in Chapter~\ref{chap:efficient-characterisation} we explored the structures of multi-time processes. Specifically, we asked in which respects can memory structures of a quantum stochastic process be sparse? And how exactly can we leverage this knowledge to reduce the both the quantum and classical overhead to learning the process? These developments led concretely to a suite of practical methods without sacrificing accuracy. Moreover, they opened up the possibility of tomographic methods that could be cognisant of correlated error in the control instruments themselves, in Chapter~\ref{chap:universal-noise}. The end result is both a formalism and an estimation procedure that completes this corner of \acs{QCVV} a self-consistent, efficent, non-Markovian characterisation method. 

Having constructed and demonstrated this formalism, we are left with a raft of open questions as to how best to apply it. 
Presuming the accuracy of characterisation, the result is an instruction manual for optimally controlling a given quantum device cognisant of its noise and its limitations. 
In Chapter~\ref{chap:NM-control}, we showcased error cancellation, \acs{DD} sequence optimization, and noise-aware two-qubit gate decompositions. 
These are agnostic to the underlying circuit, targetting correlated noise without the particulars of the computation. 
This establishes a broad framework for controlling and understanding quantum hardware independent of the underlying physics. It is our intention for this approach to be highly modular and highly versatile, so as to benefit any controlled quantum device. 

We should emphasise that although this thesis has predominantly operated within the context of non-Markovian noise, the results are a lens for the more general study of spatiotemporal quantum correlations in open quantum systems. We have shed light on this in Chapter~\ref{chap:MTP} with the initiation of the study of many-time physics.
Open quantum systems have a background far richer and broader than simply noise in non-fault-tolerant controllable quantum devices, and we anticipate a wide gamut of future directions which build off the present results outside of quantum noise.

\section{Looking forward}
This thesis has established a foundation for characterising quantum stochastic processes, leaving open many exciting prospects for future research.
The structure of quantum correlations in a many-body system is strictly richer than in classical counterparts. Moreover, we have seen the mathematical equivalence from which multi-time quantum stochastic processes can be precisely understood as a spatial many-body system~\cite{chiribella_memory_2008, Pollock2018a}. Consequently, temporal correlations inherit the same complex structure as spatial ones, which are well understood. It follows, then, that -- by some measures of the word -- multi-time processes can be as \emph{interesting} as quantum many-body states. 
There is intricate interplay between the broad scenarios in which this could be relevant. At a minimum, we have the underexplored question: ``to what extent can exotic quantum multi-time processes be manufactured in principle?''. Here, `exotic' could depend on the strength of temporal correlations, the peculiarities of their structure, or the manner in which they can transition between different phases. 
Next, we may ask, ``to what extent do exotic quantum multi-time processes exist in nature?''. Although constructing such states can shed light on unknown emergent physics, there is additionally good reason to believe that genuinely quantum dynamics play a key role in fundamental processes in nature, such as in photosynthesis~\cite{ritz2002quantum}. 
As yet, these properties are yet to be ascertained. However, with the advent of operational characterisation techniques such as \acs{PTT}, it will be possible to probe the existence of temporal quantum structures in naturally occurring processes. This has potential to elucidate the role of quantum mechanics in condensed matter, materials science, and biological systems.

These insights will be highly relevant to understanding and eliminating the effects of correlated noise on quantum information processors. 
Presently, it is unclear how both Markovian and non-Markovian noise will scale with respect to one another for increasing register sizes. However, as we have seen, non-Markovian noise can be addressed with bespoke characterisation and fed-forward control. In contrast, the only method to eliminate pre-fault-tolerant Markovian noise is through exponentially costly error mitigation methods. 
In either case, it is still an open question as to how different underlying physical errors will manifest as logical errors in a quantum error correcting context.
For example, unexpected physical errors may be propagated into non-Markovian logical errors. To understand this, we may need to turn to renormalisation group methods and construct a fully general structure for correlated memory effects at different scales of a quantum circuit: from pulse-level, to gate level, to logical level.
We saw in Chapter~\ref{chap:MTP} dynamic sampling techniques for learning mid-circuit measurement distributions. It is straightforward to imagine these in the contexts of, for example, syndrome extraction in quantum error correction. Temporal correlations between mid-circuit measurements in this context could be catastrophic to the efficacy of quantum error correcting codes. 
Chapter~\ref{chap:NM-control} showed in a general sense how one can update their control models based on characterisation priors. We anticipate that these methods will serve as a useful strategy to melding idealised quantum error correction models with the realities of intricate device noise.

Continuing away from noise as a sole motivating factor, we have also pointed out many open questions with respect to learning features of quantum stochastic processes. Multi-time features can be complex, and intractable for classical computers to simulate. There also appears to be some subtlety between learning properties of multi-time processes in comparison to states. Given a quantum computational simulation of a non-Markovian open quantum system, we might expect some quantum advantage from dynamic sampling problems in open quantum systems.
But proving this in physically realistic situations is challenging. Moreover, it invites the questions, what properties will we actually find useful to extract? And how do we efficiently learn them? Simulation of quantum systems is considered to be one of the principal applications of quantum computers. A specific focus on learning complex multi-time correlations is consequently a promising candidate for achieving useful advantage with quantum devices. Thus, we fully expect these results to appreciate into a relevant framework beyond noise and into the realm of fault-tolerant devices.

Taking this a step further, we might wonder whether a quantum computer needs to play the role of the simulation aspect at all. 
Building on the premise that quantum stochastic processes can exhibit rich quantum properties -- and that these properties can be learned with appropriate control -- the natural next step is to consider the extent to which these properties can be observed in the real world. 
Many interesting processes in nature are non-Markovian, and suspected to be classically intractable to simulate. Chapters~\ref{chap:PTT},~\ref{chap:MTP}, and~\ref{chap:efficient-characterisation} showed in theory, simulation, and in practice how a quantum sensor could be used to learn the surrounding non-Markovian environment. 
A quantum sensor placed in a complex environment could extract this information either directly or by tranducing it first to a quantum computer.
Quantum sensors serve as a natural tool for the purpose of studying open dynamics in detail, including poorly understood contexts in condensed matter physics and quantum biology~\cite{mcfadden2018origins,roman2021quantum}. Sensors are controllable quantum systems capable of detecting and responding sensitively to small changes in the environment. 
A long-hoped application of quantum computers is to transduce and process information from quantum sensors~\cite{McClean_2021}. Already, the premise of learning quantum properties with quantum algorithms has been preliminarily demonstrated in the context of quantum states~\cite{Huang2021}. Further, it has been shown that quantum algorithms may be used to achieve greater sensitivity for the task of quantum sensing~\cite{dong2022heisenberg}.
Extending these approaches naturally combines many of the discussed tenets of multi-time processes, and would serve both as a novel example of applied quantum computing, and a method to study new facets of physics in naturally occurring quantum processes.
Such an approach could be a path forward to studying naturally occurring temporal quantum correlations, which could play a role in, for example, ill-understood energy transport processes and quantum biological mechanisms.



This thesis has laid the groundwork for practical access to the nascent field of many-time physics. We have examined the problem through diverse perspectives, and developed a comprehensive and adaptable toolkit. We anticipate that this emerging area will continue to yield important insights into the behaviour of complex systems. 
We have maintained two distinct contentions: first, that temporally correlated dynamics is problematic, and we need to take every effort to remove its effects from controllable quantum hardware. But second, that temporal quantum correlations are \emph{interesting} in their own right, and warrant much further study. This constitutes the most general setting to learn about open quantum system dynamics, and we fully expect continued insight can be found through this lens.
Looking forward, it is our earnest hope that the impact and intrigue of many-body physics can be similarly matched by fruitful investigation into many-time physics.

\cleardoublepage 


\appendix

\part{Appendix} 

\chapter{Full List and Details of IBM Quantum Experiments}
\label{chap:exp-details}
\section{Hardware Specifications}
Each proper hardware experiment in this thesis takes place on an IBM Quantum superconducting transmon devices. These devices are fixed-frequency transmons with a heavy-honeycomb layout.
Table~\ref{tab:backend-data} lists calibration data of all currently available IBM Quantum backends. The daily calibrations are performed at the backend by IBM and at no point did we implement any custom gate calibrations. Unfortunately, decommissioned devices no longer have existing APIs to return data about the device itself. As a result, some information from devices used in this thesis is missing. For each backend, we list the processor type, number of qubits, quantum volume, median CNOT error, median $X$ gate error, median readout error, median $T_1$ decay time and median $T_2$ decay time. `Processor type' is named for the technological qualities that go into the different devices. The first word is a `family' and refers to the size and scale; the number is a `revision' and refers to specific design variants targeting various parts of the chip. Almost all of the devices used were from the Falcon family. Revision 4 adds on multiplexed readout; revision 5.10 uses on-chip filtering techniques to reduce qubit relaxation, as well as space-saving direct couplers; revision 5.11 uses filtering techniques to significantly speed up readout time; revision 8 enhances the coherence properties of the different qubits. We note the stark difference between various readout times -- almost an order of magnitude. This is highly relevant when aiming to use mid-circuit measurement functionality in dynamic sampling problems, and we are mindful of this specific issue.


\begin{landscape}
    \begin{table}[]
    \centering
    \resizebox{1.4\textwidth}{!}{%
    \begin{tabular}{@{}lllllllllll@{}}
    \toprule
    Backend                  & Processor type & No. qubits & QV  & CNOT error & $X$ error & Readout error & $T_1$ ($\mu$s) & $T_2$ ($\mu$s) & CNOT time (ns) & Readout time (ns) \\ \midrule
    \textit{ibmq\_montreal}  & Falcon 4       & 27         & 128 & 0.01015    & 0.0002909 & 0.0169        & 103.6          & 76.3           & 405.3          & 5202.0            \\
    \textit{ibmq\_toronto}   & Falcon 4       & 27         & 32  & 0.01149    & 0.0002798 & 0.0257        & 91.97          & 98.31          & 462.2          & 5963.0            \\
    \textit{ibmq\_kolkata}   & Falcon 5.11    & 27         & 128 & 0.0103     & 0.0002493 & 0.0119        & 111.2          & 49.41          & 451.6          & 640.0             \\
    \textit{ibmq\_mumbai}    & Falcon 5.10    & 27         & 128 & 0.008024   & 0.0002382 & 0.0203        & 117.9          & 146.9          & 433.8          & 3577.0            \\
    \textit{ibmq\_lima}      & Falcon 4       & 5          & 8   & 0.0147     & 0.0005889 & 0.0348        & 87.12          & 99.39          & 401.8          & 5913.0            \\
    \textit{ibmq\_belem}     & Falcon 4       & 5          & 16  & 0.008075   & 0.0002338 & 0.0225        & 113.1          & 117.4          & 465.8          & 6158.0            \\
    \textit{ibmq\_quito}     & Falcon 4       & 5          & 16  & 0.007669   & 0.0002824 & 0.0396        & 89.19          & 109.2          & 288.0          & 5351.0            \\
    \textit{ibmq\_guadalupe} & Falcon 4       & 16         & 32  & 0.01091    & 0.000281  & 0.01735       & 97.37          & 96.02          & 384.0          & 7111.0            \\
    \textit{ibmq\_jakarta}   & Falcon 5.11    & 7          & 16  & 0.007867   & 0.0002281 & 0.019         & 118.0          & 39.85          & 302.2          & 5518.0            \\
    \textit{ibmq\_manila}    & Falcon 5.11    & 5          & 32  & 0.005987   & 0.0002073 & 0.0196        & 177.6          & 54.17          & 344.9          & 5351.0            \\
    \textit{ibm\_hanoi}      & Falcon 5.11    & 27         & 64  & 0.006948   & 0.0001779 & 0.0094        & 161.1          & 122.2          & 346.7          & 817.8             \\
    \textit{ibm\_lagos}      & Falcon 5.11    & 7          & 32  & 0.007646   & 0.0001909 & 0.0149        & 103.6          & 72.5           & 327.1          & 789.3             \\
    \textit{ibm\_nairobi}    & Falcon 5.11    & 7          & 32  & 0.01454    & 0.0003433 & 0.0244        & 103.1          & 77.74          & 295.1          & 5561.0            \\
    \textit{ibm\_cairo}      & Falcon 5.11    & 27         & 64  & 0.01003    & 0.0002284 & 0.0136        & 115.0          & 109.8          & 316.4          & 732.4             \\
    \textit{ibm\_auckland}   & Falcon 5.11    & 27         & 64  & 0.00827    & 0.0002604 & 0.0089        & 165.8          & 126.0          & 432.0          & 757.3             \\
    \textit{ibm\_perth}      & Falcon 5.11    & 7          & 32  & 0.009231   & 0.0002885 & 0.0231        & 148.4          & 140.5          & 398.2          & 675.6             \\
    \textit{ibm\_washington} & Eagle 1        & 127        & 64  & 0.01346    & 0.0002745 & 0.0144        & 98.47          & 90.69          & 480.0          & 864.0             \\
    \textit{ibm\_oslo}       & Falcon 5.11    & 7          & 32  & 0.00885    & 0.0002241 & 0.0154        & 93.83          & 95.15          & 337.8          & 910.2             \\
    \textit{ibm\_geneva}     & Falcon 8       & 27         & 32  & 0.0479     & 0.0002604 & 0.0209        & 361.3          & 190.0          & 558.2          & 1600.0            \\ \bottomrule
    \end{tabular}%
    }
    \caption{Calibration data for all available backends used in this thesis. Error rates quoted refer to the median quantity across the device. Data accessed March 24, 2023.}
    \label{tab:backend-data}
    \end{table}
    \end{landscape}

\section{Experimental Detail and Data}
For convenience, we provide a brief summary of all the experimental results distributed throughout the thesis, in chronological order. Data and processing for process tensor fits across the thesis obtained from IBM Quantum devices can be found at \href{https://doi.org/10.26188/24295243.v1}{https://doi.org/10.26188/24295243.v1}. We also note that we conducted a significant number of experiments to test the robustness of these tools which were not included in the thesis (but is available on request). Often, this is testing the reproducibility of various results on different qubits and different devices. In the interest of keeping to space and to time, these have not been included here, but will be briefly mentioned.




\textbf{LI-PTT}

We performed \acs{LI}-\acs{PTT} in Chapter~\ref{chap:PTT}, Section~\ref{sec:LI-PTT} on \emph{ibmq\_johannesburg}, \emph{ibmq\_poughkeepsie}, \emph{ibmq\_boeblingen}, and \emph{ibmq\_valencia} between August and September 2019. Each experiment estimated a three-step process tensor with up to a basis size of 24. There were a combination of four preparation unitary operations, up to 24 unitaries in position 1, up to 24 unitaries in position 2, and then an \acs{IC} measurement. The wait time between each unitary was set to be 140 nanoseconds, or approximately the duration of two single-qubit gates.
Each experiment was run at 1600 shots, with the exception of \emph{ibmq\_boeblingen}, which was run at 4096 shots per circuit. 

These experiments also formed the basis of the non-Markovianity lower bounds reported in Figure~\ref{fig:mutual_information_circuits}.

\textbf{MLE-PTT}

The \acs{LI}-\acs{PTT} experiments were also treated to an \acs{MLE} post-processing to estimate these process tensors. This is reported in Figure~\ref{fig:RF_boxplot}. In addition, we also performed similarly structured experiments on \emph{ibmq\_manhattan} and \emph{ibmq\_bogota} in December of 2020. 

\textbf{Bootstrapping IC Instrument Sets}

In Section~\ref{sec:IC-control}, we detailed a procedure for achieving \acs{IC} control in noisy quantum devices, both with and without mid-circuit measurement capabilities. This requires the use of an ancilla qubit which may act as a spectator to the dynamics of the system. Specifically, one can generate a short interaction with the ancilla qubit, and then measure that qubit. After some local rotations on the system, this gives rise to an \acs{IC} basis which is capable of performing full \acs{PTT}. To estimate the basis, \acs{GST} is performed on both the system and ancilla qubits, to learn their effective \acs{POVM}s. Then restricted \acs{PTT} is performed to determine the effective mappings of any arbitrary unitary operation. 

In the main text, we demonstrated this procedure on \emph{ibm\_perth}, shown in Figure~\ref{fig:instrument-creation} in February of 2022. This same basis was also subsequently used in Chapter~\ref{chap:MTP} to generate the results in Figure~\ref{fig:correlated-noise-maps}. 

In total, we robustly tested this bootstrapping basis technique on \emph{ibm\_perth} and \emph{ibm\_lagos} (applied in Figure~\ref{fig:full-mpo-results}) in the main text, as well as in several omitted results including: \emph{ibmq\_casablanca}; \emph{ibmq\_sydney}; \emph{ibmq\_bogota}; multiple on \emph{ibm\_auckland}; multiple on \emph{ibm\_cairo}; and multiple on \emph{ibm\_hanoi}. In each instance, choosing all of the Clifford operations as local unitaries generated an \acs{IC} basis with condition number between 12 and 20 (cf. the analysis in Figure~\ref{fig:mlpt-analysis}). Condition number in this context seemed to be a proxy for how clean the measurement operations were: more noise decreases the linear independence of the basis elements.

\textbf{Complete MLE-PTT}

Although the results are omitted in the main text for brevity, we also performed full \acs{MLE}-\acs{PTT} to reconstruct three-step process tensors on \emph{ibmq\_casablanca}; \emph{ibmq\_sydney}; and \emph{ibmq\_bogota}. In each instance, we used our \acs{MUUB} as the local unitaries to generate the bootstrapped \acs{IC} basis. The results were then cross-validated against sequences of 100 random instruments (applied by randomising the local unitaries). The average reconstruction fidelities for each of these was $\gtrapprox 99.9\%$. This uniquely reconstructs the full multi-time statistics of various quantum stochastic processes.

\textbf{Bounding Process Properties}

When only restricted control is available -- or if instrument time is greater than the surrounding dynamics -- then only a restricted process tensor may be estimated using a complete unitary basis. For this, in Chapter~\ref{chap:MTP} we designed an algorithm that allowed lower and upper bounds to be placed on the possible process properties, as generated by the family of process tensors. 

On \emph{ibmq\_casablanca} we reconstructed six restricted process tensors with different dynamical backgrounds and an \acs{MUUB}. The number of shots used was 4096 per circuit for a total of 3000 circuits. The results are summarised in Table~\ref{tab:memory-bounds}

\textbf{Temporal Entanglement Witnessing}

In addressing the issue of incomplete estimates, in Chapter~\ref{chap:MTP} we also designed entanglement lower bounds. These lower-bounded the total amount of temporal entanglement, insofar as could be witnessed by unitary operations alone. We explored this on \emph{ibm\_cairo} and \emph{ibm\_auckland} in Figure~\ref{fig:ent-witnesses}. The first process tensor is a two-step process tensor, and the second a three-step, with interactions engineered as depicted in the figure. These were reconstructed using a \acs{MUUB} using 2048 shots per circuit. 

\textbf{Correlated Noise Characterisation}

In Chapter~\ref{chap:MTP}, Section~\ref{sec:correlated-noise}, we employed the technique of classical shadows in order to learn the correlated noise of \emph{ibm\_perth}. To do this, we used the bootstrapped \acs{IC} basis described above. With the entire single-qubit Clifford group as local operations, this basis had a condition number of 14.08. We captured the shadows using $7.68\times 10^7$ total shots split up into $300\times 2000$ circuits at 128 shots per circuit. This data was collected in April of 2022.

Note that this procedure was also repeated a second time on \emph{ibm\_perth} to ascertain robustness and consistency (with results omitted for brevity) as well as one time on \emph{ibm\_hanoi}.

\textbf{Dynamically Sampling Multi-time Processes}

In Chapter~\ref{chap:MTP}, Section~\ref{sec:dynamic-sampling}, we investigated the reconstruction of finitely correlated quantum processes using only the estimated marginals as produced by classical shadow tomography. In these instances, we again used the bootstrapped \acs{IC} basis, this time on \emph{ibm\_lagos} in April of 2022. The condition number of this basis was 17.45. We select qubit \#5 as the system on this device, qubit \#6 as the ancilla, and qubits \#0, \#1, and \#3 as the spin chain. Between each step, a random $XX+YY+ZZ$ rotation is applied, compiled down to $R_{XX}(\theta_{XX})\cdot R_{YY}(\theta_{YY})\cdot R_{ZZ}(\theta_{ZZ})$ using the usual two-qubit rotation decomposition with two CNOT gates and a single $Z$-rotation, as well as change-of-basis gates.

We collected the shadows for a 20-step process, and used the marginals to reconstruct the total finitely correlated state. Using random quantum instruments, we verified the validity of our multi-time dynamic sampling model by comparing the fidelity of predicted versus observed distributions. These results are all found in Figure~\ref{fig:full-mpo-results}. In total, we collected $23\ 400$ circuits at $1024$ shots per circuit. 

We omit the results for brevity, but note that we also repeated this experiment to check robustness and consistency with different backgrounds. We did this on \emph{ibm\_lagos}, \emph{ibm\_perth}, \emph{ibm\_hanoi}, \emph{ibm\_cairo}, and \emph{ibm\_auckland}. We were able to reproduce similarly consistent results to Figure~\ref{fig:full-mpo-results}.
Note that these devices were all chosen on account of being Falcon 5.11 processors and hence having significantly faster mid-circuit measurement speeds. 

\textbf{Multi-time Conditional Markov Order}

In Chapter~\ref{chap:efficient-characterisation}, Section~\ref{sec:MO}, we introduce the notion of Markov order to the multi-time setting, and provided a method to capture these finite memory models. This was demonstrated in Figure~\ref{fig:cmo_reconstruction}, where we estimate the breakdown of different Markov orders for four step processes on \emph{ibmq\_guadalupe} during April of 2021. For each of the ten data points, the four step process tensor was collected by estimating at most two three-step process tensors (we also estimated $\ell=1$ and $\ell=2$ models as well). These were using an \acs{MUUB} basis at 2048 shots per circuit. 

Note that the results of cancelled errors in Figure~\ref{fig:circuit_improvement_CMO} were also performed using a conditional Markov order ansatz. This time, the devices were \emph{ibmq\_montreal} and \emph{ibmq\_guadalupe} during May of 2021, and the processes were five-step.

Although the results are omitted from this thesis, we have also tested the reconstruction of processes via a conditional Markov order ansatz on: \emph{ibmq\_jakarta} ($\times 5$); \emph{ibmq\_mumbai} ($\times 5$); \emph{ibmq\_toronto} ($\times 6$); \emph{ibm\_hanoi} ($\times 10$); and \emph{ibm\_cairo} ($\times 9$). The wait times between the five steps ranged between 1~$\mu$s and 10~$\mu$s.

Each of these were five step processes, all at 2048 shots per circuit, and average reconstruction fidelities ranged from 0.976 to 0.991. Each of these characterisations were then fed forward to optimise a four-step \acs{DD} sequence, which reliably preserved state fidelities to an average between 0.89 and 0.98. 
This validates both the error-cancelling capabilities, and the capabilities of the reconstruction model.

\textbf{Tensor Network Estimates}

In Chapter~\ref{chap:efficient-characterisation}, Section~\ref{sec:tensor-networks}, we introduced the ability to reconstruct non-Markovian quantum stochastic processes as sparse tensor network models using very few numbers of circuits. We demonstrated this on IBM Quantum devices in Figure~\ref{fig:tn-ibm-benchmarks}. On \emph{ibm\_cairo} we reconstructed a 1-qubit, 18-step model; on \emph{ibm\_perth} a 2-qubit, 9 step model; on \emph{ibm\_perth} a 3-qubit, 6-step model; and on \emph{ibm\_kolkata}, a 4-qubit, 4-step model. Each process featured a wait time of 800~ns per step. Using approximately $10^7$ shots distributed among 512 circuits, we estimate the tensor network representation of each respective multi-time, multi-qubit process. The relevant data was collected in February of 2023.

\textbf{Universal Tensor Network Estimates}

Finally, we introduced self-consistency into our method by removing a prior assumptions about known models. We only benchmarked these techniques on synthetic data in the main text, but we extensively tested on IBM Quantum devices. The results have been omitted for brevity, but typically achieve a very high reconstruction fidelity ($\gtrapprox 0.9999$)This included: 2 qubits, 3 steps on \emph{ibmq\_guadalupe} ($\times 2$); 2 qubits, 3 steps on \emph{ibm\_hanoi} ($\times 4$); 1 qubits, 3 steps on \emph{ibm\_kolkata}; 1 qubit, 17 steps on \emph{ibm\_geneva}; 1 qubit, 17 steps on \emph{ibm\_guadalupe}; 3 qubits, 3 steps on \emph{ibmq\_jakarta}; 3 qubits, 9 steps on \emph{ibm\_geneva} ($\times 2$); 4 qubits, 4 steps on \emph{ibmq\_belem} ($\times 2$); 4 qubits, 4 steps on \emph{ibmq\_guadalupe}; 1 qubit, 5 steps on \emph{ibm\_brooklyn}; 1 qubit, 5 steps on \emph{ibmq\_kolkata}. In addition, we performed many 1-qubit-9-step \acs{DD} experiments, as well as 2-qubit-3-step $SU(4)$ experiments.

Most of these experiments ran 3000 circuits at 1024 shots. Partly, the aim here was to test robustness of the model across many different devices. Partly, we explored different metaparameter values for the fit, and the data was collected as we developed the technique to maturity.

\chapter{Benchmarking and Removing Markovian noise on superconducting devices}
\label{chap:Markovian}

The contents of this Appendix are based on the results of our work in Ref.~\cite{white-POST}. This work is thematically relevant to the remainder of the thesis, but not explicitly part of the same narrative. We use characterisation techniques to benchmark an IBM Quantum device and improve its gates. This provides further insight into the realistic performance of IBM Quantum devices, and the types of Markovian noise fingerprints present across different pairings of qubits. 

Specifically, we use \acs{GST} as a means by which we can improvement the calibration of two-qubit gates. We track the performance of a single CNOT gate across many weeks, and show how a single characterisation, coupled with subsequent tune-ups, can maintain an improved-fidelity quantum gate -- even in the presence of device drift. We furthermore report several two-qubit error maps of different combinations of CNOT gates.

\section{Introduction}
The nascent field of quantum computing has seen an emergence of many experimentally realised small-scale devices in recent years, most notably in superconducting qubit systems and trapped atomic spins \cite{RBK2017,Arute2019,IBM-53,intel-49,rigetti-19}. Different architectures have achieved high fidelity one and two-qubit gates, as well as the construction of multi-qubit entangled states \cite{Barends2014,Zeuner2018,PhysRevLett.117.140501,He2019,Hong2019, Kjaergaard2019,Bradley2019,Arute2019}. Despite this progress, current hardware cannot yet demonstrate large-scale topological quantum error correction below threshold, and there are many significant obstacles to overcome before qubit numbers can be scaled up to useful levels. Quantum computers presently face the challenge of imperfections in state preparation, measurement errors, and erroneous logical gates. Before improvement can be achieved, comprehensive characterisation techniques are essential in mapping where deficiencies lie. \par 
Noise on real quantum devices is challenging to understand quantitatively. In particular, it is difficult to isolate device behaviour given the tendency of noise and system parameters to \emph{drift} \cite{Klimov2018,Fogarty2015,Chow2009}. This is one of the many barriers facing the improvement of quantum hardware. Common characterisation techniques such as quantum process tomography (\ac{QPT}) \cite{PhysRevLett.78.390} and randomised benchmarking (\ac{RB}) \cite{PhysRevA.77.012307} offer an insight into the quality of a qubit, but suffer from respective self-consistency and limited information issues. Gate set tomography (\ac{GST}), introduced in \cite{gst-2013,PhysRevA.87.062119}, provides a relatively novel method in which the preparation, gate, and measurement operations can be implemented in conjunction with each other and separately characterised. The results can be highly accurate, but with the trade-off that a large number of experiments are required to provide the data. The analysis itself is also computationally demanding. As a consequence, there are relatively few examples of two-qubit \acs{GST} carried out experimentally in the literature \cite{osti_1428158,GST-improvement-china,Song2019}. Importantly, \acs{GST} puts a lens on otherwise inaccessible quantities such as the average gate infidelity, dominant noise channels, and its diamond norm. Here, we demonstrate how the comprehensive information provided by \acs{GST} can be leveraged to reduce the coherent noise in quantum operations. We consider both the diamond norm of the gate, which may be extracted from the initial estimate, and the randomised benchmarking infidelity, which may be efficiently estimated in the tune-up procedure. \par

Two-qubit gates are the most significant source of error in many quantum circuits, and so minimising their infidelity is critical to the performance of quantum algorithms. In this manuscript, we develop a method for performance optimisation seeded by tomography (POST) to consistently improve two-qubit CNOT gates based on a hybrid quantum-classical approach. 
We characterise the bare two-qubit logic gate using \acs{GST}, find the optimal corrective parameters for bookend single-qubit unitaries, and then use these as a seed to the Nelder-Mead algorithm in order to find the best improvement for a given calibration cycle.
We consider two regimes of control: corrective gates acting solely on the control qubit, and corrective gates acting both on the control and target qubits.
Following the one-time overhead of \acs{GST}, each daily optimisation is performed in a small ($<$150) number of single-length \acs{RB} experiments to overcome any drift which has occurred. 
This is similar to the use of \acs{RB} as an objective function in \cite{google-RB}, but with the incorporation of the more detailed information provided by \acs{GST}. We discuss later the practical advantages to seeding this procedure with the initial \acs{GST} estimates.
We test the method on the \emph{ibmq\_poughkeepsie} quantum device, a 20 qubit transmon device with a quantum volume of 16, CNOT error rates typically ranging between $1$ and $5$\%, and calibrations usually performed once per day. In particular, we measure the CNOT gate under consideration to have an \acs{RB} infidelity $r_{\text{CNOT}}$ of 0.0337, diamond norm $\mathcal{E}_{\Diamond\text{CNOT}}$ of 0.0683, and unitarity (as defined in \cite{noise-coherence-2015}) of 0.955. The computational costliness of computing confidence factories on two-qubit \acs{GST} means that we do not have error bars on these values. We discuss in the main body the reliability of the \acs{GST} estimates, and how the procedure is relatively insensitive to these. The coupling of the \acs{GST} seed with classical optimisation is successful at improving the gate. When tested on an experimental device, we find the POST approach is effective even weeks after the initial characterisation. 
At approximately 0.5 ms per shot, this procedure requires about 5 seconds per shot of device time for the initial \acs{GST} experiment, and less than $0.2$ seconds per shot for the subsequent tune-ups. As such, this procedure could be incorporated into the daily calibration routines of these devices.
The hybrid technique brings the gate as close as possible to its target, up to the hardware limit, but the actual effectiveness depends on the level of control afforded. Basic calibration routines tend to be amplitude optimisations in Rabi experiments -- however the calibration routine in this particular instance is opaque to us. In expanding the calibration model to a full process matrix, it appears as though a logical tune-up to bring a quantum gate as close as possible to its target can be an effective and efficient calibration technique. Further, this procedure can detect and correct for unitary errors which do not add coherently, which can be a limitation of some detection schemes \cite{unitary-RB}.
Although \acs{GST} has previously been proposed as part of a quantum error mitigation protocol in \cite{GST-improvement-china} and \cite{Endo2018}, we emphasise the need to avoid repeated application of \acs{GST} in any gate improvement techniques, owing to its extremely high experimental and computational overheads. Further, we are examining error suppression in this context rather than probabilistic error mitigation.\par 
We performed our experiments on an IBM cloud-based quantum computer with only logical level control. As a consequence, corrections to the CNOT could only be made through single qubit gate corrections, which themselves were erroneous. With control of the CNOT pulse scheme, we anticipate that the corrections are likely to be more effective.\par
In addition to the testing of the POST gate improvement scheme on a specific two-qubit case, we also conducted two-qubit \acs{GST} experiments on six separate CNOT gates as an investigation into the performance and types of noise to occur on real superconducting devices. Understanding the real noise that occurs on devices is important for several reasons: It can help inform future characterisation, which results in these procedures being less computationally expensive; it can help understand noise channels, which is important for quantum error correction \cite{OBrien2017}; and it can help identify hardware issues up on a real machine for better future implementation \cite{RBK2017}. We present these results, as well as a theoretical evaluation of the effectiveness of our technique on the additional qubit pairs.

\section{Drift-robust tune-up of coherent error}
\begin{figure*}[!t]
	\centering
	\includegraphics[width=0.95\linewidth]{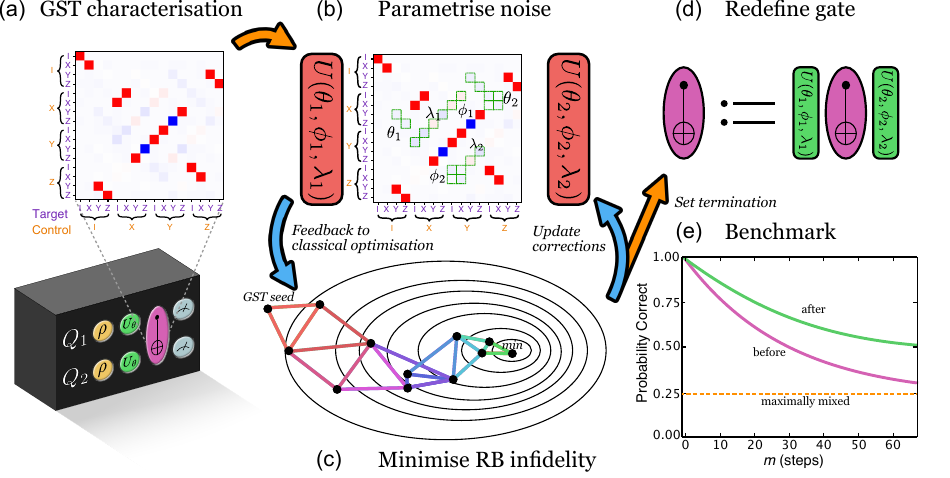}
	\caption[Overview of the POST approach for CNOT gate fidelity improvement ]{Overview of the POST approach for CNOT gate fidelity improvement. (a) Initial characterisation of the CNOT Pauli transfer matrix (PTM) using gate set tomography. (b) Gate noise is parametrised in the action of arbitrary single-qubit unitaries acting before and after the CNOT gate. The parameter values that minimise the Frobenius distance between the noisy and ideal CNOT gates are then found. (c) These values then seed a Nelder-Mead optimisation, where the objective function is the experimental infidelity of a chosen length randomised benchmarking experiment. With each iteration of POST, a new simplex is chosen and the randomised benchmarking experiment is performed for each vertex with the single qubit unitaries taking on the parameters at that point. (d) When the minimum infidelity is found, the CNOT gate is redefined to include the corrective unitaries. (e) The new gate is then fully benchmarked against the native gate, showing a revived fidelity for a much larger number of applications. }
	\label{flow-chart2}
\end{figure*}

The \acs{CPTP} map of quantum gates given by \acs{GST} highlight all Markovian errors in the operation. Examples of errors of this nature include control errors, such as axis tilt or errors in pulse shaping, or erroneous coherent rotation of the qubits due to external couplings. Our general goal in this work is to examine how best to feed this characterisation back into a quantum device with less noise. We consider only the estimate determined for the CNOT gate. The best method in which to address these noisy parameters depends on the level of control afforded in the device. Given the recent advent of cloud-based \acs{NISQ} computers, where users only have restricted control of the device, the POST protocol introduced here makes use only of additional logical operations on the qubits (although we note in the conclusion that the extension to high-level pulse control is possible). At the time of the experiments, this was the only control available to the authors, and so the only scheme examined. 
The natural CNOT gate on these devices is a cross-resonance $ZX$ interaction bookended by local rotations.
With pulse level control, provided by IBM through OpenPulse \cite{open-pulse}, we speculate that the proposed blueprint could be modified to absorb the corrections into the local pulses -- rather than applying logical corrections around it. We summarise the overall POST procedure in Figure \ref{flow-chart2}, and describe it in further detail here.\par 
GST provides a means by which errors can be identified, but it is not necessarily straightforward to then mitigate their effects. Errors occurring on two-qubit gates such as a CNOT tend to be an order of magnitude greater than those of single qubits. From the perspective of logical corrections, it is therefore optimal to address two-qubit noise with the application of single-qubit gates. \par

Consider a quantum device with an informationally complete set of two-qubit controls. This can be used to conduct a \acs{GST} experiment in the standard way on a qubit pair. The \acs{GST} analysis of a CNOT produces an estimate for the \acs{CPTP} map of the gate, designated by $\bar{G}_{CX}$. This can be decomposed into a the product of an ideal CNOT, $G_{CX}$, and some residual noise channel $G_\Lambda$ -- the inverse of which is generally unphysical \cite{inverseCP}. Previous approaches have typically treated this noise with quasiprobability decompositions \cite{Temme2017}. 
In the case of solely logical control, the nearest physical corrective $G_\Lambda^{-1}$ map may not be within the user's control-set. Importantly, given gate calibration and general hardware drift, a \acs{GST} estimate's accuracy quickly expires over time. Solely utilising \acs{GST} will require a significant overhead every time the corrections are implemented. From this, it is clear that \acs{GST} on its own faces limitations as a practical method of improving gates.\par
\begin{figure*}
	\centering
	\includegraphics[width=0.87\linewidth]{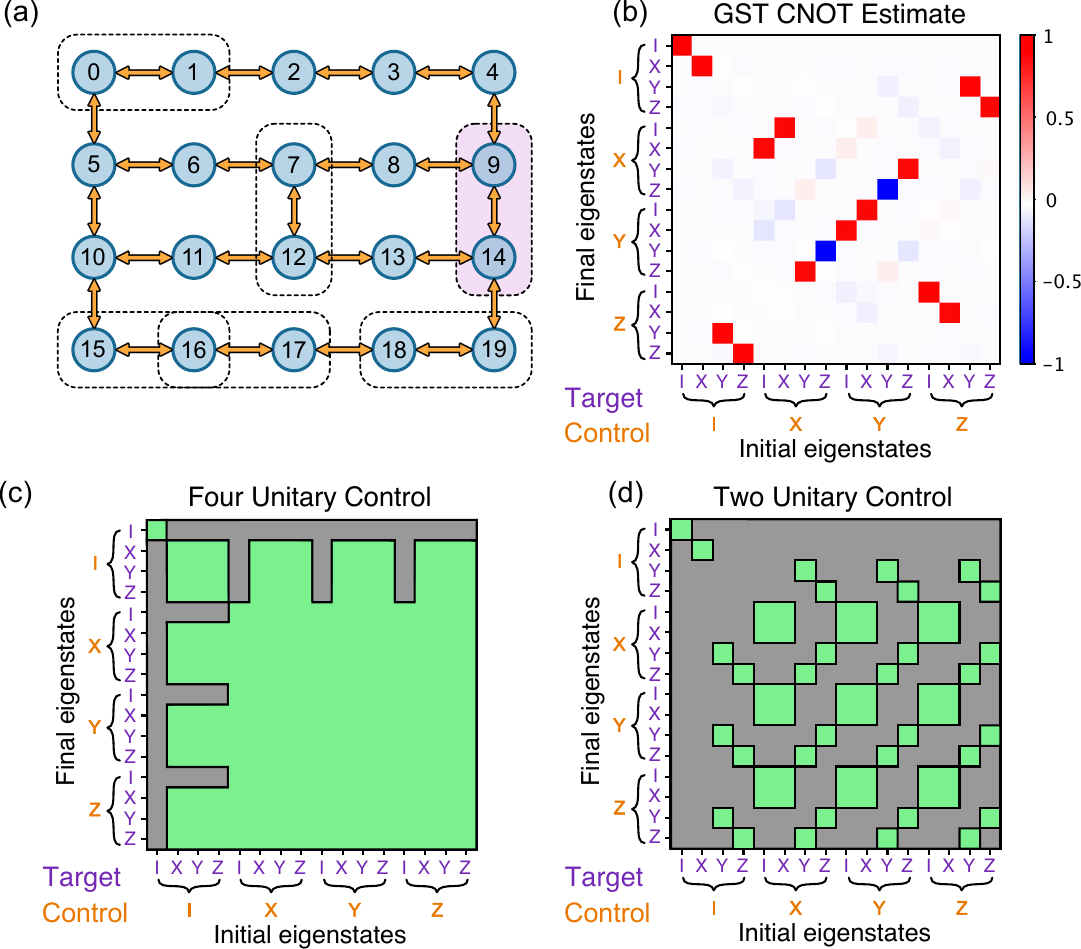}
	\caption[IBM Quantum device layout. Also depictions of addressable two-qubit components in different regimes of control.  ]{(a) The \emph{ibmq\_poughkeepsie} device layout, showing both the geometry and connectivity. Dotted lines indicate the six pairs of qubits that were characterised under \acs{GST}. Qubits 9 and 14 (highlighted in pink) were chosen as the subject for testing POST. (b) The \acs{GST} reconstructed estimate of the \acs{PTM} of the 14-9 CNOT gate under investigation. (c) The green parts of this matrix indicate the controllable parts of the overall map when combining single-qubit operations on the control and target qubits both before and after the native CNOT. Because not all elements of the \acs{PTM} can be \emph{independently} controlled, this does not imply that all error falling in the green can be corrected. Under this regime, almost all coherent noise is addressable in principle. (d) The green parts of this matrix indicate the controllable parts of the map when restricted to single-qubit operations on the control qubit before and after the native CNOT. This regime is far more restrictive than the previous for the trade-off of introducing approximately half of the amount of single-qubit error. These may also be viewed as indicators of the propagation of unitary error. }
	\label{noise-fig}
\end{figure*}

 
Without direct control of the hardware, correcting all coherent two-qubit errors will not be possible, since these corrections will contain associated errors equal or greater in magnitude than the existing ones. Further, cross-resonance errors are not controllable at this level. In this control regime we propose placing single qubit corrective gates before and after the native CNOT in order to correct as much of the local noise as possible, and then optimising over their parameters. Using a unitary parametrisation for the four correction gates $U_i$ ($i\in\{1,2,3,4\}$)
\begin{equation}
	\label{unitary-param}
	U_i(\theta_i,\phi_i,\lambda_i) = \begin{pmatrix}
		\cos(\theta_i/2) & -e^{i\lambda_i}\sin(\theta_i/2) \\
		e^{i\phi_i}\sin(\theta_i/2) & e^{i\lambda_i+i\phi_i}\cos(\theta_i/2) 
	\end{pmatrix},
\end{equation}
We propose a super-logical CNOT gate structured as 
\begin{equation}
	\label{twelve_param}
	(U_1\otimes U_2)\cdot \bar{G}_{CX} \cdot (U_3\otimes U_4),
\end{equation}
provided that the cumulative error of four single qubit gates is not greater than one two qubit gates. In the case of high single qubit error rates (or cross-talk between simultaneous gates), a similar approach can be made by applying local corrections exclusively on the control qubit (this allows for broader manipulation than corrections solely on the target qubit), at the expense of more limited noise-targeting. That is,
\begin{equation}
	\label{six_param}
	(U_1\otimes \mathbb{I})\cdot \bar{G}_{CX} \cdot (U_2\otimes \mathbb{I}).
\end{equation}
The distinction between these two levels of correction is not immediately obvious. In order to clearly see the difference between corrections on both qubits versus corrections acting solely on the control qubit, we illustrate the addressable parts of the CNOT matrix in green in Figures \ref{noise-fig}c and \ref{noise-fig}d. That is, this indicates the extra accessible dimensions of control offered in either case. Whether that control is necessary in practice will depend on where the noise manifests and the trade-off in introducing further single-qubit error. Note that that not all of elements in the green region can be necessarily independently configured. We call a matrix element `controllable' under a particular set of control operations if it is possible to change that matrix element by varying the parameters of the control. Matrix elements that are uncontrollable under a set of operations cannot be changed at all, and are marked in grey. Not all errors that fall in the green will necessarily be completely correctable. \par
The $\phi_i,\:\theta_i,\:\lambda_i$ are first selected with a simple optimisation to minimise the distance between the corrected gate and the ideal map. At this stage, any norm could be selected. For computational convenience, we chose the Frobenius norm, which also minimised the average gate infidelity,
\begin{equation}
	\frob{U_1\otimes U_2 \cdot \bar{G}_{CX} \cdot U_3\otimes U_4 - G_{CX}}.
	\label{frob-dist}
\end{equation}
If the \acs{GST} estimates of a quantum process were perfect and static with time, then this would be sufficient to have an improved CNOT gate. However, because \acs{GST} is only an estimate of a Markovian map within a (generally) non-Markovian system, a simple mathematical minimisation will not necessarily result in a physical optimisation. This could be either if the noisy parameter values were to change over time, or if the \acs{GST} estimate were slightly incorrect.
Instead, what we propose is a tune-up procedure which optimises the \emph{performance} of the gate by using \acs{GST} to identify the neighbourhood in which certain parameter values are able to be modified to improve the accuracy of the gate. As such, it is robust to the drift of different noisy parameters and does not rely on the absolute accuracy of the \acs{GST} estimate. The only assumption is that the noisy channel will remain structurally similar enough to those of the \acs{GST} estimate, that an optimisation seeded by \acs{GST} will bring us back to a better gate than the native operation in few iterations. We define our objective function as the \acs{RB} infidelity of the gate, in order to make use of the most general metric of performance. \par 

The algorithm to implement the POST procedure is as follows:
\begin{enumerate}
	\item Conduct a series of experiments given by the requirements of \acs{GST}. Use these to produce an output estimate of the gate's PTM. 
	\item Using a classical minimisation technique, find the six or twelve parameter values which numerically minimise the Frobenius distance between the super-logical and the ideal CNOT gates, given in Expression (\ref{frob-dist}). These will be the seed parameter values. The choice of the Frobenius distance is not necessarily special, but we elected to use it to make the resulting matrices as similar as possible. 
	\item Define the objective function to be the average survival probability over an ensemble of circuits for some length $m$ \acs{RB} experiment. Taking Step 2. as a newly defined CNOT gate, compute the objective function for both the native and new CNOT gates as a point of comparison. Using the parameters obtained from \acs{GST} as a seed, perform an optimisation of the CNOT gate by feeding the parameter values into the Nelder-Mead algorithm, where for each vertex of the simplex, the $m-$length \acs{RB} infidelity is computed as the objective function.
	\item Converge at some pre-defined level of change. The newly improved CNOT gate is then defined by the composition of the final single qubit unitary gates on either side of the native CNOT gate.
	\item Conduct a full \acs{RB} experiment to compare the new gate fidelity to the original. 
\end{enumerate}
The overview of the procedure is to use the \acs{GST} estimate as a seed for the Nelder-Mead optimisation algorithm, which then works to minimise the infidelity of the CNOT gate by varying the parameters given in (\ref{twelve_param}) or (\ref{six_param}). We take the infidelity measure to be a set of fixed $m$-length randomised benchmarking experiments. In a short number of iterations, this locates the optimal corrective rotations to make for a given day. This process is summarised in Figure \ref{flow-chart2}. The flexibility of the procedure is not only its robustness to drift, but generic steps (optimisation procedure, noise parametrisation, objective function) can all be chosen at the user's discretion. 
Part of the rationale in performing \acs{GST} is the ability to specifically determine the level of coherent noise in the gate from the outset, motivating whether it is possible to improve the gate at all.
Further, it provides useful information through the determination of the diamond distance between the noisy gate and the ideal.
The diamond distance is a means of assessing the distinguishability of two quantum channels. It is a worst-case error rate, taking the largest output trace distance over all possible input matrices. That is, for two quantum channels $\Phi_1$ and $\Phi_2$:
\begin{equation}
	||\Phi_1-\Phi_2||_\Diamond := \sup_{\rho} \frac{1}{2}||\Phi_1\otimes \mathbb{I}_{d^2}(\rho) - \Phi_2\otimes \mathbb{I}_{d^2}(\rho)||_1,
\end{equation}
where $||\cdot||_1$ is the trace distance, a common measure of distinguishability between two density matrices. This metric between channels is commonly used in fault tolerance calculations for quantum error-correcting codes. It is also much more sensitive to coherent error in the gate than the average gate infidelity, which is typically the figure provided from \acs{RB} curves. By performing \acs{GST}, we are able to find the global minimum for the gate infidelity, which also globally minimises the diamond distance. We discuss using our data in the next section how small differences in local minima of the infidelity can correspond to large changes in the diamond norm.

\section{Experimental Implementation}
We tested the POST framework for CNOT characterisation and improvement on the 20 qubit \emph{ibmq\_poughkeepsie} superconducting quantum device. Two qubit \acs{GST} was performed on six pairs of qubits with the gateset $\{\mathbb{I},G_{XI}\left(\frac{\pi}{2}\right), G_{IX}\left(\frac{\pi}{2}\right),G_{YI}\left(\frac{\pi}{2}\right),G_{IY}\left(\frac{\pi}{2}\right),G_{CX}\}$ up to a germ repetition of $L=4$ for a total of $10\:500$ circuits at 8190 shots each. The layout and connectivity of this device is shown in Figure \ref{noise-fig}a. We also indicate the qubits on which experiments were performed. Using the notation `control-target' to indicate the physical qubit pair used respectively as the control and target of a CNOT gate, we characterised the gates of qubits 0-1, 12-7, 14-9, 15-16, 16-17, and 18-19. In the estimation procedure, each gate was constrained to be CPTP. We then elected to test the POST procedure on the gate which had most recently been characterised, for which the control was qubit $\#14$ and target qubit $\#9$. For the germ generation and MLE steps, we used the comprehensive open source Python package \emph{pyGSTi}, introduced in \cite{pygsti}. With the tools available, we generated the required germs from our target gatesets, and conducted the analysis of our experimental data. The \acs{GST} estimate from the 14-9 qubits, used hereon in the POST tests, is shown in Figure \ref{noise-fig}b. The resulting noise maps for each additional CNOT gate can be found later in this manuscript, in Figure \ref{noise-ptms}. In the gateset estimation procedure, \emph{pyGSTi} flags violation of the Markovian process matrix model, typically indicating some form of non-Markovianity in the device. Model violation is usually summarised in the form of difference between the observed and expected maximum log-likelihood $\log(\mathcal{L})$ for a given $k$ degrees of freedom. A model-abiding $2\Delta\log(\mathcal{L})$ is expected to be distributed with mean $k$ and standard deviation $\sqrt{2k}$. For length-1 sequences where $k=737$, our data showed $2\Delta\log(\mathcal{L}) = 2861$, for length-2 where $k=6006$, $2\Delta\log(\mathcal{L}) = 15\:938$, and for length-4 where $k=26\:901$, $2\Delta\log(\mathcal{L}) = 107\:889$. This suggests that the device was not totally Markovian, and the estimation not totally perfect.

\begin{figure*}[!t]
	\centering
	\includegraphics[width=0.89\linewidth]{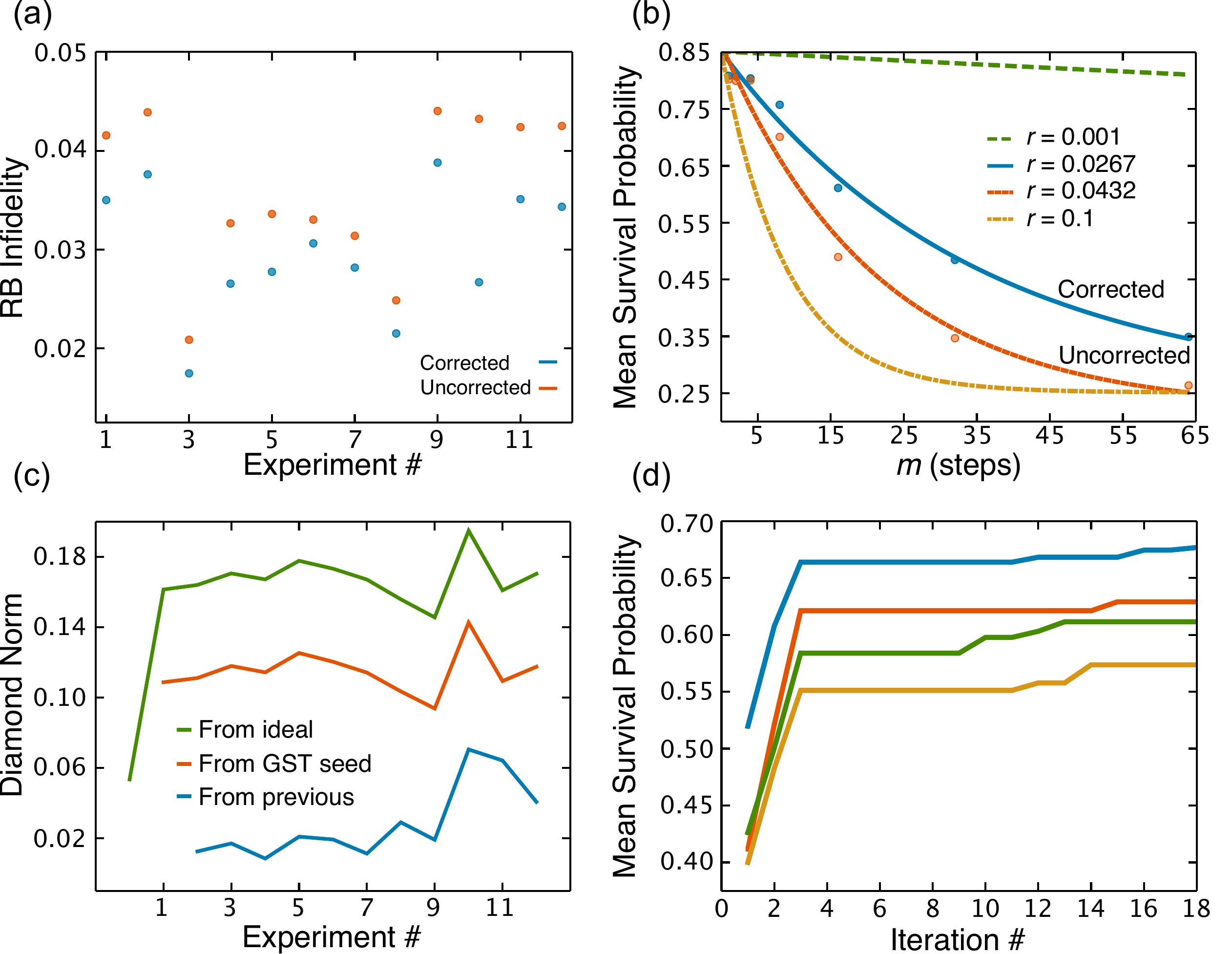}
	\caption[Experimental results tuning up an IBM Quantum CNOT gate over the course of several weeks. ]{Efficacy and robustness of the POST protocol -- experimental test on the 14-9 CNOT gate. (a) \acs{RB} values for each experiment taken over the six week period, both for the corrected and the uncorrected gates. (b) Example \acs{RB} curve showing our best improved CNOT gate decay vs. the native gate \acs{RB} experiment for experiment 10 (our most emphatic example of gate improvement). Also shown for comparison are theoretical curves corresponding to 10\% and 0.1\% error rates. (c) A comparison of the diamond norm of the corrective channel over time: firstly with respect to the ideal CNOT, secondly with respect to the initial \acs{GST} corrections, and finally with respect to each previous experiment. (d) An indication of the convergence speed of the Nelder-Mead gate optimiser for four example runs. This figure shows how the survival probability of the \acs{RB} experiment increases with each iteration. We observe an initially large increase in the fidelity of the experiment, followed by only small changes thereafter, and zero changes after 18 iterations. The sequence very quickly finds its best gate in a small number of steps.}
	\label{rb_curve}
\end{figure*}

\begin{table*}[]
	\centering
	\begin{tabular}{@{}lp{1.1cm}p{1.1cm}p{1.1cm}p{1.1cm}p{1.1cm}p{1.1cm}@{}}
		\toprule
		Date & $\theta_1$ & $\phi_1$ & $\lambda_1$ & $\theta_2$ & $\phi_2$ & $\lambda_2$ \\ \midrule
		
		31/03/19 (initial GST) & 0.046 & -1.271 & 0.480 & 0.029 & 0.480 & 0.393 \\
		02/04/19 & 0.116 & -1.234 & 0.536 & 0.116 & 0.545 & 0.403 \\
		09/04/19 & 0.130 & -1.218 & 0.461 & 0.084 & 0.565 & 0.475 \\
		24/04/19 & 0.116 & -1.234 & 0.552 & 0.119 & 0.558 & 0.415 \\
		27/04/19 & 0.148 & -1.208 & 0.502 & 0.010 & 0.530 & 0.427 \\
		29/04/19 & 0.089 & -1.228 & 0.523 & 0.071 & 0.523 & 0.436 \\
		01/05/19 & 0.218 & -1.199 & 0.552 & 0.010 & 0.552 & 0.365 \\
		03/05/19 & 0.089 & -1.228 & 0.523 & 0.071 & 0.523 & 0.436 \\
		10/05/19 & 0.096 & -1.221 & 0.530 & 0.079 & 0.530 & 0.443 \\
		13/05/19 & 0.141 & -1.310 & 0.497 & 0.123 & 0.575 & 0.488 \\* \bottomrule
		
	\end{tabular}
	\caption{Values of the corrective parameters obtained after a full improvement cycle for each day of experiments, given to the third decimal place. These correspond to the unitary parametrisation given in Equation (\ref{unitary-param}).}
	\label{param-table}
\end{table*}

\textbf{Results Summary}

The initial \acs{GST} analysis of the 14-9 CNOT gate took place on the 31st of March, 2019 and its corrective parameters used as the base vertex for the Nelder-Mead simplex method. The procedure was implemented a total of 12 times over a period of approximately six weeks, corresponding to overlap with approximately 40 different calibration cycles. Figure \ref{rb_curve}a displays a summary of the improvement shown over the native gate with each experiment run for bare \acs{RB} infidelity $r_{\text{u}}$ and corrected \acs{RB} infidelity $r_{\text{c}}$. In each case, both the corrected and uncorrected benchmarking experiments were conducted in the same job submission to avoid any bias in gate drift throughout the day, and the tune-up was often conducted shortly after the calibration of the machines. The total average improvement, which we define as $r_{\text{u}}/r_{\text{c}} - 1$ was $21.1\%$, with a notable outlier of $61.8\%$ in experiment 10. The median observed improvement was $19.1\%$. In the next section we discuss how this compares to theoretical figures based on the \acs{GST} estimates. Figure \ref{rb_curve}b is a comparison \acs{RB} curve showing the decay of an example improved gate over the native fidelity. For clarity and comparison, we also plot example curves with $10\%$ and $0.1\%$ error rates. Note that this \acs{RB} number is from the overall curve, which is composed of single and two-qubit gates. For these experiments, this partitions into $r = 3/4\cdot r_{\text{CNOT}} + 1/4\cdot r_{\text{single}}$. To reduce the total number of experiments per day, we did not compute multiple curves with different fractions of CNOT and single qubit gates. Consequently, $r_{\text{u}} /r_{\text{c}} - 1$ is really $(3/4\cdot r_{\text{u,CNOT}}+1/4\cdot r_{\text{single}})/(3/4 \cdot r_{\text{ c,CNOT}}+1/4\cdot r_{\text{single}}) - 1$, which is a lower bound for the improvement of the CNOT gate. Given that $r_{\text{single}}< r_{\text{CNOT}}$ by about an order of magnitude, we do not expect that the figure differs substantially.\par 
The minimised objective function was the average infidelity of 20 randomly sampled \acs{RB} circuits, consisting of 16 circuit layers in addition to the preparation and measurement layers. Each circuit was run at 8190 shots to minimise statistical error in the optimisation. The use of an \acs{RB} experiment as the objective function is a flexible metric and can be chosen as the user desires. In principle, context-dependence of a gate may affect the versatility of the improved gate in the sense that gates may affect their system differently depending on the circuit, however at this stage \acs{RB} curves are the most robust assessment of a gate's performance and require the fewest assumptions. \par 
Although inserting four local unitaries provides more coherent control, we found empirically on this device that this introduced more error than it eliminated. Consequently, we chose to apply corrections only on the control qubit, before and after the CNOT gate. This reduces the optimisation to 6 parameters.
In the Nelder-Mead method, a dimension 6 simplex with 7 vertices is constructed, with the base vertex given by the \acs{GST} parameters. The objective function is then evaluated for each point. In order to save on computation, we elected to omit the shrink step. After five iterations of no further improvement we would then terminate the algorithm and redefine our gate with the best point. \par
We used a relatively new form of \acs{RB} known as \emph{direct randomised benchmarking}\cite{DRB}. In a single circuit, direct \acs{RB} prepares stabiliser states, followed by $m$ randomly selected layers of gates native to that stabiliser, before finally performing a stabiliser measurement to give the success probability. A number of randomly generated circuits can then be used to provide an overall average. The utility of this over Clifford \acs{RB} is the ability to specify the occurrence of given gates. It also accommodates for future instances of the protocol where the tune-up might be efficiently computed over a larger number of qubits. Here, we randomly generated 20 \acs{RB} circuits, with CNOT gates composing on average $3/4$ of the total circuit. The average probability of success at length $m$, $P_m$
is then plotted over a series of values for $m$. These points are then fit to $P_m = A + Bp^m$ for fit parameters $A,B$, and $p$. The \acs{RB} number $r = (15/16)(1-p)$ is then the probability of an error occurring under a stochastic model. We remark here that the \acs{RB} number is not necessarily a true estimate of the average gate infidelity and can differ under, for example, gate-dependent noise. Here, our characterisation produced an initial infidelity estimate of $1.191\times10^{-2}$ for the CNOT gate, whereas the estimated \acs{RB} number was $3.369\times10^{-2}$. \acs{RB} is currently the most efficient method for estimating the quality of a gate, and the number most often cited in reporting the performance of hardware. For this reason, we use this as indicative of the gate infidelity, even though the true figures may differ. Indeed, this suggests that IBM may have better gates than their \acs{RB} curves suggest. For further details, see  \cite{PhysRevLett.119.130502,noise-coherence-2015}. \par
As previously stated, the utility of the \acs{GST} experiments is both in determining the coherence of the gate noise and in effecting a search which is placed near the global minimum of both gate infidelity and the diamond norm.
For example, using the \acs{GST} estimate we simulated the randomised benchmarking tune-up when seeded from the zero vector (that is, the case in which we optimise the \acs{RB} number without the initial \acs{GST} seed). In this case, the Nelder-Mead search terminates at a point where $r_{\text{CNOT}} = 0.0254$ and $\mathcal{E}_\Diamond=0.0426$ -- whereas the numerical search for the global minimum yields $r_{\text{CNOT}} = 0.0210$ and $\mathcal{E}_\Diamond = 0.0286$. 
Randomly seeded starts were able to find the global minimum, but only with the equivalent of between $10\:000$ and $20\:000$ \acs{RB} experiments, which is comparable to the initial number of circuits required for \acs{GST}. Crucially, this approach comes with no evidence of having reached the global minimum, which has large implications for $\mathcal{E}_\Diamond$.
\par
\textbf{Tracking the Tune-up}

It is prudent to ask how much the corrective values changed over time. We stop short of claiming that these values directly quantify drift, since a flat objective function could plausibly entail corrections whose value changed without much impact on the \acs{RB} number. However, empirically we found that the previous day's optimal corrections would typically reduce the performance of the gate, rather than improving it. It was necessary therefore to perform the optimisation, rather than simply applying the results of previous days.
Operating on the assumption that this method finds the most appropriate values for the corrective parameters on a particular day, we can use this data to loosely quantify the change in the coherent noise on a real quantum information processor (QIP). The values of the corrective parameters for each experiment are provided in Table \ref{param-table}. Moreover, the variables do not all independently affect the final map, meaning that change in the parameter values themselves might obfuscate the fact that the channel overall has not varied much. In this table, the $\theta$ values appear to change the most. These values are equivalent to $Y-$rotations, which change the populations of the control qubit. The reason for this is not entirely clear, however, inspection of the error channels in Figure \ref{noise-ptms}c shows that some of the most dominant noise occurs in the $X$ and $Z$ control qubit blocks. This could mean that they are more likely to change with time. \par 
In order to paint a more concrete picture of the effects of the parameters, we study the case of the improved gate as though it comprised three perfect gates. We then compare the diamond distance of this channel from the ideal CNOT, from the initial \acs{GST} seed, and from the previous experiment. Our results are summarised in Figure \ref{rb_curve}c. One of the key assumptions in this method was that the system and its gate noise never changed too much from the initial \acs{GST} seed that the optimiser could not easily find the best gate for the day. The distance of the corrective channel on a given day from the \acs{GST} seed supports this stance: over a six week period the mean diamond distance is 0.115, with a standard deviation of $1.2\times10^{-2}$.\par 

\textbf{Convergence Speed}

Any error suppression protocol designed to address drift in a \acs{QIP} will need to be regularly implemented as part of a tune-up procedure. It is therefore ideal that it require as few experiments as possible. In Figure \ref{rb_curve}d we present how each iteration of the Nelder-Mead optimisation increased the survival probability of the \acs{RB} experiment for four sample experiments. In each case, a substantial improvement was found in the first three iterations, beyond which we observed only small fluctuations, and zero change after 18 iterations. Each iteration requires 11 circuits to run, and so most of the improvement was found in 22 circuits, with the worst case requiring approximately 200 -- depending on when one accepts the algorithm to terminate. Assuming 0.5 milliseconds per shot, POST would take approximately nine minutes to converge in the worst case scenario. Depending on the cross-talk limitations of the device, it could then be run in parallel across all non-overlapping qubit pairs.

\section{Two-qubit gate set tomography experiments}

\begin{figure*}
	\centering
	\includegraphics[width=0.73\linewidth]{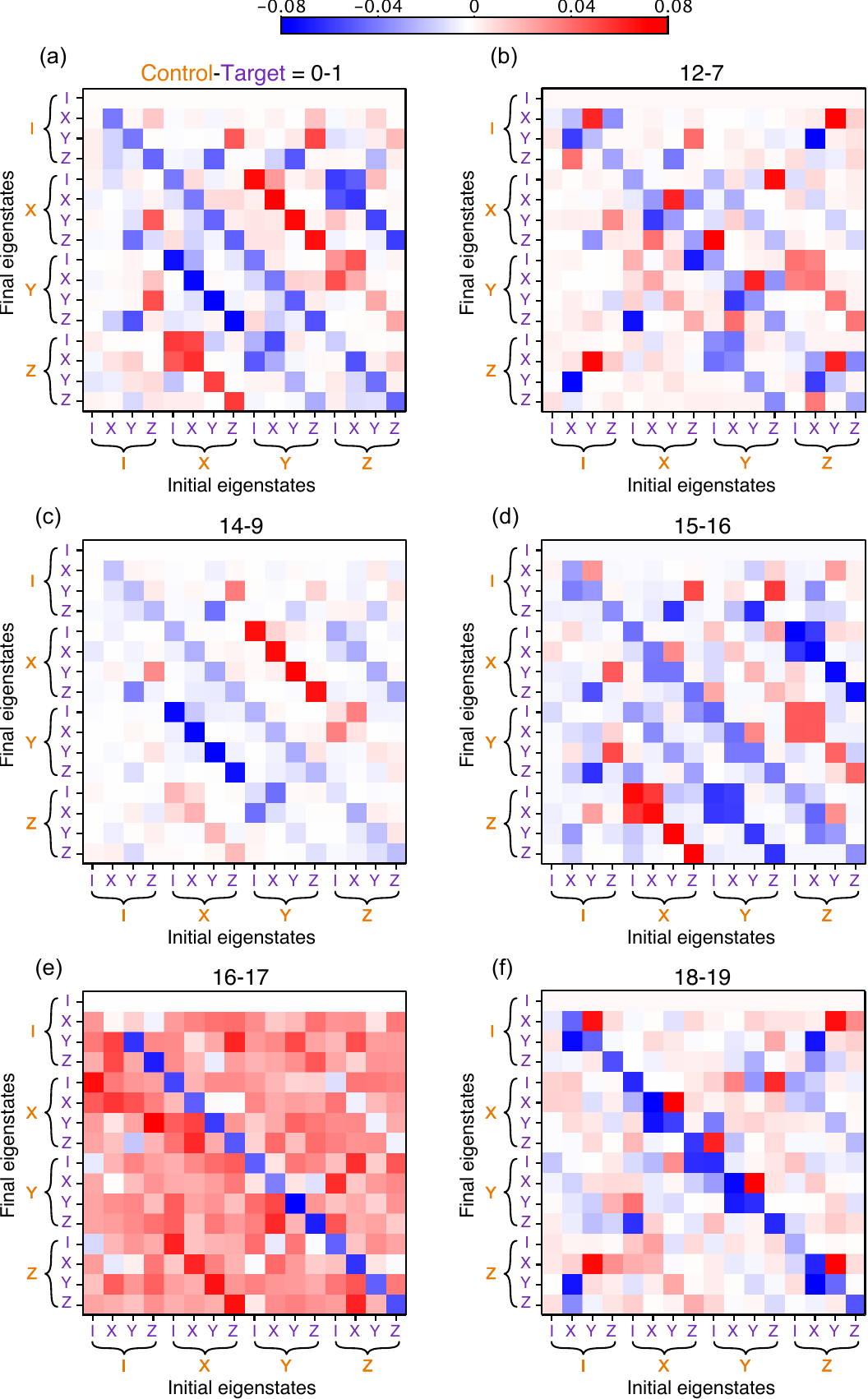}
	\caption[Gate set tomography estimates of CNOT error generators between six pairs of qubits on \emph{ibmq\_poughkeepsie} ]{The matrices here show the GST-estimated \acs{PTM} representations of the CNOT gate error generator between 6 pairs of different qubits. The control and target designations correspond to the device layout given in Figure \ref{noise-fig}. }
	\label{noise-ptms}
\end{figure*}

In addition to gate improvement on qubits 14 and 9 on the \emph{ibmq\_poughkeepsie}, we also performed \acs{GST} experiments on five other qubit CNOT pairs on the quantum device. This incorporates a further 5 experiments of between $9367$ and $20\:530$ circuits at 8190 shots each. We present this data as a case study for the noise that occurs in a real quantum device, and theoretically assess the effectiveness of the POST technique at mitigating the noise. Here, the fidelities of these gates and their structure indicate whether noise can be addressed by the single-qubit unitaries used. Figure \ref{noise-ptms} shows the noise PTMs for each of the six CNOT gates investigated. Specifically, we present this in the form of error generators $\mathcal{L}$ such that $\bar{G}_{CX}$ = $\text{e}^{\mathcal{L}} G_{CX}$. This presents the noise as though it were to occur after the ideal gate which allows for a helpful -- though not necessarily physically accurate -- picture. The control and target numbers given refer respectively to the qubits of Figure \ref{noise-fig}a acting as the control and target of the CNOT gate under characterisation. We elected to map out these qubits in order to obtain a relatively uniform sample of the full device geometry. It is instructive to compare these matrix plots with the schematics given in Figures \ref{noise-fig}c and \ref{noise-fig}d, which respectively indicate local target/control, and sole control rotations. \acs{PTM} noise whose locations are correspondingly indicated in green in the schematics can be explained as a local rotation occurring either before or after the CNOT. For example, the block-like features prominent in 0-1 and 15-16 make up the landscape of Figure \ref{noise-fig}d, suggesting a rotation of $Z-$eigenstates of the control qubit into $X-$ and $Y-$ eigenstates both before and after the CNOT. Any noise that falls outside the green of either schematic can be attributed either to decoherence or cross-resonance errors.\par
We also use this data to estimate the effectiveness of the POST procedure on the other qubit-pairs. Using the \acs{GST} estimates for each CNOT gate, we compute its \acs{RB} number as well as its diamond distance to the ideal case. We then investigate how these numbers are ideally minimised both in the case with corrective unitaries on the target qubit, and where all coherent noise is removed. This provides an insight into the location of most of the noise in these gates, as well as a partition of the gate infidelity into coherent and stochastic error. Interestingly, in most cases (barring the 12-7 pair) there is little difference between the best ideal gate using two single-qubit corrective unitaries, and where a perfect $U(4)$ gate is used to remove all coherent noise. A summary of the data is provided in Table \ref{tab:theoretical-improvements}. \par 
Pulse sequences for the implementation of a CNOT gate typically consist of a local pulse to each qubit, as well as an additional cross-resonance pulse coupling the two. We would expect implementing POST with absorbed corrective rotations into the native CNOT pulse sequence would see a much larger increase in fidelity.
\begin{table*}[]
	\centering
	\begin{tabular}{@{}lllllll@{}}
		\toprule 
		\begin{tabular}[c]{@{}l@{}}Qubit pair \\ (control-target)\end{tabular} & $r_{\text{CNOT}}$ & $\mathcal{E}_{\Diamond \text{CNOT}}$ & \begin{tabular}[c]{@{}l@{}}POST \\ minimum $r_{\text{CNOT}}$\end{tabular} & \begin{tabular}[c]{@{}l@{}}POST \\minimum $\mathcal{E}_{\Diamond \text{ CNOT}}$\end{tabular} & \begin{tabular}[c]{@{}l@{}} Stochastic $r_{\text{CNOT}}$\end{tabular} & \\ \midrule
		0-1 & $2.28\times 10^{-2}$ & $4.74\times 10^{-2}$ & $1.79\times 10^{-2}$ & $2.44\times 10^{-2}$ & $1.78\times 10^{-2}$ \\
		12-7 & $4.52\times 10^{-2}$ & $9.02\times 10^{-2}$ & $4.16\times 10^{-2}$ & $7.77\times 10^{-2}$ & $2.88\times 10^{-2}$ \\
		14-9 & $3.37\times 10^{-2}$ & $6.83\times 10^{-2}$ & $2.10\times 10^{-2}$ & $2.86\times 10^{-2}$ & $2.07\times 10^{-2}$ \\
		15-16 & $3.55\times 10^{-2}$ & $8.10\times 10^{-2}$ & $2.29\times 10^{-2}$ & $3.79\times 10^{-2}$ & $2.21\times 10^{-2}$  \\
		16-17 & $5.56\times 10^{-2}$ & $8.09\times 10^{-2}$ & $5.26\times 10^{-2}$ & $7.70\times 10^{-2}$ & $5.09\times 10^{-2}$  \\
		18-19 & $2.20\times 10^{-2}$ & $4.39\times 10^{-2}$ & $2.16\times 10^{-2}$ & $4.21\times 10^{-2}$ & $1.94\times 10^{-2}$ \\ \bottomrule
	\end{tabular}
	\caption{
		For each \acs{GST} analysis of the qubit pairs, we present the \acs{RB} number of the CNOT gate, as well as its diamond distance from the ideal CNOT. Further, we show the same figures in the idealised scenario where perfect single qubit corrective unitaries are employed on the control qubit to remove coherent error. In the final column, we have included the predicted \acs{RB} number for each gate with the complete removal of coherent noise. The diamond norm of almost every entry stands to be significantly improved with the removal of coherent error.}
	\label{tab:theoretical-improvements}
\end{table*}

\section{Discussion}
The transition from mathematical maps to physical operations is not always a seamless one. Besides errors in the \acs{GST} characterisation, absent a good method of characterising non-Markovian behaviour, assumptions must be made of weak system-environment correlations, composability of operations, and minimal cross-talk between qubits. The emergence of unexpected behaviour from quantum systems means that in-principle operational improvements, such as the direct application of corrections from \acs{GST} estimates, cannot always be relied upon. We have presented a general quantum-classical hybrid method which uses the real-life performance of the gate as the feed-forward for corrective updates. The success of the procedure is therefore self-fulfilling. \par 
randomised benchmarking is a robust method of measuring the Markovian fidelity of a given operation. However, in a system with environmental back-action or context dependent gates, it is not clear whether the situation will always be so simple as transplanting a redefined gate into a quantum circuit and seeing an increased fidelity in this new context. The markedly better performance of quantum algorithms consisting of these redefined gates remains to be demonstrated and will be the subject of future work. In particular, the POST algorithm would be easily adapted to any characterisation technique more inclusive of non-Markovian behaviour. \par
Developing high-fidelity gate hardware is imperative for the field of quantum computing to achieve its ambitious aims. An underrated measure of device quality, however, is \emph{consistency} -- the ability to achieve reported minimal error rates again and again despite gradual changes in device parameters and system-environment correlations. In this work we presented a consistent method that combines an initial overhead of gate set tomography with a classical optimisation algorithm that delivers an improved two-qubit gate in relatively few experiments. We emphasise that although POST was tested on an IBM Quantum device, it is applicable to any hardware with logical-level control. Furthermore, the method is adaptable to any level of control. The key aspect is identifying noisy parameters from the \acs{GST} estimate using the afforded set of device controls. In particular, we would expect to see significantly better results with pulse-level control wherein instead of separately implementing the corrective unitaries, they would be absorbed into modifying the CNOT pulse, there would be minimal additional gate errors introduced, or increase in depth.





\cleardoublepage

\label{app:bibliography} 

\manualmark 
\markboth{\spacedlowsmallcaps{\bibname}}{\spacedlowsmallcaps{\bibname}} 
\refstepcounter{dummy}

\addtocontents{toc}{\protect\vspace{\beforebibskip}} 
\addcontentsline{toc}{chapter}{\tocEntry{\bibname}}

\printbibliography


\end{document}